\documentclass
[
    twoside,                 % The thesis is formatted like a book. That is, odd and even pages are handled differently.
    openright,               % Starts a new chapter on an odd page number (right side).
    cleardoublepage = empty, % Clear pages inserted in order to have new chapters appear on odd pages are formatted with an empty style.
    fontsize = 12 pt,        % The size of the font.
    american,                % Support for American English.
    captions = tableheading, % Places the correct amount of space when the caption of a table is above the table.
    numbers = noenddot,      % Does not use a period at the end of numbered titles, such as sections or figures.
    footheight = 1 pt,      % Defines the height of the foot. Due to the line, it needs extra height.
]
{scrbook}

\newcommand{\specialannex}[3][]{%
  \cleardoublepage              % o \clearpage si no usas 'book'
  \phantomsection               % necesario para hyperref
  \refstepcounter{chapter}      % avanza capítulo y actualiza referencias
  \if\relax\detokenize{#1}\relax
    \addcontentsline{toc}{chapter}{\thechapter\; #2}%
  \else
    \addcontentsline{toc}{chapter}{\thechapter\; #1}%
  \fi
  \thispagestyle{empty}
  \vspace*{\fill}
  \begin{center}
    {\fontsize{36pt}{44pt}\selectfont \bfseries \textcolor{stroke1}{#2}} \\[1em]
    {\color{stroke1}\rule{0.6\linewidth}{2pt}} \\[1em]
    \if\relax\detokenize{#3}\relax
    \else
      {\normalsize #3}
    \fi
  \end{center}
  \vspace*{\fill}
  \clearpage
}

\newif\ifprintVersion   % Defines a binary variable that signals whether the document is prepared for physical or digital print.
\newif\ifprofessionalPrint % Defines a binary variable that signals whether the print will be done by a professional printing service that requests extra margin for page cutting and is not bound to paper formats like A4.
\newif\iffancyTheorems  % Defines a binary variable that signals whether theorems are formatted in the classical style or in a new format that better suits the overall flavor of this thesis.
\newif\ifboldNumberSets % Defines a binary variable that signals whether the variables for number sets (like N or R) should be in bold. If not, they are in blackboard bold instead.

\printVersionfalse
\professionalPrintfalse
\fancyTheoremstrue
\boldNumberSetstrue

\newcommand*{\printTitle}{}
\newcommand*{\myTitle}[1]{\renewcommand*{\printTitle}{#1}}

\newcommand*{\printAuthor}{}
\newcommand*{\myName}[1]{\renewcommand*{\printAuthor}{#1}}

\newcommand*{\printSubject}{}
\newcommand*{\mySubject}[1]{\renewcommand*{\printSubject}{#1}}

\newcommand*{\printKeywords}{}
\newcommand*{\myKeywords}[1]{\renewcommand*{\printKeywords}{#1}}

\newcommand*{\printTopParagraph}{\null}

\newlength{\extraborderlength}
\newcommand*{\extraBorder}[1]{\setlength{\extraborderlength}{#1}}

\newlength{\mybindingcorrection}
\newcommand*{\bindingCorrection}[1]{\setlength{\mybindingcorrection}{#1}} % Contains commands that are used for certain information that is printed.
\usepackage{comment}
\usepackage{amsmath}
\usepackage{tikz}
\usepackage{adjustbox}
\usepackage{multirow}
\usepackage[final]{pdfpages}
\usepackage{afterpage}
\newcommand\blankpage{%
    \null
    \thispagestyle{empty}%
    \newpage}
\usetikzlibrary{shapes.geometric, arrows}

\DeclareMathOperator{\sech}{sech}
\DeclareMathOperator{\arctanh}{arctanh}
\DeclareMathOperator{\arccosh}{arccosh}
\tikzstyle{block1} = [rectangle, minimum width=3cm, minimum height=1cm, text centered, draw=black, fill=red!30]
\tikzstyle{block2} = [rectangle, minimum width=3cm, minimum height=1cm, text centered, draw=black, fill=green!30]
\tikzstyle{block3} = [rectangle, minimum width=3cm, minimum height=1cm, text centered, draw=black, fill=yellow!30]
\tikzstyle{block4} = [rectangle, minimum width=2.5cm, minimum height=1cm, text centered, draw=black, fill=blue!30]
\tikzstyle{block5} = [rectangle, minimum width=2.5cm, minimum height=1cm, text centered, draw=black, fill=orange!30]
\tikzstyle{arrow} = [thick,->,>=stealth]
\tikzstyle{arrow} = [thick,->,>=stealth]
    \extraBorder{3mm}

    \bindingCorrection{6 mm}
    \myTitle{Evolution and scattering of excited topological solitons: Interaction between internal modes}

    \myName{David Miguélez Caballero}

    \mySubject{PhD thesis}

    \myKeywords{PhD thesis | Topological soliton | Internal mode | Vortices | Kinks | Abelian-Higgs model}

\usepackage[utf8]{inputenc} % Allows to input UTF8 characters.
\usepackage[T1]{fontenc}    % Allows to print special characters correctly.
\usepackage
[
    ngerman,         % German is used for the German abstract.
    main = american, % This is the main language of the thesis.
]
{babel}                     % Is responsible for sensible hyphenations.

\input glyphtounicode
\usepackage{calc} % Makes it easer to do math with TeX measurements.

\newlength{\myparindent}
\newlength{\myparskip}
\setfootnoterule{0 cm}

\deffootnote[1.2 em]{1.2 em}{0 em}{\makebox[1.4 em][l]{\textbf{\thefootnotemark}}}

\makeatletter%
    \@removefromreset{footnote}{chapter}%
\makeatother

\usepackage[dvipsnames]{xcolor} % Allows it to define colors. The option says that common names can be used.

\definecolor{stroke1}{HTML}{2574A9} % This color is used as the standard color to highlight things.

\colorlet{captionlabel}{black}
\colorlet{footerpagenr}{black}
\colorlet{footerchapter}{stroke1}
\colorlet{footerchaptername}{black}
\colorlet{footersection}{stroke1}
\colorlet{footersectionname}{black}
\colorlet{chapternumber}{stroke1}

\newlength{\mypaperwidth}
\newlength{\mypaperheight}
\newlength{\mybodywidth}
\newlength{\mybodyheight}
\newlength{\myoutermargin}
\ifprintVersion
    \ifprofessionalPrint
    \else
    \fi
\else
\fi

\newlength{\mytopmargin}
\ifprintVersion
    \ifprofessionalPrint
    \fi
\fi

\newlength{\myinnermargin}
\ifprintVersion
    \ifprofessionalPrint
    \fi
\fi

\newlength{\mybottommargin}
\ifprintVersion
    \ifprofessionalPrint
    \fi
\fi

\newcommand{\goldenratio}{1.618}

\newlength{\myheadsep} % Distance from the header to the body.
\ifprintVersion
    \ifprofessionalPrint
    \fi
\fi

\newlength{\myfootskip} % Distance from the body to the footer.
\ifprintVersion
    \ifprofessionalPrint
    \fi
\fi

\newlength{\mymargininnersep} % Distance between the margin and the body.
\newlength{\mymarginoutersep} % Distance between the margin and the paper border.
\ifprintVersion
    \ifprofessionalPrint
    \fi
\fi

\newlength{\mymarginwidth} % Width of the margin.
\newlength{\mymarginwidthwithinnersep} % Width of the margin.
\usepackage
[
    \ifprintVersion
        \ifprofessionalPrint
            paperwidth = \mypaperwidth + 2\extraborderlength + \mybindingcorrection,
            paperheight = \mypaperheight + 2\extraborderlength,
        \else
            paperwidth = \mypaperwidth,
            paperheight = \mypaperheight,
        \fi
    \else
        paperwidth = \mypaperwidth,
        paperheight = \mypaperheight,
    \fi
    textwidth = \mybodywidth,
    textheight = \mybodyheight,
    outer = \myoutermargin,
    inner= \myoutermargin,
    top = \mytopmargin,
    headsep = \myheadsep,
    footskip = \myfootskip,
    marginparsep = \mymargininnersep,
    marginparwidth = \mymarginwidth,
]
{geometry} % Used in order to define the dimensions of the page and its layout.

\usepackage
[
]
{scrlayer-scrpage} % Allows to adjust the definitions of the head and foot of a page.

\clearpairofpagestyles

\KOMAoptions
{%
    headwidth = \textwidth + \mymarginwidthwithinnersep,%
    footwidth = \myoutermargin : \textwidth,%
}

\automark[chapter]{chapter}
\automark*[section]{}

\lehead%
{%
}
\cehead{\hspace*{\mymarginwidthwithinnersep}\parbox{\textwidth}{\raggedright\leftmark}}

\rohead%
{%
    \Ifstr{\rightmark}{\leftmark}%
    {%
    }%
    {%
    }%
}
\cohead{\hspace*{-\mymarginwidthwithinnersep}\parbox{\textwidth}{\raggedleft\rightmark}}

\lefoot*%
{%
    \vspace*{1 ex}%
    {\color{stroke1}\rule{\myoutermargin - \mymargininnersep}{0.5 mm}}\\
    \begin{minipage}[b]{\myoutermargin - \mymargininnersep}%
        \raggedleft\normalfont\color{footerpagenr}\textbf{\thepage}%
    \end{minipage}%
}
\rofoot*%
{%
    {\color{stroke1}\rule{\myoutermargin - \mymargininnersep}{0.5 mm}}\\
    \begin{minipage}[b]{\myoutermargin - \mymargininnersep}%
        \raggedright\normalfont\color{footerpagenr}\textbf{\thepage}%
    \end{minipage}%
}

\usepackage{caption}
\usepackage{changepage}
\usepackage{tikz} % Used in order to draw the stylistic elements.

\newlength{\mytmpa}
\newlength{\mytmpb}

\renewcommand*{\partlineswithprefixformat}[3]%
{%
    #2
    \thispagestyle{empty}
    \ifprintVersion
        \setlength{\mytmpa}{0.618\mypaperwidth + \mybindingcorrection + \extraborderlength}%
        \setlength{\mytmpb}{0.382\mypaperheight + \extraborderlength}%
    \else
        \setlength{\mytmpa}{0.618\mypaperwidth}%
        \setlength{\mytmpb}{0.382\mypaperheight}%
    \fi
    \begin{tikzpicture}[overlay, remember picture]%
        \node [inner sep = 0, outer sep = 0, anchor = north] at (current page.north west)%
        {%
            \begin{tikzpicture}[overlay, remember picture]%
            \draw[color = stroke1, line width = 0.7 mm] (\mytmpa, 0) -- (\mytmpa, -\mytmpb);%
            \end{tikzpicture}%
        };%
        \node (align) [align = right, below = \mytmpb - 2 ex, inner sep = 0, outer sep = 0, anchor = north west] at (current page.north west)%
        {%
            \hspace{\mytmpa}\hspace{0.5 em}\partname\ \thepart\\[1 ex]
            \color{stroke1}#3%
        };%
    \end{tikzpicture}%
}
\RedeclareSectionCommand%
[%
    font = \normalfont\Huge\sffamily,
    prefixfont = \normalfont\Huge\sffamily,
]
{part}

\usepackage{etoolbox}

\newbool{chapterHasANumber}
\newbool{chapterHasAStar}
\renewcommand*{\chapterlinesformat}[3]%
{%
    \Ifnumbered{#1}{\setbool{chapterHasANumber}{true}}{\setbool{chapterHasANumber}{false}}%
    \Ifstr{#2}{}{\setbool{chapterHasAStar}{true}}{\setbool{chapterHasAStar}{false}}%
    \ifboolexpr{bool{chapterHasANumber} and not bool{chapterHasAStar}}%
    {%
        \begin{tikzpicture}[overlay, remember picture]%
            \node [right = \myinnermargin, below = \mytopmargin, inner sep = 0, outer sep = 0, anchor = north west] (numbernode) at (current page.north west)%
            {%
                \hspace{\myinnermargin}%
                \sffamily\fontsize{60}{60}\selectfont%
                \color{chapternumber}%
                \thechapter%
            };%
            \node [inner sep = 0, outer sep = 0, anchor = north west] at (numbernode.south west)%
            {%
                \begin{tikzpicture}[overlay, remember picture]%
                    \draw[color = stroke1, line width = 0.7 mm] (\myinnermargin, -1 ex) -- (\paperwidth, -1 ex);%
                \end{tikzpicture}%
            };%
            \node (align) [text width = \textwidth - 2 cm, align = right, right = \myinnermargin + \mybodywidth, inner sep = 0, outer sep = 0, anchor = east] at (numbernode.west)%
            {%
                #3%
            };%
        \end{tikzpicture}%
    }%
    {%
        \begin{tikzpicture}[overlay, remember picture]%
            \node [right = \myinnermargin, below = \mytopmargin, inner sep = 0, outer sep = 0, anchor = north west] (numbernode) at (current page.north west)%
            {%
                \hspace{\myinnermargin}%
                \sffamily\fontsize{60}{60}\selectfont%
                \color{white}%
                \thechapter%
            };%
            \node [inner sep = 0, outer sep = 0, anchor = north west] at (numbernode.south west)%
            {%
                \begin{tikzpicture}[overlay, remember picture]%
                    \draw[color = stroke1, line width = 0.7 mm] (\myinnermargin, -1 ex) -- (\paperwidth, -1 ex);%
                \end{tikzpicture}%
            };%
            \node (align) [align = left, right = \myinnermargin, inner sep = 0, outer sep = 0, anchor = south west] at (numbernode.south west)%
            {%
                #3%
            };%
        \end{tikzpicture}%
    }%
}
\RedeclareSectionCommand%
[%
    font = \color{stroke1}\normalfont\huge\sffamily,
    afterskip = 20 pt,
]
{chapter}

\BeforeStartingTOC[toc]{\pagestyle{plain}}
\AfterStartingTOC{\thispagestyle{plain}}                        % Contains commands that define the general format and layout of the thesis.
\usepackage
[
    babel = true, % Enables language-specific tuning.
]
{microtype}           % Uses the text space more efficiently.
\usepackage{csquotes} % Uses the correct quotes according to the current language.

\usepackage{amsmath}
\usepackage{amssymb}
\usepackage{amsthm}
\usepackage{thmtools}
\usepackage{mathtools}
\usepackage{thm-restate}
\usepackage{dsfont}        % Yields far better blackboard-bold letters than \mathbb. Use \mathds in order to write such letters.
\usepackage{braceMnSymbol} % Adjusts overbraces and underbraces such that longer versions are put together seamlessly.

\usepackage
[
    ttscale = 0.85, % Scales the typewriter font.
]
{libertine} % The main font used in this thesis.
\usepackage
[
    libertine,    % Changes the math font to libertine (the main font).
    slantedGreek, % Makes all greek letters italic by default. If you want to use an upright greek letter, use ›\up‹ immediately followed by the letter’s name. For example, \upGamma displays an upright uppercase gamma.
    vvarbb,       % Changes the \mathbb font to another font. However, \mathbb remains ugly and should not be used. Use \mathds instead.
    libaltvw,     % Uses different characters for v und w that look far better than the default ones.
]
{newtxmath} % The main math font of this thesis. It fits well with the main font.
\usepackage{url} % Responsible for URL formatting.
\usepackage{bm}  % Allows to use sensible bold letters in math mode. This package has to go after the font packages. Otherwise it does not work correctly!

\usepackage{graphicx} % The standard package for including graphics into your document.
\usepackage
[
    subrefformat = simple, % Formats the label of the \subref command without parentheses.
    labelformat = simple,  % Formats the mark of a subfigure without parentheses.
]
{subcaption}         % Enables it to have subfigures inside of a single figure.
\usepackage{wrapfig} % Allows to put figures next to text.

\columnsep = \mymargininnersep
\usepackage{array}     % Improves the way that tables can be formatted.
\usepackage{booktabs}  % Adds lines (called ›rules‹) that can be used in tables and improves spacing.
\usepackage{longtable} % Allows to make tables that span multiple pages.
\usepackage{pdflscape} % Allows to change a page into landscape. This is handy if a table is very wide.

\usepackage{enumitem} % Adds tons of useful features to enumeration environments.

\usepackage
[
    ruled,         % Creates lines at the top and at the bottom. Further, the caption is now above the algorithm.
    vlined,        % Shows the scope of a statement spanning multiple lines via a small vertical bar. Thus, no closing tags are needed.
    linesnumbered, % Shows line numbers.
]
{algorithm2e} % Allows to write pseudocode.

\usepackage{cite}
\usepackage{xspace}   % Adds the functionality that a space after a command will be shown as a space in the output.
\usepackage
[
    shortcuts, % Allows to use short symbols for non-breaking hyphens and dashes instead of lengthy commands.
]
{extdash}             % Adds non-breaking hyphens and dashes.
\usepackage{setspace} % Allows to easily chnage the spacing inside of the document.

\usepackage{xparse}    % Is used in order to define reasonable commands.
\usepackage{footnote}  % Allows it to extend the environments footnotes can be used in. It is said that this package is in conflict with ›hyperref‹. I did not note any troubles. However, if something is fishy, it is probably best to not use this package.
\usepackage{afterpage} % Adds the \afterpage command, which specifies that the provided argument shall be processed after the current page is finished.
\usepackage
[
    textsize = scriptsize, % Determines the text size of the TODO note.
]
{todonotes}            % Adds TODO notes to the document. These are small text areas inside of the margin of a page.

\usepackage
[
    bookmarks = true,                 % Generates boodmarks for the PDF.
    bookmarksopen = false,            % The bookmarks are closed by default.
    bookmarksnumbered = true,         % The bookmarks use the numbers of the corresponding headline.
    pdfstartpage = 1,                 % The first page seen when opening the PDF.
    pdftitle = {{\printTitle}},       % The PDF’s title in the meta data.
    pdfauthor = {{\printAuthor}},     % The PDF’s author name in the meta data.
    pdfsubject = {{\printSubject}},   % The PDF’s subject in the meta data.
    pdfkeywords = {{\printKeywords}}, % The PDF’s keywords in the meta data.
    breaklinks = true,                % Allows it to break links.
    \ifprintVersion
        hidelinks,                    % In the printed version, links are not highlighted, as they are not clickable.
    \else
    colorlinks = true,            % The text of hyperlinks is colored instead of having a colored box around it.
    allcolors = stroke1,          % Every hyperlink uses the same color. If you want to change specific colors, use the commands below.
    \fi
]
{hyperref} % The standard package that is used for creating hyperlinks inside of a document.

\usepackage
[
    noabbrev,   % The words in front of the labels are not abbreviated.
    nameinlink, % Extends the link of a reference to the word in front of it.
]
{cleveref} % This package must be included after ›hyperref‹. It creates clever references that know what they refer to.     % Contains the packages that this template provides.
\makeatletter
    \def\IfEmptyTF#1%
    {%
        \if\relax\detokenize{#1}\relax%
            \expandafter\@firstoftwo%
        \else%
            \expandafter\@secondoftwo%
        \fi%
    }
\makeatother

\NewDocumentCommand{\mathOrText}{m}
{%
    \ensuremath{#1}\xspace%
}

\let\originalleft\left
\let\originalright\right
\renewcommand{\left}{\mathopen{}\mathclose\bgroup\originalleft}
\renewcommand{\right}{\aftergroup\egroup\originalright}

\makeatletter
    \DeclareRobustCommand{\bfseries}%
    {%
        \not@math@alphabet\bfseries\mathbf%
        \fontseries\bfdefault\selectfont%
        \boldmath%
    }
\makeatother

\xspaceaddexceptions{]\}}

\allowdisplaybreaks

\crefname{ineq}{inequality}{inequalities}
\creflabelformat{ineq}{#2{\upshape(#1)}#3} 

\crefname{term}{term}{terms}
\creflabelformat{term}{#2{\upshape(#1)}#3}

\let\oldfootnote\footnote

\newlength{\spaceBeforeFootnote} % Denotes the space before the footnote mark in em.
\newlength{\spaceAfterFootnote}  % Denotes the space after the footnote mark in em.

\RenewDocumentCommand{\footnote}{o o o m}%
{%
    \IfNoValueTF{#1}%
    {%
        \oldfootnote{#4}%
    }%
    {%
        \setlength{\spaceBeforeFootnote}{\IfEmptyTF{#1}{0}{#1} em}%
        \IfNoValueTF{#2}%
        {%
            \hspace*{\spaceBeforeFootnote}\oldfootnote{#4}%
        }%
        {%
            \setlength{\spaceAfterFootnote}{\IfEmptyTF{#2}{0}{#2} em}%
            \hspace*{\spaceBeforeFootnote}\IfNoValueTF{#3}{\oldfootnote{#4}}{\oldfootnote[#3]{#4}}\hspace*{\spaceAfterFootnote}%
        }%
    }%
}

\makesavenoteenv{figure}
\makesavenoteenv{table}
\makesavenoteenv{tabular}

\iffancyTheorems
    \declaretheoremstyle
    [
        spaceabove = \topsep,
        spacebelow = \topsep,
        headfont = \bfseries,
        headformat = \textcolor{stroke1}{$\blacktriangleright$} \NAME~\NUMBER \NOTE,
        notefont = \bfseries,
        notebraces = {(}{)},
        bodyfont = \normalfont,
        postheadspace = 0.5 em,
        qed = \textcolor{stroke1}{\bfseries$\blacktriangleleft$},
    ]
    {myTheoremStyle}
    
    \declaretheorem
    [
        style = myTheoremStyle,
        name = Conjecture,
        numberwithin = chapter,
    ]
    {conjecture}
    \declaretheorem
    [
        style = myTheoremStyle,
        name = Proposition,
        sharenumber = conjecture,
    ]
    {proposition}
    \declaretheorem
    [
        style = myTheoremStyle,
        name = Claim,
        sharenumber = conjecture,
    ]
    {claim}
    \declaretheorem
    [
        style = myTheoremStyle,
        name = Lemma,
        sharenumber = conjecture,
    ]
    {lemma}
    \declaretheorem
    [
        style = myTheoremStyle,
        name = Corollary,
        sharenumber = conjecture,
    ]
    {corollary}
    \declaretheorem
    [
        style = myTheoremStyle,
        name = Theorem,
        sharenumber = conjecture,
    ]
    {theorem}
    \declaretheorem
    [
        style = myTheoremStyle,
        name = Definition,
        sharenumber = conjecture,
    ]
    {definition}
    \declaretheorem
    [
        style = myTheoremStyle,
        name = Example,
        sharenumber = conjecture,
    ]
    {example}
    \declaretheorem
    [
        style = myTheoremStyle,
        name = Remark,
        sharenumber = conjecture,
    ]
    {remark}
\else
    \theoremstyle{plain}
    
    \newtheorem{conjecture}{Conjecture}[chapter]
    \newtheorem{proposition}[conjecture]{Proposition}
    
    \newtheorem{lemma}[conjecture]{Lemma}
    
    \newtheorem{theorem}[conjecture]{Theorem}
    \newtheorem{definition}[conjecture]{Definition}

\fi

\NewDocumentCommand{\functionTemplate}{m m m m o}%
{%
    \IfNoValueTF{#5}%
    {%
        \mathOrText{#1\left#2{#4}\right#3}%
    }%
    {%
        \mathOrText{#1#5#2{#4}#5#3}%
    }%
}

\newcommand*{\leftBracketType}{(}
\newcommand*{\rightBracketType}{)}

\NewDocumentCommand{\createFunction}{m m o o}%
{%
    \renewcommand*{\leftBracketType}{\IfNoValueTF{#3}{(}{#3}}%
    \renewcommand*{\rightBracketType}{\IfNoValueTF{#4}{)}{#4}}%
    \NewDocumentCommand{#1}{o o}%
    {%
        \IfNoValueTF{##1}%
        {%
            \mathOrText{#2}%
        }%
        {%
            \functionTemplate{#2}{\leftBracketType}{\rightBracketType}{##1}[##2]%
        }%
    }%
}

\DeclareDocumentCommand{\probabilisticFunctionTemplate}{m m O{} o}
{%
    \functionTemplate{#1}%
    {\lbrack}%
    {\rbrack}%
    {#2\IfEmptyTF{#3}{}{\ \IfNoValueTF{#4}{\left}{#4}\vert\ \vphantom{#2}#3\IfNoValueTF{#4}{\right.}{}}}%
    [#4]%
}

\ifboldNumberSets

    \newcommand*{\indicatorFunctionSymbol}{\mathbf{1}}
\else

    \newcommand*{\indicatorFunctionSymbol}{\mathds{1}}
\fi

\RenewDocumentCommand{\Pr}{m O{} o}%
{%
    \probabilisticFunctionTemplate{\mathrm{Pr}}{#1}[#2][#3]%
}

\NewDocumentCommand{\E}{m O{} o}%
{%
    \probabilisticFunctionTemplate{\mathrm{E}}{#1}[#2][#3]%
}

\NewDocumentCommand{\Var}{m O{} o}%
{%
    \probabilisticFunctionTemplate{\mathrm{Var}}{#1}[#2][#3]%
}

\DeclareDocumentCommand{\bigO}{m o}%
{%
    \functionTemplate{\mathrm{O}}{(}{)}{#1}[#2]%
}

\DeclareDocumentCommand{\smallO}{m o}%
{%
    \functionTemplate{\mathrm{o}}{(}{)}{#1}[#2]%
}

\DeclareDocumentCommand{\bigTheta}{m o}%
{%
    \functionTemplate{\upTheta}{(}{)}{#1}[#2]%
}

\DeclareDocumentCommand{\bigOmega}{m o}%
{%
    \functionTemplate{\upOmega}{(}{)}{#1}[#2]%
}

\DeclareDocumentCommand{\smallOmega}{m o}%
{%
    \functionTemplate{\upomega}{(}{)}{#1}[#2]%
}

\DeclareDocumentCommand{\eulerE}{o}%
{%
    \mathOrText{\mathrm{e}\IfNoValueTF{#1}{}{^{#1}}}%
}

\DeclareDocumentCommand{\poly}{m o}%
{%
    \functionTemplate{\mathrm{poly}}{(}{)}{#1}[#2]%
}

\createFunction{\id}{\mathrm{id}}

\NewDocumentCommand{\ind}{m o o}%
{%
    \IfNoValueTF{#2}%
    {%
        \mathOrText{\indicatorFunctionSymbol_{#1}}%
    }%
    {%
        \functionTemplate{\indicatorFunctionSymbol_{#1}}{(}{)}{#2}[#3]%
    }%
}

\DeclareDocumentCommand{\dom}{m o}%
{%
    \functionTemplate{\mathrm{dom}}{(}{)}{#1}[#2]%
}

\DeclareDocumentCommand{\rng}{m o}%
{%
    \functionTemplate{\mathrm{rng}}{(}{)}{#1}[#2]%
}

\DeclareDocumentCommand{\d}{o}%
{%
    \mathrm{d}\IfNoValueTF{#1}{}{^{#1}}%
}

\DeclareDocumentCommand{\set}{m m o}%
{%
    \mathOrText{\IfNoValueTF{#3}{\left}{#3}\{#1\ \IfNoValueTF{#3}{\left}{#3}\vert\ \vphantom{#1}#2\IfNoValueTF{#3}{\right.}{}\IfNoValueTF{#3}{\right}{#3}\}}%
}      % Contains newly defined commands useful for mathematics.
\begin{document}

    \frontmatter
    \includepdf[pages=1]{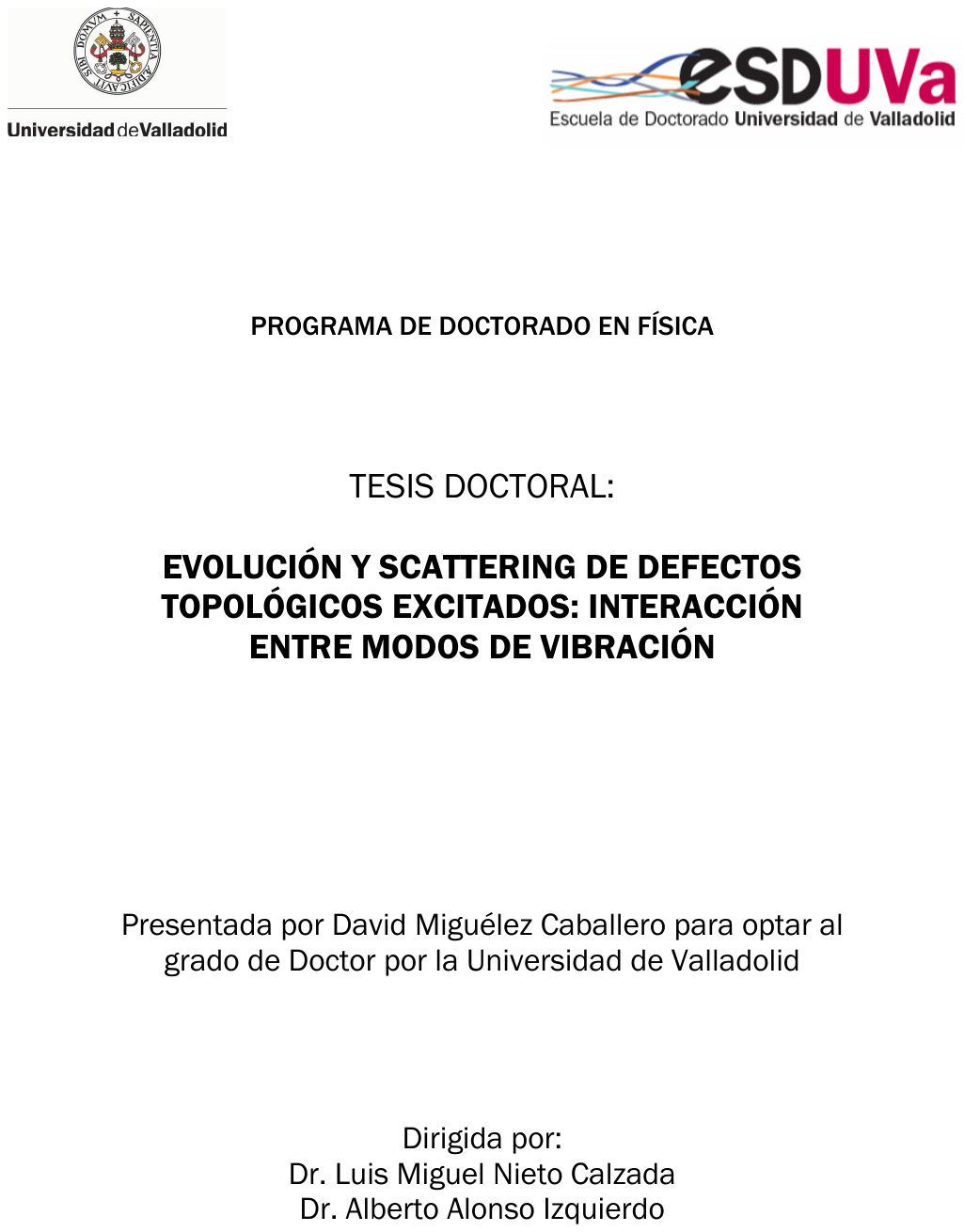}
\clearpage
\thispagestyle{empty}
\newpage
\mbox{~}
\clearpage
\newpage
\thispagestyle{empty}
\clearpage
    \includepdf[pages=2]{PortadaTesis1.pdf}

\clearpage
\thispagestyle{empty}
\newpage
\mbox{~}
\clearpage
\newpage
\thispagestyle{empty}
\clearpage

\clearpage
\thispagestyle{empty}
\newpage
\mbox{~}

\vspace*{\fill}
\begin{center}
    ``\textit{Everything has a beginning and an end. Life is just a cycle of starts and
stops. There are ends we don’t desire, but they’re inevitable, we have to
face them. It’s what being human is all about.}''\\ Jet Black, Cowboy Bebop
\end{center}
\vspace*{\fill}
\clearpage
\newpage
\thispagestyle{empty}
\clearpage

       % \selectlanguage{american}
    \addchap{Acknowledgments}
    % Here you can write whom you want to thank.

La presentación de esta tesis pone fin a una etapa compleja, llena de lecciones, de fallos y de aciertos. Durante estos años he vivido innumerables experiencias que no habría podido disfrutar si nunca me hubiese atrevido a iniciar un doctorado.  Este espacio está dedicado a todos aquellos que han mostrado su apoyo y cariño durante este proceso.

En primer lugar me gustaría agradecer a mis directores Luis Miguel Nieto Calzada y Alberto Alonso Izquierdo por acogerme al inicio de mi doctorado. A ellos les tengo que agradecer su paciencia, sus consejos, su apoyo y dedicación. Sin ellos nunca habría podido  iniciarme en el mundo de la investigación y  culminar esta tesis doctoral.

En segundo lugar he de agradecer  a Sergio N. su compañía no solo durante el doctorado, si no también durante nuestra estancia en Cracovia y durante todos los congresos a los que he asistido. Gracias a él he vivido un sinfín de innumerables experiencias que se han transformado en innumerables anécdotas y buenos recuerdos. 

Gracias también a Sergio S., Albert, Paz y Samane por su apoyo y camaradería. Nunca olvidaré todos los buenos momentos y experiencias que hemos vivido juntos durante todos estos años. Gracias por estar a mi lado en los peores y mejores momentos que he vivido en esta etapa de mi vida. 

I would like to thank Andrzej,  Kasia and Tom for their hospitality. I am deeply grateful for the enriching scientific discussions and their support during my stay in Kraków, as well as for the good times we shared together.

También me gustaría dedicar unas palabras para dar gracias a mis amigos por estar a mi lado durante tantos años, tanto en los mejores momentos como en los peores. Gracias a Judit, Patri, Nerea, Dani y Rubén por todos los planes, escapadas y cervezas que hemos compartido juntos.

Por último, me gustaría agradecer a toda mi familia todo el apoyo brindado desde que comencé  el grado hace ya más de 7 años. 

\vspace{2cm}

Esta tesis ha sido financiada por un contrato predoctoral concedido por la Junta de Castilla y León (Orden EDU 1868/2022 de 19 de diciembre). También agradezco a los proyectos ``\textit{New developments in mathematical modelling of quantum phenomena}''  (PID2020-113406GB-I00), \textit{Q-CAYLE} (PRTRC17.I1), ``\textit{Dynamics of Topological Defects: New Analytical and Numerical Developments with Applications }'' (PID2023-148409NB-I00) y ``\textit{Laboratory of disruptive and interdisciplinary sciences}'' (CLU-2023-1-05) por el apoyo financiero ofrecido para la asistencia a congresos. Asimismo, los primeros meses de realización tesis también ha sido financiada a través de un contrato asociado al proyecto Q-CAYLE.

\newpage
    \pagestyle{plain}

    \addchap{Abstract}

This thesis presents an extensive analysis of the behavior of topological solitons when one or more of their internal modes are activated. The first part of this manuscript is devoted to the study of the simplest topological solitons in $(1+1)$ dimensions: kinks. Specifically, we investigate how these solutions emit radiation when one of their internal modes is initially excited, within the framework of a particular two-component scalar field theory—the double $\phi^4$ model. The simplest kink solution in this theory exhibits a complex internal mode structure that depends on a coupling constant appearing in the potential governing the dynamics. We will show how the amplitude and frequency of the emitted radiation are affected by changes in this coupling constant.

In addition to static kink configurations, we also examine the dynamics of wobbling kink/antikink scattering when the kinks possess more than one internal mode. To this end, we study kink/antikink collisions in the context of the simplest kink solution arising in the Montonen–Sarker–Trullinger– Bishop (MSTB) model. This analysis provides valuable insight into the resonant energy exchange mechanism, which enables the transfer of energy between internal modes and the translational mode.

The second part of this thesis focuses on excited vortex solutions in $(2+1)$ dimensions. In particular, we begin with a detailed study of the internal mode structure associated with vortex solutions in the Abelian-Higgs model. We demonstrate how the problem can be significantly simplified by choosing an appropriate angular dependence for the eigenfunctions. Furthermore, we investigate the radiation emitted by a vortex with winding number $n=1$ when its internal mode is initially activated. To achieve this, we extend the analytical techniques used in $(1+1)$ dimensions to field theories defined in two spatial dimensions. This enables us to compute the radiation amplitude, its frequency, and the decay of the internal mode amplitude due to energy loss via radiation.

All analytical results are contrasted with data from numerical simulations, allowing us to shed light to the validity of the analytical methods employed.

\newpage

\pagestyle{plain}

    \selectlanguage{american}
    \addchap{Resumen}

% This file should contain the Spanish abstract.
Esta tesis contiene un extenso an\'alisis en el cual se estudia en detalle el comportamiento de solitones topológicos una vez estos han sido excitados mediante uno o m\'as de sus modos internos. En la  primera parte de este manuscrito estudiaremos los solitones topológicos más sencillos definidos en $(1+1)$ dimensiones, los kinks. En concreto, analizaremos la emisión de radiación por parte de estas soluciones una vez uno de sus modos internos ha sido inicialmente excitados en el contexto del \textit{modelo} \textit{doble} $\phi^4$, la cual es una teoría de campos escalar de dos componentes. El kink más sencillo que posee esta teoría posee una compleja estructura interna la cual, a su vez, depende de la constante de acople entre campos definida por el potencial que define el modelo. Analizaremos en detalle  la dependencia de la amplitud de radiación y de su frecuencia de la ya mencionada constante de acople entre ambos campos.

Además de abordar el estudio de soluciones tipo kink estáticas, también analizaremos la dinámica asociada a procesos de scattering kink/antikink cuando dichos kinks poseen más de un modo interno de vibración. Para llevar a cabo esta tarea,  estudiaremos colisiones entre kinks excitados en el contexto del \textit{modelo} \textit{Montonen-Sarker-Trullinger-Bishop (MSTB)}. Este análisis nos permitirá entender mejor el mecanismo responsable de la transferencia de energía entre modos internos que permite redistribuir la energía entre los modos internos y el modo traslacional. 

La segunda parte de esta tesis está dedicada al estudio de vórtices excitados, los cuales están definidos en $(2+1)$ dimensiones. En particular, estudiaremos en detalle la estructura interna asociada a soluciones tipo vórtice en el \textit{modelo de Higgs abeliano}. En concreto, veremos como este análisis puede ser simplificado una vez se ha encontrado la dependencia angular correcta para autofunciones correspondientes. Adicionalmente, también nos centraremos en el estudio de la emisión de radiación por parte de vórtices excitados con carga topológica $n=1$. Esta tarea se llevará a cabo extendiendo las técnicas desarrolladas para el estudio de este tipo de fenómenos en $(1+1)$ dimensiones. Todo ello nos permitirá obtener la amplitud de radiación, su frecuencia y el decaimiento  de la amplitud del modo interno.

Todas las técnicas analíticas se contrastarán con los datos  extraídos de  las  correspondientes simulaciones numéricas, lo cual, a su vez, nos permitirá entender mejor el rango de validez de dichas técnicas analíticas. 

\newpage
    \pagestyle{plain}

    \addchap{List of Publications}
    The work presented in this thesis has given rise to the following publications:
\begin{itemize}
    \item A. Alonso-Izquierdo, D. Miguélez-Caballero and L.M. Nieto, \textit{Wobbling kinks and shape mode interactions in a coupled two-component} $\phi^4$ \textit{theory}, \href{https://doi.org/10.1016/j.chaos.2023.114373}{Chaos Solitons Fractals \textbf{178}, 114373 (2024)}. 

    \item A. Alonso-Izquierdo, J.J. Blanco-Pillado, D. Miguélez-Caballero, S. Navarro-Obregón and J. Queiruga,  \textit{Excited Abelian-Higgs vortices: Decay rate and radiation emission}, \href{ https://doi.org/10.1103/PhysRevD.110.065009}{Phys. Rev. D \textbf{110}, 065009  (2024).}

    \item A. Alonso-Izquierdo and D. Miguélez-Caballero, \textit{Dissecting normal modes of vibration on vortices in Ginzburg-Landau superconductors}, \href{https://doi.org/10.1103/PhysRevD.110.125026}{Phys. Rev. D \textbf{110} , 125026 (2024).}

    \item A. Alonso-Izquierdo, D. Miguélez-Caballero and L.M. Nieto, \textit{Scattering between orthogonally wobbling kink}, \href{https://doi.org/10.1016/j.physd.2024.134438}{Physica D \textbf{471}, 134438 (2025)}. 

    \item A. Alonso-Izquierdo, D. Miguélez-Caballero and J. Queiruga, \textit{Spectral structure of fluctuation around $n$-vortices in the Abelian-Higgs model}, \href{https://doi.org/10.48550/arXiv.2505.05039}{ arXiv:2505.05039 (2025)}  \textbf{ (pending of acceptance)}.
    
\end{itemize}

\newpage
    \pagestyle{plain}

   % \setuptoc{toc}{totoc}
    \tableofcontents

    \pagestyle{headings}
    \mainmatter

    \chapter{Introduction}\label{Intro0}
    % Here you introduce your topic to the reader.
\section{Topological solitons: origin and history}

Topological solitons have played an essential role over the last few decades. They have been significant in several areas, such as superconductivity and cosmology \cite{Manton2004, Vilenkin2000, Shnir2018, Kevrekidis2019, Davydov1985, Rebbi1984, Vachaspaty2006, Manton2022}. The first soliton ever studied dates back to 1834, when John Scott Russell observed a solitary water wave in a canal. Years later, in a report to the British Association \cite{Russell1845}, he described this wave as a ``\textit{large solitary elevation, a rounded, smooth, and well-defined heap of water, which continued its course along the channel apparently without change of form or diminution of speed}”.

Despite this early observation, the phenomenon remained largely overlooked for many years. Even Airy and Stokes claimed that the existence of such a wave was mathematically impossible and violated the laws of hydrodynamics \cite{Kevrekidis2019}. Nevertheless, in 1895, Korteweg and de Vries published a paper in which they mathematically proved what Russell had observed half a century earlier \cite{Korteweg1895}. They showed that the phenomenon was governed by what has since become known as the KdV equation.

%Many years later, in 1953, Fermi, Pasta, Ulam and Tsingou conducted the FPUT experiment, in which they simulated the dynamics of an elastic string with fixed ends in which they had added nonlinear terms. They expected the energy to be distributed equally among all the vibrational modes of the string, but, instead, they found a quasi-periodic behaviour in which the energy was redistributed passing through different modes of vibration. This problem reamain unsolved until  twelve years later,    Kruskal and Zabusky discoverd the connection between the FPUT lattice system and the KdV equation. In their paper, they presented the results of several simulations in which soliton solutions in such a manner that when they collide they conserved its initial velocity and profile. Due to their findings, they decided to baptize these solutions as  ``\textit{solitons}''. 

%So far, we have talked about the origin of the study of solitons, but how can we define them? We can define a soliton as a solution to a non-linear partial differential equation that behaves as a wave-like solution, whose profile remains constant and travels with constant velocity remaining unperturbed. This definition is not entirely precise since the soliton profile can be perturbed due to the presence of an impurity; we will explain this in more detail later.

Many years later, in 1953, Fermi, Pasta, Ulam and Tsingou conducted the FPUT experiment, in which they simulated the dynamics of an elastic string with fixed ends, to which they had added nonlinear terms. They expected the energy to be distributed equally among all the vibrational modes of the string; however, they found instead a quasi-periodic behavior in which the energy was redistributed, passing through different modes of vibration \cite{Shnir2018}.

The results were published after Fermi's death in 1955 \cite{Fermi1955}. This problem remained unsolved until twelve years later, when Kruskal and Zabusky discovered the connection between the FPUT lattice system and the KdV equation. In their paper, they presented the results of several simulations showing soliton solutions that, when colliding, preserved their initial velocity and profile. As a result of their findings, they decided to name these solutions “solitons” \cite{Zabusky1965}.

So far, we have discussed the origins of the study of solitons. But how can we define them? A soliton can be defined as a solution to a nonlinear partial differential equation that behaves like a wave: its profile remains unchanged while propagating at a constant velocity, unperturbed by dispersion. This definition is not entirely precise, since it is now well-known that the soliton profile can be affected by the presence of impurities. For instance, these kinds of models have been constructed in the case of vortices \cite{Tong2014,Ashcroft2020,Bazeia2024} and solitons that arise in scalar field theories \cite{Adam2019, AlonsoIzquierdo2023}.

So far, we have discussed the basic concept of a soliton, but the main focus of this thesis is the analysis of \textit{topological solitons}. A possible definition for these objects is the following:

\begin{definition}[Topological soliton]
A topological soliton is a solution to a nonlinear field theory that satisfies the following conditions:
\begin{itemize}
    \item The theory possesses a quantized conserved quantity, known as the \textbf{topological charge}, which is closely related to the topology of the problem under consideration.

    \item The energy density is localized around a specific point in space, referred to as the \textbf{center} of the soliton.

    \item In the absence of impurities, this solution can be Lorentz-boosted, propagating with constant velocity while maintaining its shape.
\end{itemize}
\end{definition}

It should be noted that the existence of a topological charge does not necessarily imply the stability of the soliton. For instance, a vortex with winding number \( n = 2 \) above the Bogomol'nyi–Prasad–Sommerfield (BPS) limit constitutes an unstable solution, which decays into two \( n = 1 \) vortices that repel each other \cite{Manton2004}.

The study of these objects dates back to the 1950s and 1960s, when physicists and mathematicians began investigating nonlinear classical field theories. Historically, the first example of a topological soliton was found in the Ginzburg--Landau superconductivity model. This model, defined in two spatial dimensions, features a complex scalar field \(\phi\), whose modulus represents the density of electrons present on the surface of the superconductor. In turn, this field is coupled with an electromagnetic field. The model was first proposed in 1950 by V. L. Ginzburg and L. D. Landau\footnote{A translation of the original paper can be found in \cite{Ginzburg2009}.} \cite{Ginzburg1950} , but the first vortex solutions were discovered in 1957 by Abrikosov \cite{Abrikosov1957}. Nevertheless, these solutions were found without taking into account the topology of the problem. For this reason, the skyrmions present in the Skyrme model are often considered the earliest identified topological solitons.

The Skyrme model is a nonlinear field theory defined in \((3+1)\) dimensions, originally proposed by Tony Skyrme in 1961 as a model for baryons \cite{Skyrme1961, Skyrme1962}. The soliton solutions arising in this model were identified as baryons, with the corresponding topological charge identified as the baryon number. This model gained significant popularity in the 1980s, when Witten showed that Skyrme's picture of protons and neutrons made sense from the perspective of quantum chromodynamics \cite{Witten1983,Witten1983b}. Subsequently, solutions with baryon number \(2\) and higher were discovered \cite{Kopeliovich1987, Verbaarschot1987, Braaten1990}, and corrections to the model were added, making it of great importance in nuclear physics to this day \cite{Manton2022}.

As time passed, other topological solitons were discovered, such as kinks, baby skyrmions, monopoles, and instantons, arising in various areas of physics and becoming increasingly important over time \cite{Manton2004,Shnir2018}.

Another remarkable development in the study of these objects is the discovery of the Bogomol'nyi equations. In a seminal paper published in 1976, Bogomol'nyi presented several examples of how to reduce the second-order differential equations associated with certain nonlinear field theories to a set of first-order time-independent equations \cite{Bogomolnyi1976}. This was achieved by cleverly rearranging the terms of the energy density such that, if these first-order equations are satisfied, the energy of the field configuration is simply a multiple of the topological charge. This bound is known as the Bogomol'nyi--Prasad--Sommerfield (BPS) bound. Concurrently, Prasad and Sommerfield arrived at reduced first-order differential equations for dyons and monopoles, albeit through a direct derivation from the field equations \cite{Prasad1975}.

It is important to note that these equations allow for finding soliton solutions in a much simpler way than solving the full-field equations. Nevertheless, there exist field theories supporting solitons for which this method cannot be applied; for example, for Abelian-Higgs vortices and 't Hooft monopoles this trick can only be applied in a certain limit\footnote{Papers \cite{Zabusky1965, Skyrme1962, Witten1983b , Abrikosov1957, Bogomolnyi1976, Prasad1975, Rubinstein1970} can also be found as reprints in reference \cite{Rebbi1984}. }
 \cite{Bogomolnyi1976, Prasad1975}. 
This discovery also opened the door to studying the small fluctuation operator associated with static soliton solutions in a simpler manner. This is because the BPS equations can be used to factorize the second-order operator. In this thesis, we will analyze in detail how this can be achieved in the context of vortices.

\subsection{ A brief history of kinks}

Kinks are the simplest topological solitons that can be found in a \((1+1)\)-dimensional scalar field theory. When discussing kink solutions, two models stand out as the first in which solutions of this kind were discovered: the sine-Gordon model and the \(\phi^4\) model. Both support the existence of topological solitons, but the main difference between them lies in the fact that the former is an integrable theory, whereas the latter is non-integrable. In other words, the sine-Gordon model possesses an infinite number of conserved quantities. This is the main reason why exact multi-kink solutions can be constructed in the sine-Gordon model \cite{Shnir2018, Kevrekidis2019}.

On the other hand, the \(\phi^4\) model exhibits unique properties that are absent in any integrable theory. One of these is the presence of internal (or vibrational) modes in the spectrum of the second-order fluctuation operator around the kink solution. Another is the emergence of a fractal structure in kink–antikink collisions for varying initial velocities. When plotting the final velocity of the individual kinks as a function of their initial velocity, one observes the appearance of multi-bounce windows, with self-similar patterns emerging between them. We will discuss this phenomenon in more detail later in this chapter and in Chapters~\ref{Intro1} and \ref{Chap2}.

Historically, the sine-Gordon model first appeared in the 19\textsuperscript{th} century, being applied in various contexts such as the study of surfaces with constant negative curvature and the modeling of dislocations in solids \cite{Eisenhart1960, Seeger1951, Seeger1953}. Skyrme and Perring also analyzed this model in the context of constructing particles from mesonic wave packets \cite{Perring1962}. In 1973, Ablowitz, Kaup, Newell and Segur solved the equation \cite{Ablowitz1973} using an inverse scattering transform \cite{Zakharov1972}, although Seeger, Donth, and Kochendorfer had already obtained analytical multi-kink solutions in 1950 using Bäcklund transformations \cite{Seeger1953}. These multi-soliton solutions were later rediscovered by Hirota in 1972 using an alternative approach \cite{Hirota1972}. As a final remark, the name \textit{sine-Gordon} was coined by Martin Kruskal and Norman Zabusky as a pun on the Klein–Gordon equation (see seventh footnote in \cite{Rubinstein1970}).

The \(\phi^4\) model has been widely used in condensed matter physics \cite{Bishop1978}, to study phase transitions \cite{Krumhansl1975}, chains of atoms \cite{Ishibashi1972}, polyacetylene \cite{Rice1979} and superconductors \cite{Falk1984}. In the 1970s and 1980s, the scientific community began to question whether analytical breather-like and multi-kink solutions could be found in other field theories such as the \(\phi^4\) model. However, as previously mentioned, the lack of integrability in this theory prevents the existence of such analytical solutions. As a result, several authors began investigating kink–antikink scattering phenomena in this model. Due to the limited computational resources available at the time, only a few simulations could be performed, with each author reporting different behaviors \cite{Aubry1976,Ablowitz1979,Kudryavtsev1975,Makhankov1978,Sugiyama1979,Campbell1983}. In some studies, the annihilation of kinks forming a breather-like configuration—later termed an \textit{oscillon}—was observed \cite{Kudryavtsev1975, Makhankov1978}. Many of these findings were subsequently corroborated by other researchers. In fact, Ablowitz, Kruskal, and Ladik confirmed several of these phenomena in a study initiated in 1972 and published seven years later \cite{Ablowitz1979}.

%Other authors noted that, after the first collision, the kinks could attract and collide multiple times before escaping each other’s interaction

\begin{figure}[htb]
    \centering
    \includegraphics[width=1\linewidth]{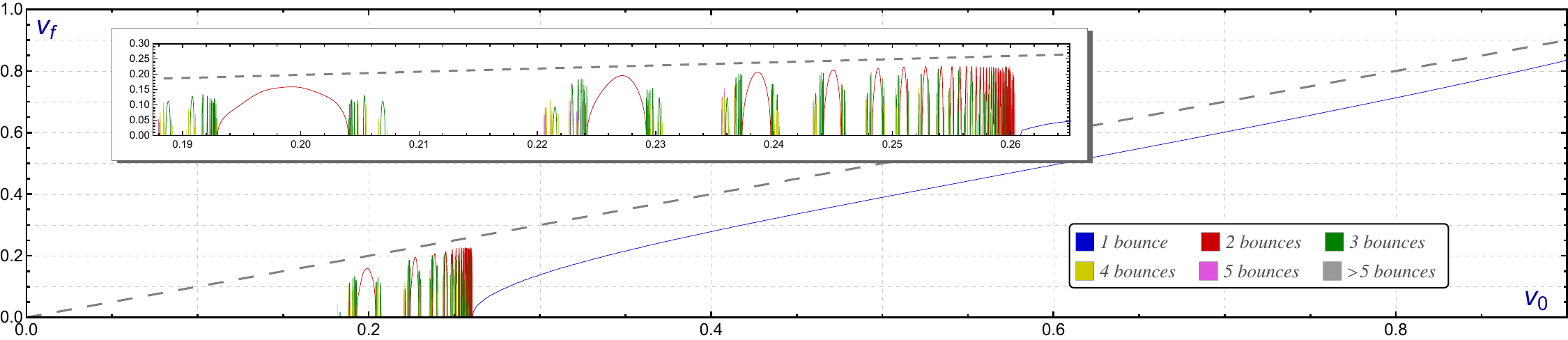}
    \caption{\textit{Kink-antikink velocity diagram for kinks in the $\phi^4$ model. The final velocity $v_f$ of the scattered kinks as a function of the initial velocity $v_0$ has been plotted. The color code shown in the graphs indicates the number of bounces suffered by the kink-antikink pair before moving appart. In initial velocity ranges where no final velocity is shown, a bion is assumed to form. The resonance window where various bounces can be observed has been expanded inside each graphics to better show the fractal pattern. For the sake of comparison the dashed grey line indicates the elastic scenario $v_0 = v_f$.}}
    \label{FigI1:KAKCollision}
\end{figure}

In 1983, Campbell, Schonfeld and Wingate published a seminal paper in which they conducted a large number of simulations varying the initial velocities of the colliding kinks \cite{Campbell1983b}. They identified that for initial velocities \(v > 0.258\), the kinks collided inelastically and escaped, whereas for velocities \(v < 0.188\), the kinks were annihilated, forming an oscillon. Additionally, they observed that, unlike the sine-Gordon breather, this bound state decays slowly in amplitude due to radiation emission. In the intermediate regime, they reported a fractal structure in the final velocity diagram, where alternating regions of reflection and oscillon formation appeared (see Figure~\ref{FigI1:KAKCollision}). This phenomenon was qualitatively explained in this work. The underlying mechanism was identified as the \textit{resonant energy transfer mechanism}, which allows energy exchange between the internal and translational modes of the kinks. This mechanism was later more rigorously explain in \cite{Goodman2005, Goodman2007}.  We will explore this mechanism in detail in Chapter~\ref{Chap2}, in the context of the MSTB model.

Another important milestone was the development of the Collective Coordinate Method (CCM), introduced by Sugiyama in 1979 \cite{Sugiyama1979}, inspired in the collective coordinates methods  used in particle physics \cite{Gervais1975}. In his work, Sugiyama proposed a technique to reduce the dynamics of kink–antikink interactions to a mechanical Lagrangian system depending on only three dynamical variables—namely, the kink position and the amplitudes of their internal modes. However, Takyi and Weigel later discovered a typographical error in Sugiyama’s original calculations. Once corrected, they showed that the dynamics predicted by the CCM did not fully agree with results from numerical simulations \cite{Weigel2014, Takyi2016}. Nonetheless, N.S.~Manton, K.~Oles, T.~Romanczukiewicz, and A.~Wereszczynski demonstrated that if the metric in the reduced system is appropriately reparametrized, a singularity is avoided, and the resulting CCM accurately reproduces the numerical results \cite{Manton2021}.

As a final remark on the \(\phi^4\) model, it is worth mentioning that in 1997, Manton proved that a static kink with an initially excited internal mode can radiate energy \cite{Manton1997}. This result was independently confirmed in 2009 by Oxtoby and Barashenkov using a different analytical approach \cite{Barashenkov2009, Barashenkov2009b}.

In recent decades, many researchers have extended the study of kink solutions to \(N\)-component field theories. These models generalize the results obtained in simpler settings and provide a deeper understanding of phenomena such as the resonant energy transfer mechanism. Furthermore, these models allow for the existence of \textit{non-topological} kinks that interpolate between the same vacuum states of the potential\footnote{Formally speaking, these non-topological kinks are indeed \textit{sphalerons}—unstable, non-topological soliton solutions that interpolate between identical vacua of the theory under consideration.} \cite{AlonsoIzquierdo2019, Halavanau2012, Halcrow2024, AlonsoIzquierdo2021}.%

%\color{red} Mencionar origen, sine gordon $\phi^4$, colisiones, diferencia entre modelos y desarrollo de modelos en dos componentes (MSTB)

\subsection{ A brief history of vortices}

The beginning of the study of vortices dates back to 1950, when Ginzburg and Landau published a foundational paper introducing a theory of superconductivity. In their formulation, a complex scalar field \(\phi\) represented the density of Cooper pairs within the superconducting material \cite{Ginzburg1950}. Their theory accounted for the existence of two distinct types of superconductors: Type I and Type II. Type I superconductors expel external magnetic fields entirely, and they lose their superconducting properties when subjected to a sufficiently strong magnetic field. On the other hand, Type II superconductors do not completely expel the magnetic field. When an external magnetic field is applied, part of it penetrates the material through quantized tubes of magnetic flux, giving rise to vortices. This unique property allows for the existence of stable vortex configurations in Type II superconductors \cite{Ginzburg2009}.

In 1957, Abrikosov published a paper in which he found vortex solutions within the Ginzburg– Landau theory of superconductivity an in which he introduced the aforemention classification of types of superconductor \cite{Abrikosov1957}. As previously mentioned, however, Abrikosov did not consider the topological nature of these solutions. The equations he derived were valid only for vortices with winding number \(n = 1\), although he did find that the magnetic flux in the presence of \(N\) vortices was quantized. Although initially met with skepticism, the existence of vortices was experimentally confirmed in 1967 \cite{Essmann1967}. Abrikosov and Ginzburg were later awarded the Nobel Prize in 2003 for their groundbreaking contributions to the theory of superconductivity. In the following decades, numerous experimental studies confirmed the existence of vortices in various superconducting materials, including the formation of vortex lattices \cite{Suderow2014, Blatter1994, ErikGoa2001, Vinnikov1988, Vinnikov2000, Harada1996}.

In 1964, P.W.~Higgs introduced the Abelian-Higgs model as an example of a relativistic field theory in which gauge fields acquire mass through coupling to scalar fields \cite{Higgs1964}. This model can be seen as a relativistic generalization of the Ginzburg–Landau theory. In 1973, Nielsen and Olesen studied vortex solutions in the Abelian-Higgs model in a manner similar to Abrikosov's earlier work, but this time paying attention to the model's topological structure. In the same work, they introduced the concept of a \textit{string} by extending the vortex solution to three spatial dimensions, visualized as a stack of planar vortices forming a one-dimensional object \cite{Nielsen1973}. Furthermore, they established a connection between these strings and the Nambu–Goto action, which provides a simplified method for analyzing their dynamics \cite{Vilenkin2000}. In 1976, Kibble proposed that such topological defects may have formed during phase transitions in the early universe, therefore coining the concept of \textit{cosmic strings} \cite{Kibble1976}.

It is also worth noting that similar vortex structures arise in superfluids, as these systems are governed by equations analogous to those in the Ginzburg–Landau theory of superconductivity \cite{Hatsuda1994, Thuneberg1987}.

 As will be discussed in Chapter~\ref{Intro2}, the Abelian-Higgs model depends on a real parameter \(\lambda\) that defines two distinct regimes. The interaction between vortices in these regimes attracted considerable attention in the early studies of the model. Despite the fact that vortices only exist in Type II superconductors, Berger and Chen rigorously established that there exist solutions for every value greater than zero of the constant $\lambda$ that couples the scalar and electromagnetic fields \cite{Berger1989}. In 1980, Jaffe and Taubes conjectured that vortices with vorticity \(n = 1\) attract each other when \(\lambda < 1\) and repel when \(\lambda > 1\) \cite{Jaffe1980}. Bogomol'nyi had previously provided an argument supporting the repulsive interaction for \(\lambda > 1\) \cite{Bogomolnyi1976}. In 1999, Gustafson and Sigal rigorously established the criterion for attraction and repulsion of vortices, confirming the conjectures of the aforementioned authors \cite{Gustafson2000}. This task was accomplished through the analysis of a small perturbation operator associated with static vortex solutions.  Meanwhile, Jacobs and Rebbi computed the interaction energy between vortices as a function of their separation for various values of \(\lambda\), finding that in the BPS limit \(\lambda = 1\), the energy is independent of their separation \cite{Jacobs1979}. 

The long-range forces between vortices have also been extensively studied in recent decades using various analytical techniques. Notable contributions in this area include the works of Bettencourt and Rivers \cite{Bettencourt1995}, as well as Speight \cite{Speight1997} and Tong \cite{Tong2002}.

An important contribution was made by Weinberg, who studied the BPS limit of the Abelian-Higgs model and uncovered the supersymmetric structure underlying the small fluctuation operator associated with static vortex solutions. He employed this structure to show that, in the BPS limit, a $n$-vortex possesses \(2n\) zero modes\footnote{We will analyze the internal mode structure and the consequences of this supersymmetric structure in detail in Chapter~\ref{Chap3}.}\cite{Weinberg1979}.%

Weinberg later extended his analysis to similar models, such as the Abelian-Chern–Simons model \cite{Jackiw1990,Jackiw1990b}. In fact, numerous generalizations and modifications of the Abelian-Higgs model have been explored by different authors \cite{Peterson2015, Jatkar2000, Adhikari2018, AlonsoIzquierdo2022, AlonsoIzquierdo2020}, including the introduction of magnetic and electric impurities into the Lagrangian density \cite{Tong2014,Ashcroft2020,Bazeia2024, Zhang2015, Han2016, Cockburn2017}.

\begin{figure}[htb]
    \centering

    % Fila 1
    \begin{subfigure}[b]{0.325\textwidth}
        \centering
        \includegraphics[width=\textwidth]{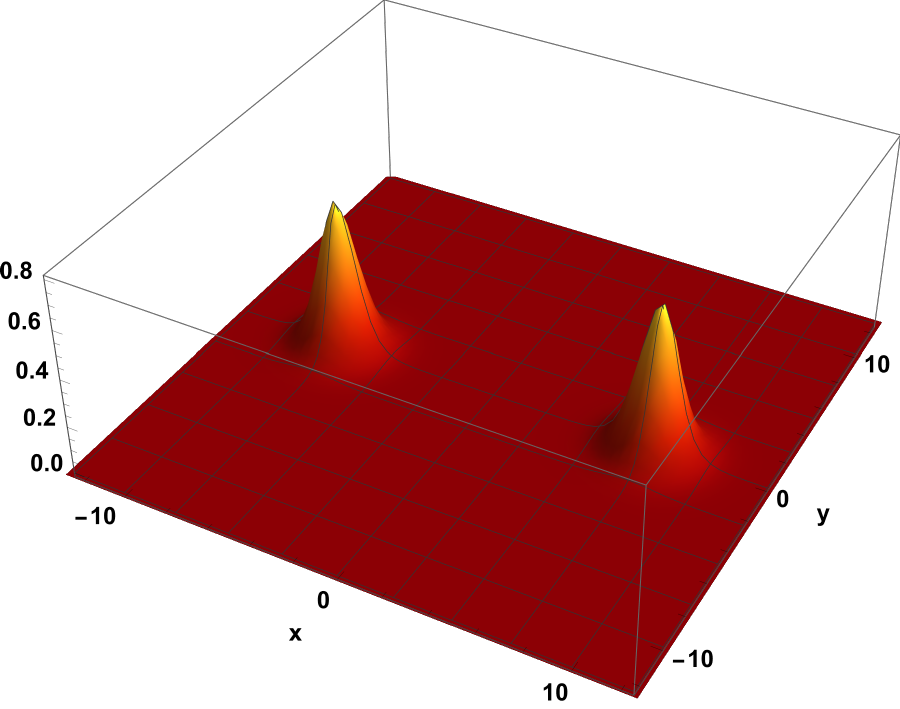}
    \end{subfigure}
    \hfill
    \begin{subfigure}[b]{0.325\textwidth}
        \centering
        \includegraphics[width=\textwidth]{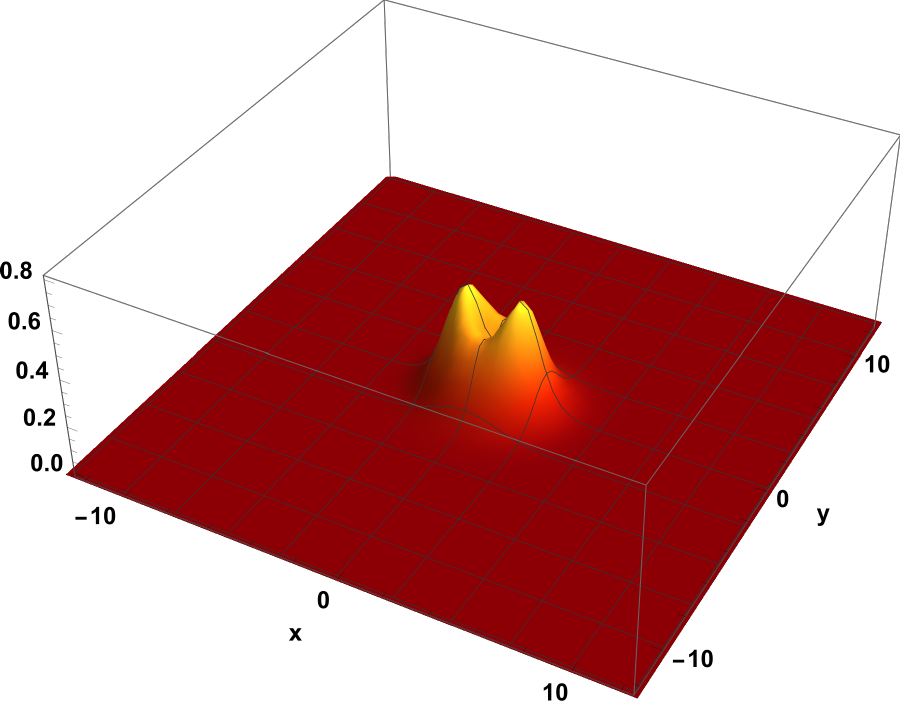}
    \end{subfigure}
    \hfill
    \begin{subfigure}[b]{0.325\textwidth}
        \centering
        \includegraphics[width=\textwidth]{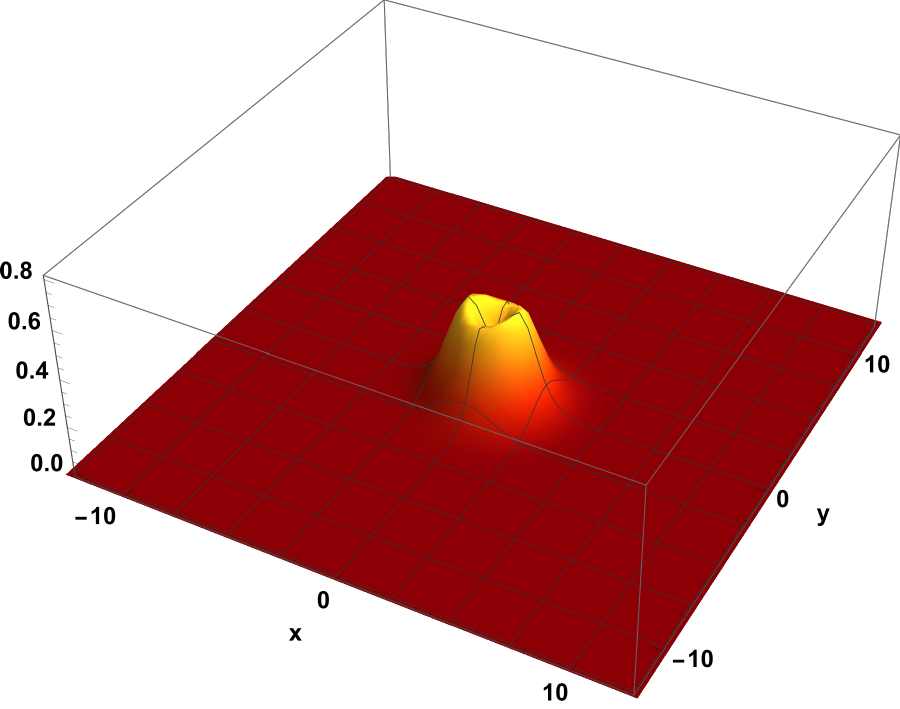}
    \end{subfigure}

    \vspace{0.5cm}

    % Fila 2
    \begin{subfigure}[b]{0.325\textwidth}
        \centering
        \includegraphics[width=\textwidth]{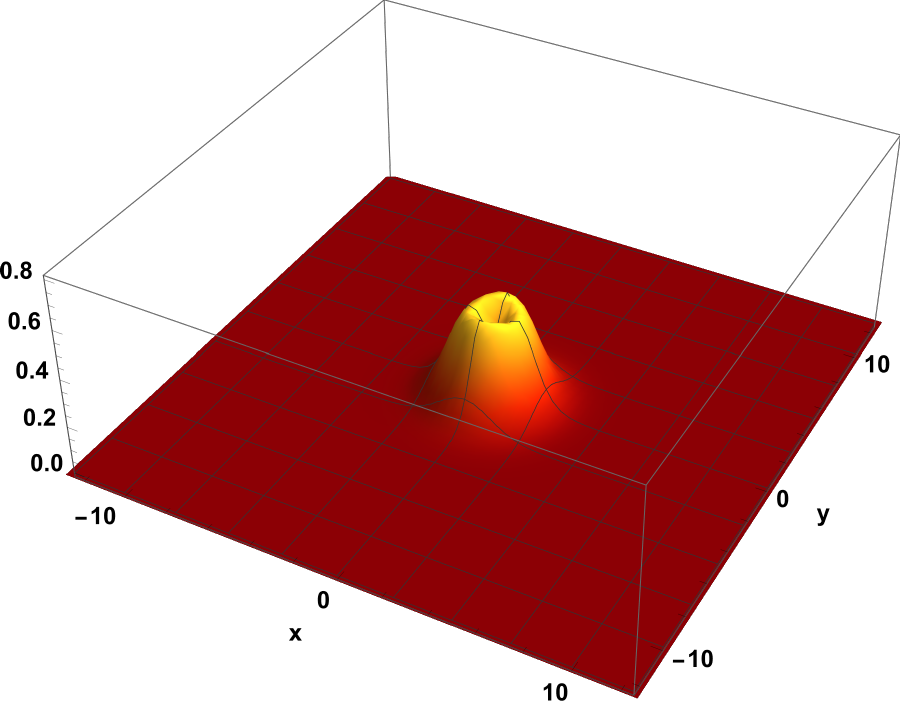}
    \end{subfigure}
    \hfill
    \begin{subfigure}[b]{0.325\textwidth}
        \centering
        \includegraphics[width=\textwidth]{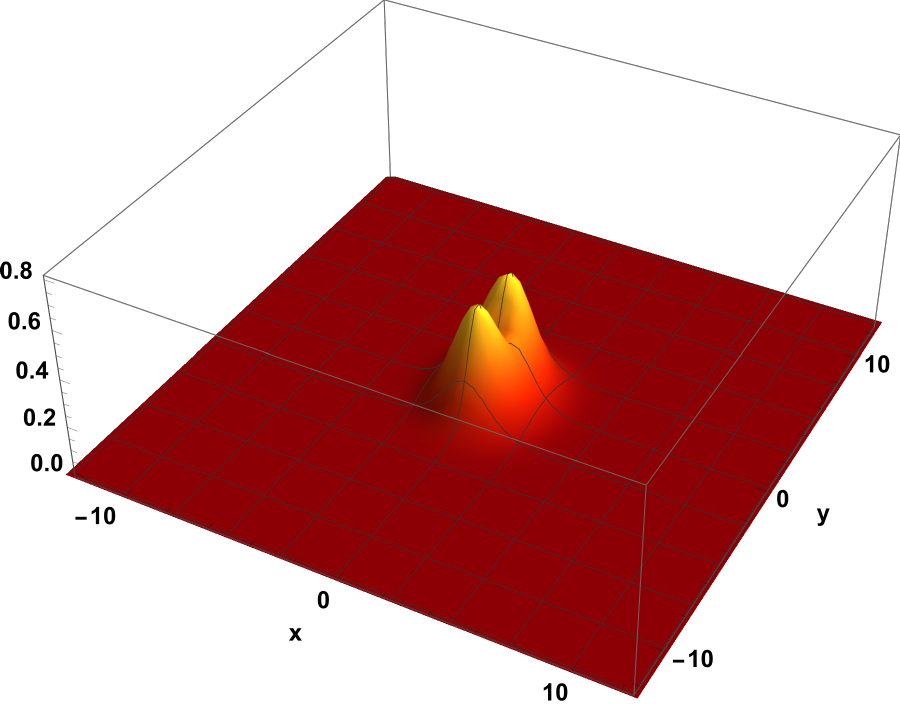}
    \end{subfigure}
    \hfill
    \begin{subfigure}[b]{0.325\textwidth}
        \centering
        \includegraphics[width=\textwidth]{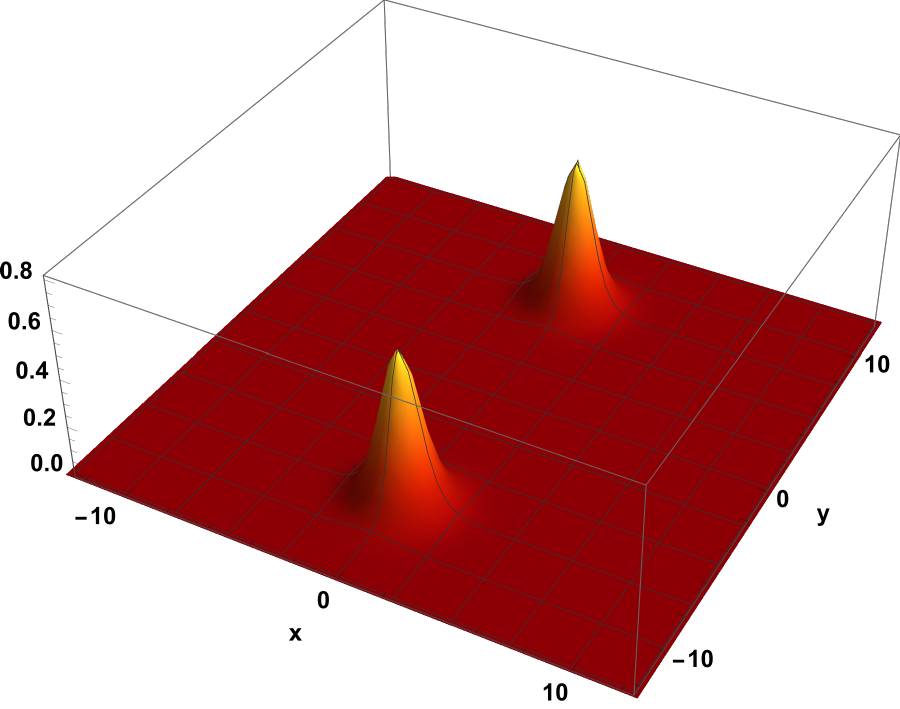}
    \end{subfigure}

    \caption{\textit{Energy density plots at increasing times during the right-angle scattering
of two vortices in a head-on collision in the BPS limit at a velocity $v=0.1c$.}}
    \label{FigI1:VortexCollision}
\end{figure}

Several additional phenomena related to vortex interactions have also been studied. In the 1980s, numerical simulations revealed that in head-on collisions, the scattering angle of two vortices is always \(90^\circ\) \cite{Ruback1988, Moriarty1988, Matzner1988, Shellard1988} (see Figure \ref{FigI1:VortexCollision}). This counterintuitive result was later given an analytical explanation by Rosenzweig, Srivastava, Abdewahid, Burzlaff, MacKenzie and Arthur \cite{Rosenzweig1991, Abdelwahid1994, MacKenzie1995, Arthur1996}. Moreover, the dynamics of vortex collisions can also be described through the moduli space approximation derived by Samols \cite{Samols1992} (see Chapter~\ref{Intro2}), who based his study on the analysis of the BPS equations carried out by Taubes \cite{Taubes1980,Taubes1980b} and Strachan's study of the moduli space approximation for an integrable model of vortices in the hyperbolic plane \cite{Strachan1992}.

It has also been discovered that when internal modes of vortices are excited prior to collision, a fractal structure emerges in the scattering outcomes, reminiscent of the resonance windows observed in kink–antikink collisions \cite{Krusch2024}.

Finally, it is worth highlighting that the moduli space approximation has been recently extended to study the dynamics of solitary moving vortices beyond the BPS limit \cite{MiguelezCaballero2025}.

%{\color{red} plot energía distancia diferentes regímenes }

%{\color{red} plot scattering noventa grados}

%{\color{red} weinberg, otras teorías con vórtices, Abelian Higgs, interaccion entre vórtices colisiones, colisiones vórtices excitados, fuerzas entre vórtices (M. Speight), cuerdas cósmicas, superfluidos }

\section{Objectives and structure }

The main objective of this thesis is to study the internal mode structure associated with kinks and vortices, and to analyze in detail the interaction between internal modes and radiation emission when such modes are initially activated. To this end, the numerical and analytical techniques that have been implemented will be presented in Section~\ref{Intro1.3}.

This thesis is divided into two main parts. In Part~\ref{Part1}, we discuss some classical results regarding kinks. These results focus on general aspects such as how to find these solutions, how to compute their topological charge, their internal mode structure, the corresponding BPS equations, and related properties.

Once the basic features characterizing these solutions have been introduced, we present the results concerning  \cite{AlonsoIzquierdo2024} and \cite{AlonsoIzquierdo2025} in Chapters \ref{Chap1} and \ref{Chap2}, respectively. In Chapter~\ref{Chap1}, we analyze in detail the simplest solution arising in the double \(\phi^4\) model. Specifically, we first obtain its shape modes analytically, and then study the dynamics up to second order in perturbation theory when one of these modes is excited.

In Chapter~\ref{Chap2}, kink–antikink scattering phenomena in the MSTB model are investigated. In order to gain a deeper understanding of the resonant energy exchange mechanism, the internal mode associated with the second field component will be excited prior to the collision. This setup enables us to explore the behavior of the system as the corresponding eigenfrequency is varied. In particular, we will study the system's response in cases where resonance conditions between different frequencies occur.

Part~\ref{Part2} is devoted to the review of the basic properties of Abelian-Higgs vortices and the presentation of the results related to \cite{AlonsoIzquierdo2024b}, \cite{AlonsoIzquierdo2025b}, and \cite{AlonsoIzquierdo2024c}. In Chapter~\ref{Chap3}, we describe the internal mode structure of a vortex with winding number \(n\), both inside and outside of the BPS limit. This analysis is facilitated by an appropriate choice of angular dependence for the internal modes, which also allows for a classification of the modes based on their behavior near the vortex core.

Following this, Chapter~\ref{Chap4} focuses on the dynamics of a vortex with topological charge \(n = 1\) when its only internal mode is excited. A detailed analysis of the radiation emitted by the system will also be conducted.

Appendix \ref{appen} contains details about the numerical scheme used in Chapters \ref{Chap3} and \ref{Chap4} Additionally, Appendices \ref{Eppen1}-\ref{Eppen4} contain the original papers on which Chapters \ref{Chap1}, \ref{Chap2}, \ref{Chap3} and \ref{Chap4} are based.

The manuscript concludes with a summary of the main results and a discussion of possible directions for future research based on the work developed in this thesis.

\section{Methodology }\label{Intro1.3}

In the present work, we have employed a combination of analytical and numerical techniques. Analytical techniques are mainly based on perturbative methods, which allow us to study the structure of the internal mode associated with static solutions and to analyze the behavior of the system up to the second order. This last aspect is essential for investigating the coupling between internal modes and the emission of radiation when at least one of these modes is initially excited. These techniques are based on the approach introduced by Manton and Merabet in 1997 \cite{Manton1997} and play a key role in Chapters~\ref{Chap1} and~\ref{Chap4}.

Regarding the numerical techniques, in Chapters~\ref{Chap1}, \ref{Chap2}, and~\ref{Chap4}, the field equations governing the dynamics of each model have been discretized to perform simulations under various scenarios. Specifically, in Chapters~\ref{Chap1} and~\ref{Chap2}, a fourth-order algorithm designed to efficiently handle Klein–Gordon-type equations has been implemented. A detailed description of this numerical method can be found in \cite{AlonsoIzquierdo2021}. In Chapter~\ref{Chap1}, a static kink perturbed through one of its internal modes is simulated in different configurations. To investigate radiation emission and the decay of internal mode amplitudes, this algorithm is supplemented by the implementation of the Fast Fourier Transform (FFT), which allows the radiation amplitudes to be computed numerically.

In Chapter~\ref{Chap2}, the same algorithm is used to simulate kink–antikink collisions at different initial velocities. The FFT is again employed to extract the amplitude of each vibrational mode following the final collision.

In Chapters~\ref{Chap3} and~\ref{Chap4}, a gradient flow method is used to obtain vortex profiles (see Section 2.8 and 7.7.2 of \cite{Manton2004}). This involves reformulating the field equations by exploiting the symmetry of the system and replacing the second-order time derivatives with first-order ones. An initial ansatz approximating the vortex configuration is allowed to evolve using a second-order finite-difference scheme until convergence to the true vortex profile is reached. Regarding the vibrational spectra studied in Chapter~\ref{Chap3}, the small perturbation problem is discretized, yielding a matrix formulation that can be diagonalized to extract the internal modes and their corresponding eigenvalues.

As for the numerical techniques employed in Chapter~\ref{Chap4}, the equations used to compute the vortex profiles are extended to simulate the dynamics of an excited vortex. All simulations were carried out using the programming language \texttt{C++}, supplemented with the \textit{Armadillo}, \textit{OpenMP}, and \textit{FFTW} libraries.

Due to the large number of simulations required in Chapter~\ref{Chap2}, the supercomputer \textit{Caléndula}, located at \textit{SCAYLE} (Supercomputación Castilla y León), was used to perform the necessary computations. All the analytical and numerical methods mentioned here will be described in detail throughout the following chapters.

%{\color{red} conectar con los solitones topologicos hablar de skyrmiones, despues el modelo de abelian higgs, toeria de ginzburg landau y luego kinks. Tammbien mencionar colisiones kink antikink}

%Its study dates back to the 1960s and 1970s when physicists and mathematicians began to study nonlinear classical field theories. When these theories were first studied particle like-solutions were found, whose energy density was concentrated towards a fixed point in space. 

    \part{Topological kinks}\label{Part1}

    \chapter{Topological kinks: foundations and their dynamics}\label{Intro1}

%In this Chapter we will analyze the basics of topological kinks. We will study how these solutions can be obtained starting from the BPS equations, their stability in terms of scaling arguments, and the definition of topological charge in this context. 

%Additionally, we will also talk  about  the internal mode structure associated with these solutions and how to simplify this problem making use of the supersymmetric structure that can be derived from the BPS equations. 

%Finally, we will generalize all these results in order to study kink solutions arising in two-component field theories. All these results will gain relevance when we analyze the dynamics associated with kink solutions arising in Part \ref{Part2}. 

%In this Chapter we will only examine non integrable field theories, but most of the results here obtained can be extended to study integrable models such us the sine-Gordon model.
%{\color{red} plots centro simulacion KAK y AKK $\phi^6$ y $\phi^4$ ? Cambiar $\phi\to\phi,\psi\to\psi$ en seccion 2.6?

%Hacer imulaciones inestabilidad kinks   y MSTB?
%}
In this chapter, we will explore the fundamentals  of topological kinks. We begin by deriving these solutions from the BPS equations and analyzing their stability via scaling arguments. The notion of topological charge in this context will also be presented and discussed in detail.

Furthermore, we will examine the internal mode structure associated with kink solutions, and generalize it to study the appearance of spectral walls in the context of kink-antikink configurations.

We will then  analyze the radiation emission of an excited kink in the $\phi^4$ field theory. The results here obtained will play a key role when analyzing excited kinks in Chapter \ref{Chap1} and excited Abelian-Higgs vortices in Chapter \ref{Chap4}

Finally, we will extend these results to investigate kink solutions in two-component scalar field theories. These insights will become particularly relevant when studying the dynamics of kink configurations in Chapters \ref{Chap1} and \ref{Chap2}.

Although we will restrict ourselves in this chapter to non-integrable field theories, many of the concepts and methods presented here are applicable to integrable models as well, including the sine-Gordon model.

\section{Stability based on scaling arguments: Derrick's theorem}\label{DerrickTheoremSec}

Let us consider a Lagrangian density with the following structure 
\begin{equation}\label{eqI2:LagDens}
    \mathcal{L}=\frac{1}{2}\partial_\mu \phi\, \partial^\mu\phi - U(\phi),
\end{equation}
 where the Minkowski metric is taken in the form $g_{\mu\nu} = \mathrm{diag}\{1, -1, -1,\dots\}$ and where the scalar field $\phi$ is assumed to be defined in $(n+1)$-dimensional spacetime. 
 
 If $\phi$ is a static solution to the time independent field equation, then,  the total energy of the field configuration is 
\begin{equation}\label{eqI2:StaticEn}
    E=\int \left[\frac{1}{2}(\nabla\phi)^2+U(\phi)\right] dx^n.
\end{equation}

Let $\phi$ be a solution to the Euler-Lagrange equation derived from  the Lagrangian density \eqref{eqI2:LagDens}. Now, let us define $\phi_\lambda(\mathbf{x})=\phi(\lambda\,\mathbf{x})$ where $\lambda$ is an arbitrary real constant. Let us also define the energy functionals 
\begin{equation*}
    I_1 = \int \frac{1}{2} (\nabla \phi)^2\, d^n x, \qquad I_2 = \int U(\phi)\, d^n x.
\end{equation*}

Consequently, the total energy of the rescaled field $\phi_\lambda$ becomes
\begin{equation*}
    E_\lambda =\int \left[\frac{1}{2}(\nabla\phi_\lambda)^2+U(\phi_\lambda)\right] dx^n=\lambda^{2-n}I_1+ \lambda^{-n} I_2.
\end{equation*}

For $\phi$ to be a stable solution, we must have that,  for $\lambda=1$, $\frac{d \,E_\lambda}{d\, \lambda}=0$ and $\frac{d^2 \,E_\lambda}{d^2\, \lambda}>0$. In our case, we have that
\begin{equation*}
    \frac{d \,E_\lambda}{d\, \lambda}=(2-n)\,\lambda^{1-n}I_1-n\,\lambda^{-n-1}I_2, \qquad  \frac{d^2 \,E_\lambda}{d\, \lambda^2}=(2-n)(1-n)\, \lambda^{-n}I_1+n(n+1)\,\lambda^{-n-2}I_2.
\end{equation*}

Assuming that both $I_1$ and $I_2$ are  positive quantities, then the condition $\frac{d \,E_\lambda}{d\, \lambda}=0$ only holds for $d=1$. In this case, 
\begin{equation}
    \lambda=\sqrt{\frac{I_2}{I_1}}=1 \quad \text{and} \quad \frac{d^2 \,E_\lambda}{d^2\, \lambda}>0. \label{eqI2:DerrickCond}
\end{equation}
For higher dimensions, it is straightforward to prove that $E_{\lambda}(\lambda)$ does not have any minimum for $n>1$. 

%All exposed so far can be stated as follows:
%\begin{theorem}[Derrick's Theorem]
  %  Consider the Langrian density \eqref{eqI2:LagDens} and suppose a finite energy field configuration $\phi$ that is solution to the corresponding field equations which is not the vacuum. If $d>1$, the function $E_\lambda$ has no minima, and therefore, there are not other static solutions other than the vacuum.
%\end{theorem}

%In Chapter \ref{Intro2}, we will examine how this theorem is applied to the case of Abelian-Higgs vortices as a consequence of the altered structure of the Lagrangian density.

All that has been exposed so far can be stated as follows:

\begin{theorem}[Derrick's Theorem]
    Consider the Lagrangian density \eqref{eqI2:LagDens} and suppose a finite energy field configuration $\phi$ that is a solution to the corresponding field equations and is not the vacuum. If $n > 1$, the function $E_\lambda$ has no minima, and therefore, there are no static solutions other than the vacuum.
\end{theorem}

In Chapter \ref{Intro2}, we will examine how this theorem applies to the case of Abelian-Higgs vortices as a consequence of the modified structure of the Lagrangian density.

%---------------------------------------------------------------------------------------------
\section{BPS equations, vacuum structure, kink solutions and topological charge}

In the present section and Sections \ref{intmode}-\ref{I2Manton}, the  field theory under consideration will possess the structure defined by the Lagrangian density \eqref{eqI2:LagDens} from which we derive the Euler-Lagrange equation
\begin{equation}\label{eqI2:FieldEq}
    \partial_\mu \partial^\mu \phi+\frac{d\, U}{d\, \phi}=0,
\end{equation}
defined in $1+1$ dimensions.

%Additionally, we assume that that $U(\phi)$ is a real and positive function that possesses at least two degenerate global minima in such a way that $U(\phi_{min})=0$\footnote{This requireent can always be fulfilled by only adding the right constant to the potential $U(\phi)$. }. Moreover, the Mikowski metric $g_{\mu\nu}=\mathrm{diag}\{1,-1\}$. 

Additionally, we assume that $U(\phi)$ is a real, positive function possessing at least two degenerate global minima such that $U(\phi_{\min})=0$\footnote{This requirement can always be fulfilled by adding an appropriate constant to the potential $U(\phi)$.}. Moreover, the Minkowski metric is taken as $g_{\mu\nu} = \mathrm{diag}\{1, -1\}$.

In this case the potential and kinetic energy is given, respectively,  by 
\begin{equation}\label{eqI2:PotKin}
    V=\int^{\infty}_{-\infty}\left[\frac{1}{2}\phi'^2+U(\phi)\right]dx,\qquad T=\frac{1}{2}\int^{\infty}_{-\infty}\dot{\phi}^2 dx,
\end{equation}
where $\phi'=\frac{\partial\, \phi}{\partial\, x}$, $\dot{\phi}=\frac{\partial\, \phi}{\partial\, t}$.
Let us now denote the set of vacua as follows
\begin{equation*}
    \mathcal{V}=\left\{\phi_0,\, \mathrm{such\,\,\, that}\,\, \phi_0\neq\phi(x,t)\rightarrow \dot{\phi_0}=\phi_0'=0 \mathrm{\,\,and\,\,}U(\phi_0)=0 \right\}.
\end{equation*}

%Kink solutions must interpolate between these vacua, so this set must contain at least two elements that are different from each other, in other words $\lim_{x\rightarrow\pm\infty}\phi=\phi_\pm$, being $\phi_+\neq\phi_-$ and $\phi_+,\phi_-\in\mathcal{V}$. If $\phi_+=\phi_-$, then we could continuous deformate the solution in such a way that $\phi(x)=\phi_+=\phi_-$. If the vacua are different, then, we cannot deformate one of the vaccums into the other, which means that the field configuration is stable.

Kink solutions must interpolate between these vacua, so the set of vacua $\mathcal{V}$ must contain at least two distinct elements. In other words, the field must satisfy
\[
\lim_{x \to \pm \infty} \phi(x) = \phi_\pm,
\]
where $\phi_+ \neq \phi_-$ and $\phi_\pm \in \mathcal{V}$. If $\phi_+ = \phi_-$, then the solution could be continuously deformed into the trivial vacuum solution $\phi(x) = \phi_+ = \phi_-$. However, if the vacua are different from each other, it is not possible to continuously deform one vacuum into the other, which implies that the field configuration is stable.

Indeed, there exists a non-topological soliton that interpolates between the same vacua in this type of theory: the sphaleron. A distinctive feature of this solution is its instability, when analyzing its internal mode structure, since at least an unstable mode always arises. For more detailed information about the sphaleron references \cite{Manton2004,Manton1984} provide a suitable background.

There also exist stable kink solutions that interpolate between the same vacuum. These kinks are referred to as \textit{vrochosons} in the literature. Such solutions arise in multicomponent field theories where the target space is defined over the surface of a torus. For more details, see \cite{BalseyroTesis, AlonsoIzquierdo2022b, AlonsoIzquierdo2023b}.

It is also possible to find metastable, non-topological kinks in multicomponent field theories whose target space is $\mathbb{R}^n$. This type of solution is commonly referred to as a \textit{lavion} in the literature \cite{Halcrow2023,AlonsoIzquierdo2020z}.

Alternatively, in the so-called \textit{singular models}, where the potential defining the theory exhibits at least one divergence, it is possible to find stable, non-topological kinks whose orbits enclose the aforementioned divergence \cite{AlonsoIzquierdo2007}.

The question now is: How can we extract the kink solutions starting from the Lagrangian density \eqref{eqI2:LagDens}?  To address this question, let us write the energy  of a static solution \eqref{eqI2:StaticEn} in a more convenient way:
\begin{equation}\label{eqI2:BPS1}
    E=\int^{\infty}_{-\infty} \left[\frac{1}{2}\phi'^2+U(\phi) \right]dx= \int^{\infty}_{-\infty} \left(\frac{1}{\sqrt{2}}\phi'\mp \sqrt{U(\phi)} \right)^2dx\, \pm\,\int^{\infty}_{-\infty} \sqrt{2\,U(\phi)}\phi'dx.
\end{equation}

Introducing the smooth superpotential function  $W(\phi)$ such that \begin{equation}\label{suppppp}
    U(\phi)=\frac{1}{2}\left(\frac{d W}{d\phi}\right)^2,
\end{equation}
the last part of \eqref{eqI2:BPS1} can reformulated as
\begin{equation*}
    \pm\int^{\infty}_{-\infty} \sqrt{2\,U(\phi)}\phi'dx=\pm\int^{\phi_+}_{\phi_-}\sqrt{2 U(\phi)}\, d\phi=|W(\phi_+)-W(\phi_-)|.
\end{equation*}

If the equation 
\begin{equation}\label{eqI2:BPS}
    \phi'=\pm\sqrt{2\,U(\phi)}
\end{equation}
is fulfilled, then the energy \eqref{eqI2:BPS1} of the field configuration is just 
\begin{equation*}
    E=|W(\phi_+)-W(\phi_-)|.
\end{equation*}

Equation \eqref{eqI2:BPS} is the BPS equation of the theory. This equation has to be integrated taking into account the asymptotic behavior of kink solutions in order to find its analytical formula. This trick can also be applied to find kinks in multicomponent field theories, however, in those cases, it is not as straightforward as in the single-component scenario, and sometimes it is not even possible to be able to write down the BPS equations. We will address this problem in Section \ref{Intro2.6}.

In equation \eqref{eqI2:BPS}, the sign $+$ is for the kink solution, whereas the sign $-$ corresponds to the antikink solution. 

It is straightforward to prove that the static field equation can be derived from \eqref{eqI2:BPS} and that if we apply a Lorenz boost to the kink solution, it is still a solution to the field equation. 

Also note that, as a consequence of  Derrick's theorem, condition \eqref{eqI2:DerrickCond} must be fulfilled, which leads to
\begin{equation*}
    \int^\infty_{-\infty} U(\phi)\,dx=\frac{1}{2} \int^\infty_{-\infty} \phi'^2\,dx.
\end{equation*}

As a final remark to this section, we will define  the topological charge of any field configuration regarding the Lagrangian density \eqref{eqI2:LagDens} as
\begin{equation}\label{eqI2:TopCharge}
    N=\frac{1}{2}\int^{\infty}_{-\infty}\phi'(x) \,dx=\frac{\phi(\infty)-\phi(-\infty)}{2}.
\end{equation}

This quantity is always conserved in any scenario, such as, for example, kink-antikink collisions.
%\vspace{\fill}

\section{Internal mode structure}\label{intmode}

When analyzing the dynamics of these solutions, their mode structure is crucial for understanding the interaction between kinks.

The most straightforward way to compute the internal mode structure is by using the following ansatz as a solution to the field equation:
\begin{equation*}
    \Tilde{\phi}(x,t)=\phi(x)+\epsilon\,e^{i\omega t}\,\eta(x),
\end{equation*}
where the parameter $\epsilon$ is supposed to be a small real parameter and where $\phi(x)$ is the kink solution.

Plugging this ansatz into the field equation \eqref{eqI2:FieldEq} leads to the following spectral problem,
\begin{equation}\label{eqI2:specProb}
    \mathcal{H}\,\eta=\omega^2\eta,
\end{equation}
where
\begin{equation*}
    \mathcal{H}=-\frac{d^2}{d^2 x}+\left.\frac{\partial^2 U}{\partial\phi^2}\right|_{\phi=\phi(x)}.
\end{equation*}

As it can be appreciated, spectral problem \eqref{eqI2:specProb} is a Schrödinger-like equation.  The limits 
\[\lim_{x\to \pm\infty}\left.\frac{\partial^2 U}{\partial\phi^2} \right|_{\phi=\phi(x)}=m_{\pm}\,,\]
define the mass thresholds of the spectral problem. In theories such as the $\phi^4$ model $m_+=m_-$, but there exists other models, like the $\phi^6$, where this property does not hold. Depending of the possible values of the eigenfrequencies we can have four possible types of internal modes:
\begin{itemize}

    \item \textbf{Unstable modes $\mathbf{(\omega^2<0)}$:} This modes imply the instability of the kink solution. Depending on the theory under consideration the solution would decay into the vacuum or into a less energetic kink solution.

    \item \textbf{Zero modes or translational modes $(\mathbf{\omega^2=0)}$:} One of the modes that arises in the resolution of  \eqref{eqI2:specProb} is the zero mode of the translational mode of the kink. The existence of this eigenmode is due to the translational invariance of the theory, in other words
\begin{equation*}
    \phi(x+\epsilon)\approx\phi(x)+\epsilon\,\frac{d\,\phi(x)}{d\,x}+\mathcal{O}(\epsilon^2)=\phi(x)+\epsilon\,\eta_0(x) +\mathcal{O}(\epsilon^2).
\end{equation*}

It can be easily proved that $\eta_0(x)$ is a solution to \eqref{eqI2:specProb} with $\omega^2=0$. 

    \item \textbf{Shape mode or discrete modes $\mathbf{(0<\omega^2<\mathrm{min}(m_+,m_-))}$:} These modes vanish at infinity. In other words, they are nonzero only in a region near the kink center.

      \item   \textbf{Semi-propagating modes $ (\mathrm{max}(m_+,m_-)>\omega^2 >\mathrm{min}(m_+,m_-))$: } These modes behave as radiation only for $x\to-\infty$ if $m_+>m_-$, and for $x\to\infty$ if  $m_+<m_-$. In other words,
      \begin{equation*}
          \lim_{x\to-\infty}\eta(x)=C e^{-i k_- x}, \quad  \lim_{x\to\infty}\eta(x)=0, \quad \text{if} \quad m_+>m_-,
      \end{equation*}
      and
        \begin{equation*}
          \lim_{x\to-\infty}\eta(x)=0, \quad  \lim_{x\to\infty}\eta(x)=C e^{-i k_+ x}, \quad \text{if} \quad m_+<m_-.
      \end{equation*} 
where $k_\pm=\sqrt{\omega^2-m_\pm}$.

    \item \textbf{Pure radiation modes $\mathbf{(\omega^2>\mathrm{max}(m_+,m_-))}$:} They are not localized, at infinity they behave as a traveling wave, in other words
    \begin{equation*}
        \lim_{x\to\pm\infty}\eta(x)=C_{\pm} e^{\pm i k_\pm x},
    \end{equation*}
    where $k_{\pm}=\sqrt{\omega^2-m_\pm}$.

\end{itemize}

The spectral problem \eqref{eqI2:specProb} can also be used as an approximation to obtain the modes corresponding to kink/antikink  configurations. Although this approach is not entirely correct, since a kink/antikink configuration is not a static solution to the field equations, there exists an attractive force between both kinks and a spectra derived only for static solutions is used. Despite this limitation, this method proves useful for identifying spectral walls: localized spatial regions where a kink can either bounce back, become trapped, or pass through without interference, depending on the amplitude of the kink’s internal mode. These spectral walls are localized and are marked by   the separation between kink and antikink for which one  of the discrete modes merges into the continuum \cite{Adam2019b,Dorey2011}. We will address this phenomenon in the following section for the kinks arising in the $\phi^4$ and $\phi^6$ models.

%\vspace{\fill}

%\section{Two simple models: the $\phi^4$ and $\phi^6$ models}
\section{Two simple field theories: the \texorpdfstring{$\phi^4$}{phi4} and \texorpdfstring{$\phi^6$}{phi6} models}
%In this section, the simplest  field theories that support kink solutions will be studied in detail. The kink solutions will be found and  some details about kink-antinkink configurations and spectral walls will also be given. Specifically, some of the results related to the $\phi^4$ model will be generalized in the study of the MSTB and double $\phi^4$ models, which will be presented in Chapter \ref{Intro1.7}.

In this section, the simplest field theories that support kink solutions will be studied in detail. The kink solutions will be explicitly constructed, and some aspects of kink–antikink configurations and spectral walls will also be discussed. Specifically, some of the results related to the $\phi^4$ model will be generalized in the study of the MSTB and double $\phi^4$ models, which will be presented in Section~\ref{Intro1.7}.

\subsection{The \texorpdfstring{$\phi^4$}{phi4} model}\label{Intro2.4.1}

%This field theory is governed by the Lagrangian density \eqref{eqI2:Lag} with potential 
%\begin{equation}\label{eqI2:PotPhi4}
 %   U(\phi)= \frac{1}{2}(\phi^2-1)^2.
%\end{equation}

%Therefore, the superpotential of the theory and the BPS equation are, respectively,
%\begin{equation*}
 %   W(\phi)=\phi-\frac{\phi^3}{3}, \qquad \frac{d\,\phi}{d\,x}=\pm(1-\phi^2).
%\end{equation*}

%Note that \eqref{eqI2:PotPhi4} has two degenerate vacua in $\phi=\pm1$, implying that both the kink and the antikink must interpolate between these two field  values. 

%Integrating the BPS equation, the next relation is found 
%\begin{equation*}
 %   x-x_0=\pm \int \frac{1}{1-\phi^2}d\, \phi=\pm \mathrm{arctanh}(\phi),
%\end{equation*}
%which, in turn, implies that 
%\begin{equation}\label{eqI2:KAK}
 %   \phi_{K}(x)=\tanh(x-x_0),\qquad \phi_{-K}(x)=-\tanh(x-x_0).
%\end{equation}

%Functions \eqref{eqI2:KAK} are the kink and antikink solutions of this theory. Using \eqref{eqI2:PotKin} we find that the energy density of these field configurations is 
%\begin{equation}
 %   \mathcal{E}(x)=\mathrm{sech}^4(x-x_0),
%\end{equation}
%and the total energy is 
%\begin{equation}
 %   E=\int^{\infty}_{-\infty}  \mathcal{E}(x) d\,x
%=\frac{4}{3}.
%\end{equation}

This field theory is governed by the Lagrangian density \eqref{eqI2:LagDens} with potential
\begin{equation}\label{eqI2:PotPhi4}
    U(\phi)= \frac{1}{2}(\phi^2 - 1)^2.
\end{equation}

Therefore, the superpotential of the theory and the BPS equation are, respectively,
\begin{equation}\label{nosequeponer}
    W(\phi) = \phi - \frac{\phi^3}{3}, \qquad \frac{d\phi}{dx} = \pm(1 - \phi^2).
\end{equation}

Note that \eqref{eqI2:PotPhi4} has two degenerate vacua at $\phi = \pm 1$, implying that both the kink and the antikink must interpolate between these two field values.

Integrating the BPS equation \eqref{nosequeponer} yields
\begin{equation*}
    x - x_0 = \pm \int \frac{1}{1 - \phi^2} \, d\phi = \pm \mathrm{arctanh}(\phi),
\end{equation*}
which, in turn, implies that
\begin{equation}\label{eqI2:KAK}
    \phi_K(x) = \tanh(x - x_0), \qquad \phi_{-K}(x) = -\tanh(x - x_0).
\end{equation}

The functions in \eqref{eqI2:KAK} are the kink ($\phi_K$) and antikink ($\phi_{-K}$)  solutions of this theory. Therefore, from \eqref{eqI2:StaticEn} we find that the energy density of these field is 
\begin{equation}
    \mathcal{E}(x) = \mathrm{sech}^4(x - x_0),
\end{equation}
and the total energy is
\begin{equation}
    E = \int_{-\infty}^{\infty} \mathcal{E}(x) \, dx = \frac{4}{3}.
\end{equation}

\begin{figure}[h!]
    \centering
    \begin{subfigure}{0.49\textwidth}
        \centering
        \includegraphics[width=\linewidth]{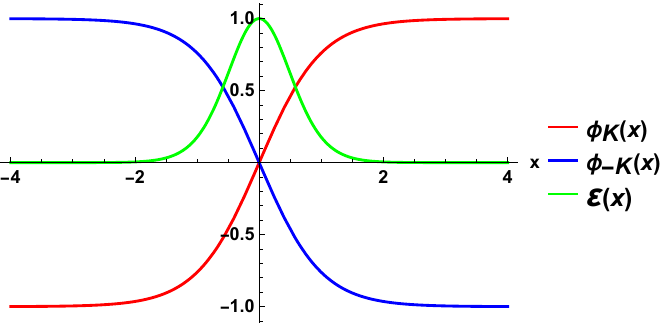}
    \end{subfigure}
    \begin{subfigure}{0.49\textwidth}
        \centering
        \includegraphics[width=\linewidth]{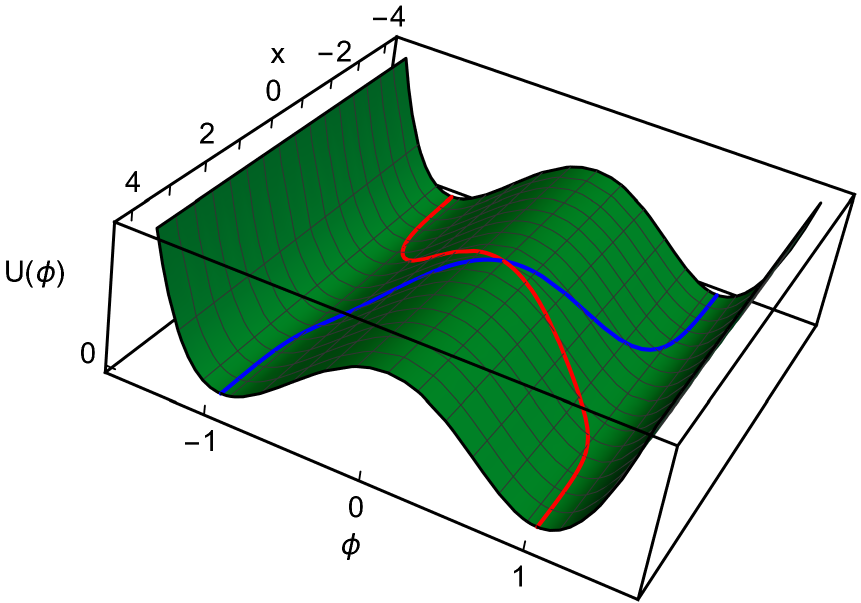}
    \end{subfigure}
    \caption{\textit{Left: kink solutions arising in the $\phi^4$ model and their energy density. Right: kink solutions plotted over the potential \eqref{eqI2:PotPhi4} (right).}}
    \label{figI2:Phi4kinks}
\end{figure}

As it can be appreciated in Figure \ref{figI2:Phi4kinks}, both solutions interpolate between the vacua of the potential \eqref{eqI2:PotPhi4}. Additionally, the energy density distribution is localized near the center of the kink.

Taking into account \eqref{eqI2:TopCharge}, the topological charge is $N = 1$ for the kink and $N = -1$ for the antikink. Moreover, for a kink–antikink configuration, this charge is just $N = 0$.

For this model, we have
\begin{equation}
    m_+ = m_- = m = \lim_{x \to \pm \infty}\left. \frac{\partial^2 U}{\partial \phi^2} \right|_{\phi=\phi(x)} = 4,
\end{equation}
which implies that when studying the small vibration operator, no semi-propagating modes will be found. The spectral problem then reads
\begin{equation}\label{eqI2:PerProbPhi4}
    -\frac{d^2 \eta(x)}{d x^2} + \left(4 - 6 \,\mathrm{sech}^2 x\right) \eta(x) = \omega^2 \eta(x),
\end{equation}
which is a Schrödinger-like equation with a modified Pösch–Teller potential well. The detailed solution to this equation can be found in \cite{Rosen1932, Flugge1971, Morse1953}.

The internal mode structure of this system consists of:
\begin{itemize}
    \item \textbf{Zero mode:} As previously stated, this mode is associated with the frequency $\omega^2 = 0$ and is given by
    \begin{equation*}
        \eta_0(x) = \mathrm{sech}^2(x).
    \end{equation*}
    
    \item \textbf{Shape mode or discrete mode:} A localized mode with frequency $\omega^2 = 3$ arises. Its profile is
    \begin{equation}
        \eta_D(x) = \mathrm{sech}(x)\tanh(x).
    \end{equation}
    
    \item \textbf{Radiation modes:} There is also a continuum of modes associated with frequencies $\omega^2 = 4 + q^2$. The eigenfunctions take the form
    \begin{equation}\label{eqI2:CCradiation}
        \eta_q(x) = \left(3 \tanh^2(x) - 3i\, q \tanh(x) - 1 - q^2\right)e^{i q x}.
    \end{equation}
    It can be verified that the complex conjugate of \eqref{eqI2:CCradiation} is also a solution to \eqref{eqI2:PerProbPhi4}. This second solution can also be obtained by performing the change $q \to -q$. The Wronskian corresponding to these two solutions is
    \begin{equation*}
        W_q = \eta_q \,\eta_{-q}' - \eta_q'\, \eta_{-q} = -2i q\, (q^2 + 1)(q^2 + 4).
    \end{equation*}
\end{itemize}

\begin{figure}[h!]
    \centering  
    \includegraphics[width=0.52\linewidth]{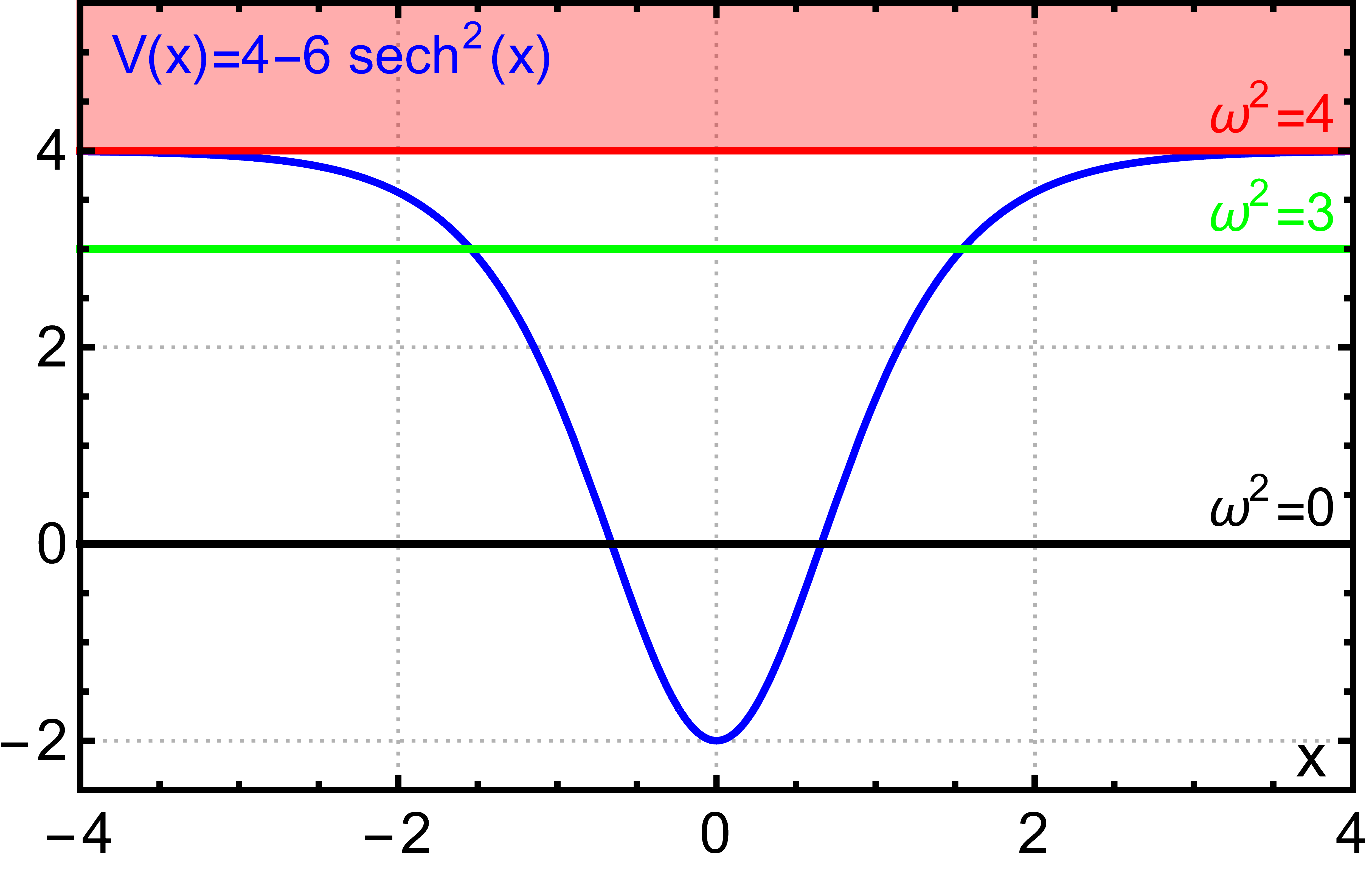}
    \caption{\textit{Potential well corresponding to the spectral problem \eqref{eqI2:PerProbPhi4} and the allowed eigenvalues. The shaded area represents the continuum starting at $\omega^2=4$.}}
    \label{figI2:PTPhi4}
\end{figure}

In Figure \ref{figI2:PTPhi4}, the spectrum corresponding to the allowed vibration modes are plotted over the potential of the spectral problem \eqref{eqI2:PerProbPhi4}.

\begin{figure}[h!]
    \centering
    \includegraphics[width=0.52\linewidth]{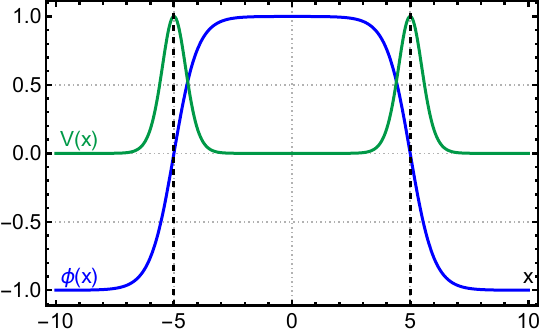}
    \caption{\textit{Kink–antikink pair \eqref{eqI2:KAKPhi4} with $d=5$, and its corresponding potential energy $U(x)$.}}
    \label{figI2:KAKPair}
\end{figure}

As previously stated, a similar spectral problem can be constructed to analyze the so-called spectral walls associated with kink–antikink configurations. For a kink–antikink pair with a large separation, the following ansatz can be employed:
\begin{equation}\label{eqI2:KAKPhi4}
    \phi(x) = \phi_K(x+d) + \phi_{-K}(x-d) - 1 = \tanh(x + d) - \tanh(x - d) - 1,
\end{equation}
 where $d$ is the distance from each of the kink centers to the axis origin.  
 The ansatz \eqref{eqI2:KAKPhi4} is not an exact solution of the field equation \eqref{eqI2:FieldEq}, but for $d \gg 1$ it serves as a good approximation to the true solution. Note also that the topological charge corresponding to this field configuration is $N = 0$. A plot of this field for $d=5$ is shown in Figure \ref{figI2:KAKPair}.

By plugging \eqref{eqI2:KAKPhi4} into the general spectral problem \eqref{eqI2:specProb}, we obtain the following equation:
\begin{equation}\label{eqI2:SpecProbKAKPhi4}
    -\frac{d^2 \eta}{d x^2} + \left[6\left(\tanh(x+d) - \tanh(x-d) - 1 \right)^2 - 2\right] \eta = \omega^2 \eta.
\end{equation}

\begin{figure}[h!]
    \centering
    \includegraphics[width=0.52\linewidth]{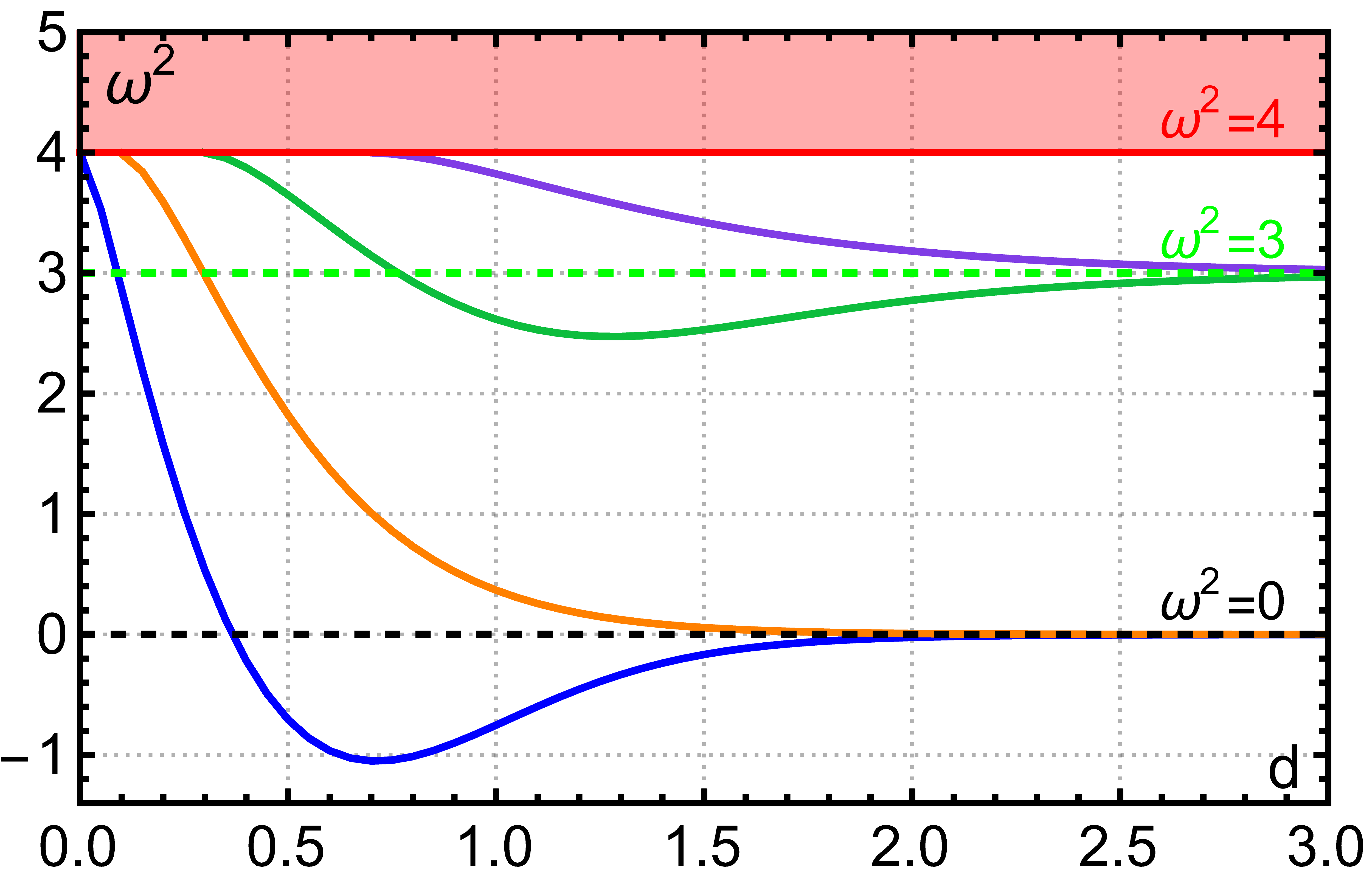}
    \caption{\textit{Spectrum for kink–antikink configurations from \eqref{eqI2:SpecProbKAKPhi4} in the $\phi^4$ model as a function of the separation $d$.}}
    \label{figI2:SpecKAKPhi4}
\end{figure}

Unlike the previous spectral problem, this one  must be solved numerically. The resulting spectrum as a function of the distance $d$ is presented in Figure \ref{figI2:SpecKAKPhi4}.

This figure shows that three modes merge into the continuum, indicating the existence of three distinct spectral walls approximately located at $d \approx 0.1,\ 0.3,\ 0.7$. If the internal mode amplitudes of both kinks are suitably tuned, one would observe that the kinks become trapped at these distances. Each spectral wall corresponds to a different required amplitude. The wall located at $d \approx 0.1$ is more subtle and harder to resolve. In this regime, interkink forces become significant, which, in turn, implies that the ansatz \eqref{eqI2:KAKPhi4} becomes less accurate \cite{Adam2020}.

Furthermore, as $d \gg 1$, the eigenvalues of the spectral problem \eqref{eqI2:SpecProbKAKPhi4} asymptotically approach those of the single-kink fluctuation spectrum.

\begin{figure}[h!]
    \centering
    \begin{subfigure}{0.49\textwidth}
        \centering
        \includegraphics[width=\linewidth]{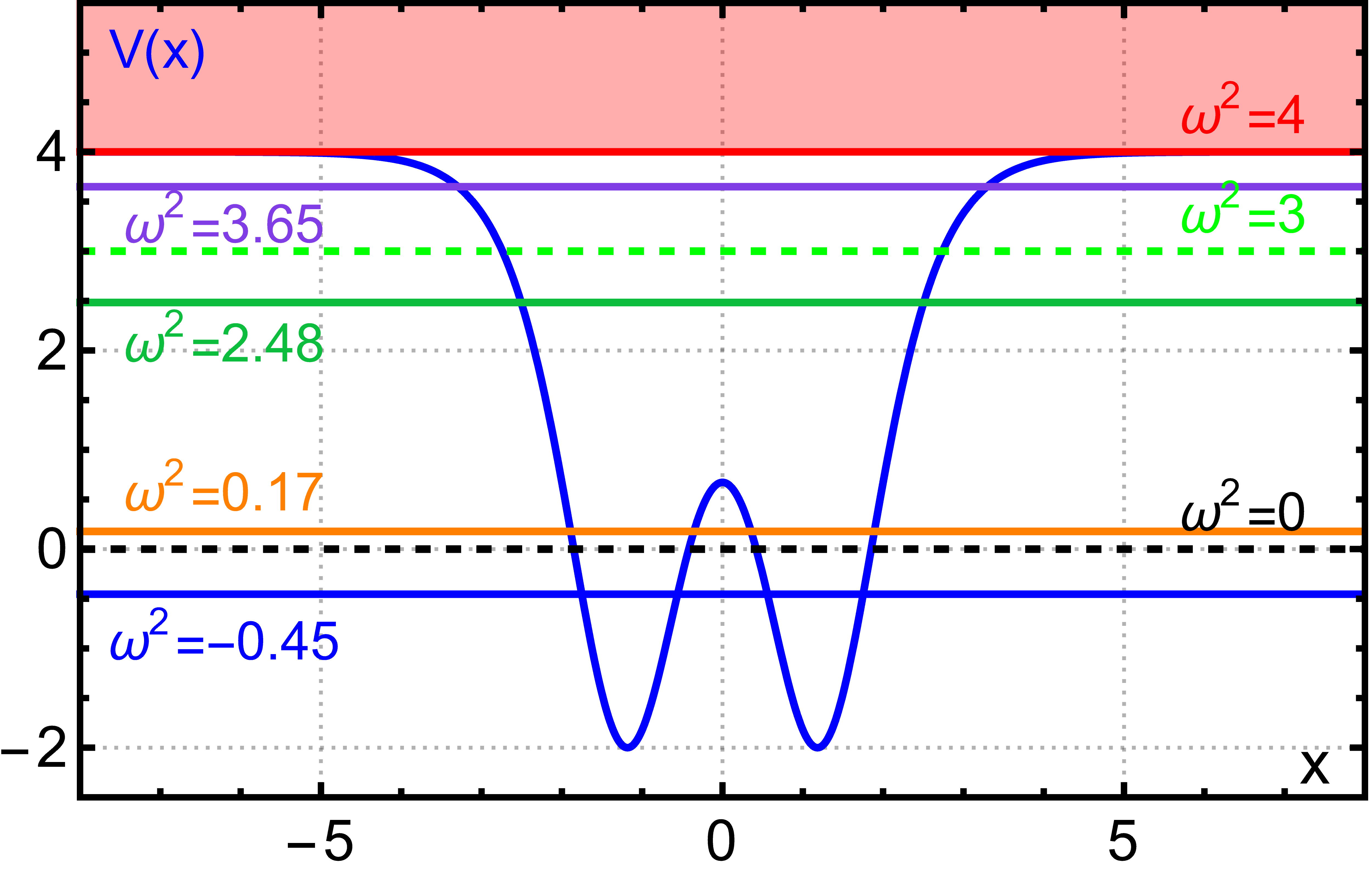}
    \end{subfigure}
    \begin{subfigure}{0.49\textwidth}
        \centering
        \includegraphics[width=\linewidth]{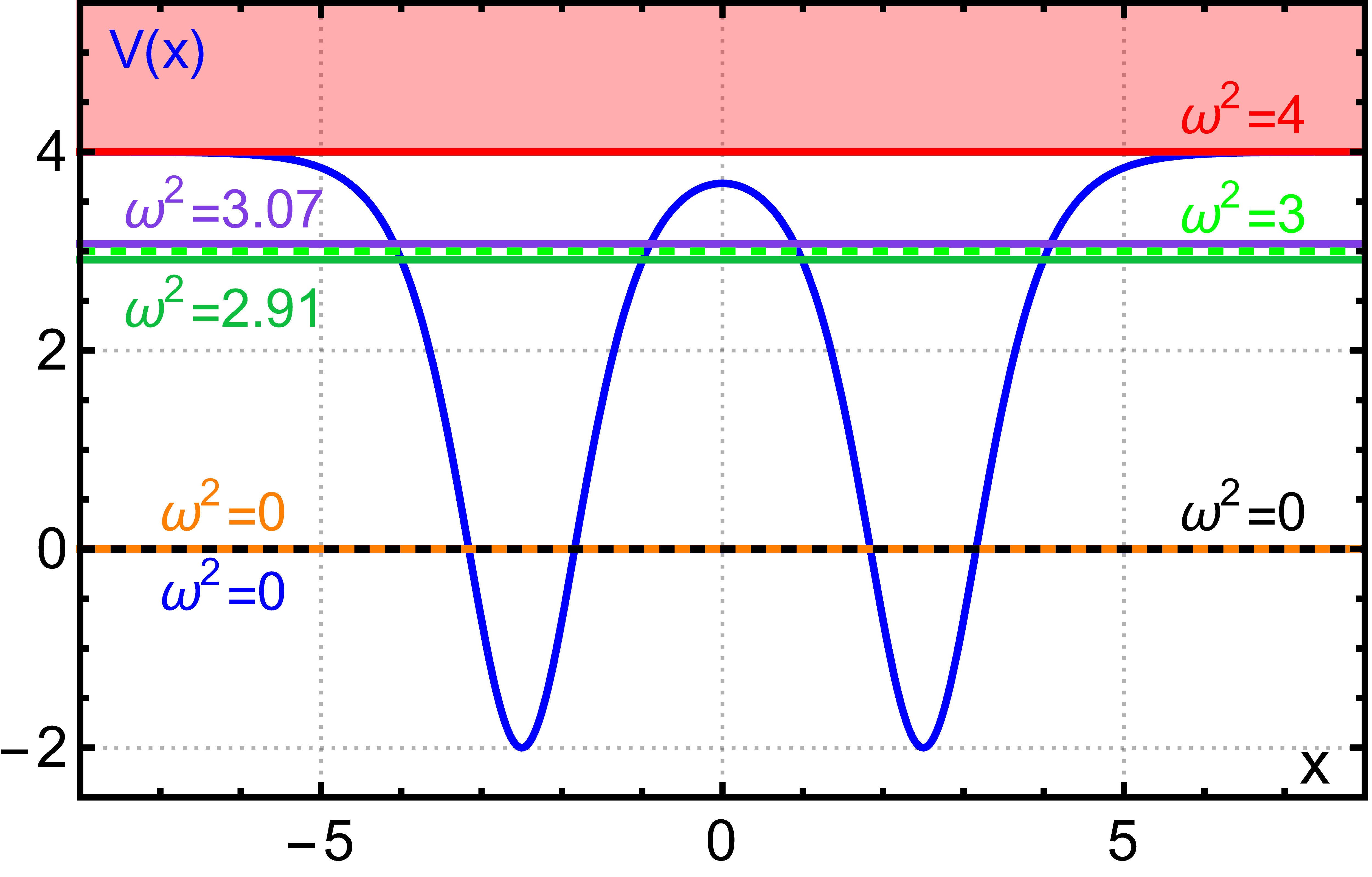}
    \end{subfigure}
    \caption{\textit{Potential well for the spectral problem \eqref{eqI2:SpecProbKAKPhi4} for $d = 1.2$ (left) and $d = 2.5$ (right). The resulting eigenvalues (solid lines) and the eigenvalues from the single-kink problem \eqref{eqI2:PerProbPhi4} are also plotted for comparison.}}
    \label{figI2:PpotentialKAK}
\end{figure}

In Figure \ref{figI2:PpotentialKAK}, the potential well from \eqref{eqI2:SpecProbKAKPhi4} is plotted for $d = 1.2$ and $d = 2.5$. As the distance $d$ increases, the eigenvalues tend toward those obtained from the fluctuation operator of a single kink. Moreover, the potential begins to resemble two separate copies of the potential from \eqref{eqI2:PerProbPhi4}.

\begin{figure}[h!]
    \centering
    \begin{subfigure}{0.49\textwidth}
        \centering
        \includegraphics[width=\linewidth]{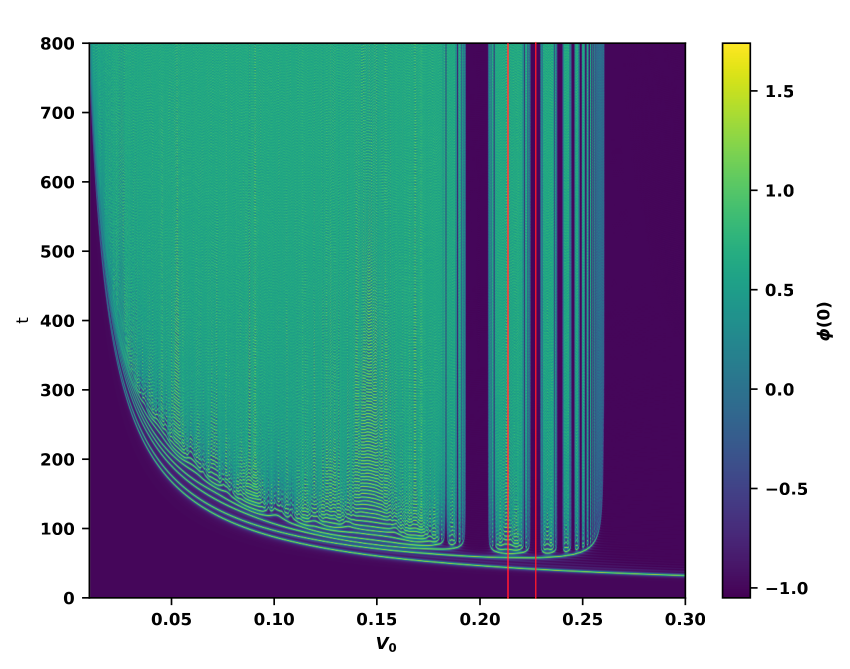}
    \end{subfigure}
    \begin{subfigure}{0.49\textwidth}
        \centering
        \includegraphics[width=\linewidth]{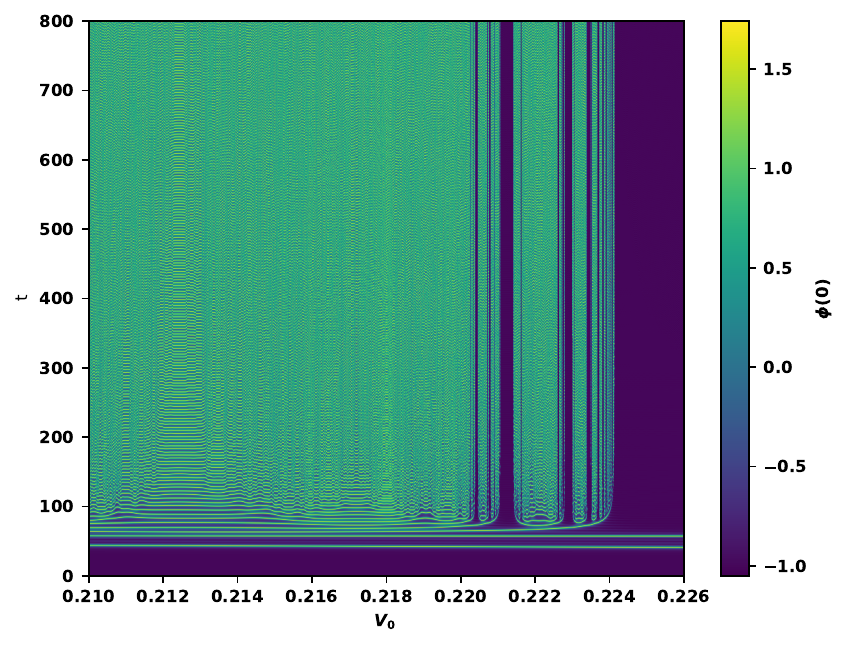}
    \end{subfigure}
    \caption{\textit{Center of the field configuration during antikink–kink collisions in the $\phi^4$ model as a function of time and the initial velocity of the kink–antikink pair ($v_0$). The right plot shows a zoomed-in view of the region outlined in red in the left plot.}}
    \label{figI2:FractalPhi4}
\end{figure}

To conclude the discussion of the $\phi^4$ model, Figure \ref{figI2:FractalPhi4} shows a plot of the center of the field $\phi$ during kink–antikink scattering as a function of the initial velocity of the kink–antikink pair. As can be seen, when the plot is zoomed in, the same pattern reappears, clearly indicating the existence of a fractal structure in kink–antikink collisions in this model.
%{\color{red} meter plot potencial posvh teller+ modos internos+ plot modos internos.}
%{\color{red}hablar configuraciones KAK y spectral walls, meter plot KAK en este modelo, plot potencial para }
\subsection{The \texorpdfstring{$\phi^6$}{phi6} model}
\begin{comment}
    In this case, this model is governed by the Lagrangian density,
\begin{equation*}
   \mathcal{L}=\frac{1}{2}\partial_{tt}\phi-\frac{1}{2}\partial_{xx}\phi-\frac{1}{2}\phi^2(\phi^2-1)^2.
\end{equation*}

Now the potential 
\begin{equation}
    U(\phi)=\frac{1}{2}\phi^2(\phi^2-1)^2,
\end{equation}
possesses three degenerate vacua. 

Consequently, the superpotential can be expressed as 
\begin{equation}
    W(\phi)=\pm \frac{\phi}{2}\mp\frac{\phi^4}{4}.
\end{equation}
and the BPS equation is just
\begin{equation}\label{eqI2:PotPhi6}
    \frac{d\, \phi}{d\, x}=\pm\phi(1-\phi^2).
\end{equation}
\end{comment}
In this case, the model is governed by the Lagrangian density
\begin{equation*}
   \mathcal{L} = \frac{1}{2} (\partial_t \phi)^2 - \frac{1}{2} (\partial_x \phi)^2 - \frac{1}{2} \phi^2(\phi^2 - 1)^2.
\end{equation*}

The associated potential is
\begin{equation}\label{potphi6}
    U(\phi) = \frac{1}{2} \phi^2(\phi^2 - 1)^2,
\end{equation}
which has three degenerate vacua located at $\phi = 0$ and $\phi = \pm1$.

Consequently, the superpotential can be written as
\begin{equation}
    W(\phi) = \pm \left( \frac{\phi}{2} - \frac{\phi^4}{4} \right),
\end{equation}
and the associated BPS equation \eqref{eqI2:BPS} takes the form
\begin{equation}\label{eqI2:PotPhi6}
    \frac{d\phi}{dx} = \pm \phi (1 - \phi^2).
\end{equation}

\begin{comment}
    A simple integration gives two kink solutions and two  antikink solutions, which are
\begin{equation}
    \begin{matrix}
        \phi_{1,K}(x)=\phi_{(0,1)}(x)=\sqrt{\frac{1+\tanh(x)}{2}},  &\phi_{1,-K}(x)=\phi_{(1,0)}(x)=\sqrt{\frac{1-\tanh(x)}{2}},\\
         \phi_{2,K}(x)=\phi_{(-1,0)}(x)=-\sqrt{\frac{1-\tanh(x)}{2}},  &\phi_{2,-K}(x)=\phi_{(0,-1)}(x)=-\sqrt{\frac{1+\tanh(x)}{2}}.
    \end{matrix}
\end{equation}
\end{comment}
 A simple integration gives the two kink solutions
\begin{equation}
    \begin{matrix}
        \phi_{1,K}(x)=\phi_{(0,1)}(x)=\sqrt{\frac{1+\tanh(x)}{2}},  &\phi_{2,K}(x)=\phi_{(-1,0)}(x)=-\sqrt{\frac{1-\tanh(x)}{2}},
    \end{matrix}
\end{equation}
and the two corresponding antikinks
\begin{equation}
    \begin{matrix}
       \phi_{1,-K}(x)=\phi_{(1,0)}(x)=\sqrt{\frac{1-\tanh(x)}{2}},  &\phi_{2,-K}(x)=\phi_{(0,-1)}(x)=-\sqrt{\frac{1+\tanh(x)}{2}}.
    \end{matrix}
\end{equation}

\begin{figure}[h!]
    \centering
    \begin{subfigure}{0.49\textwidth}
        \centering
        \includegraphics[width=\linewidth]{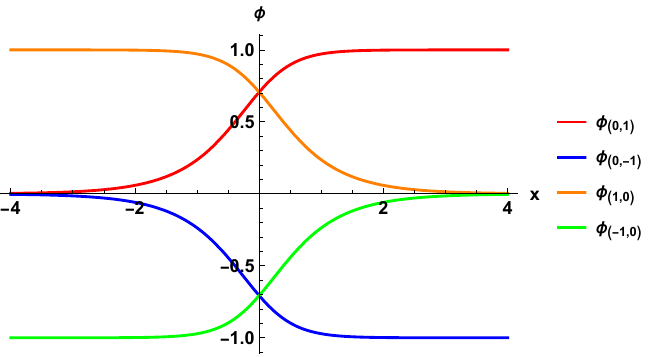}
    \end{subfigure}
    %\hfill
    \begin{subfigure}{0.49\textwidth}
        \centering
        \includegraphics[width=\linewidth]{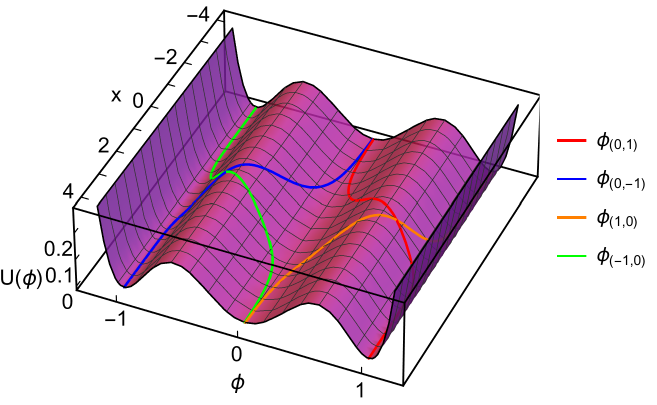}
    \end{subfigure}
    \caption{\textit{Kink solutions arising in the $\phi^6$ model  (left) and kink solutions plotted over the potential \eqref{potphi6} (right).}}
    \label{figI2:Phi6kinks}
\end{figure}

These solutions are displayed in Figure \ref{figI2:Phi6kinks}.

 Let us now consider the small fluctuation operator around the kink solutions $\phi_{(0,1)}$ and $\phi_{(0,-1)}$. The fluctuation operator can be written as\footnote{Note that for the kinks $\phi_{(1,0)}$ and $\phi_{(-1,0)}$, the corresponding spectral problem can be obtained by simply performing the change $x \to -x$ in the potential of \eqref{eqI2:PerProblPhi6}.}
\begin{equation}\label{eqI2:PerProblPhi6}
    \left[-\frac{d^2}{dx^2} - \frac{5}{4} + \frac{3}{2} \tanh(x) + \frac{15}{4} \tanh^2(x) \right] \eta = \omega^2 \eta,
\end{equation}
which implies
\begin{equation}
    m_+ = \left.\lim_{x \to \infty} \frac{\partial^2 U}{\partial \phi^2}\right|_{\phi=\phi(x)} = 4, \qquad    
    m_- = \lim_{x \to -\infty} \left.\frac{\partial^2 U}{\partial \phi^2}\right|_{\phi=\phi(x)}= 1.
\end{equation}
Hence, unlike the case of the $\phi^4$ model, we now have $m_+ \neq m_-$.

The spectral structure of \eqref{eqI2:PerProblPhi6} is as follows \cite{Shnir2018, Lohe1979}:
\begin{itemize}
    \item \textbf{Zero mode:} As previously stated, this mode corresponds to $\omega^2 = 0$ and is given by
    \begin{equation*}
        \eta_0(x) = e^{2x} \left( \frac{1}{1 + e^{2x}} \right)^{3/2}.
    \end{equation*}
    
    \item \textbf{Semi-propagating modes:} A continuum of semibound modes appears for $m_+ > \omega^2 > m_-$. These can be expressed as
    \begin{equation}
        \eta(x) = \frac{e^{-(k_1 + i\, k_2)x / 2}}{(e^x + e^{-x})^{\alpha}} 
        \,{}_2F_1\left( \alpha - \frac{3}{2},\, \alpha + \frac{5}{2},\, k_1 + 1;\, \frac{1}{1 + e^{2x}} \right),
    \end{equation}
    where
    \begin{equation*}
        k_1 = \sqrt{\omega^2 - 1}, \quad 
        k_2 = \left| \sqrt{\omega^2 - 4} \right|, \quad 
        \alpha = \frac{k_1 - i\, k_2}{2},
    \end{equation*}
    and  where ${}_2F_1$ is the hypergeometric function.
    
    Asymptotically, these functions behave as
    \begin{equation*}
        \eta(x \to \infty) \to 0, \qquad 
        \eta(x \to -\infty) \to C_{k_1} e^{i k_1 x} + C_{k_1}^* e^{-i k_1 x}.
    \end{equation*}

    \item \textbf{Radiation modes:} A continuum spectrum arises for $\omega^2 > m_+$, and the corresponding eigenfunctions take the form
    \begin{equation}
        \eta(x) = \frac{e^{-i(k_2 - k_1)x / 2}}{(e^x + e^{-x})^{\alpha}} 
        \,{}_2F_1\left( \alpha - \frac{3}{2},\, \alpha + \frac{5}{2},\, -i k_1 + 1;\, \frac{1}{1 + e^{2x}} \right).
    \end{equation}
    These modes resemble radiation waves in both asymptotic regions $x \to \pm \infty$.
\end{itemize}

\begin{figure}[h!]
    \centering  \includegraphics[width=0.55\linewidth]{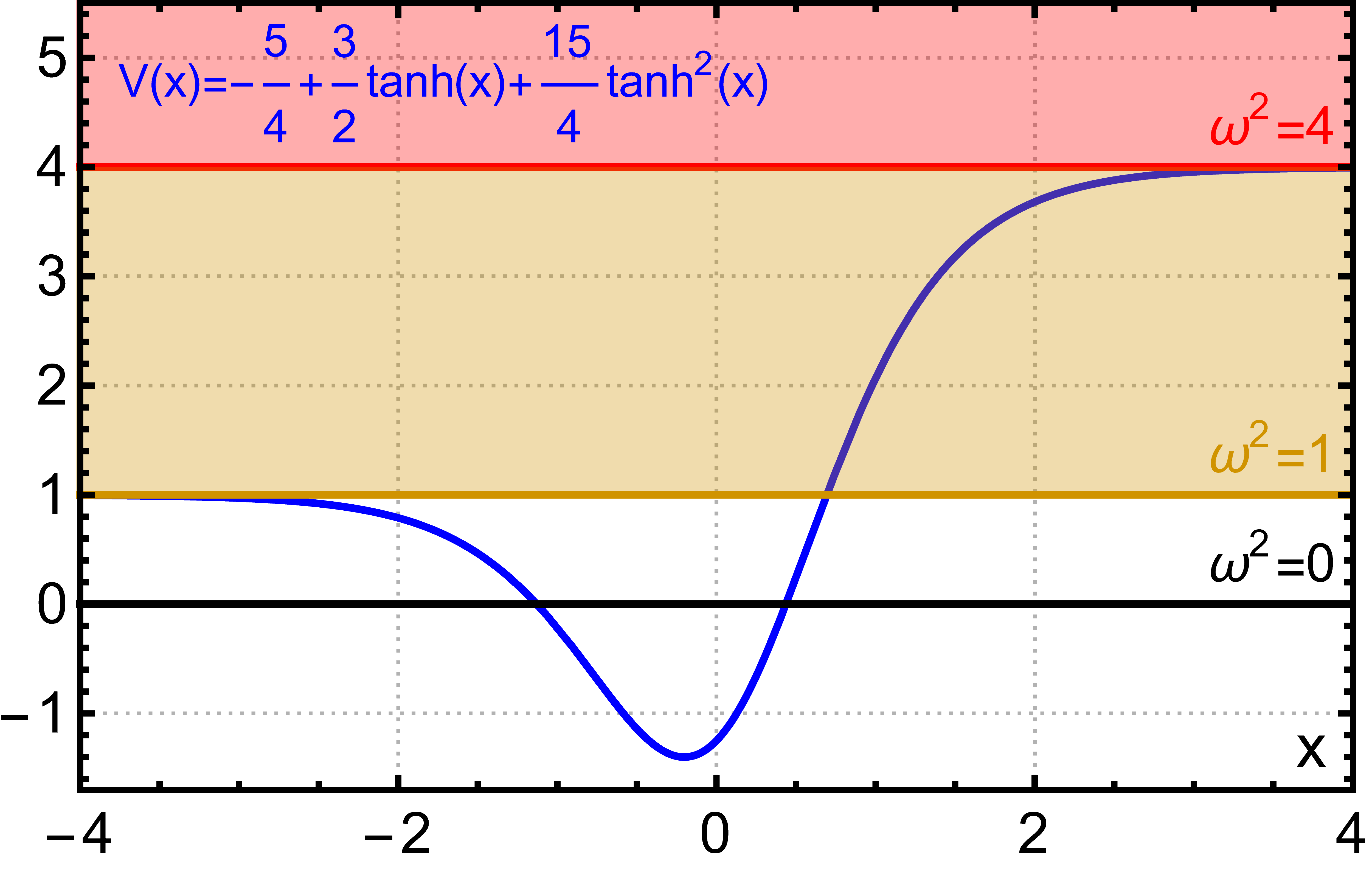}
    \caption{\textit{Potential well corresponding to the spectral problem \eqref{eqI2:PerProblPhi6} and the allowed eigenvalues. The shaded areas represents the continuum arsing in $\omega^2=1,4$.}  }
    \label{figI2:PTPhi6}
\end{figure}

As can be seen, in this case there are no bound modes. A plot of the eigenfrequencies described above is shown in Figure~\ref{figI2:PTPhi6}.

\begin{figure}[h!]
    \centering
    \begin{subfigure}{0.49\textwidth}
        \centering
        \includegraphics[width=\linewidth]{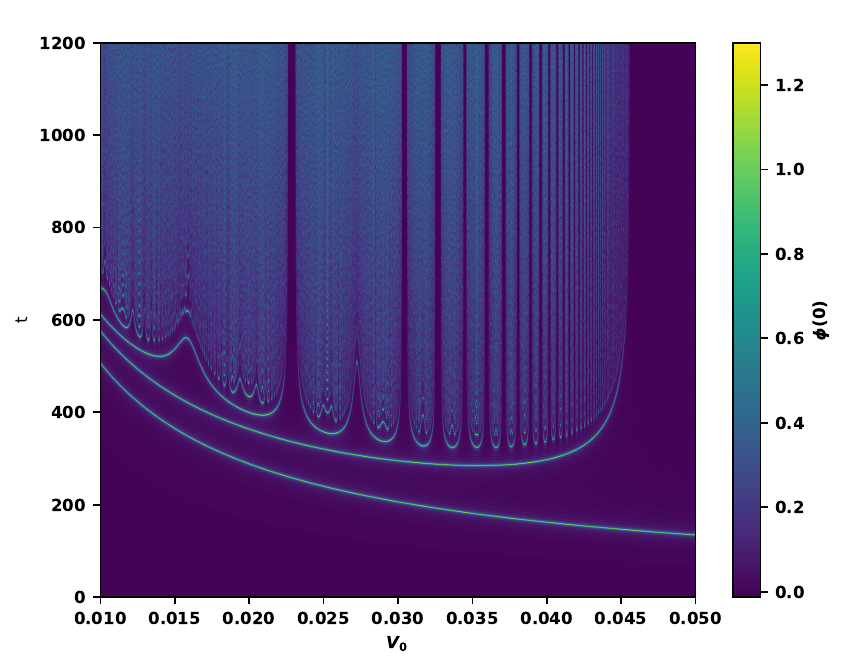}
    \end{subfigure}
    \begin{subfigure}{0.49\textwidth}
        \centering
        \includegraphics[width=\linewidth]{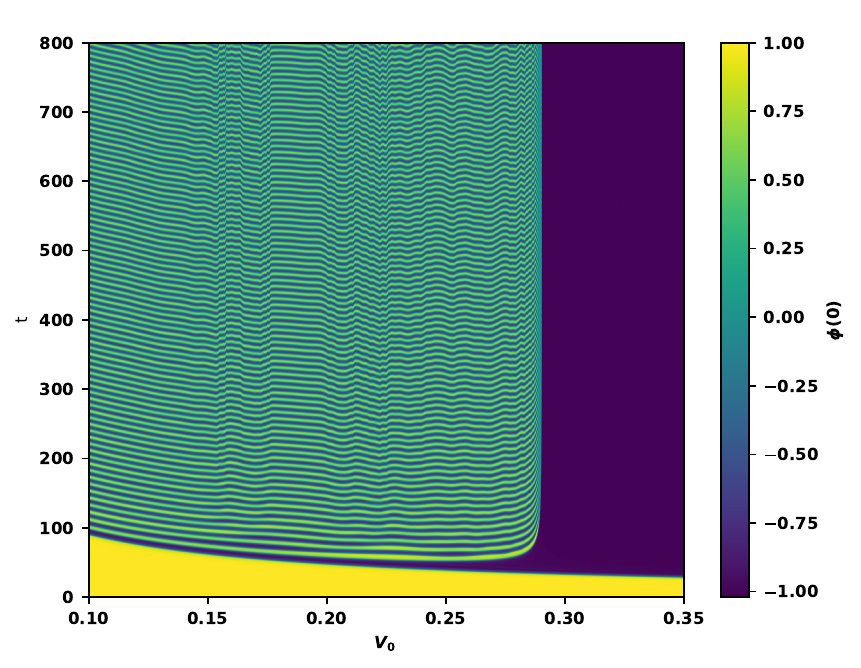}
    \end{subfigure}
    \caption{\textit{Center of the field configuration during antikink–kink (left) and kink-antikink (right) collisions in the $\phi^6$ model as a function of time and the initial velocity of the kink–antikink pair ($v_0$).}}
    \label{figI2:FractalPhi6}
\end{figure}

Another notable property of the kink solutions in this model is that kinks and antikinks are no longer related by the transformation $x \to -x$. As a consequence, several features observed in the $\phi^4$ model no longer hold. For instance, in the $\phi^4$ model, kink/antikink collisions yielded the same results regardless of the ordering of the pair. However, in the $\phi^6$ model, a fractal structure emerges only in antikink/kink collisions. This phenomenon can be better understood by looking at Figure \ref{figI2:FractalPhi6}. In it, it can be clearly be seen that only a fractal structure similar to that found for the $\phi^4$ model can be found only for antikink-kink scattering phenomena. 

\begin{figure}[h!]
    \centering
    \begin{subfigure}{0.49\textwidth}
        \centering
        \includegraphics[width=\linewidth]{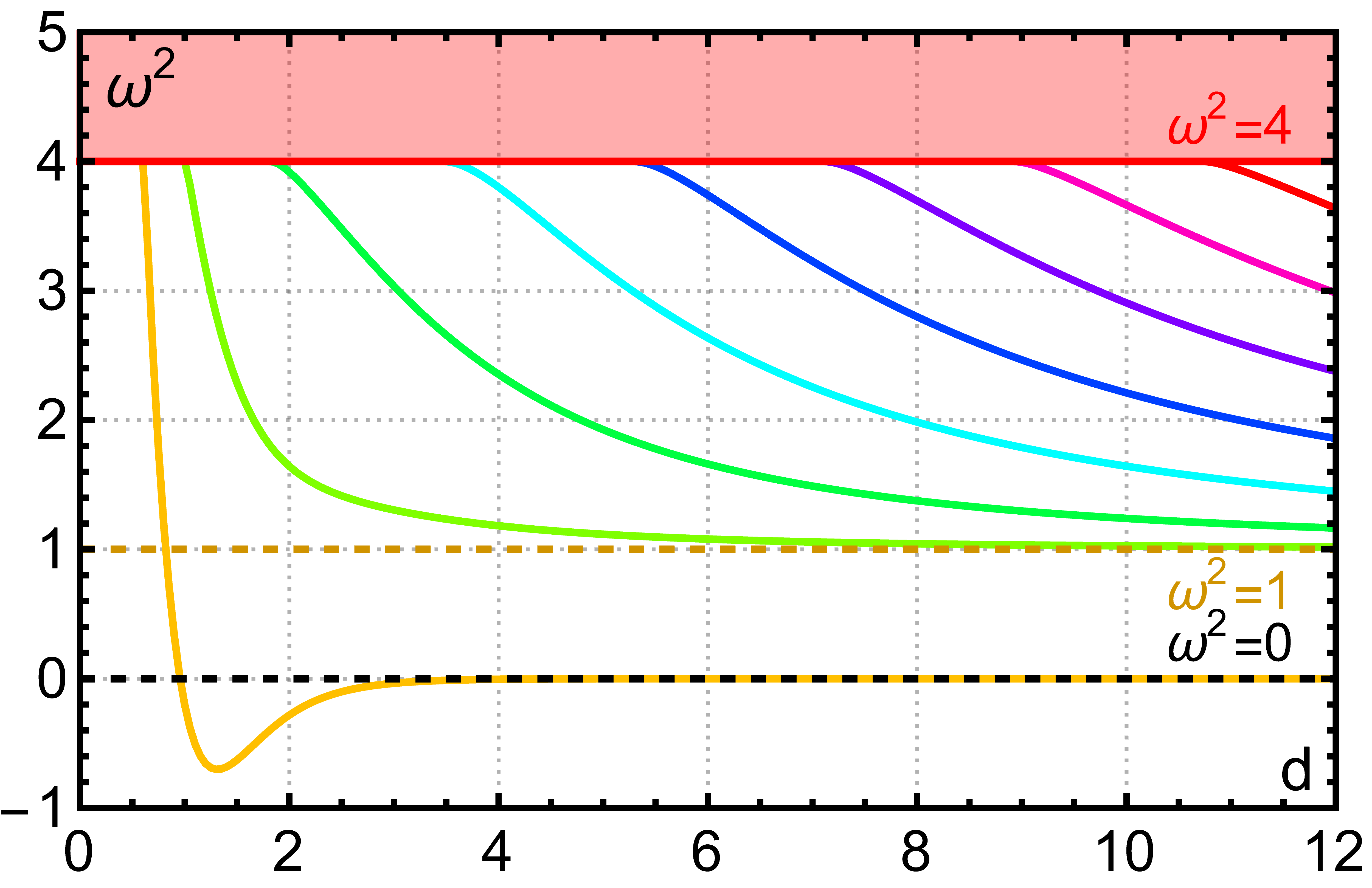}
    \end{subfigure}
    \begin{subfigure}{0.49\textwidth}
        \centering
        \includegraphics[width=\linewidth]{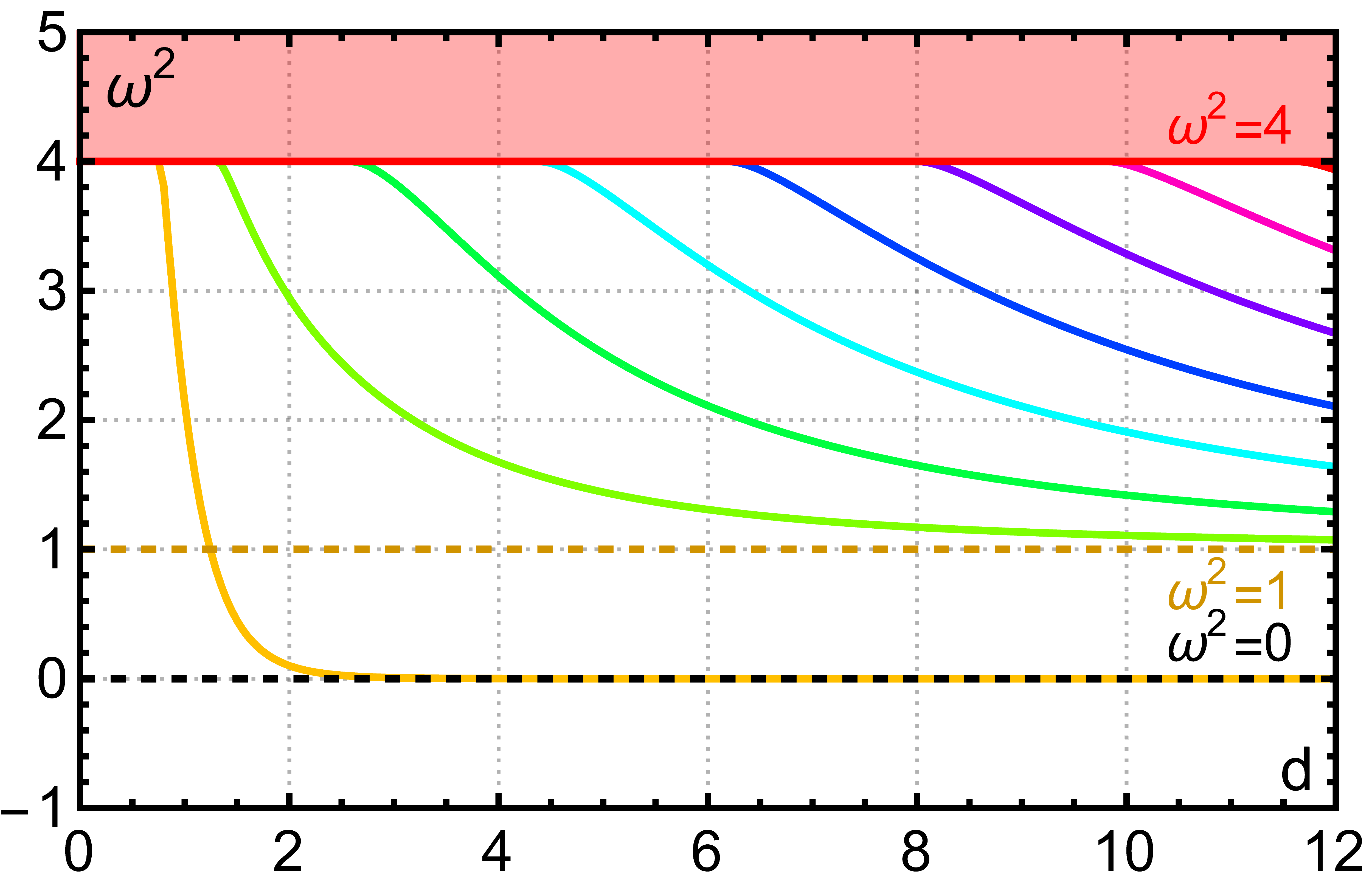}
    \end{subfigure}
    \caption{\textit{Even (left) and odd (right) modes arising from the solution of \eqref{eqI2:SpecProbSpecWallPhi6}.}}
    \label{figI2:SpectralWallsPhi6}
\end{figure}

The origin of this phenomenon lies in the absence of bound modes in kink/antikink configurations. In contrast, bound modes do exist in antikink/kink configurations (see Figure~\ref{figI2:SpectralWallsPhi6}). These modes allow the transfer of energy between internal and translational degrees of freedom, giving rise to the well-known fractal structure observed in the scattering process.

In other words, for the spectral problem
\begin{equation}\label{eqI2:SpecProbSpecWallPhi6}
    \left[-\frac{d^2}{dx^2} + \phi_{-KK}(x) - 4\phi_{-KK}^3(x) + 3\phi_{-KK}^5(x) \right]\eta = \omega^2 \eta,
\end{equation}
where
\begin{equation*}
    \phi_{-KK}(x) = \phi_{(1,0)}(x+d) + \phi_{(0,1)}(x-d),
\end{equation*}
a plethora of bound modes emerges. These modes facilitate energy exchange during collisions, explaining the fractal structure that appears when varying the initial velocity of the antikink/kink pair. The corresponding spectrum, obtained from the resolution of \eqref{eqI2:SpecProbSpecWallPhi6}, is shown in Figure~\ref{figI2:SpectralWallsPhi6} \cite{Dorey2011, Adam2022, Adam2022b}.

In this case, the spectral wall phenomenon is also present, as the bound modes shown in Figure~\ref{figI2:SpectralWallsPhi6}  merge with the continuum when $d$ is decreased.

%\section{Forces between kinks and antikinks}

\section{Wobbling kinks and radiation emission}\label{I2Manton}

%Once we have introduced the simplest models in which kink solutions can be found, let us now address the radiation emission associated to a wobbling kink or wobbler in the $\phi^4$ model. For this purpose, we will introduce the Manton and Merabet method \footnote{A generalization of this method can be found in \cite{NavarroObregon2023}. } \cite{Manton1997}.

Having introduced the simplest models that support kink solutions, we now turn our attention to the radiation emitted by a wobbling kink (or wobbler), in the $\phi^4$ model. In order  to analyze this phenomenon, we will employ the method developed by Manton and Merabet in \cite{Manton1997}. A generalization of this method can be found in \cite{NavarroObregon2023}.

%This method will prove to be of great importance throughout this thesis, since will be also implemented in the study of wobbling kinks in the double $\phi^4$ model in Chapter \ref{Chap1} and, it will be also extended in order to analyze the radiation emission of excited Abelian-Higgs vortices in Chapter \ref{Chap4}. 

 This method will prove to be of great importance throughout this thesis, as it will also be applied to the study of wobbling kinks in the \textit{double} $\phi^4$ model in Chapter \ref{Chap1}, and extended to analyze the radiation emission of excited Abelian-Higgs vortices in Chapter \ref{Chap4}.

%Firstly, let us assume the following expansion of the field
%\begin{equation}\label{eqI2:AnsRad}
 %   \phi(x,t)=\phi_K(x)+a(t)\,\eta_D(x)+\eta(x,t)=\tanh(x)+a(t)\mathrm{sech}(x)\tanh (x)+\eta(x,t),
%\end{equation}
%where we are implicitly assuming that $a(t)$ is a function that dictates the temporal evolution of the internal mode amplitude. Moreover, $\eta(x,t)$ is the spatial and temporal function that dictates the radiation emission of the system. 

Firstly, let us assume the following expansion of the field:
\begin{equation}\label{eqI2:AnsRad}
    \phi(x,t) = \phi_K(x) + a(t)\,\eta_D(x) + \eta(x,t) = \tanh(x) + a(t)\,\mathrm{sech}(x)\tanh(x) + \eta(x,t),
\end{equation}
where we implicitly assume that \( a(t) \) is a function that governs the time evolution of the internal mode amplitude. Moreover, \( \eta(x,t) \) describes the  radiation emitted by the wobbler.

Plugging \eqref{eqI2:AnsRad} into the field equation \eqref{eqI2:FieldEq} leads to
\begin{equation}\label{eqI2:FirstExpansion}
    \Ddot{\eta}-\eta''+(6\phi_K^2-2)\eta+ (\Ddot{a}+3\, a)\eta_D=-6(\eta +\phi_K )\eta_D^2a^2-6(\eta+2\phi_K)\eta a\eta_D-2a^3\eta_D^3-6\phi_K\eta^2-2\eta^3.
\end{equation}

By construction, if $a_0$ is the initial wobbling amplitude, the function $\eta(x,t)$ is assumed to be of order $\mathcal{O}(a_0^2)$. Taking this into account, let us now truncate \eqref{eqI2:FirstExpansion} up to order $\mathcal{O}(a_0^2)$. Then, terms proportional to $\eta^2$, $\eta^3$, $\eta \,a^2$ and $a^3$ will be neglected resulting into
\begin{equation}\label{eqI2:SecondExpansion}
    \Ddot{\eta}-\eta''+(6\phi_K^2-2)\eta+ (\Ddot{a}+3\, a)\eta_D=-6\,\phi_K \,\eta_D^2\,a^2.
\end{equation}

Now, assuming that $\eta$ and $\eta_D$ are orthogonal in the following way
\begin{equation*}
    \int_{-\infty}^{\infty}\eta(x,t)\eta_D(x) dx=0,
\end{equation*}
if equation \eqref{eqI2:SecondExpansion} is projected over the shape mode $\eta_D$, the next second order ordinary differential equation is obtained
\begin{equation}\label{eq2}
    \Ddot{a}(t)+3a(t)+\frac{9\pi}{16}a(t)^2=0.
\end{equation}

Thus, at linear order 
\begin{equation}\label{eqI2:vibAmp}
    a(t)\approx a_0 \cos(\sqrt{3}t).
\end{equation}

Plugging \eqref{eqI2:vibAmp} and \eqref{eq2} into \eqref{eqI2:SecondExpansion} we find 
\begin{equation}\label{eqI2:secOrEq}
    \Ddot{\eta} - \eta'' + (6\phi_K^2 - 2)\eta = 3a_0^2\left(\frac{3\pi}{32}\eta_D - \phi_K\,\eta_D^2\right)\left(\cos(2\sqrt{3}\,t) + 1\right).
\end{equation}

In \eqref{eqI2:secOrEq} the time-independent terms located on the right side of the equation will not be taken into account, since the response of $\eta$ to those terms would be time independent, and we are only interested in terms that can carry away energy. Consequently, the equation to be solved is 
\begin{equation}\label{eqI2:Fulleq}
     \Ddot{\eta}-\eta''+(6\phi_K^2-2)\eta=3a_0^2\left(\frac{3\pi}{32}\eta_D-\,\phi_K \,\eta_D^2\,\right)\cos(2\sqrt{3}t)=f(x) \cos(2\sqrt{3}t),
\end{equation}
where
\begin{equation*}
    f(x)=a_0^2\,\mathrm{sech}(x)\tanh(x)\left(\frac{9\pi}{32}-3\mathrm{sech}(x)\tanh^2(x)\right).
\end{equation*}

Setting $\eta(x,t)=e^{i2\sqrt{3}t}\eta(x)$, we obtain 
\begin{equation}\label{eqI2:secOrEq2}
     -\eta''+(6\phi_K^2-2-\omega^2)\eta=f(x),
\end{equation}
where $\omega=2\sqrt{3}$.

The solutions to the homogeneous part of \eqref{eqI2:secOrEq2} are functions \eqref{eqI2:CCradiation} where now $q=2\sqrt{2}$.

A particular solution to \eqref{eqI2:secOrEq2} can be written as 
\begin{equation}\label{eqI2:partsol}
    \eta(x)=-\frac{1}{W_q}\left(\eta_{-q}(x)\int^x_{-\infty} f(y)\eta_q(y)dy +\eta_{q}(x)\int^\infty_{x} f(y)\eta_{-q}(y)dy \right),
\end{equation}
where $W_q=\eta_q\,\eta_{-q}'-\eta_q'\,\eta_{-q}=-2iq\,(q^2+1)(q^2+4)$.

Asymptotically, we can eliminate the second part of \eqref{eqI2:partsol}, allowing us to rewrite this solution as
\begin{equation}\label{eqI2:imRad}
\eta(x,t)\xrightarrow{x\rightarrow\infty}\frac{\int^{\infty}_{-\infty}\eta_q(y)f(y) dy}{2 i q(2-q^2-3iq)}\,e^{i 2\sqrt{3}t-i q x}.
\end{equation}

Since the inhomogeneous term in \eqref{eqI2:Fulleq} is proportional to $\cos(2\sqrt{3}t)$, then, we must take the real part in \eqref{eqI2:imRad}, which leads to 
\begin{equation} \label{eqI2:RadPhi4}
\eta(x,t)\xrightarrow{x\rightarrow\infty}\frac{\pi q(q^2-2)}{32 \sinh(\pi q/2)}\sqrt{\frac{q^2+4}{q^2+1}} a_0^2 \cos(2\sqrt{3 }t-q x+ \delta(q)).
\end{equation}

Furthermore, we can conclude that the radiation amplitude emitted by the wobbler is
\begin{equation*}
    R= \frac{\pi q(q^2-2)}{32 \sinh(\pi q/2)}\sqrt{\frac{q^2+4}{q^2+1}} a_0^2=0.0453\, a_0^2,
\end{equation*}
where $a_0$ is the initial wobbling amplitude. 

%
%As a matter of fact, as the energy stored in the internal mode vibrations is radiated away, the wobbling amplitude  $a_0$ must decay. In order to analytically compute this decay, let us consider that the internal mode behaves as an harmonic oscillator in each spatial point, which translates into that the associated energy density is  energy
%\begin{equation}
 %   \mathcal{E}=\frac{1}{2}a_0^2 \omega^2 \eta_D^2,
%\end{equation}

%Thus, the energy associated to these vibrations is just 
%\begin{equation}\label{eqI2:Energy}
 %   E=a_0^2.
%\end{equation}

%On the other hand, the radiation emitted by the system can be written as 
%\begin{equation}\label{eqI2:Radiation}
 %   \langle R \rangle=\frac{d\, E}{d\,t}=-R^2 (2\omega) \,q=-0.0201 \, a_0^4.
%\end{equation}

%Putting together \eqref{eqI2:Energy} and \eqref{eqI2:Radiation} the next second order ordinary differential equation is obtained
%\begin{equation*}
 %   \frac{d a_0^2}{d t}=-0.0201 \,a_0^4,
%\end{equation*}
%whose solution is 
%\begin{equation}\label{eqI2:MantonLaw}
 %   a_0(t)=\frac{a_0(0)}{\sqrt{1+0.0201 \,a_0(0)^2\, t}}.
%\end{equation}
%

As a matter of fact, as the energy stored in the internal mode vibrations is radiated away, the wobbling amplitude $a_0$ must decay. In order to analytically compute this decay, let us consider that the internal mode behaves as a harmonic oscillator at each spatial point, which implies that the associated energy density is given by
\begin{equation}
    \mathcal{E} = \frac{1}{2} a_0^2 \omega^2 \eta_D^2\,.
\end{equation}

Thus, the energy associated to these vibrations is just 
\begin{equation}\label{eqI2:Energy}
    E = a_0^2.
\end{equation}

On the other hand, the radiation emitted by the system can be written as 
\begin{equation}\label{eqI2:Radiation}
    \left\langle R \right\rangle = \frac{d E}{dt} = -R^2 (2\omega)\, q = -0.0201\, a_0^4.
\end{equation}

Putting together \eqref{eqI2:Energy} and \eqref{eqI2:Radiation}, the following second-order ordinary differential equation is obtained:
\begin{equation*}
    \frac{d a_0^2}{dt} = -0.0201\, a_0^4,
\end{equation*}
whose solution is
\begin{equation}\label{eqI2:MantonLaw}
    a_0(t) = \frac{a_0(0)}{\sqrt{1 + 0.0201\, a_0(0)^2\, t}}.
\end{equation}

%In Figure \ref{figI2:MantonLaw} a comparison between the numerical simulation and the theoretical decay law can be appreciated.

%All explained so far can be extended to study the radiation emission of wobbler in multicomponent field theories or complex topological solitons, as vortices. These two scenearios will be addressed in Chapters \ref{Chap1} and \ref{Chap4}. Another perturbative approach to analyze this scenerio can be found in \cite{Barashenkov2009, Barashenkov2009b, AlonsoIzquierdo2023c} This method allows to analyze wobblers up to fourth order in perturbation theory and the results that can be extracted match with the ones here displayed, but for more complex theories extracting analytical information is much more complicated. Both methods have also been applied when studying wobblers in the MSTB model. 

\begin{figure}[h!]
    \centering
    \includegraphics[width=0.6\linewidth]{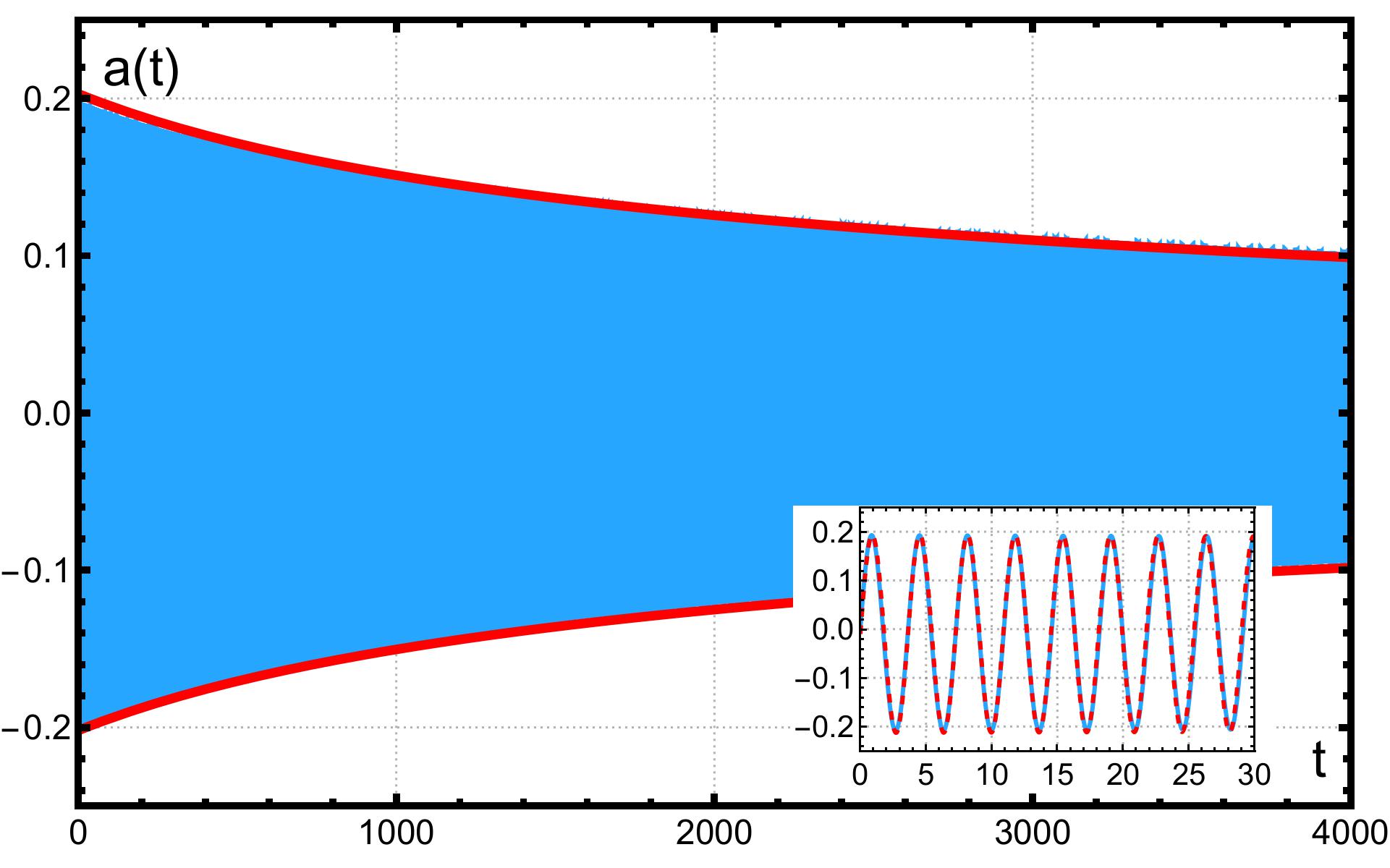}
    \caption{\textit{Numerical evolution of the amplitude of a wobbling kink in the $\phi^4$ model. The blue lines correspond with the results extracted from a numerical simulation and the red ones correspond with the analytical decay law \eqref{eqI2:MantonLaw}. }}
    \label{figI2:MantonLaw}
\end{figure}

In Figure \ref{figI2:MantonLaw}, a comparison between the numerical simulation and the theoretical decay law can be appreciated.

All the results presented so far can be extended to study the radiation emission of wobblers in multicomponent field theories or in complex topological solitons, such as vortices. These two scenarios will be addressed in Chapters \ref{Chap1} and \ref{Chap4}. 

Another perturbative approach to analyze this scenario can be found in \cite{Barashenkov2009, Barashenkov2009b, AlonsoIzquierdo2023c}. This method allows for the analysis of wobblers up to fourth order in perturbation theory. The results obtained match those presented here; however, for more complex theories, extracting analytical information becomes significantly more challenging. Both methods have also been applied in the study of wobblers in the MSTB model \cite{AlonsoIzquierdo2023c}.

The general steps to apply the method used here for studying excited solitons can be summarized as follows:
\begin{enumerate}
    \item First, we construct an expansion of the field consisting of the sum of the static soliton solution, an internal mode multiplied by a time-dependent function (the internal mode amplitude), and an initially unknown function that depends on both spatial and temporal coordinates that will be called $\eta(x,t)$. This function is assumed to be orthogonal to the shape modes.

    \item We plug this ansatz into the field equations and neglect higher order terms. \label{Step2}

    \item The resulting equations are projected onto the internal modes, yielding a set of second-order ordinary differential equations. 

    \item These equations are solved under the assumption that the amplitudes of the initially activated modes can be written as
    \begin{equation*}
        a_i(t) \approx a_{i,0} \cos{(\omega_i \, t)}.
    \end{equation*}

    \item The approximate solutions to the aforementioned system of ordinary differential equations are substituted into the truncated expansion from Step~\ref{Step2}. \label{Step5}

    \item From the resulting equations for $\eta(x,t)$, we neglect the time-independent terms coming from the inhomogeneous part. Then, the different vibrational terms are solved separately using an ansatz of the type
    \begin{equation*}
        \eta(x,t)=\sum \eta_{\Tilde{\omega}_i}(x)e^{i\Tilde{\omega}_i t}.
    \end{equation*}

    \item The resulting equations are solved asymptotically in order to find the radiation terms, such that
    \begin{equation*}
        \eta_{\Tilde{\omega}_i}(x,t) \xrightarrow{x\rightarrow\infty} R_i \,e^{i \Tilde{\omega}_i t - i q_i x}.
    \end{equation*}

    \item Frequencies $\Tilde{\omega}_i$ naturally arise from Step~\ref{Step5}. In some cases, these terms do not correspond to radiative modes because $q_i \notin \mathbb{R}$.

    \item Finally, an analytical decay law can be derived if the internal mode vibrations are assumed to behave as a harmonic oscillator and the energy radiated away in each period is considered. This leads to a first-order differential equation governing the decay of the wobbling amplitude.
\end{enumerate}

\section{Kinks in two component scalar field theories} \label{Intro2.6}
In this section, we study two-component scalar field theories defined by the Lagrangian density
\begin{equation}\label{eqI2:LagDensTwoFields}
    \mathcal{L} = \frac{1}{2} (\partial_{t} \phi)^2 - \frac{1}{2} (\partial_{x} \phi)^2 + \frac{1}{2} (\partial_{t} \psi)^2 - \frac{1}{2} (\partial_{x} \psi)^2 - U(\phi,\psi),
\end{equation}
where $U(\phi,\psi)$ is a potential function that possesses at least two degenerate vacua.

The analytical methods discussed so far can be extended to the study of multicomponent scalar field theories. However, in this case, finding kink solutions is more involved than in single-component scalar field theories. The main difficulty lies in constructing a superpotential function $W(\phi,\psi)$ that fulfills the relation 
\begin{equation}
    U(\phi,\psi) = \frac{1}{2} \left( \frac{\partial W}{\partial \phi} \right)^2 + \frac{1}{2} \left( \frac{\partial W}{\partial \psi} \right)^2.
\end{equation}

If such a function can be constructed, then the BPS equations can be written down and take the form
\begin{equation}
    \frac{d \phi}{d x} = \pm \frac{\partial W}{\partial \phi}, \qquad \frac{d \psi}{d x} = \pm \frac{\partial W}{\partial \psi},
\end{equation}
which greatly simplifies the task of finding kink solutions.

Another method for obtaining these solutions is the \textit{trial orbit method} introduced by Rajaraman \cite{Rajaraman1982}, which  can be summarized as follows:
\begin{enumerate}
    \item A family of orbits $g(\phi,\psi) = 0$ is proposed, depending on a set of undetermined coefficients.
    \item This relation is plugged into the following identity:
    \begin{equation}\label{eqI2:Orbits}
        \left( \frac{\partial g}{\partial \phi} \right)^2 \left( \int \frac{\partial U(\phi,\psi)}{\partial \phi} d\phi + A_1 \right) = 
        \left( \frac{\partial g}{\partial \psi} \right)^2 \left( \int \frac{\partial U(\phi,\psi)}{\partial \psi} d\psi + A_2 \right).
    \end{equation}
    \item The undetermined coefficients are fixed by matching both sides of \eqref{eqI2:Orbits}, giving rise to the kink orbits in the $(\phi,\psi)$-plane.
\end{enumerate}

An example of the application of this method can be found in \cite{Rajaraman1982}.

In some cases, depending on the form of the potential $U(\phi,\psi)$, it is  possible to construct the superpotential explicitly. A classification of such potentials is provided in Reference \cite{AlonsoIzquierdo2001}.

Regarding the internal mode structure associated with these new kink solutions, the operator governing small perturbations now takes the form
\begin{equation}
    \mathcal{H} =
    \begin{pmatrix}
        -\frac{d^2}{d x^2} + \frac{\partial^2 U(\phi,\psi)}{\partial \phi^2} & 
        \frac{\partial^2 U(\phi,\psi)}{\partial \phi \partial \psi} \\
        \frac{\partial^2 U(\phi,\psi)}{\partial \phi \partial \psi} & 
        -\frac{d^2}{d x^2} + \frac{\partial^2 U(\phi,\psi)}{\partial \psi^2}
    \end{pmatrix},
\end{equation}
which, in turn, implies that the associated spectral problem consists of two coupled Schrödinger-like differential equations.

In the next section, we will study two examples of kink solutions in two-component field theories. In the first case, the MSTB model, we will derive the corresponding BPS equations and obtain the full internal mode structure. In the second case, we will briefly discuss the kink solutions arising in the model and how to obtain them numerically the ones that cannot be found analytically.

\subsection{Two examples: The MSTB and double \texorpdfstring{$\phi^4$}{phi4} models}\label{Intro1.7}

\subsubsection{The MSTB model}

This model is governed by the Lagrangian density \eqref{eqI2:LagDensTwoFields}, with the potential 
\begin{equation}\label{eqI2:PotMSTB}
    U(\phi,\psi) = \frac{1}{2}(\phi^2 + \psi^2 - 1)^2 + \frac{\sigma^2}{2} \psi^2,
\end{equation}
where $\sigma$ is a real positive coupling constant.

The presence of the last term in \eqref{eqI2:PotMSTB} indicates that this model represents a deformation of the $O(2)$ linear sigma model.

The corresponding field equations are
\begin{eqnarray}
    \partial_{tt} \phi - \partial_{xx} \phi \!\!\!\! &=&  \!\!\!\! 2\phi(1 - \phi^2 - \psi^2),\\
    \partial_{tt} \psi - \partial_{xx} \psi  \!\!\!\!  &=&  \!\!\!\!  2\psi\left(1 - \phi^2 - \psi^2 - \frac{\sigma^2}{2}\right).
\end{eqnarray}

Consequently, the set of vacua in this model is given by
\begin{equation}
    \mathcal{V} = \left\{ A_- =  \left( 
\begin{array}{c}
-1 \\ %[0.5ex]
0
\end{array}
\right) ,\,\, A_+ = \left( 
\begin{array}{c}
 1 \\ %[0.5ex]
0
\end{array}
\right) \right\}.
\end{equation}

Additionally, if $\sigma^2 \in (0, 2]$, the potential \eqref{eqI2:PotMSTB} also exhibits a local maximum at $(0,0)$ and two saddle points at $(0, \pm\sqrt{1 - \sigma^2/2})$. For $\sigma \in [2, \infty)$, the saddle points disappear, and the local maximum at the origin becomes a saddle point.

For reasons that will become clear later, it is convenient to use elliptic coordinates. Specifically,
\begin{equation}
    \phi = \frac{u \, v}{\sigma}, \qquad \psi = \pm\frac{1}{\sigma} \sqrt{(u^2 - \sigma^2)(\sigma^2 - v^2)},
\end{equation}
where $u \in [\sigma, \infty)$ and $v\in[-\sigma,\sigma]$.

Rewriting the energy in terms of these new coordinates gives as a result
\begin{equation}
    E = \int_{-\infty}^{\infty} \left[
    \frac{1}{2} \frac{u^2 - v^2}{u^2 - \sigma^2} \left( \frac{d u}{dx} \right)^2 + 
    \frac{1}{2} \frac{u^2 - v^2}{\sigma^2 - v^2} \left( \frac{d v}{dx} \right)^2 + 
    \frac{(u^2 - 1)^2 (u^2 - \sigma^2) + (1 - v^2)^2 (\sigma^2 - v^2)}{2(u^2 - v^2)}
    \right] dx,
\end{equation}
which, in turn, can be rearranged as
{\small
\begin{equation}
    E = \int_{-\infty}^{\infty} \left[
    \frac{1}{2} \frac{u^2 - v^2}{u^2 - \sigma^2} \left( \frac{d u}{d x} - (-1)^a \frac{u^2 - \sigma^2}{u^2 - v^2} (1 - u^2) \right)^2 + 
    \frac{1}{2} \frac{u^2 - v^2}{\sigma^2 - v^2} \left( \frac{d v}{d x} - (-1)^b \frac{\sigma^2 - v^2}{u^2 - v^2} (1 - v^2) \right)^2 
    \right] dx + |T|,
\end{equation}
}
where
\begin{equation}
    |T| = \int_{-\infty}^{\infty} \left| \frac{d u}{dx} (1 - u^2) \right| dx + \int_{-\infty}^{\infty} \left| \frac{d v}{dx} (1 - v^2) \right| dx.
\end{equation}

The BPS bound is saturated when the following BPS equations are satisfied:
\begin{eqnarray}
    \frac{d u}{d x} \!\!\!\!&=& \!\!\!\!(-1)^a \frac{u^2 - \sigma^2}{u^2 - v^2} (1 - u^2), \label{eqI2:BPSu} \\
    \frac{d v}{d x} \!\!\!\!&=& \!\!\!\!(-1)^b \frac{\sigma^2 - v^2}{u^2 - v^2} (1 - v^2), \label{eqI2:BPSv}
\end{eqnarray}
where $a, b \in \{0, 1\}$.

With the BPS equations at hand, kink solutions can now be obtained more straightforwardly. These solutions can be classified into two categories: \textit{topological} and \textit{non-topological}. Topological kinks interpolate between the vacua $A_+$ and $A_-$. Despite the fact that they connect distinct vacua, we will see that under certain conditions, they can still be unstable.

\begin{itemize}
    \item \textbf{One null component topological kinks:}
    In this case, two distinct regimes must be considered:
    \begin{itemize}
        \item \textbf{Regime 1 ($\boldsymbol{\sigma > 1}$):} The vacuum points correspond to $u_{\pm} = \sigma$, $v = \pm 1$. It is straightforward to verify that setting $u(x) = \sigma$ satisfies equation \eqref{eqI2:BPSu} identically. Integrating equation \eqref{eqI2:BPSv} for $v \in (-1,1)$ yields $v(x) = (-1)^a \tanh(x - x_0)$. Thus, the kink solution reads
        \begin{equation}\label{eqI2: K1}
            \phi_{K1} = \left((-1)^a \tanh(x - x_0),\, 0\right)^T.
        \end{equation}

        Consequently, the BPS bound for this regime is \begin{equation}
        |T_1| = \int_{-1}^{1} (1 - v^2)\,dv = \frac{4}{3}.%, \qquad 
      %  |T_2| = \int_{-\sigma}^{\sigma} (1 - u^2)\,du = 2\sigma\left(1 - \frac{\sigma^2}{3}\right).
    \end{equation}
        
        \item \textbf{Regime 2 ($\boldsymbol{0 < \sigma < 1}$):} Since $u \in [\sigma, \infty)$ and $v \in [-\sigma, \sigma]$, the vacua are now located at $u_\pm = 1$ and $v = \pm \sigma$. The solution can be constructed following the next steps: First, we fix $v = -\sigma$ and integrate \eqref{eqI2:BPSu} over the interval $\sigma < u < 1$; then, we fix $u = \sigma$ and integrate \eqref{eqI2:BPSv} over $-\sigma < v < \sigma$; finally, we fix $v = \sigma$ and integrate \eqref{eqI2:BPSu} again over $\sigma < u < 1$. As in Regime 1, the resulting kink is given by \eqref{eqI2: K1}. Here, the BPS bound is given by
         \begin{equation}
        %|T_1| = \int_{-1}^{1} (1 - v^2)\,dv = \frac{4}{3}.%, \qquad 
        |T_2| = \int_{-\sigma}^{\sigma} (1 - u^2)\,du = 2\sigma\left(1 - \frac{\sigma^2}{3}\right).
    \end{equation}
        
    \end{itemize}

   % The BPS bounds for each regime can be calculated as
    %\begin{equation}
     %   |T_1| = \int_{-1}^{1} (1 - v^2)\,dv = \frac{4}{3}, \qquad 
      %  |T_2| = \int_{-\sigma}^{\sigma} (1 - u^2)\,du = 2\sigma\left(1 - \frac{\sigma^2}{3}\right).
    %\end{equation}
    In both regimes, the energy associated with the kink solution \eqref{eqI2: K1} is $E = \frac{4}{3}$. This implies that in Regime 2, the kink is not a BPS configuration, since its energy is greater than the energy of the corresponding BPS bound. Therefore, it is unstable and the field configuration will decay into a field profile that does saturate the BPS bound.

    The stability analysis of this kink solution will be addressed in Chapter~\ref{Chap2}. A graphical representation of this solution is shown in Figure~\ref{figI2:K1PlotMSTB}.
    \begin{figure}[h!]
        \centering
        \includegraphics[width=0.6\linewidth]{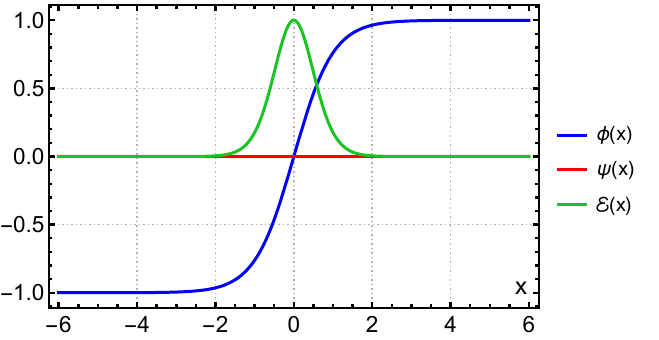}
        \caption{\textit{Plot of both field components and the energy density of $\phi_{K1}(x)$.}}
        \label{figI2:K1PlotMSTB}
    \end{figure}

    \item \textbf{Non-null component topological kinks:}
    In Regime 2, by fixing $u = 1$ and integrating  \eqref{eqI2:BPSv} over the interval $-\sigma < v < \sigma$, we find $$v(x) = (-1)^b \sigma \tanh(\sigma(x - x_0)),$$ which yields the kink
    \begin{equation}\label{eqI2:MSTBk2}
        \phi_{K2} = \left(q  \tanh(\sigma(x - x_0)),\, \lambda \sqrt{1 - \sigma^2}\,\mathrm{sech}(\sigma(x - x_0))\right)^\intercal,
    \end{equation}
where $q,\lambda=\pm1$.
    As is evident from \eqref{eqI2:MSTBk2}, this kink only exists in Regime 2.
    Furthermore, the energy of this configuration exactly matches the BPS bound $|T_2|$, implying that this solution is stable. Figure~\ref{figI2:K2PlotMSTB} shows a representative plot of this solution.
    
    \begin{figure}[h!]
        \centering
        \includegraphics[width=0.6\linewidth]{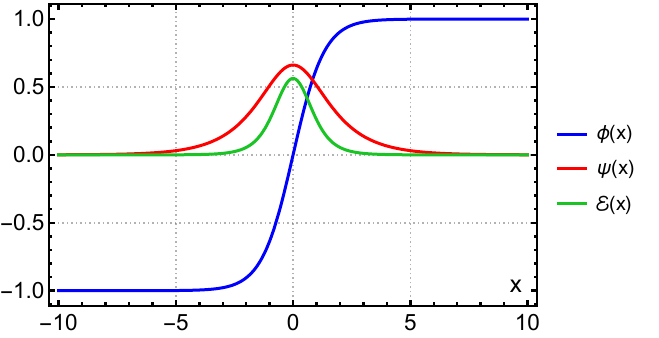}
        \caption{\textit{Plot of both field components and the energy density of $\phi_{K2}(x)$ for $\sigma = 0.75$.}}
        \label{figI2:K2PlotMSTB}
    \end{figure}

    \item \textbf{Non-topological kinks:}
   In Regime 2, it can be shown that there exist two families of non-topological kink solutions, given by
    \begin{equation}\label{eqI1:NonTopKink}
        \phi_{K3} = \left(
        (-1)^a \frac{\sigma_- \cosh(\sigma_+ x_+) - \sigma_+ \cosh(\sigma_- x_-)}{\sigma_- \cosh(\sigma_+ x_+) + \sigma_+ \cosh(\sigma_- x_-)},\,
        \frac{2\sigma_+ \sigma_- \sinh(x - x_0)}{\sigma_- \cosh(\sigma_+ x_+) + \sigma_+ \cosh(\sigma_- x_-)}
        \right)^\intercal,
    \end{equation}
    where $\sigma_\pm = 1  \pm \sigma$, $x_\pm = x - x_0 - \gamma \sigma(\sigma \mp 1)$, and $\gamma \in \mathbb{R}$.

    The energy associated with this solution can be written as the sum of the energies of the previously described topological kinks:
    \begin{equation*}
        E = E_{K1} + E_{K2} = \frac{4}{3} + 2\sigma\left(1 - \frac{\sigma^2}{3}\right).
    \end{equation*}

    \begin{figure}[h!]
        \centering
        \begin{subfigure}{0.49\textwidth}
            \centering
            \includegraphics[width=\linewidth]{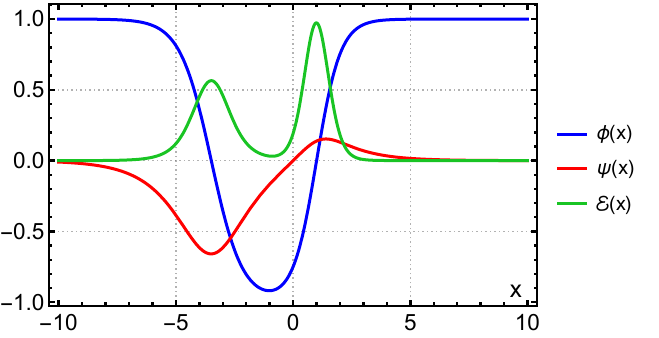}
        \end{subfigure}
        \begin{subfigure}{0.49\textwidth}
            \centering
            \includegraphics[width=\linewidth]{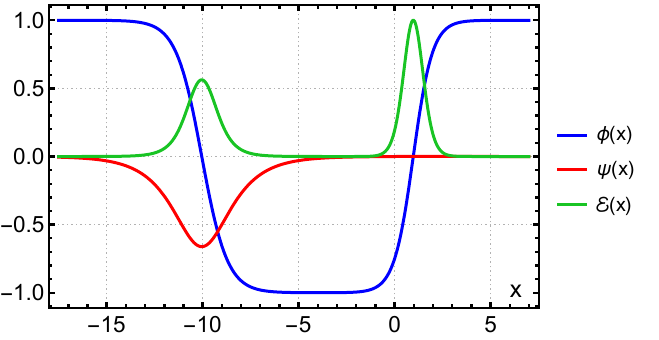}
        \end{subfigure}
        \caption{\textit{Plots of both field components and the energy density of $\phi_{K3}(x)$ for $\gamma = 5$ (left) and $\gamma = 20$ (right), with $\sigma = 0.75$.}}
        \label{figI2:K3PlotMSTB}
    \end{figure}

    Moreover, as shown in Figure~\ref{figI2:K3PlotMSTB}, for sufficiently large $\gamma$, the kink profile of $\phi_{K3}$ resembles a superposition of the kinks $\phi_{K1}$ and $\phi_{K2}$.

\end{itemize}

\begin{figure}[h!]
    \centering
    \includegraphics[width=0.5\linewidth]{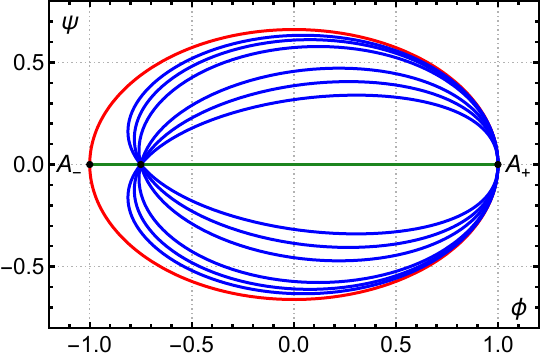}
    \caption{\textit{Kink orbits in the $(\phi, \psi)$ plane for $\phi_{K1}$ (green), $\phi_{K2}$ (red), and $\phi_{K3}$ (blue) for $\sigma=0.75$.}}
    \label{figI2:MSTBOrbits}
\end{figure}

Figure~\ref{figI2:MSTBOrbits} displays the kink orbits for all the solutions arising in this model. As can be seen, the kink $\phi_{K3}(x)$ does not interpolate between distinct vacua, which implies that it is unstable. These kinks are, in fact, sphalerons: unstable solutions that asymptotically return to the same vacuum \cite{Manton2004}. Finally, it is worth noting that in Regime 2, the one null component topological kink decays into a non-null component topological kink when perturbed.

\subsubsection{The double $\phi^4$ model}

Now, the energy density is governed by the potential
\begin{equation}\label{eqI2:DoblePhi4}
    U(\phi,\psi)=\frac{1}{2}(\phi^2-1)^2+\frac{1}{2}(\psi^2-1)^2+\kappa \,\phi^2 \psi^2-\frac{1}{2},
\end{equation}
which is completely symmetric  in the interchange of $\phi$ for $\psi$. As can be seen, this theory consist on two separated copies of the $\phi^4$ model coupled by means of the term $\kappa\,\phi^2\psi^2$. The field equations that govern the evolution of both fields,  which can be  calculated from the Lagrangian density \eqref{eqI2:LagDensTwoFields}, are
\begin{eqnarray}
    \partial_{tt}\phi-\partial_{xx}\phi+2\phi(\phi^2-1+\kappa\,\psi^2)\!\!\!\!&=&\!\!\!\!0,\label{eqI2:FieldEqDouble1}\\
    \partial_{tt}\psi-\partial_{xx}\psi+2\psi(\psi^2-1+\kappa\,\phi^2)\!\!\!\!&=&\!\!\!\!0\label{eqI2:FieldEqDouble2},
\end{eqnarray}
where $\kappa>0$.

Depending on the value of $\kappa$ there exist two different regimes:
\begin{itemize}
    \item \textbf{Regime 1} \textbf{($0<\kappa<1$):}  The set of vacua is 
\begin{equation}
\mathcal{V}_{\kappa<1}=\left\{ 
% \begin{pmatrix}
% \phi  \\ 
%\psi
% \end{pmatrix}
% =
\frac1{\sqrt{1+\kappa}} 
 \left( 
\begin{array}{c}
(-1)^a  \\ %[0.5ex]
(-1)^b
\end{array}
\right) 
, \  a,b=0,1\right\}.
\label{eqI2:vacua4<1}
\end{equation}

Indeed, for this set to be strictly vacua, it is necessary to add to the potential \eqref{eqI2:DoblePhi4} the term $\frac{1-\kappa}{2(1+\kappa)}$, so the potential would be $U^*(\phi,\psi)=U(\phi,\psi)+\frac{1-\kappa}{2(1+\kappa)}$.

  \item \textbf{Regime 2} \textbf{($\kappa>1$):} The four vacua solutions are slightly different than in the previous case \eqref{eqI2:vacua4<1}:
  \begin{equation}
\mathcal{V}_{\kappa>1}=
\left\{ 
\left( 
\begin{array}{c}
(-1)^a \\
0 
\end{array}
\right) 
, \ \ 
\left( 
\begin{array}{c}
0 \\
(-1)^b
\end{array}
\right) 
, \quad  a,b=0,1\right\}.
\label{eqI2:vacua4>1}
\end{equation}

\end{itemize}

As the set of vacua changes, the possible kink solutions must also change. As in both regimes there are four different vacua, there exist at least $12$ kink/antinkink topological kink solutions. For this specific potential, the BPS equations cannot be written down. Nevertheless, some of these kink can be obtained exploring the field equations \eqref{eqI2:FieldEqDouble1}-\eqref{eqI2:FieldEqDouble2}:

\begin{itemize}
    \item  \textbf{Regime 1} \textbf{($0<\kappa<1$):} 
    The kink structure for  $\kappa<1$ has already been studied in \cite{Halavanau2012}, where the authors investigate the kink scattering characteristics in this model. 
%As for this other regime, the vacua structure changes, so the kink structure also changes,
The discrete symmetries of the Lagrangian density  can be used to find that the equation that governs both components of the kink is 
\begin{equation}
     -\partial_{xx} \phi^\kappa_K + 2 \phi^\kappa_K \left( (1+\kappa)(\phi_K^\kappa)^2 -1 \right)=0,
\end{equation}
whose solutions, except for an irrelevant translation in the $x$ coordinate, are $ \phi_K^\kappa(x)= \pm  \frac{ \phi_K(x)}{\sqrt{1+\kappa}}$. Here, $\phi_K(x)$ satisfies the well-known field equation of the $\phi^4$ model 
\begin{equation}\label{fi4}
     -\partial_{xx} \phi_K + 2 \phi_K (\phi_K^2 -1 )=0,
\end{equation}
whose solutions are $ \phi_K(x)=\pm \tanh (x)$.
%In what follows, without loss of generality, we will assume that each kink is centered at the origin of the axis, in other words, $x_0=0$.
Therefore, there are two pairs of different kinks, in such a way that each pair joins two non-adjacent vacua of those we have in \eqref{eqI2:vacua4<1}, one of them in one direction and the other in the opposite direction, and these four kinks $K(x)$ are of the form
\begin{equation}
    K^{(a,b)} (x)=\frac{\tanh x }{\sqrt{1+\kappa}} \left(
    \begin{array}{c}
    (-1)^a\\ 
    (-1)^b 
    \end{array}
    \right), \quad a,b=0,1.
\end{equation}

There exist other kinks that interpolate between non-consecutive vacua in this regime. However, they have to be obtained numerically. We will call them $K_{2,\kappa<1}(x)$. In Figure \ref{figI2:kinkNumK<1} a plot for one of these kinks can be found. 
\begin{figure}[h!]
    \centering
    \begin{subfigure}{0.6\textwidth}
        \centering
        \includegraphics[width=\linewidth]{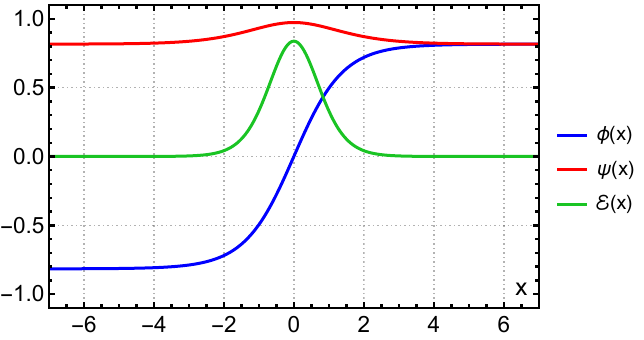}
    \end{subfigure}
    %\hfill
    \caption{ \textit{Plot for one of the kink solutions $K_{2,\kappa<1}(x)$ and its corresponding energy density for $\kappa=0.5$.}}
    \label{figI2:kinkNumK<1}
\end{figure}
Additionally, in Figure \ref{figI2:kinkOrbitk<1} we can find the plot for all kink orbits arising in this regime  over the  $(\phi,\psi)$-plane.
\begin{figure}[h!]
    \centering
    \begin{subfigure}{0.43\textwidth}
        \centering
        \includegraphics[width=\linewidth]{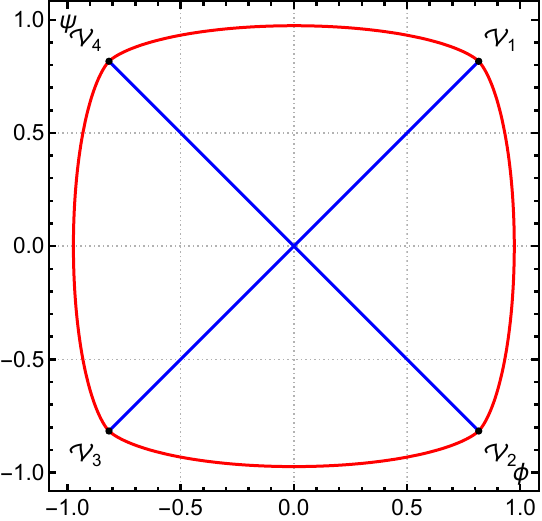}
    \end{subfigure}
    %\hfill
    \caption{ \textit{Kink orbits for kinks $K^{(a,b)}(x)$ (blue) and $K_{2,\kappa<1}(x)$ (red) for $\kappa=0.5$. }}
    \label{figI2:kinkOrbitk<1}
\end{figure}

    \item \textbf{Regime 2} \textbf{($\kappa>1$):}
    It is now possible to find the kink solutions  by canceling one of its components since, by doing this, from the field equations \eqref{eqI2:FieldEqDouble1}--\eqref{eqI2:FieldEqDouble2} we can deduce that the evolution of the non-zero component of the field is again described by the $\phi^4$ field equation,
%\begin{equation*}
%     -\partial_{xx} \phi_K + 2 \phi_K (\phi_K^2 -1 )=0,
%\end{equation*}
whose solutions have been already given in Section \ref{Intro2.4.1}.
Taking into account all the previous comments, we can finally infer that for $\kappa>1$ the kinks  in this case are given by the two pairs
\begin{eqnarray}
&& K^{(\pm)}_{1} (x)= 
 \left(
    \begin{array}{c}
\pm \phi_K(x) \\
0
    \end{array}
    \right)=
     \left(
    \begin{array}{c}
\pm \tanh x \\
0
    \end{array}
    \right)
    ,\label{Kink1}\\ [1ex]
&& 
K^{(\pm)}_{2} (x)= 
 \left(
    \begin{array}{c}
0  \\
\pm\phi_K(x)
    \end{array}
    \right)
=
 \left(
    \begin{array}{c}
0  \\
\pm \tanh x
    \end{array}
    \right).
    \label{Kink2}
\end{eqnarray}
As before, each pair of kinks joins two non-contiguous vacua that we have in \eqref{eqI2:vacua4>1}, one of them in one direction and the other in the opposite. For example, $K^{(+)}_{1} (x)$ connects 
$\begin{pmatrix}
-1 \\
0
    \end{pmatrix}
$
with 
$\begin{pmatrix}
1 \\
0
    \end{pmatrix}
$
as $x$ goes from $-\infty$ to $+\infty$, and $K^{(-)}_{1} (x)$ connects the same vacua but in reverse order. We will analyze more in detail these solutions in Chapter \ref{Chap1}.

On the other hand, there exist also kinks that interpolate contiguous vacua. These solutions cannot be obtained analytically, but they can be obtained numerically. From now on, these kinks will be called $K_3(x)$ kinks. In Figure \ref{figI2:K3DoblePhi4} a plot for these solutions can be found.

\begin{figure}[htb]
    \centering
    \begin{subfigure}{0.6\textwidth}
        \centering
        \includegraphics[width=\linewidth]{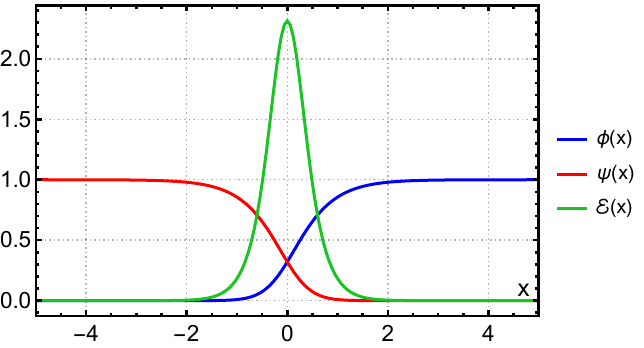}
    \end{subfigure}
    %\hfill
    \caption{ \textit{Plot for one of the kink solutions $K_3$ and its corresponding energy density for $\kappa=15$.}}
    \label{figI2:K3DoblePhi4}
\end{figure}

In Figure \ref{figI2:KOrbitKmay1} we can find the kink orbits for all kink solutions arising in this regime.
\begin{figure}[h!]
    \centering
    \begin{subfigure}{0.45\textwidth}
        \centering
        \includegraphics[width=\linewidth]{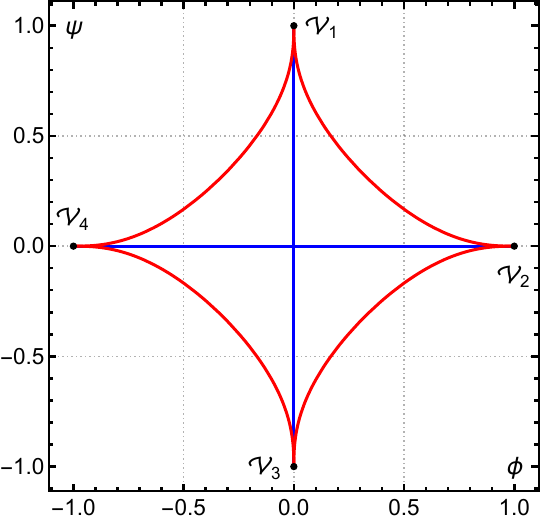}
    \end{subfigure}
    %\hfill
    \caption{ \textit{Kink orbits for kinks $K_1(x)$, $K_2(x)$ (blue) and $K_3(x)$ (red) for $\kappa=15$. }}
    \label{figI2:KOrbitKmay1}
\end{figure}

As we will see in Chapter \ref{Chap1}, kinks $K_1$ and $K_2$ are unstable for $1<\kappa<3$. In this scenario, these kinks decay into pairs of kinks $K_3$. In Figure \ref{figI2:kinkDoblePhi4evol} we have depicted the evolution of one of these kinks for $\kappa=2$, the orbit of the field configuration has been also depicted. As it can be seen,  the final configuration interpolates between the vacua $\mathcal{V}_4$ and $\mathcal{V}_2$ passing through $\mathcal{V}_3$. It can also be appreciated that the kink disintegrates into a pair of $K_3$ kinks\footnote{Depending on how the initial profile is perturbed the final kink orbit can go through $\mathcal{V}_2$ instead of $\mathcal{V}_3$.}.

\begin{figure}[htb]
    \centering
    \begin{subfigure}{0.49\textwidth}
        \centering
        \includegraphics[width=\linewidth]{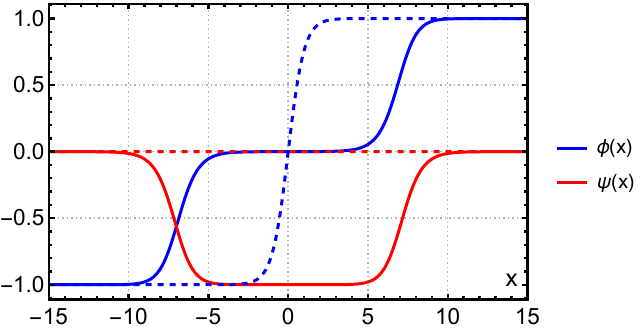}
    \end{subfigure}
    %\hfill
    \begin{subfigure}{0.3\textwidth}
        \centering
        \includegraphics[width=\linewidth]{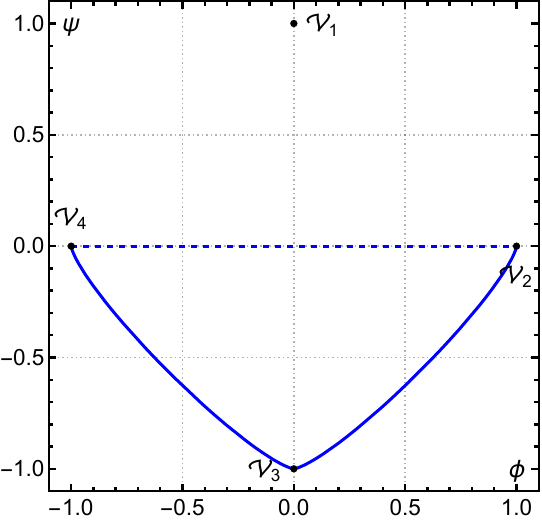}
    \end{subfigure}
    \caption{\textit{ On the left, initial kink configuration (solid lines) and configuration after a few time steps (dashed lines). On the right, their corresponding field orbits over the $(\phi,\psi)$-plane (right).}}
    \label{figI2:kinkDoblePhi4evol}
\end{figure}

\end{itemize}

%\section{Other aspects}

    \chapter[  Wobblers and internal mode interactions in the double \texorpdfstring{$\phi^4$}{phi4} model]{\LARGE Wobblers and internal mode interactions in the double \texorpdfstring{$\phi^4$}{phi4} model}\label{Chap1}

This chapter is an adaptation from Reference \cite{AlonsoIzquierdo2024}: 

\begin{figure}[htb]
    \centering
   \includegraphics[width=1\linewidth]{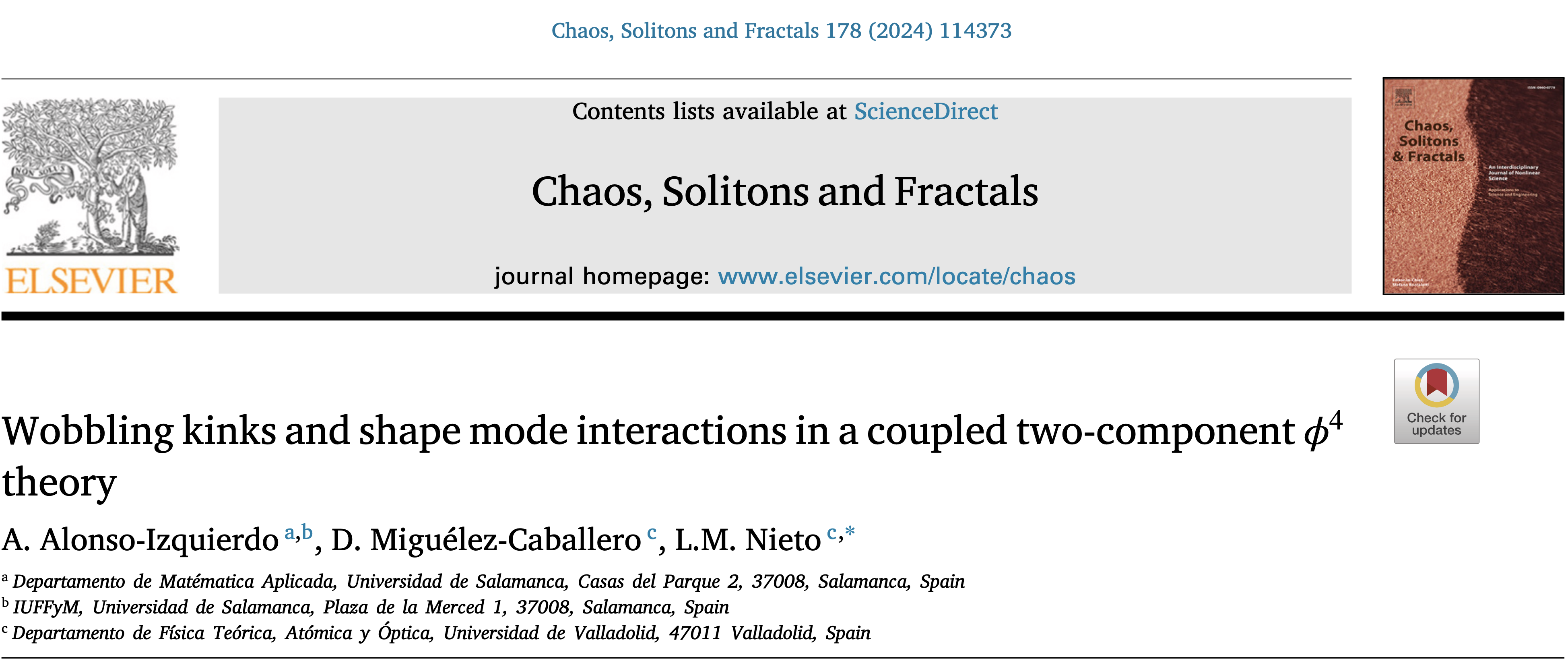}
\end{figure}

%----------------------------------Introduction-------------------------

\section{Introduction}

%It is well known that topological defects have played a key role in the last decades to understand nonlinear phenomena in many areas, such as condensed matter \cite{Bishop1978, Rajaraman1982,Dauxois2006,CuevasMaraver2014},  cosmology \cite{Kibble1976, Vachaspati2006,Vilenkin1994},  superconductivity \cite{Buzea1998} or quantum communications \cite{Dymarsky2021,Buican2023}. 

As previously stated in Chapters \ref{Intro0} and \ref{Intro1}, among all the existing range of topological defects, the kinks, which arise when considering nonlinear scalar field theories, are undoubtedly the simplest.
In this context, the most famous theory is the $\phi^4$ model, for which kink scattering and collisions have been extensively investigated due to the complex pattern of collisions in which a fractal pattern arises \cite{Goodman2005, AlonsoIzquierdo2021b,AlonsoIzquierdo2021c,Mohammadi2022,Kevrekidis2019b}. 
This same type of phenomena has also been observed when studying other more complex scalar models, such as theories with other polynomial potentials such as $\phi^6$ \cite{Springer2019,Dorey2011,Marjaneh2017,Gani2014} or $\phi^8$ \cite{Belendryasova2018, Bazeia2023, Gani2015}, theories with two-component scalar fields \cite{AlonsoIzquierdo2021,AlonsoIzquierdo2023, Katsura2013, Bazeia1995, Shifman1998, Ashcroft2016, AlonsoIzquierdo2019, Aguirre2020, AlonsoIzquierdo2002, Halavanau2012}, or more complex ones \cite{Blinov2022, Christov2021, Askari2020, Takyi2023, Mohammadi2021}. 

In addition to numerical analysis methods,  analytical tools have been developed to study and understand the  physical behavior of these fascinating solutions to the $\phi^6$ and $\phi^8$ models. Some of these techniques have already been discussed in Chapter \ref{Chap1}. There exist other analytical tools developed with the same purpose we have not discussed in thesis. An example of what we just said can be found in the use of the  moduli space approximation \cite{Sugiyama1978, NavarroObregon2023, Manton2004, Adam2022c, Adam2023, Takyi2016} to reduce the degrees of freedom of the system.  These methods have successfully  replicated phenomena observed in kink solutions, such as radiation emission or oscillon formation and have even been extended to study vortices in the Abelian-Higgs model \cite{MiguelezCaballero2025}.

When kink scattering phenomena are studied, among the possible final configurations of the system, it is possible to find \textit{wobbling kinks} or \textit{wobblers}, which  consists of a kink solution where one of its vibration modes has been triggered due to the energy transfer mechanism between its shape modes and the kinetic energy of the kink. In the previous chapter, we discussed this scenario in the context of the $\phi^4$ model.  It was found  \cite{Barashenkov2019, Barashenkov2009,Manton1997} that a wobbling kink in the $\phi^4$ model emits radiation with twice the frequency associated with its shape mode, which also causes a decay in the wobbling amplitude due to the loss of energy in the form of radiation.   
As already mentioned, these perturbative theories have also been implemented to study the evolution of wobbling kinks in two-component scalar field theories, such as the MSTB model \cite{AlonsoIzquierdo2023c}. 
In the present chapter, the theory under study will be the \textit{double $\phi^4$} field theory \cite{Halavanau2012}. The kink structure was analyzed in detail in Section \ref{Intro1.7}. The kink solution under study will be identical in this chapter and in Chapter \ref{Chap2}. The main difference is that, as the Lagrangian density governing the dynamics of the kink is different in both cases, the internal mode structure is completely different. Indeed, the shape mode structure arising in this model will be much more complex than in the MSTB field theory, since the number of shape modes corresponding to the second field component depends on the coupling constant between the two copies of the $\phi^4$ model.

This chapter is structured as follows: in Section~\ref{C1:Section2}  the linear stability of the kink solutions under study will be analyzed, which will allow us to find the corresponding vibration eigenfunctions and eigenfrequencies. 
In Section~\ref{C1:Section3} the perturbative approach introduced by Manton and Merabet in~\cite{Manton1997} will be used to describe the behavior of a kink when one of its shape modes is activated. In Section~\ref{C1:Section4} all the analytical results found in the preceding chapters will be compared with those obtained by numerical simulations. The chapter ends with  some concluding remarks.

%----------------------------------Model-------------------------
\section{ Kink solutions under study and their linear stability
%The two-component coupled $\phi^4$ model: kink solutions, linear stability and shape modes 
}\label{C1:Section2}

\begin{comment}
As already mentioned above, in the following Chapters we will deal with a two-component real scalar field theory, consisting of two separate copies of a $\phi^4$ model, coupled by means of a cross term $\kappa \phi^2 \psi^2$, where $\kappa$ is  a real positive parameter.  Thus, the dynamics of this physical system is governed by the Lagrangian density
\begin{equation}\label{C1:LagrangianDensity}
    \mathcal{L}=\frac{1}{2}\partial_\mu \phi \partial^\mu \phi+\frac{1}{2}\partial_\mu \psi \partial^\mu \psi- U(\phi, \psi),
\end{equation}
where the potential $U(\phi,\psi)$ is given by 
\begin{equation}\label{C1:Potential}
    U(\phi, \psi)=\frac{1}{2}(\phi^2-1)^2+\frac{1}{2}(\psi^2-1)^2+ \kappa \phi^2 \psi^2 -\frac{1}{2} ,
\end{equation}
which is completely symmetric  in the interchange of $\phi$ for $\psi$.
In equations \eqref{C1:LagrangianDensity}--\eqref{C1:Potential} we assume that $\phi$ and $\psi$ are real scalar fields and the Minkowski metric is taken in the usual form:  $g_{\mu,\nu}=\text{diag} \{ 1,-1\}$. 
In general, the field equations that govern the evolution of both fields,  which can be  calculated from the Lagrangian density \eqref{C1:LagrangianDensity}, are
\begin{eqnarray}
  &&  \partial_{tt} \phi -\partial_{xx} \phi + 2 \phi (\phi^2 -1 + \kappa \psi^2)=0, \label{C1:FieldEqn1} \\ %[1ex]
   && \partial_{tt} \psi -\partial_{xx} \psi + 2 \psi (\psi^2 -1 + \kappa\phi^2)=0. \label{eqI2:FieldEqDouble2}
\end{eqnarray}
\end{comment}

We recall that  the scalar field theory under study is governed by the Lagrangian density 
\begin{equation}\label{C1:LagrangianDensity}
    \mathcal{L}=\frac{1}{2}\partial_\mu \phi \partial^\mu \phi+\frac{1}{2}\partial_\mu \psi \partial^\mu \psi- U(\phi, \psi),
\end{equation}
where the potential $U(\phi,\psi)$ is given by 
\begin{equation}\label{C1:Potential}
    U(\phi, \psi)=\frac{1}{2}(\phi^2-1)^2+\frac{1}{2}(\psi^2-1)^2+ \kappa \phi^2 \psi^2 -\frac{1}{2} ,
\end{equation}
which is completely symmetric  in the interchange of $\phi$ for $\psi$.
In the field equations  \eqref{C1:LagrangianDensity}--\eqref{C1:Potential} we assume that $\phi$ and $\psi$ are real scalar fields and the Minkowski metric is taken in the usual form:  $g_{\mu,\nu}=\text{diag} \{ 1,-1\}$.

The vacua structure and the kink solutions that arise in this model have been already discussed in Section \ref{Intro1.7}. In this chapter we will only analyze the analytical kinks that arise in the regime $\kappa>1$. In other words, we will only study the two pair of kinks 
\begin{eqnarray}
 K^{(\pm)}_{1} (x) \!\!&\!\! = \!\!&\!\!
 \left(
    \begin{array}{c}
\pm \phi_K(x) \\
0
    \end{array}
    \right)=
     \left(
    \begin{array}{c}
\pm \tanh x \\
0
    \end{array}
    \right)
    ,\label{C1:Kink1}
    \\ %[1ex] 
K^{(\pm)}_{2} (x) \!\!&\!\! =  \!\!&\!\!
 \left(
    \begin{array}{c}
0  \\
\pm\phi_K(x)
    \end{array}
    \right)
=
 \left(
    \begin{array}{c}
0  \\
\pm \tanh x
    \end{array}
    \right).
    \label{C1:Kink2}
\end{eqnarray}

\begin{figure}[htb] 
         \includegraphics[width=0.49\textwidth]{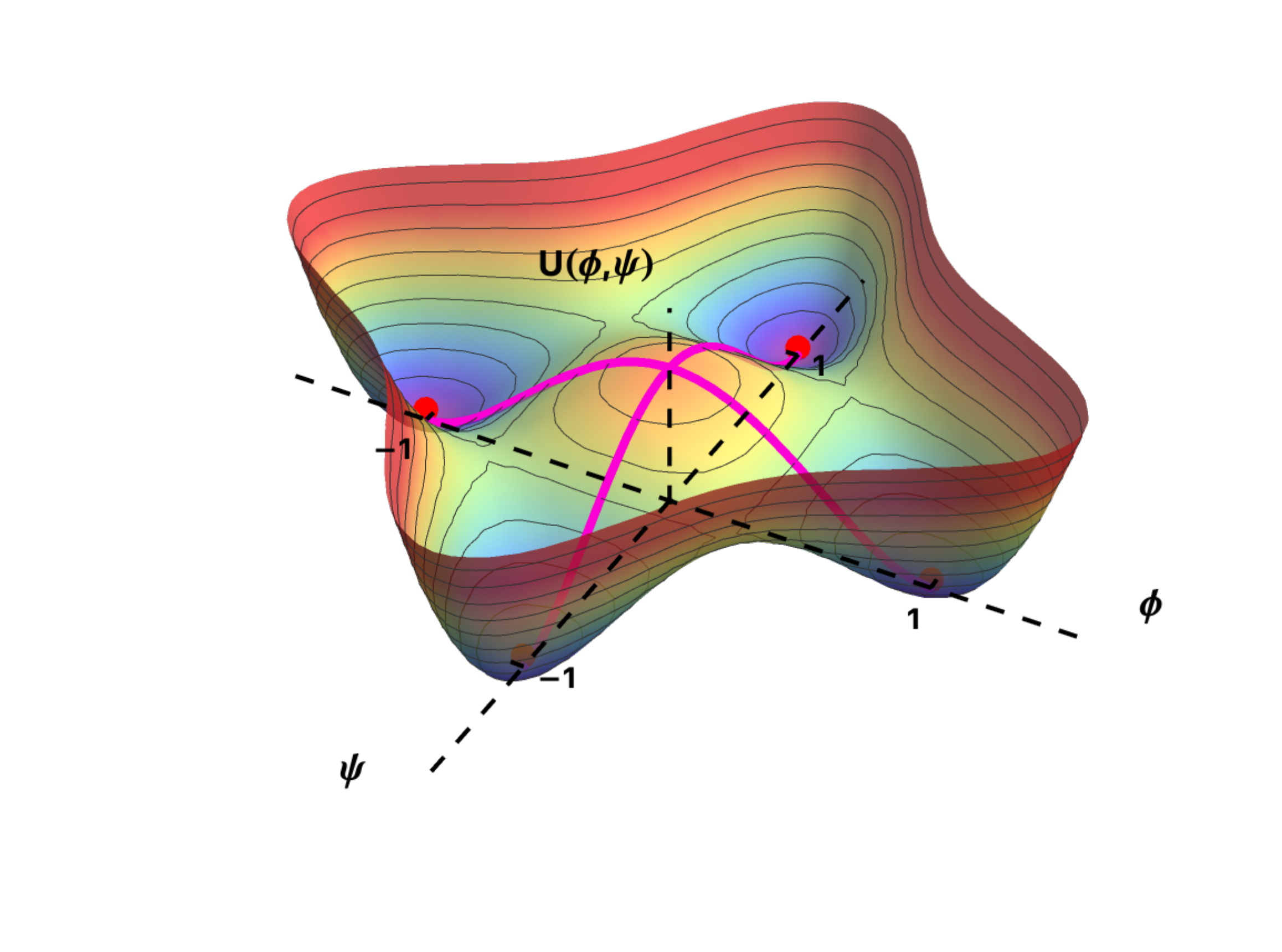}
         \includegraphics[width=0.49\textwidth]{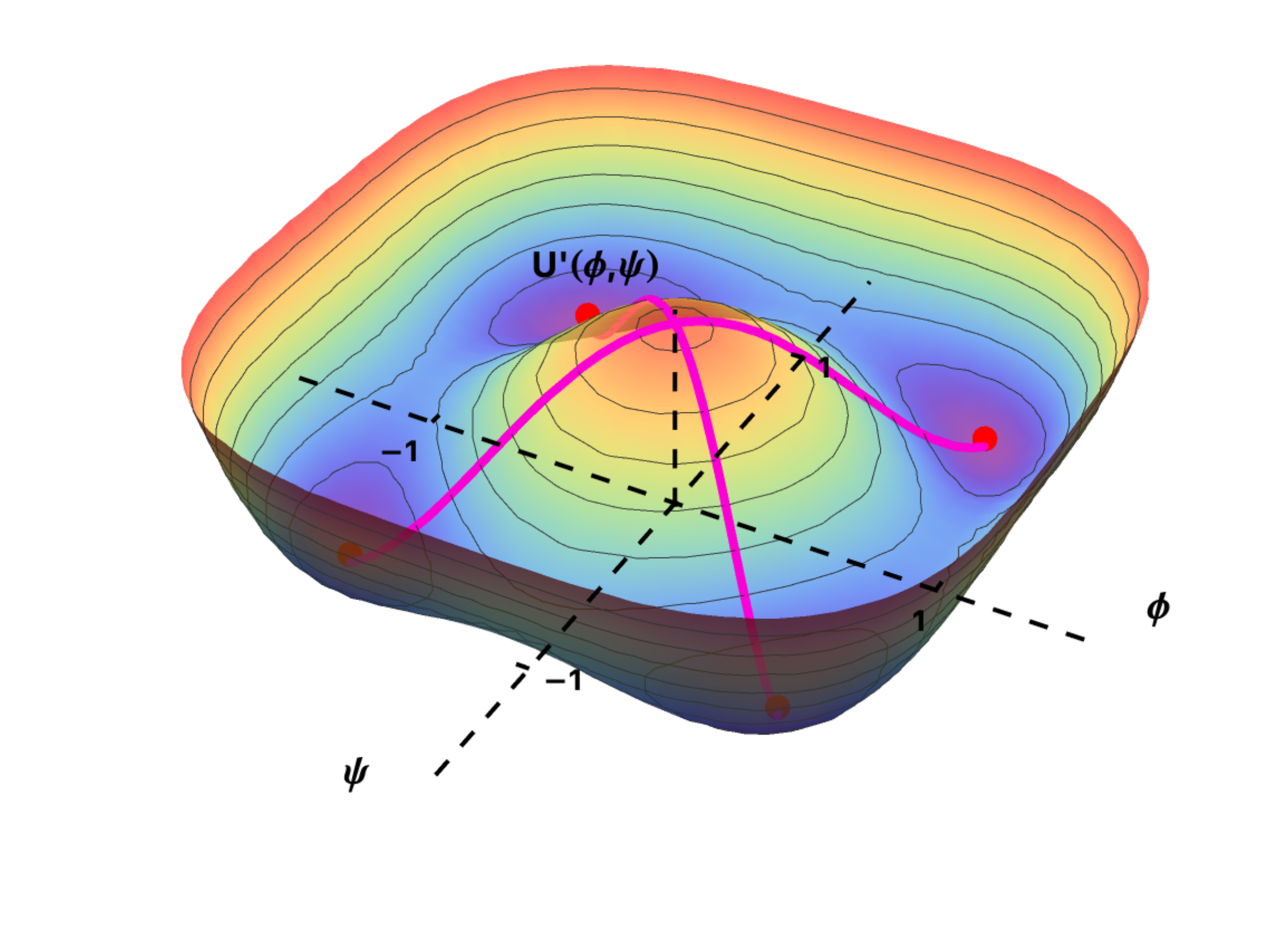}
\caption{\textit{Plot of the potential $U(\phi,\psi)$ given by \eqref{C1:Potential} for $\kappa=4$ (upper plot) and for the shifted potential $U^*(\phi,\psi)=U(\phi,\psi)+\frac{1-\kappa}{2(1+\kappa)}$ for $\kappa=0.5$ (bottom plot). The corresponding kinks (pink curves) and the minima (red dots) are shown in each case.}}
\label{C1:Fig:Potential}
\end{figure}

As previously said, these kinks connect non consecutive vacua.  In fact, all analytical calculations that will be developed can also be performed with the kink solutions 
\begin{equation}\label{C1:otherKink}
    K^{(a,b)} (x)=\frac{\tanh x }{\sqrt{1+\kappa}} \left(
    \begin{array}{c}
    (-1)^a\\ 
    (-1)^b 
    \end{array}
    \right), \quad a,b=0,1,
\end{equation}
 but all the results would be the same as those corresponding to the kink \eqref{C1:Kink1}-\eqref{C1:Kink2}.
To clarify as much as possible the problem we are dealing with, in Figure~\ref{C1:Fig:Potential} we show  graphs of the potential for two values of $\kappa$. 
On the first graph of the figure the four vacua \eqref{eqI2:vacua4>1} of the potentials $U(\phi,\psi)$ in \eqref{C1:Potential} are shown as the minima (red dots) of the potential for $\kappa>1$, and the kinks \eqref{C1:Kink1}--\eqref{C1:Kink2} are  the pink curves connecting two of the non-adjacent vacua.
On the second graph there is a plot of the situation that occurs when $\kappa<1$, with the potential $U^*(\phi,\psi)$ appropriately shifted to have the four vacua \eqref{eqI2:vacua4<1} (red dots), showing again the kinks by pink lines joining two of the non-contiguous vacua.

\subsection{Linear stability and shape modes when \texorpdfstring{$\kappa>1$}{kappa>1} }

Since  we will focus on the analysis of the regime $\kappa>1$, we will now discuss in detail the linear stability of the kinks \eqref{C1:Kink1}--\eqref{C1:Kink2}, as well as the shape modes and the corresponding vibrational frequencies that arise in this kind of study. 
First of all, to perform the linear stability analysis, a perturbation in the kink solution $K(x)$ of the form
\begin{equation}\label{C1:complexperturbation}
    \widetilde{K}(x,t)=K(x)+ a \, e^{i\omega t} F(x)= \begin{pmatrix}
\phi(x)  \\ 
\psi(x)
    \end{pmatrix}
+ a \, e^{i \omega t} \begin{pmatrix}
\overline{\eta} (x)  \\ 
\widehat{\eta} (x)
    \end{pmatrix}
\end{equation}
is inserted into the field equations \eqref{eqI2:FieldEqDouble1}--\eqref{eqI2:FieldEqDouble2}, where $\phi(x)$ and $\psi(x)$ are the two components of the static solution  of the model field equations and  $a$ is a small real parameter. This leads to the spectral problem 
\begin{equation}
    \mathcal{H}
\begin{pmatrix}
\overline{\eta} (x)  \\ 
\widehat{\eta} (x)
    \end{pmatrix}    = 
    \omega^2 
\begin{pmatrix}
\overline{\eta} (x)  \\ 
\widehat{\eta} (x)
    \end{pmatrix}    ,
\end{equation}
where
\begin{equation}
    \mathcal{H}=
    \begin{pmatrix}
          -\dfrac{d^2}{d x^2}+ 6 \phi(x)^2 -2 +2 \kappa\, \psi(x)^2 &  4 \kappa\, \phi(x) \psi(x)  \\
           \hskip-2cm    4 \kappa \, \phi(x) \psi(x)  &  \hskip-2cm -\dfrac{d^2}{d x^2}+ 6 \psi(x)^2 -2 +2 \kappa \,\phi(x)^2  \\
    \end{pmatrix}
    .
\end{equation}
Therefore, to study the kink stability of the solution \eqref{C1:Kink1} in the regime $\kappa>1$, the spectral problem to be solved is essentially the following:
\begin{equation}\label{C1:SpectralProblem}
      \begin{pmatrix}
          -\dfrac{d^2}{d x^2}+ 6 \tanh^2 x -2  & 0  \\
               \hskip-1.7cm 0 &   \hskip-1.7cm -\dfrac{d^2}{d x^2}  +2 \kappa \tanh^2 x -2 \\
    \end{pmatrix}
    \begin{pmatrix}
    \overline{\eta}(x) \\
    \widehat{\eta}(x) 
    \end{pmatrix}
    =
    \omega^2 
    \begin{pmatrix}
    \overline{\eta}(x) \\
    \widehat{\eta}(x) 
    \end{pmatrix}
    .
\end{equation}
From \eqref{C1:SpectralProblem} it can be  seen that the matrix-operator $\mathcal{H}$ is diagonal which, in turn, implies that this problem consists of two decoupled Schr\"odinger-like  equations with P\"oschl-Teller potential wells $\mathcal{H}_{11}=-\frac{d^2}{d x^2}+ 6 \tanh^2 x -2 $ and $\mathcal{H}_{22}=-\frac{d^2}{d x^2}  +2 \kappa \tanh^2 x -2 $. The solution to this type of equations has been widely studied and can be found, for example, in \cite{Flugge1971,Morse1953,Morse1933}. 
Since the orbit of the kink solution \eqref{C1:Kink1} lies  on the $\phi$-axis, the fluctuations corresponding to the first  component of the field will be called \textit{longitudinal eigenmodes} and those corresponding to the second component of the field will be called \textit{orthogonal eigenmodes}. Solving the spectral problem \eqref{C1:SpectralProblem}, the following eigenmodes and eigenfrequencies can be found.
\begin{itemize}
\item 
\textbf{Longitudinal eigenmodes:} 
    There are two eigenmodes, one associated with $\omega=0$, called  \textit{zero mode} or \textit{translational eigenmode}, and another associated with a frequency $\overline{\omega}=\sqrt{3}$, which is known  as \textit{longitudinal shape mode}. 
The formulas for these two modes can be expressed as follows \cite{Vachaspati2006,Shnir2018,Manton1997,Barashenkov2009,Barashenkov2019,Springer2019,AlonsoIzquierdo2023c}:
    \begin{eqnarray}
        \overline{F}_{0} (x) \!\!&\!=\!&\!\!         \begin{pmatrix}
    \overline{\eta}_0 (x) \\
    0
    \end{pmatrix}
= \begin{pmatrix}
    \sech^2 x \\
    0
    \end{pmatrix},
    \\ % [1ex]
        \overline{F}_{\!\!\sqrt{3}}(x)  \!\!&\!=\!&\!\!    \begin{pmatrix}
    \overline{\eta}_D(x) \\
    0
    \end{pmatrix} 
=  \begin{pmatrix}
    \sech x\tanh x \\
    0
    \end{pmatrix} ,
    \label{C1:LongitudinalShapeMode}
    \end{eqnarray}
where the subindex in $F(x)$ indicates the corresponding frequency.
    In  addition to these   vibration eigenfunctions, there are also continuous  modes associated with the frequencies
        \begin{equation}\label{C1:Continuousfrequencylongitudinal}
\overline{\omega}^c_{\bar{q}}=\sqrt{4+\bar{q}^2}, \quad \bar{q}\in\mathbb{R}.
    \end{equation}
In this case, these eigenmodes take the form 
    \begin{equation}\label{C1:ContinuousEigenfunctionFirstField}
        \overline{F}_{\sqrt{4+\bar{q}^2}} \,(x)=
        \begin{pmatrix}
    \overline{\eta}_{\bar{q}}(x) \\
    0
    \end{pmatrix} ,
    \end{equation}
    with $ \overline{\eta}_{\bar{q}}(x)  = ( -1-\bar{q}^2+ 3 \, \tanh^2 x -3 i \bar{q} \tanh x) \, e^{i \bar{q} x}$.
It is easy to verify that the complex conjugate of the previous function is also a solution of \eqref{C1:SpectralProblem}, independent of the first one, which is also obtained with the change $ \bar{q}\to - \bar{q}$. Therefore, this second solution will be denoted as $\overline{\eta}_{-\bar{q}}$ and it can be verified that the Wronskian associated to these two functions is        
\begin{equation}\label{C1:WronskianoFirstField}
        \overline{W}_{\bar{q}}=\overline{\eta}_{\bar{q}}(x)\, \overline{\eta}\,'_{-\bar{q}}(x)-\overline{\eta}\,'_{\bar{q}}(x)\, \overline{\eta}_{-\bar{q}}(x)=-2 i \bar{q}(\bar{q}^2+1)(\bar{q}^2+4),
\end{equation}
where the prime in \eqref{C1:WronskianoFirstField} stands for the derivative with respect to the variable $x$.

As a final remark in this section, let us note that from \eqref{C1:Continuousfrequencylongitudinal} it follows that the continuous spectrum in frequencies begins for $\bar{q}=0$, $\overline{\omega}^c_{0}=2$, that is to say, that the longitudinal discrete spectrum will necessarily be contained in the interval $[0,2 ]$. The internal mode structure found here is essentially the same that the one for the kink in the $\phi^4$ model. This problem was also addressed in Section \ref{Intro2.4.1}.

\item 
\textbf{Orthogonal eigenmodes:} 
The number of possible shape modes depends on the value of the parameter $\kappa$ that appears in the differential equation corresponding to the second component of \eqref{C1:SpectralProblem}, where the differential operator $\mathcal{H}_{22}$ appears. More precisely, there exists a number $n_{max}\geq 0$ that is given by the largest integer that verifies the following inequality \cite{Flugge1971}
    \begin{equation}\label{C1:ConditionNumberOrthogonalShapeModes}
             \kappa>{n_{max}(n_{max}+1)}/{2},
    \end{equation}
    and which determines the number of discrete eigenmodes ($n_{max}+1$) corresponding to this P\"oschl-Teller type equation.
The eigenfrequencies corresponding to these modes are determined by the expression
\begin{equation}\label{C1:FrequencyOrthogonalModes}
    \widehat{\omega}_n=\sqrt{(2n+1)\rho-n^2-n-\tfrac{5}{2}} , \quad n=0,1,\dots, n_{max},
\end{equation}
where $\rho=\sqrt{2\kappa+\frac{1}{4}}$. 
 Notice that the above equation is telling us something extremely important: that the kink is unstable when $\kappa<3$, since in that case it happens that $\widehat{\omega}^2_0<0$. Therefore, in the following we will focus exclusively on values of $\kappa>3$.

The associated eigenfunctions to these discrete modes \eqref{C1:FrequencyOrthogonalModes} are \cite{Flugge1971}
    \begin{equation}\label{C1:OrthogonalSahpeMode0}
            \widehat{F}_{\widehat{\omega}_n}  (x)=
            \begin{pmatrix}
    0 \\
    \widehat{\eta}_{D,n}(x)
    \end{pmatrix} 
    , \quad n=0,1,\dots, n_{max},
    \end{equation}
    with
    \begin{equation*} \hskip-0.5cm
     \widehat{\eta}_{D,n}(x) = (\sech x)^{\,\rho-n-\frac{1}{2}} {}_2F_1  \Bigl( -n,2\rho-n;\rho-n+1/2;\frac{1-\tanh{x}}2 \Bigr),
    \end{equation*}
 being ${}_2F_1 (a,b;c;z)$ the well known hypergeometric function.
Using the following relation (see \cite{NIST2010})
\begin{equation}
 {}_2F_1 \Bigl(-n,n+2\lambda,\lambda+1/2, \frac{1-x}2 \Bigr)  = \frac{n!}{(2 \lambda)_n} 
     \sum_{m=0}^{\lfloor n/2 \rfloor }(-1)^m \frac{(\lambda)_{n-m}}{m!(n-2m)!}(2x)^{n-2m},
     \label{C1:HypergeometricRelation}
\end{equation}
where $(\lambda)_n= \Gamma(\lambda+n)/\Gamma(\lambda)$ represents the Pochhammer symbol and $ \lfloor x \rfloor$ denotes the integer part of $x$, 
the orthogonal eigenmodes \eqref{C1:OrthogonalSahpeMode0} can be written as
    \begin{equation}
     \widehat{\eta}_{D,n}(x) =
    (\sech x)^{\,\rho-n-\frac{1}{2}} \frac{n!}{(2\rho-2n)_n} 
         \sum_{m=0}^{\lfloor n/2 \rfloor}(-1)^m \frac{(\rho-n)_{n-m}}{m!(n-2m)!}(2\tanh x)^{n-2m}.
    \label{C1:OrthogonalSahpeMode}
\end{equation}
Note that the orthogonal modes $ \widehat{\eta}_{D,n}  (x) $ with even or odd $n$ are even and odd functions, respectively.

On the other hand, the orthogonal continuous spectrum is composed of the following frequencies
\begin{equation}\label{C1:Continuousfrequencyorthogonal}
 \widehat{\omega}^c_{\hat{q}}=\sqrt{\hat{q}^2+2\kappa-2}, \quad \hat{q}\geq 0,
\end{equation}
being the corresponding eigenfunctions
\begin{equation}\label{C1:ContinuousEigenfunctionSecondField}
    \widehat{F}_{\sqrt{\hat{q}^2+2\kappa-2}} \, (x)=
    \begin{pmatrix}
    0 \\
    \widehat{\eta}_{\hat{q}}(x)
    \end{pmatrix} 
\end{equation}
    with
    \begin{eqnarray*} \hskip-0.5cm
     \widehat{\eta}_{\hat{q}}(x) \!\!&\!\!=\!\!&\!\! 
      {}_2F_1  \left(  \frac12-\rho,\frac12+\rho; 1-i \hat{q} ; \frac{1-\tanh x}2  \right) \, e^{i \hat{q} x} .
\end{eqnarray*}
Similar to what happens in the longitudinal case, a second linearly independent solution is obtained by changing $\hat{q}\to - \hat{q}$, so that the Wronskian for these two eigenfunctions can be written in this case as
\begin{equation}\label{C1:WronskianoSecondField}
    \widehat{W}_{\hat{q}}=\widehat{\eta}_{\hat{q}}(x)\, \widehat{\eta}\,'_{-\hat{q}}(x)-\widehat{\eta}\,'_{\hat{q}}(x)\, \widehat{\eta}_{-\hat{q}}(x)=-2 i \hat{q}.
\end{equation}
\vspace{-1cm}

Note that from \eqref{C1:Continuousfrequencyorthogonal} it follows that the continuous frequency spectrum begins for $\hat{q}=0$, that is, given a value of $\kappa>3$, the discrete orthogonal spectrum will necessarily be contained in the interval $[0,\sqrt{2\kappa-2} ]$.

\end{itemize}

%----------------------------------Perturbation theory-------------------------
\section{Interaction between vibrational modes: Perturbative approach}\label{C1:Section3}

As  shown in Section \ref{I2Manton},  in some nonlinear models  when we initially excite a shape mode, this  mode couples with the rest of  eigenmodes of the system,  causing  the emission of  radiation  \cite{AlonsoIzquierdo2023c, AlonsoIzquierdo2024c}. 
The reason of this phenomenon is the nonlinearity associated with the generalised Klein-Gordon equations  that dictate the behavior of the physical system. 
An example of this can be found  in the $\phi^4$ model, in which, when the shape mode associated with this theory is initially triggered, radiation is found with twice the frequency of the vibrational mode. 

As we mentioned before, in the present chapter we are going to study how the kink \eqref{C1:Kink1} evolves when an orthogonal shape mode is initially excited. 
The situation in which only the non-zero frequency longitudinal eigenmode is excited reduces the analysis to the study of  a one-component $\phi^4$ model, and this has been extensively investigated by several authors \cite{Manton1997, Barashenkov2009, Barashenkov2019}. 
In light of all this, if the initial configuration takes the form 
\begin{equation}\label{C1:InitialHypothesis}
    \widetilde{K}(x,t)= K(x)+ a_0 \sin(\widehat{\omega}_n t)\, \widehat{F}_{\widehat{\omega}_n}(x),
\end{equation}
where $a_0$ is a small parameter and ${\widehat{\omega}_n}$ is one of the possible values given in \eqref{C1:FrequencyOrthogonalModes}, hopefully at least some of the other modes will be also triggered. Note that \eqref{C1:InitialHypothesis} is similar to \eqref{C1:complexperturbation}, only now we are not looking for complex  but real solutions.

In this section we will address this situation and to do so we will use the perturbative approach that Manton and Merabet first introduced in \cite{Manton1997}. In other words, we will follow the method exposed in Section \ref{I2Manton} and assume the following expansion for field components:
\begin{eqnarray}
\label{C1:MantonExpansion1}
\phi(x,t) \!\!&\!\!=\!\!&\!\! \phi_K(x)+ \overline{a}(t)\,\overline{\eta}_D(x)+\overline{\eta}(x,t),  \\
\psi(x,t) \!\!&\!\!=\!\!&\!\! \sum_p\widehat{a}_p(t)\, \widehat{\eta}_{D, p}(x)+\widehat{\eta}(x,t), \label{C1:MantonExpansion2}
\end{eqnarray}
where $\overline{\eta}_D$, $\widehat{\eta}_{D,p}$ and $\phi_K(x)$ are defined in equations \eqref{C1:LongitudinalShapeMode}, \eqref{C1:OrthogonalSahpeMode} and \eqref{C1:Kink1}. 
The time dependent functions $\overline{a}(t)$ and $\widehat{a}_p(t)$ describe the  evolution of the amplitudes corresponding to each shape mode (longitudinal or transversal), the sum going from $p=0$ to $p=n$, where $n$ is the largest natural number for which the condition \eqref{C1:ConditionNumberOrthogonalShapeModes} is satisfied for a specific value of the coupling constant $\kappa$. 
By construction, $a_0$ is the small parameter in our perturbation approach.  
Furthermore, the functions $\overline{\eta}(x,t)$ and $\widehat{\eta}(x,t)$ are the space and time dependent functions that will dictate the behavior of the radiation found when $x\rightarrow\infty$. For the sake of simplicity,  the dependency of the functions mentioned above will be omitted in subsequent calculations. 
Therefore, if we now plug \eqref{C1:MantonExpansion1}--\eqref{C1:MantonExpansion2} into the field equations     \eqref{eqI2:FieldEqDouble1} and \eqref{eqI2:FieldEqDouble2}
find for the first and second field components, respectively,
{
\begin{eqnarray}
    \label{C1:FirstFieldExpansionManton}
   && \hspace{-0.6cm}(\overline{a}_{tt}+\overline{\omega}^2\, \overline{a} \, ) \,\overline{\eta}_D-\overline{\eta}_{xx}+\overline{\eta}_{tt}+2\,\overline{\eta}^3 + 6\, \overline{a}\,\overline{\eta}^2\,\overline{\eta}_D+6\, \overline{a}^2\,\overline{\eta}\,\overline{\eta}_D^2 +2\, \overline{a}^3 \,\overline{{\eta}}_D^3+6\, \overline{\eta}^2\,\phi_K+12\, \overline{a}\,\overline{\eta}\,\overline{\eta}_D\, \phi_K 
    \\ %[0.5ex]
   && \hspace{-0.6cm}\quad + 6\,\overline{a}^2\, \overline{\eta}^2_D \,\phi_K + 6\,\overline{\eta}\,\phi_K^2-2\,\overline{\eta}  +2\, \kappa \left(\phi_K+\overline{a}\,\overline{\eta}_D +\overline{\eta} \right) 
   \Bigl(  \sum_{p,r} \widehat{a}_p\,\widehat{a}_r\,\widehat{\eta}_{D ,p}\,\widehat{\eta}_{D,r}+ 2 \sum_p \widehat{a}_p\,\widehat{\eta}_{D, p}\,\widehat{\eta}+\widehat{\eta}^2 \Bigr)=0,
   \nonumber
    \\    %[1ex]    
       && \hspace{1.3cm} \sum_p \Bigl( (\widehat{a}_{p})_{tt} + \widehat{\omega}_p^2 \, \widehat{a}_p \Bigr) \widehat{\eta}_{D, p}+\widehat{\eta}_{tt}-\widehat{\eta}_{xx}+2\sum_{p,r,s} \widehat{a}_p \,\widehat{a}_r\,\widehat{a}_s\, \widehat{\eta}_{D,p}\,\widehat{\eta}_{D,r}\,\widehat{\eta}_{D,s}  
         \nonumber\\
         && \hspace{1.3cm}
        + 6 \sum_{p,r} \widehat{a}_p \,\widehat{a}_r\,\widehat{\eta}_{D, p}\,\widehat{\eta}_{D, r} \,\widehat{\eta}  + 6 \sum_p \widehat{a}_p\, \widehat{\eta}_{D, p}\,\widehat{\eta}^2+2 \widehat{\eta}^3-2\widehat{\eta} + 2\kappa\,\widehat{\eta}\,\phi_K^2 \label{C1:SecondFieldExpansionManton} \\   
         && \hspace{1.3cm}  +2 \kappa \Bigl(
         \sum_p \widehat{a}_p\,\widehat{\eta}_{D,p}+\widehat{\eta} \Bigr)\Bigl( \overline{\eta}^2+ 2 \overline{a}\,\overline{\eta}_D\, \overline{\eta}+ \overline{a}^2 \,\overline{\eta}_D^2+ 2 \overline{\eta}\,\phi_K+2 \overline{a}\,\overline{\eta}_D\,\phi_K \Bigr)  =0,
         \nonumber
    \end{eqnarray}
    }
where $\overline{\omega}=\sqrt{3}$ and $\widehat{\omega}_p$ is given by  \eqref{C1:FrequencyOrthogonalModes}. 
By physical reasons, all the functions $\eta(x,t)$ and $a(t)$ (with a bar or with a hat on top) are small quantities. 
Then, the terms $\eta^2$, $\eta^3$, $\eta^2 a$, $\,a^3$\dots  can be neglected in the formulas \eqref{C1:FirstFieldExpansionManton} and \eqref{C1:SecondFieldExpansionManton}, which leads us to the following truncated expansion for the field component equations:
   \begin{eqnarray}
&&  \hspace{-1.25cm }      (\overline{a}_{tt}+\overline{\omega}^2 \,\overline{a} )\,\overline{\eta}_D +\overline{\eta}_{tt}-\overline{\eta}_{xx}-2\overline{\eta}+6\,\overline{\eta}\,\phi_K^2+6\overline{a}^2 \,\overline{\eta}_D^2\,\phi_K+2\kappa\,\phi_K \Bigl( \sum_p \widehat{a}_p \widehat{\eta}_{D, p} \Bigr)^2\approx 0,
        \label{C1:FirstFieldExpansionMantonTruncated}
        \\  %[0.5ex]
&&  \hspace{-1.25cm }      \sum_p \Bigl(  (\widehat{a}_{p})_{tt} + \widehat{\omega}_p^2\,  \widehat{a}_p \Bigr)\, \widehat{\eta}_{D, p}+\widehat{\eta}_{tt}-\widehat{\eta}_{xx}-2 \widehat{\eta}+2\kappa\,\widehat{\eta}       \,\phi_K^2+4 \kappa \,\overline{a}\,\overline{\eta}_D\,\phi_K \sum_p \widehat{a}_p\,
        \widehat{\eta}_{D, p}\approx 0.
        \label{C1:SecondFieldExpansionMantonTruncated}
    \end{eqnarray}
If we now project the formula \eqref{C1:FirstFieldExpansionMantonTruncated} onto the longitudinal shape mode $\overline{\eta}_D(x)$, we find the following relation:
    \begin{equation}
             \Bigl( \overline{a}_{tt}+\overline{\omega}^2 \, \overline{a}\Bigr) \overline{C}+ 6 \, \overline{a}^2 \, \overline{V}+ \sum_{p,r} \widehat{a}_p\, \widehat{a}_r\, \widehat{B}_{p r}=0,
             \label{C1:ODEAmplitudesFirstComponent}
    \end{equation}
    where $ \overline{C} = {2}/{3}$, $\overline{V}= {\pi}/{16}$ and 
    \begin{equation}\label{C1:Bjm}
        \widehat{B}_{p r}= 2 \kappa \int^{\infty}_{-\infty} \overline{\eta}_D(x)\ \phi_K(x)\ \widehat{\eta}_{D, p}(x)\ \widehat{\eta}_{D, r}(x)\ dx.
    \end{equation}
Note that given the parities of the functions in the integrand of \eqref{C1:Bjm}, if $p+r$ is an odd number, then $\widehat{B}_{p r}=0$.
    
Similarly, we can project the relation \eqref{C1:SecondFieldExpansionMantonTruncated} onto $\widehat{\eta}_{D, m}$, which leads to
\begin{equation}
    \Bigl(   (\widehat{a}_{m})_{tt}   + \widehat{\omega}_m^2 \,  \widehat{a}_{m} \Bigr) \widehat{C}_{m} +2 \sum_{p} \overline{a}\, \widehat{a}_p \,\widehat{B}_{p m}=0,
    \label{C1:ODEAmplitudesSecondComponent}
\end{equation}
$m=0,1, \dots n$, where 
\begin{equation}\label{C1:CDj}
 \widehat{C}_{m} =\int^{\infty}_{-\infty} \widehat{\eta}_{D, m}^2(x)\ dx.
\end{equation} 
When necessary, the numbers $\widehat{B}_{p r}$ in \eqref{C1:Bjm} and $\widehat{C}_{ m}$ in \eqref{C1:CDj} should be evaluated numerically, since the presence of the functions $\widehat{\eta}_{D,n}$ in the corresponding integrals makes it impossible to evaluate them analytically.

The expressions \eqref{C1:ODEAmplitudesFirstComponent} and \eqref{C1:ODEAmplitudesSecondComponent} form a system of $n+1$ ordinary nonlinear differential equations that must be determined according to the shape modes that are initially triggered (the initial conditions). Plugging \eqref{C1:ODEAmplitudesFirstComponent} and \eqref{C1:ODEAmplitudesSecondComponent} into \eqref{C1:FirstFieldExpansionMantonTruncated} and \eqref{C1:SecondFieldExpansionMantonTruncated},  we find 
\begin{adjustwidth}{-0.75cm}{0cm}
\begin{eqnarray}\label{C1:EquationRadiationFirstField}
 \overline{\eta}_{tt}-\overline{\eta}_{xx}+
         ( 6 \phi_K^2-2 ) \overline{\eta}  \!\!&\!\! \hspace{-0.0cm}=\!\!&\hspace{0cm}\!\!   \hspace{-0.0cm}\overline{a}^2 \Bigl( -6\,\overline{\eta}_D^2\,\phi_K+\frac{\displaystyle6\,  \overline{V}}{\displaystyle\overline{C}}\,\overline{\eta}_D \Bigr)+\sum_{p,r} \widehat{a}_p\, \widehat{a}_r\,\Bigl(\displaystyle \frac{\widehat{B}_{p r}}{\displaystyle\overline{C}}\,\overline{\eta}_D-2\,\kappa\, \phi_K \,\widehat{\eta}_{D, p}\,\widehat{\eta}_{D, r}\Bigr),
\\ 
\label{C1:EquationRadiationSecondField}
             \widehat{\eta}_{tt}-\widehat{\eta}_{xx}+  ( 2\kappa\, \phi_K^2-2  ) \widehat{\eta}
             \!\!&\!\!=\!\!&\!\! 2\,\overline{a} \, \Bigl(\sum_{p,r}\frac{\displaystyle  \widehat{a}_r\, \widehat{B}_{pr}\,\widehat{\eta}_{D, r}}{\displaystyle\widehat{C}_{p}}-2 \kappa \sum_p\,\overline{\eta}_D \,\phi_K \,\widehat{a}_p \,\widehat{\eta}_{D, p}\Bigr),
\end{eqnarray}
\end{adjustwidth}
for the first and second components. 
These are the key equations that we will need to analyze next.

\subsection{Evolution of the system when only an orthogonal shape mode is initially activated}

Next, we want to study how the systems evolves in case we exclusively trigger the $j$-th orthogonal shape mode at $t=0$. It can be seen from \eqref{C1:ODEAmplitudesFirstComponent}--\eqref{C1:ODEAmplitudesSecondComponent} that this excitation also activates the rest of the shape modes, but their corresponding amplitudes will be much smaller. From this reasoning and from the differential equation \eqref{C1:ODEAmplitudesSecondComponent}, it is logical to assume that the amplitude associated with the $j$-th mode can be approximated as
\begin{equation}\label{C1:OthogonalAmplitudeEvolution}
    \widehat{a}_j(t)\approx a_0 \sin (\widehat{\omega}_j t),
\end{equation}
where ${\widehat{\omega}_j}$ is fixed and is one of the possible values given in \eqref{C1:FrequencyOrthogonalModes}. 
If we now plug \eqref{C1:OthogonalAmplitudeEvolution} into \eqref{C1:ODEAmplitudesFirstComponent}, we neglect terms of the form $\overline{a}^2$ and $\widehat{a}_p\, \widehat{a}_r$ with $ p ,r\neq j$ (because they are of order $\mathcal{O}(a_0^4)$) and then we solve the resulting differential equation taking into account the initial conditions
\begin{equation}\label{C1:IntialConditions}
    \overline{a}_t(0)=\overline{a}(0)=0, \quad  \widehat{a}_m(0)=\widehat{a}_{m,t}(0)=0 \quad  \mathrm{ with} \quad  m\neq j, 
\end{equation}
then it turns out that the evolution of $\overline{a}(t)$ can be described as
\begin{equation}\label{C1:LongitudinalAmplitudeEvolution}
        \overline{a}(t)\approx\frac{a_0^2\, \widehat{B}_{jj}\left( 4 \widehat{\omega}_j^2 -\overline{\omega}^2+\overline{\omega}^2 \cos (2 \widehat{\omega}_j t)-4\widehat{\omega}_j^2 \cos(\overline{\omega}t)\right)}{2\overline{C}\,  \overline{\omega}^2 \left( \overline{\omega}^2-4 \widehat{\omega}_j^2\right)}.
\end{equation}
The initial conditions \eqref{C1:IntialConditions} have been taken because only the $j$-th mode is activated and none of the others.

On the other hand, the rest of the amplitudes $\widehat{a}_m$ can be estimated by solving the differential equations \eqref{C1:ODEAmplitudesSecondComponent}, neglecting in them the terms $\overline{a}\; \widehat{a }_p$, with $p\neq j$, with the initial conditions \eqref{C1:IntialConditions}, which leads to {
\begin{eqnarray}\label{C1:OrthogonalAmplitudes}
&&\widehat{a}_m(t) = \widehat{a}_{\widehat{\omega}_j} \sin(\widehat{\omega}_j t)+ \widehat{a}_{3\widehat{\omega}_j} \sin(3\widehat{\omega}_j t)+ \widehat{a}_{\widehat{\omega}_m} \sin(\widehat{\omega}_m t)\\
&&\hspace{3.5cm}+\widehat{a}_{\widehat{\omega}_j+\overline{\omega}} \sin((\widehat{\omega}_j+\overline{\omega})t)+ \widehat{a}_{\widehat{\omega}_j-\overline{\omega}} \sin((\widehat{\omega}_j-\overline{\omega})t),\nonumber
\end{eqnarray}
}
where the amplitudes associated with each of the frequencies that appear in the expression \eqref{C1:OrthogonalAmplitudes} are
{\small
\begin{eqnarray}
     \widehat{a}_{\widehat{\omega}_j} \!\!&\!\!=\!\!&\!\! \frac{a_0^3 \, \widehat{B}_{jm} \, \widehat{B}_{jj} \left( 3 \overline{\omega}^2-8 \widehat{\omega}_j^2\right)}{2\overline{C} \, \widehat{C}_m \left( \widehat{\omega}_j^2-\widehat{\omega}_m^2\right) \overline{\omega}^2 \left( 4\widehat{\omega}_j^2-\overline{\omega}^2\right)}, \label{C1:OrthogonalAmplitudes1}
    \quad
    \widehat{a}_{3\widehat{\omega}_j}  =  \frac{-a_0^3 \,  \widehat{B}_{jm} \, \widehat{B}_{jj}}{2\overline{C} \, \widehat{C}_m \left( 9\widehat{\omega}_j^2-\widehat{\omega}_m^2 \right) \left( 4\widehat{\omega}_j^2-\overline{\omega}^2 \right)},
%    \label{C1:OrthogonalAmplitudes2} 
    \\
      \widehat{a}_{\widehat{\omega}_m}  \!\!&\!\!=\!\!&\!\! \frac{- 4 a_0^3 \,  \widehat{B}_{jm} \, \widehat{B}_{jj}  \, \widehat{\omega}_j^3 \left( 3\widehat{\omega}_j^2+5\widehat{\omega}_j^2-3\overline{\omega}^2 \right)}{\overline{C} \, \widehat{C}_m  \, \widehat{\omega}_m \left( 9\widehat{\omega}_j^4-10\widehat{\omega}_j^2 \, \widehat{\omega}_m^2+\widehat{\omega}_m^4 \right) \left( \widehat{\omega}_j^4+\left( \widehat{\omega}_m^2-\overline{\omega}^2 \right)^2 -2\widehat{\omega}_j^2 \left( \widehat{\omega}_m^2+\overline{\omega}^2 \right) \right) }, 
    \label{C1:OrthogonalAmplitudes3}
    \\
      \widehat{a}_{\widehat{\omega}_j+\overline{\omega}}  \!\!&\!=\!\!&\!\! \frac{-2 a_0^3 \, \widehat{B}_{jm} \, \widehat{B}_{jj}  \, \widehat{\omega}_j^2 }{\overline{C} \, \widehat{C}_m  \, \overline{\omega}^2  ( \widehat{\omega}_j-\widehat{\omega}_m+\overline{\omega}  )  ( \widehat{\omega}_j+\widehat{\omega}_m+\overline{\omega}  )  ( \overline{\omega}^2-4\widehat{\omega}_j^2  )}, 
    \label{C1:OrthogonalAmplitudes4}
    \\
    \widehat{a}_{\widehat{\omega}_j-\overline{\omega}}   \!\!&\!=\!\!&\!\! \frac{-2 a_0^3 \, \widehat{B}_{jm} \, \widehat{B}_{jj} \, \widehat{\omega}_j^2 }{\overline{C}  \widehat{C}_m  \overline{\omega}^2  ( \overline{\omega}^2-4\widehat{\omega}_j^2  )  ( \widehat{\omega}_j^2-\widehat{\omega}_m^2 -2\widehat{\omega}_j  \overline{\omega}+\overline{\omega}^2  )}.
    \label{C1:OrthogonalAmplitudes5}
\end{eqnarray}
}
From these formulas and from the parity of the shape modes, it can be shown that $\widehat{a}_m$ is zero when we consider shape modes with different parities. 
This is because  the integrand of \eqref{C1:Bjm} is odd  when $j+m$ is not an even number. This phenomenon will be studied in detail in Section~\ref{C1:Section4.4}, where we will compare these results with those obtained by numerical simulations.

When we substitute \eqref{C1:OrthogonalAmplitudes1}--\eqref{C1:OrthogonalAmplitudes5} into \eqref{C1:OrthogonalAmplitudes}, the resulting amplitudes lead to terms of order $\mathcal{O}(a_0^4)$ in  \eqref{C1:EquationRadiationFirstField}--\eqref{C1:EquationRadiationSecondField}, which will be ignored because we are only considering quantities up to the order $\mathcal{O}(a_0^3)$. In other words, the differential equations \eqref{C1:EquationRadiationFirstField} and \eqref{C1:EquationRadiationSecondField} can be approximated up to the order indicated by the following ones
\begin{eqnarray}
 \hskip-0.7cm    \overline{\eta}_{tt}-\overline{\eta}_{xx}+  ( 6 \phi_K^2-2 )\overline{\eta} \!\!&\!\! =\!\!&\!\!  \widehat{a}_j^2 \left[ \frac{\displaystyle \widehat{B}_{jj}}{\displaystyle \overline{C}}\overline{\eta}_D -
    2 \kappa  \phi_K  \widehat{\eta}_{D, j}^2\right],
    \label{C1:EquationRadiationFirstField2} 
    \\ %[1ex]
 \hskip-0.7cm        \widehat{\eta}_{tt}-\widehat{\eta}_{xx}+  ( 2 \kappa  \phi_K^2-2  ) \widehat{\eta}  \!\!&\! =\!&\!\! 2 \overline{a} \,\widehat{a}_j \left[\frac{  \widehat{B}_{jj}}{\widehat{C}_{ j}}  \widehat{\eta}_{D, j}-
    2  \kappa  \overline{\eta}_D  \phi_K    \widehat{\eta}_{D, j}\right]. \label{C1:EquationRadiationSecondField2}
\end{eqnarray}
Taking into account that from \eqref{C1:OthogonalAmplitudeEvolution}
\begin{equation}\label{C1:shapeModeAmplitudeSquare}
     \widehat{a}_j^2(t)=\frac{a_0^2}{2} \left( 1-\cos(2 \widehat{\omega}_j t )\right),
\end{equation}
and that the time-independent part of  \eqref{C1:shapeModeAmplitudeSquare} causes a time-independent response of  $\overline{\eta}$ that carries zero energy, then it is possible to rewrite \eqref{C1:EquationRadiationFirstField2} as
\begin{equation}\label{C1:EquationRadiationFirstField3}
     \overline{\eta}_{tt}-\overline{\eta}_{xx}+ \left( 6 \phi_K^2-2 \right)\,\overline{\eta}=f(x)\ e^{i2\widehat{\omega}_j t},
\end{equation}
where
\begin{equation}
    f(x)= -\frac{a_0^2}{2}\left(\frac{\widehat{B}_{jj}}{\overline{C}}\,\overline{\eta}_D-2\kappa\, \phi_K\, \widehat{\eta}_{D, j}^2\right).
\end{equation}
It is important to clarify that, to simplify the subsequent calculations as much as possible, we have taken imaginary exponentials in \eqref{C1:EquationRadiationFirstField3} instead of sines or cosines. In fact, as in the case of the $\phi^4$ wobbler, this does not affect the final analytical result at all since the relevant results can be retrieved simply by taking the real part of the final result.
On the other hand, from \eqref{C1:OthogonalAmplitudeEvolution} and \eqref{C1:LongitudinalAmplitudeEvolution} it can be obtained that 
{
\begin{equation}
     \overline{a}(t)\, \widehat{a}_j (t)\hspace{-0.1cm}=\hspace{-0.1cm}\frac{\displaystyle a_0^3 \, \widehat{B}_{jj}\left[4 \widehat{\omega}_j^2 \left( \sin{((\widehat{\omega}_j-\overline{\omega})t)}\hspace{-0.1cm}+\hspace{-0.1cm}\sin{((\widehat{\omega}_j+\overline{\omega})t)} \right) \hspace{-0.1cm}+\hspace{-0.1cm} \left( 3 \overline{\omega}^2-8 \widehat{\omega}_j^2  \right)\sin(\widehat{\omega}_j t )\hspace{-0.1cm}-\hspace{-0.1cm}\overline{\omega}^2 \sin(3\widehat{\omega}_j t) \right]}{\displaystyle 4\overline{C}  \, \overline{\omega}^2 \left(4 \widehat{\omega}_j^2-\overline{\omega}^2 \right)}.
\end{equation}
}
Clearly, in the expression we have just found, four relevant frequencies naturally appear, which are
\begin{equation}\label{C1:PossibleFrequencies}
    \omega_1=\widehat{\omega}_j,\quad \omega_2=3\widehat{\omega}_j,\quad  \omega_3=\widehat{\omega}_j+\overline{\omega},\quad   \text{and} \quad  \omega_4=|\widehat{\omega}_j-\overline{\omega}|.
\end{equation}
Following the same procedure used with the equation of the first field component $\overline{\eta}$, the formula \eqref{C1:EquationRadiationSecondField2} can be rewritten as
\begin{equation}\label{C1:EquationRadiationSecondField3}
     \widehat{\eta}_{tt}-\widehat{\eta}_{xx}+ \left( 2 \kappa \, \phi_K^2-2 \right)\,\widehat{\eta}=\sum_{\ell=1}^4 g_\ell(x) \ e^{i \omega_\ell t},
\end{equation}
where
\begin{eqnarray*}
\hskip-0.7cm  &&    g_1(x)  =  \frac{ a_0^3 \, \widehat{B}_{jj}\left( 3\overline{\omega}^2-8\widehat{\omega}_j^2 \right)}{2\overline{C}\,\overline{\omega}^2\left( 4\widehat{\omega}_j^2-\overline{\omega}^2 \right)}\left(\frac{  \widehat{B}_{jj}}{\widehat{C}_{j}}\,\widehat{\eta}_{D, j}-2 \kappa \,\overline{\eta}_D \,\phi_K \, \widehat{\eta}_{D, j}\right),\\
     && g_2(x)   =  \frac{- a_0^3\, \widehat{B}_{jj}}{2\overline{C} \left(4\widehat{\omega}_j^2-\overline{\omega}^2 \right)}\left(\frac{  \widehat{B}_{jj}}{\widehat{C}_{j}}\,\widehat{\eta}_{D, j}-2 \kappa \,\overline{\eta}_D \, \phi_K \, \widehat{\eta}_{D, j}\right),
      \\
 \hskip-0.7cm  &&     g_3(x)=g_4(x)   =  \frac{2 a_0^3\, \widehat{B}_{jj}\,\widehat{\omega}_j^2}{\overline{C}\,\overline{\omega}^2\left( 4\widehat{\omega}_j^2-\overline{\omega}^2 \right)}\left(\frac{  \widehat{B}_{jj}}{\widehat{C}_{j}}\,\widehat{\eta}_{D,j}-2 \kappa \,\overline{\eta}_D\, \phi_K  \,\widehat{\eta}_{D, j}\right).
\end{eqnarray*}
Under these circumstances, \eqref{C1:EquationRadiationFirstField3} and \eqref{C1:EquationRadiationSecondField3} can be solved if we separate the spatial and temporal part of the two $\eta(x,t)$-functions as follows
\begin{equation}
\overline{\eta}(x,t)= \overline{\eta}_{2\widehat{\omega}_j}(x)\ e^{i\, 2\widehat{\omega}_j t},
\quad
    \widehat{\eta}(x,t)= \sum_{\ell=1}^4\widehat{\eta}_{\omega_\ell} (x)\ e^{i \omega_\ell t},
\end{equation}
which  will lead to the following non-homogeneous linear ordinary differential equations ($\ell=1,2,3,4$):
\begin{eqnarray}
 -\overline{\eta}''_{2\widehat{\omega}_j}(x)+ \left( 6  \phi_K^2-2-4\widehat{\omega}_j^2 \right)\,\overline{\eta}_{2\widehat{\omega}_j}(x) \!\!&\!\!=\!\!&\!\!  f(x),
 \label{C1:EquationRadiationFirstField4}
 \\
    -\widehat{\eta}''_{\omega_\ell}(x)+ \left( 2 \kappa \, \phi_K^2-2-\omega_\ell^2 \right) 
    \widehat{\eta}_{\omega_\ell}(x) \!\!&\!\!=\!\!&\!\!  g_\ell(x),
    \label{C1:EquationRadiationSecondField4} 
\end{eqnarray}
If we now take into account the dispersion relations  for a longitudinal channel \eqref{C1:Continuousfrequencylongitudinal} of frequency $ 2\widehat{\omega}_j$ and the four orthogonal modes \eqref{C1:Continuousfrequencyorthogonal}, which are the $\omega_\ell$ given in equation \eqref{C1:PossibleFrequencies},
\begin{equation}
\bar{q}=\sqrt{4 \widehat{\omega}^2_j-4} , \qquad \widehat{q } _ \ell=\sqrt {\omega^2_\ell+2-2\kappa},
\end{equation}
 as well as the homogeneous solutions of \eqref{C1:EquationRadiationFirstField4} and \eqref{C1:EquationRadiationSecondField4}, which correspond to the expressions \eqref{C1:ContinuousEigenfunctionFirstField} and \eqref{C1:ContinuousEigenfunctionSecondField}, then the  solutions to the inhomogeneous equations \eqref{C1:EquationRadiationFirstField4}--\eqref{C1:EquationRadiationSecondField4} are given by the following functions:
\begin{eqnarray}
    \overline{\eta}_{2 \widehat{\omega}_j} \!\!&\!\!=\!\!&\!\! -\frac{  \overline{\eta}_{-{\bar{q}}}(x)}{\overline{W}_{\bar{q}}}\int_{-\infty}^{x}\overline{\eta}_{{\bar{q}}} (y)\, f(y)\, dy 
    -\frac{ \overline{\eta}_{{\bar{q}}} (x)}{\overline{W}_{\bar{q}}}\int_{x}^{\infty}\overline{\eta}_{-{\bar{q}}} (y)\, f(y)\, dy,\label{C1:SolutionRadiationFirstField}  \\[1ex]
    \widehat{\eta}_{\omega_\ell}  \!\!&\!\!=\!\!&\!\!  -\frac{ \widehat{\eta}_{-{\widehat{q}_\ell}}(x)}{\widehat{W}_{\widehat{q}_\ell}}\int_{-\infty}^{x}\widehat{\eta}_{{\widehat{q}_\ell}} (y)\, g_\ell(y)\, dy  -\frac{ \widehat{\eta}_{{\widehat{q}_\ell}}(x)}{\widehat{W}_{\widehat{q}_\ell}}\int_{x}^{\infty}\widehat{\eta}_{-{\widehat{q}_\ell}}(y)\, g_\ell(y)\, dy,  \label{C1:SolutionRadiationSecondField}
\end{eqnarray}
where $\overline{W}_{\bar{q}}$ and $\widehat{W}_{\widehat{q}_\ell}$ are given by \eqref{C1:WronskianoFirstField} and \eqref{C1:WronskianoSecondField}. The asymptotic behavior of \eqref{C1:SolutionRadiationFirstField} and \eqref{C1:SolutionRadiationSecondField} is
\begin{eqnarray}
     \overline{\eta}_{2 \widehat{\omega}_j}&\xrightarrow{x\rightarrow\infty}& \frac{i\left(\int_{-\infty}^{\infty }\overline{\eta}_{{\overline{q}}}(y)f(y)dy\right)}{2(\overline{q}+i)(\overline{q}+2i)} e^{-i\overline{q}x}\label{C1:FinalRadiationFirstField},\\ 
    \widehat{\eta}_{\omega_\ell}&\xrightarrow{x\rightarrow\infty}& \frac{\left(\int_{-\infty}^{\infty}\widehat{\eta}_{{\widehat{q}_\ell}}(y)g_\ell(y)dy\right)}{2 i \widehat{q}_\ell }\,  e^{-i \widehat{q}_\ell x},\label{C1:FinalRadiationSecondField}
\end{eqnarray}
which provides us with the amplitudes of the radiation that travels in the longitudinal and orthogonal channels respectively.
Unfortunately, the functions \eqref{C1:FinalRadiationFirstField} and \eqref{C1:FinalRadiationSecondField} cannot be calculated analytically since the integrals that are present in these formulas involve mixed hypergeometric and hyperbolic functions. Note that it can be seen that the amount of radiation propagated in the longitudinal channel is going to be greater than that emitted in the orthogonal channel because the term \eqref{C1:FinalRadiationFirstField} is proportional to $a_0^2$, while the term \eqref{C1:FinalRadiationSecondField} is proportional to $a_0^3$, being  $a_0$ a small parameter.

Now that the possible radiation frequencies are known, we have to find out which of them are capable of producing radiation.
In fact, for this to happen both $2\widehat{\omega}_j$ and $\omega_{\ell}$  in \eqref{C1:PossibleFrequencies} have to lie within the continuous vibration spectra of the components of the first and second fields, respectively \cite{AlonsoIzquierdo2023c}.
In other words, $\bar{q}$ and $\widehat{q}_\ell$ must both be positive real quantities. This can be verified from \eqref{C1:FinalRadiationFirstField} and \eqref{C1:FinalRadiationSecondField}, since, if the aforementioned dispersion relations were imaginary, then this would lead to divergences in the solutions \eqref{C1:SolutionRadiationFirstField}--\eqref{C1:SolutionRadiationSecondField} when we are far from the center of the kink.

\begin{figure}[htb]
\centering
 \includegraphics[width=0.49\textwidth]{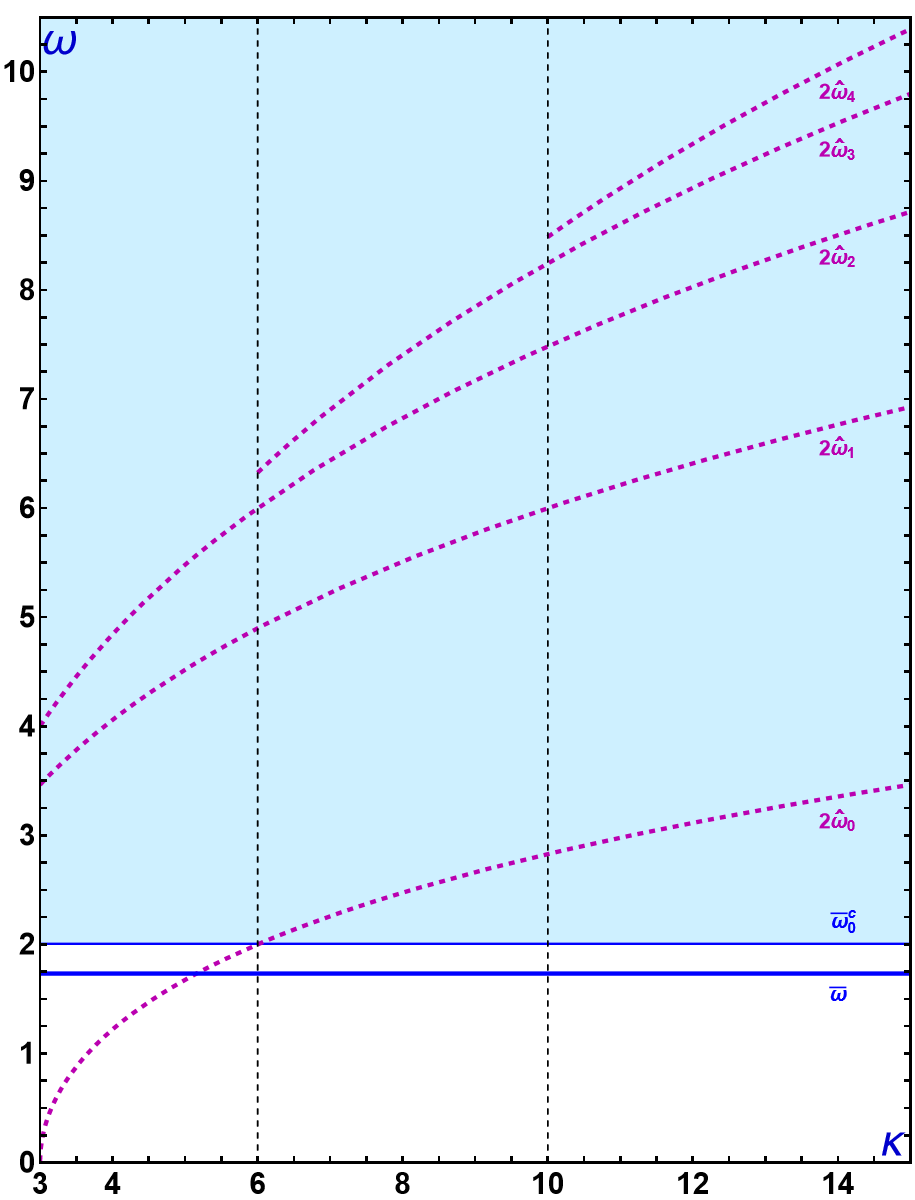}
\caption{\textit{Longitudinal eigenfrequencies of the operator $\mathcal{H}_{11}$ in \eqref{C1:SpectralProblem} as a function of the coupling constant $\kappa$. The black dashed lines are the values of $\kappa$ for which a new orthogonal mode arises in the spectrum of $\mathcal{H}_{22}$. The purple lines represent the lowest frequencies that can be excited by the coupling with the orthogonal fluctuations, which can be realized as radiation when plunged into the continuous spectrum (blue area).}}
                           \vspace{-0.4cm}
\label{C1:Fig:LongitudinalEspectrum}
\end{figure}

Below we are going to graphically illustrate part of the analytical results that we have obtained so far, to help us better understand the solutions to the problem that we are analyzing. 
 Thus, Figure~\ref{C1:Fig:LongitudinalEspectrum} shows the eigenfrequencies found in Section~\ref{C1:Section2} for the longitudinal fluctuations of $\mathcal{H}_{11}$ in \eqref{C1:SpectralProblem}, from which we can infer that if we initially activate only $\widehat{\eta}_{D,0} $, then we can only find radiation with frequency $2 \widehat{\omega}_0 $ in the regime $\kappa>6 $ for the longitudinal channel, since for $\kappa<6$ it happens that $2\widehat{\omega} _0< \overline{\omega}^c_0$.
 For higher modes, the frequencies $2\widehat{\omega}_n$ are always embedded in the continuous part of the spectrum $\mathcal{H}_{22}$.

\begin{figure}[htb]
         \centering
         \includegraphics[width=0.45\textwidth]{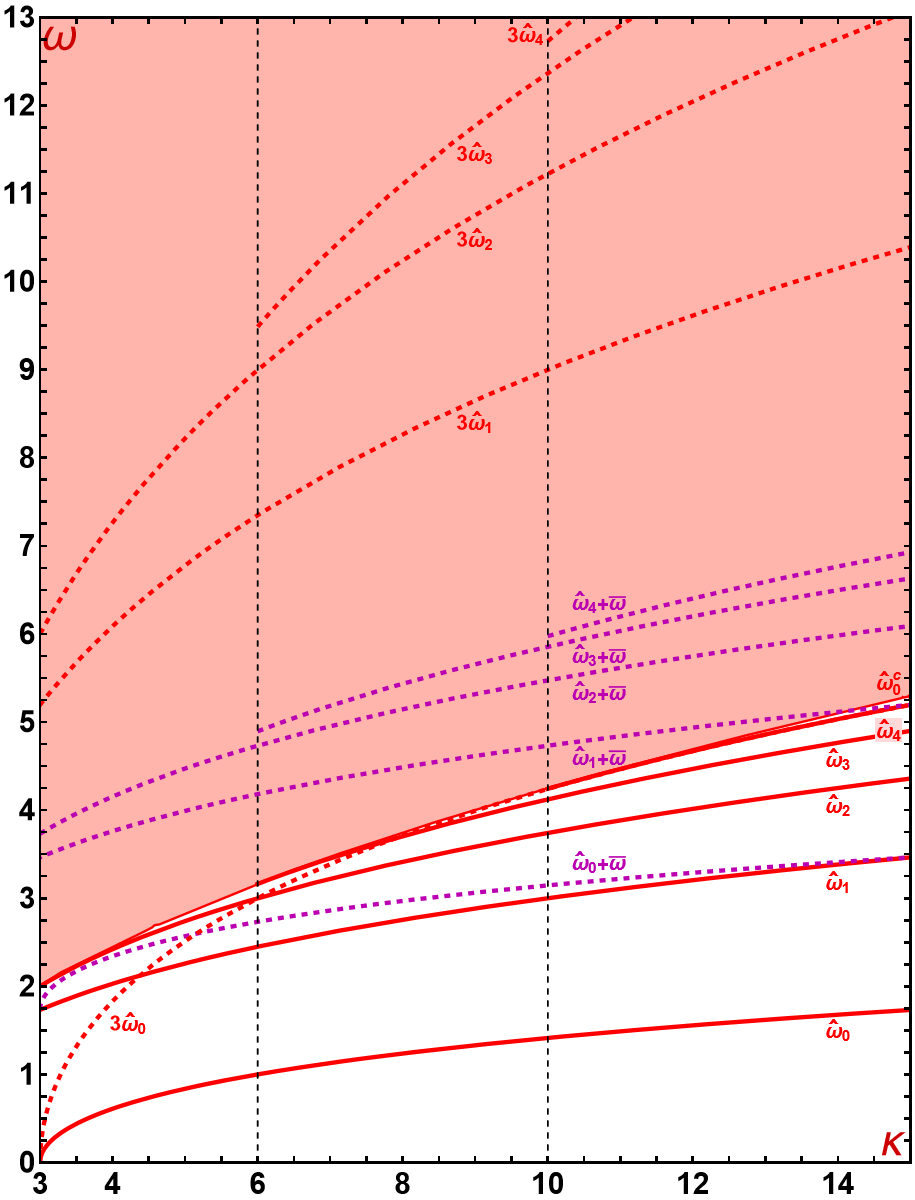}
         \qquad
         \includegraphics[width=0.45\textwidth]{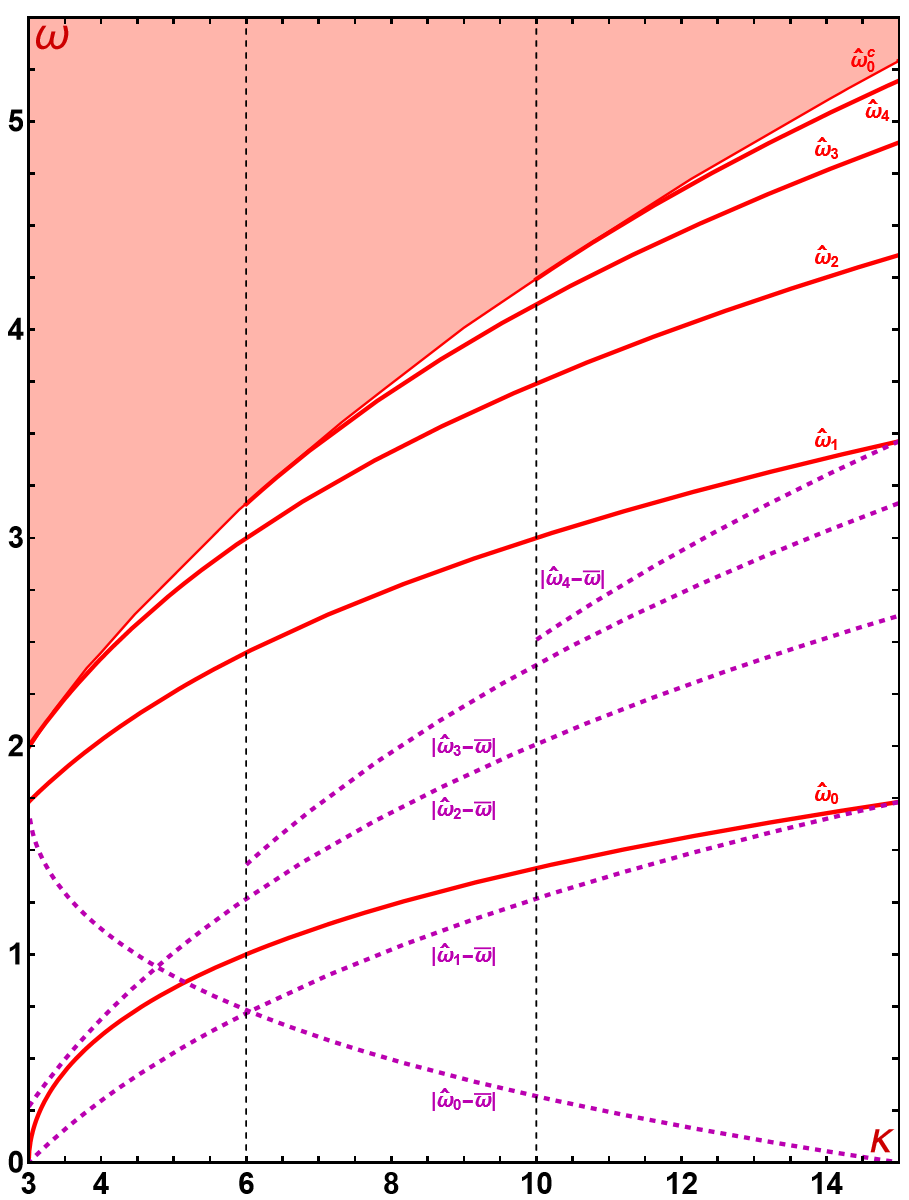}
\caption{\textit{On the left is part of the spectrum of the operator $\mathcal{H}_{22}$ that involves the orthogonal fluctuations as a function of the coupling constant $\kappa$. The graph on the right shows a zoom of the aforementioned spectrum near the threshold value $\widehat{\omega}^c_0$.
The black dashed vertical lines are the values of $\kappa$ for which a new orthogonal shape mode arises. 
The red and purple dotted lines represent the lowest frequencies that can be excited by coupling with longitudinal fluctuations, which can be realized as radiation when immersed in the continuous spectrum (red area).}}
                           \vspace{-0.5cm}
     \label{C1:Fig:OrthogonalSpectrum}
\end{figure}

The two graphs in Figure~\ref{C1:Fig:OrthogonalSpectrum} show the spectrum of the orthogonal operator $\mathcal{H}_{22}$, in addition to the frequencies $\omega_1,\dots\omega_4$ defined in \eqref{C1:PossibleFrequencies}. 
As we have already said, only the frequencies embedded in the continuous spectrum will be able to produce radiation. Following this reasoning, as we can see in the second of the drawings in Figure~\ref{C1:Fig:OrthogonalSpectrum}, the frequencies $|\widehat{\omega}_i-\overline{\omega}|$ are not embedded in the continuous part of the orthogonal channel spectrum, which implies that no radiation associated with these frequencies will be found. 
On the other hand, from the first of the drawings in Figure~\ref{C1:Fig:OrthogonalSpectrum} it can be inferred that $3\widehat{\omega}_0$ will not produce radiation in the second field component, although this frequency is ``almost" embedded in the continuous part of the spectrum around $\kappa=10$. 
Also note that $3\widehat{\omega}_0$ coincides with $\widehat{\omega}_4$ for $\kappa>10$.  The rest of the frequencies $3\widehat{\omega}_i$ are part of the radiation spectrum, which implies that radiation associated with these frequencies can be detected. On the other hand,  the frequencies $\widehat{\omega_i}+\overline{\omega}$ (except for $\widehat{\omega_0}+\overline{\omega}$)  are contained in the continuous spectrum only for a range of values of the coupling constant $\kappa$. For example, it can be shown that $\widehat{\omega}_1+\overline{\omega}>\widehat{\omega}^c_0$ when $3<\kappa<14.14$ and $\widehat{\omega}_2+\overline{\omega}>\widehat{\omega}^c_0$ when $3<\kappa<24.93$.

\begin{figure}[htb]
     \centering
         \includegraphics[width=0.55\textwidth]{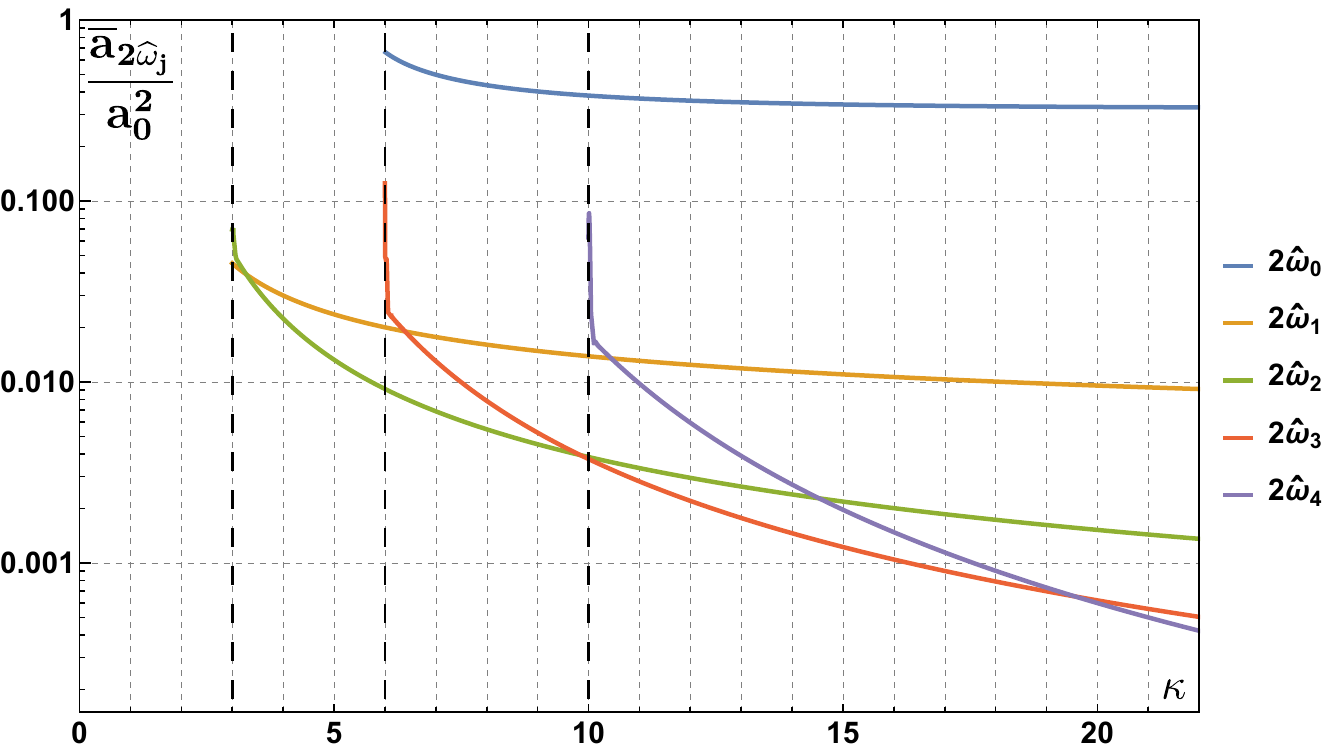}
                  \vspace{-0.3cm}
         \caption{\textit{Radiation amplitudes  associated with the frequencies  $ 2\widehat{\omega}_0,\dots 2\widehat{\omega}_4$ in the longitudinal  channel. The vertical dashed lines indicate the values of $\kappa$ for which a new shape mode arises in the spectrum of $\mathcal{H}_{22}$.}}
         \label{C1:Fig:LongitudinalAmplitudes}
\end{figure}

\begin{figure}[htb]
     \centering
         \includegraphics[width=0.55\textwidth]{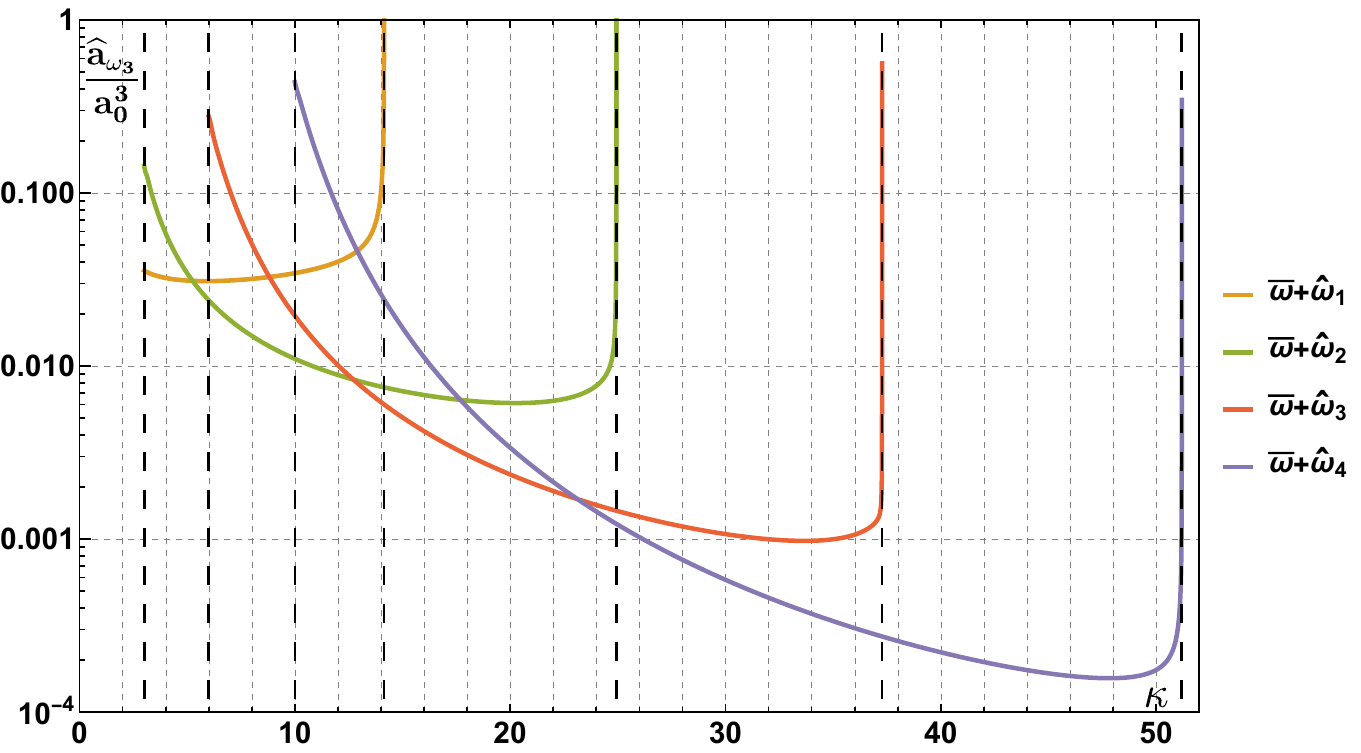}
                  \vspace{-0.3cm}
         \caption{\textit{Radiation amplitudes  associated with the frequencies $ \overline{\omega}+\widehat{\omega}_j$ ($j=1,\dots,4$) in the orthogonal channel. The dashed lines indicate the values of $\kappa$ for which a new shape mode arises in the spectrum of $\mathcal{H}_{22}$ and for which a radiation term disappears.}}
         \label{C1:Fig:OrthogonallAmplitudes1}
\end{figure}

\begin{figure}[htb]
     \centering
         \includegraphics[width=0.55\textwidth]{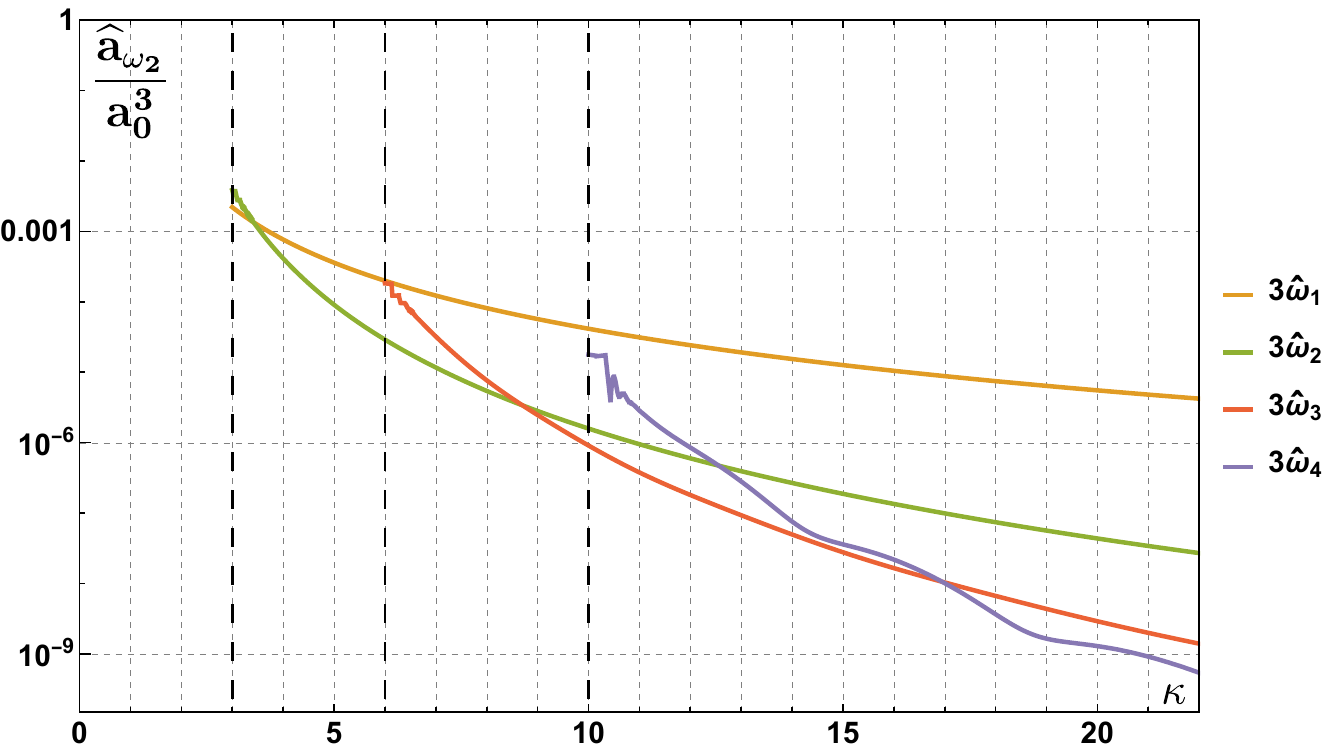}
         \vspace{-0.3cm}
         \caption{\textit{Radiation amplitudes  associated with the frequencies $ 3\widehat{\omega}_1,\dots 3\widehat{\omega}_4$ in the orthogonal  channel. The dashed lines indicate the values of $\kappa$ for which a new shape mode arises in the spectrum of $\mathcal{H}_{22}$.}}
         \label{C1:Fig:OrthogonallAmplitudes2}
\end{figure}

Next we focus on obtaining the radiation amplitudes associated with each frequency as a function of the coupling constant $\kappa$, depending on which orthogonal mode is initially activated. For this we must use \eqref{C1:FinalRadiationFirstField}--\eqref{C1:FinalRadiationSecondField}, which must necessarily be evaluated numerically, showing the results in Figures~\ref{C1:Fig:LongitudinalAmplitudes}--\ref{C1:Fig:OrthogonallAmplitudes2}.
More specifically, the real part of the result obtained in \eqref{C1:FinalRadiationFirstField} must be taken for the longitudinal channel amplitudes and the imaginary part of the result found in \eqref{C1:FinalRadiationSecondField} for the orthogonal channel amplitudes.

In Figure~\ref{C1:Fig:LongitudinalAmplitudes} we show the behavior of longitudinal radiation amplitudes when the first five orthogonal modes are activated separately.

As we can see, as $\kappa$ grows, the radiation amplitudes get smaller, which can be explained by the fact that higher frequencies are more difficult to trigger, and as $\kappa$ grows, the gap between $2\widehat {\omega}_i$ and the threshold $\overline{\omega}^c_0$ also increases.
Furthermore, it can be seen that the radiation emitted when we excite $\widehat{\eta}_{D,0}$ is much bigger than when we trigger  higher shape modes. 
In addition to all this, it is important to point out that, for large values of $\kappa$, the amplitudes associated with the higher modes are smaller than those corresponding to the first modes. 
For example, for $\kappa>20$, $\overline{a}_{2\widehat{\omega}_0}>\overline{a}_{2\widehat{\omega}_1}>\overline{a}_{2\widehat{\omega}_2}>\overline{a}_{2\widehat{\omega}_3}>\overline{a}_{2\widehat{\omega}_4}$. 
Another notable phenomenon is that the radiation associated with the frequency $2\widehat{\omega}_0$ begins when $\kappa=6$. 
As mentioned above, this is because when $\kappa<6$ the aforementioned frequency is not embedded in the continuous part of the spectrum of the longitudinal channel (see Figure \ref{C1:Fig:LongitudinalEspectrum}).

On the other hand, in Figure \ref{C1:Fig:OrthogonallAmplitudes1} we can see the graphs corresponding to the amplitudes of the orthogonal radiation emitted in $\overline{\omega}+\widehat{\omega}_j$, $j=1,\dots,4$, as a function of $\kappa$.
In this case, these frequencies only emit radiation in a certain range of values of $\kappa$.
In other words, the coupling between $\widehat{\eta}_{D,j}$ and $\overline{\eta}_D$ only produces radiation at $\overline{\omega}+\widehat{\omega}_j $ for a value of $\kappa$ that is greater than the minimum value for which the orthogonal shape mode arises and less than the critical value for which this frequency no longer belongs to the continuous part of the spectrum of the second channel.

Note also that the amplitude of the radiation diverges near the value of $\kappa$ where there is a resonance between $\overline{\omega}+\widehat{\omega}_i$ and the threshold value of $\widehat{\omega}_c$. 
These resonance structures must be addressed by other analytical methods due to the fact that these limits are outside the range of validity where our perturbative approach works well.
Furthermore, the first orthogonal shape mode cannot trigger radiation in the second field component because this frequency is always below the continuous spectrum.

Finally,  Figure \ref{C1:Fig:OrthogonallAmplitudes2} contains the graphs of the amplitudes corresponding to the radiation emitted at frequencies $3\widehat{\omega}_1,\dots 3\widehat{\omega}_4$. 
Note that, as in Figure \ref{C1:Fig:LongitudinalAmplitudes}, for large values of $\kappa$ the amplitudes associated with the higher modes are smaller that those corresponding to the first shape modes.  In fact, in this case the radiation amplitudes are much smaller than in the previous cases. 
This phenomenon can be explained taking into account that the higher frequencies are more difficult to excite, as can be seen in Figures \ref{C1:Fig:LongitudinalEspectrum} and \ref{C1:Fig:OrthogonalSpectrum}: the frequencies $3\widehat{\omega }_j $ are much larger than $2\widehat{\omega}_j$ and $\widehat{\omega}_j+\overline{\omega}$.
Note also that the orthogonal channel radiation terms  are proportional to $a_0^3$, making them much smaller than the radiation propagated in the longitudinal channel, which is proportional to $a_0^2$.

In the next section we will compare all the analytical results that we have just developed with those obtained through numerical simulations.

%----------------------------------Numerical Results: Comparison with the analitical ones------------------------

\section{Numerical analysis}\label{C1:Section4} %: Comparison with the analytical ones

Once we have developed the perturbative method for the problem we are analyzing in the preceding section, it now seems reasonable to compare the results obtained there with those that arise when the field equations \eqref{eqI2:FieldEqDouble1}--\eqref{eqI2:FieldEqDouble2} are solved numerically. To carry out these simulations, the aforementioned nonlinear partial differential equations have been discretized using an explicit fourth-order finite difference algorithm implemented with fourth-order Mur boundary conditions \cite{AlonsoIzquierdo2021} in the spatial interval $x\in (- 100,100)$ for a time $0<t<1200$. The initial configuration is determined by \eqref{C1:InitialHypothesis}, that is, the same one used in the perturbation approach developed in Section~\ref{C1:Section3}.
Specifically, the simulations have been run for initial configurations for which one of the first three orthogonal eigenmodes has been excited and for various initial amplitudes. To study the radiation emitted by the wobbling kink and its internal vibration, the Fast Fourier Transform algorithm has been implemented at several points on the real axis to obtain the spectral data. 
This analysis has been carried out at points far from both the center of the kink ($x_B$) and the points where the shape modes have their maxima. 
For $\overline{\eta}$ the maximum is 
\begin{equation}
x_M=\ln(1+\sqrt{2}),
\end{equation}
and  for the first three orthogonal modes $\widehat{\eta}_{D,0}$, $\widehat{\eta}_{D,1}$ and $\widehat{\eta}_{D,2}$ the maxima are, respectively,
\begin{eqnarray}
x_{M0}\!\!&\!\! = \!\!&\!\!x_0=0,
\\ 
x_{M1 }\!\!&\!\! = \!\!&\!\!\arctanh\left(\sqrt{\frac{2}{\sqrt{8\kappa+1}-1}}\, \right), 
\\
x_{M2}\!\!&\!\! = \!\!&\!\!\frac{1} {2}\arccosh \left(\frac{-5+4\kappa+\sqrt{1+8\kappa}}{4(2+\kappa-\sqrt{1+8\kappa})}\,\right),
\end{eqnarray}

For ease of presentation, this section will be organized as follows: in Section~\ref{C1:Section4.1} we will check whether the assumption of a constant amplitude associated with the excited orthogonal shape mode is true in the perturbative regime.  
In Section~\ref{C1:Section4.2} we will discuss how the initially triggered shape mode couples with longitudinal vibration mode. 
In Section~\ref{C1:Section4.3} the radiation emitted by the kink along the longitudinal and orthogonal channels will be analyzed. 
In Section~\ref{C1:Section4.4} a study of the coupling between orthogonal modes will be addressed. Finally, in Section \ref{C1:Section4.5} we will present an analytical explanation of the energy loss of the first orthogonal shape mode when the initial amplitude is increased.

\subsection{Hypothesis validation (first order in  \texorpdfstring{$a_0$}{a0})}\label{C1:Section4.1}

\begin{figure}[htb]
     \centering
         \includegraphics[width=0.47\textwidth]{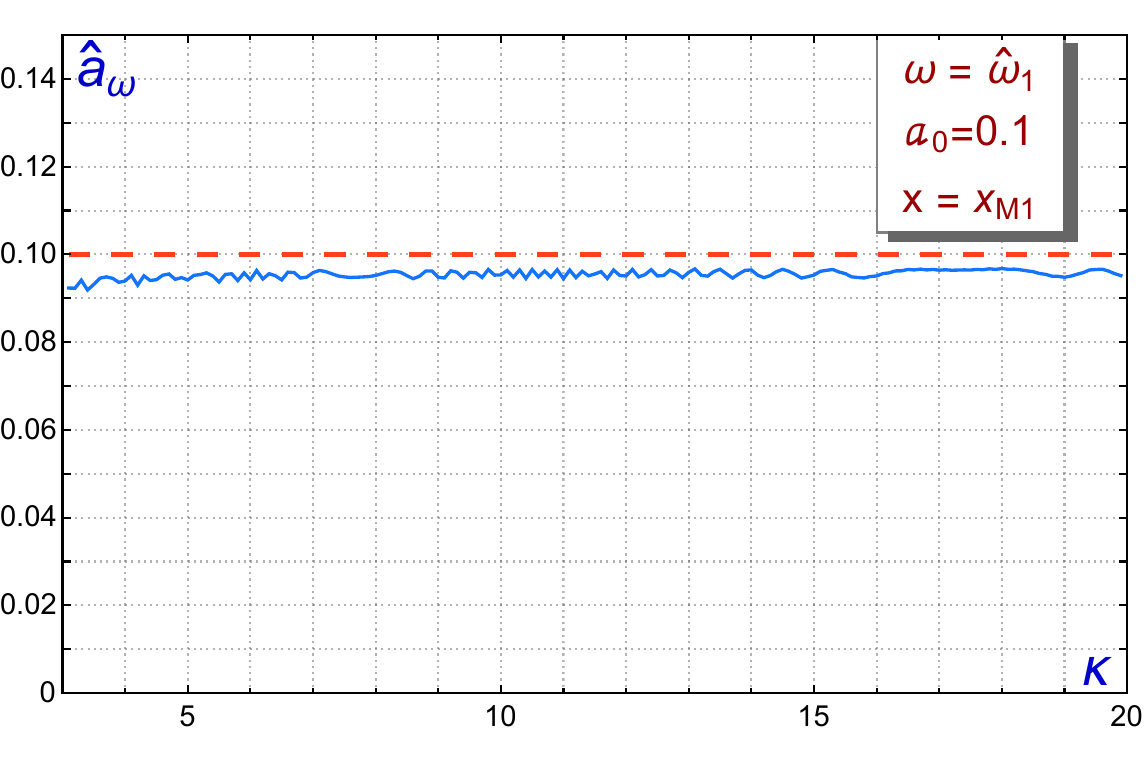}\qquad
         \includegraphics[width=0.47\textwidth]{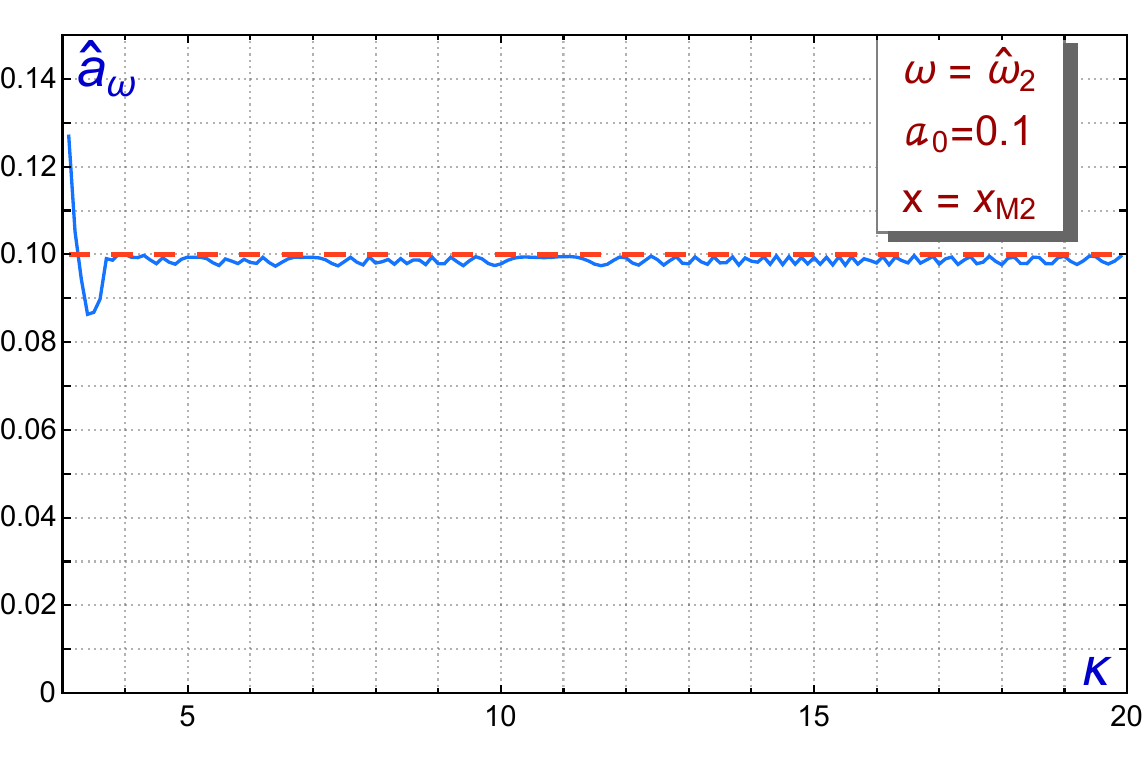}
      %                     \vspace{-0.3cm}
     \caption{\textit{Vibration amplitudes (blue curve) associated with  $\widehat{\eta}_{D,1}$ (first drawing) and $\widehat{\eta}_{D,2}$ (second drawing) as a function of $\kappa$  when the second and third shape modes are initially activated. The red dashed line corresponds to the analytical hypothesis \eqref{C1:OthogonalAmplitudeEvolution}.}}
     \label{C1:Fig:Mode0-1-2-OrthogonalAmplitude}
\end{figure}

In Section~\ref{C1:Section2} it was assumed that, when an orthogonal mode is activated, its associated amplitude remains essentially constant \eqref{C1:OthogonalAmplitudeEvolution}.
In Figure \ref{C1:Fig:Mode0-1-2-OrthogonalAmplitude} it can be seen that this hypothesis agrees quite well with the numerical results for various typical values, specifically  for $\widehat{\eta}_{D,1}$ (first drawing) and $\widehat {\eta}_{D,2}$ (second drawing) with $a_0\approx \mathcal{O}(0.1)$.

\begin{figure}[htb]
     \centering
     \includegraphics[width=0.47\textwidth]{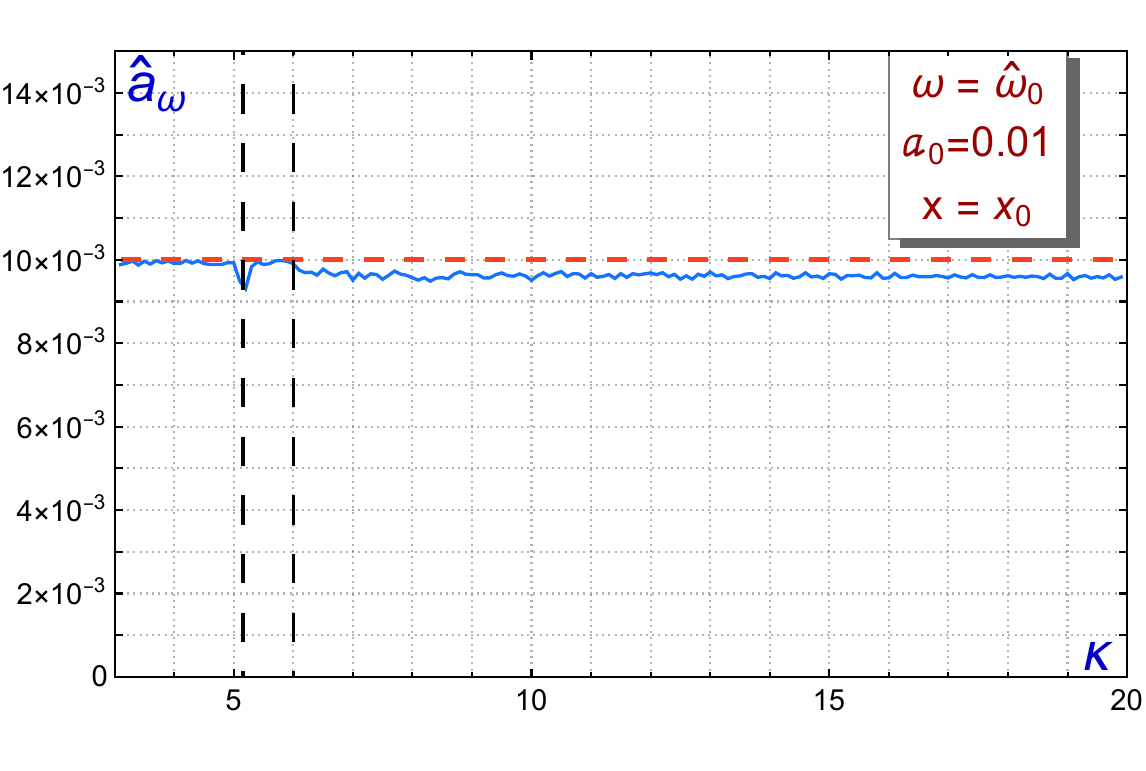}\qquad
         \includegraphics[width=0.47\textwidth]{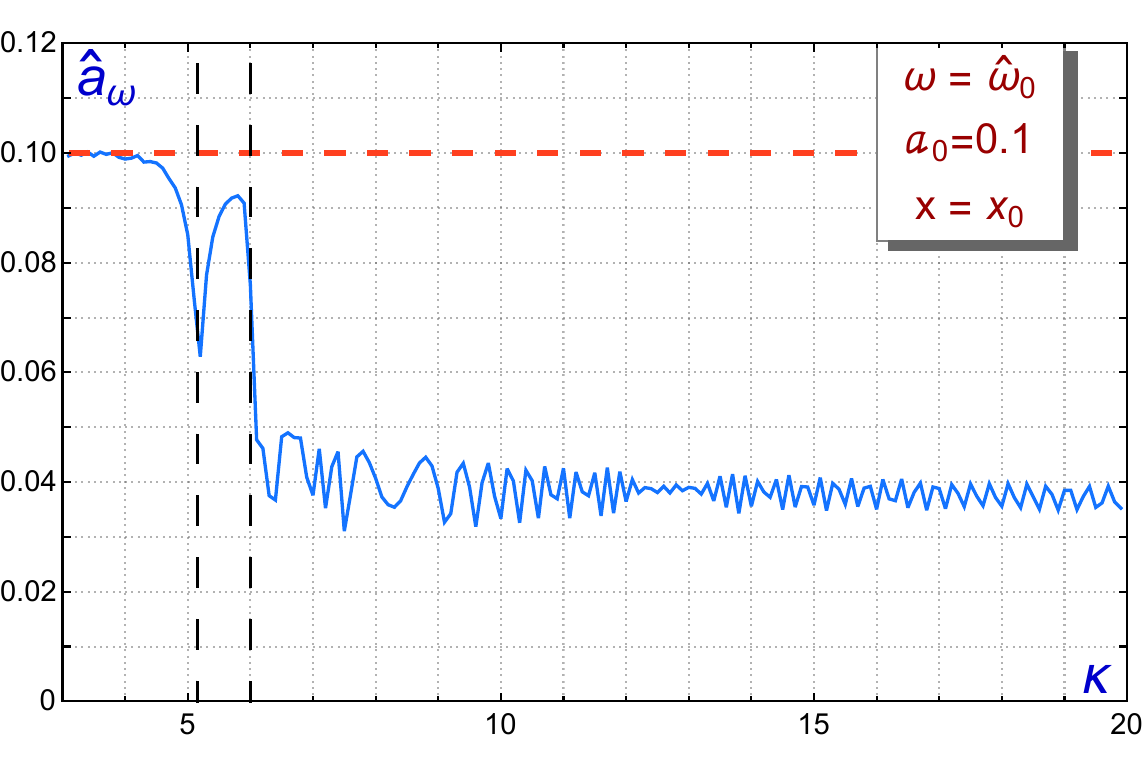}
 \qquad
          \includegraphics[width=0.47\textwidth]{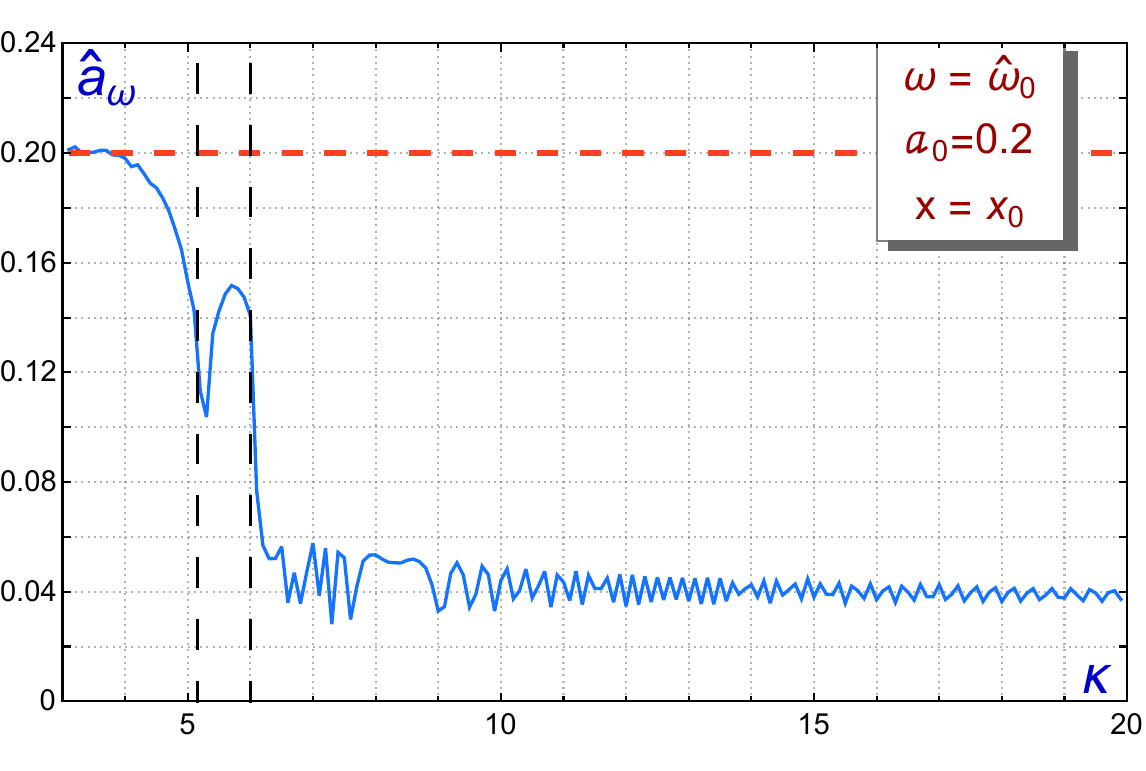}
                            %\vspace{-0.3cm}
     \caption{\textit{Vibration amplitudes associated with $\widehat{\eta}_{D,0}$ as a function of $\kappa$,  for $a_0=0.01, 0.1$ and $0.2$.}}
     \label{C1:Fig:Mode0rthogonalAmplitude}
\end{figure}

However, in Figure~\ref{C1:Fig:Mode0rthogonalAmplitude}  it can be seen that this hypothesis works fine for amplitudes of order $a_0\approx\mathcal{O }(0.01)$ for $\widehat{\eta}_{D,0}$, but fails when considering the simulations  with $a_0\approx\mathcal{O }(0.1)$.

Indeed, a large decrease in the initial amplitude can be observed for $\kappa>6$, which is the regime in which the kink is capable of emitting radiation with frequency $2\widehat{\omega}_0$. This phenomenon can be explained if we take into account that for large values of $t$ part of the vibration energy is dissipated in the form of radiation. 
In fact, the amplitude of the radiation emitted when $\widehat{\eta}_{D,0}$ is excited is much larger than when higher shape modes are activated (see Figure~\ref{C1:Fig:LongitudinalEspectrum}). This is also the reason why this decay is much smaller when we consider the simulations performed for $\widehat{\eta}_{D,1}$ and $\widehat{\eta}_{D,2}$ (see Figure~\ref{C1:Fig:Mode0-1-2-OrthogonalAmplitude}).
Since the radiation emitted when we consider $\widehat{\eta}_{D,1}$ is greater than that emitted when we consider  
$\widehat{\eta}_{D,2}$, then the observed decrease in $ a_0$ in this last case will be less than for the second shape mode (see  Figure~\ref{C1:Fig:Mode0-1-2-OrthogonalAmplitude}). A decay law for this amplitude will be discussed in Section~\ref{C1:Section4.5} taking into account the radiation emitted by the wobbling kink.
In addition to all that has been mentioned above, analyzing the Figures~\ref{C1:Fig:LongitudinalEspectrum} and \ref{C1:Fig:Mode0rthogonalAmplitude}, an additional decrease in the values of $a_0$ can be observed  for $\kappa\approx 5.15$, which is the value for which $2\widehat{\omega}_0=\overline{\omega}$.

\subsection{Amplitude of the longitudinal shape mode (second order in \texorpdfstring{$a_0$}{a0})}\label{C1:Section4.2}

\begin{figure}[htb]
     \centering
         \includegraphics[width=0.47\textwidth]{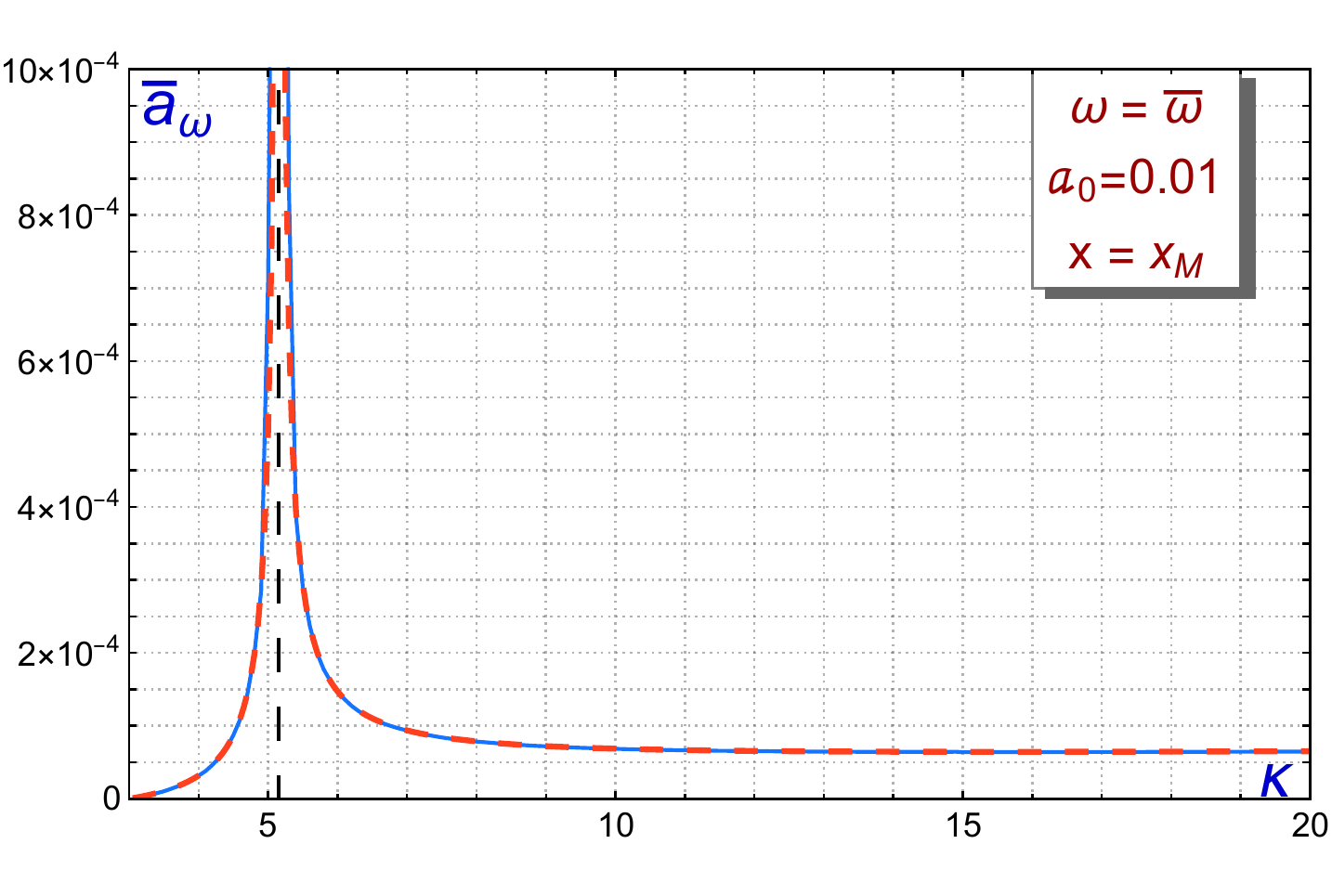}
\qquad
         \includegraphics[width=0.47\textwidth]{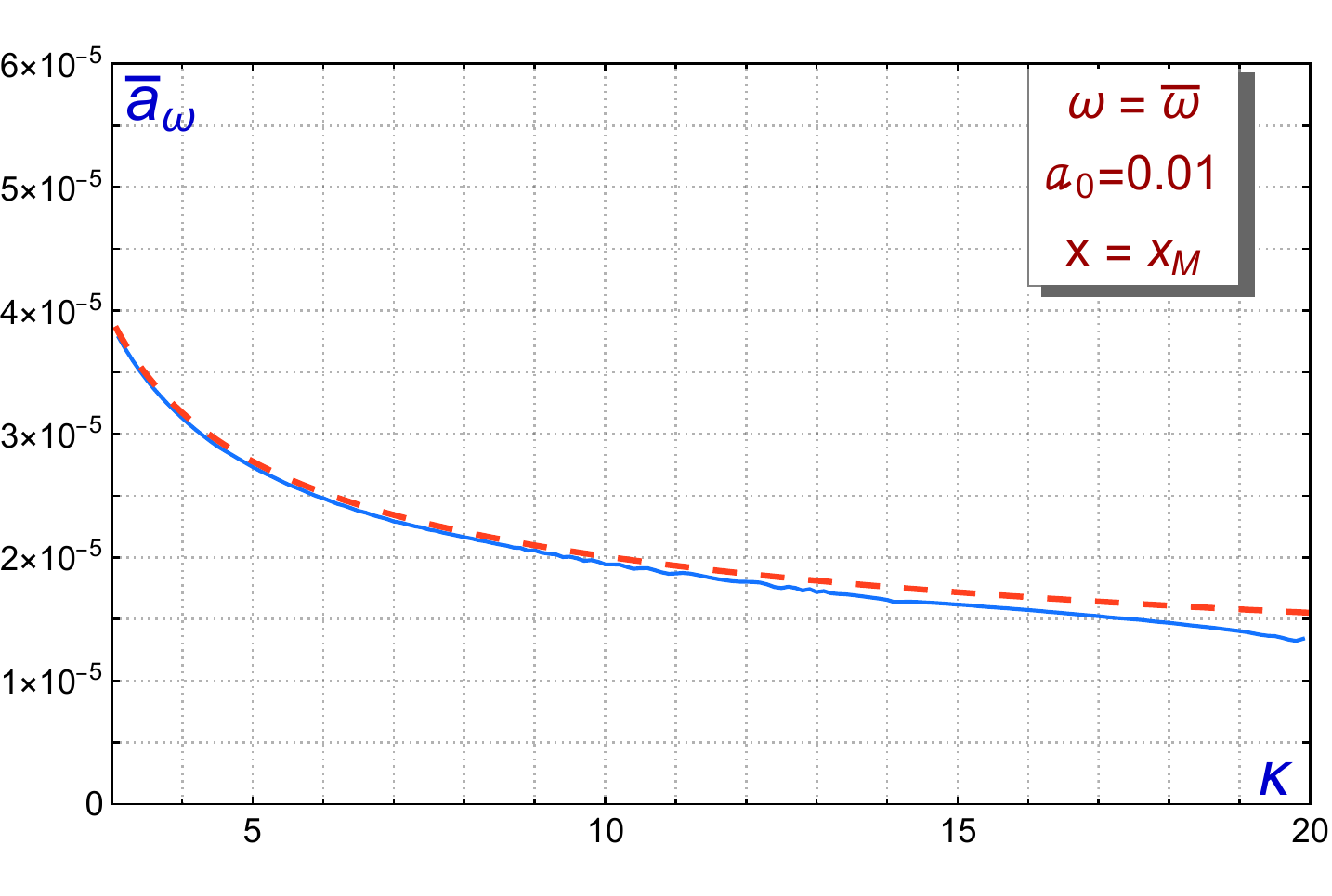}
\qquad
         \includegraphics[width=0.47\textwidth]{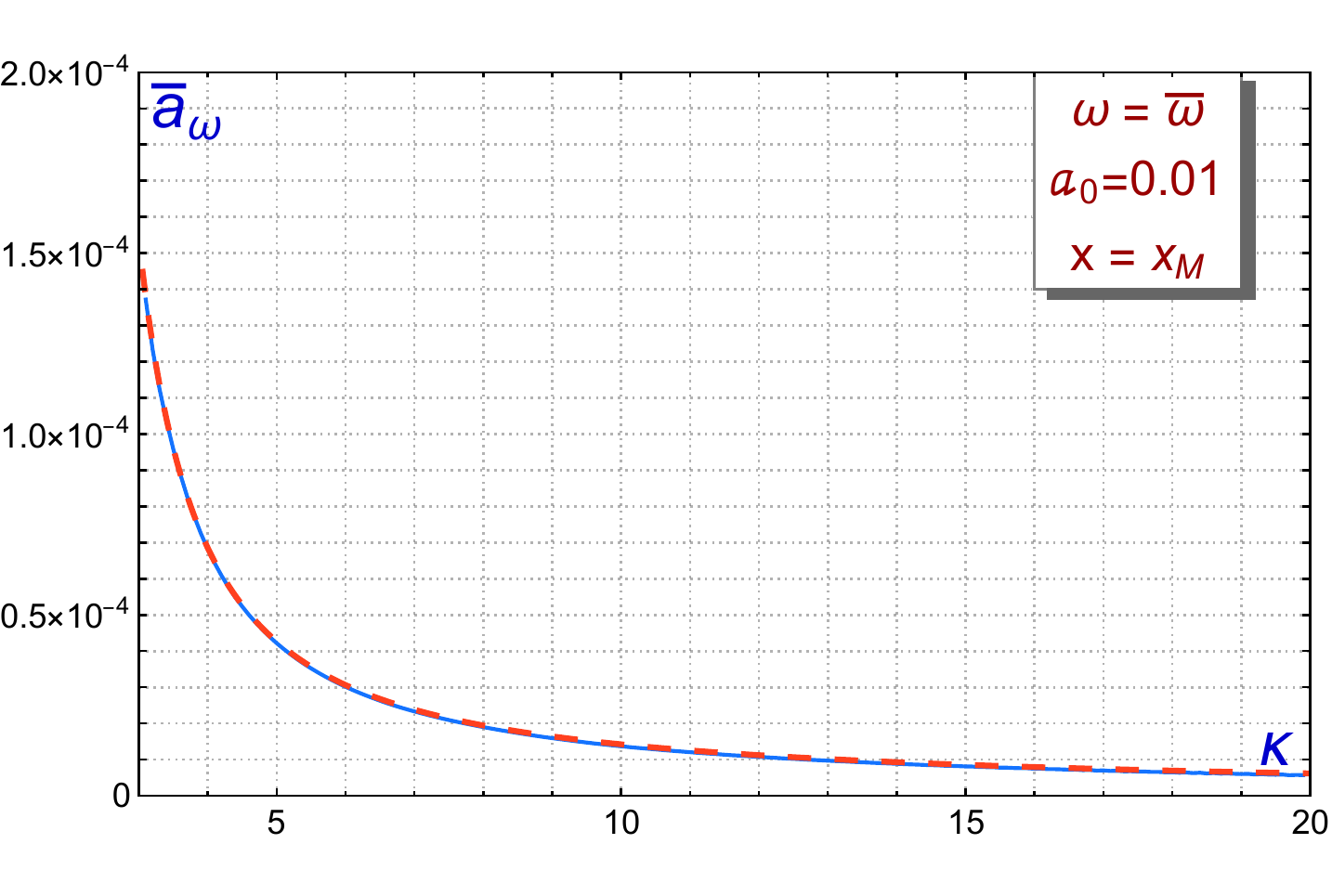}
                           %\vspace{-0.3cm}
     \caption{\textit{Vibration amplitudes (blue curves) of the longitudinal mode with frequency $\overline{\omega}$ in \eqref{C1:LongitudinalAmplitudeEvolution} when $\widehat{\eta}_{D,0}$ (first drawing), $\widehat {\eta}_{D,1}$ (second drawing) and $\widehat{\eta}_{D,2}$ (third drawing) are triggered, always with $a_0=0.01$. The red dashed lines correspond to the analytical estimate.}}
     \label{C1:Fig:Mode0-1-2-LongitudinalAmplitude}
\end{figure}

In Figure~\ref{C1:Fig:Mode0-1-2-LongitudinalAmplitude} it can be seen that the analytical estimate made in Section~\ref{C1:Section2} is fully consistent with the numerical simulations we have developed.    

It is worth mentioning that in the first drawing of Figure~\ref{C1:Fig:Mode0-1-2-LongitudinalAmplitude} it is possible to observe a resonance for the value $\kappa=\frac{165}{32}\approx5.15$, which coincides with the decrease in the amplitude of the shape mode observed in the previous section, specifically in the first drawing of Figure~\ref{C1:Fig:Mode0-1-2-OrthogonalAmplitude} and in Figure~\ref{C1:Fig:Mode0rthogonalAmplitude}.
This means that a large amount of energy  is transferred from the shape mode to the longitudinal mode for this particular value of the coupling constant, where it happens that $2\widehat{\omega}_0=\overline{\omega}$. 
In fact, we will see in Section~\ref{C1:Section4.4} that part of this energy is also transferred to $\widehat{\eta}_{D,2}$ when $\kappa=6$.

It can also be seen that, for large values of $\kappa$, when considering higher shape modes, the amplitude of the longitudinal shape mode becomes smaller and smaller.
Another remarkable phenomenon is that, for $\kappa\gg1$, the amplitude of the shape mode in \eqref{C1:LongitudinalAmplitudeEvolution} can be approximated as 
\begin{equation}
   \overline{a}_{\overline{\omega}}\approx \frac{a_0^2\, \widehat{B}_{jj}}{2\, \overline{C}\ \overline{\omega}^2}
=\frac{a_0^2\, \widehat{B}_{jj}}{4}.
\end{equation}

\subsection{Radiation amplitudes (second-third order in \texorpdfstring{$a_0$}{a0})}\label{C1:Section4.3}

\begin{figure}[htb]
     \centering
         \includegraphics[width=0.47\textwidth]{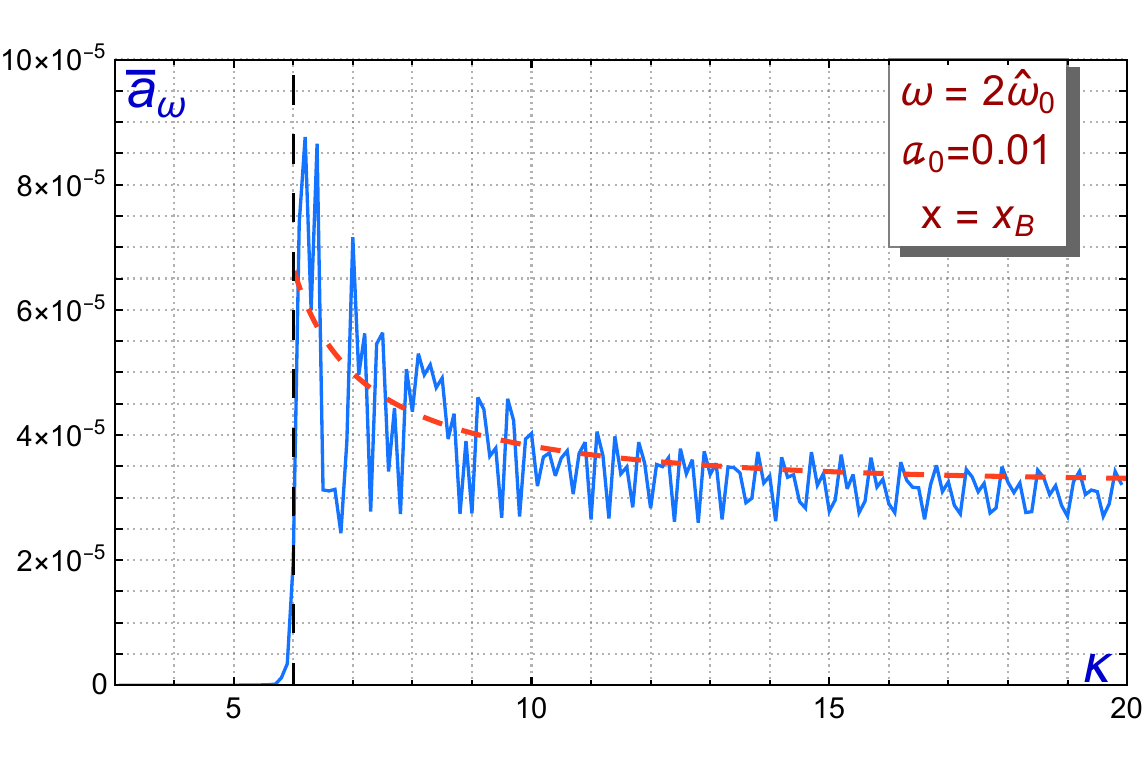} 
         \includegraphics[width=0.47\textwidth]{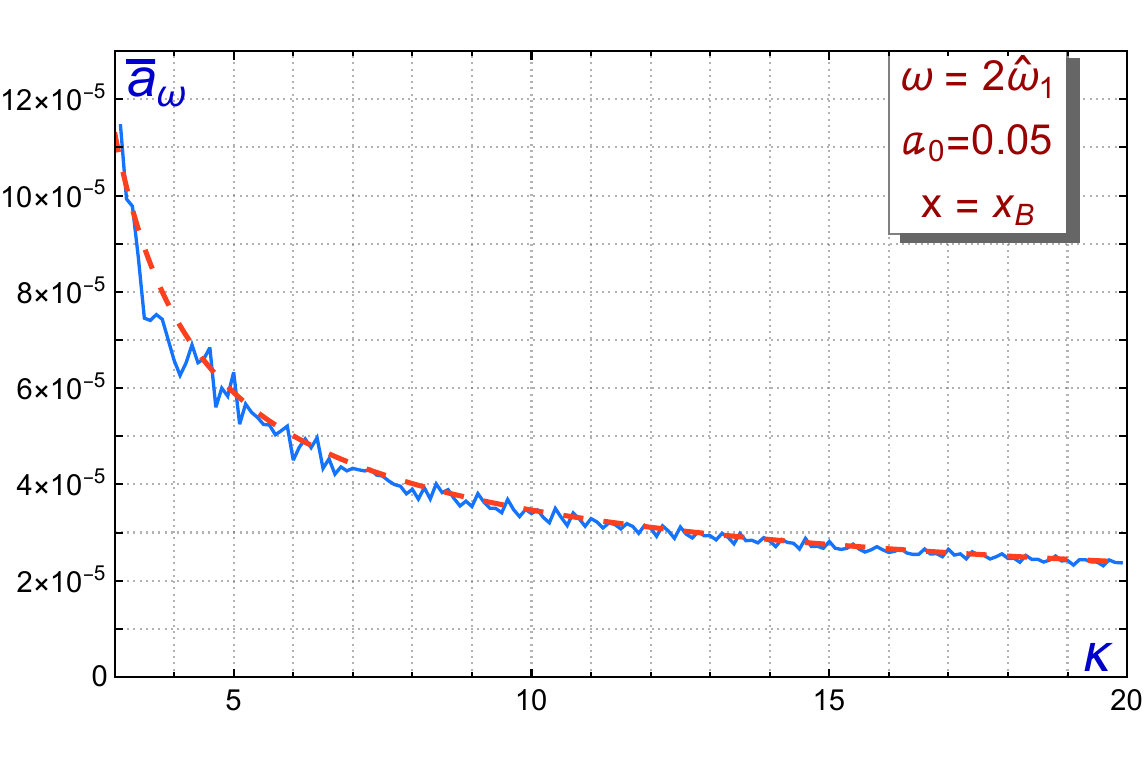} 
         \includegraphics[width=0.47\textwidth]{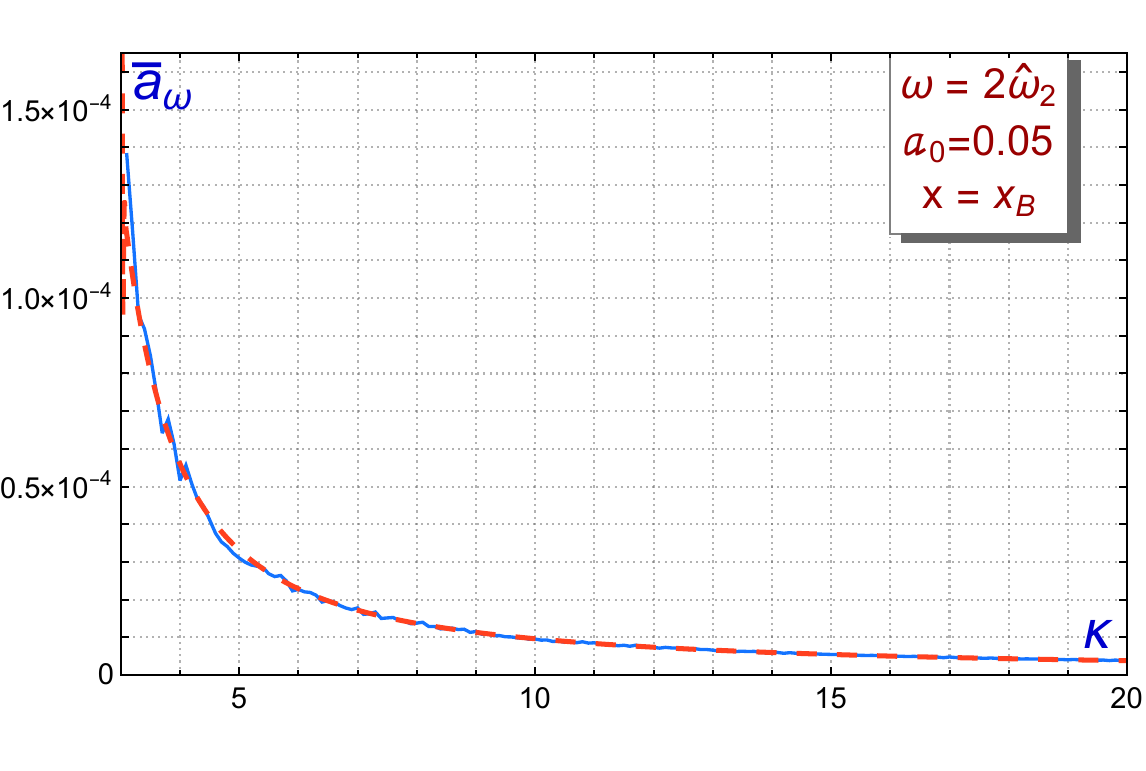}                  \vspace{-0.3cm}
                  %\vspace{-0.3cm}
     \caption{\textit{Radiation amplitudes (blue curves) emitted   at the frequency $2\widehat{\omega}_j$ by the kink when we trigger $\widehat{\eta}_{D,0}$ (first drawing), $\widehat{\eta}_{D,1}$ (second drawing) and $\widehat{\eta}_{D,2}$ (third  drawing). 
     The red dotted lines represent the analytical prediction \eqref{C1:FinalRadiationFirstField}.}}
     \label{C1:Fig:RadiationLongitudinalNumeric}
\end{figure}

In Section~\ref{C1:Section3} we show  that the kink is capable of emitting radiation  when we trigger at least one of its shape modes. 
In this section we will compare  the  radiation amplitudes obtained by numerical simulations with the theoretical predictions for the longitudinal  and  orthogonal channels  separately. 
Firstly we will focus on studying the behavior of longitudinal radiation.
Note that in the first drawing of  Figure~\ref{C1:Fig:RadiationLongitudinalNumeric} it can be seen that the excited kink cannot emit radiation with a frequency $2\widehat{\omega}_0$ when $\kappa<6$. 
This phenomenon is due to the  fact that the frequency $2\widehat{\omega}_0$ is only embedded into the continuous spectrum of the longitudinal mode when $\kappa>6$ (see Figure~\ref{C1:Fig:LongitudinalEspectrum}). 
On the other hand, since $2\widehat{\omega}_1$ and $2\widehat{\omega}_2$ are always included in the continuous spectrum, radiation associated with these frequencies will always be found, as can be seen in the second and third drawings of Figure~\ref{C1:Fig:RadiationLongitudinalNumeric}, in which it can also be seen that the numerical results coincide very well with the analytical ones.
Note also that the highest radiation amplitude is the one associated with $\widehat{\eta}_{D,0}$ and that, when  considering higher shape modes, the radiation amplitudes corresponding to the frequency $2\widehat{\omega}_j$ become smaller, as predicted in Section~\ref{C1:Section3}.

\begin{figure}[htb]
     \centering
         \includegraphics[width=0.47\textwidth]{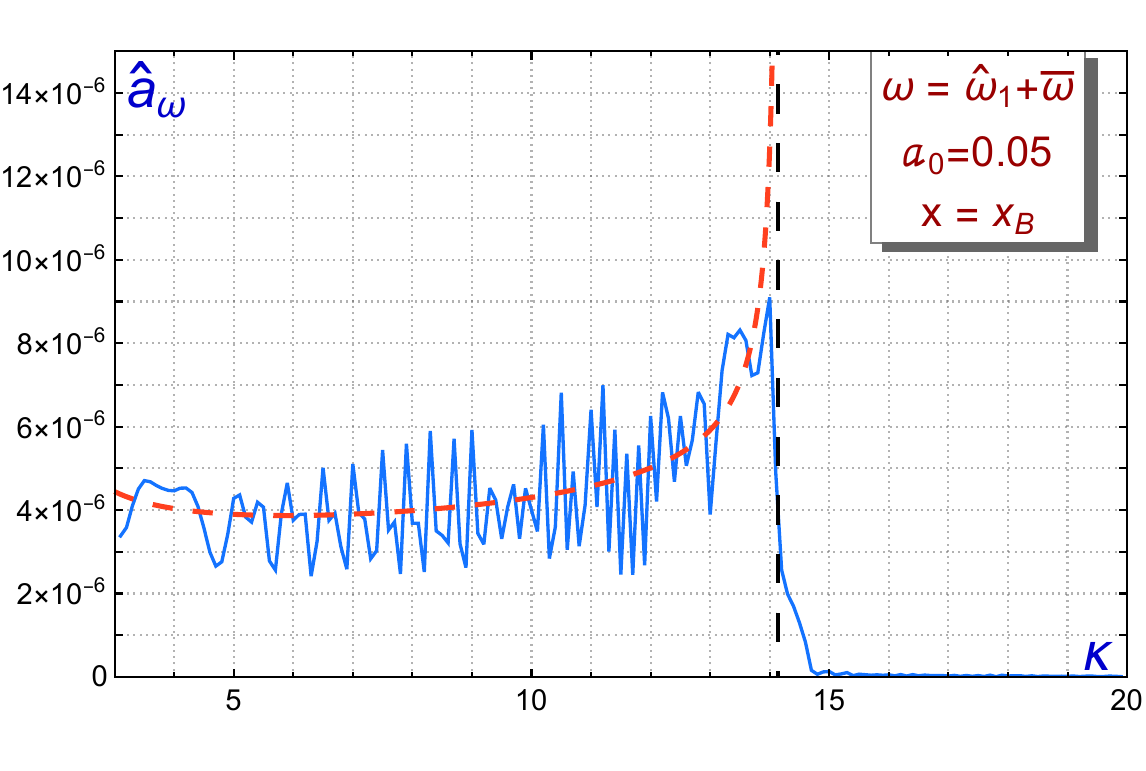}
         \qquad
         \includegraphics[width=0.47\textwidth]{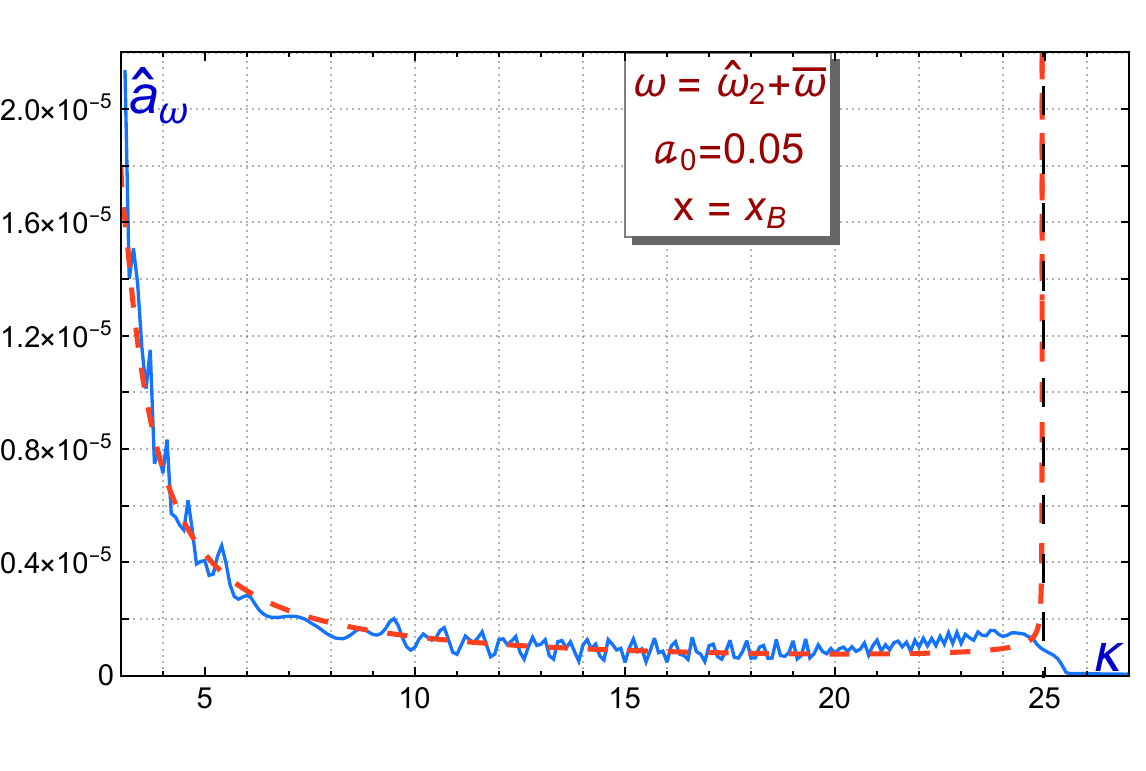}
                           %\vspace{-0.3cm}
    \caption{\textit{Radiation amplitudes (blue curve) emitted  at frequency $\widehat{\omega}_j+\overline{\omega}$ by the kink  when we trigger the orthogonal shape modes $\widehat{\eta}_{D,1}$ (first drawing) and  $\widehat{\eta}_{D,2}$ (second drawing). 
    The red dotted line represents the analytical prediction of formula \eqref{C1:FinalRadiationSecondField}.}}
     \label{C1:Fig:RadiationOrthogonalNumeric}
\end{figure}

The behavior of orthogonal radiation is shown in Figure \ref{C1:Fig:RadiationOrthogonalNumeric}.
From the simulations performed for $\widehat{\eta}_{D,1}$, shown in the first drawing of Figure~\ref{C1:Fig:RadiationOrthogonalNumeric}, it can be seen that no radiation is found in the orthogonal channel for $\kappa>14.14$ at frequency $\widehat{\omega}_1+\overline{\omega}$.
In this range, the aforementioned  frequency  is less than the threshold  value of $\widehat{\omega}^c_0$. 
It can also be observed the existence of a resonance close to this particular value of $\kappa$, which agrees with the result obtained with the equation \eqref{C1:FinalRadiationSecondField}.
Furthermore, on the second drawing of Figure~\ref{C1:Fig:RadiationOrthogonalNumeric} it can be observed that a similar phenomenon occurs for the simulation performed with $\widehat{\eta}_{D,2}$: the kink stops emitting radiation in the range $\kappa >24.93$, which agrees with the analytical prediction made in Section~\ref{C1:Section3}.
Finally, it is worth mentioning that no radiation was found for $\widehat{\omega}_0+\overline{\omega}$ because this frequency is less than  $\widehat{\omega}_0^c$.

The graphs corresponding to the radiation emitted at $\omega=3\widehat{\omega}_j$ will not be shown, since from Figure~\ref{C1:Fig:OrthogonallAmplitudes2} it can be seen that these magnitudes are too small, which makes it difficult to  observe them adequately from the data extracted from numerical simulations. 
This fact can be explained because higher frequencies are more difficult to excite and higher nonlinearities end up exciting frequencies close to the continuous frequency threshold.

\vspace{-0.2cm}
\subsection{Amplitude of other orthogonal shape modes (third order in \texorpdfstring{$a_0$}{a0})}\label{C1:Section4.4}

\begin{figure}[htb]
     \centering
         \includegraphics[width=0.49\textwidth]{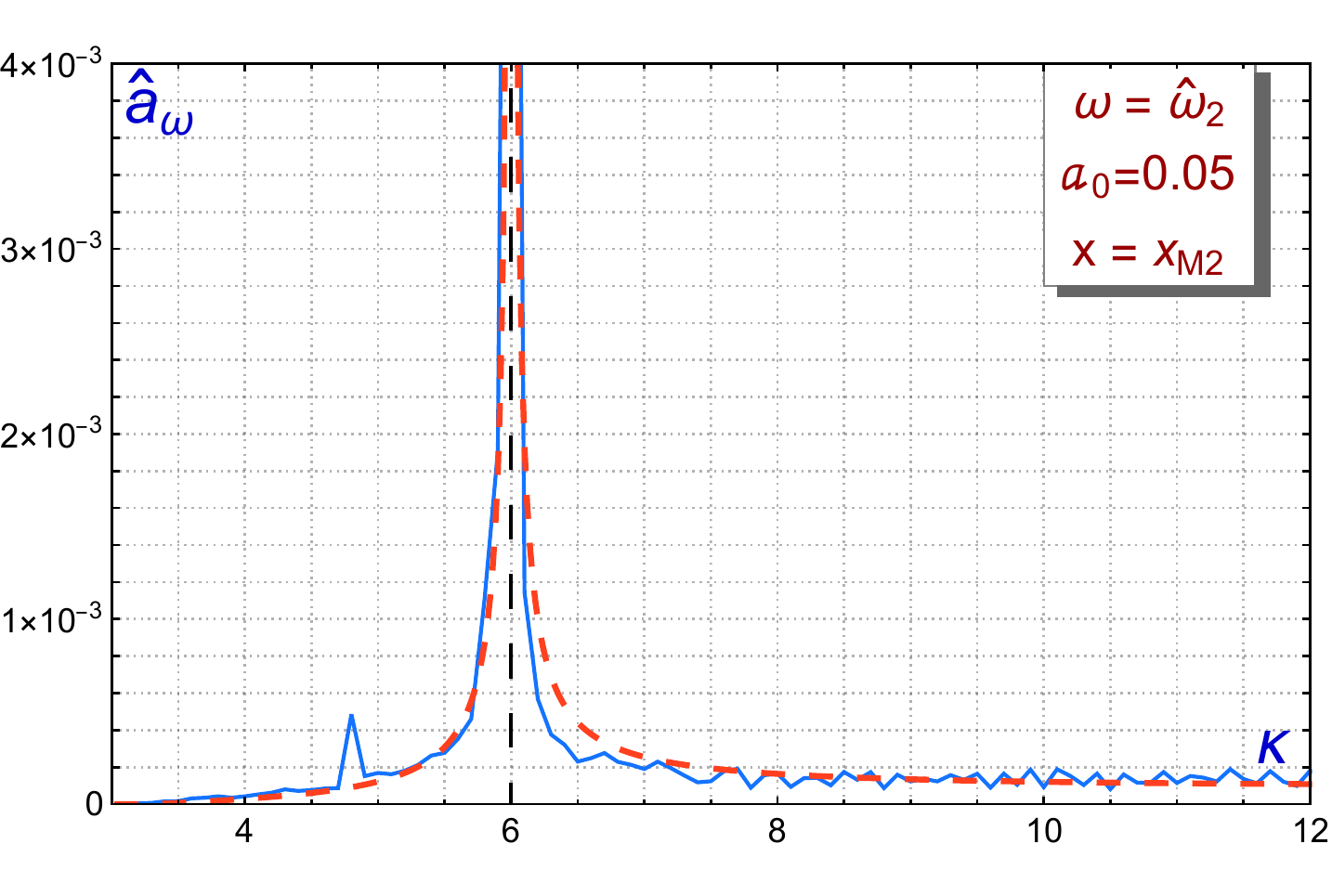}  
         \includegraphics[width=0.49\textwidth]{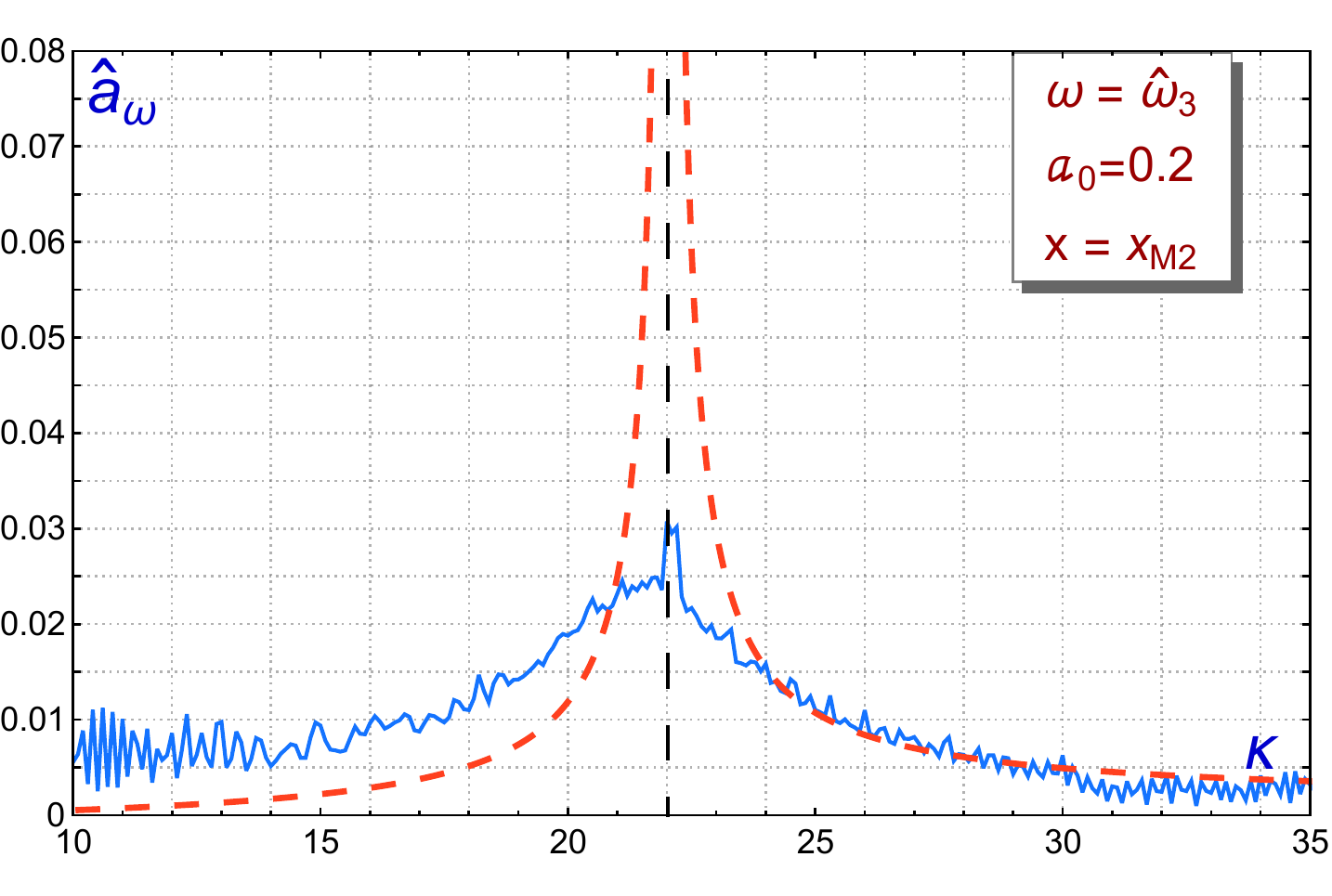} 
         \includegraphics[width=0.49\textwidth]{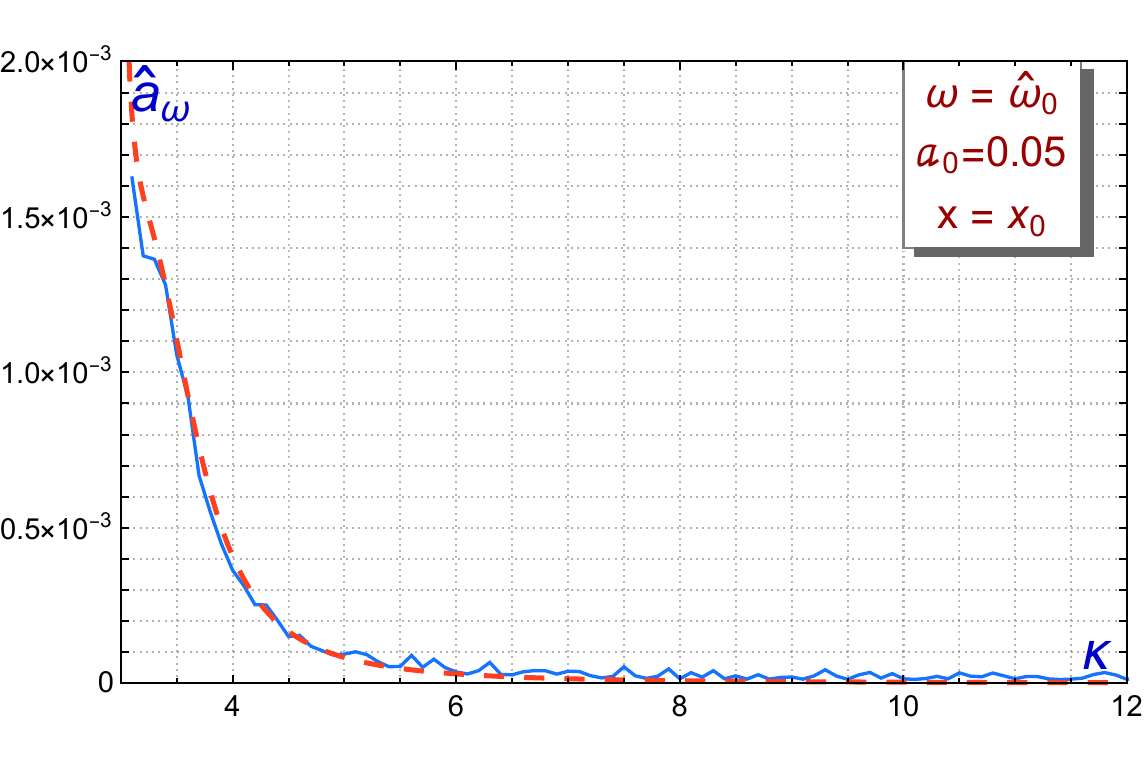}
                         %  \vspace{-0.3cm}

     \caption{\textit{Numerical vibration amplitudes (blue curves) associated with excitation of the shape modes $\widehat{\eta}_{D,2}$ (first drawing), $\widehat{\eta}_{D,3}$ (second drawing) and $\widehat{\eta}_{D,0}$ (third drawing) when $\widehat{\eta}_{D,0}$, $\widehat{\eta}_{D,1}$ and $\widehat{\eta}_{D,2}$ were initially excited, respectively. 
     The red dashed lines correspond to the analytical prediction \eqref{C1:OrthogonalAmplitudes3}.}}
     \label{C1:Fig:CouplingModes}
\end{figure}

In Section~\ref{C1:Section3} it was found that exciting a certain orthogonal mode also activates all other orthogonal modes that have the same parity.
This means that if, for example, $\widehat{\eta}_{D,0}$ is excited, then $\widehat{\eta}_{D,2}, \widehat{\eta}_{ D, 4}, \widehat{\eta}_{D,6},\dots$ will have a non-zero amplitude.
In fact, the analytical expression that describes this event is given by  \eqref{C1:OrthogonalAmplitudes3}.
In this section we will analyze some examples of this phenomenon.
\vspace{0.1cm}
In Figure~\ref{C1:Fig:CouplingModes} the theoretical (red curves) and numerical (blue curves) results obtained for the amplitudes associated with $\widehat{\eta}_{D,2}$, $\widehat{\eta}_{D,3 } $ and $ \widehat {\eta}_{D,0}$, as a function of the coupling constant $\kappa$, are shown together. It is obvious that, depending on the case, the similarity between both types of results is better or worse.

     It is important to point out that several divergences appear in the graphs, which are due to different resonances between the frequencies involved in the calculation carried out to obtain the behavior of these amplitudes.
     For example, in the first drawing in Figure \ref{C1:Fig:CouplingModes} there is a divergence at $\kappa=6$ that can be explained by the fact that, for this specific value of the coupling constant, there is a resonance between $ \widehat{\omega}_2$ and $3\widehat{\omega}_0$, as can be seen on the left side of Figure~\ref{C1:Fig:OrthogonalSpectrum}.

On the other hand, in the second of the graphs in Figure~\ref{C1:Fig:CouplingModes} it is possible to identify the amplitude associated with $\widehat{\eta}_{D,3}$ when $\widehat{\eta}_{D,1}$ is triggered.  This amplitude has been plotted for $\kappa>10$, which is the range where this orthogonal shape mode arises. 
In this case we can notice the presence of a divergence at $\kappa\approx 22.02$, which is due to a resonance between $\widehat{\omega}_3$ and $\widehat{\omega}_1+\overline{\omega}$.
 Finally, in the third drawing of Figure~\ref{C1:Fig:CouplingModes} it can be seen that we cannot appreciate any type of coupling between frequencies, but it can be verified that there is a divergence at $\kappa=3$, which is the value for which $\widehat{\omega}_0=0$.

\subsection{Decay law for the orthogonal  mode amplitude }\label{C1:Section4.5}

 In Section~\ref{C1:Section4.1} it was shown that if we set $\widehat{\eta}_{D,0}$ with $a_0\approx\mathcal{O}(0.01)$, then $a_0$ remains constant, as assumed in \eqref{C1:OthogonalAmplitudeEvolution}.
 However, this is not true when considering higher values of the initial amplitude and the  $\kappa>6$ regime.
In this section, this energy loss will be quantified taking into account the radiation emitted by the wobbling kink  for this specific case.
 
Next we will show that a decay law can be calculated analytically using reasoning similar to that used in \cite{Manton1997} and \cite{AlonsoIzquierdo2023c}.
In Section~\ref{C1:Section3} we assumed that if we initially trigger $\widehat{\eta}_{D,0}$, its corresponding amplitude can be approximated as
\begin{equation}
    (\widehat{a}_{0})_{tt}+ \widehat{\omega}_0^2\  \widehat{a}_0\approx 0,
\end{equation}
which implies that the first orthogonal shape mode behaves as a harmonic oscillator at each point on the real axis. 
Thus, the  energy density  can be written as 
\begin{equation}\label{C1:EnergyDensity}
        \mathcal{E}= \frac{1}{2}\ \widehat{\omega}_0^2\ a_0^2\ \widehat{\eta}_0^2.
\end{equation}
If we now integrate \eqref{C1:EnergyDensity} over the real axis, the total energy stored in this vibration mode will be 
\begin{equation}\label{C1:TotalEnergy}
        E=\int^{\infty}_{-\infty} \mathcal{E}\ dx=\frac{1}{2}\ \widehat{\omega}_0^2\ a_0^2\ \widehat{C}_{0},
\end{equation}
where $\widehat{C}_{0}$ is defined in \eqref{C1:CDj}.

On the other hand, the total average power emitted in a period by the plane wave $\eta=A \sin(\omega t- q x+\delta)$ in both parts of the real axis is $\langle P \rangle =\frac{d E}{d t}=- A^2\, \omega\, q$.
For this mode there is only one radiation term for $\kappa>6$, its frequency being $2\widehat{\omega}_0$.
If we rewrite the amplitude described by \eqref{C1:FinalRadiationFirstField} as
\begin{equation}
    \overline{A}_{2\widehat{\omega}_0}=a_0^2\  \overline{A}'_{2\widehat{\omega}_0}
\end{equation}
then, the radiated power emitted by the wobbling kink is
\begin{equation}\label{C1:RadiatedEnergy}
    \langle P\rangle=\frac{d E}{d t}=- (a_0^2\,  \overline{A}'_{2\widehat{\omega}_0})^2\, (2\, \widehat{\omega}_0)\, \overline{q},
\end{equation}
 where $\overline{q}=\sqrt{4\, \widehat{\omega}_0^2-4}$. 
Taking into account the equations \eqref{C1:TotalEnergy} and \eqref{C1:RadiatedEnergy}, we arrive at the differential equation
\begin{equation}
        \frac{1}{2}\, \widehat{\omega}_0^2\  \widehat{C}_{0}\ \frac{d a_0^2(t)}{ d t}\approx- 2\widehat{\omega}_0\, \overline{A}_{2\widehat{\omega}_0}'^{2}\,  \overline{q}\, a_0^4(t), 
\end{equation}
whose solution is 
\begin{equation}\label{C1:DecayLawMode0}
             a_0(t)\approx\frac{a_0(0)}{\sqrt{1+ t\left(\dfrac{4\, \overline{q}\, a_0(0)^2\, 
             \overline{A}_{2\widehat{\omega}_0}'^{2} }{\widehat{C}_{0}\, \widehat{\omega}_0}\right)}} .
\end{equation}
For $\kappa>6$, $a_0(0)=0.2$ and $800<t<1000$, \eqref{C1:DecayLawMode0} predicts that $a_0\approx 0.05-0.04$, which is the same range of values obtained through numerical simulations in Section~\ref{C1:Section4.1} and is represented in the second drawing of Figure~\ref{C1:Fig:Mode0rthogonalAmplitude}.
A completely similar calculation can be done for higher modes, but in these cases the radiation amplitudes are much smaller than for $\widehat{\eta}_{D,0}$, which implies that the decay in $a_0(t)$ will be almost insignificant.

%----------------------------------Concluding   Remarks------------------------------------------------
\section{Concluding remarks}

Kink solutions in the $\phi^4$ model have been widely used in the literature to explain numerous natural phenomena whose origin is based on the presence of nonlinear terms in the model that describes the physical system. 
This model implies only a scalar field that can describe a given physical quantity. In this chapter we have proposed the study of a more general system that involves two scalar fields. The proposed system is a natural generalization of the $\phi^4$ model with two copies of its potential coupled with a cross term of the type $\kappa \phi^2 \psi^2$. 
This model, like the $\phi^4$ model, has kink solutions, which can now be perturbed by both longitudinal and orthogonal fluctuations, giving rise to new types of wobblers. It is important to study the evolution of these new solutions (which can appear spontaneously  in any physical system through phase transitions or a thermal bath) since they can critically affect the dynamics of the system and give rise to new behaviors.

Throughout the present chapter, we have studied in detail the behavior of a wobbling kink and how the excited eigenmode couples with the rest of the shape modes in the context of the aforementioned  two-component scalar field theory.
For this case, it was found that the shape mode structure depends on the value of the coupling constant between both field components. 
In addition to this, we also found that the wobbler emits radiation with frequency $2\widehat{\omega}_j$ in the longitudinal channel, and with frequencies $3\widehat{\omega}_j$ and $\widehat{\omega}_j+\overline{\omega}$ in the orthogonal channel, where $\widehat{\omega}_j$ and $\overline{\omega}$ are the frequencies associated with the  orthogonal shape mode and  longitudinal mode triggered. 
This differs from what is known for the $\phi^4$ model, where the wobbling kinks only have one radiation term that emits with frequency $2\overline{\omega}$. 
The value of these amplitudes also depends on the coupling constant $\kappa$, a parameter that also determines whether a frequency is embedded in the continuous spectrum and, therefore, whether the kink is capable of emitting radiation with that frequency. 

We can see a clear example of this when $\widehat{\eta}_{D,0}$ is activated since in this case, if $\kappa<6$, there is no frequency radiation $2\widehat{\omega}_0$, in contrast to what happens when higher orthogonal modes are excited. Another example can be found when analyzing the frequencies $\overline{\omega}+\widehat{\omega}_j$, since these terms are only included in the continuous part of the spectrum of the orthogonal channel for a range of values of $ \kappa$.

In addition to what has already been mentioned, the coupling mechanism between shape modes has also been studied, which allowed us to find that the triggered shape mode also couples with shape modes that have the same parity and not only with the longitudinal one.
We could also observe some divergences in both  shape mode amplitudes and radiation amplitudes, due to resonances between frequencies in the vibrational spectrum of the small second-order kink fluctuation.

Another notable phenomenon that appears among the results of this chapter is the decay in the wobbling amplitude due to the loss of energy in the form of radiation.
This energy loss was of great importance when the first orthogonal mode was studied, since when this mode is triggered the emitted radiation is much greater than when higher eigenmodes are considered.

The analitical tools discussed here will be generalized in order to study the radiation emitted by Abelian-Higgs vortices in Chapter \ref{Chap4}.

%As a future line of research that serves as a natural continuation of this work, we consider the possibility of using the techniques presented here to study the behavior of vortices in the Abelian-Higgs model that have been triggered by one of its excited states, something that may be physically very relevant, given the ubiquity of this type of systems in various physical applications. Work in this direction is in progress. 

    \chapter{ Scattering between wobblers in the MSTB model}\label{Chap2}
    
This chapter is an adaptation from Reference \cite{AlonsoIzquierdo2025}: 

\begin{figure}[h!]
    \centering
    \includegraphics[width=1\linewidth]{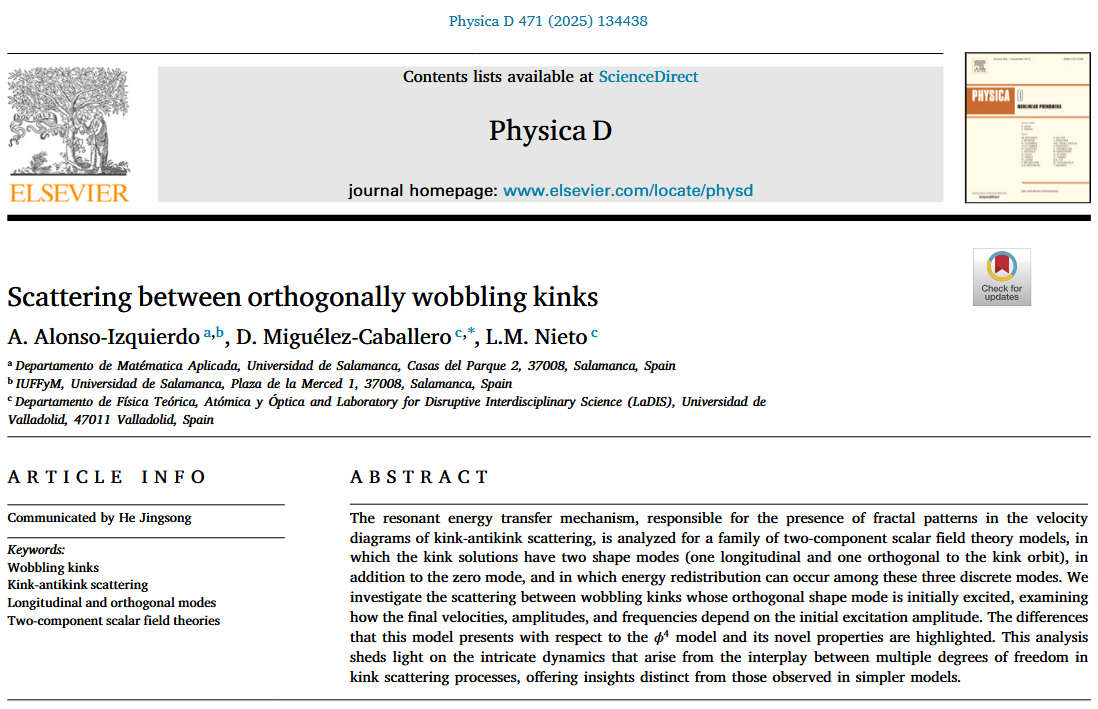}
\end{figure}

\section{Introduction}

%The investigation of kink scattering has garnered significant attention in recent years due to its remarkable characteristics such as the presence of fractal structures in the velocity diagrams showing the final velocity of the scattered kinks as a function of the colliding velocity. Initially investigated in seminal references \cite{Campbell1983,Sugiyama1979,Anninos1991,Peyrard1983,Campbell1986,Kudryavtsev1975}, the collision dynamics between kinks and antikinks in the $\phi^4$  model and deformed sine-Gordon models  unveiled two scattering channels: bion formation and kink reflection. 

In Chapter \ref{Intro1} we addressed kink-antikink collisions both in the $\phi^4$ and $\phi^6$ models unveiling two scattering channels: bion formation and kink reflection.
Bion formation occurs when kinks collide and bounce back repeatedly, emitting radiation with each impact, while kink reflection involves a finite number of collisions before the kinks move apart. These channels dominate for different ranges of initial collision velocity, with bion formation prevalent at low velocities and kink reflection at higher velocities.
A particularly intriguing phenomenon arises during the transition between these regimes, where bion formation and kink reflection intertwine infinitely, giving rise to a fractal structure within the velocity diagram. 
Understanding these intricate dynamics has significant implications for various physical applications, shedding light on nonlinear phenomena fostered by the presence of such topological defects \cite{Manton2004, Shnir2018, Rajaraman1982, Vilenkin2000, Vachaspati2006, Rebbi1984, Kevrekidis2019, Dauxois2006,Belova1997}.

As previously said, the emergence of a fractal structure in the velocity diagram depicting kink scattering within the $\phi^4$ model stems from the presence of an internal vibrational mode, known as the shape mode, associated with kink solutions. 
This mode, together with the zero mode (or translational mode), allows the existence of the resonant energy transfer mechanism, which facilitates the redistribution of energy between kinetic and vibrational modes during kink collisions. 
Normally, in a scattering event, kinks and antikinks approach each other and collide, transferring some of the kinetic energy to the shape mode. 
This turns them into wobblers (vibrationally excited kinks) when they try to separate. 
If the kinetic energy of the wobblers is not enough, they can get closer and collide again. 
This cycle can continue indefinitely or cease after a finite number of collisions. In the latter scenario, enough vibrational energy is converted back to kinetic energy in the zero mode, enabling the wobblers to separate. 
This intricate mechanism and its associated phenomena have been widely investigated through various models, revealing profound complexity \cite{NavarroObregon2023,Adam2022c,Adam2023,Long2024,Campos2023,Adam2019,Adam2021,Weigel2013,Gani2014,Gani2015,Demirkaya2017}. 
It is evident that the natural scenario in a physical system where topological defects can be found is that they are vibrating (i.e., with their shape modes excited), either due to collisions between them or due to the formation of topological defects after a phase transition. 
Therefore, from a physical point of view, the most interesting scenario lies in the investigation of scattering processes between wobbling kinks. 
The analysis of scattering between wobblers within the $\phi^4$ model has been widely discussed in previous works \cite{AlonsoIzquierdo2021b, AlonsoIzquierdo2022}. 
Through the numerical analysis of the scattering solutions derived from the Klein-Gordon partial differential equation, these investigations aim to shed light on the resonant energy transfer mechanism. It is worth noting that studies on wobblers scattering in the double sine-Gordon model have also been carried out in \cite{Gani2018,Gani2019,Campos2021b}, contributing to a broader understanding of similar phenomena across different models \cite{AlonsoIzquierdo2019, Mohammadi2022,Halavanau2012, Dorey2011, Dorey2023, Bazeia2023, Campos2022, AlonsoIzquierdo2021, AlonsoIzquierdo2018, Belendryasova2019, AlonsoIzquierdo2020b,Hahne2024,AlonsoIzquierdo2018b,Christov2019,Christov2019b,Campos2021}. The resonant energy transfer mechanism has been recently demonstrated to occur in vortex scattering within the Abelian-Higgs model in (2+1)-dimensions, leading to the emergence of fractal patterns \cite{Krusch2024}. A collective coordinate model explaining this behavior is detailed in \cite{AlonsoIzquierdo2024d}. This discovery underscores the significance of investigating topological defects in (1+1)-dimensions, as many phenomena observed in this lower-dimensional context persist in higher dimensions. Notably, the phenomenon of spectral walls observed in kink dynamics \cite{Adam2019b} has also been identified in vortex scattering \cite{AlonsoIzquierdo2024e}.

In the present chapter, the main goal is to investigate the scattering between wobblers within a one-parametric family of relativistic (1+1)-dimensional two-component scalar field theories known as the MSTB model, which has already been studied in detail in Section \ref{Intro2.6} and is a natural generalization of the $\phi^4$.
%which is a natural generalization of the $\phi^4$ model in this context. This system arises as a deformation of the $O(2)$ linear sigma model where the term breaking the symmetry includes the coupling constant $\sigma$. 
This model has been the focus of study by many researchers for decades \cite{Sarker1976, Subbaswamy1980b, Subbaswamy1980, Magyari1984, Ito1985, Ito1985b, Montonen1976, Rajaraman1975, AlonsoIzquierdo2008, Rajaraman1979, AlonsoIzquierdo2000}. 
The search for two-component kink solutions has been explored in references \cite{AlonsoIzquierdo2019,Sarker1976, Subbaswamy1980b, Subbaswamy1980, Magyari1984, Ito1985, Ito1985b}, while the generalization of such models is considered in \cite{Rajaraman1979, AlonsoIzquierdo2000,AlonsoIzquierdo2008}. 
These works have proved the existence of topological kinks and two families of non-topological kinks for $\sigma < 1$. 
However, when $\sigma \geq 1$, only topological kinks persist. The investigation of the quantum corrections to these kinks is discussed in \cite{Montonen1976, Rajaraman1975}, while the scattering behavior of these unexcited kinks is analyzed in \cite{AlonsoIzquierdo2019}.  
The existence of wobbling kinks in this model was described in \cite{AlonsoIzquierdo2023c}, where it was shown that the simplest kink solutions that arise in this model have two shape modes, which allows their vibration in two different channels: one longitudinal and  another orthogonal to the kink orbit. 
Through perturbation theory, the radiation emission channels and the attenuation of the shape mode amplitudes were analytically identified. 
Therefore, the next natural step in studying the MSTB model is to investigate scattering between wobblers. In particular, we focus on the investigation of the scattering of wobbling kinks or wobblers emerging in the MSTB model in the regime $\sigma>1$. 
Several reasons justify the interest in carrying out this study. First, the MSTB model has been widely addressed in the literature, as mentioned earlier. Exploring the dynamics of topological defects in this model carries an implicit interest. 
Second, the study of scattering between wobblers has been limited to a few models, all of which involve a single scalar field. 
This restriction poses a limited scenario, exemplified by cases like the $\phi^4$ model, in which the kink possesses only one shape mode, causing the wobblers to vibrate longitudinally along their trajectory in internal space. 
However, 
the MSTB model stands out for its unique features: the kink exhibits two shape modes, allowing the wobblers to vibrate both longitudinally and orthogonally to their orbit. 
This distinctive feature adds complexity and richness to the scattering dynamics in the MSTB model, warranting further exploration. This constitutes the fundamental reason for this research.
Note that now the resonant energy transfer mechanism can be activated, leading to energy redistribution among three discrete modes: the zero (or translational) mode (responsible for the kink motion), and the longitudinal and orthogonal shape modes (responsible for the kink vibration in each component of the field). 
This expansion in the number of modes involved introduces a new layer of complexity to the scattering dynamics, potentially yielding novel phenomena not observed in models with fewer degrees of freedom.

The organization of this paper is as follows. Section \ref{SecC2:section2} introduces the MSTB model along with its kink solutions. The stability of these solutions and their vibrational modes are described. Section \ref{SecC2:section3} outlines the numerical setup used for simulating collisions of wobbling kinks. Section \ref{SecC2:section4} presents detailed numerical results, analyzing the behavior of the fractal patterns observed in the velocity diagrams resulting from the scattering between wobbling kinks, and exploring their dependency on both the coupling constant $\sigma$ and the initial excitation amplitude. By analyzing the final amplitudes and frequencies of the shape modes, the behavior of the resonant energy transfer mechanism in such models is examined. Finally, Section \ref{SecC2:section5} provides the conclusions of the present chapter.

\section{The MSTB model: kink solutions, linear stability and shape modes}\label{SecC2:section2}

In this chapter we deal with the MSTB model, previously described in Section \ref{Intro2.6}. As previously stated, the  dynamics of this theory is governed by the Lagrangian density 
\begin{equation}\label{EqC2:LagrangianDensity}
    \mathcal{L}=\frac{1}{2} \, \partial_\mu \phi \, \partial^{\mu} \phi+\frac{1}{2}\, \partial_\mu \psi \, \partial^\mu \psi- U(\phi, \psi),
\end{equation}
where the potential $U(\phi,\psi)$ is given by the fourth degree algebraic expression
\begin{equation}\label{EqC2:Potential}
    U(\phi,\psi)=\frac{1}{2}(\phi^2+\psi^2-1)^2+\frac{\sigma^2}{2}\psi^2.
\end{equation}
In \eqref{EqC2:LagrangianDensity} and \eqref{EqC2:Potential}, $\phi$ and $\psi$ denote real scalar fields, and the Minkowski metric is represented as $g_{\mu,\nu}=\text{diag} \{ 1,-1\}$. 
%Note that the potential term \eqref{EqC2:Potential} introduces a coupling constant $\sigma^2$, which is a real positive parameter. This family of models is usually referred to as the MSTB model and has been extensively investigated in the literature, as indicated in the Introduction. 
Our focus here lies in analyzing the interaction between discrete modes (longitudinal and orthogonal) during collisions between topological defects in this model. The main objective of this chapter is to understand how the mechanism of resonant energy transfer is influenced when the kinks include an orthogonal vibration channel. The reason for choosing the MSTB model for this study is twofold. The first is because it is a natural generalization of the $\phi^4$ model,  in which the energy transfer mechanism between modes was first studied. In fact, this two-component scalar field theory incorporates the $\phi^4$ model, which implies that the so-called $\phi^4$-kink arises as a solution of this model when the second field is null, as already addressed in Section \ref{Intro2.6}. This can be directly observed from the field equations of this model:
\begin{eqnarray} 
  \frac{\partial^2 \phi}{\partial t^2} -\frac{\partial^2 \phi}{\partial x^2} =2 \phi \left(1-\phi^2 -\psi^2\right), 
\qquad    \frac{\partial^2 \psi}{\partial t^2} -\frac{\partial^2 \psi}{\partial x^2} = 2 \psi \left(1-\phi^2 -\psi^2-\frac{\sigma^2}{2}\right). \label{EqC2:FieldEqn2}
\end{eqnarray} 
%It can be verified that the following static kink/antikink solutions satisfy the field equations (\ref{EqC2:FieldEqn2}) of the model:
In other words, the kink solution under study will be
\begin{equation}\label{EqC2:KinkSol}
    K^{(\pm)}(x)=(\phi(x),\psi(x))^\intercal=(\pm \tanh(x-x_0),0)^\intercal .
\end{equation}

\vspace{-0.2cm}

\begin{figure}[h!]
    \centering
    \includegraphics[width=0.63\linewidth]{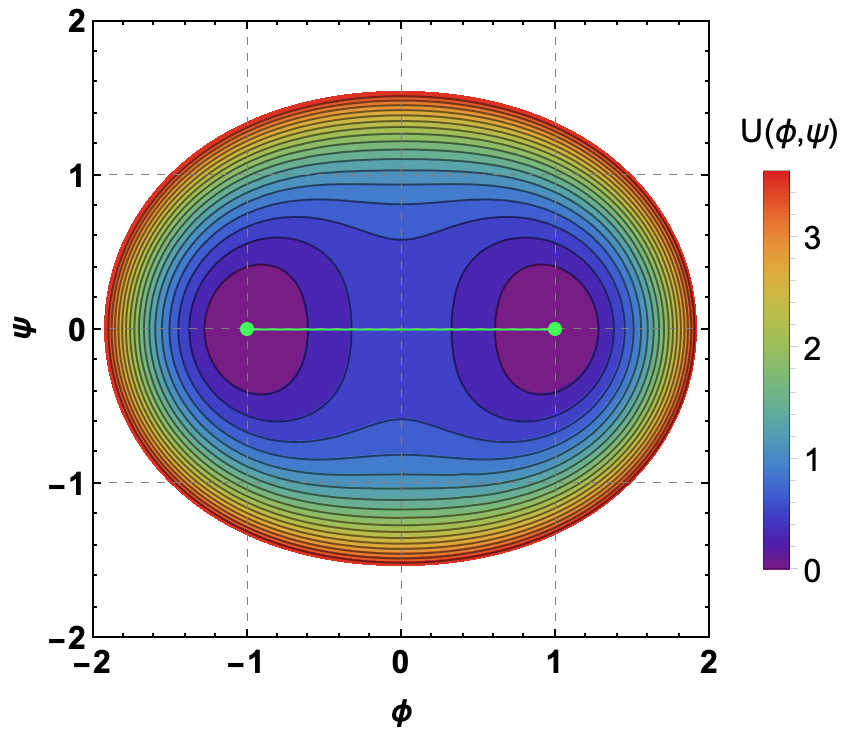}
    \caption{\textit{Contour density plot of the potential \eqref{EqC2:Potential} for $\sigma=1.5$. The green line represents the orbit of the kink solutions \eqref{EqC2:KinkSol}.}}
    \label{FigC2:PlotPotential}
\end{figure}

The kink solution that will be analyzed is formally identical to the one studied in Chapter \ref{Chap1}, but, as the potential in this case is completely different, then the dynamics associated with this solution will change. In Figure~\ref{FigC2:PlotPotential}, the kink orbit has been depicted on the MSTB potential (\ref{EqC2:Potential}). Note that both kink solutions (\ref{EqC2:KinkSol}) interpolate between the two vacua or minimum points of \eqref{EqC2:Potential}. 

\vspace{-0.2cm}

For future use, it is  worth noting that these solutions can be transformed into traveling kinks 
\begin{equation}\label{EqC2:KinkSolLorentz}
    K^{(\pm)}(x,t)=\left(\pm \tanh \frac{x-x_0-v_0 t}{\sqrt{1-v_0^2}} \, , \, 0\right)^\intercal,
\end{equation}
simply by applying a Lorentz boost, where $v_0$ is the kink velocity and $-1<v_0<1$. 
The second reason that makes the MSTB model a natural candidate to develop the study we have embarked on is that the kink solution (\ref{EqC2:KinkSol}) can vibrate both longitudinally and orthogonally with respect to the kink orbit, in the same way as the kink presented in Chapter \ref{Chap1}, and these channels include only one eigenstate each. 
Specifically, the second-order small kink fluctuation operator involves only one  longitudinal vibration mode (the so-called shape mode, already present in the $\phi^4$ model).
But we also have an orthogonal vibration mode, that is, as it will be explained later, the second order fluctuation operator consists of two decoupled Schr\"odinger-type equations whose corresponding solution will provide us with a single mode shape each, contrary to what occurred in the double $\phi^4$ model in which the number of shape modes for the second field component for the simplest kink depended on the value of the coupling constant $\kappa$.
This scenario provides a natural framework to analyze the energy transfer mechanism between longitudinal and orthogonal modes. This statement is justified in the following way:  substituting a linearly perturbed kink
\begin{equation}\label{EqC2:KinkPlusPertubation}
    \widetilde{K}(x,t;\omega, a)=K^{(\pm)}(x)+ a\, e^{i \omega t} F(x)=
    \begin{pmatrix}
\pm \tanh(x)  \\ 
0
    \end{pmatrix}
    + a\, e^{i \omega t}
    \begin{pmatrix}
\overline{\eta}(x) \\ 
\widehat{\eta}(x)
    \end{pmatrix}
    ,
\end{equation}
into the field equations  \eqref{EqC2:FieldEqn2} leads to the spectral problem
\begin{equation}\label{EqC2:SpectralProblem}
    \mathcal{H}
\begin{pmatrix}
\overline{\eta} (x)  \\ 
\widehat{\eta} (x)
    \end{pmatrix}    = 
    \omega^2 
\begin{pmatrix}
\overline{\eta} (x)  \\ 
\widehat{\eta} (x)
    \end{pmatrix}    ,
\end{equation}
where ${\cal H}$ is given by the matrix differential operator
\begin{equation}\label{EqC2:HMatrix}
    \mathcal{H}=
    \begin{pmatrix}
    -\dfrac{d^2}{d x^2}+ 4-6 \sech^2 x  & 0  \\
    0 &-\dfrac{d^2}{d x^2}  +\sigma^2 -2 \sech^2 x  \\
    \end{pmatrix}
    .
\end{equation}
From \eqref{EqC2:HMatrix} it is clear that the longitudinal modes (vibrating along the kink orbit) 
\begin{equation}\label{EqC2:LongFluc}
\overline{F}_{\omega}(x)=
    \begin{pmatrix}
\overline{\eta} (x)  \\ 
0
    \end{pmatrix} 
    ,
\end{equation}
are decoupled from the orthogonal modes (vibrating orthogonally with respect to the kink orbit)
\begin{equation}\label{EqC2:OrthoFluc}
\widehat{F}_{\omega}(x)=
    \begin{pmatrix}
 0 \\ 
\widehat{\eta} (x) 
    \end{pmatrix} 
    .
\end{equation}
In addition to this, the longitudinal and orthogonal components $\overline{\eta} (x)$ and $\widehat{\eta} (x)$ are respectively determined as the eigenfunctions of the Schr\"odinger-like operators 
\[
\mathcal{H}_{11}=-\dfrac{d^2}{d x^2}+ 4-6 \sech^2 x \hspace{0.5cm} \mbox{and} \hspace{0.5cm} \mathcal{H}_{22}=-\dfrac{d^2}{d x^2}  +\sigma^2 -2 \sech^2 x\, ,
\]
 with P\"oschl-Teller potential wells  \cite{Flugge1971,Morse1953,Morse1933}. Therefore, the eigenfunctions associated with the operator \eqref{EqC2:HMatrix} are described as:
\begin{itemize}
\item \textit{Longitudinal eigenmodes:} The fluctuations of the kink solution correspond to those found in the $\phi^4$ model with a single component \cite{Vachaspati2006, Shnir2018, Manton1997, Barashenkov2009, Barashenkov2009b, AlonsoIzquierdo2023}. Therefore, the discrete eigenfrequencies are given by $\omega=0$ and $\overline{\omega}=\sqrt{3}$, whose eigenfunctions are respectively the \textit{zero mode}  
\begin{equation}\label{EqC2:ZeroMode}
     \overline{F}_{0} (x) =        
     \begin{pmatrix}
    \overline{\eta}_0 (x) \\
    0
    \end{pmatrix}
= \begin{pmatrix}
    \sech^2 x \\
    0
    \end{pmatrix},
\end{equation}
and the \textit{longitudinal shape mode}
\begin{equation}\label{EqC2:LongitudinalShapeMode}
     \overline{F}_{\sqrt{3}} (x) =        
     \begin{pmatrix}
    \overline{\eta}(x) \\
    0
    \end{pmatrix}
= \begin{pmatrix}
    \sech x \tanh x \\
    0
    \end{pmatrix}.
\end{equation}
Additionally, a continuous spectrum arises on the threshold value $\overline{\omega}_{0}^c=2$ with frequency $\overline{\omega}_{\overline{q}}^c=\sqrt{4+\overline{q}^2}$, where $\overline{q}$ is a real positive parameter. The corresponding eigenfunctions for this case can be written as
    \begin{equation}\label{EqC2:ContinuousLongitudinalMode}
        \overline{F}_{\sqrt{4+\bar{q}^2}} \,(x)=
        \begin{pmatrix}
    \overline{\eta}_{\bar{q}}(x) \\
    0
    \end{pmatrix} 
=
    \begin{pmatrix}
    ( -1-\bar{q}^2+ 3 \, \tanh^2 x -3 i \bar{q} \tanh x) \, e^{i \bar{q} x} \\
    0
    \end{pmatrix} . 
    \end{equation}

    It is important to note that this eigenmode structure for this field component is identical to that found for the longitudinal component of the kink solution presented in Chapter \ref{Chap1} and the $\phi^4$ kink in Chapter \ref{Intro1}.
\item \textit{Orthogonal eigenmodes:} Only one discrete mode appears, whose frequency $\widehat{\omega}=\sqrt{\sigma^2-1}$ depends on the coupling constant $\sigma$, while the now-called \textit{orthogonal shape mode} is determined as
\begin{equation}\label{EqC2:OrthogonalsShapeMode}
     \widehat{F}_{\sqrt{\sigma^2-1}} (x) =        
     \begin{pmatrix}
    0 \\
    \widehat{\eta} (x)
    \end{pmatrix}
= \begin{pmatrix}
    0 \\
     \sech x
    \end{pmatrix}.
\end{equation}
Note that these fluctuations describe perturbations which are orthogonal to the kink orbit, in the direction of the second field $\psi$. It is clear from the expression of the eigenvalue $\widehat{\omega}$ that the kink (\ref{EqC2:KinkSol}) is unstable when $\sigma < 1$. 
As mentioned in Section \ref{Intro2.6}, in this regime, the kink (\ref{EqC2:KinkSol}) decays into the kink
\begin{equation*}
    K^*(x)=(q \tanh(\sigma(x-x_0)),\lambda \sqrt{1-\sigma^2}\sech(\sigma(x-x_0)))^\intercal,
\end{equation*}
where $q,\lambda=\pm1$.
However, for the regime $\sigma > 1$, the kink is stable and the orthogonal shape mode corresponds to a fluctuation in the second field (which  was initially zero), describing a vibration of the kink orbit, which we can visualize like the movement of a violin string.
For this reason, the scattering of orthogonally wobbling kinks addressed in this article will  be restricted to the range $\sigma>1$. 
Along with the discrete mode (\ref{EqC2:OrthogonalsShapeMode}) there is also a continuous spectrum with eigenfunctions 
\begin{equation}\label{EqC2:ContnuousOrthogonalMode}
     \widehat{F}_{\sqrt{\sigma^2-1}} (x) =        
     \begin{pmatrix}
    0 \\
    \widehat{\eta}_{\widehat{q}} (x)
    \end{pmatrix}
= \begin{pmatrix}
    0 \\
    (\,\widehat{q}+ i \tanh x)\, e^{i \widehat{q} x}
    \end{pmatrix}, \qquad \widehat{\omega}_{\widehat{q}}^c=\sqrt{\sigma^2+\widehat{q}^2}.
\end{equation}
\end{itemize}

\begin{figure}[h!p]
         \centering
         \includegraphics[width=0.48\textwidth]{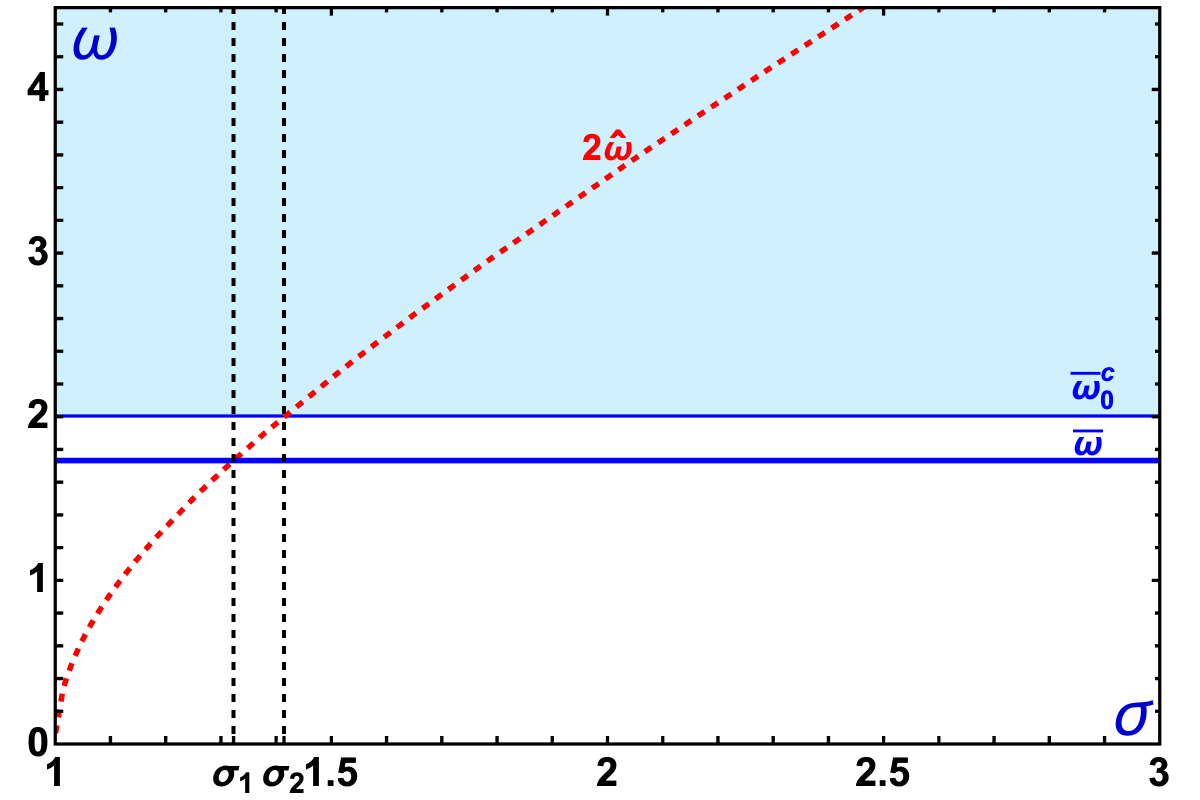}
         \hfill
         \includegraphics[width=0.48\textwidth]{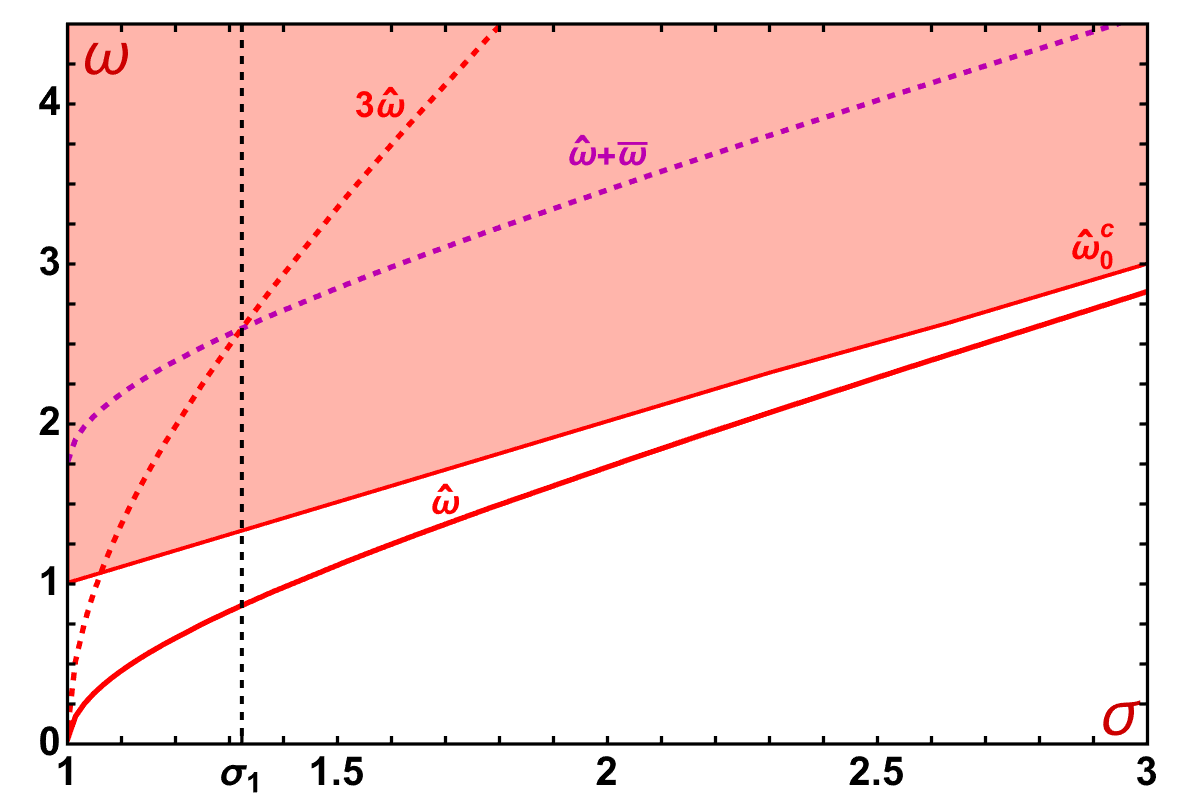}
\caption{\textit{Spectrum of the longitudinal (left) and orthogonal (right) fluctuations of the operator \eqref{EqC2:HMatrix}.
The blue and red areas coincide with the continuous eigenvalues for the first and second field components. The dashed colored lines represent the radiation frequencies that arise when $\widehat{\eta}$ is initially triggered, while the black dashed vertical lines represent the values at which there is resonance between two frequencies.}}
     \label{FigC2:VibrationSpectrum}
\end{figure}

The spectrum of the kink fluctuation operator (\ref{EqC2:HMatrix}) as a function of the coupling constant $\sigma$ is shown in Figure~\ref{FigC2:VibrationSpectrum}, in which the graph on the left represents the spectrum of longitudinal fluctuations, while the one on the right corresponds to orthogonal fluctuations. The discrete eigenfrequencies are represented by solid lines, while the continuous spectrum appears as a shaded region. 

Note that the longitudinal spectrum (left plot) does not depend on $\sigma$, exhibiting a single longitudinal eigenfrequency denoted as $\overline{\omega}=\sqrt{3}$ and a continuous spectrum emerging in the threshold value $\overline{\omega}_0^c= 2$. 
On the other hand, the orthogonal spectrum changes depending on the value of $\sigma$, presenting a single orthogonal frequency $\widehat{\omega}=\sqrt{\sigma^2 -1 }$, and a continuum that emerges from the value $\widehat{\omega}_0^c = \sigma$. 
As a first conclusion, it should be noted that the orthogonal shape mode is easily excited when the coupling constant is close to the value 1. 
In particular, it can be observed that in the range $\sigma\in [1,2]$, the orthogonal eigenfrequency is lower than the longitudinal one and, therefore, more prone to excitation. 
When $\sigma >2$, the situation is reversed and it will be easier to excite the longitudinal mode. 
Since these regimes are expected to lead to different collision behavior of the wobblers, simulations with  values of $\sigma$  both lower and higher than $2$ (typically the values $\sigma=1.2$, $\sqrt{7}/2$, $\sqrt{2} $, $1.5$ and $ 2.5$) will be carried out in Section~\ref{SecC2:section4}.
Additionally, certain harmonics of the discrete eigenfrequencies are represented by dashed curves in Figure~\ref{FigC2:VibrationSpectrum}. These frequencies can be excited because of nonlinear terms of the theory at higher orders and play an essential role in radiation phenomena. This type of process where a kink is initially excited by the orthogonal shape mode $\widehat{\eta}$ has been analyzed in \cite{AlonsoIzquierdo2023} by means of a perturbation approach. It was found that the orthogonally wobbling kink is able to emit radiation with frequency $2\widehat{\omega}$ in the longitudinal channel.  Note that the longitudinal radiation captures the information coming from the orthogonal channel. On the other hand, orthogonal radiation (traveling in the second field) is emitted with frequencies $3\widehat{\omega}$ and $\widehat{\omega}+\overline{\omega}$. Typically, the radiation emission remains small when the initial amplitude of the orthogonal shape mode is reasonably small. However, for certain values of the coupling constant $\sigma$, resonance phenomena occur in which this radiation emission can increase by one or two orders of magnitude. These special model parameter values can be visualized in Figure~\ref{FigC2:VibrationSpectrum} by the coincidence of certain eigenfrequencies, and are listed below:
\begin{itemize}
    \item 
    $\mathbf{\sigma_1=\frac{\sqrt{7}}{2}} \approx 1.32288$\textbf{:}
    As seen in Figure~\ref{FigC2:VibrationSpectrum} (left), a resonance occurs when twice the eigenfrequency of the orthogonal shape mode $2\widehat{\omega}$ coincides with the longitudinal discrete eigenfrequency $\overline{\omega}$ for the model parameter $\sigma_1$. 
    In this case,  a second resonance coincidentally arises between $3\widehat{\omega}$ and $\widehat{\omega}+\overline{\omega}$ in the orthogonal fluctuation spectrum, see Figure~\ref{FigC2:VibrationSpectrum} (right). 
    In this scenario, there is a significant loss of energy from the orthogonal shape mode in favor of the longitudinal mode, accompanied by a pronounced emission of radiation in the orthogonal channel with frequencies $\overline{\omega} + \widehat{\omega}$ and $3\widehat{\omega}$ \cite{AlonsoIzquierdo2023}.  
    Obviously, this phenomenon can have relevant effects on the resonant energy transfer mechanism, causing wobbler collisions to exhibit singular behaviors within the values of the coupling constant $\sigma$. In other words, as the wobblers approach each other, a significant portion of the energy from the orthogonal shape mode is transferred to the longitudinal shape mode  and also radiated away. Consequently, for this $\sigma$ value, the collision dynamics is predominantly governed by the scattering of longitudinally wobbling kinks. 
    Under these circumstances, we can anticipate that behavior is similar to the wobbler scattering observed in the $\phi^4$ model.
    Deviations from the patterns observed  in this case provide valuable information about the interaction between the orthogonal shape mode and other modes within the resonant energy transfer mechanism.
    
    \item 
    $\mathbf{\sigma_2=\sqrt{2}} \approx 1.41421$\textbf{:} 
    In this case, the frequency $2\widehat{\omega}$ begins to form part of the continuous longitudinal spectrum (see Figure~\ref{FigC2:VibrationSpectrum} left). In fact, this value is characterized by the coincidence between the values $2\widehat{\omega}$ and $\overline{\omega}_0^c$. This implies that the quadratic longitudinal radiation source is activated from this point. 
    As a consequence, the radiation emission in the longitudinal channel with frequency $2\widehat{\omega}$ is very significant (relative to other values of $\sigma$). 
    This implies that a large amount of energy is extracted from the orthogonal shape mode in the form of longitudinal radiation. 
Again, in this situation some peculiar behaviors could appear in the scattering between wobblers.    However, despite this phenomenon, the energy loss resulting from radiation emission is not considered significant enough to alter the wobbler scattering characteristics for this particular value.
\end{itemize}

For the analysis of kink scattering in this model, we will take into account previous observations regarding the relevant values of the coupling constant $\sigma$, with the aim of performing simulations at various values of the model parameter that can lead to different behaviors. 
We have already mentioned the values $\sigma=1.5$ and $\sigma=2.5$ as representative of possible different behaviors. 
To these two values we will also add $\sigma=1.2$, for which the emission of radiation in the longitudinal channel with frequency $2\widehat{\omega}$ is suppressed. 
Additionally, studies of the scattering of orthogonally wobbling kinks will be carried out in the special cases $\sigma_1=\sqrt{7}/2$ and $\sigma_2=\sqrt{2}$ mentioned above.

\section{Numerical setup for the scattering between wobblers in the MSTB model}\label{SecC2:section3}

In this chapter we are analyzing  collisions between a kink and antikink \eqref{EqC2:KinkSol}, with their orthogonal shape modes initially activated. 
It is well-known that the scattering between wobbling kinks in the $\phi^4$ model exhibits complex fractal structures depending on the vibration amplitude of its only shape mode \cite{AlonsoIzquierdo2021c,AlonsoIzquierdo2021b}. 
These fractal structures arise due to the resonant energy transfer mechanism between the zero and shape discrete modes. In this model both of these eigenfluctuations are embedded in the longitudinal channel. 
Our goal now is to numerically investigate whether this is a necessary condition to trigger the resonant energy transfer mechanism or whether it can also be initiated by orthogonal eigenmodes. 
The initial numerical setup will consist of a well-separated orthogonally wobbling kink/antikink configuration whose centers are initially located at points $x_0$ and $-x_0$ (with $x_0 \gg 1$) and travel respectively with velocities $v_0$ and $-v_0$.

Therefore, the initial configuration for the numerical simulations can be characterized as 
\vspace{-0.15cm}
{\small
\begin{equation}\label{EqC2:InitialConditionSimulation}
   K^{(\pm)}_{W K}(x,t,v_0,x_0,\sigma,a_0) \cup K^{(\mp)}_{WK}(x,t,-v_0,-x_0,\sigma,a_0)=\left\lbrace
   \begin{array}{cc}
       K^{(\pm)}\left(  \Tilde{x}_{K^{\pm}}\right)+\widehat{a}_0 \, \sin (\widehat{\omega} t)\, \widehat{F}_{\widehat{\omega}} \left(  \Tilde{x}_{K^{\pm}}\right) , \quad  \mathrm{if} \, x<0, \vspace{0.3cm} \\ 
        K^{(\mp)}\left(\Tilde{x}_{K^{\mp}}\right)+ \widehat{a}_0\, \sin( \widehat{\omega} t)\, \widehat{F}_{\widehat{\omega}} \left(\Tilde{x}_{K^{\mp}}\right) ,  \quad \mathrm{if}\, x>0,
   \end{array}
   \right.
\end{equation}}
where
\begin{equation*}
    \Tilde{x}_{K^{\pm}}=\frac{x+x_0-v_0 t}{\sqrt{1-v_0^2}}, \qquad \Tilde{x}_{K^{\mp}}=\frac{x-x_0+v_0 t}{\sqrt{1-v_0^2}}. 
\end{equation*}
Here, $\widehat{a}_0$ is the initial amplitude of the orthogonal shape mode, which measures the excitation of the wobbling kinks. 
Note that \eqref{EqC2:InitialConditionSimulation} determines well-behaved initial conditions because the single kinks in this model are exponentially localized. 
It is expected that during the scattering process there will be an energy transfer between the orthogonal, longitudinal and translational modes, together with a small radiation emission in both field components. 
This energy loss causes  the wobbling amplitude $\widehat{a}$ to decay following the theoretical law  \cite{AlonsoIzquierdo2023}
\begin{equation}\label{EqC2:DecayLaw}
    \widehat{a}(t)^2\approx \frac{\widehat{a}(0)^2}{1+ \widehat{\omega}  \, \, \widehat{c}\,\,  \widehat{a}(0)^2\, t},
\end{equation}
where $\widehat{c}$ is a constant that  depends on the radiation amplitudes and $\widehat{\omega}$ and, therefore, also on $\sigma$. Generally, this phenomenon only plays an important  role when $\widehat{a}(0)>0.1$. In a scattering event, once the collision occurs, a part of the energy could be transferred to the longitudinal eigenmode. 
In this case, the resulting longitudinally wobbling kinks behave approximately  like those of  $\phi^4$, that is, the kink would emit radiation in the longitudinal channel whose frequency would  be $2\overline{\omega}$. 
If this happens, the wobbling amplitude of this internal mode will also decay  following the formula 
\begin{equation}
        \overline{a}(t)^2\approx \frac{\overline{a}(0)^2}{1+ \overline{\omega} \, \,\overline{c} \,\,  \overline{a}(0)^2\, t},
\end{equation}
as described in Section \ref{I2Manton} \cite{Manton1997, Barashenkov2009,Barashenkov2009b}. 
As before, this decrease is only relevant only for amplitudes $\overline{a}(0) > 0.1$. This phenomenon implies that the initial amplitudes of the shape modes of the colliding kinks will decrease as they approach each other. 
When the initial amplitudes of the  shape modes involved are small, the described phenomenon can be considered negligible and these magnitudes are suitable parameters to describe the scattering processes described in this article.

Based on all the results presented above, after the collision of the wobbling kinks and in the case that they are reflected, it is expected that the kinks will move away with a certain final velocity $v_f$, with both longitudinal and orthogonal shape modes excited, that is, with a configuration of the form:
\vspace{-0.3cm}
%\begin{small}
\begin{equation}
   K^{(\pm)}_{W K,f}(x,t,v_0,x_0,\sigma,a_0) \cup K^{(\mp)}_{WK,f}(x,t,-v_0,-x_0,\sigma,a_0)=\left\lbrace
 \!\!  \begin{array}{cc}
       K^{(\pm)}\left(  \Tilde{x}_{K^{\pm},f}\right)+\overline{a}_f  \sin (\overline{\omega} t+\delta)\, \overline{F}_{\overline{\omega}}\left(  \Tilde{x}_{K^{\pm},f}\right)\\+\widehat{a}_f  \sin (\widehat{\omega} t)\, \widehat{F}_{\widehat{\omega}} \left(  \Tilde{x}_{K^{\pm},f}\right) , \,   \mathrm{if} \, x<0, \vspace{0.3cm} \\ 
        K^{(\mp)}\left(\Tilde{x}_{K^{\mp},f}\right)+\overline{a}_f \sin (\overline{\omega} t+\delta)\, \overline{F}_{\overline{\omega}} \left(  \Tilde{x}_{K^{\mp},f}\right)\\+\widehat{a}_f  \sin (\widehat{\omega} t)\, \widehat{F}_{\widehat{\omega}} \left(  \Tilde{x}_{K^{\mp},f}\right)  ,  \,  \mathrm{if}\, x>0,
   \end{array}
   \right.
\end{equation}
%\end{small}
where
\begin{equation}\label{EqC2:InitialConditionSimulation2}
    \Tilde{x}_{K^{\pm},f}=\frac{x+v_f t}{\sqrt{1-v_f^2}}, \qquad \Tilde{x}_{K^{\mp},f}=\frac{x-v_f t}{\sqrt{1-v_f^2}}. 
\end{equation}

 Once the initial conditions of our problem have been described, the kink scattering processes are simulated by discretizing equations \eqref{EqC2:FieldEqn2} using a fourth-order algorithm designed specifically  to handle general nonlinear systems of two coupled Klein-Gordon partial differential equations, as described in \cite{AlonsoIzquierdo2021}. 
 To prevent the radiation emitted by the system from being reflected from the spatial grid boundaries, Mur absorbing boundary conditions were implemented at $x=100$ and $x=-100$. 
 Simulations were performed for various values of the coupling constant $\sigma \in [1.2,2.5]$ over a range of initial velocities $v_0\in\left[0.1, 0.9\right]$ with increments of $\Delta v_0=10^{-5}$. For velocity ranges where the kinks experience only a rebound, the step size was increased to $\Delta v_0=10^{-3}$. 
 In each of these cases the simulations have been carried out changing the initial orthogonal amplitude $\widehat{a}_0$ with steps $\Delta \widehat{a}_0 = 0.02$ in the range $\widehat{a}_0 \in (0,0.2)$. 
 The total number of simulations carried out to obtain the results presented in this article amounts to approximately $10^6$, which have been carried out by using the \textit{Cal\'endula} Supercomputer integrated into the high-performance computing resources of the Castilla y Le\'on Supercomputing Center (SCAYLE). All simulations were performed within the spatial interval $x\in[-100,100]$ with a step size of $\Delta x=0.005$. 
 Initially, the kink/antikink centers are separated by a distance $d=2x_0=15$. 
 When kinks are reflected, their final  velocities are measured numerically. In addition to this, the amplitudes of the orthogonal and longitudinal shape modes are also estimated by using a fast Fourier transform algorithm, implemented at the comoving kink centers $\widehat{x}_M=x_C$ and the respective points
\begin{equation*}
    \overline{x}_M=x_C\pm \sqrt{1-v_0^2} \, \arccosh\sqrt{2},
\end{equation*}
where the discrete longitudinal fluctuations present their maximum elongation. { Another point could be chosen to measure the amplitude corresponding to each mode, but the numerical error resulting from this analysis would be greater.}

It is also  worth mentioning that the initial conditions proposed in \eqref{EqC2:InitialConditionSimulation} are symmetric with respect to the origin of the spatial coordinate. 
That is, the kink and the antikink approach each other at the same velocity (this does not imply a loss of generality since it is simply assumed that we fix the reference system with the origin located at the center of mass) and their shape modes have been excited with the same amplitude and phase. 
As demonstrated in \cite{AlonsoIzquierdo2023, AlonsoIzquierdo2024}, this scenario provides the most extreme phenomena that can arise in kink collisions, since at the moment of impact a constructive interference occurs, amplifying the redistribution of energy between the  vibrational and translational modes. 
Since the numerical algorithm used to discretize the field equations (\ref{EqC2:FieldEqn2}) also preserves this symmetry, the scattering results can be extracted directly from one of the traveling wobblers involved in the simulations.

\vspace{-0.45cm}
\section{Numerical results for the scattering between wobblers in the MSTB model}\label{SecC2:section4}

Now that we have detailed the implementation of the numerical simulations, we proceed to present the main results derived from the data obtained from the simulations. 
The parameter $\sigma$ emerges as pivotal to understanding the energy transfer between internal modes and significantly influences the velocity diagrams of the scattering process. In fact, $\sigma$ plays a crucial role in the appearance of new single-bounce windows in these diagrams. Indeed, as it will be explained later, increasing the value of $\sigma$ is equivalent to increasing the energy needed to trigger the orthogonal eigenmode. This implies that, for example, for the same value of  $\widehat{a}_0$ the critical velocity where the one-bounce tail arises will be smaller as $\sigma$ is increased,  since more energy  can be transferred to the translational eigenmode.
Furthermore, the amplitude of the orthogonal shape mode holds considerable importance: a larger $\widehat{a}_0$ facilitates greater energy transfer to the longitudinal shape mode and the translational one, facilitating separation of the wobblers. The main characteristics of these scattering processes are delineated in the following subsections. 
Firstly, Section \ref{SecC2:section4.1} delves into the analysis of the velocity diagrams, elucidating their behavior in each of the cases. 
Next, in Section \ref{SecC2:section4.2}, we investigate the resonant energy transfer mechanism by examining the final wobbling amplitudes (both longitudinal and orthogonal), which provides information on the strength of shape mode excitations at the end of the process.

\subsection{Properties of the velocity diagrams for the scattering between wobblers in the MSTB model}\label{SecC2:section4.1}

In this subsection we will describe the characteristics of the velocity diagrams observed in the scattering between wobblers within the MSTB model, when the orthogonal shape mode is initially excited. We will classify these scenarios according to the strength of the initial excitation into three distinct regimes. 
We will first examine the regime in which wobblers are weakly excited, typically for $\widehat{a}_0<0.05$. 
Next, we will investigate the regime characterized by moderate excitations, within the range $0.05<\widehat{a}_0<0.1$. 
Finally, we will explore the strong excitation regime, identified by $\widehat{a}_0>0.1$. 
Through this analysis we aim to provide a comprehensive understanding of the behavior exhibited by the velocity diagrams (and the fractal structures included in them) when varying the value of the initial excitation and also the coupling constant $\sigma$, as we have already indicated in Section \ref{SecC2:section2}.

\begin{figure}[h!]
\centering
\begin{subfigure}{1\textwidth}
    \includegraphics[width=1.0\linewidth]{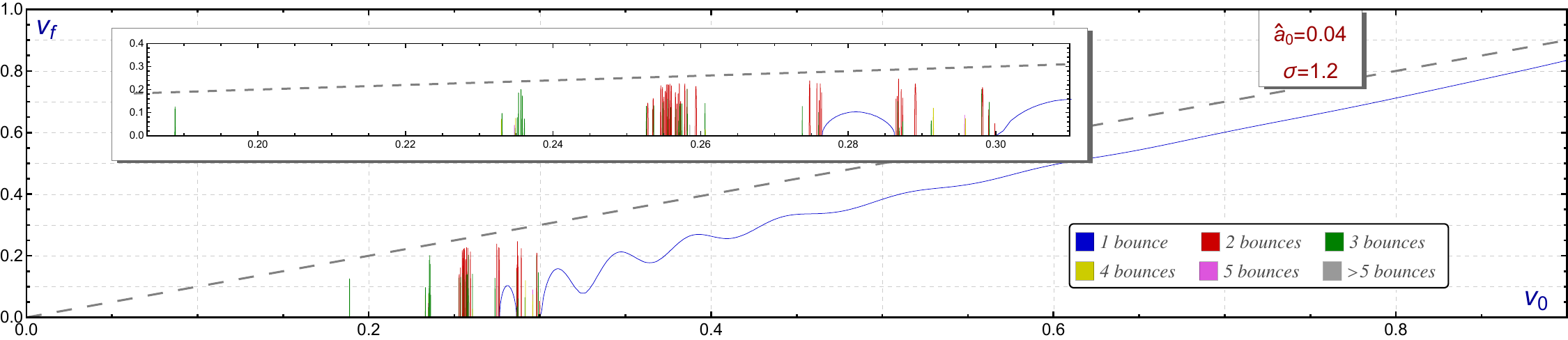}
\end{subfigure}
\hfill
\begin{subfigure}{1\textwidth}    \includegraphics[width=1.0\linewidth]{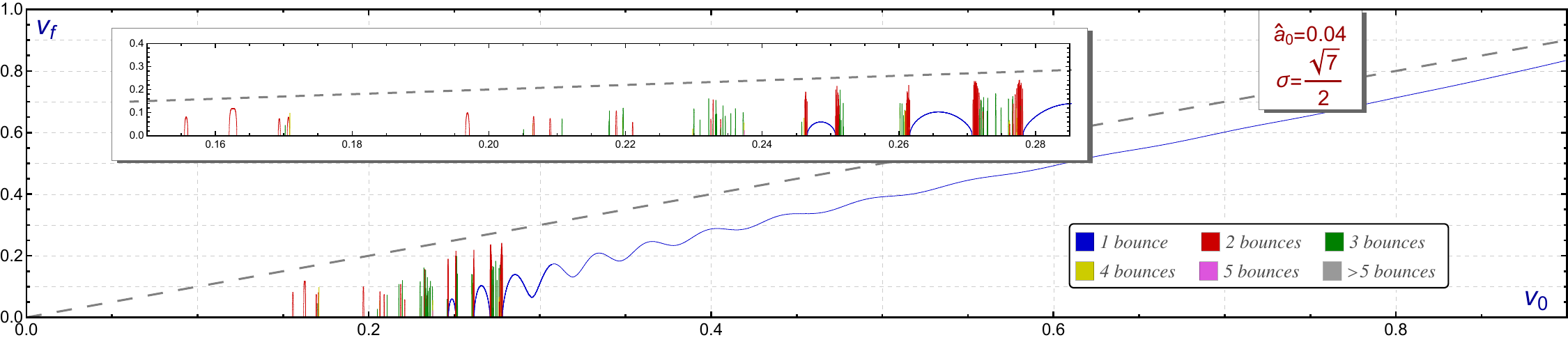}
\end{subfigure}
\hfill
\begin{subfigure}{1\textwidth}
    \includegraphics[width=1.0\linewidth]{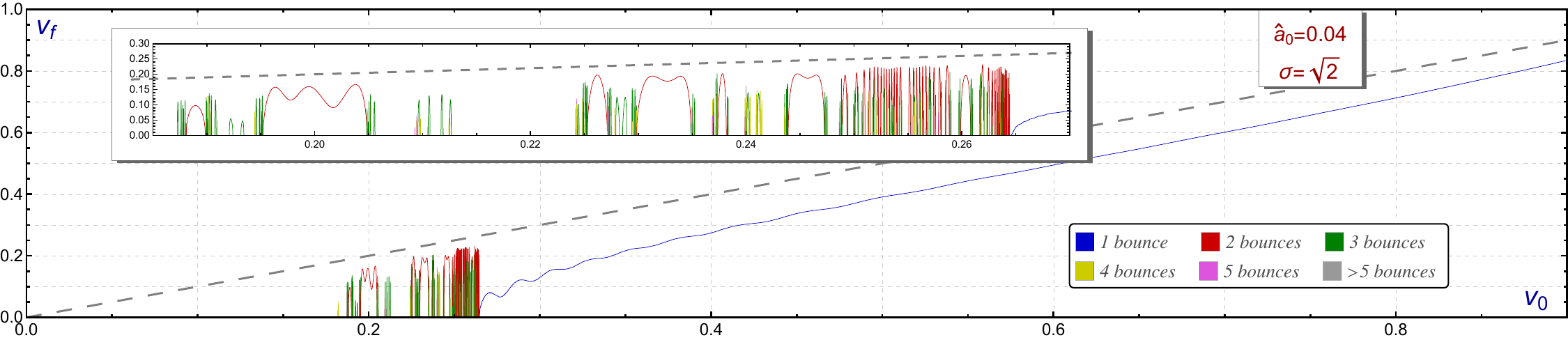}
\end{subfigure}
\hfill
\begin{subfigure}{1\textwidth}
    \includegraphics[width=1.0\linewidth]{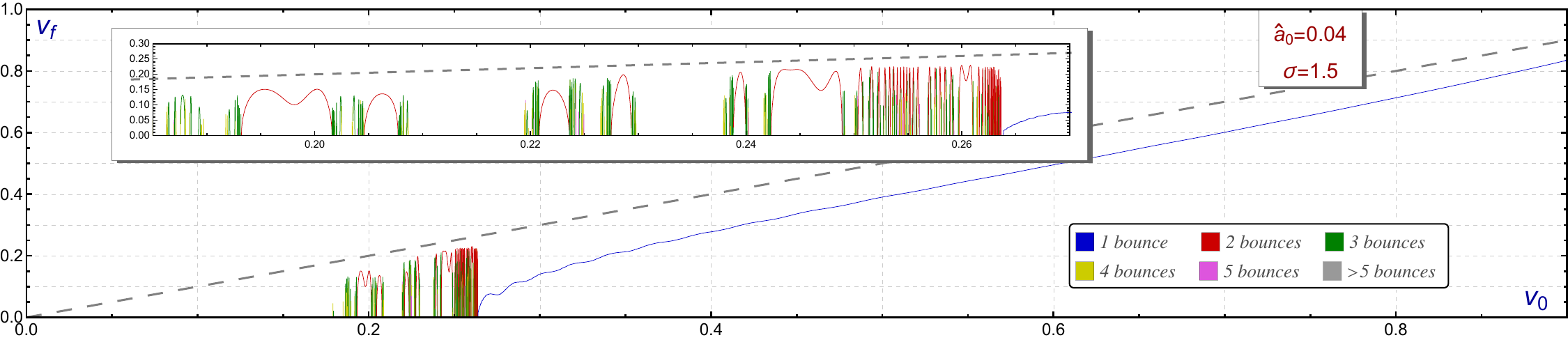}
\end{subfigure}
\hfill
\begin{subfigure}{1\textwidth}
    \includegraphics[width=1.0\linewidth]{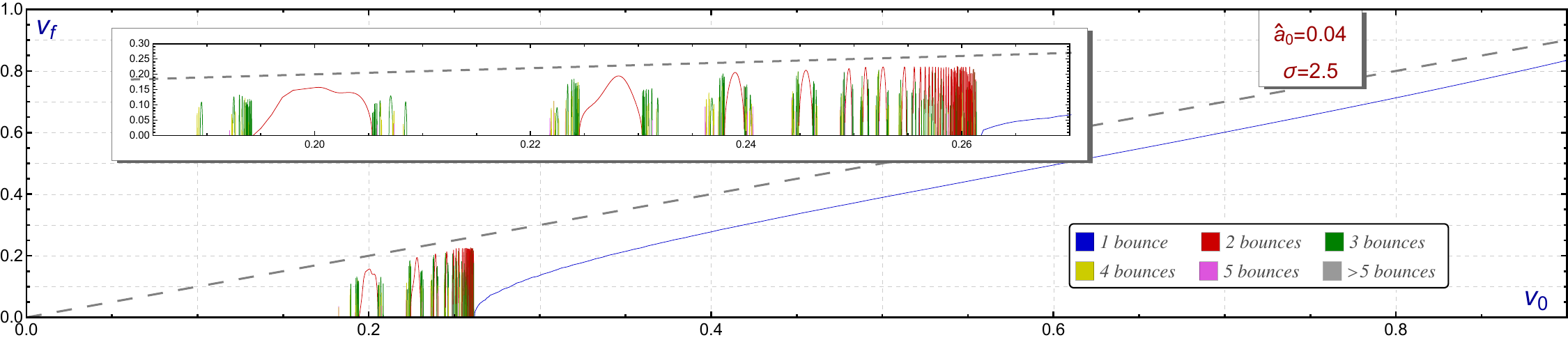}
\end{subfigure}
        
  \caption{\textit{Final velocity $v_f$ of the scattered kinks as a function of the initial velocity $v_0$. The color code shown in the graphs indicates the number of bounces suffered by the kink-antikink pair before moving appart. 
  In initial velocity ranges where no final velocity is shown, a bion is assumed to form.
  The resonance window where various bounces can be observed has been expanded inside each graphics to better show the fractal pattern. For the sake of comparison the dashed grey line indicates the elastic scenario $v_0=v_f$.}}
    \label{FigC2:LowAmplitudeViVf}
\end{figure}

\subsubsection{{Velocity diagrams for the scattering between weakly wobbling kinks}}
In Figure~\ref{FigC2:LowAmplitudeViVf}, the velocity diagram for a low initial wobbling amplitude $\widehat{a}_0 = 0.04$ for the orthogonal mode is displayed for different values of $\sigma$. 
%Specifically, the values chosen  are $\sigma=1.2$, $\sqrt{7}/2$, $\sqrt{2}$,  $1.5$ and $2.5$, which allows us to describe the entire range of behaviors of the velocity diagrams in this scenario. When the values of $\sigma$ are close to $1$, the resulting fractal structure appears relatively simple, with only a few windows where the wobblers collide more than once. 
Specifically, the values chosen  are $\sigma=1.2$, $\sqrt{7}/2$, $\sqrt{2}$,  $1.5$ and $2.5$, which allows us to describe the entire range of behaviors of the velocity diagrams in this scenario.
{ When $\sigma$ values are close to 1, the resulting fractal structure appears relatively simple, annihilation phenomena predominate over kink-antikink scattering. As shown in  Figure~\ref{FigC2:LowAmplitudeViVf} , the $n$-bounce windows are restricted to small intervals of $v_0$. Nevertheless, additional $n$-bounce reflection windows would emerge with increased resolution. Moreover, the absence of an evident self-similar pattern indicates that chaotic behavior is  more pronounced in other cases.}
The resonant energy transfer mechanism redistributes energy between three modes: two vibrational modes (longitudinal and orthogonal) and the zero mode. The chances of gaining energy by the zero mode are less than in the case where there is only one shape mode. 
Consequently, the number of velocity windows in the resonance interval where the kinetic energy of the wobblers is large enough  to allow them to escape must be smaller compared to the $\phi^4$ model, where only one of these massive modes is present. 
It is important to note that, for the same reason, the one-bounce tail (represented in blue in the upper plots of Figure~\ref{FigC2:LowAmplitudeViVf}) exhibits small oscillations, indicating that part of the kinetic energy has been trapped by one of the shape modes.

The previous effect is diminished in the singular case $\sigma_1=\sqrt{7}/2$. As mentioned above, the resonance occurring under the present circumstances  implies that the energy accumulated by the orthogonal shape mode is partially discharged into the longitudinal mode before the collision occurs. 
Scattering can therefore be described in this case as the collision between two longitudinally wobbling kinks. 
Thus, the velocity diagrams found are similar to those observed in the collision between wobblers in the $\phi^4$ model. Note, for example, that the oscillations of the one-bounce tail are more pronounced than in the case $\sigma=1.2$.

As $\sigma$ increases, the fractal structure undergoes substantial changes, becoming increasingly complex. For sufficiently large values of this coupling constant $\sigma$, the diagrams stabilize and resemble the velocity diagram observed for kink/antikink scattering in the $\phi^4$ field theory. 
Furthermore, the critical velocity at which the one-bounce tail emerges is approximately the same, $v_{cr}\approx0.26$ \cite{Campbell1983b, AlonsoIzquierdo2021b}. This suggests that when the initial orthogonal amplitude $\widehat{a}_0$ is very small, the shape mode retains its energy without significant transfer to other modes. As a result, the resulting scattering process resembles that of unexcited kinks. 
This phenomenon can be attributed to the fact that the value of $ \widehat{\omega}$ increases as the model parameter $\sigma$ increases. Consequently, exciting the orthogonal mode during collisions becomes much more challenging. Certainly, this behavior changes when the amplitude $\widehat{a}_0$ reaches a certain threshold. 
This suggests that the orthogonal shape mode must be sufficiently excited to initiate the resonant energy transfer mechanism. More discussion on this point will be provided in Section \ref{SecC2:section4.2} when the behavior of the amplitude of the internal modes after the collision will be discussed.

Another remarkable phenomenon is that the critical velocity $v_{cr}$, at which the one-bounce tail arises, decreases as $\sigma$ increases. As the energy of the orthogonal mode increases with $\sigma$ this also implies that more energy can be transferred to the zero mode allowing the kink-antinkink pair to escape more easily. As it will be shown below, this phenomenon can also be observed for higher values of $\widehat{a}_0$.

\subsubsection{{Velocity diagrams for the scattering between moderate wobbling kinks}} \label{SecC2:section4.1.2}
When the initial amplitude of the orthogonal shape mode is increased, the behavior of the velocity diagrams and the fractal pattern that appears in them exhibit novel characteristics with respect to the previous situation, which, in turn, depends on the value of $\sigma$. 
It is worth noting that in the case of the $\phi^4$ model, the study of wobbling kink collisions led to the identification of new phenomena such as the splitting of the one-bounce tail into isolated one-bounce windows, along with dividing $n$-bounce windows into those with a higher number of bounces. 
These phenomena are also present in the MSTB model with distinctive characteristics. Recall that at the critical value $\sigma_1=\sqrt{7}/2$, the collision of wobblers occurs with a maximum longitudinal wobbling amplitude compared to other $\sigma$ values. 
Consequently, the influence of the orthogonal shape mode decreases, allowing the characteristic features observed in the scattering of wobblers in the $\phi^4$ model to emerge more prominently and distinctively. For other values of $\sigma$, the orthogonal shape mode assumes a more important role in the dynamics. 
For example, when $\sigma<\sigma_1$, the orthogonal frequency is low, which makes the orthogonal shape mode easily excited. 
This implies that the appearance of $n$-bounce windows within the fractal structure  decreases. On the other hand, when $\sigma>\sigma_1$, the number of isolated one-bounce windows decreases, although the complexity of the fractal pattern involving $n$-windows grows and the resonance interval reduces with increasing values of $\sigma$. 
All of these characteristics can be observed in  Figure~\ref{FigC2:VFinalVIncDifferentSgmaA008}, where the velocity diagrams are shown for an initial amplitude $\widehat{a}_0=0.08$ for the values of the model parameter $\sigma=1.2$, $\sigma=\sqrt{7}/2$, $\sigma=\sqrt{2}$ and $\sigma=2.5$. 
One of the most notable behaviors occurs in the resonance interval, where the splitting of the $n$-bounce windows into others of the same type occurs, the gap being filled by windows with a greater number of bounces. Additionally, isolated one-bounce windows appear that were not present before. 
Another surprising behavior is that the one-bounce tail begins to exhibit oscillations that, when the initial vibration amplitude is  large enough, lead to the reappearance of isolated one-bounce windows. 
These processes were also present in the scattering of wobbling kinks in other models such as the $\phi^4$ \cite{AlonsoIzquierdo2021} and double sine-Gordon models \cite{Campos2021b}, although, as we will see in the present scenario, these phenomena show novel features, which we describe in more detail at the following points:

%\vspace{0.3cm}

\begin{itemize}
\item \textit{Emergence of one-bounce windows in the resonance regime:} 
As a general rule, when $\sigma$ is not close to the critical value $\sigma_1=\sqrt{7}/{2}$, only a few one-bounce windows arising via this mechanism will be observed in the velocity diagrams. In Figure~\ref{FigC2:OneBounceWindowSigma1.2}, the velocity plots for one-bounce scattering events are depicted as a function of the initial amplitude of the orthogonal shape mode  for $\sigma=1.2$ within the range $\widehat{a}_0\in[0,0.2]$ using a color-coded scheme. 
In it, red represents the case in which the kinks are not excited, while blue corresponds to the maximum value of the amplitude within the selected range $\widehat{a}_0\in [0,0.2]$. From this figure it is clear that there are only four windows present under the critical velocity  ${v_0 \approx 0.27}$ (associated with unexcited kink-antikink scattering) for relatively high values of $\,\widehat{a}_0$, and their heights are relatively small. {  This feature is also present in the scattering between wobbling kinks in the $\phi^4$ and double sine-Gordon models. In the latter, the kink solution also possesses an internal frequency that depends on the model's coupling constant. In both models, since there is only one internal mode, more energy can be transferred to the translational mode, which explains why this mechanism plays a more prominent role in these models.}
\vspace{0.2cm}
 \begin{figure}[h!]
\centering
\begin{subfigure}{1\textwidth}
    \includegraphics[width=1.0\linewidth]{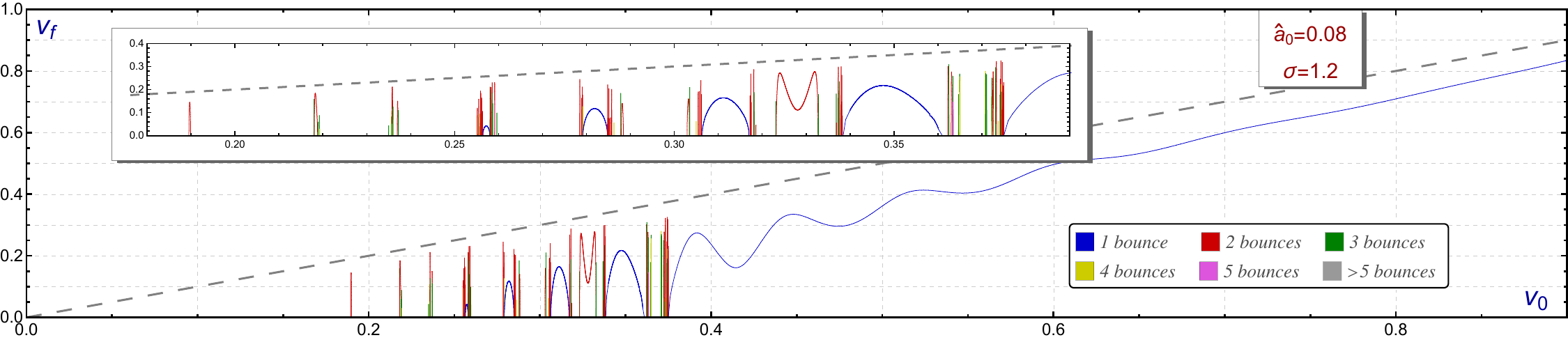}
\end{subfigure}
\hfill
\begin{subfigure}{1\textwidth}    \includegraphics[width=1.0\linewidth]{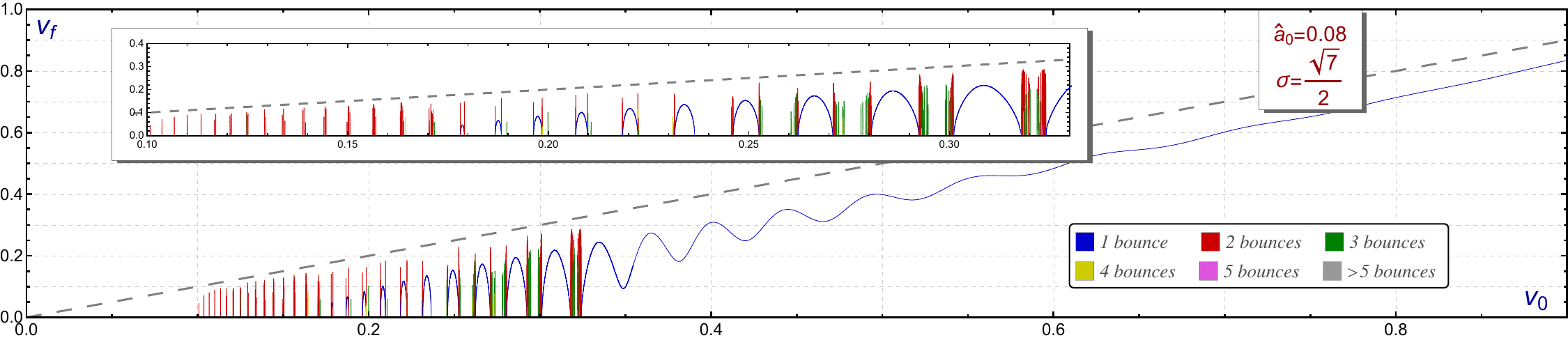}
\end{subfigure}
\hfill
\begin{subfigure}{1\textwidth}
    \includegraphics[width=1.0\linewidth]{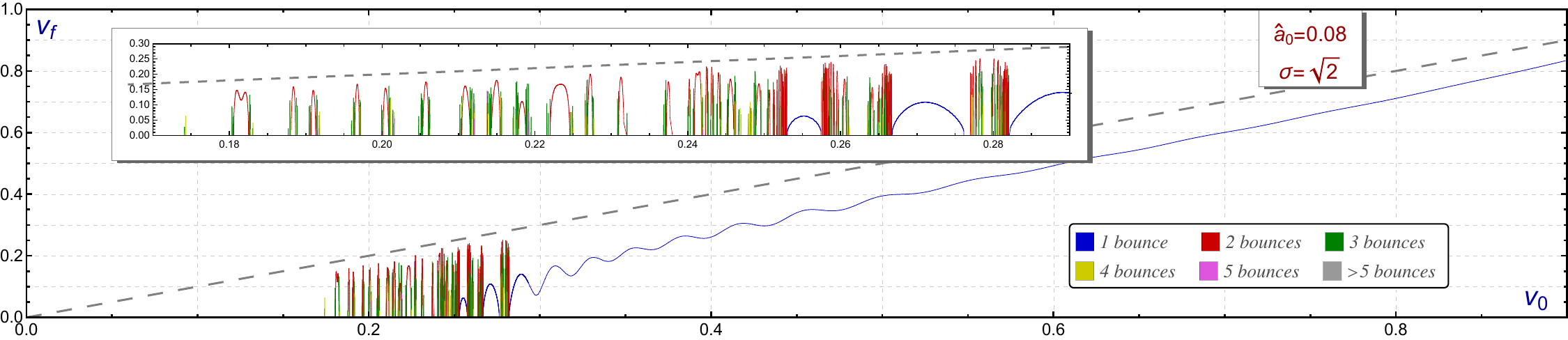}
\end{subfigure}
\hfill
\begin{subfigure}{1\textwidth}
    \includegraphics[width=1.0\linewidth]{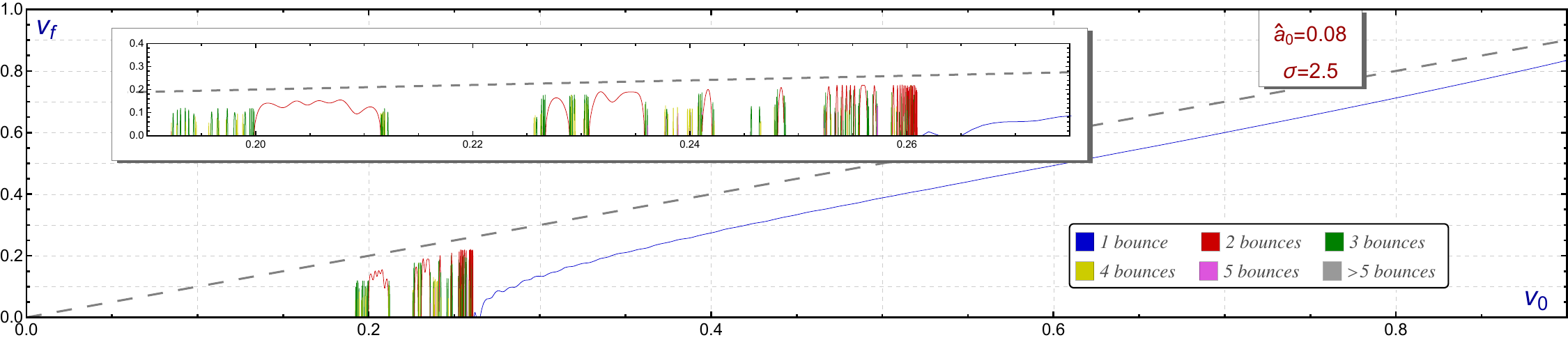}
\end{subfigure}
        
   \caption{\textit{Final velocity $v_f$ of the scattered kinks as a function of the initial velocity. The color code shown in the graphs indicates the number of bounces suffered by the kink-antikink pair before moving apart. 
  In initial velocity ranges where no final velocity is shown, a bion is assumed to form.
  The resonance window where various bounces  can be observed has been expanded. For the sake of comparison the dashed grey line indicates the elastic scenario $v_0=v_f$. }}
    \label{FigC2:VFinalVIncDifferentSgmaA008}
\end{figure}

\begin{figure}[htb]
    \centering
    \includegraphics[width=0.99\linewidth]{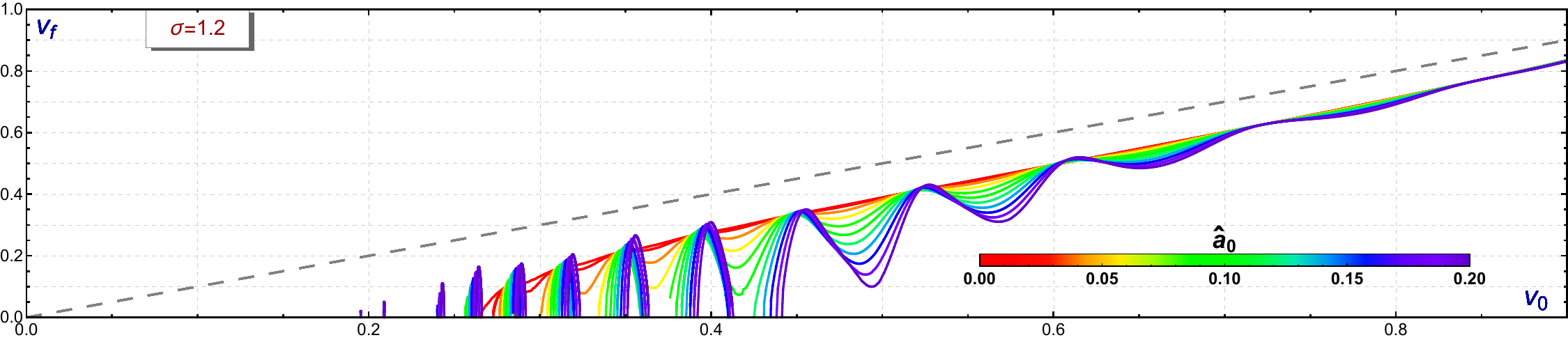}
    \vspace{0.25cm}
    \caption{ \textit{Velocity diagram for one-bounce scattering events for different values of the initial wobbling amplitude, $\widehat{a}_0\in[0,0.2]$.}}
    \label{FigC2:OneBounceWindowSigma1.2}
\end{figure}

\vspace{0.3cm}

The presence of these windows can be explained by considering that the larger $\widehat{a}_0$ is, the more energy the kinks possess before the collision. Consequently, more energy can be transferred to the translational mode, allowing the kinks to escape more easily after a collision for relatively low values of $v_0$. This phenomenon can be better understood by referring to Figure~\ref{FigC2:ZoomSigma1.2SeveralA}, which represents a sequence of velocity diagrams for closely spaced initial amplitude values $\widehat{a}_0$ for the case of $\sigma=1.2$. 
It can be seen that the fractal structure around $v_0\approx0.255$ is compressed when the amplitude of the orthogonal mode increases. When this amplitude is high enough, a new one-bounce window emerges in the same place where the two-bounce windows existed.

\begin{figure}[h!]
    \centering
    \includegraphics[trim={3cm 0 3cm 0},width=0.99\linewidth]{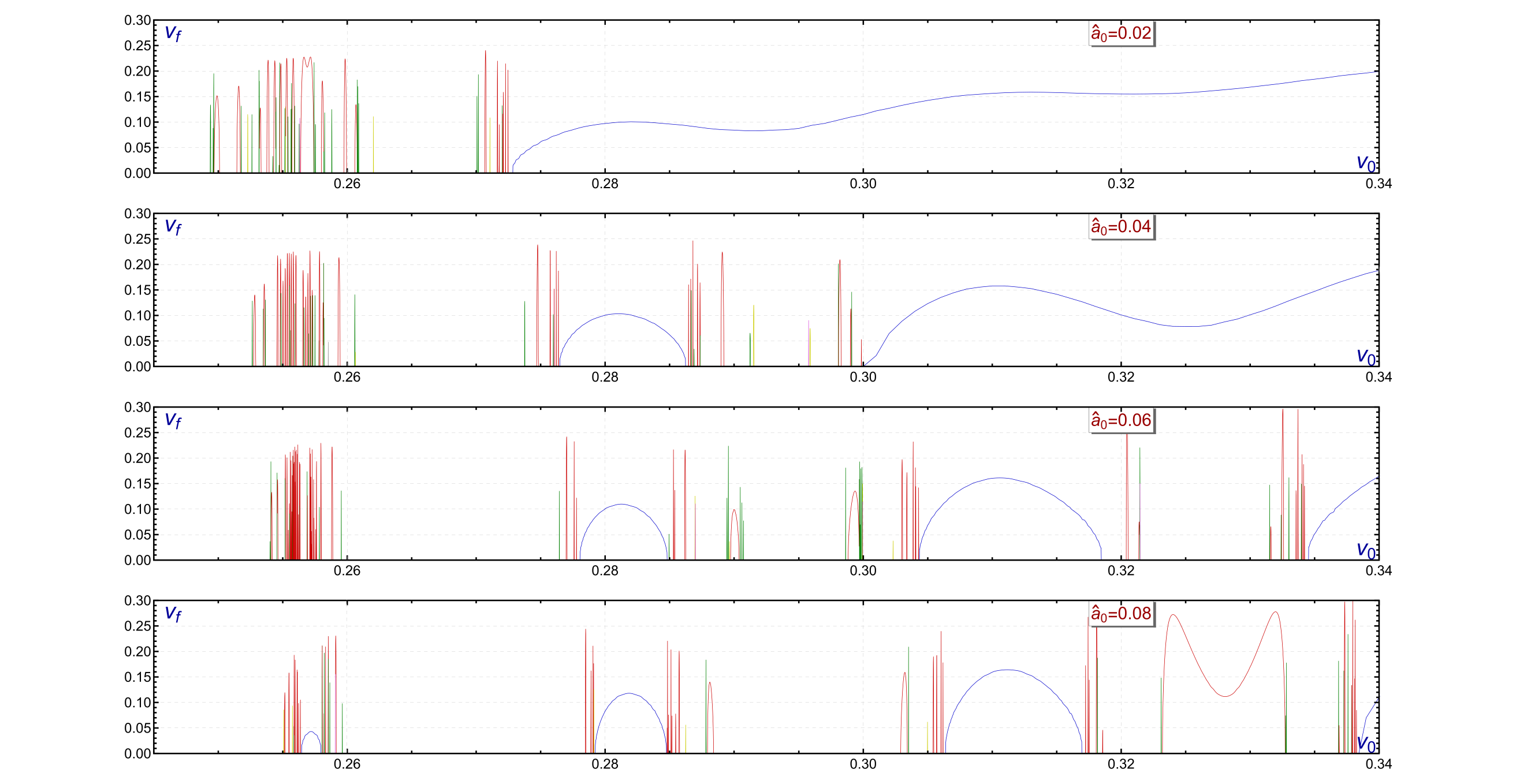}
        \vspace{-0.25cm}
    \caption{\textit{Velocity diagrams for the case $\sigma=1.2$ in the initial velocity range $v_0\in[0.25,0.34]$ for several values of excitation amplitude $\widehat{a}_0$. This graphics illustrates the window splitting mechanism. The color code used to indicate the total number of bounces is the same as in Figure~\ref{FigC2:LowAmplitudeViVf}.} }
    \label{FigC2:ZoomSigma1.2SeveralA}
\end{figure}
%\vspace{-0.25cm}
\item\textit{One-bounce reflection tail splitting:} 
As shown in  Figures~\ref{FigC2:VFinalVIncDifferentSgmaA008} and \ref{FigC2:OneBounceWindowSigma1.2}, the one-bounce tail (plotted in blue) begins to oscillate. In fact, for weakly excited wobbling kinks, this phenomenon is already noticeable in the graph associated with $\sigma=1.2$ in Figure~\ref{FigC2:LowAmplitudeViVf}. 
With an increase in the amplitude of the orthogonal mode, these oscillations can extend to the $v_f=0$ axis, resulting in isolated one-bounce windows. 
This behavior is more pronounced for values of $\sigma$ close to 1, as seen in Figure~\ref{FigC2:VFinalVIncDifferentSgmaA008}. 
This process is clearly illustrated in Figure~\ref{FigC2:ZoomSigma1.2SeveralA}, where the one-bounce tail is recurrently split into multiple isolated one-bounce windows. In the sequence represented, one can observe the formation of such a window around $v_0\approx 0.28$ in the second graph, while a new window appears in the third graph for $v_0\approx0.31$. Furthermore, as $\widehat{a}_0$ increases, these new windows begin to shift to the right side of the graph. 
%This phenomenon was not observed in the study of scattering between wobblers in the $\phi^4$ and double sine-Gordon models \cite{AlonsoIzquierdo2021b,Campos2021b}. 
{  This mechanism was also observed in studies of scattering between wobblers in the $\phi^4$ and double sine-Gordon models \cite{AlonsoIzquierdo2021b, Campos2021b}; however, in these models, the new one-bounce windows did not shift as the amplitude of the internal mode increased.}
For example, considering the velocity diagram for $\sigma=1.2$ shown in Figure~\ref{FigC2:ZoomSigma1.2SeveralA}, it is evident that the initial point of the second one-bounce window is shifted from $v_0\approx 0.277$ when $\widehat{a}_0=0.04$ to $v_0\approx 0.279$ when $\widehat{a}_0=0.08$. Indeed, this phenomenon is similar to the aforementioned squeezing of the fractal structure that gave as a result new one-bounce windows in the resonant interval of the diagram. The main difference between both phenomena is that now no new one-bounce windows are created; instead, the existing window is compressed to fill the gap between one-bounce windows with $n$-bounce windows. Note in Figure~\ref{FigC2:VFinalVIncDifferentSgmaA008} that  splitting  the one-bounce tail results in significantly more isolated one-bounce windows for small values of $\sigma$ than for higher values of $\sigma$.

\item\textit{$n$-bounce window splitting:} 
Another mechanism that contributes to the increasing complexity of the fractal structure of the velocity diagrams as the initial amplitude increases is the splitting of $n$-bounce windows, where one of these windows is divided into others of the same type. 
This phenomenon already appeared in the $\phi^4$ model.  In this case, increasing the amplitude of the single shape mode results in the generation of more two-bounce windows, which appear progressively from left to right in the corresponding velocity diagram. Figure~\ref{FigC2:EvolutionTwoBounce} illustrates the evolution of a single two-bounce window  as $\widehat{a}_0$ goes from $0.02$ to $0.12$ for the kink-antikink scattering in the MSTB model for $\sigma=2.5$. 
It can be seen that as the orthogonal wobbling amplitude increases, a new two-bounce window emerges on the left side of the original window between $\widehat{a}_0=0.06$ and $\widehat{a}_0=0.08$. The gap between these two regions is filled with scattering events exhibiting three and four bounces. 
Subsequently, for $0.10<\widehat{a}_0<0.12$, another window is created, this time on the right side of the plot. This phenomenon can be explained by considering that, for large values of $\sigma$, the frequency of the orthogonal mode is greater than that corresponding to the longitudinal mode. 
Consequently, the vibrational energy of the wobblers increases, providing more energy to transfer to translational mode. This allows the wobbling kink-antikink pair to generate new windows on either side of the original ones as $\widehat{a}_0$ increases.

\end{itemize}

\begin{figure}[h!]
    \centering
    \includegraphics[width=0.84\linewidth]{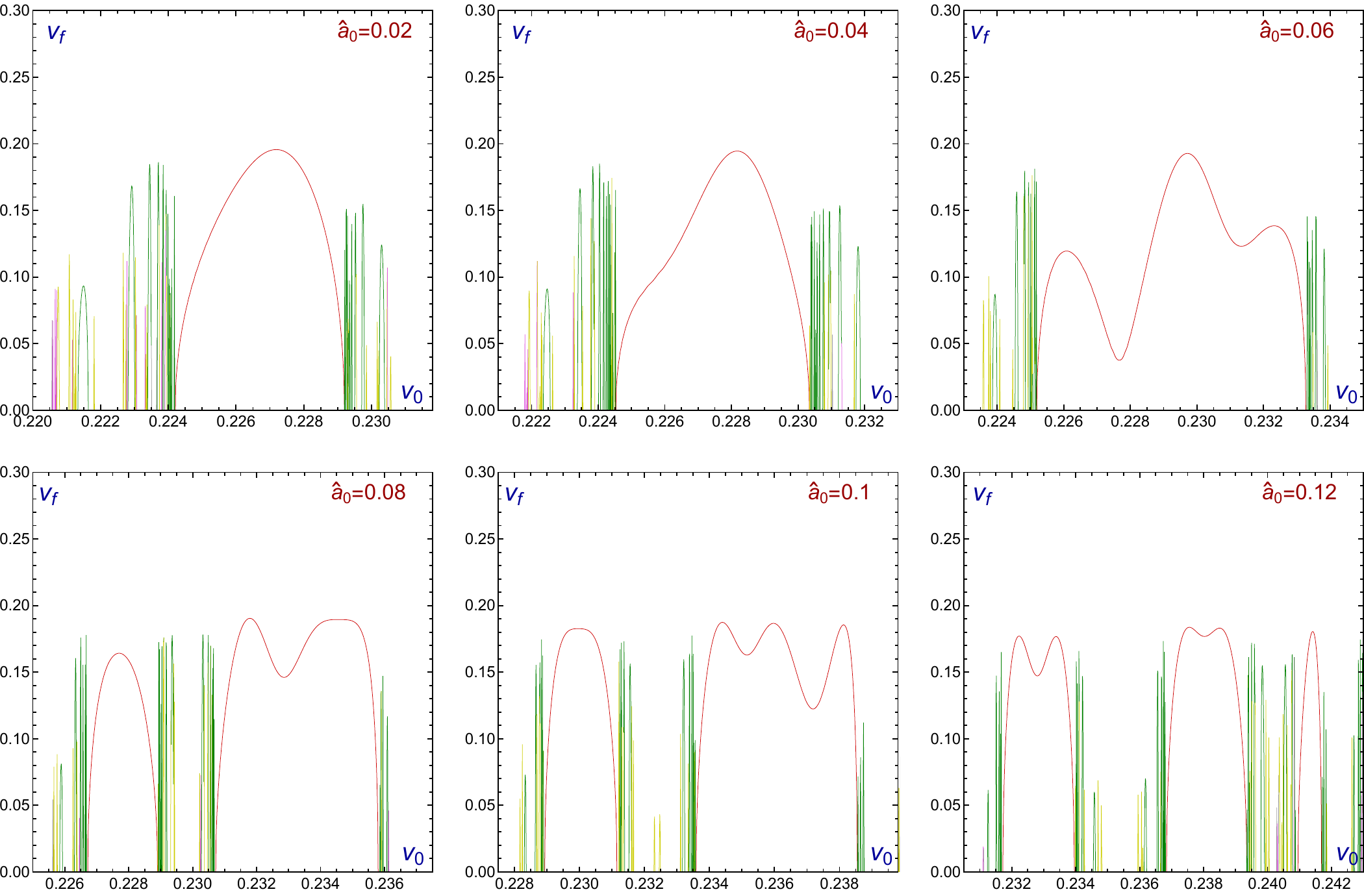}
    \vspace{-0.2cm}
    \caption{\textit{Evolution of a two-bounce window found in the velocity diagram for $\sigma=2.5$ as the value of $\widehat{a}_0$ increases from $\widehat{a}_0=0.02$ to $\widehat{a}_0=0.12$. The color code used to indicate the total number of bounces is the same as in Figure~\ref{FigC2:LowAmplitudeViVf}.}}
    \label{FigC2:EvolutionTwoBounce}
\end{figure}

\subsubsection{Velocity diagrams for the scattering between strongly wobbling kinks}  \label{SecC2:section4.1.3}

\begin{figure}[h!]
\centering
\begin{subfigure}{1\textwidth}
    \includegraphics[width=1.0\linewidth]{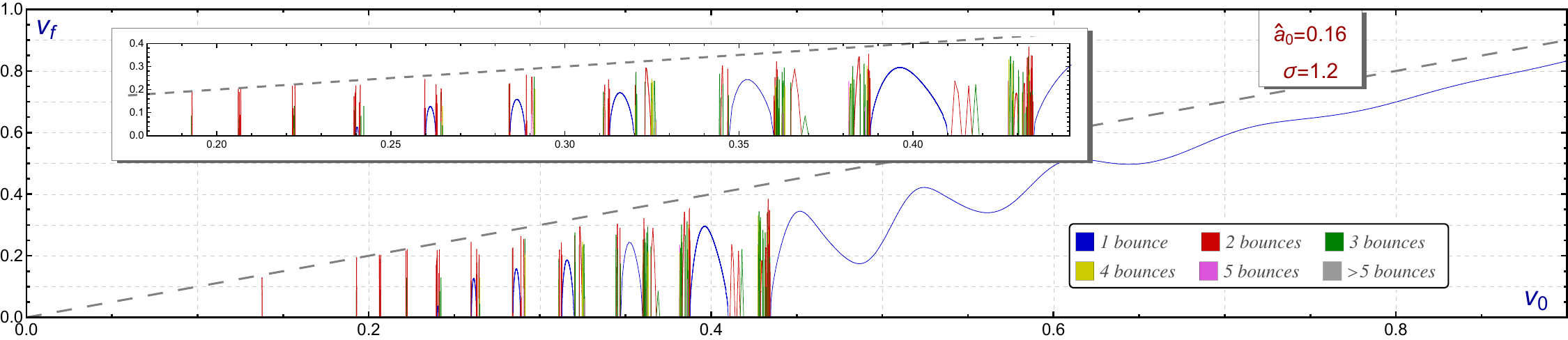}
\end{subfigure}
\hfill
\begin{subfigure}{1\textwidth}
    \includegraphics[width=1.0\linewidth]{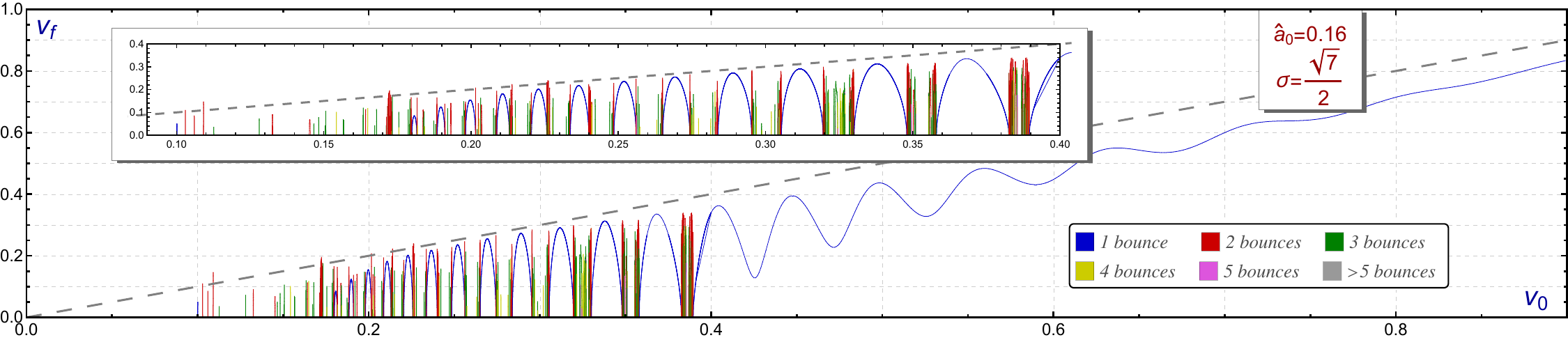}
\end{subfigure}
\hfill
\begin{subfigure}{1\textwidth}
    \includegraphics[width=1.0\linewidth]{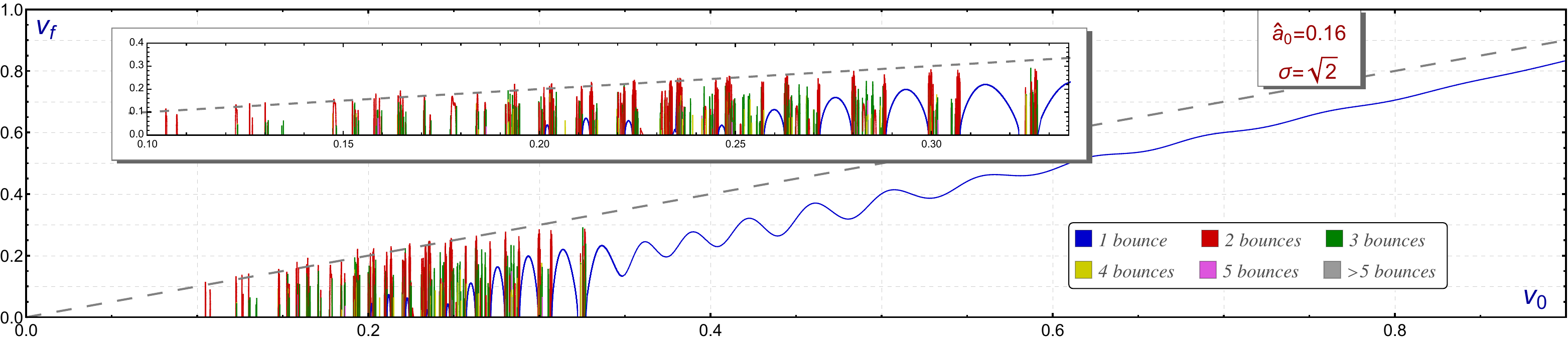}
\end{subfigure}
\hfill
\begin{subfigure}{1\textwidth}
    \includegraphics[width=1.0\linewidth]{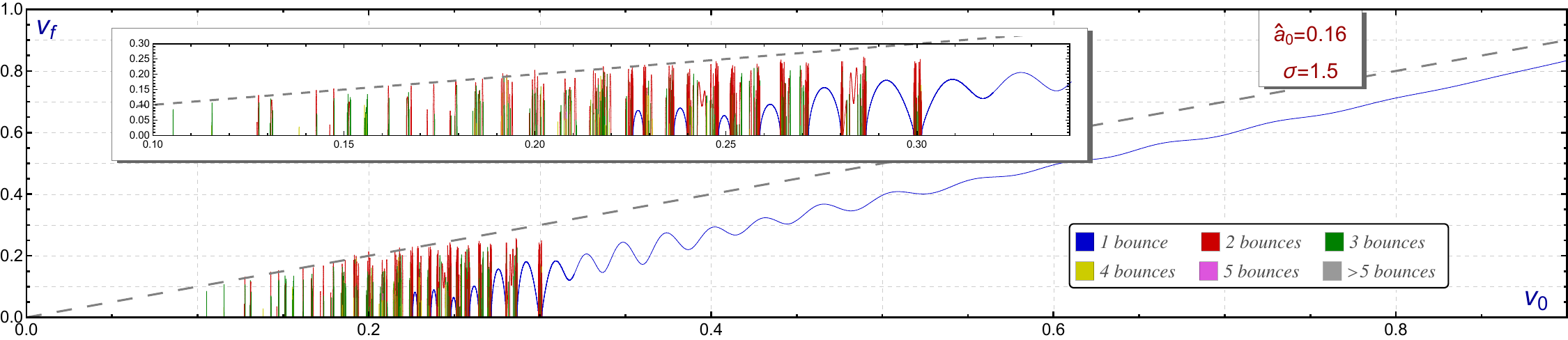}
\end{subfigure}
\hfill
\begin{subfigure}{1\textwidth}
    \includegraphics[width=1.0\linewidth]{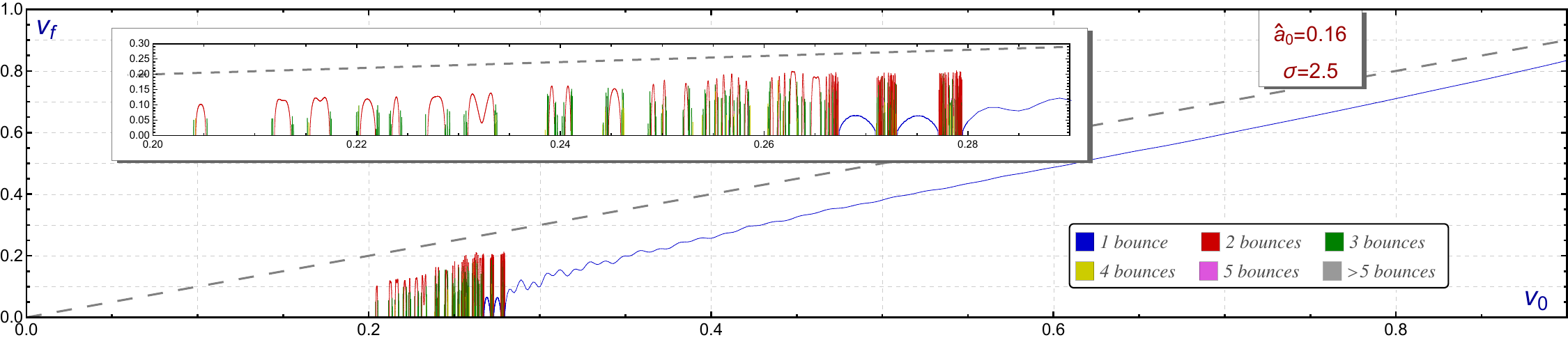}
\end{subfigure}
  \caption{\textit{Final velocity $v_f$ of the scattered kinks as a function of the initial velocity. The color code shown in the graphs indicates the number of bounces suffered by the kink-antikink pair before moving apart. In initial velocities ranges where no final velocity is shown, a bion is assumed to form. The resonance window where various bounces can be observed has been expanded. For the sake of comparison tThe dashed grey line indicates the elastic scenario $v_0=v_f$. }    }
    \label{FigC2:LowAmplitudeViVf16}
\end{figure}

Let us analyze the case where the scattering of orthogonally wobbling kinks occurs with a large initial amplitude $\widehat{a}_0$. In Figure~\ref{FigC2:LowAmplitudeViVf16}, the velocity diagrams for the case in which $\widehat{a}_0=0.16$ are depicted. Now, the behaviors described above are much more pronounced. For example, for small values of $\sigma$, the prevalence of isolated one-bounce windows is observed, which for the case of $\sigma=1.2$ occupies the range between $v_0\in [0.19,0.405]$, and the one-bounce tail presents significant oscillations.

In the singular case $\sigma_1={\sqrt{7}}/{2}$, we observe a highly organized structure in the fractal pattern within the velocity diagram. Isolated one-bounce windows emerge in a repeating pattern, decreasing in both height and width as the initial velocity $v_0$ decreases. As previously mentioned, this is related to the fact that the resonant energy mechanism is practically governed in this case by only two modes (the zero mode and the longitudinal one), resembling the behavior observed in the $\phi^4$ model. It is worth noting that the peak of these windows corresponds closely to the velocity associated with the elastic scenario. 

For the cases where $\sigma=\sqrt{2}$ and $1.5$, the fractal structure is highly complex, also showing isolated one-bounce windows along with others with a greater number of bounces, all densely packed. Note that the regularity seen in the case of $\sigma=\sigma_1$ is completely lost, replaced by a sequence of isolated one-bounce windows that alternate between higher and lower heights. This suggests a higher level of chaotic behavior compared to that observed in the $\phi^4$ model, attributable to the presence of an additional shape mode. 
Similarly, it can be seen in  Figure~\ref{FigC2:LowAmplitudeViVf16} how the oscillations of the one-bounce tail decrease as the value of $\sigma$ increases, as seen in the cases of $\sigma=1.5$ and $\sigma=2.5$. Finally, in the case where $\sigma=2.5$, the presence of isolated one-bounce windows is practically suppressed, and the fractal structure is compressed into a much smaller region than that presented in the previously mentioned cases. 
Additionally, it is worth mentioning that the behavior observed in the scattering of wobbling kinks in the $\phi^4$ model, where the kinks transferred a significant amount of energy from the vibrational modes to the zero modes, allowing final velocities of the kinks greater than the initial ones, does not usually appear in the present case. The reason is clearly due to the presence of two shape modes in this case, both accumulating energy, and it is difficult for both to simultaneously transfer their energy to the zero mode to recreate the situation discussed above.

\subsection{Analysis of the shape mode amplitudes} \label{SecC2:section4.2}

\begin{figure}[h!]
\centering
\begin{subfigure}{1\textwidth}
    \includegraphics[width=1.0\linewidth]{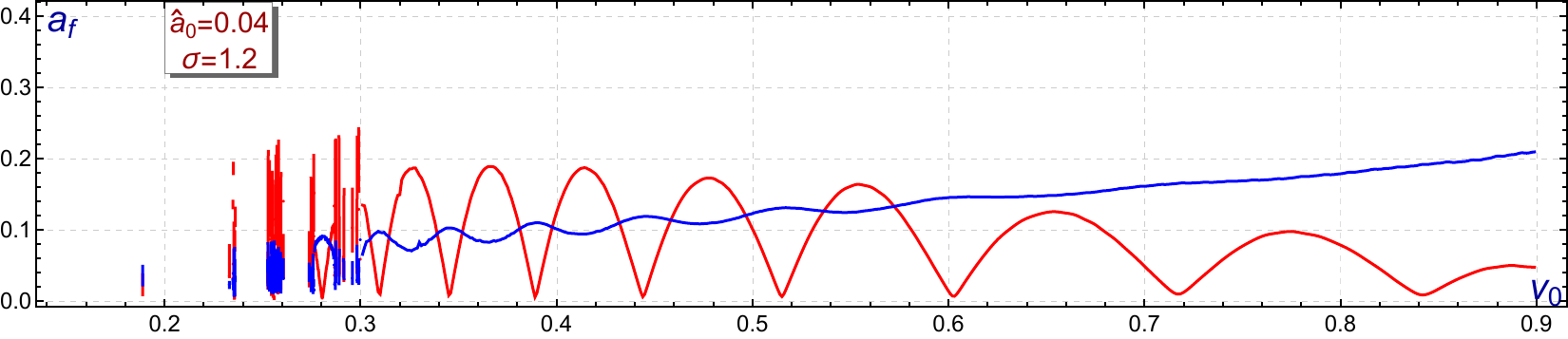}
\end{subfigure}
\hfill
\begin{subfigure}{1\textwidth}
    \includegraphics[width=1.0\linewidth]{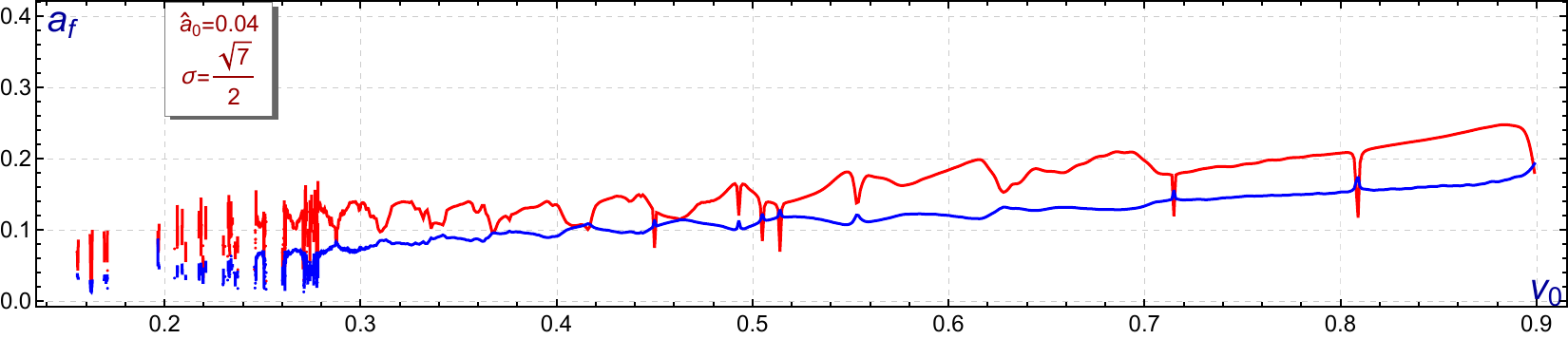}
\end{subfigure}
\hfill
\begin{subfigure}{1\textwidth}
    \includegraphics[width=1.0\linewidth]{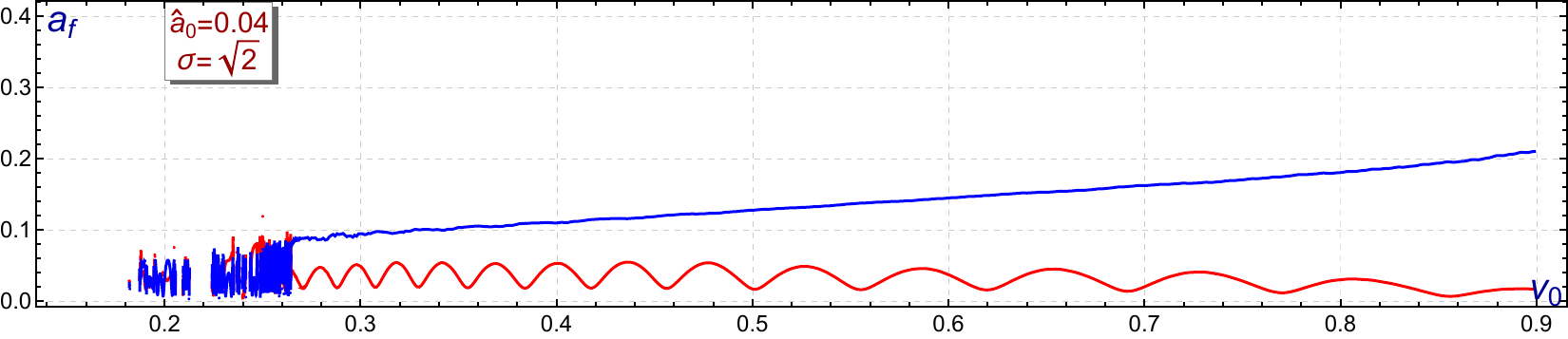}
\end{subfigure}
\hfill
\begin{subfigure}{1\textwidth}
    \includegraphics[width=1.0\linewidth]{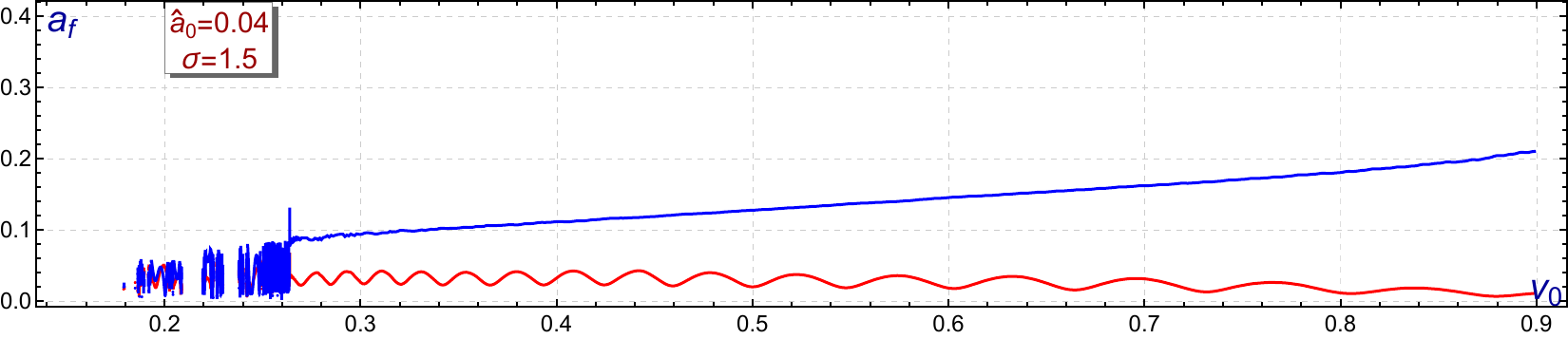}
\end{subfigure}
\hfill
\begin{subfigure}{1\textwidth}
    \includegraphics[width=1.0\linewidth]{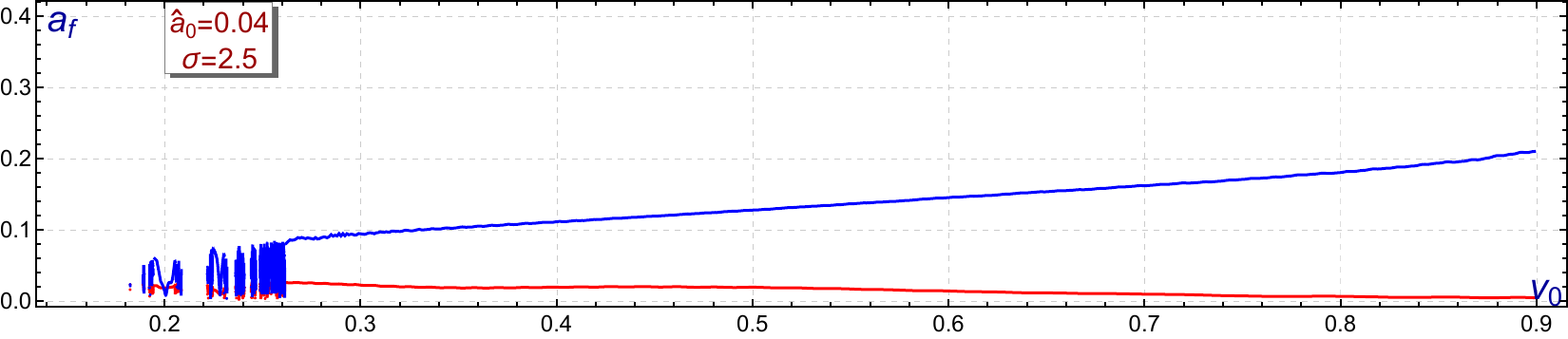}
\end{subfigure}

  \caption{\textit{Final wobbling amplitudes of the scattered kinks after the last collision as a function of $v_0$ for various values of $\sigma$ and initial wobbling amplitude $\widehat{a}_0=0.04$. The blue line represents $\overline{a}_f$ (the final longitudinal wobbling amplitude) and the red line represents $\widehat{a}_f$ (the final orthogonal wobbling amplitude).}}
    \label{FigC2:AmplitudeInternalModesVsViSeveralSigma}
\end{figure}

In this section, we will analyze the behavior of the final vibration amplitudes in both the longitudinal and orthogonal channels. This analysis provides valuable information on how  the resonant energy transfer mechanism works in this model. 
To accomplish this task, the final amplitudes of both shape modes (longitudinal and orthogonal) have been plotted as a function of initial velocity for various values of the amplitude $\widehat{a}_0$ of the orthogonal shape mode and the coupling constant $\sigma$. 
Figure~\ref{FigC2:AmplitudeInternalModesVsViSeveralSigma} illustrates the behavior of these amplitudes with the initial amplitude value set to $\widehat{a}_0=0.04$, while the model parameter $\sigma$ takes various values: $\sigma=1.2, \sqrt{7}/2, \sqrt{2}, 1.5$, and $2.5$. 
In the graphs shown in Figure~\ref{FigC2:AmplitudeInternalModesVsViSeveralSigma}, the amplitude of the longitudinal shape mode is represented in blue, while that of the orthogonal shape mode is represented in red. The behavior shown in this figure is similar for other values of the amplitude $\widehat{a}_0$. 

It is evident that the dispersion process depends largely on the value of the constant $\sigma$. Analogous to the results obtained previously, we can identify two significant regimes. If the value of $\sigma$ is close to 1 and below the resonance value associated with $\sigma=\sigma_1$, the effect of the resonant energy transfer mechanism becomes particularly pronounced.
There are values of the collision velocity $v_0$ for which the final amplitude of the orthogonal shape mode is almost suppressed, making its value almost disappear. These values are aligned with the peaks of the longitudinal shape mode amplitude. On the other hand, the minima of the longitudinal shape mode amplitude correspond to the maxima of the orthogonal mode amplitude. 
This clearly indicates the energy transfer between these two modes. It is worth noting that the oscillations of the former mode are smaller than those of the latter. This difference arises because the eigenvalue of the orthogonal shape mode is smaller than that of the longitudinal mode, which means that the orthogonal shape mode is less energetic than the longitudinal mode.

\begin{figure}[h!]
      \centering
	   \begin{subfigure}{1\linewidth}
		\includegraphics[width=1.0\linewidth]{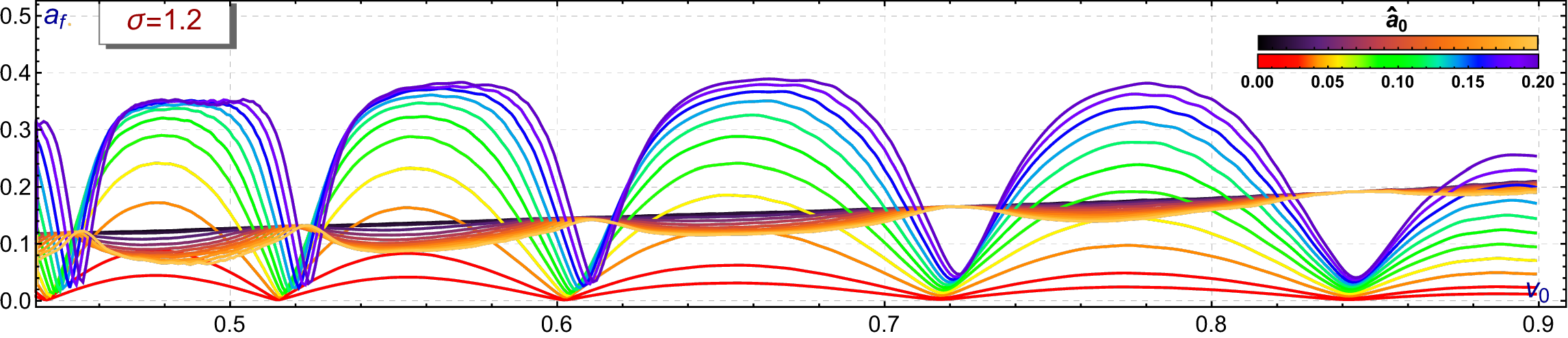}
		\label{FigC2:FinalAmplitudesVsV0Sigma12}
	   \end{subfigure}
	   \begin{subfigure}{1\linewidth}
      \vspace{-0.4cm}
		\includegraphics[width=1.0\linewidth]{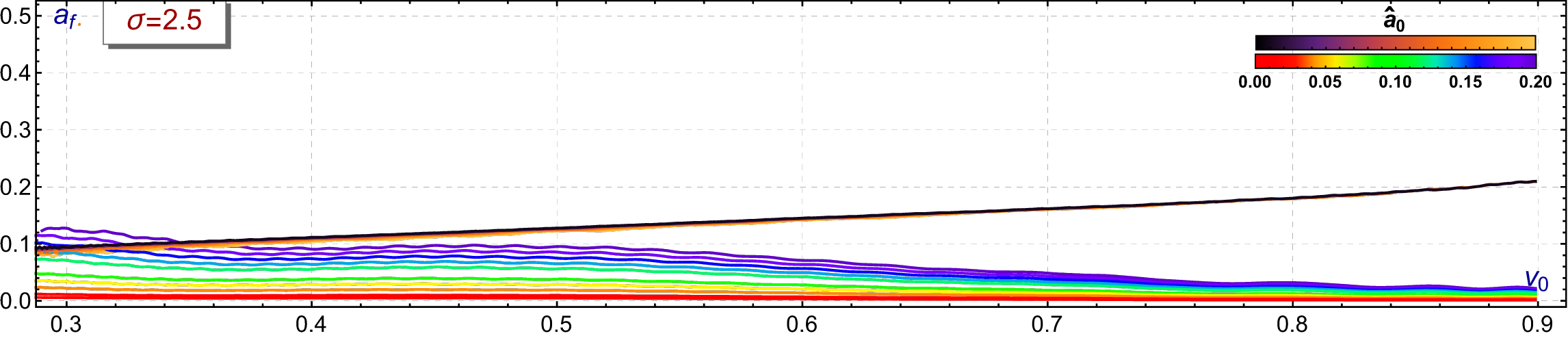}
		\label{FigC2:FinalAmplitudesVsV0Sigma25}
	    \end{subfigure}   
        \vspace{-1.1cm}
	\caption{\textit{Final amplitudes of both shape modes after the scattering as a function of the initial velocity $v_0$ for various values of $\widehat{a}_0\in [0,0.20]$ and for $\sigma=1.2$ and $2.5$. The first legend located in the upper right corner of the figure corresponds to the amplitude of the longitudinal shape mode while the second corresponds to the amplitude associated with the  orthogonal mode.}}
	\label{FigC2:FinalAmplitudesVsV0Sigma12-25}
\end{figure}

In the other regime that we mentioned, for values of $\sigma$ greater than $\sigma_1=\sqrt{7}/2$, we observe a similar behavior in the amplitude of the orthogonal shape mode, which oscillates as a function of the initial velocity. However, these oscillations have a smaller amplitude and decrease as the coupling constant $\sigma$ increases. 
In fact, for $\sigma=2.5$, these fluctuations are almost imperceptible. Additionally, it should be noted that, as a general trend, the frequency of these oscillations increases as the value of $\sigma$ increases, which is related to the fact that the frequency of the orthogonal shape mode also increases (see \cite{AlonsoIzquierdo2023}). 
On the other hand, the amplitude of the longitudinal shape mode is not influenced by the orthogonal shape mode. In this case, the excitation of this mode can be attributed almost entirely to the collision of the kinks, or more specifically, to the interaction with the zero modes of these solutions. This behavior is best observed in Figure~\ref{FigC2:FinalAmplitudesVsV0Sigma12-25}, where the evolution of the amplitudes corresponding to both shape modes for $\sigma=1.2$ and $\sigma=2.5$ as a function of $v_0$ and $\widehat{a}_0$ can be observed. 
As shown in the top graph of Figure~\ref{FigC2:FinalAmplitudesVsV0Sigma12-25}, for low values of $\sigma$ increasing the value of the initial amplitude $\widehat{a}_0$ results in a greater  final amplitude of orthogonal shape mode $\widehat{\eta}$ after the last collision. 
Simultaneously, as the maxima corresponding to $\widehat{a}_f$ increase, the minima of $\overline{a}_f$ decrease. Note that there is an energy loss in the longitudinal shape mode compared to the case where the orthogonal shape mode is not initially excited. 
Another important aspect to highlight is that the nodes of the oscillations exhibited by the shape modes depend on the initial amplitude. As seen in Figure~\ref{FigC2:FinalAmplitudesVsV0Sigma12-25}, the nodes shift slightly to the right as the value of $\widehat{a}_0$ increases.
On the other hand, as shown in the lower graph of Figure~\ref{FigC2:FinalAmplitudesVsV0Sigma12-25}, the behavior of the longitudinal shape mode amplitude does not depend on the initial wobbling amplitude $\widehat{a}_0$ when $\sigma$ is high enough. In fact, the wobbling amplitude of the orthogonal mode stops oscillating and its value decreases as the kinks collide with greater velocity.

\begin{figure}[h!]
\centering
\begin{subfigure}{1\textwidth}
    \includegraphics[width=1.0\linewidth]{Images/Chap2/Section4.2/AmplitudeInternalModesVsViS13A004.pdf}
\end{subfigure}
\hfill
\begin{subfigure}{1\textwidth}
    \includegraphics[width=1.0\linewidth]{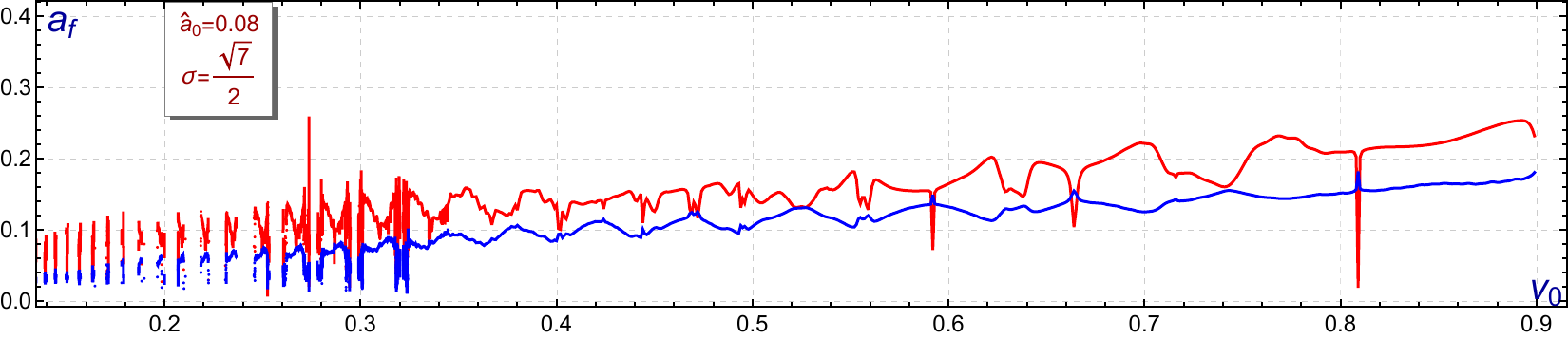}
\end{subfigure}
\hfill
\begin{subfigure}{1\textwidth}
    \includegraphics[width=1.0\linewidth]{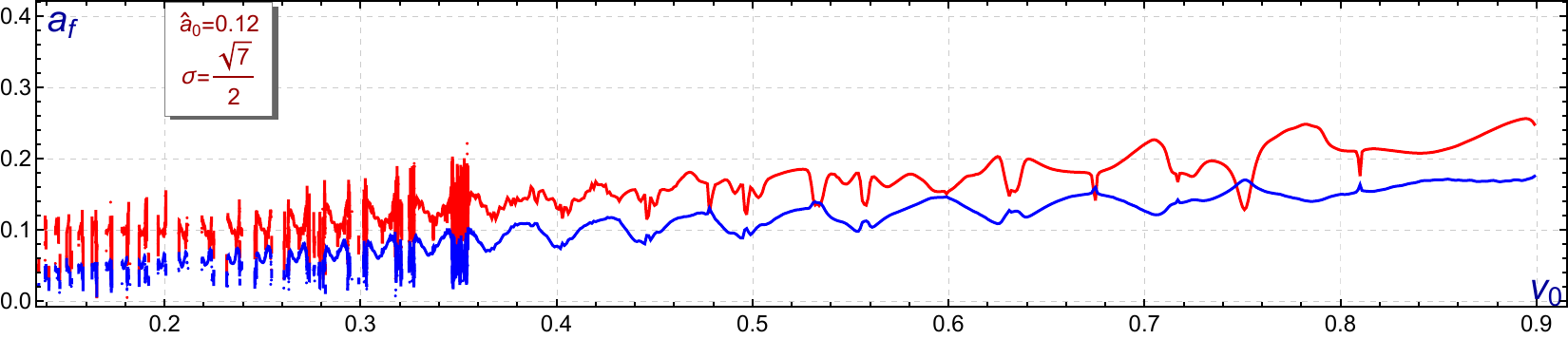}
\end{subfigure}
\caption{\textit{Final wobbling amplitudes for the scattered kinks after the last collision as a function of $v_0$ for $\sigma=\sqrt{7}/2$ and wobbling amplitude $\widehat{a}_0=0.04$. The blue line represents $\overline{a}_f$ (the final longitudinal wobbling amplitude) and the red line represents $\widehat{a}_f$ (the final orthogonal wobbling amplitude).}}
    \label{FigC2:AmplitudeInternalModesVsViSeveralA0S13}
\end{figure}

However, the behavior of the system is completely different for the special value $\sigma_1=\sqrt{7}/2$, where a resonance occurs between the frequencies $\overline{\omega}$ and $2\widehat{\omega}$. In this case, the graphical representation of the amplitudes is more irregular than in the other cases and, surprisingly, the final amplitude of both shape modes does not depend on the initial wobbling amplitude $\widehat{a}_0$, being $\widehat{a}_f$ almost always slightly larger than the amplitude corresponding to the longitudinal eigenmode (see Figure~\ref{FigC2:AmplitudeInternalModesVsViSeveralA0S13}). 
{  The mechanism that seems to explain the scattering of the initially excited orthogonally wobbling kinks in this case is the following: as explained in \cite{AlonsoIzquierdo2023}, when the orthogonal mode is initially triggered with a certain amplitude a significant amount of energy will be transferred to the longitudinal eigenmode due to the aforementioned resonance. In turn, this explains why the velocity diagrams are similar to those found  for the $\phi^4$ model, where there is no second component. After the last collision, the longitudinal mode  is recharged via both the resonance and the redistribution of energy in the scattering process. Once this phenomenon takes place, the longitudinal eigenmode has enough energy to excite the orthogonal one once again until both amplitudes reach an equilibrium state.  This phenomenon can be appreciated in Figure \ref{FigC2:AmplitudeInternalModesVsViSeveralA0S13}. Here, it can be seen that the pattern of both shape mode amplitudes remains consistent as the initial wobbling amplitude increases. The lack of smoothness observed in the figure can be attributed to the chaotic behavior of the system for this specific value of $\sigma$.

Since both modes have been excited, the energy transfer channel from the orthogonal to the longitudinal mode is suppressed.
However, it should be noted that for this critical value of $\sigma$, there is also strong radiation emission in the orthogonal channel with frequencies $\omega=\overline{\omega}+ \widehat{\omega}$ and $3\widehat{\omega}$, resulting in an increase in decay in the shape mode chain. It is important to consider that this process is quite complex, and its description through numerical schemes may depend on the $t_{\mathrm{ max}}$ used in the simulations, which justifies the irregular behavior of the graphs shown for this value. 
Furthermore, the reason behind $\widehat{a}_f>\overline{a}_f$ can be explained by considering that the longitudinal eigenvalue is larger than the orthogonal one, which makes the orthogonal mode easier to excite. In fact, it can be seen in Figure~\ref{FigC2:AmplitudeInternalModesVsViSeveralA0S13} that in the one bounce tail regime $\frac{\overline{a}_f}{\widehat{a}_f}\approx\frac{\overline{\omega}}{\widehat{\omega}}$.}

\section{Concluding remarks}\label{SecC2:section5}

In this chapter, we have investigated the interaction between the translational mode and the shape modes in a two-component scalar field theory in the context of kink-antikink scattering processes when only the mode corresponding to the second field component is initially activated. 
The energy transfer mechanism between modes has already been discussed, both analytically and numerically in \cite{AlonsoIzquierdo2023}, but only for a static kink whose orthogonal mode had initially been  triggered. The results presented here are therefore a natural continuation of the aforementioned paper, shedding light on how the resonant energy mechanism  affects the fractal structure found in the velocity diagrams. 
The presence of two different shape modes makes the energy transfer mechanism more complex than when wobbler collisions were studied in the $\phi^4$ model. %Since there are now two different discrete eigenmodes instead of just one, there is less  kinetic energy available for the translational mode, which explains why the escape velocity of the kink-antikink pair after the last collision $v_f$ is never greater than the initial kink velocity collision $v_0$. 
%This is in contrasts to what was found in the $\phi^4$ model, since in the latter, when the amplitude of the only shape mode was triggered with enough amplitude, $v_f$ could be larger than $v_0$. 
{  With two distinct discrete eigenmodes now present instead of just one, less kinetic energy is available for the translational mode. This explains why the escape velocity of the kink-antikink pair after the final collision, $v_f$, only exceeds the initial kink collision velocity, $v_0$, within a few specific windows.
This differs from the results observed in the $\phi^4$ model, where sufficiently triggering the amplitude of the single shape mode enables the existence of larger windows where $v_f$ exceeds $v_i$. }

Another notable phenomenon found in this study is the changing of the velocity diagrams as $\widehat{a}_0$ and $\sigma$ change their value. For low values of $\sigma$, when $\widehat{a}_0$ increases, one-bounce windows were created in the resonant part of the velocity diagram. 
However, for high values of $\sigma$ and a small value for $\widehat{a}_0$ the velocity diagram resembles the behavior found for an unexcited kink in the $\phi^4$ field theory. This phenomenon does not manifest itself for higher values of the initial orthogonal amplitude, since in the latter case the resonant part of the diagram becomes chaotic and the amplitude of the  oscillations of the one-bounce regime decreases.    

The system response changes dramatically  when the value of $\sigma$ is set to $\sqrt{7}/2$. Under these conditions, the orthogonal mode discharges most of its energy into the longitudinal mode and  radiation modes. This implies that the longitudinal mode is activated  with a certain amplitude before the scattering process. Such circumstance clarifies the similarity between the velocity diagrams for the resonance value $\sigma_1=\sqrt{7}/2$ and the velocity diagrams coming from the wobbler scattering in the $\phi^4$ model \cite{AlonsoIzquierdo2021b,AlonsoIzquierdo2022}.

In addition to what has already been explained, it is worth highlighting the correspondence between the maxima and minima of the final wobbling amplitudes explained in detail in Section~\ref{SecC2:section4.2}, which has shed light on the energy transfer mechanism between modes which, as has being shown, strongly depends on the value of $\sigma$ since the orthogonal mode can only be triggered again after the scattering process for values of $\sigma\approx 1$.

As a future line of research that serves as a natural continuation of the study presented here, the numerical techniques presented here can be generalized to study the scattering of excited topological defects defined in higher dimensions. Among the innumerable possibilities that are contemplated in this regard, one of them is generalize the results found for the excited vortex scattering process in the Abelian-Higgs model recently presented  in  \cite{AlonsoIzquierdo2024d, Krusch2024} studying the same situation in generalized models that support the existence of vortices. Another potential avenue of exploration considered involves the study  of  kink-antikink  collisions in two-component scalar field models whose kink solutions possess more than three internal modes. In this regard, a suitable candidate for this research would be the two-component coupled $\phi^4$ model \cite{Halavanau2012, AlonsoIzquierdo2024}, in which the internal structure and the possible number of eigenmodes depend on the value of the coupling constant that the potential defining the model depends on.

    \part{Abelian-Higgs vortices}\label{Part2}

    \chapter{Introduction to Abelian-Higgs vortices}\label{Intro2}
    %In this chapter, we will introduce the Abelian-Higgs model. As mentioned in Chapter \ref{Intro0}, this field theory  takes on particular importance in areas such as cosmology, superfluidity and superconductivity.  We will explore the fundamentals of this model and how to obtain the vortex solutions. 

%We will also explore some details related to the behaviour of the profile functions near the vortex center and the infinity. We will also briefly discuss the interactions between vortices both outside and inside the BPS limit of the theory.   Some details of the relation between Abelian-Higgs vortices and global vortices will also be given. Some details about vortex scattering phenomena will also be given as well as a short discussion about the moduli space approximation. 

%This chapter will lay the foundation for the study of the internal mode structure of these solutions and its radiation emission when one of these modes is initially triggered. These aforementioned results will be discussed in Chapters \ref{Chap3} and \ref{Chap4} respectively.

In this chapter, we introduce the Abelian-Higgs model. As mentioned in Chapter \ref{Intro0}, this field theory is particularly important in areas such as cosmology, superfluidity, and superconductivity. We will explore the fundamentals of the model and the procedure for obtaining vortex solutions.

We will also examine the behavior of the profile functions near the vortex center and at infinity. Additionally, we will  discuss the interactions between vortices, both outside and within the BPS limit of the theory. The relationship between Abelian-Higgs vortices and global vortices will be also addressed. Moreover, the spectral wall phenomenon for Abelian-Higgs vortices in the BPS limit will be discussed.

This chapter lays the foundation for studying the internal mode structure of these solutions and their associated radiation emission when one of these modes is initially excited. These results will be discussed in Chapters \ref{Chap3} and \ref{Chap4}, respectively.

\section{From global vortices to Abelian-Higgs vortices}

\subsection{Global vortices}
\begin{comment}
  Let us start with a simpler model defined in only $(2+1)$ dimensions depending only on a complex scalar field. This theory is ruled by the Lagrangian density
\begin{equation}\label{eqI3:LAGdENS}
    \mathcal{L}=\frac{1}{2}\overline{\partial_\mu \phi}\, \partial^\mu\phi-\frac{\lambda}{8}(1-\overline{\phi}\,\phi)^2,
\end{equation}
where the Minkowski metric is taken in the form $g_{\mu\nu}=\mathrm{diag}\{1,-1,-1\}$ and $\lambda$ is a real positive constant.

Therefore, the corresponding field equation is 
\begin{equation}\label{eqI3:globalFieldEq}
    \partial_{tt}\phi-\nabla^2\phi=\frac{\lambda}{2}(1-\overline{\phi}\,\phi)\phi.
\end{equation}

The first thing to notice is that the vacuum manifold $\mathcal{V}$ arising from \eqref{eqI3:LAGdENS} is the circle $\phi=e^{i\chi}$ where $\chi$ is an arbitrary phase. By plugging $\phi=e^{i\chi}$ into \eqref{eqI3:globalFieldEq} we inmediatly see that $\nabla^2\chi=\nabla\chi\nabla\chi=0$, which implies that $\chi$ must be constant. Another remarkable fact about this is  theory by Derrick's theorem, which was introduced in Chapter \ref{DerrickTheoremSec}, there are no static solutions to \eqref{eqI3:globalFieldEq} with finite energy.   
\end{comment}

Let us start with a  model defined in  $(2+1)$ dimensions and depending solely on a complex scalar field $\phi$ involving a $U(1)$ symmetry in the internal target space. This theory is governed by the Lagrangian density
\begin{equation}\label{eqI3:LAGdENS}
\mathcal{L}=\frac{1}{2}\overline{\partial_\mu \Phi}\, \partial^\mu\Phi - \frac{\lambda}{8}(1 - \overline{\Phi}\,\Phi)^2,
\end{equation}
where the Minkowski metric is taken in the form $g_{\mu\nu} = \mathrm{diag}\{1, -1, -1\}$, and $\lambda$ is a real positive constant.

The corresponding field equation is therefore
\begin{equation}\label{eqI3:globalFieldEq}
\partial_{tt}\Phi - \nabla^2\Phi = \frac{\lambda}{2}(1 - \overline{\Phi}\,\Phi)\Phi.
\end{equation}

The first thing to notice is that the vacuum manifold $\mathcal{V}$ arising from \eqref{eqI3:LAGdENS} is the circle $\Phi = e^{i\chi}$, where $\chi$ is an arbitrary phase. By plugging $\Phi = e^{i\chi}$ into \eqref{eqI3:globalFieldEq}, we immediately see that $\nabla^2\chi = \nabla\chi \cdot \nabla\chi = 0$, which implies that $\chi$ must be constant.

Another remarkable fact about this theory is that, by Derrick's theorem (introduced in Section \ref{DerrickTheoremSec}), there are no static solutions to \eqref{eqI3:globalFieldEq} with finite energy.
In order to demonstrate this, let us first rewrite the static energy using polar coordinates:
\begin{equation}
    x_1 = r\cos\theta, \quad x_2 = r\sin\theta.
\end{equation}
This allows us to express the energy as
\begin{equation}\label{eqIe:PotPolarsGlobal}
    V = \frac{1}{2} \int_0^\infty \int_0^{2\pi} \left( \partial_r \overline{\Phi}\, \partial_r \Phi + \frac{1}{r^2} \partial_\theta \overline{\Phi}\, \partial_\theta \Phi + \frac{\lambda}{4}(1 - \overline{\Phi} \Phi)^2 \right) r\, dr\, d\theta.
\end{equation}

Let us now consider a field configuration $\Phi(r,\theta)$ whose corresponding energy density vanishes rapidly as $r \to \infty$. From \eqref{eqIe:PotPolarsGlobal}, it is easy to see that as $r \to \infty$, $|\Phi| \to 1$ and $\partial_r \Phi \to 0$, which allows us to write
\[
\lim_{r \to \infty} \Phi(r, \theta) = \Phi^\infty(\theta) = e^{i\chi^\infty(\theta)}.
\]

Necessarily, this must be a map from the circle at infinity $S^1_\infty$ to the vacuum manifold $S^1$. Since $\Phi^\infty(\theta)$ is single-valued, the phase $\chi^\infty(\theta)$ must satisfy
\[
\chi^\infty(2\pi) = \chi^\infty(0) + 2\pi n,
\]
where $n$ is an integer. This integer is commonly referred to as the \textit{winding number} or \textit{topological charge} of the field.

Considering a sufficiently large circle of radius $r_\infty$, the only contribution to the energy outside this circle is
\begin{equation}\label{eqI3:EnergyDivergence}
    \frac{1}{2} \int_{r_\infty}^{\infty} \int_0^{2\pi} \frac{(\partial_\theta \chi^\infty)^2}{r} \, dr\, d\theta.
\end{equation}

From \eqref{eqI3:EnergyDivergence}, it is evident that the energy \eqref{eqIe:PotPolarsGlobal} diverges logarithmically,  unless
\[
\int_0^{2\pi} (\partial_\theta \chi^\infty)^2\, d\theta = 0,
\]
which only occurs when $n = 0$. This, in turn, corresponds to the trivial vacuum solution $\Phi = e^{i\chi^\infty}$, where $\chi^\infty$ is a constant. All this implies that there are no any finite energy solution with $n=1,2,3,\dots$

Although finite-energy field configurations do not exist in this theory, we can still find vortex solutions. To do so, we assume the field takes the form
\[
\Phi(r, \theta) = f_n(r)e^{i n\theta},
\]
which leads to the following equation for the profile function $f_n(r)$\footnote{Since $\lambda$ corresponds to a rescaling of $\Phi$, we set $\lambda = 2$ without loss of generality.}:
\begin{equation}\label{eqI3:GVortexEq}
    \frac{d^2 f_n}{d r^2} + \frac{1}{r} \frac{d f_n}{d r} - \frac{n^2}{r^2} f_n + (1 - f_n^2)f_n = 0.
\end{equation}

\begin{figure}[h!]
    \centering
    \includegraphics[width=0.67\linewidth]{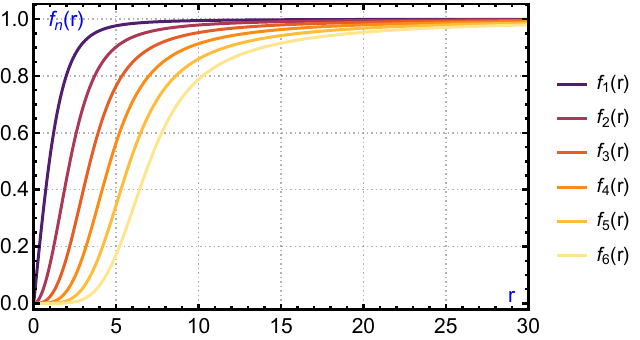}
    \caption{\textit{Global vortex profile functions for different values of $n$.}}
    \label{figI3:GlobalVortexProfiles}
\end{figure}

As it can be appreciated, \eqref{eqI3:GVortexEq} is a non-linear second-order ordinary differential equation that must be solved numerically. Nevertheless, it can be shown that near the vortex origin $f_n(r) \sim r^n$, and the asymptotic behaviour for large values of $r$ is
\[
f_n(r) \sim 1 - \frac{n^2}{2 r^2} - \frac{n^2(n^2+8)}{8 r^4}.
\]
The profile functions $f_n(r)$ are depicted in Figure \ref{figI3:GlobalVortexProfiles} for different values of the topological charge $n$. As can be seen, the behavior of the functions $f_n(r)$ near the origin matches the expected analytical behavior.

Now that we have obtained the vortex solutions from this model, let us briefly discuss their stability and internal mode structure.

To do so, we assume that the perturbations take the form
\begin{equation}
\Phi(r,\theta,t) = \left(f_n(r) + \epsilon \left( s^r_m(r)\cos(m\theta) + i\, s^i_m(r)\sin(m\theta) \right) \right)e^{i n \theta},
\end{equation}
where $\epsilon$ is a small real parameter.

The corresponding spectral problem takes the form
\begin{equation}\label{eqI3:SpecProbGVortex}
    \mathcal{H}
    \begin{pmatrix}
        s^r_m(r)\\
        s^i_m(r)
    \end{pmatrix}
    = \omega^2
    \begin{pmatrix}
        s^r_m(r)\\
        s^i_m(r)
    \end{pmatrix},
\end{equation}
where
\begin{equation}
    \mathcal{H} = 
    \begin{pmatrix}
        -\frac{d^2}{dr^2} - \frac{1}{r}\frac{d}{dr} + \frac{m^2 + n^2}{r^2} + (3f_n(r)^2 - 1) & \frac{2mn}{r^2} \\
        \frac{2mn}{r^2} & -\frac{d^2}{dr^2} - \frac{1}{r}\frac{d}{dr} + \frac{m^2 + n^2}{r^2} + (f_n(r)^2 - 1)
    \end{pmatrix}.
\end{equation}

At large $r$, the operator asymptotically becomes
\begin{equation}
    \mathcal{H}_\infty = 
    \begin{pmatrix}
        -\frac{d^2}{dr^2} - \frac{1}{r}\frac{d}{dr} + 2 & 0 \\
        0 & -\frac{d^2}{dr^2} - \frac{1}{r}\frac{d}{dr}
    \end{pmatrix},
\end{equation}
which implies that there are two distinct mass thresholds: one at $\omega_r^2 = 2$ and another at $\omega_i^2 = 0$.

In the simplest case ($m = 0$, $n = 1$), the spectral problem reduces to solving
\begin{equation}
    -\frac{d^2 s^r_m(r)}{dr^2} - \frac{1}{r} \frac{d s^r_m(r)}{dr} + U(r) s^r_m(r) = \omega^2 s^r_m(r),
\end{equation}
with the effective potential
\begin{equation}
    U(r) = \frac{1}{r^2} + (3f_n(r)^2 - 1).
\end{equation}

\begin{figure}[h!]
    \centering
    \begin{subfigure}{0.49\textwidth}
        \centering
        \includegraphics[width=\linewidth]{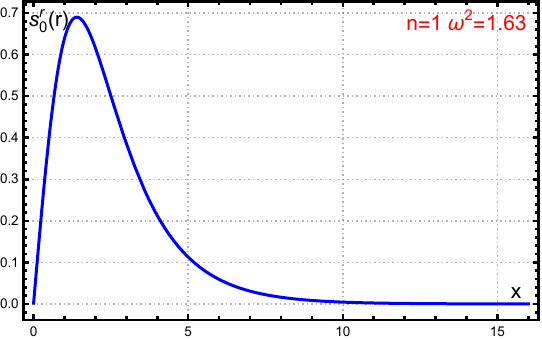}
    \end{subfigure}
    \begin{subfigure}{0.49\textwidth}
        \centering
        \includegraphics[width=\linewidth]{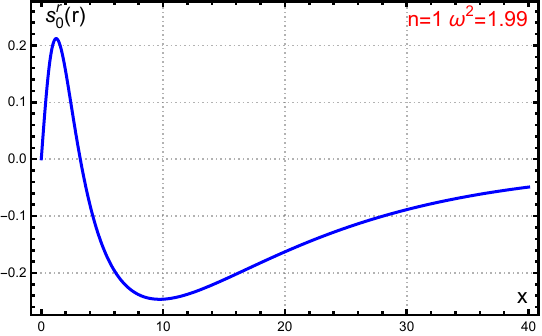}
    \end{subfigure}
    \caption{\textit{First two internal bound modes for a global vortex with $n=1$.}}
    \label{figI3:GVInternalModesn1}
\end{figure}

At large distances, the potential behaves as
\begin{equation}
    U(r) \approx 1 - \frac{2}{r^2},
\end{equation}
suggesting a Coulomb-like decay. Thus, an infinite number of internal modes exist, with frequencies lying below the threshold $\omega^2 = 2$. In Figure \ref{figI3:GVInternalModesn1}, the first two bound modes for a global vortex with winding number $n = 1$ are shown. Note that the second mode, being close to the continuum threshold, is more spread out and less localized around the vortex core.

\begin{figure}[h!]
    \centering
    \includegraphics[width=0.6\linewidth]{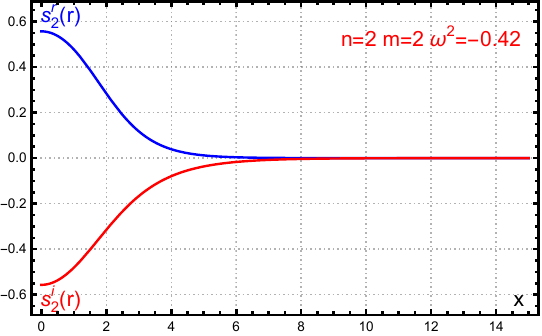}
    \caption{\textit{Unstable mode profile for a global vortex with $n=2$.}}
    \label{figI3:UnstableModen2m2}
\end{figure}

Moreover, for $n > 1$, the static vortex configuration becomes unstable, and several unstable modes arise in the resolution of \eqref{eqI3:SpecProbGVortex}. In Figure \ref{figI3:UnstableModen2m2}, an unstable mode for a vortex with winding number $n=2$ is depicted.

As a final remark, the zero modes can be computed similarly to the case of kink solutions. They are given by
\begin{eqnarray}
    \Phi(x + \epsilon, y) \!\!\!\!&\approx&\!\!\!\! \Phi(x, y) + \epsilon \frac{\partial \Phi(x, y)}{\partial x} = \Phi(x, y) + \epsilon\, s_{1,x}(x, y), \label{eqI3:ZeroMode1} \\
    \Phi(x, y + \epsilon) \!\!\!\!&\approx& \!\!\!\!\Phi(x, y) + \epsilon \frac{\partial \Phi(x, y)}{\partial y} = \Phi(x, y) + \epsilon\, s_{1,y}(x, y), \label{eqI3:ZeroMode2}
\end{eqnarray}
where
\begin{eqnarray}
    s_{1,x}(r, \theta) \!\!\!\!&=& \!\!\!\!\frac{e^{i n \theta}}{r} \left(-i n f_n(r) \sin(\theta) + r \cos(\theta) f_n'(r)\right), \\
    s_{1,y}(r, \theta) \!\!\!\!&=&\!\!\!\! \frac{e^{i n \theta}}{r} \left(i n f_n(r) \cos(\theta) + r \sin(\theta) f_n'(r)\right).
\end{eqnarray}

As we shall see in Chapter~\ref{Chap3}, the calculation of zero and internal modes for Abelian-Higgs vortices is significantly more complex. For instance, the zero modes cannot be obtained using \eqref{eqI3:ZeroMode1}–\eqref{eqI3:ZeroMode2}. Instead, they must satisfy the Gauss law and all imposed gauge conditions.

For more details about the internal mode structure of global vortices, references~\cite{Goodband1995,BlancoPillado2021} provide a suitable background.

\subsection{Static finite energy vortex configurations  and the Abelian-Higgs model}

As previously stated, global vortices are static field configurations whose energy diverges. The reason behind this divergence is Derrick's theorem. The question now is: is it possible to find a model in which vortices do exist and have finite energy? 

To address this question, let us consider the Lagrangian density \footnote{The scalar field $\Phi$ is usually referred to as the Higgs field.}
\begin{equation}\label{eqI3:LagDens}
    \mathcal{L} = -\frac{1}{4} F_{\mu\nu} F^{\mu\nu} + \frac{1}{2} D_{\mu} \Phi\, \overline{D^{\mu} \Phi} - \frac{\lambda}{8} (1 - \Phi\, \overline{\Phi})^2,
\end{equation}
where $D_\mu = \partial_\mu - i A_\mu$, $F_{\mu\nu}$ is the electromagnetic field tensor defined as $F_{\mu\nu} = \partial_\mu A_\nu - \partial_\nu A_\mu$, and the Minkowski metric is taken as $g_{\mu\nu} = \mathrm{diag}\{1, -1, -1, \}$. Furthermore, we assume that \eqref{eqI3:LagDens} is defined in two spatial dimensions.

Assuming the temporal gauge $A_0 = 0$, the static energy is given by
\begin{equation}\label{eqI3:StaticEnergy}
    E = \frac{1}{2} \int \left(B^2 + \overline{D_i \Phi} \, D_i \Phi + \frac{\lambda}{4} (1 - \overline{\Phi} \Phi)^2 \right) d^2x,
\end{equation}
where $F_{12} =B = \partial_1 A_2 - \partial_2 A_1$  is the magnetic field.

Now, rescaling the spatial coordinates as $x \to \mu x$ implies that the $1$-form potential $A$ transforms as \cite{Manton2004}
\begin{equation*}
    A^{(\mu)}(\mathbf{x}) = \mu \, A(\mu \mathbf{x}).
\end{equation*}
Consequently, the components of the electromagnetic field scale as
\begin{equation*}
    F^{(\mu)}(\mathbf{x}) = \mu^2 F(\mu \mathbf{x}),
\end{equation*}
and the covariant derivative is now
\begin{equation*}
    D^{(\mu)} \Phi^{(\mu)}(\mathbf{x}) = \mu \, D \Phi(\mu \mathbf{x}).
\end{equation*}

Let us define the following energy contributions:
\begin{equation*}
    E_4 = \frac{1}{2} \int B^2 \, d^2x, \quad 
    E_2 = \frac{1}{2} \int \overline{D_i \Phi} \, D_i \Phi \, d^2x, \quad 
    E_0 = \frac{\lambda}{8} \int (1 - \Phi\, \overline{\Phi})^2 \, d^2x.
\end{equation*}
Then, under the rescaling, the energy becomes
\begin{equation}
    E_\mu = \mu^2 E_4 + E_2 + \mu^{-2} E_0.
\end{equation}

This function has a minimum at
\begin{equation}
    \mu = \left(\frac{E_0}{E_4} \right)^{1/4}.
\end{equation}

This indicates that static finite-energy configurations for the Lagrangian density \eqref{eqI3:LagDens} do exist. Furthermore, since the minimum must occur at $\mu = 1$ for a true static solution of the original system, the following relation must be satisfied:
\begin{equation}
    \frac{\lambda}{4} \int (1 - \Phi\, \overline{\Phi})^2 \, d^2x = \int B^2 \, d^2x.
\end{equation}

\section{Topological charge and static vortex solutions}\label{SecVortexSol}

Once we have introduced the Abelian-Higgs model, let us now discuss the topology associated with the configuration space of finite energy static solutions. To begin, we rewrite the static energy associated with \eqref{eqI3:LagDens} in polar coordinates, assuming the change of variables\footnote{In Chapters \ref{Chap3} and \ref{Chap4} we will change this definition for $A_r$ and $A_\theta$ and, instead, we will define these functions as 
\begin{equation*}
    A_r = A_1 \cos\theta + A_2 \sin\theta, \quad 
    A_\theta = -A_1  \sin\theta + A_2 \cos\theta.
\end{equation*}}
\begin{equation}
    A_r = A_1 \cos\theta + A_2 \sin\theta, \quad 
    A_\theta = -A_1 r \sin\theta + A_2 r \cos\theta,
\end{equation}
which allows us to express the static energy as
\begin{equation}
    E = \frac{1}{2} \int_{0}^{\infty} \int_0^{2\pi} \left( \frac{F_{r\theta}^2}{r^2} + \overline{D_r \Phi} D_r \Phi + \frac{\overline{D_\theta \Phi} D_\theta \Phi}{r^2} + \frac{\lambda}{4}(1 - \overline{\Phi} \Phi)^2 \right) r \, dr \, d\theta,
\end{equation}
where $D_r = \partial_r - i A_r$, $D_\theta = \partial_\theta - i A_\theta$, and $F_{r\theta} = \partial_r A_\theta - \partial_\theta A_r$.

Let us now assume the ``radial gauge'' $A_r = 0$. For the energy to be finite, the field $\Phi$ must tend to a vacuum configuration for large values of $r$, i.e.,
\begin{equation*}
    \Phi(r, \theta) \sim  e^{i \chi(r, \theta)}.
\end{equation*}
Moreover, since $D_r \Phi$ must vanish as $r \to \infty$ in this gauge, we have
\begin{equation}
    D_r \Phi = i \,  (\partial_r \chi(r,\theta) - A_r(r,\theta))e^{i\chi(r,\theta)} \Rightarrow \partial_r \chi(r,\theta) = A_r(r,\theta) = 0 \quad \Rightarrow \quad \chi(r,\theta) = \chi(\theta).
\end{equation}

On the other hand, to ensure the energy remains finite, $D_\theta \Phi \to 0$ and $F_{r\theta} \to 0$ as $r \to \infty$. This implies:
\begin{equation}
    \lim_{r \to \infty} A_\theta(r,\theta) = A_\theta^\infty(\theta), \quad \partial_\theta \chi^\infty(\theta) = A_\theta^\infty(\theta).
\end{equation}

Hence, the asymptotic field $\Phi^\infty$ defines a map from the circle at infinity $S_\infty^1$ to the vacuum manifold $S^1$, characterized by a ``winding number'' given by
\begin{equation}
    n = \frac{1}{2\pi} \int_0^{2\pi} \partial_\theta \chi^\infty(\theta) \, d\theta = \frac{\chi^\infty(2\pi) - \chi^\infty(0)}{2\pi},
\end{equation}
which corresponds to the topological charge of the field configuration.

Furthermore, this winding number is directly related to the total magnetic flux via Stokes' theorem:
\begin{equation}
    n = \frac{1}{2\pi} \int_0^{2\pi} \partial_\theta \chi^\infty(\theta) \, d\theta = \frac{1}{2\pi} \int_0^{2\pi} A_\theta^\infty(\theta) \, d\theta = \frac{1}{2\pi} \int B \, d^2x.
\end{equation}

Now that we have introduced the concept of topological charge in the Abelian-Higgs model, let us turn to the question of how to obtain vortex solutions explicitly. The field equations governing the theory are:
\begin{align}
    D_\mu D^\mu \Phi - \frac{\lambda}{2}(1 - \overline{\Phi} \Phi)\Phi &= 0, \label{eqI3:FE1} \\
    \partial_\mu F^{\mu \nu} + \frac{i}{2} \left( \overline{\Phi} \, D^\nu \Phi - \Phi \, \overline{D^\nu \Phi} \right) &= 0. \label{eqI3:FE2}
\end{align}

If we now set the gauge conditions $A_r = A_0 = 0$ and assume the following ansatz:
\begin{equation}\label{eqI3:ansatzVortex}
    \Phi(r,\theta) = e^{i n \theta} f_n(r), \quad  A_\theta(r) = n \beta_n(r),
\end{equation}
then, the field equations \eqref{eqI3:FE1}--\eqref{eqI3:FE2} reduce to the following system of ordinary differential equations for the profile functions:
\begin{align}
    \frac{d^2 f_n}{d r^2} + \frac{1}{r} \frac{d f_n}{d r} - \frac{n^2}{r^2}(1 - \beta_n)^2 f_n + \frac{\lambda}{2}(1 - f_n^2) f_n &= 0, \label{eqI3:ProfEq1} \\
    \frac{d^2 \beta_n}{d r^2} - \frac{1}{r} \frac{d \beta_n}{d r} + f_n^2 (1 - \beta_n) &= 0. \label{eqI3:ProfEq2}
\end{align}

Regularity at the origin and finiteness of the energy impose the following boundary conditions:
\begin{equation}\label{eqI3:AssymptoticVortex}
    f_n(0) = 0, \quad f_n(\infty) = 1, \quad \beta_n(0) = 0, \quad \beta_n(\infty) = 1.
\end{equation}

It can be shown from equations \eqref{eqI3:ProfEq1}--\eqref{eqI3:ProfEq2} that near the vortex center, the profile functions admit the power series expansions \cite{Manton2004, Jaffe1980}
\begin{equation}
    f_n(r) = r^n \sum_{i=0}^\infty d_i r^i, \quad \beta_n(r) = r^2 \sum_{i=0}^\infty c_i r^i,
\end{equation}
with $d_{2i+1} = c_{2i+1} = 0$ for $i \in \mathbb{N}$ and $c_2 = c_4 = \dots = c_{2n-2} = 0$. Therefore, these become:
\begin{equation}\label{eqI3:PowerExpansion}
    f_n(r) = r^n \sum_{i=0}^\infty d_{2i} \, r^{2i}, \quad 
    \beta_n(r) = r^2 \left( c_0 + \sum_{i=n}^\infty c_{2i} \, r^{2i} \right).
\end{equation}

On the other hand, the asymptotic behavior as $r \to \infty$ can be studied by introducing small fluctuations around the vacuum \cite{Plohr1981,Perivolaropoulos1993,Berger1989}:
\begin{equation}
    f_n(r) = 1 - g_{f}(r), \quad \beta_n(r) = 1 - g_{\beta}(r),
\end{equation}
and linearizing the resulting equations. This yields:
\begin{align}
    \frac{d^2 g_{f}}{d r^2} + \frac{1}{r} \frac{d g_{f}}{d r} - \lambda \, g_{f} &= 0, \\
    \frac{d^2 g_{\beta}}{d r^2} - \frac{1}{r} \frac{d g_{\beta}}{d r} - n \, g_{\beta} &= 0.
\end{align}

Thus, the asymptotic behavior of the profile functions is given by:
\begin{align}
    f_n(r) &\approx 1 - \frac{A_{f_n}}{2\pi} K_0(\sqrt{\lambda} r), \\
    \beta_n(r) &\approx 1 - \frac{A_{\beta_n}}{2\pi} r K_1(\sqrt{n} r),
\end{align}
where $A_{f_n}$ and $A_{\beta_n}$ are constants that must be determined numerically, and $K_0$, $K_1$ are modified Bessel functions of the second kind.

However, for large values of the coupling constant $\lambda$, the linearization above is no longer valid, and nonlinear terms that were previously neglected become relevant.

\begin{figure}[h!]
    \centering
    \begin{subfigure}{0.495\textwidth}
        \includegraphics[width=\linewidth]{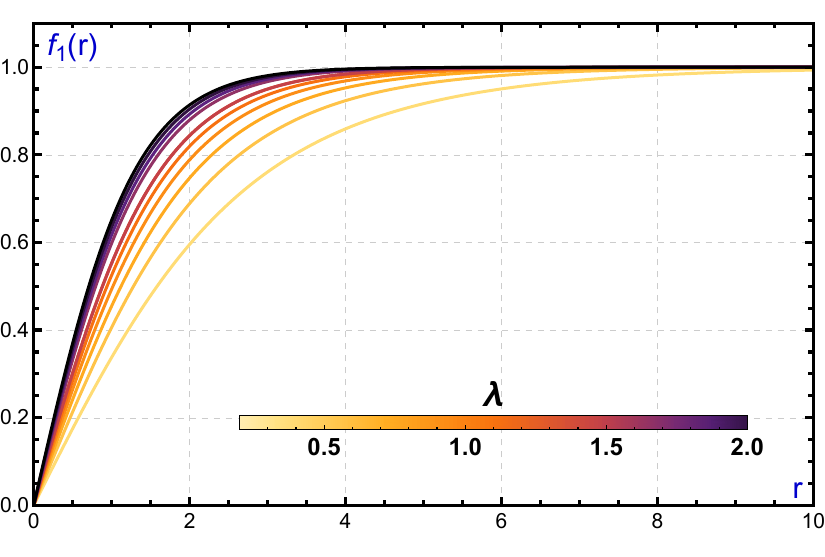}
    \end{subfigure}
    \begin{subfigure}{0.495\textwidth}
        \includegraphics[width=\linewidth]{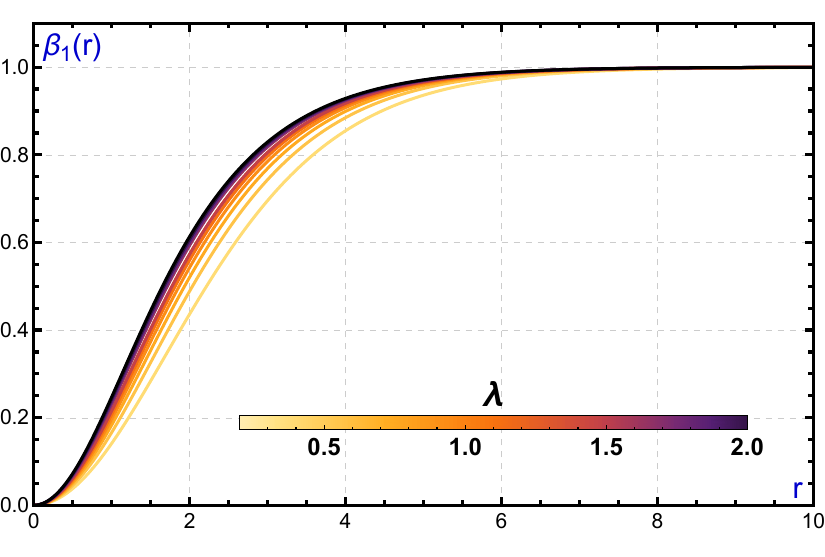}
    \end{subfigure}
    \begin{subfigure}{0.495\textwidth}
        \includegraphics[width=\linewidth]{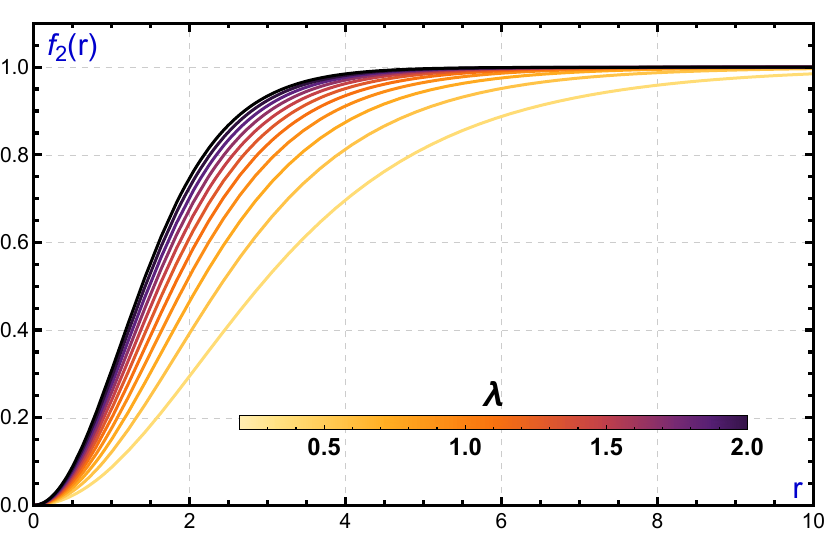}
    \end{subfigure}
    \begin{subfigure}{0.495\textwidth}
        \includegraphics[width=\linewidth]{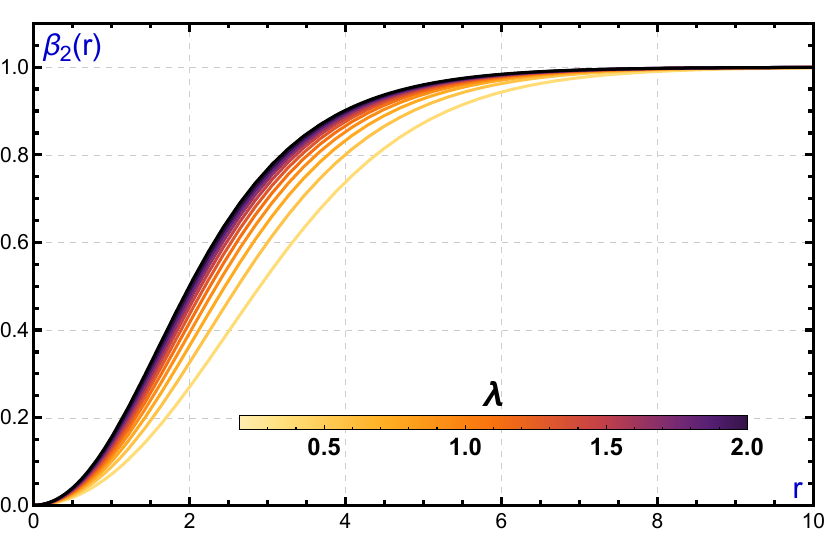}
    \end{subfigure}
    %\vspace{-0.3cm}
    \caption{\textit{Vortex profiles for different values of $\lambda$.}}
    \label{figI3:VortexProfiles}
\end{figure}

\begin{figure}[h!]
    \centering
    \begin{subfigure}{0.495\textwidth}
        \includegraphics[width=\linewidth]{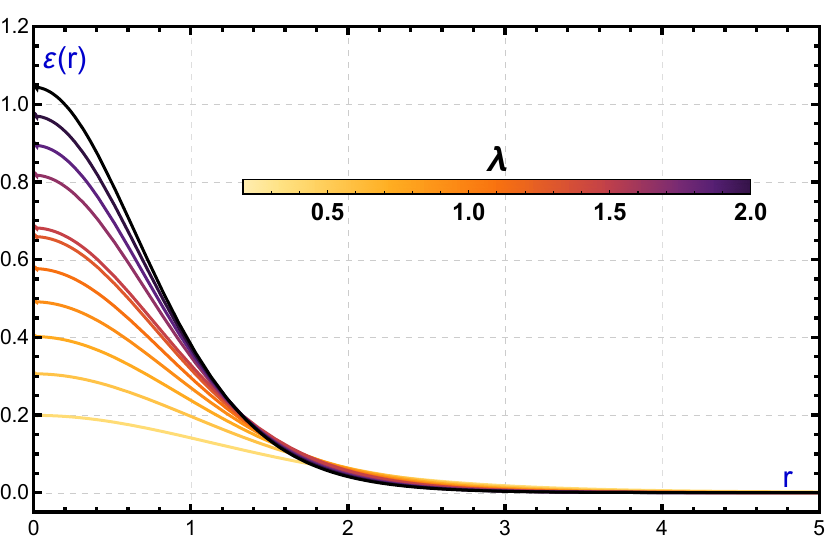}
    \end{subfigure}
    \begin{subfigure}{0.495\textwidth}
        \includegraphics[width=\linewidth]{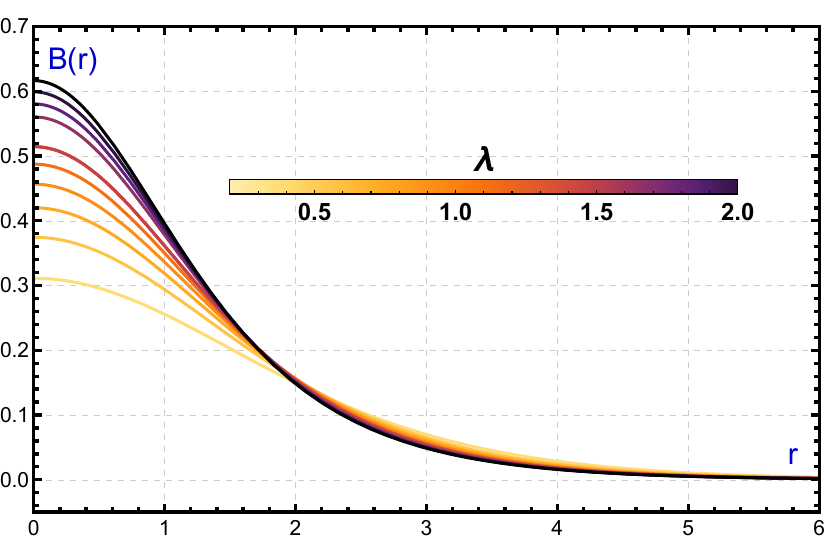}
    \end{subfigure}
    \begin{subfigure}{0.495\textwidth}
        \includegraphics[width=\linewidth]{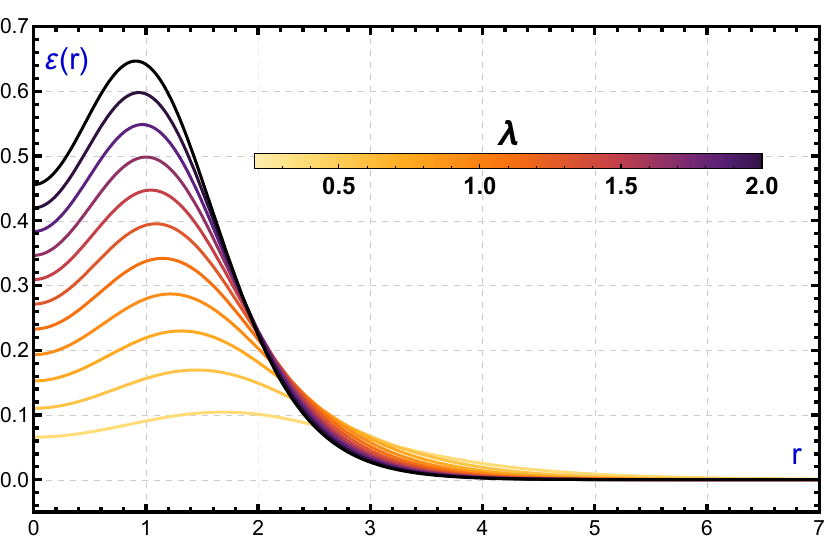}
    \end{subfigure}
    \begin{subfigure}{0.495\textwidth}
        \includegraphics[width=\linewidth]{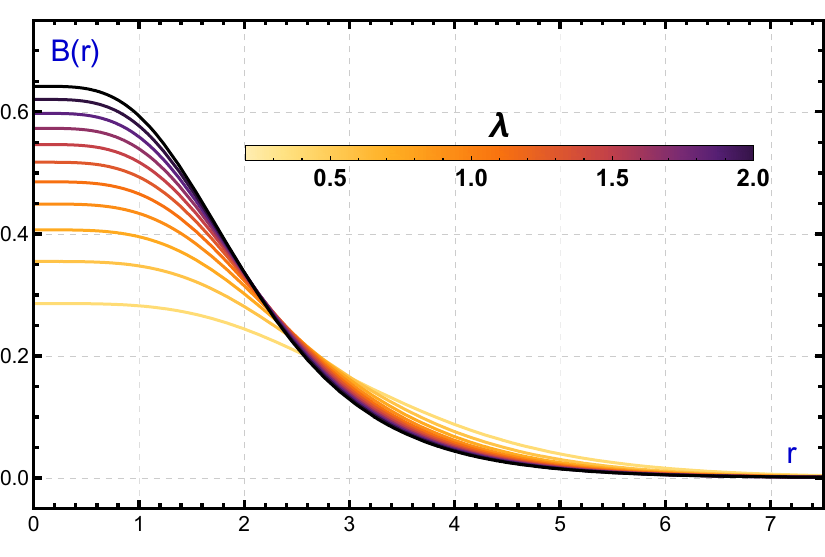}
    \end{subfigure}
   % \vspace{-0.3cm}
    \caption{\textit{Energy density and magnetic field profiles for vortices with $n=1$ (top) and $n=2$ (bottom) for several values of $\lambda$.}}
    \label{figI3:VortexEnergyDensAndB}
\end{figure}

Figure \ref{figI3:VortexProfiles} shows the vortex profiles with topological charges $n=1$ and $n=2$ for several values of the coupling constant $\lambda$. The corresponding energy density and magnetic field profiles are displayed in Figure \ref{figI3:VortexEnergyDensAndB}. Near the vortex center, the scalar field profile for $n=1$ behaves linearly in $r$, while for $n=2$ the behavior is quadratic, i.e., $f_2(r) \sim d_0 r^2$, as expected from the power expansion \eqref{eqI3:PowerExpansion}.

As can be seen in Figure \ref{figI3:VortexEnergyDensAndB}, the energy density for $n=2$ vortices is nearly constant near the origin. This is due to the quadratic behavior of the scalar field profile $f_2(r)$ around $r=0$.

\begin{figure}[h!]
    \centering
    \includegraphics[width=0.65\linewidth]{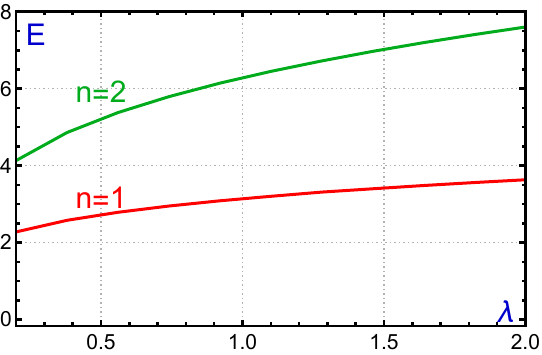}
    \caption{\textit{Energy of vortex solutions as a function of $\lambda$ for $n=1$ and $n=2$.}}
    \label{figI3:energyvslambda}
\end{figure}

To conclude this section, the dependence of the vortex energy on the coupling constant $\lambda$ is illustrated in Figure \ref{figI3:energyvslambda}. As $\lambda$ increases, the energy of the vortex configuration also increases. In fact, the energy diverges in the limit $\lambda \to \infty$. This is because, in that limit, the scalar field decouples from the electromagnetic field, and Abelian-Higgs vortices become effectively global vortices. This transition can be intuitively observed by comparing Figures \ref{figI3:GlobalVortexProfiles} and \ref{figI3:VortexProfiles}: for both $n=1$ and $n=2$, the scalar field profiles increasingly resemble those of global vortices as $\lambda$ grows.

Indeed, in Chapter~\ref{Chap4}, we will show how bound modes in global vortices can be obtained as the limiting case of the perturbation spectrum of Abelian-Higgs vortices.

\vspace{-0.5cm}
\section{The BPS regime}\label{SecBPSlIMIT}

In Chapter~\ref{Intro1} we studied how kink solutions can be obtained from a set of first-order ordinary differential equations instead of solving the full field equations. The natural question now is: can we do something similar for vortex solutions?

The answer is yes, but only for a specific value of the coupling constant $\lambda$. This restriction will have important implications when discussing  interactions between vortices.

Let us consider the static energy for $\lambda = 1$, that is,
\begin{equation}
    E = \frac{1}{2} \int \left(B^2 + \overline{D_i \Phi}\, D_i \Phi + \frac{1}{4}(1 - \overline{\Phi}\, \Phi)^2 \right)\, d^2x.
\end{equation}

This expression can be rewritten as
\begin{equation}\label{eqI3:BPS11}
\begin{aligned}
    E = \frac{1}{2} \int \Big[ & \left(B - \frac{1}{2}(1 - \overline{\Phi}\, \Phi)\right)^2 
    + (\overline{D_1 \Phi} - i \overline{D_2 \Phi})(D_1 \Phi + i D_2 \Phi) \\
    & + B - i \left(\partial_1(\overline{\Phi} D_2 \Phi) - \partial_2(\overline{\Phi} D_1 \Phi)\right) \Big]\, d^2x.
\end{aligned}
\end{equation}

Applying Stokes' theorem to the last two terms in~\eqref{eqI3:BPS11}, we get
\begin{equation}
    \frac{1}{2} \int \left[B - i\left(\partial_1(\overline{\Phi} D_2 \Phi) - \partial_2(\overline{\Phi} D_1 \Phi)\right)\right] d^2x = \pi n,
\end{equation}
since $D_i \Phi$ vanishes in the limit $r \to \infty$.

Thus, the energy becomes
\begin{equation}
    E = \frac{1}{2} \int \left[\left(B - \frac{1}{2}(1 - \overline{\Phi}\, \Phi)\right)^2 + (\overline{D_1 \Phi} - i \overline{D_2 \Phi})(D_1 \Phi + i D_2 \Phi) \right] d^2x + \pi n.
\end{equation}

The energy bound $E = \pi n$ is saturated when the field configuration satisfies the following first-order equations:
\begin{align}
    D_1 \Phi + i D_2 \Phi &= 0, \label{eqI3:BPS1}\\
    B - \frac{1}{2}(1 - \overline{\Phi}\, \Phi) &= 0. \label{eqI3:BPS2}
\end{align}
Equations \eqref{eqI3:BPS1}-\eqref{eqI3:BPS2}  are known as the \textit{BPS equations} for the Abelian-Higgs model.

For antivortices (i.e., $n < 0$), the corresponding BPS equations take the form:
\begin{align}
    D_1 \Phi - i D_2 \Phi &= 0, \label{eqI3:BPS3}\\
    B + \frac{1}{2}(1 - \overline{\Phi}\, \Phi) &= 0, \label{eqI3:BPS4}
\end{align}
which follows from the fact that the sign of $n$ is reversed under the transformation $(x_1,x_2)\to(x_1,-x_2)$.

Assuming the ansatz~\eqref{eqI3:ansatzVortex} for a static $n$-vortex solution, the BPS equations~\eqref{eqI3:BPS1}–\eqref{eqI3:BPS2} reduce to the radial form:
\begin{align}
    \frac{d f_n}{dr} &= \frac{n}{r} f_n (1 - \beta_n), \\
    \frac{d \beta_n}{dr} &= \frac{r}{2n}(1 - f_n^2).
\end{align}

An important consequence of the BPS structure for this limit is that the energy of a vortex is simply $E = \pi |n|$. Moreover, in this limit, vortices neither attract nor repel each other. In fact, the equations~\eqref{eqI3:BPS1}–\eqref{eqI3:BPS2} allow for the existence of multi-vortex solutions \cite{Jaffe1980,Taubes1980,Taubes1980b}.

Let us start  by writing  the Higgs field in polar form:
\begin{equation}
    \Phi = |\Phi| e^{i\chi}.
\end{equation}

Defining $h = \log |\Phi|^2$, equation~\eqref{eqI3:BPS1} yields:
\begin{equation}
    A_1 = \frac{1}{2} \partial_2 h + \partial_1 \chi, \quad A_2 = -\frac{1}{2} \partial_1 h + \partial_2 \chi,
\end{equation}
from which we infer that the magnetic field can be written as
\begin{equation}
    B = -\frac{1}{2} \nabla^2 h.
\end{equation}

Plugging all this into~\eqref{eqI3:BPS2}, the following partial differential equation is obtained for $h$:
\begin{equation}\label{eqI3:SimpBPS}
    \nabla^2 h + 1 - e^{h} = 0.
\end{equation}

Equation~\eqref{eqI3:SimpBPS} is valid everywhere except at the vortex centers, i.e. the places where $\Phi$ is zero, where we would have a logarithmic divergence.  

In order to fix this issue and take into account these singularities, we add delta functions to~\eqref{eqI3:SimpBPS} that act as vortex centers, in such a way that the final equation becomes
\begin{equation}\label{eqI3:EqH}
     \nabla^2 h + 1 - e^{h} = 4\pi \sum_{i=1}^N N_i\, \delta^2(x - X_i),
\end{equation}
where $X_i$ and $N_i$ are the vortex positions and their winding numbers, respectively.

To check the validity of this ansatz, let us consider the behavior of a vortex with winding number $n$ near its center. As previously discussed, near the vortex center:
\begin{equation}
    \Phi \sim |x - X|^n \quad \Rightarrow \quad h \sim 2n \log|x - X|.
\end{equation}

On the other hand, near the vortex center,
\begin{equation}
    \nabla^2 h = 4\pi n\, \delta^2(x - X) \quad \Rightarrow \quad h \sim 2n \log|x - X|.
\end{equation}

It is possible to take advantage of~\eqref{eqI3:EqH} to numerically construct multi-vortex configurations by solving it, assuming that it is possible to take a small circle of radius $\epsilon$ around each vortex center in which the solution is approximately $h \approx 2n \log \epsilon$. As $|\Phi| \to 1$ at infinity, then in this limit $h \to 0$.

As previously mentioned, this method is only valid in the BPS limit. In general, the vortex profiles can be obtained numerically by solving the gradient flow equations \cite{AlonsoIzquierdo2024c}:
\begin{align}
   \frac{\partial f_n}{\partial t} &= \frac{\partial^2 f_n}{\partial r^2} + \frac{1}{r} \frac{\partial f_n}{\partial r} - \frac{n}{r^2}(1 - \beta_n)^2 f_n + \frac{\lambda}{2}(1 - f_n^2) f_n, \label{eqI3:ProfEqGF1}\\
   \frac{\partial \beta_n}{\partial t} &= \frac{\partial^2 \beta_n}{\partial r^2} - \frac{1}{r} \frac{\partial \beta_n}{\partial r} + n(1 - \beta_n) f_n^2. \label{eqI3:ProfEqGF12}
\end{align}

Then, if the vortices are well separated, the following ansatz, known as the Abrikosov's product ansatz \cite{Manton2004}, can be used to build the field configuration:
\begin{equation}\label{eqI3:AbrikosovAnsatz}
    \Phi = \prod_{i=1}^N \Phi^{(i)}(x - X_i), \quad A_j = \sum_{i=1}^N A_j^{(i)}(x - X_i),
\end{equation}
where $X_i$ denotes the positions of the vortices. 

If the vortices are located at short distances, another possibility is to use the splitting modes of the vortices.  We will discuss this in more detail in Chapter~\ref{Chap3}.

Another possibility is to construct an ansatz for the field configuration that depends on a real parameter, and minimize the static energy~\eqref{eqI3:StaticEnergy} with respect to it. For example, as proposed in \cite{AlonsoIzquierdo2024f}, for two vortices separated by a distance $2d$, the following ansatz can be used:
\begin{align*}
    \Phi(x_1, x_2, d, \alpha) &= \widehat{\Phi}(x_1, x_2, d, \alpha) \, \widehat{\Phi}(x_1, x_2, -d, \alpha), \\
    A_1(x_1, x_2, d, \alpha) &= \widehat{A}_1(x_1, x_2, d, \alpha) + \widehat{A}_1(x_1, x_2, -d, \alpha), \\
    A_2(x_1, x_2, d, \alpha) &= \widehat{A}_2(x_1, x_2, d, \alpha) + \widehat{A}_2(x_1, x_2, -d, \alpha),
\end{align*}
where
\begin{align}
    \widehat{\Phi}(x_1, x_2, d, \alpha) &= \left( \alpha f_1(\overline{r}) + (1 - \alpha) \sqrt{f_2(\overline{r})} \right) e^{i\overline{\theta}}, \label{eqI3:AnsatzMultivortex1} \\
    \widehat{A}_1(x_1, x_2, d, \alpha) &= -\left( \alpha \beta_1(\overline{r}) + (1 - \alpha) \beta_2(\overline{r}) \right) \frac{\sin \overline{\theta}}{\overline{r}}, \\
    \widehat{A}_2(x_1, x_2, d, \alpha) &= \left( \alpha \beta_1(\overline{r}) + (1 - \alpha) \beta_2(\overline{r}) \right) \frac{\cos \overline{\theta}}{\overline{r}}, \label{eqI3:AnsatzMultivortex3}
\end{align}
and where $\overline{r} = \sqrt{(x - d)^2 + y^2}$ and $\overline{\theta} = \arctan\left( \frac{y}{x - d} \right)$.

As can be inferred from the ansatz~\eqref{eqI3:AnsatzMultivortex1}--\eqref{eqI3:AnsatzMultivortex3}, for $d = 0$, we must have $\alpha = 0$, since for this specific value the field configuration corresponds to a two-vortex. On the other hand, for $d \gg 1$, $\alpha$ must tend to one, since in that case we recover the Abrikosov ansatz~\eqref{eqI3:AbrikosovAnsatz}, which is valid for widely separated vortices. This ansatz is valid both in the BPS limit and beyond it.

\section{Interactions between vortices}\label{SecIntBetVort}

Now that we have studied the BPS limit in detail, let us focus on the study of vortex interactions beyond the BPS regime. In the self-dual limit, there is no interaction between static vortices (or antivortices); however, once we move away from this limit, forces between vortices begin to arise.  However the presence of vortices and antivortices at the same time gives place  to the presence of forces. 

In Table \ref{tabI3:tableinteractions}, a summary of the interactions between vortices is shown:

\begin{table}[h!]
\centering
\begin{tabular}{@{}|>{\bfseries}c||c|c|c|@{}}
\hline
\textbf{} & \textbf{Vortex|Vortex} & \textbf{Vortex|Antivortex} & \textbf{Antivortex|Antivortex} \\
\hline\hline
\textbf{$\mathbf{\lambda<1}$} & Attract & Attract & Attract \\
\hline
\textbf{$\mathbf{\lambda=1}$} & No interaction & Attract & No interaction \\
\hline
\textbf{$\mathbf{\lambda>1}$} & Repel & Attract & Repel \\
\hline
\end{tabular}
\vspace{0.2cm}
\caption{\textit{Behaviour of the interactions between vortices/antivortices in the Abelian-Higgs model.}}
\label{tabI3:tableinteractions}
\end{table}

As shown in Table~\ref{tabI3:tableinteractions}, vortex–antivortex pairs always attract. In contrast, the behavior of vortex–vortex interactions depends strongly on the value of $\lambda$.
\vspace{0.15cm}

The long-range force between these objects has been computed by several authors \cite{Bettencourt1995,Speight1997,Tong2002}. For two vortices with topological charge $n=1$, the interaction energy is given by
\begin{equation}
    E_{int}(d) = -\frac{A_{f_n}^2}{2\pi}K_0(\sqrt{\lambda}\,d) + \frac{A_{\beta_n}^2}{2\pi}K_0(d),
\end{equation}
\vspace{0.15cm}
which implies that the force is
\vspace{0.15cm}
\begin{equation}\label{eqI3:ForceLongRange}
    F_{int}(d) = -\frac{A_{f_n}^2\sqrt{\lambda}}{2\pi}K_1(\sqrt{\lambda}\,d) + \frac{A_{\beta_n}^2}{2\pi}K_1(d).
\end{equation}

\vspace{0.15cm}
From \eqref{eqI3:ForceLongRange}, we conclude that the scalar field contributes with an attractive force, while the magnetic field contributes with a repulsive one. For $\lambda < 1$, the scalar contribution dominates, and the total force is attractive. On the other hand, in the regime $\lambda > 1$, the magnetic contribution dominates, resulting in repulsion. At the BPS limit ($\lambda = 1$), we have $A_{f_n} = A_{\beta_n}$, and the total force vanishes. This type of analysis has also been applied to study interactions between vortices and magnetic impurities in modified versions of the Abelian-Higgs model \cite{Ashcroft2020}.
\vspace{0.15cm}

Another aspect that will be discussed in more detail in Chapter~\ref{Chap3} is the stability of $n$-vortex configurations with $n>1$. In the regime $\lambda > 1$, these configurations are unstable. For example, a vortex with $n=2$ eventually splits into two $n=1$ vortices. This instability is checked by means of the presence of negative modes in the small fluctuation operator. In contrast, for $\lambda < 1$, these modes become stable. A rigorous study of the stability of these solutions was made by \cite{Gustafson2000}.
\vspace{0.15cm}

\begin{figure}[h!]
    \centering
    \begin{subfigure}{0.49\textwidth}
        \centering
        \includegraphics[width=\linewidth]{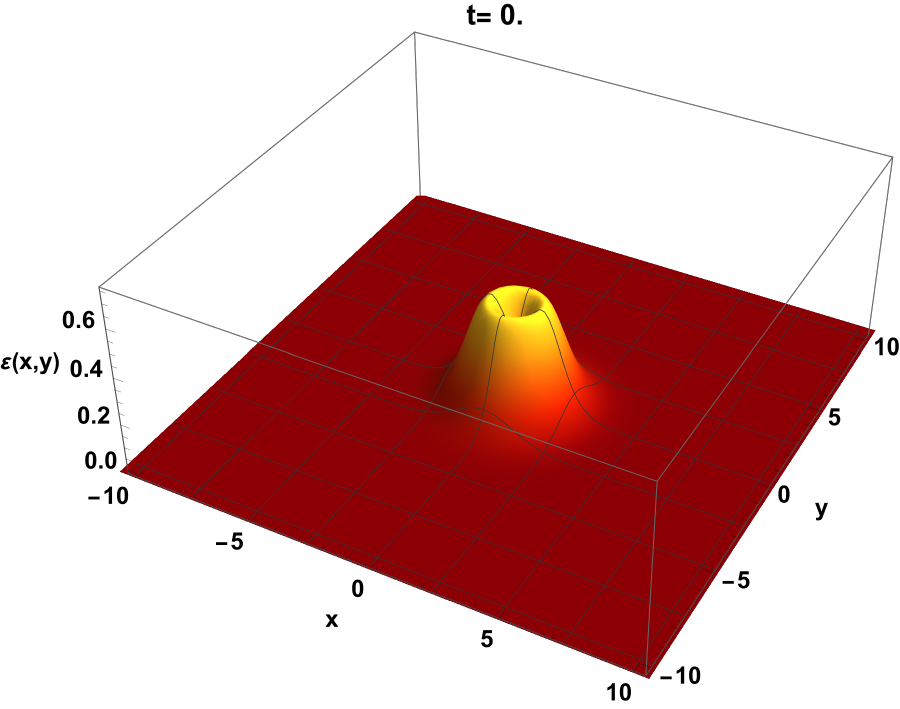}
    \end{subfigure}
    \begin{subfigure}{0.49\textwidth}
        \centering
        \includegraphics[width=\linewidth]{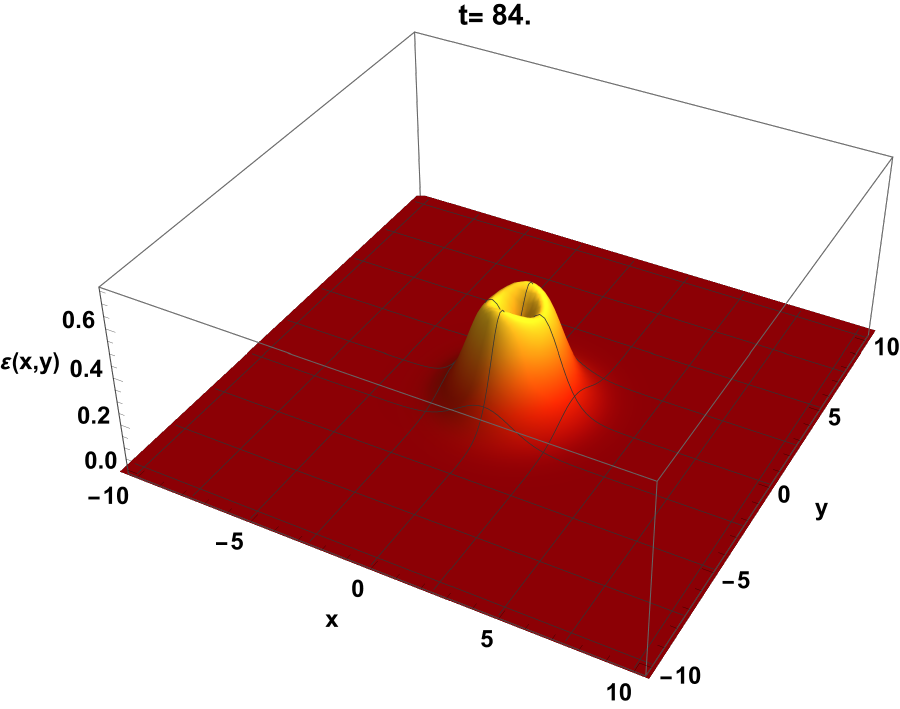}
    \end{subfigure}
    \begin{subfigure}{0.49\textwidth}
        \centering
        \includegraphics[width=\linewidth]{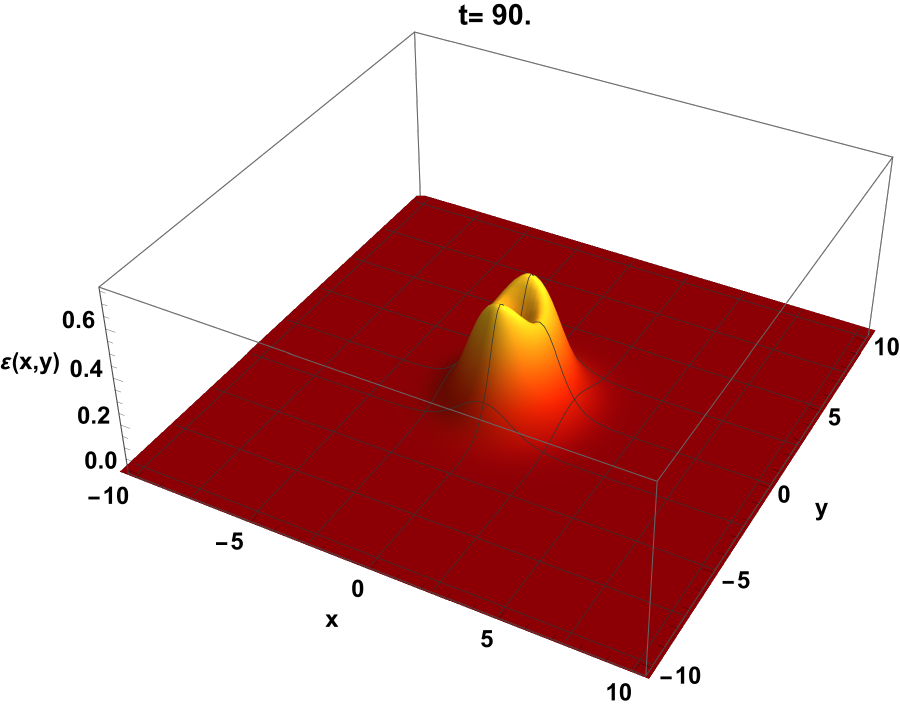}
    \end{subfigure}
    \begin{subfigure}{0.49\textwidth}
        \centering
        \includegraphics[width=\linewidth]{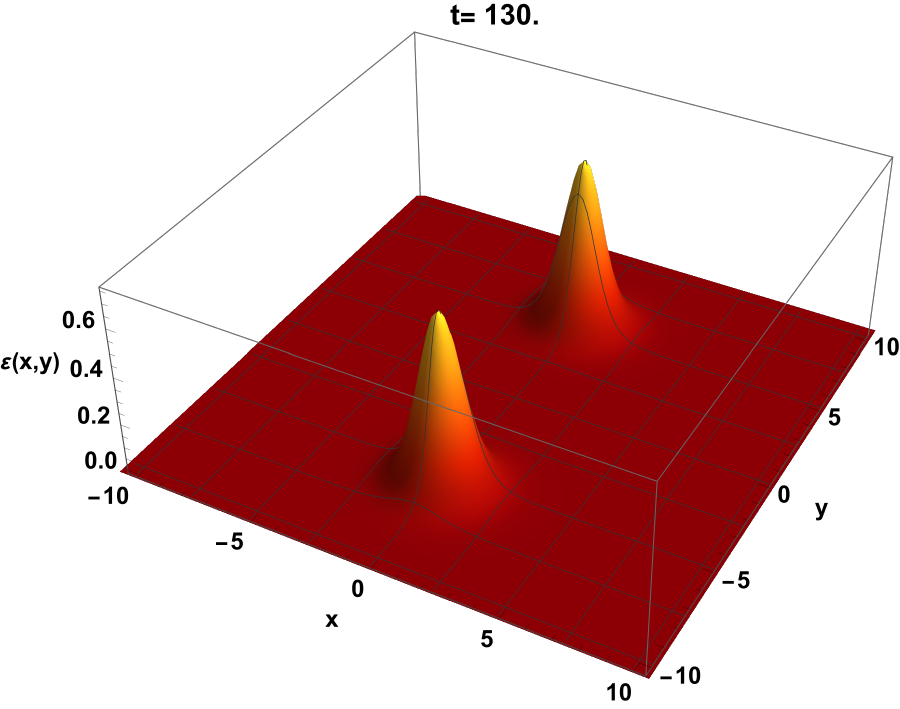}
    \end{subfigure}
    \vspace{0.35cm}
    \caption{\textit{Snapshots of the energy density during the splitting of a two-vortex configuration for $\lambda=1.2$.}}
    \label{figI3:SplittingVortex}
\end{figure}

In Figure~\ref{figI3:SplittingVortex}, the splitting of a two-vortex configuration into two $n=1$ vortices can be seen. Initially, the configuration appears stable, but over time, one of the unstable modes activates because of numerical oscillations and causes the two-vortex to split, with the two resulting one-vortices repelling each other.

\vspace{0.15cm}

\begin{figure}[htb]
    \centering
    % Row 1
    \begin{subfigure}[b]{0.325\textwidth}
        \includegraphics[width=\textwidth]{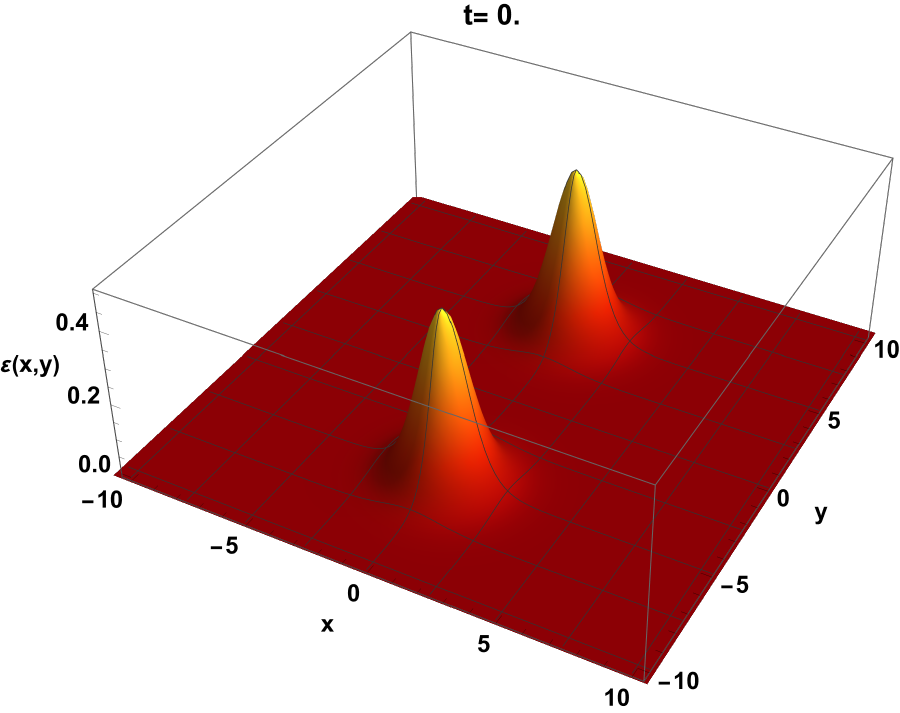}
    \end{subfigure}
    \hfill
    \begin{subfigure}[b]{0.325\textwidth}
        \includegraphics[width=\textwidth]{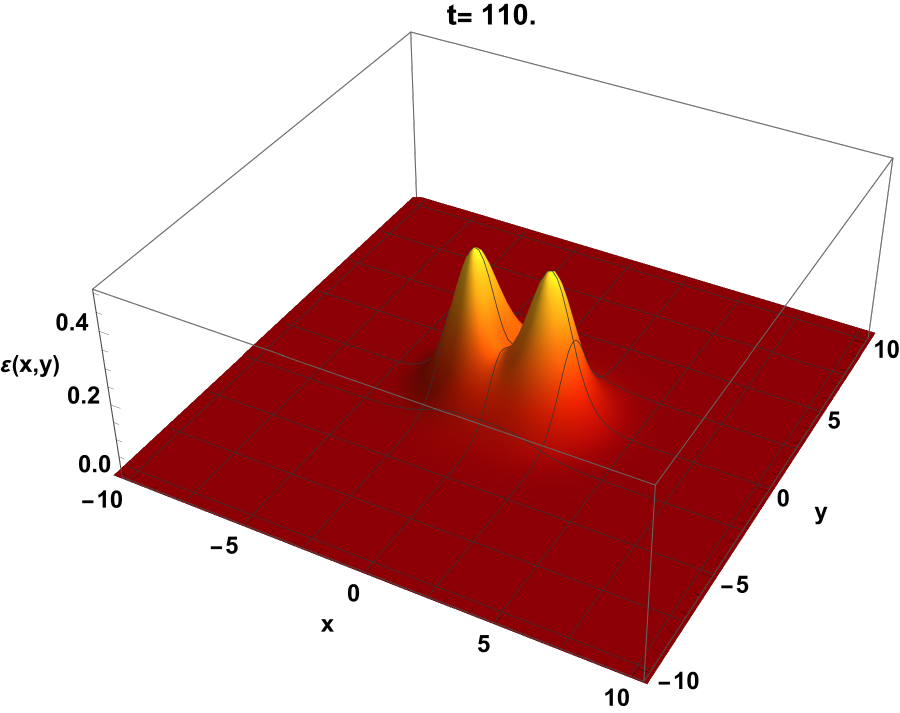}
    \end{subfigure}
    \hfill
    \begin{subfigure}[b]{0.325\textwidth}
        \includegraphics[width=\textwidth]{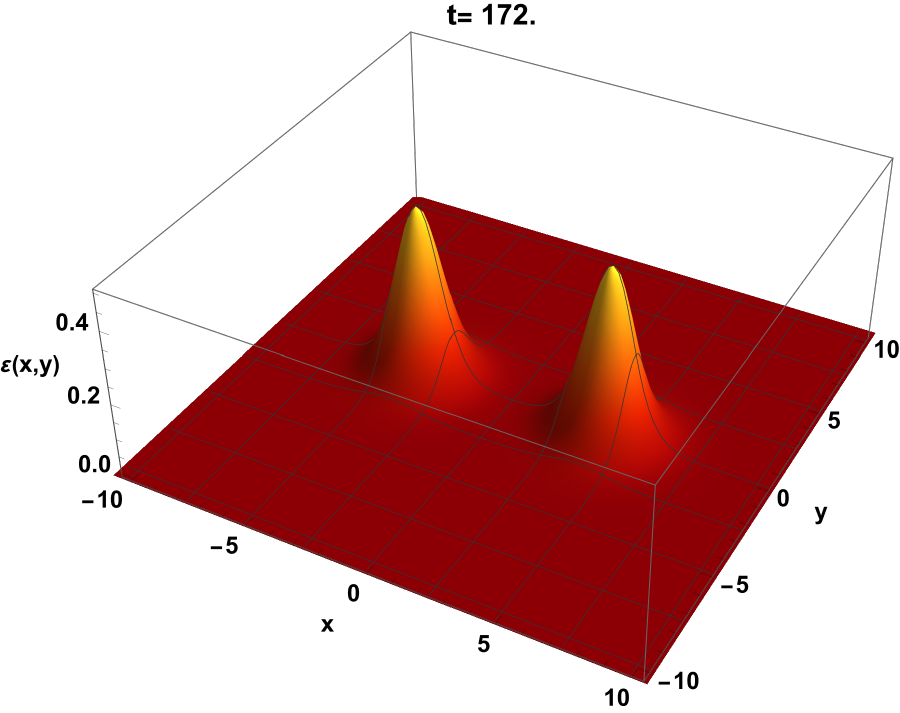}
    \end{subfigure}

    \vspace{0.5cm}

    % Row 2
    \begin{subfigure}[b]{0.325\textwidth}
        \includegraphics[width=\textwidth]{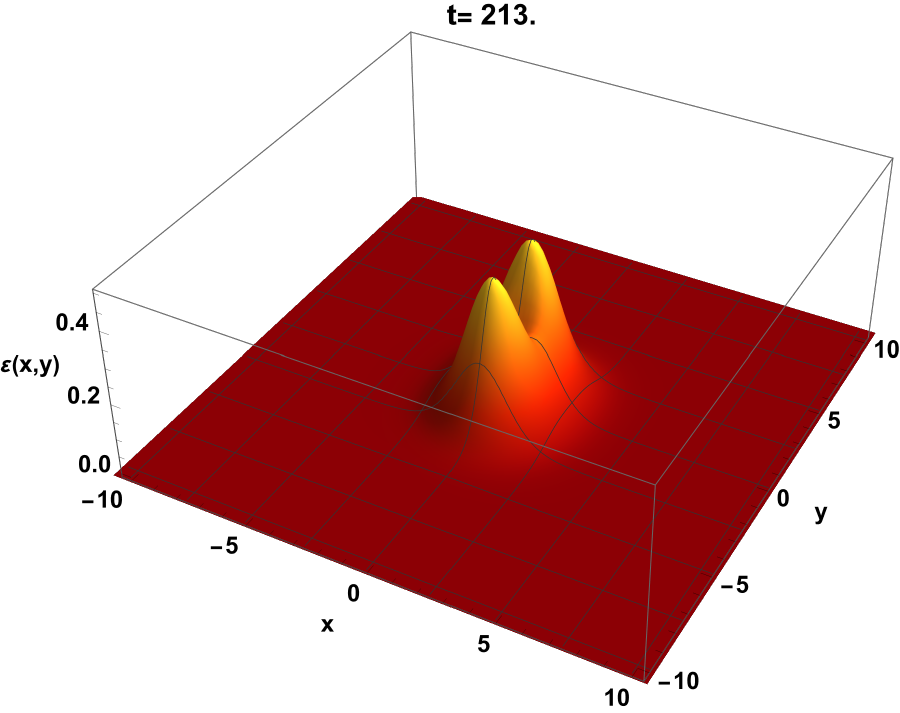}
    \end{subfigure}
    \hfill
    \begin{subfigure}[b]{0.325\textwidth}
        \includegraphics[width=\textwidth]{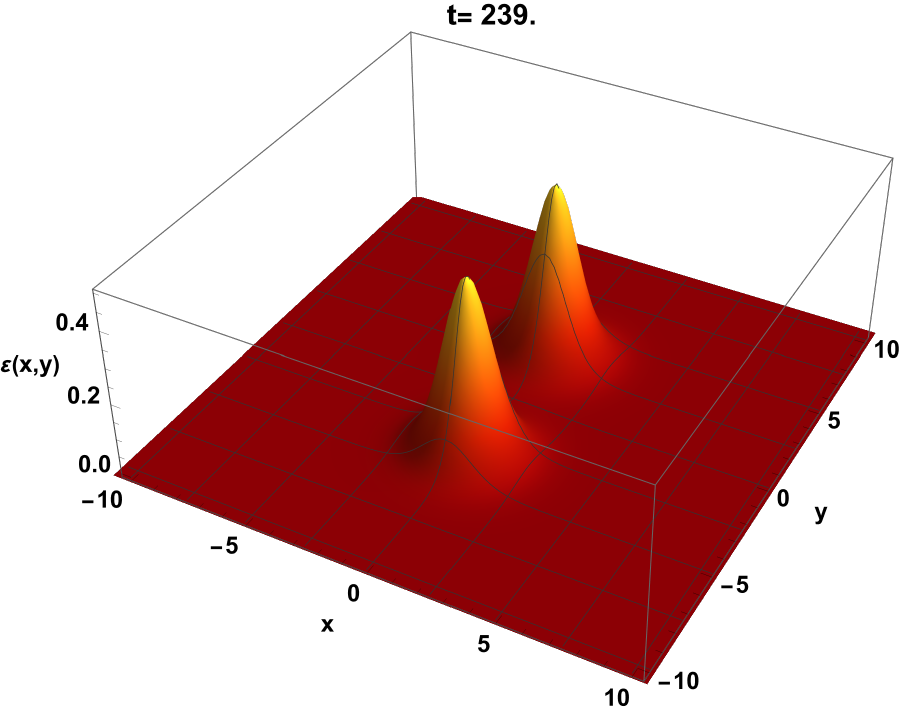}
    \end{subfigure}
    \hfill
    \begin{subfigure}[b]{0.325\textwidth}
        \includegraphics[width=\textwidth]{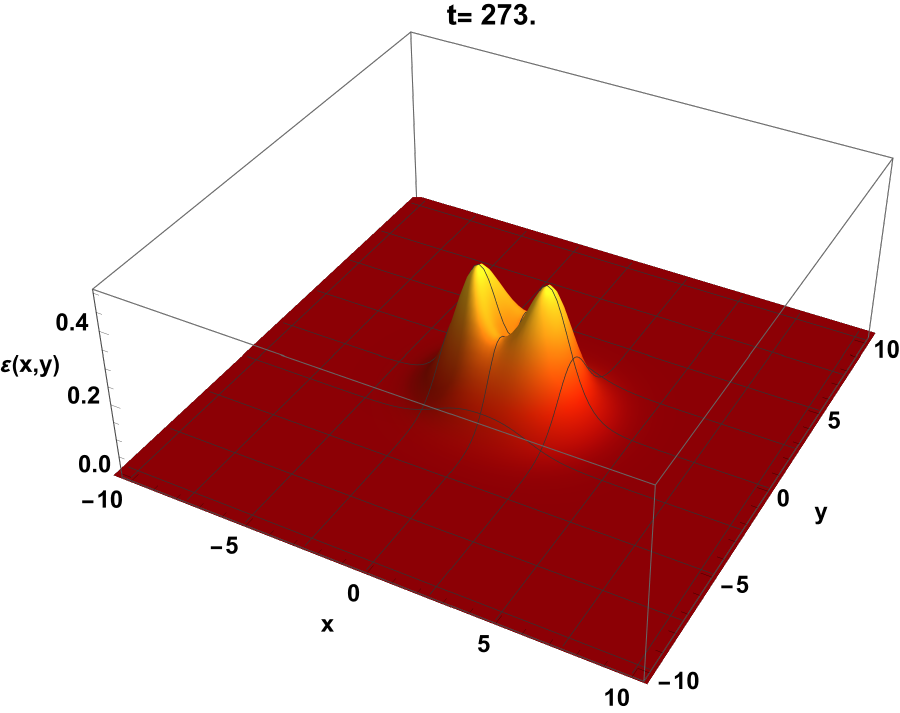}
    \end{subfigure}
\vspace{0.35cm}
    \caption{\textit{Snapshots of the energy density of two initially separated $n=1$ vortices for $\lambda = 0.7$.}}
    \label{FigI3:VortexAttraction}
\end{figure}

In Figure~\ref{FigI3:VortexAttraction}, we show snapshots of a simulation with $\lambda = 0.7$ where two initially static vortices are placed at a certain distance. As vortices attract in this regime, they accelerate and scatter off each other. As mentioned in Chapter~\ref{Intro0}, this head-on collision results in a $90^\circ$ scattering process. After scattering, the vortices move apart, but the mutual attraction pulls them together again, resulting in another collision. This process repeats, gradually radiating energy via internal modes, and eventually forming a stable two-vortex  at the center of the simulation box. The mechanism of radiation through internal modes will be studied in Chapter~\ref{Chap4}.

\vspace{0.15cm}

\begin{figure}[h!]
    \centering
    % Row 1
    \begin{subfigure}[b]{0.325\textwidth}
        \includegraphics[width=\textwidth]{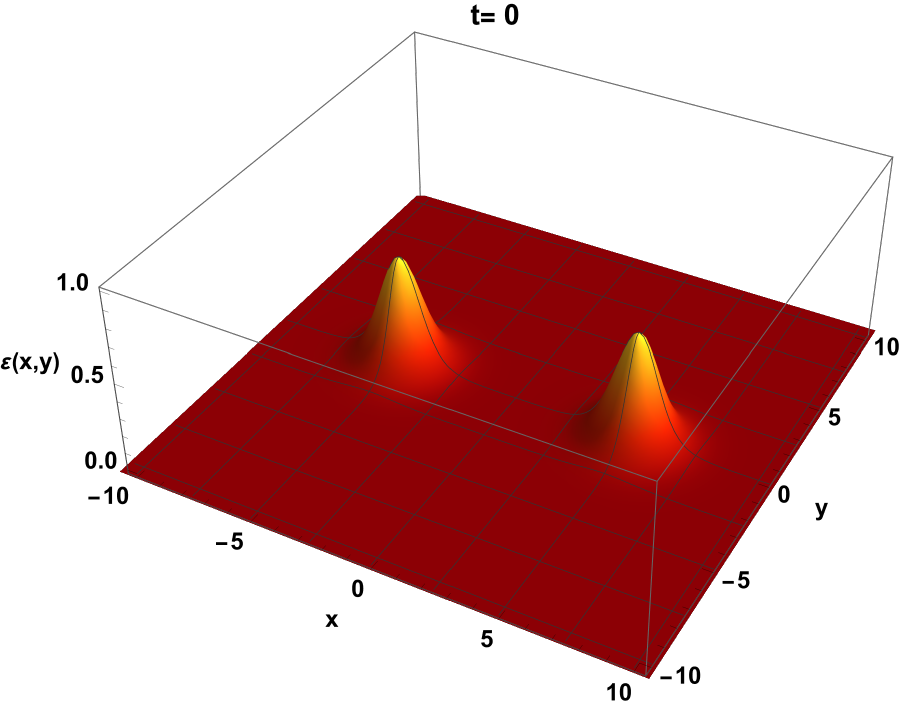}
    \end{subfigure}
    \hfill
    \begin{subfigure}[b]{0.325\textwidth}
        \includegraphics[width=\textwidth]{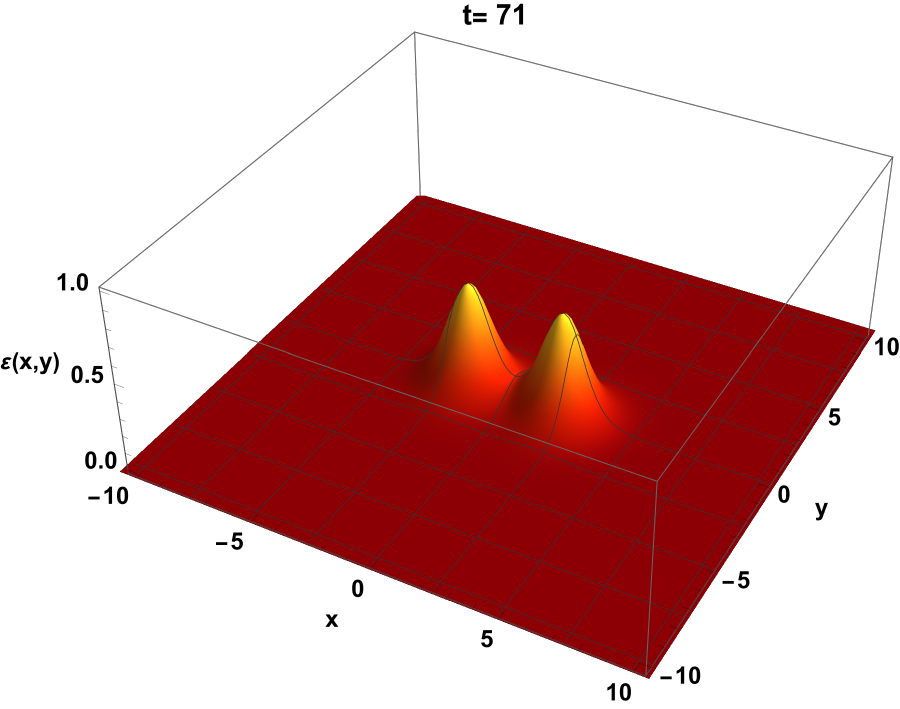}
    \end{subfigure}
    \hfill
    \begin{subfigure}[b]{0.325\textwidth}
        \includegraphics[width=\textwidth]{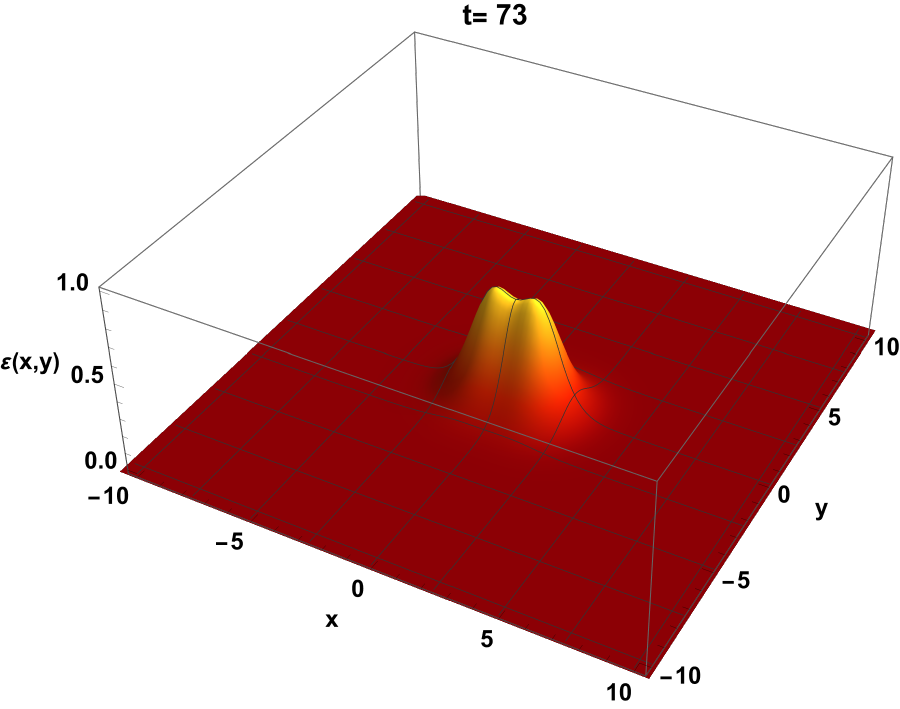}
    \end{subfigure}

    \vspace{0.5cm}

    % Row 2
    \begin{subfigure}[b]{0.325\textwidth}
        \includegraphics[width=\textwidth]{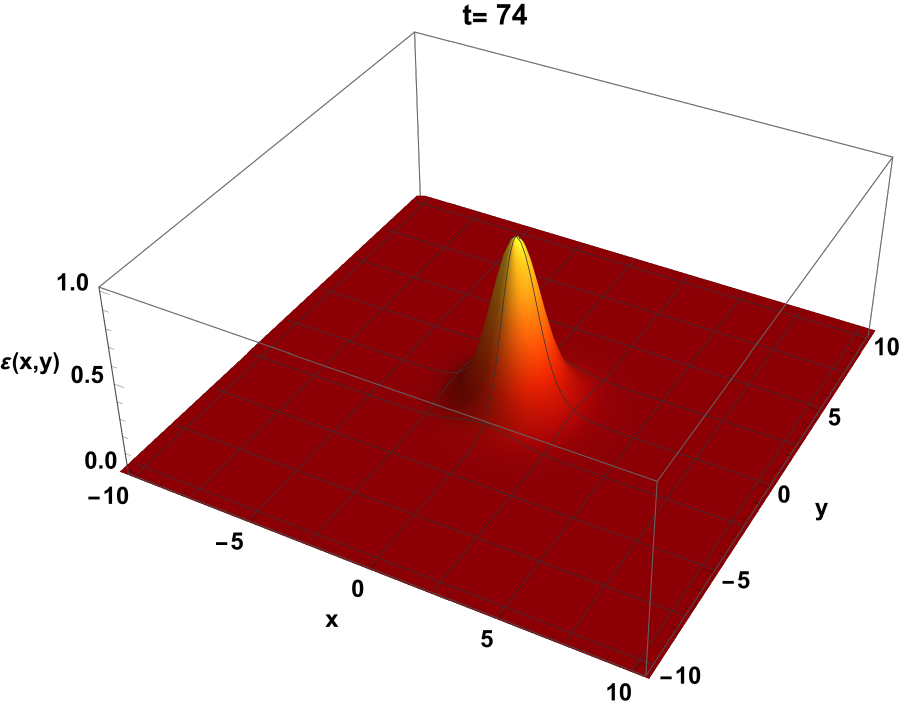}
    \end{subfigure}
    \hfill
    \begin{subfigure}[b]{0.325\textwidth}
        \includegraphics[width=\textwidth]{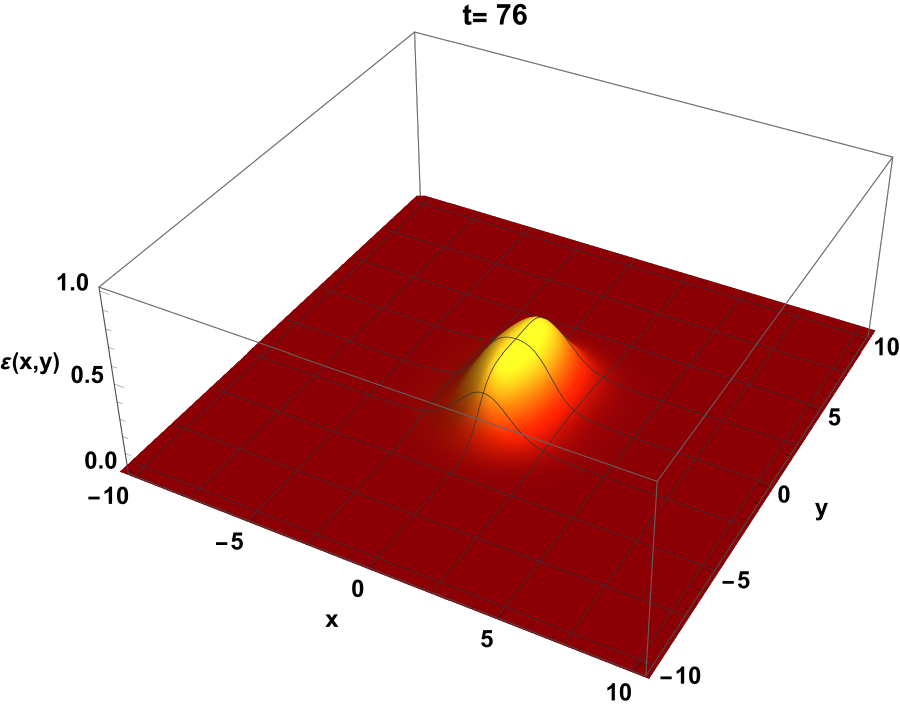}
    \end{subfigure}
    \hfill
    \begin{subfigure}[b]{0.325\textwidth}
        \includegraphics[width=\textwidth]{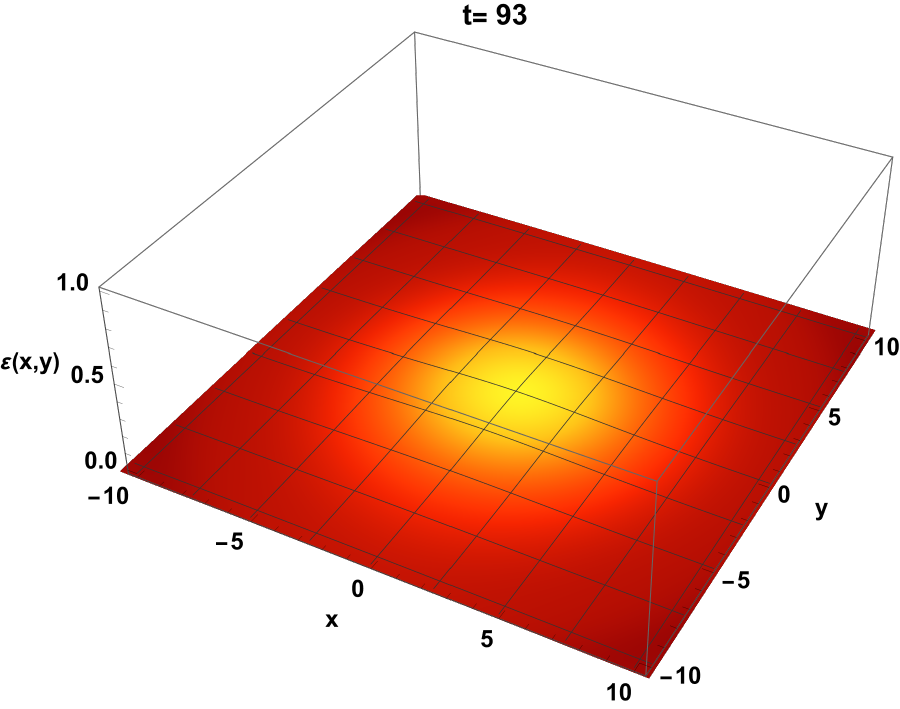}
    \end{subfigure}

    \caption{\textit{Snapshots of the energy density during vortex–antivortex annihilation for $\lambda=1$.}}
    \label{FigI3:VortexKaboomLambda1}
\end{figure}

Regarding vortex–antivortex scattering: when a vortex and an antivortex are initially at rest at some distance, they eventually attract and annihilate, releasing their energy as radiation. This behavior occurs in both the BPS case ($\lambda=1$) and for $\lambda \ne 1$. Figure~\ref{FigI3:VortexKaboomLambda1} illustrates this process in the BPS limit.
\vspace{0.15cm}

\begin{figure}[h!]
    \centering
    % Row 1
    \begin{subfigure}[b]{0.245\textwidth}
        \includegraphics[width=\textwidth]{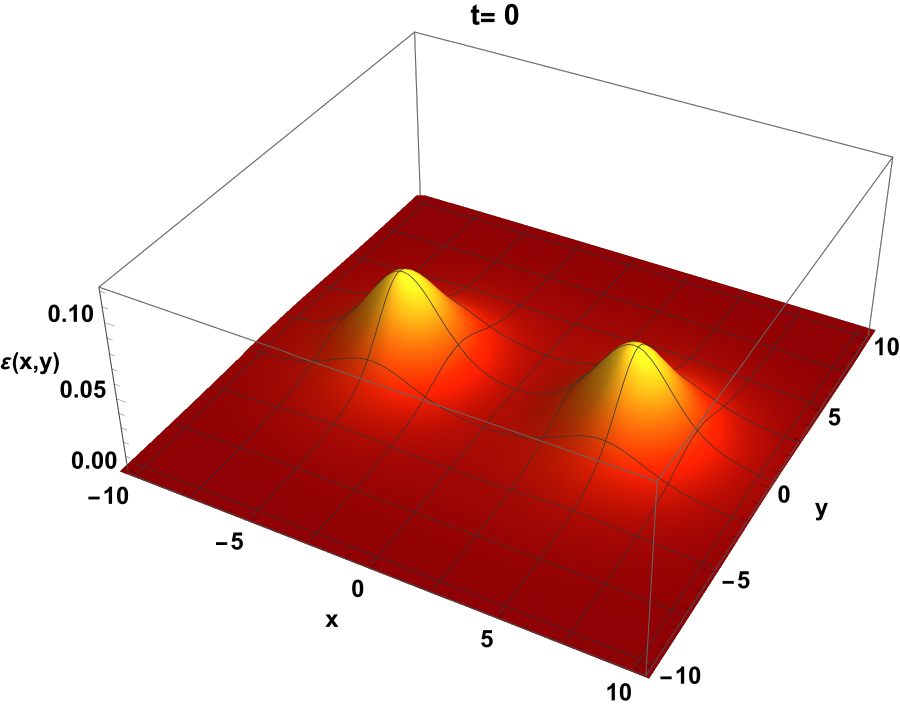}
    \end{subfigure}
    \hfill
    \begin{subfigure}[b]{0.245\textwidth}
        \includegraphics[width=\textwidth]{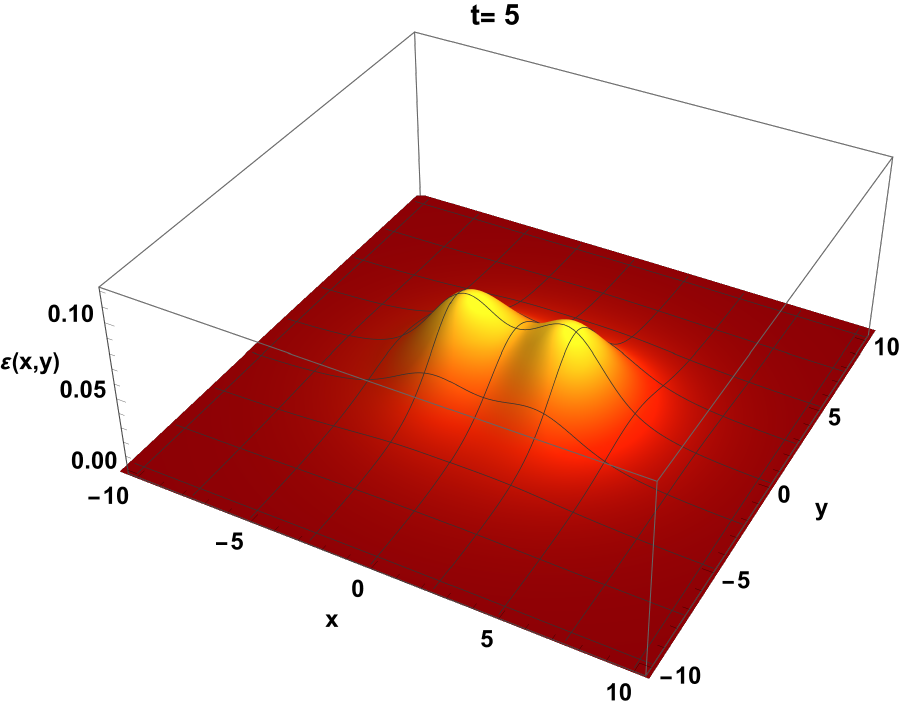}
    \end{subfigure}
    \hfill
    \begin{subfigure}[b]{0.245\textwidth}
        \includegraphics[width=\textwidth]{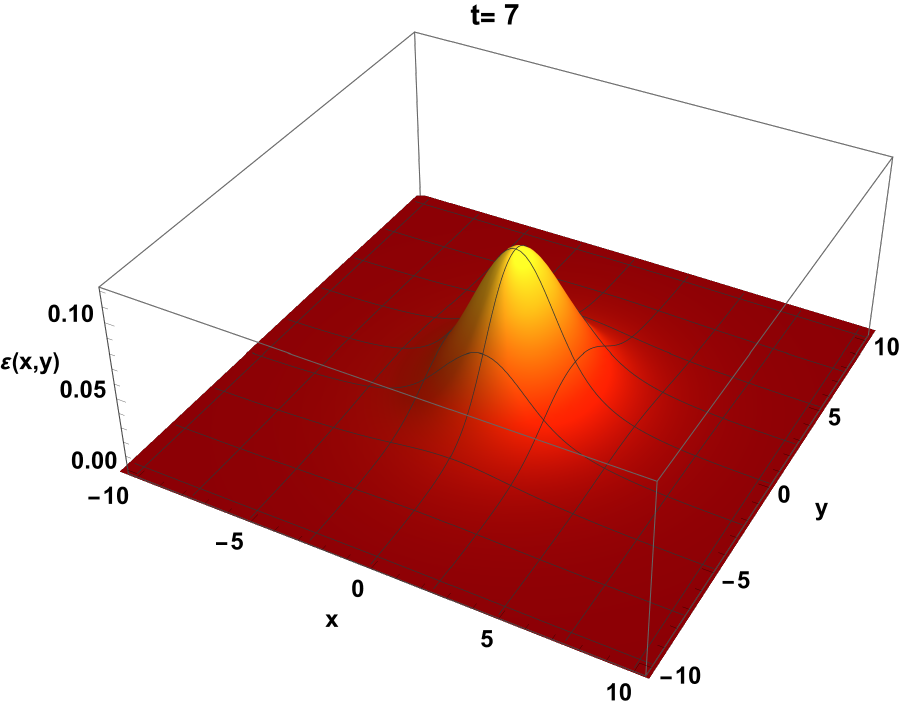}
    \end{subfigure}
    \hfill
    \begin{subfigure}[b]{0.245\textwidth}
        \includegraphics[width=\textwidth]{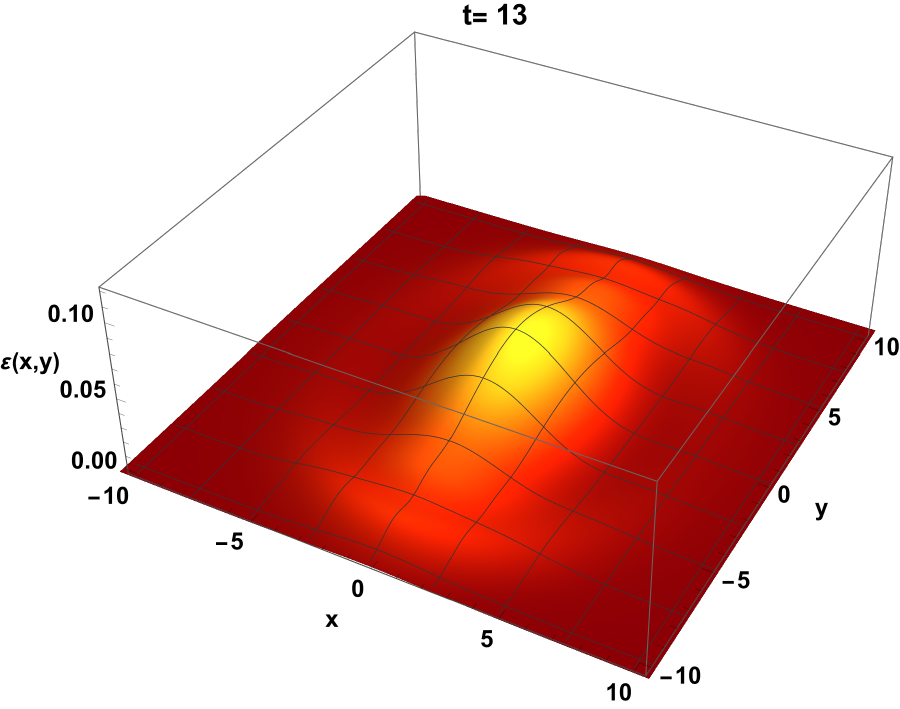}
    \end{subfigure}

    \vspace{0.5cm}

    % Row 2
    \begin{subfigure}[b]{0.245\textwidth}
        \includegraphics[width=\textwidth]{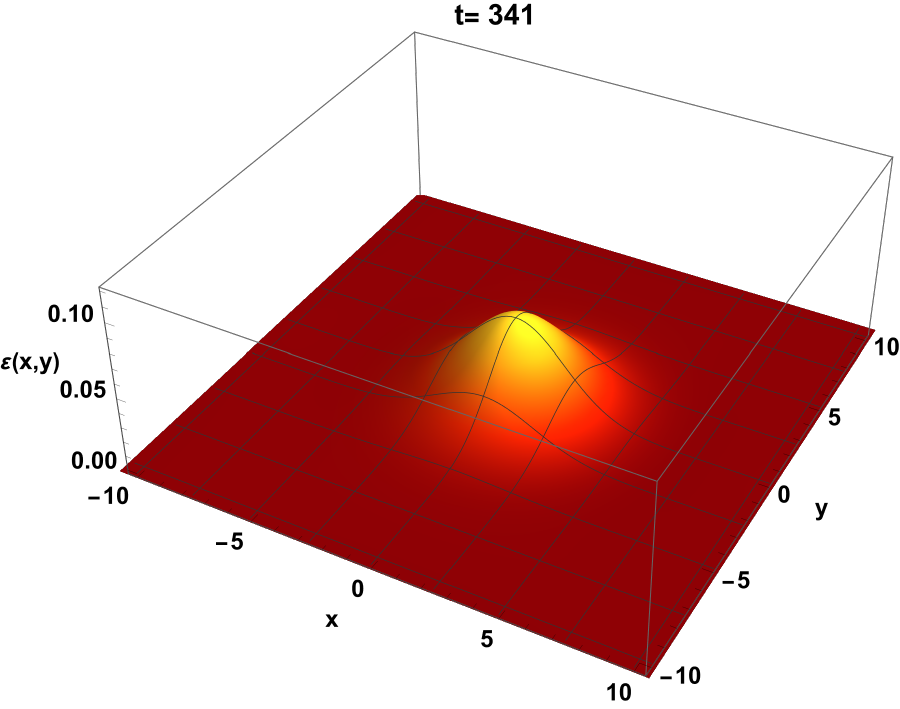}
    \end{subfigure}
    \hfill
    \begin{subfigure}[b]{0.245\textwidth}
        \includegraphics[width=\textwidth]{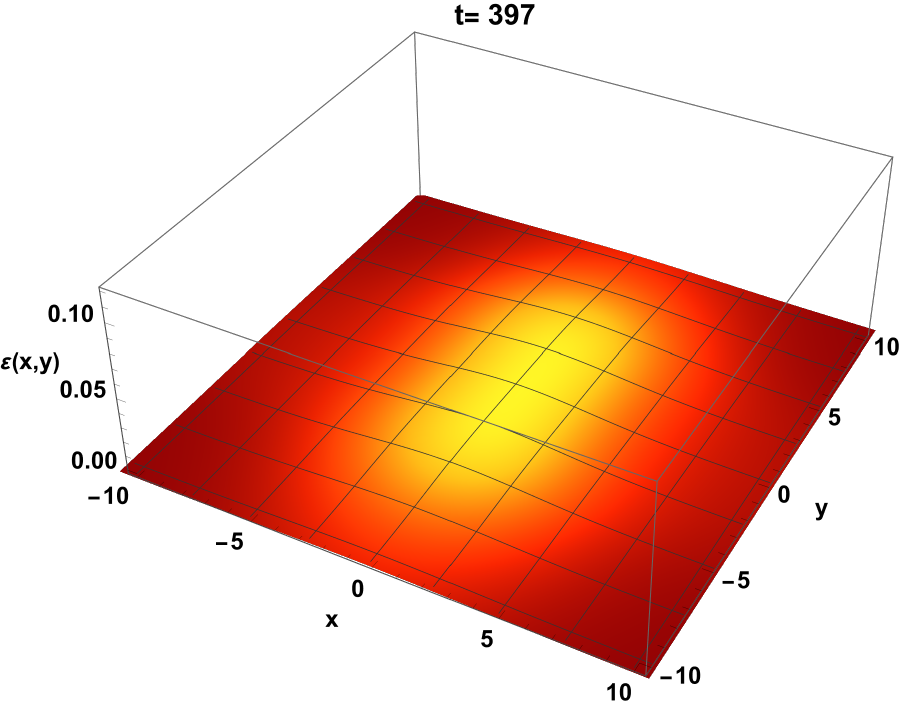}
    \end{subfigure}
    \hfill
    \begin{subfigure}[b]{0.245\textwidth}
        \includegraphics[width=\textwidth]{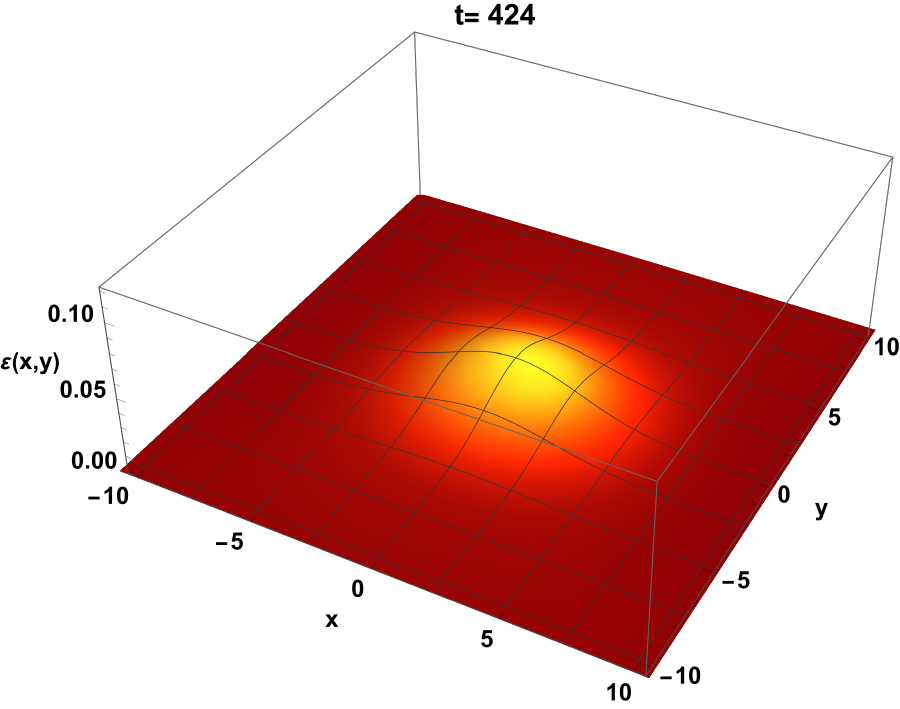}
    \end{subfigure}
    \hfill
    \begin{subfigure}[b]{0.245\textwidth}
        \includegraphics[width=\textwidth]{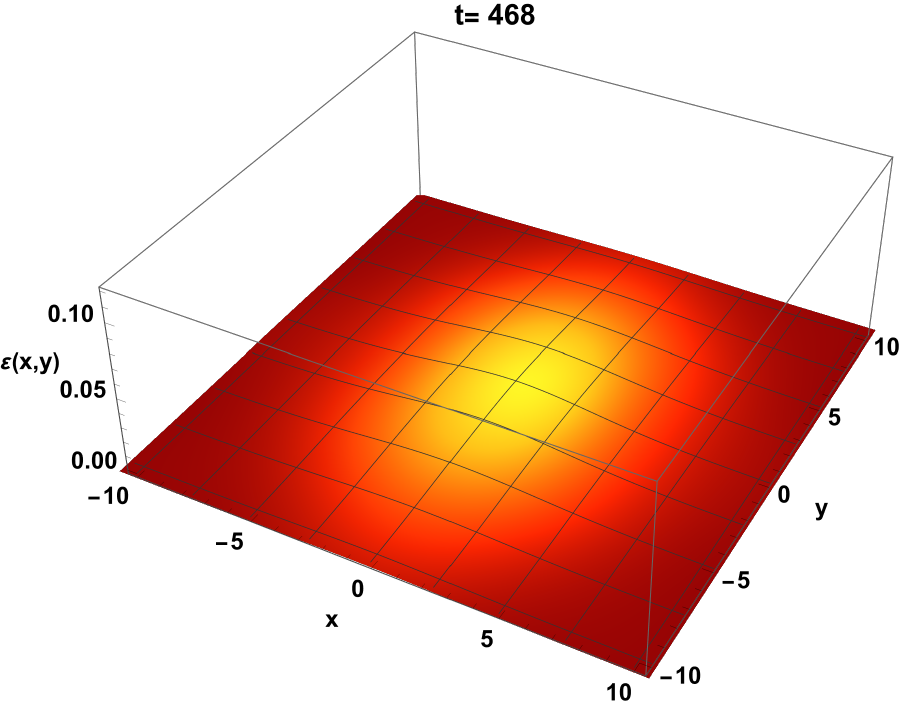}
    \end{subfigure}
    \caption{\textit{Snapshots of the energy density showing the formation and vibration of an oscillon for $\lambda=0.05$.}}
    \label{FigI3:VortexOscillonLambda0.05}
\end{figure}

However, for very small values of $\lambda$, the annihilation results in the formation of an oscillon. These results were first established by {\cite{Gleiser2007}}. Figure~\ref{FigI3:VortexOscillonLambda0.05} illustrates this phenomenon in more detail for $\lambda=0.05$. Initially, the vortex–antivortex pair attracts and annihilates, emitting radiation. Once the field configuration stabilizes, an oscillon remains at the center of the simulation box. Snapshots in the lower row of Figure~\ref{FigI3:VortexOscillonLambda0.05} show the vibration of this object: when the amplitude is maximal, the energy density is concentrated at the origin; when it is minimal, the energy density spreads out. 

%\vspace{-0.1cm}

High-speed vortex–antivortex scattering has also been explored. Interestingly, an oscillon always forms for $\lambda\ll1$, unlike the case of kink–antikink collisions in $(1+1)$-dimensional scalar field theories, where kinks may bounce back without annihilating.
\vspace{-0.1cm}

Nevertheless, a similar fractal-like structure can appear in vortex scattering involving excited BPS vortices \cite{Krusch2024}. If the relative phase between the two vortices is varied, the vortices can undergo multiple bounces before escaping. This phenomenon was reproduced in \cite{AlonsoIzquierdo2024d} using a collective coordinate approach that incorporates internal vibrational modes.

\section{Vortices and spectral walls }\label{spectralwallVort}
In Chapter \ref{Intro1}, we analyzed the spectral wall phenomenon for kinks in both the $\phi^4$ and $\phi^6$ models. Nevertheless, this phenomenon can also be observed in the case of Abelian-Higgs vortices. In this context, the study of internal modes is considerably more complex and will be addressed in detail in Chapter \ref{Chap3}. Nevertheless, we anticipate some key aspects here in order to discuss the emergence of spectral walls in the BPS limit.

As mentioned in Chapter \ref{Chap1}, the BPS equations can be used to derive the small fluctuation operator. Furthermore, they allow us to decompose this operator as follows\cite{Weinberg1979,AlonsoIzquierdo2016,AlonsoIzquierdo2016b}:
\begin{equation}
    \mathcal{H}^+\xi=\mathcal{D}^\dagger\mathcal{D}\,\xi=\omega^2\,\xi,
\end{equation}
where
\begin{equation}
    \xi(x_1,x_2)=\left(a_1(x_1,x_2)\,\, a_2(x_1,x_2)\,\, \varphi_1(x_1,x_2)\,\,\varphi_2(x_1,x_2)\right),
\end{equation}
with $a_i(x_1,x_2)$ and $\varphi_i(x_1,x_2)$ being the perturbations of $A_i$ and the real and imaginary parts of $\Phi$, respectively.

The key point here is that the following operators are supersymmetric partners:
\begin{equation}
     \mathcal{H}^+=\mathcal{D}^\dagger\mathcal{D},\quad  \mathcal{H}^-=\mathcal{D}\mathcal{D}^\dagger,
\end{equation}
which implies that the spectrum of internal modes can be extracted from the study of $\mathcal{H}^-$, since both operators are isospectral.

For BPS Abelian-Higgs vortices, using the BPS equations \eqref{eqI3:BPS1}--\eqref{eqI3:BPS2} together with the background gauge condition $\partial_k a_k -(\psi_1\,\varphi_2-\psi_2\,\varphi_1)=0$, the kernel operator $\mathcal{D}$ takes the form
\begin{eqnarray}\label{eqI3:OperatorD}
    \mathcal{D}=\begin{pmatrix}
        -\partial_2 & \partial_1 &\psi_1 & \psi_2 \\
        \partial_1 & \partial_2 &-\psi_2 & \psi_1 \\ 
        \psi_1 & -\psi_2 &  -\partial_2+V_1 & -\partial_1-V_2\\
        \psi_2 & \psi_1 &  \partial_1+V_2 & -\partial_2+V_1
    \end{pmatrix}\,,
\end{eqnarray}
where $V_1$, $V_2$, $\psi_1$, and $\psi_2$ correspond to a static field configuration solving the BPS equations. This allows us to construct the operator
\begin{equation*}
    \mathcal{H}^-=\begin{pmatrix}
        -\nabla^2+|\psi|^2 & 0 &0 &0\\
        0 &  -\nabla^2+|\psi|^2 &0 &0\\
        0 & 0& -\nabla^2+\frac{1}{2}(|\psi|^2+1)+V_kV_k & -2V_k\partial_k-\partial_kV_k\\
        0 &0 & 2V_k\partial_k+\partial_kV_k & -\nabla^2+\frac{1}{2}(|\psi|^2+1)+V_kV_k 
    \end{pmatrix}\,.
\end{equation*}

Thus, to obtain the discrete spectrum, it is sufficient to solve the eigenvalue problem
\begin{equation}\label{eqI3:ReducedEigenvalueProblem}
    (-\nabla^2+|\psi|^2)a_1=\omega^2 a_1.
\end{equation}

The corresponding eigenmode of the operator $\mathcal{H}^+$ is then given by
\begin{equation}
    \xi_+=\frac{\mathcal{D}^{\dagger}\xi_-}{\omega}=\frac{1}{\omega}\left(\partial_2 a_1\,\, -\partial_1 a_1\,\, \psi_1\,a_1\,\,\psi_2\,a_1\right).
\end{equation}

\begin{figure}[h!]
    \centering
    \includegraphics[width=0.48\linewidth]{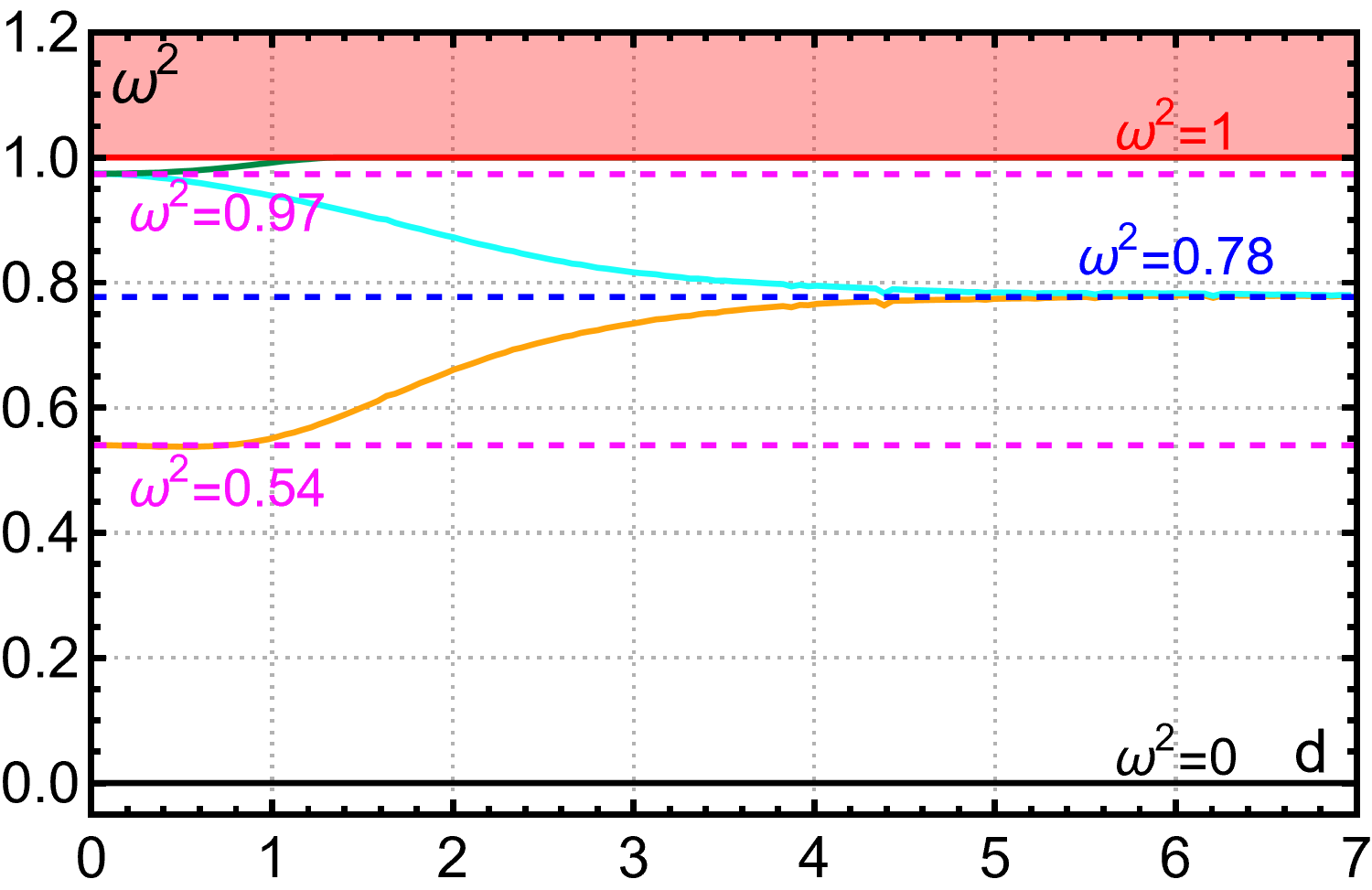}
    \vspace{-0.3cm}
    \caption{\textit{Eigenvalues obtained from solving the spectral problem \eqref{eqI3:ReducedEigenvalueProblem} for two unit-charge vortices separated by a distance $2d$. The pink dashed lines correspond to the discrete eigenfrequencies of a two-vortex configuration, while the blue dashed line indicates the eigenfrequency of a single vortex.}}
    \label{figI3:SpectralWall}
\end{figure}

To compute the spectrum for two vortices each with charge $n=1$ separated a distance $2d$, we can use the method addressed at the end of Section \ref{SecBPSlIMIT} to obtain $|\psi|$ for the field configuration and solve the eigenvalue problem \eqref{eqI3:ReducedEigenvalueProblem}. The  eigenfrequencies  found using this procedure as a function of the vortex separation are shown in Figure \ref{figI3:SpectralWall}. Here, for $d \ll 1$, the eigenvalues match those of a  two-vortex configuration. Conversely, for $d \gg 1$, the eigenvalues approach those of a single vortex, in a similar way that occurred when studying kink-antikink configurations in $1+1$ scalar field theories \cite{AlonsoIzquierdo2024f}.

Crucially, we observe a mode crossing into the continuum, signaling the presence of a spectral wall in the range  $d \in\left[ 1.6,2.2\right]$. As described in \cite{AlonsoIzquierdo2024e}, when vortices collide and the amplitude of internal excitations is sufficiently large, the interaction becomes repulsive. However, if the amplitude is finely tuned, the vortices may scatter and become trapped near the spectral wall at $d \in\left[ 1.6,2.2\right]$. If the amplitude is below the critical threshold, the vortices may undergo multiple bounces before eventually escaping each other’s influence.

\begin{figure}[h!]
    \centering
    \includegraphics[width=0.62\linewidth]{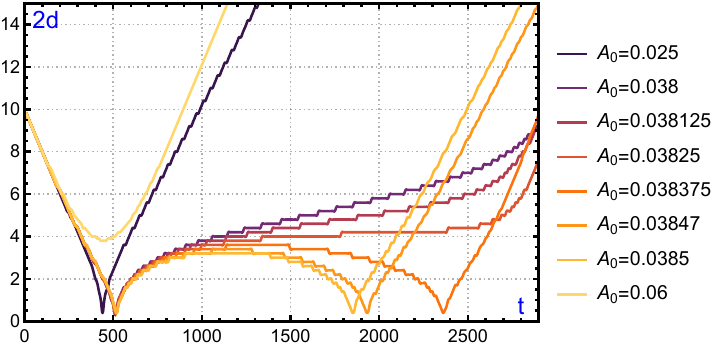}
    \vspace{-0.3cm}
    \caption{\textit{Distance between vortices for scattering at $v=0.01$ as a function of time for different values of the internal mode amplitude.}}
    \label{figI3:SpecWallAH}
\end{figure}

In Figure \ref{figI3:SpecWallAH}, it can be appreciated the results of a series of simulations in which both vortices have been scattered with an initial velocity $v=0.01$ whose internal modes have been initially excited in opposite phase. As it can be appreciated, if the amplitude is large, both vortices repel without colliding. Conversely, if the amplitude of the internal modes is very small, then, a regular $90^\circ$ scattering event occur. Nevertheless, if the amplitude is well tuned, then, vortices scatter and then freeze at a fixed distance. A similar analyisis to that one presented here have also been carried out for three vortices configurations \cite{AlonsoIzquierdo2025c}.

    \chapter{ Internal mode structure }\label{Chap3}
    This chapter is an adaptation from References \cite{AlonsoIzquierdo2024b,AlonsoIzquierdo2025b}: 

\begin{figure}[htb]
    \centering
   \includegraphics[width=1\linewidth]{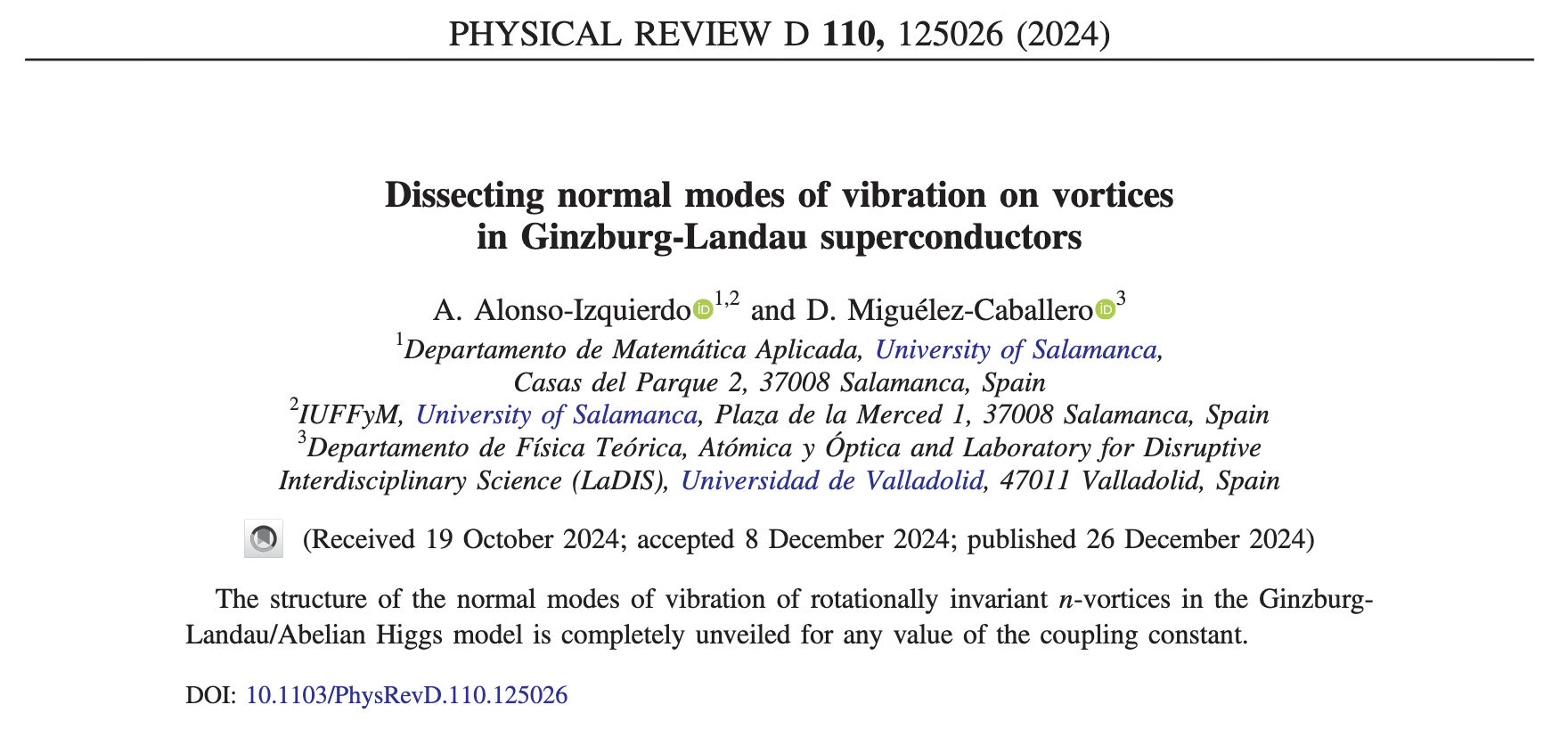}
\end{figure}
%\vspace{-0.4cm}
\begin{figure}[htb]
    \centering
   \includegraphics[width=1\linewidth]{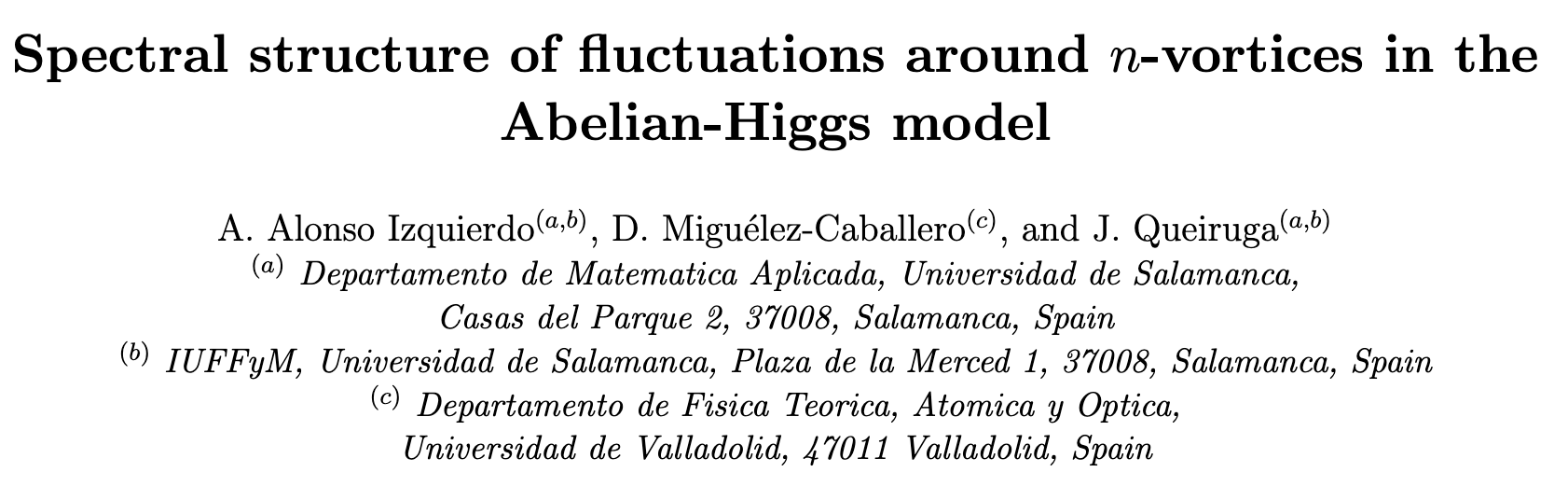}
\end{figure}
\vspace{-0.8cm}
\begin{figure}[htb]
    \centering
   \includegraphics[width=0.4\linewidth]{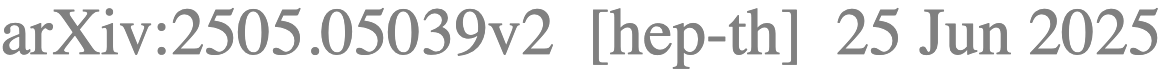}
\end{figure}

\section{Introduction}

As previously stated in Chapters \ref{Intro0} and \ref{Intro2}, vortices are two dimensional solitons whose charge is exponentially localized in $\mathbb{R}^2$. Their topological charge is determined by the homotopy group $\pi_1(\mathbb{S}_1)=\mathbb{Z}$, and as a consequence, they may exist in multivortex configurations with any (positive or negative) charge. In their local version, they are described by a complex field (usually called  Higgs field) and a $U(1)$ gauge field responsible for the local gauge symmetry. Vortex solutions arising in the Abelian-Higgs model are fundamental in many areas of physics from condensed matter \cite{Abrikosov1957, superfluid-vortex, Nielsen1973} to cosmology \cite{Vilenkin2000}. In the usual formulation of the Abelian-Higgs model, the functional form of the self-interacting potential of the Higgs field determines the vacuum structure of the model (as analized in Section \ref{SecVortexSol}), and therefore, the topological structure of the solitons. But more interestingly, as stated in Section \ref{SecIntBetVort}, the strength of the potential, characterized by a self-coupling constant, $\lambda$, has important consequences for higher charge solitons. There is a special value of $\lambda=\lambda_c$ (usually called critical, BPS or self-dual) at which individual charge one vortices do not interact statically \cite{Manton2004, Jaffe1980}. This means that, at this value, the static intersolitonic forces vanish. Below $\lambda_c$ individual vortices attract, as it can be appreciated in Figure \ref{FigI3:VortexAttraction}. As a consequence, at sufficiently large times, a configuration of $N$ vortices located at different points in space will eventually coalesce and form a charge $N$ vortex. On the other hand, for $\lambda>\lambda_c$ and at sufficiently large times, the field configuration will consist of a set of charge one vortices moving at (possibly) different velocities, as we saw in Figure \ref{figI3:SplittingVortex}. 

The BPS Abelian-Higgs model deserves special attention. At $\lambda=\lambda_c$ vortices satisfy the set of first order static equations defined by equations \eqref{eqI3:BPS1}-\eqref{eqI3:BPS2}, the so-called BPS equations which saturate the energy in each topological sector \cite{Bogomolnyi1976, Prasad1975, Taubes1980, Taubes1980b}. In addition, the low velocity dynamics is dictated by the kinetic terms in the action. This gives rise to an effective manifold, called the moduli space, whose geodesics correspond to trajectories of the interacting vortices \cite{manton1981, Samols1992}. The vortex scattering has been thoroughly studied in the literature \cite{Ruback1988, Shellard1988, Matzner1981,Moriarty1988,Arthur1996} and it has become clear recently that the moduli space picture can be improved by adding possible internal excitations \cite{Krusch2024, AlonsoIzquierdo2024, AlonsoIzquierdo2024e}. Internal excitations are linear perturbations on top of the vortex, and, for the charge one case, they have been studied in the literature \cite{AlonsoIzquierdo2016,AlonsoIzquierdo2024b, AlonsoIzquierdo2016b, Goodband1995, Kojo2007, AlonsoIzquierdo2024f}. In an excited vortex scattering the internal modes may move through the spectrum changing their frequency, as we briefly introduced in Section \ref{spectralwallVort}.
%This generates effective forces that may lead to a relevant deviation form the moduli space trajectories. In the non-BPS case, the internal modes may play also a relevant role, in this case as competitors of the intersolitonic forces.
Recently, it has been suggested that the vibrational modes of cosmic strings—spatially extended counterparts of the vortices studied in this work—could act as sources of gravitational waves \cite{Blanco-Pillado2024}. Although such applications go beyond the immediate scope of this chapter, they highlight the broader relevance of a detailed understanding of the fluctuation spectrum in the Abelian-Higgs model.

The analysis of the spectral problem for the second-order small fluctuation operator associated with a vortex solution is often essential to understanding its dynamical  properties. For instance, their stability is ensured if the eigenvalues are non-negative. However, the significance of eigenfluctuations extends beyond stability. Recent studies have demonstrated that the dynamics of two self-dual vortices are drastically altered when they are excited \cite{AlonsoIzquierdo2024f,AlonsoIzquierdo2024d}. In the absence of excitation, BPS vortices do not interact. However, when they vibrate in phase, an attractive force emerges; conversely, when they vibrate out of phase, a repulsive force arises. Additionally, a resonant energy transfer mechanism can occur during vortex collisions, facilitating energy exchange between different eigenmodes and leading to chaotic vortex dynamics \cite{Krusch2024}. 
%Other works emphasize the role of these vibration modes in the context of topological defects in cosmology \cite{Arodz1,Arodz2,Kojo,kop}. 
In particular, interesting effects of massive states on vortex cores have also been identified in the superfluid phase of the isotope ${}^3{\mathrm{ He}}$ \cite{Kopnin1991}. Moreover, bound modes play a key role in estimating quantum corrections to the mass of topological defects \cite{Rajaraman1982}.

%Despite the importance of the vibrational modes, the full spectral structure of the rotationally invariant $n$-vortex beyond the critical point $\lambda_c = 1$ has remained largely unexplored. For $\lambda_c \neq 1$, the supersymmetry structure present at the critical value is lost, necessitating a different mathematical approach to address this problem. In this chapter, we present a comprehensive description of the spectral structure of the vortex fluctuation operator as a function of the coupling constant $\lambda$. One application of this study is in understanding vortex dynamics in the Abelian-Higss model when $\lambda\neq 1$.
%It is well known that in Type I superconductors ($\lambda < 1$), vortices attract each other, whereas in Type II superconductors ($\lambda > 1$), vortices repel. However, as discussed earlier, this interaction scheme changes drastically when the vortices are in an excited state. Excited vortices in Type I materials could exhibit repulsive behavior, while vibrations could stabilize $n$-vortices in Type II superconductors.

It is worth mentioning that an $n$-vortex possesses $2n$ zero modes in the BPS limit. This result was rigorously established by E. Weinberg \cite{Weinberg1979} through a generalization of the index theorem for elliptic operators. In addition to these zero modes, the second-order small fluctuation operator associated with the self-dual vortices includes massive discrete modes. Derrick-type bound states, where the vortex size oscillates periodically, were first identified by Goodman and Hindmarsh \cite{Goodband1995}. These modes exhibit the same angular dependence as the static $n$-vortex solution. Other eigenfluctuations, however, do not follow this form. Indeed, the complete set of normal modes for the rotationally invariant $n$-vortex at the critical value $\lambda_c=1$ was unveiled in \cite{AlonsoIzquierdo2016,AlonsoIzquierdo2016b}.
%, where the hidden supersymmetry of the spectral problem associated with the self-dual vortex fluctuation operator was exploited.
For the 1-vortex, it was shown that there are two zero modes, one Derrick-type mode with frequency $\omega^2 = 0.777476$, and a continuous spectrum starting at the threshold value $\omega^2=1$. In contrast, the rotationally invariant 2-vortex exhibits a richer structure, involving four zero modes, one Derrick-type mode with frequency $\omega^2 = 0.53859$, a doubly degenerate shape mode with $\omega^2 = 0.97303$, and a continuous spectrum beginning at $\omega^2=1$. As the vorticity increases, the spectrum becomes more intricate.

This chapter constitutes a natural continuation of two previously mentioned studies \cite{AlonsoIzquierdo2016, AlonsoIzquierdo2016b}, which investigate the Abelian-Higgs model at critical coupling ($\lambda_c= 1$). 
At this special point, powerful supersymmetric techniques allow for an elegant and efficient spectral analysis. In contrast, the present work addresses the general (non-BPS) case $\lambda_c \neq 1$, where such techniques are no longer applicable, and the spectral problem must be approached from first principles. The study of the excitation spectrum of vortex solitons in the Abelian-Higgs model has a rich history, beginning with the foundational work of Goodband and Hindmarsh in 1995 \cite{Goodband1995}, who offered only a partial spectrum. Other subsequent investigations \cite{Arodz1991, Arodz1996, Kojo2007} addressed specific limits or configurations. To the best of our knowledge, our contribution is the first complete and systematic investigation of the full excitation spectrum of vortices away from the BPS limit. Our results yield a comprehensive classification of the eigenmodes and their properties for arbitrary values of the coupling constant, resolving a decades-old problem. In this sense, the present article concludes the research program initiated in the aforementioned works.

This chapter is organized as follows: in Section \ref{sec:intro} we present the model and compute the spectral equations. In Section \ref{bps} the review the main properties of the spectrum of self-dual vortices. In Section \ref{non-bps} we analyzed in detail the spectral structure of non-BPS vortices. In Section \ref{num_spec} we compute numerically the frequencies and profiles of the first internal modes. Finally, Section \ref{conclusions} is devoted to our conclusions and further discussion. We also add one appendix (\ref{appen}) with numerical details. 

%%%%%%%%%%%%%%%%%%%%%%%%%%%%%%%%%%%%%%%%
%%%%%%%%%%%%%%%%%%%%%%%%%%%%%%%%%%%%%%%%
%%%%%%%%%%%%%%%%%%%%%%%%%%%%%%%%%%%%%%%%
%%%%%%%%%%%%%%%%%%%%%%%%%%%%%%%%%%%%%%%%
%%%%%%%%%%%%%%%%%%%%%%%%%%%%%%%%%%%%%%%%
%%%%%%%%%%%%%%%%%%%%%%%%%%%%%%%%%%%%%%%%

\vspace{0.2cm}
\section{An introduction to the small perturbation problem}\label{sec:intro}

In this section, we will introduce the small perturbation operator fro which we will obtain the internal mode structure associated with vortex solution in the Abelian-Higgs model. 
\vspace{0.2cm}

From now on, for a static rotationally invariant $n$-vortex solution we will denote the scalar and vector components as
\vspace{0.2cm}
\[
\psi(\vec{x};n)=\psi_1(\vec{x};n) + i \, \psi_2(\vec{x};n) \hspace{00cm},\hspace{0.5cm} V(\vec{x};n)=(V_1(\vec{x};n),V_2(\vec{x};n)) \hspace{0.5cm}\mbox{with}\hspace{0.5cm} \vec{x}=(x_1,x_2) ,
\]
which will be assembled into a four real component column $\Psi(r,\theta) \in {\cal C}^\infty(\mathbb{R}^2) \oplus \mathbb{R}^4$ in the form
\begin{equation}
\Psi(r,\theta)=\left(\begin{array}{c} V_1(\vec{x};n) \\ V_2(\vec{x};n) \\ \psi_1(\vec{x};n) \\ \psi_2(\vec{x};n) \end{array}  \right) =  \left(\begin{array}{c}- \frac{n\beta_n(r)}{r} \sin \theta \\ \frac{n\beta_n(r)}{r} \cos \theta  \\ f_n(r) \cos (n\theta) \\ f_n(r) \sin (n\theta) \end{array}  \right) \label{vortexcolumn}.
\end{equation}
At this point, we will define certain quantities that contain information about the angular dependence of the vector and scalar components functions $F(r,\theta)\in {\cal C}^\infty(\mathbb{R}^2) \oplus \mathbb{R}^4$ of interest, such as solutions, fluctuations, etc., as well as their radial behavior near the origin. The latter is particularly relevant as it allows us to identify the presence of singularities by simply analyzing the form of these quantities. For this reason, we will call them \textit{characteristics of the function} under study:

\vspace{-0.3cm}

\begin{definition} [Angular and radial characteristics] The \textit{angular characteristic} ${\mathrm{ ch}}_\theta[F]$ of a function $F(r,\theta)\in {\cal C}^\infty(\mathbb{R}^2) \oplus \mathbb{R}^4$ is defined as
\[
{\mathrm{ ch}}_\theta[F]  = {\alpha_1;\dots;\alpha_q \choose \beta_1;\dots;\beta_q },
\]
where the set of numbers in the rows specifies the angular dependence of the function $F$. Specifically, the vector components of $F$ depend only on a linear combination of sine and cosine functions with arguments $\alpha_1 \theta, \dots, \alpha_q\theta$ while the scalar component depends only on the angles $\beta_1 \theta, \dots, \beta_q\theta$. The \textit{angular characteristics number} is defined as $\#{\mathrm{ ch}}_\theta[F] = \max\{ p,q\}$. 

On the other hand, the \textit{radial characteristic} ${\mathrm{ch}}_r[F]$,
\[
{\mathrm{ ch}}_r[F]  = {\ell_1 \choose \ell_2} ,
\]
determines the radial dependence of the function in the neighborhood of $r=0$. In particular, the vector components of the function $F$ behave as $r^{\ell_1}$, whereas the scalar components behave as $r^{\ell_2}$.
\end{definition}
\vspace{-0.3cm}

For example, the angular and radial characteristics of the $n$-vortex solution $\Psi(r,\theta)$ are
\begin{equation}
{\mathrm{ ch}}_\theta[\Psi] = {1\choose n} \hspace{0.5cm}\mbox{and}\hspace{0.5cm} {\mathrm{ ch}}_r[\Psi] = {2\choose n} \label{chvortex},
\end{equation}
which is directly obtained from (\ref{vortexcolumn}) and the power expansion around the origin \eqref{eqI3:PowerExpansion}.

%\vspace{0.2cm}

The main goal of this chapter is the study of the spectrum of normal modes associated with $n$-vortices. In order to achieve this, the dynamics of small fluctuations $\xi(\vec{x})$ around the static $n$-vortex (\ref{vortexcolumn}) will be analyzed. Consequently, the perturbed $n$-vortex, denoted collectively by $\widetilde{\Psi}(\vec{x},n)$, is expressed as
\begin{equation}
\widetilde{\Psi}(\vec{x},n) = \Psi(\vec{x},n) + \epsilon\, \xi(\vec{x})=\left(\begin{array}{c} V_1(\vec{x};n) \\ V_2(\vec{x};n) \\ \psi_1(\vec{x};n) \\ \psi_2(\vec{x};n)\end{array}  \right) + \epsilon\,  \left(\begin{array}{c} a_1(\vec{x}) \\ a_2(\vec{x})  \\ \varphi_1(\vec{x}) \\ \varphi_2(\vec{x}) \end{array}  \right) \label{perturvortexcolumn},
\end{equation}
where 
\[
\xi(\vec{x})=\left( \begin{array}{c c c c}a_1(\vec{x}), & a_2(\vec{x}), & \varphi_1(\vec{x}) & \varphi_2(\vec{x}), \end{array} \right)^\intercal ,
\]
denote the fluctuation column. As mentioned in Chapter \ref{Intro2}, to discard pure (non-physical) gauge fluctuations, the so called \textit{background gauge}
\begin{equation}
	\partial_1 a_1( \vec{x}) + \partial_2 a_2( \vec{x}) - (\,\psi_1( \vec{x})\, \varphi_2( \vec{x})-\psi_2( \vec{x})\,\varphi_1( \vec{x})\,)=0  , 
	\label{backgroundgauge}
\end{equation}
is imposed as the gauge fixing condition on the fluctuation modes. With this set-up the normal modes of an $n$-vortex solution $\Psi(\vec{x},n)$ are determined by the spectral condition 
\begin{equation}
{\cal H}^+ \xi_\nu(\vec{x}) =\omega_\nu^2 \, \xi_\nu(\vec{x}) , \label{spectralproblem}
\end{equation}
where $\nu$ is a label in either the discrete or the continuous spectrum useful to enumerate the eigenfunctions and eigenvalues, and ${\cal H}^+$ is the second-order small fluctuation operator
{\scriptsize \begin{equation}
	{\cal H}^+= \left( \begin{array}{cccc}
	-\Delta + |\psi|^2 & 0 & -2 \widetilde{D}_1 \psi_2 & 2 \widetilde{D}_1 \psi_1 \\
	0 & -\Delta +|\psi|^2 & -2 \widetilde{D}_2 \psi_2 & 2 \widetilde{D}_2 \psi_1 \\
	-2 \widetilde{D}_1 \psi_2 & -2 \widetilde{D}_2\psi_2 & -\Delta + \frac{3\lambda}{2} \psi_1^2 + (1+\frac{\lambda}{2})\psi_2^2 -\frac{\lambda}{2}  +V_kV_k & -2 V_k \partial_k + (\lambda-1)\psi_1\psi_2\\
	2\widetilde{D}_1\psi_1 & 2 \widetilde{D}_2 \psi_1 & 2V_k \partial_k + (\lambda-1)\psi_1\psi_2 & -\Delta + (1+\frac{\lambda}{2})\psi_1^2 +\frac{3\lambda}{2} \psi_2^2  -\frac{\lambda}{2} + V_kV_k
	\end{array} \right)\label{hessianoperator}
   ,
	\end{equation}}
where $\widetilde{D}_i\psi_j = \partial_i\psi_j+\epsilon^{jk} V_i \psi_k$. The Sturm-Liouville eigenvalue problem (\ref{spectralproblem}) is obtained by linearizing the field equations (in the background gauge) around the vortices.

The fluctuation eigenfunctions $\xi(\vec{x})$ belong in general to a rigged Hilbert space, such that there exist square integrable eigenfunctions $\xi_\nu(\vec{x})\in L^2(\mathbb{R}^2)\oplus \mathbb{R}^4$ belonging to the discrete spectrum, for which the norm $\|\xi(\vec{x})\|$ is bounded:
\begin{equation}
	\|\xi(\vec{x})\|^2  = \int_{\mathbb{R}^2} d^2x \Big[ (a_1(\vec{x}))^2 + (a_2(\vec{x}))^2 + (\varphi_1(\vec{x}))^2 + (\varphi_2(\vec{x}))^2 \Big] < +\infty \, ,
	\label{normalization}
\end{equation}
together with scattering (unbounded) eigenfunctions $\xi_\nu(\vec{x})$ with $\nu$ ranging in a dense set. 

Note that the operator (\ref{hessianoperator}) resulting for the asymptotic values (\ref{eqI3:AssymptoticVortex}) of the vortices exhibits a triply degenerate spectrum emerging at the threshold value $\omega_c^2 = 1$ together with once degenerate spectrum on the value $\omega_c^2= \lambda$. As a consequence, the continuous spectrum of the operator (\ref{hessianoperator}) begins at $\omega_c^2=\lambda$ when $\lambda <1$, and at the value $\omega_c^2=1$ when $\lambda \geq 1$. From a technical perspective, eigenfunctions corresponding to eigenvalues in the range $\lambda < \omega_\nu^2 < 1$ consist of a localized gauge field component and a non-localized scalar field component (which behave as a scattering mode). These type of modes is usually called quasi-bound modes. This behavior is reversed in the range $1<\omega_\nu^2 < \lambda$.

%%%%%%%%%%%%%%%%%%%%%%%%
%%%%%%%%%%%%%%%%%%%%%%%%
%%%%%%%%%%%%%%%%%%%%%%%%
%%%%%%%%%%%%%%%%%%%%%%%%
%%%%%%%%%%%%%%%%%%%%%%%%

\section{Spectral structure for the fluctuations around self-dual vortices}

\label{bps}

In this section, our objective is to determine the complete structure of the normal modes for rotationally invariant $n$-vortices in the Abelian-Higgs model for arbitrary values of the coupling constant $\lambda$. This analysis is entirely new, with the exception of the specific case $\lambda=1$ (self-dual vortices), which was thoroughly investigated in \cite{AlonsoIzquierdo2016, AlonsoIzquierdo2016b}. In this section, we briefly summarize the results presented in those works. The motivation for this summary is twofold: (i) The general results obtained in this chapter must reproduce those of the aforementioned works when the coupling constant takes the specific value $\lambda=1$; and (ii) it will become evident that the mathematical techniques employed in \cite{AlonsoIzquierdo2016, AlonsoIzquierdo2016b} are applicable exclusively to self-dual vortices, corresponding to the special case $\lambda=1$. This limitation underscores the necessity of developing new methods to extend the analysis beyond the self-dual regime. This new approach is presented in Section \ref{non-bps}.

As previously mentioned in Section \ref{SecBPSlIMIT}-\ref{spectralwallVort}, we will describe the spectral structure of self-dual vortices, which satisfy the BPS first-order differential equations
\begin{equation}
\widetilde{D}_1\psi_1 \mp \, \widetilde{D}_2 \psi_2 =0, \hspace{0.5cm} \hspace{0.5cm} \widetilde{D}_1\psi_2 \pm \, \widetilde{D}_2 \psi_1 =0, \hspace{0.5cm} \hspace{0.5cm} F_{12}\mp \frac{1}{2} (1-|\psi|^2 )=0 .\label{bpsequation}
\end{equation}
\begin{comment}
Equations (\ref{bpsequation}) are derived using a Bogomolny decomposition of the energy functional. Substituting the ansatz (\ref{ansatz}) into the first-order PDE system (\ref{bpsequation}) leads to the following system of first-order ODEs
\begin{equation}
\frac{df_n(r)}{d r}=\frac{n}{r} f_n(r) [1-\beta_n(r)]\hspace{0.0cm},\hspace{0.5cm} \frac{d\beta_n(r)}{d r}=\frac{r}{2n}[1-f_n^2(r)] \, . \label{ode1}
\end{equation}
\end{comment}
As briefly explained in Section \ref{spectralwallVort}, in the aforementioned papers the spectral problem in the self-dual/BPS case was solved by exploiting a hidden supersymmetric structure involving the second order vortex fluctuation operator (\ref{hessianoperator}), which can be factorized as ${\cal H}^+= {\cal D}^\dagger \, {\cal D}$, where ${\cal D}$ is the first order differential matrix operator 
\[
{\cal D} = \left( \begin{array}{cccc}
-\partial_2 & \partial_1 & \psi_1 & \psi_2 \\
-\partial_1 & -\partial_2 & -\psi_2 & \psi_1 \\
\psi_1 & -\psi_2 & -\partial_2 + V_1 & -\partial_1 -V_2 \\
\psi_2 & \psi_1 & \partial_1+V_2 & -\partial_2 + V_1
\end{array} \right) \,,
\]
obtained from the BPS equations (\ref{bpsequation}) and the background gauge condition (\ref{backgroundgauge}). It can be shown that the intertwined supersymmetric partner operator ${\cal H}^-= {\cal D} \, {\cal D}^\dagger$ has a block diagonal structure. This property can be exploited to investigate the fluctuation spectral problem of the BPS vortex and the appearance of spectral walls in multi-vortex field configurations \cite{AlonsoIzquierdo2025c}  (as addressed in Section \ref{spectralwallVort}). Recall that our goal in this chapter is to study this problem for general (non-BPS) vortices for arbitrary values of $\lambda$, where this supersymmetric approach is no longer applicable.

The spectral structure identified in the BPS limit reveals that two apparently different types of eigenfunctions emerge in the spectral problem for self-dual vortices:

\begin{itemize}
	\item \textbf{Zero modes}: It is well known that at the critical value $\lambda=1$, vortices can be located at any point without experiencing any forces between them. For charge-$n$ vortices, this freedom in the positions of the vortex centers suggests the existence of $2|n|$ linearly independent zero modes (modes with vanishing eigenvalue $\omega^2_\mu=0$). This was proved by E. Weinberg in \cite{Weinberg1979} using a generalization of the index theorem for elliptic operators and the supersymmetric structure mentioned earlier. Following the notation used in \cite{Weinberg1979}, these $2|n|$ zero modes associated with the rotationally invariant BPS $n$-vortex have the structure
    \vspace{-0.2cm}
\begin{eqnarray}
\xi_0(\vec{x},n,k)&=& r^{n-k-1} \left( \begin{array}{c} h_{nk}(r) \, \sin[(n-k-1)\theta] \\ h_{nk}(r) \, \cos[(n-k-1)\theta] \\  -\frac{h_{nk}'(r)}{f_n(r)} \, \cos(k\theta) \\ - \frac{h_{nk}'(r)}{f_n(r)} \, \sin(k\theta) \end{array} \right)\,  , \label{bpszeromode4}
\\
\xi_0^\perp(\vec{x},n,k)&=& r^{n-k-1} \left( \begin{array}{c} h_{nk}(r) \, \cos[(n-k-1)\theta] \\ -h_{nk}(r) \, \sin[(n-k-1)\theta] \\  - \frac{h_{nk}'(r)}{f_n(r)} \, \sin(k\theta) \\  \frac{h_{nk}'(r)}{f_n(r)} \, \cos(k\theta)  \end{array} \right) \, , \label{bpszeromode41}
\end{eqnarray}
where $k=0,1,2,\dots,n-1$ and the radial form factor $h_n(r)$ satisfies the second-order ODE
\begin{equation}
-r \, h_{nk}''(r)+[1+2k-2n\,\beta_n(r)]\,h_{nk}'(r) + r \,f_n^2(r)\, h_{nk}(r)=0, \label{ode5}
\end{equation}
with the boundary conditions $h_{nk}(0)\neq 0$ and $\lim_{r\rightarrow \infty} h_{nk}(r) =0$. The eigenfunctions (\ref{bpszeromode4}) and (\ref{bpszeromode41}) are degenerate, although they are mutually orthogonal with respect to the interior product defined by the norm \eqref{normalization}.

\vspace{-0.2cm}

The angular characteristic of the zero modes is
\begin{equation}
{\mathrm{ ch}}_\theta[\xi_0(\vec{x},n,k)] = {\mathrm{ ch}}_\theta[\xi_0^\perp(\vec{x},n,k)] = {n-k-1\choose k}, \label{angchbpszeromode}
\end{equation}
which means that only one trigonometric function enters in the expression of the components, that is,
$\# {\mathrm{ ch}}_\theta[\xi_0(\vec{x},n,k)] =  \# {\mathrm{ ch} }_\theta[\xi_0^\perp(\vec{x},n,k)] = 1$. 

\vspace{-0.2cm}

Since $ h_{nk}(r) = c_0^{(n,k)} + c_{2k+2}^{(n,k)} r^{2k+2} + \dots$ near $r=0$, 
\begin{equation}
{\mathrm{ ch}}_r[\xi_0(\vec{x},n,k)] = {\mathrm{ ch}}_r[\xi_0^\perp(\vec{x},n,k)] = {n-k-1\choose k}.\label{radchbpszeromode}
\end{equation}
%\vspace{-0.4cm}
Note that the \textit{radial characteristic} ${\mathrm{ ch}}_r[\xi_0(\vec{x},n,k)]$ directly provides the constraints on the values of $k$. Regularity conditions require the components of this characteristic to be non-negative, otherwise a singularity would emerge at the origin. This means in this case that $k\in \{0,1,\dots, n-1\}$, as previously indicated. 
\vspace{-0.1cm}
In addition to these $2 \vert n \vert $ orthogonal zero modes, a continuous spectrum of scattering modes emerges for $\omega^2_\mu\geq1$.

\item \textbf{Shape modes}: Eigenfunctions belonging to the strictly positive spectrum of ${\cal H}^+$ arise with the form:
{\small
\begin{eqnarray}
\hspace{-0.4cm} \xi_\lambda(\vec{x},n,k)&=& \left( \begin{array}{c} \sin \theta \cos (k\theta) \frac{d v_{nk}(r)}{d r} - \frac{k}{r} \, v_{nk}(r) \cos \theta \sin(k\theta) \\ -\cos \theta \cos (k\theta) \frac{d v_{nk}(r)}{d r} - \frac{k}{r} \, v_{nk}(r) \, \sin \theta \sin(k\theta) \\ f_n(r) \, v_{nk}(r)\, \cos(n\theta)\, \cos(k\theta) \\ f_n(r)\, v_{nk}(r)\, \sin(n\theta) \,\cos(k\theta)
 \end{array} \right) \hspace{0.0cm}\, , \hspace{0.3cm} k=0,1,2,\dots\, , \hspace{0.2cm} \label{bpsexcitedmode1}
\end{eqnarray}
\begin{eqnarray}
\hspace{-0.5cm} \chi_\lambda(\vec{x},n,k)&=& \left( \begin{array}{c} \sin \theta \sin (k\theta) \frac{d v_{nk}(r)}{d r} + \frac{k}{r} \, v_{nk}(r) \cos \theta \cos(k\theta) \\ -\cos \theta \sin (k\theta) \frac{d v_{nk}(r)}{d r} + \frac{k}{r} \, v_{nk}(r) \, \sin \theta \cos(k\theta) \\ f_n(r) \, v_{nk}(r)\, \cos(n\theta)\, \sin(k\theta) \\ f_n(r)\, v_{nk}(r)\, \sin(n\theta) \,\sin(k\theta)
 \end{array} \right) \hspace{0.0cm}\,, \hspace{0.3cm} k=1,2,\dots \, ,\hspace{0.2cm} \label{bpsexcitedmode2}
\end{eqnarray}}
\hspace{-0.175cm}where the radial form factors $v_{nk}(r)$ can be determined as a solution of the 1D Sturm-Liouville problem
\begin{equation}
-\frac{d^2 v_{nk}(r)}{d r^2} -\frac{1}{r} \frac{d v_{nk}(r)}{d r} + \Big[ f_n^2(r)-\omega_\lambda^2 + \frac{k^2}{r^2} \Big] v_{nk}(r)=0  .
\label{ode55}
\end{equation}
For $k\neq 0$ the eigenfunctions $\xi_\lambda(\vec{x},n,k)$ and $\chi_\lambda(\vec{x},n,k)$ are degenerate, although linearly independent. The number of shape modes is non trivially related to the vortex charge $n$ (contrary to the zero modes) and depends on the strength of the potential well arising in the spectral problem (\ref{ode55}).

Since $v_{nk}(r) = v_0^{(n,k)} r^k + v_2^{(n,k)} r^{k+2} + \dots$ near $r=0$, the angular characteristics of the shape modes (\ref{bpsexcitedmode1}) and (\ref{bpsexcitedmode2}) become
\begin{equation}
{\mathrm{ch}}_\theta[\xi(\vec{x},n,k)] = {\mathrm{ch}}_\theta[\chi(\vec{x},n,k)] = {k-1;k+1\choose n-k;n+k} \hspace{0.5cm} \mbox{for} \hspace{0.5cm} k=1,2,\dots, \label{angchbpsshapemode}
\end{equation}
while the radial characteristics are
\begin{equation}
{\mathrm{ch}}_r[\xi(\vec{x},n,k)] = {\mathrm{ch}}_r[\chi(\vec{x},n,k)] = {k-1\choose n+k}  \hspace{0.5cm} \mbox{for} \hspace{0.5cm} k=1,2,\dots.  \label{radchbpsshapemode}
\end{equation}
The expression (\ref{angchbpsshapemode}) implies that the eigenfunctions describing these shape modes exhibit a more complex angular dependence than the zero modes, which can be described by the value of the angular characteristic number, $\# {\mathrm{ch}}_\theta[\xi(\vec{x},n,k)]  = \# {\mathrm{ch}}_\theta[\chi(\vec{x},n,k)] = 2$. In general, the vector and scalar contribution of the shape modes has  angular dependence of the forms $(k\pm 1)\theta$ and $(n\pm k)\theta$ respectively.

The double angular dependence is broken for $k=0$, where we find 
\begin{equation}
{\mathrm{ch}}_\theta[\xi(\vec{x},n,0)]  = {1\choose n} \hspace{0.5cm}\mbox{and} \hspace{0.5cm} {\mathrm{ch}}_r [\xi(\vec{x},n,0)] = {1\choose n}. \label{chbpsshapemode0}
\end{equation}
In this case we recover the dependence on only one angle for the components of the eigenfunction, as in the case of the zero modes. It is worth mentioning that for $k=0$ the angular characteristics (\ref{chbpsshapemode0}) of the shape mode (\ref{bpsexcitedmode1}) with $k=0$ coincides with that of the charge-$n$ vortex (\ref{chvortex}). This implies in particular that the $k=0$ modes are the so-called Derrick-type modes, i.e., the $k=0$ modes only affect the vortex size but not its symmetry. Note that this is the only shape mode that is not doubly degenerate, see (\ref{bpsexcitedmode1}) and (\ref{bpsexcitedmode2}).
\end{itemize}

In summary, the spectral structure for rotationally invariant $n$-vortices in the self-dual case comprises a non-degenerate Derrick-type mode with positive eigenvalue (responsible for size oscillations of the $n$-vortex); $2 \vert n \vert$ zero modes (which changes the vortex centers) and a charge-dependent number of doubly degenerate shape modes with positive eigenvalues (describing normal modes with non-trivial multipolar structure with respect to the $n$-vortex solution). 
%The information obtained in the self-dual case can be used to contrast the results in the non-BPS case, meaning that we should be able to reproduce these outcomes when $\lambda=1$ in our general analysis. 

%%%%%%%%%%%%%%%%%%%%%%%%
%%%%%%%%%%%%%%%%%%%%%%%%
%%%%%%%%%%%%%%%%%%%%%%%%
%%%%%%%%%%%%%%%%%%%%%%%%
%%%%%%%%%%%%%%%%%%%%%%%%
\section{The spectral structure for the fluctuations around non-self-dual vortices}\label{corepaperinternalmodes}

\label{non-bps}

In this section we aim to describe the spectral structure of the non-BPS vortices. At the same time, we seek to develop a unified understanding of the different types of normal modes of vibration analyzed in Section \ref{bps}. The first step is to identify the angular dependence of the eigenfunctions, that is, their angular characteristics. One could guess that beyond the self-dual case the angular dependence of the eigenfunctions would become more complex than in the case of BPS vortices. However, as we show in the following lemma, the angular characteristics number is still 2:

%\vspace{0.1cm}

%\noindent \textbf{Lemma:}
\begin{lemma}\label{lemma6.2}
 The spectral problem (\ref{spectralproblem}) associated to the non self-dual rotationally invariant $n$-vortex admits Derrick-type eigenfunctions together with generic eigenmodes $\xi_\nu(\vec{x})$ whose angular characteristics are
\begin{equation}
{\mathrm{ch}}_\theta [\xi_\nu(\vec{x},n)] = {\overline{k}-1 \,;\,\overline{k}+1 \choose n-\overline{k}\,;\,n+\overline{k}} \hspace{0.5cm} \mbox{for} \hspace{0.5cm} \overline{k}=1,2,\dots \label{angthnonbpsvortex}
\end{equation}
\end{lemma}

\begin{proof}
    The form of (\ref{angthnonbpsvortex}) indicates that the expression of the vector and scalar components of the ${\cal H}^+$-eigenfunctions depends only on two different angles. For this reason, it is enough to show that the ansatz 
\begin{equation}
    \xi_\nu(r,\theta) = \left( \begin{array}{c} \overline{g}_1(r) \, \sin (a \,\theta)  + \overline{g}_2(r) \sin (b \,\theta)  \\ \overline{g}_1(r) \, \cos (a \,\theta)  + \overline{g}_2(r) \cos (b \,\theta)  \\  \overline{t}_1(r) \, \cos (c \,\theta)  + \overline{t}_2(r) \cos (d \,\theta) \\  \overline{t}_1(r) \, \sin (c \,\theta)  + \overline{t}_2(r) \sin (d \,\theta) \end{array} \right)\,, \label{genericform01}
\end{equation}
is compatible with the spectral equation ${\cal H}^+ \xi_\nu = \omega_\nu^2 \xi_\nu$ associated to the operator (\ref{hessianoperator}) for some choice of the parameters $a,b,c,d$ in (\ref{genericform01}). Because of the continuity of the eigenfunctions these parameters must be integers, that is, $a,b,c,d \in \mathbb{Z}$. If we substitute (\ref{genericform01}) into (\ref{spectralproblem}) the angular characteristics of both sides of the spectral equation can be obtained. The left-hand side verifies that
\[
{\mathrm{ch}}_\theta [{\cal H}^+ \xi_\nu] = { a \, ; \, b \,  ; \, 1+c-n\, ; \, 1-c+n\, ;\, 1+d-n\, ;\, 1-d+n \choose c\, ;\, d\, ;\, 1+a-n\, ;\, 1+a+n\, ;\, 1+b-n\, ;\, 1+b+n\, ;\, c-2n\, ;\, d-2n}\,,
\]
while the right-hand side leads to
\[
{\mathrm{ch}}_\theta [\omega_\nu^2 \xi_\nu] = { a\, ;\,  b \choose c\, ;\, d}\,.
\]
This means that we end up with four equations to determine the radial profiles $\overline{g}_i(r)$ and $\overline{t}_i(r)$ with $i=1,2$, which involve sums of trigonometric functions with different angular dependencies. Specifically, the first and second equation include the angles $a\theta$, $b\theta$, $(1+c-n)\theta$, $(1-c+n)\theta$, $(1+d-n)\theta$, $(1-d+n)\theta$ while the third and the fourth ones involve the angles $c\theta$, $d\theta$, $(1+a-n)\theta$, $(1+a+n)\theta$, $(1+b-n)\theta$, $(1+b+n)\theta$, $(c-2n\theta)$ and $(d-2n)\theta$. Thus, $\#{\mathrm{ch}}_\theta [{\cal H}^+ \xi_\nu]=8$. Fortunately, there is a particular choice of the parameters
\begin{equation}
b=-2-a \hspace{0.0cm},\hspace{0.5cm} c=n-1-a  \hspace{0.0cm},\hspace{0.5cm} d= 1+n+a ,\label{constraint01}
\end{equation}
which reduces these dependencies to only two independent trigonometric functions for each equation, leading then to $\#{\mathrm{ch}}_\theta [{\cal H}^+ \xi_\nu]=2$ \footnote{Indeed, the choice of parameters  $b=-2-a ,\, d=n-1-a, \,c= 1+n+a$ is also a valid choice to solve the spectral problem, but leads to the same results as the choice \eqref{constraint01}.}. For the sake of simplicity, it is convenient to choose $a=\overline{k}-1$, such that the relation (\ref{constraint01}) implies that the fluctuations (\ref{genericform01}) follows the form
\begin{equation}
    \xi_\nu(\vec{x},n,\overline{k}) = \left( \begin{array}{c} -\overline{g}_1(r) \, \sin [(1-\overline{k}) \,\theta]  - \overline{g}_2(r) \sin [(1+\overline{k}) \,\theta]   \\  \overline{g}_1(r) \, \cos [(1-\overline{k}) \,\theta]  + \overline{g}_2(r) \cos [(1+\overline{k}) \,\theta]  \\  \overline{t}_1(r) \, \cos [(n-\overline{k}) \,\theta]  + \overline{t}_2(r) \cos [(n+\overline{k}) \,\theta]\\ \overline{t}_1(r) \, \sin [(n-\overline{k}) \,\theta]  + \overline{t}_2(r) \sin [(n+\overline{k}) \,\theta] \end{array} \right) \label{genericform02}\,.
\end{equation}
In this way, it is direct to prove that these eigenfunctions exhibit the symmetry $\overline{k} \rightarrow - \overline{k}$, in such a way that we can restrict our study to the case $\overline{k} \geq 0$. This justifies the relation (\ref{angthnonbpsvortex}) introduced in the Lemma \ref{lemma6.2}. The expression (\ref{genericform02}) when plugged into the spectral equation leads to four ordinary differential equations for the four radial functions $\overline{g}_i(r)$ and $\overline{t}_i(r)$ with $i=1,2$. The explicit form of these differential equations will be shown below. Recall that (\ref{genericform02}) provides the admissible eigenfunctions, but new requirements will be imposed on them, for example, to guarantee the regularity conditions at the origin or the fulfillment of the \textit{background gauge condition} in order to remove non-physical degrees of freedom. 

Note also that $\overline{k}=0$ is a special case, because the two trigonometric functions that enter each of the fluctuation components of (\ref{genericform02}) are equal, and the angular characteristic number is reduced to 1, that is, $\# {\mathrm{ch}}_\theta[\xi_\nu(\vec{x},n,0) ] = 1$. In this case, the eigenfunction (\ref{genericform02}) becomes 
\begin{equation}
    \xi_\nu(\vec{x},n,0) = \left( \begin{array}{c} -\widetilde{g}(r) \, \sin [\theta] \\  \widetilde{g}(r) \, \cos [\theta]  \\  \widetilde{t}(r) \, \cos [n \,\theta] \\ \widetilde{t}(r) \, \sin [n \,\theta] \end{array} \right)\,, \label{genericform03}
\end{equation}
where
$$
\widetilde{g}(r)=\overline{g}_1(r)+ \overline{g}_2(r), \qquad \widetilde{t}(r)=t_1(r)+t_2(r).
$$
These eigenfunctions preserve the angular form of $n$-vortices, which means that they are Derrick-type modes.

The same argument employed in this proof can be applied to a different choice of (\ref{genericform01}), which leads to the generic eigenfunctions
\begin{equation}
    \chi_\nu(\vec{x},n,\overline{k}) = \left( \begin{array}{c} -\overline{g}_1(r) \, \cos [(1-\overline{k}) \,\theta] + \overline{g}_2(r) \cos [(1+\overline{k}) \,\theta]   \\  -\overline{g}_1(r) \, \sin [(1-\overline{k}) \,\theta]  + \overline{g}_2(r) \sin [(1+\overline{k}) \,\theta]  \\  -\overline{t}_2(r) \, \sin [(n-\overline{k}) \,\theta]  + \overline{t}_1(r) \sin [(n+\overline{k}) \,\theta]\\ -\overline{t}_2(r) \, \cos [(n-\overline{k}) \,\theta]  - \overline{t}_1(r) \cos [(n+\overline{k}) \,\theta] \end{array} \right)\,, \label{genericform02b}
\end{equation}
which are orthogonal to (\ref{genericform02}). The radial functions $\overline{g}_i(r)$ and $\overline{t}_i(r)$ with $i=1,2$ in (\ref{genericform02b}) verify the same differential equations as those included in (\ref{genericform02}).
\end{proof}
%\noindent \textbf{Proof:} 

\vspace{0.2cm}

One important remark is that the generic form (\ref{genericform02}) and  (\ref{genericform02b}) unifies the two different types of eigenmodes described in the self-dual regime in Section \ref{bps}: 
\begin{enumerate}
\item The zero modes follow the pattern (\ref{genericform02}) (or (\ref{genericform02b})) by taking $\overline{k}=n-k$. With this choice, the angular momentum parameter $\overline{k}$ used in (\ref{genericform02}) ranges the values $\overline{k}=1,2,\dots,n-1,n$. Notice that the zero modes in the self-dual case have an \textit{angular characteristics number} equals to 1, $\#{\mathrm{ch}}_\theta[\xi_0(\vec{x})]=1$, but this is no longer possible for the general case. These modes acquire a new angular dependence when $\lambda\neq 1$.

\item On the other hand, the shape modes are recovered by simply taking $\overline{k}=k$ with $\overline{k}=0,1,\dots$, with $\overline{k}=0$ designating the Derrick-type mode. Comparing  (\ref{genericform02}) with (\ref{bpsexcitedmode1}) the relation between the radial functions is given by $\overline{g}_1(r) = -\frac{1}{2} (v_{nk}'(r) + \frac{k v_{nk}(r)}{r})$, $\overline{g}_2(r)= -\frac{1}{2} (v_{nk}'(r) - \frac{k v_{nk}(r)}{r})$ and $\overline{t}_1(r)=\overline{t}_2(r)=\frac{1}{2} f_n(r) v_{nk}(r)$.
\end{enumerate}

As mentioned previously, by substituting the expressions (\ref{genericform02}) or (\ref{genericform03}) into the general problem (\ref{spectralproblem}) we end up with four second-order ordinary differential equations for the radial profile functions $\overline{g}_i(r)$ and $\overline{t}_i(r)$, $i=1,2$. We distinguish two different cases, $\overline{k}=0$ and $\overline{k}>0$:

\vspace{0.2cm}

\noindent \textsc{Derrick-type modes:} In the first case, $\overline{k}=0$, the reduced 1D eigenvalue problem is given by
{\begin{eqnarray}
	&&\hspace{-0.75cm} - \frac{d^2 \widetilde{g}(r)}{dr^2} - \frac{1}{r}\frac{d\widetilde{g}(r)}{dr} + \Big( \frac{1}{r^2} + f_n^2(r) - \omega_n^2 \Big) \widetilde{g}(r) - \frac{2n}{r} (1-\beta_n(r)) f_n(r) \widetilde{t}(r) =0 ,\label{radialedo01} \\
	&& \hspace{-0.75cm}- \frac{d^2 \widetilde{t}(r)}{dr^2} - \frac{1}{r} \frac{d\widetilde{t}(r)}{dr} + \Big( \frac{n^2 (1-\beta_n(r))^2}{r^2} + \frac{3 \lambda}{2} f_n^2(r) - \frac{\lambda}{2} - \omega_n^2 \Big) \widetilde{t}(r) - \frac{2n}{r} (1-\beta_n(r)) f_n(r) \widetilde{g}(r) =0 .\nonumber
\end{eqnarray}
}

%\vspace{0.2cm}

\noindent \textsc{Multipolar modes:} In the general case, $\overline{k}>0$ the radial profiles of the eigenfunctions (\ref{genericform02}) are determined by
{\footnotesize \begin{eqnarray}
	&&\hspace{-0.75cm}- \frac{d^2 \overline{g}_1(r)}{dr^2} - \frac{1}{r} \frac{d\overline{g}_1(r)}{dr}+ \Big[ \frac{(\overline{k}-1)^2}{r^2} +f_n^2(r) \Big] \overline{g}_1(r) - \Big[ \frac{n (1-\beta_n(r))f_n(r)}{r} + f_n'(r) \Big] \overline{t}_1(r) + \nonumber\\
	&&  \hspace{0.5cm}\hspace{-0.75cm} + \Big[ - \frac{n(1-\beta_n(r))f_n(r)}{r} + f_n'(r) \Big] \overline{t}_2(r) = \omega_n^2 \overline{g}_1(r) \hspace{0.0cm}, \nonumber \\
	&&- \frac{d^2 \overline{g}_2(r)}{dr^2} - \frac{1}{r} \frac{d\overline{g}_2(r)}{dr}+ \Big[ \frac{(\overline{k}+1)^2}{r^2} +f_n^2(r) \Big] \overline{g}_2(r) + \Big[ -\frac{n (1-\beta_n(r))f_n(r)}{r} + f_n'(r) \Big] \overline{t}_1(r) - \nonumber \\
	&&  \hspace{0.5cm}\hspace{-0.75cm} - \Big[ \frac{n(1-\beta_n(r))f_n(r)}{r} + f_n'(r) \Big] \overline{t}_2(r) = \omega_n^2 \overline{g}_2(r) \hspace{0.0cm}, \label{radialedo02} \\
	&&\hspace{-0.75cm}- \frac{d^2 \overline{t}_1(r)}{dr^2} - \frac{1}{r} \frac{d\overline{t}_1(r)}{dr}+ \Big[ \frac{(\overline{k}-n(1-\beta_n(r)))^2}{r^2} -\frac{\lambda}{2} + (\lambda+\frac{1}{2})f_n^2(r) \Big] \overline{t}_1(r) - \Big[ \frac{n (1-\beta_n(r))f_n(r)}{r} + f_n'(r) \Big] \overline{g}_1(r) + \nonumber \\
	&&  \hspace{-0.75cm}\hspace{0.5cm} + \Big[ - \frac{n(1-\beta_n(r))f_n(r)}{r} + f_n'(r) \Big] \overline{g}_2(r) + \frac{1}{2} (\lambda -1) f_n^2(r) \overline{t}_2(r) = \omega_n^2 \overline{t}_1 (r)\hspace{0.0cm}, \nonumber \\
	&&\hspace{-0.75cm}- \frac{d^2 \overline{t}_2(r)}{dr^2} - \frac{1}{r} \frac{d\overline{t}_2(r)}{dr}+ \Big[ \frac{(\overline{k}+n(1-\beta_n(r)))^2}{r^2} -\frac{\lambda}{2} + (\lambda+\frac{1}{2})f_n^2(r) \Big] \overline{t}_2(r) + \Big[- \frac{n (1-\beta_n(r))f_n(r)}{r} + f_n'(r) \Big] \overline{g}_1(r) - \nonumber \\
	&&  \hspace{-0.75cm}\hspace{0.5cm} - \Big[ \frac{n(1-\beta_n(r))f_n(r)}{r} + f_n'(r) \Big] \overline{g}_2(r) + \frac{1}{2} (\lambda -1) f_n^2(r) \overline{t}_1(r) = \omega_n^2 \overline{t}_2(r) \hspace{0.0cm}. \nonumber
\end{eqnarray}}

At this point, we restrict the admissible eigenfunctions (\ref{genericform02}) to those satisfying the \textit{background gauge condition}, removing all the non-physical degrees of freedom. This is a very delicate point because it may add a new condition on the radial spectral problems (\ref{radialedo01}) and (\ref{radialedo02}), which could lead to inconsistencies. Here, it is not the case as we prove in the following proposition:

%\vspace{0.2cm}

\begin{proposition}
    
 The admissible eigenfunctions of the fluctuation spectral problem (\ref{spectralproblem}) associated with the non self-dual rotationally invariant $n$-vortices without superfluous gauge degrees of freedom are characterized as:

\vspace{0.1cm}
\begin{enumerate}
\item The Derrick-type modes arise in the form
\begin{equation}
	\xi_\nu(\vec{x},n,0) = \left( \begin{array}{c} \widetilde{v}(r) \, \sin \theta  \\ - \widetilde{v}(r) \,\cos \theta  \\ \widetilde{u}(r) \, \cos(n\theta) \\ \widetilde{u}(r) \, \sin(n\theta) 
	\end{array} \right)\,,  \label{genericform03b}
\end{equation}
where the radial functions $\zeta_\nu (r)= (\widetilde{v}(r),\widetilde{u}(r) )$ satisfy the reduced spectral problem $\overline{\cal H}_0 \zeta_\nu (x) = \omega_\nu^2 \zeta_\nu(x)$ where
\begin{equation}
\overline{\cal H}_0  = \left( \begin{array}{cc} - \frac{d^2}{dr^2} - \frac{1}{r} \frac{d}{dr} + \frac{1}{r^2} + f_n^2(r)  &  \frac{2n}{r} (1-\beta_n(r)) f_n(r) \\ \frac{2n}{r} (1-\beta_n(r)) f_n(r) & - \frac{d^2}{dr^2}- \frac{1}{r} \frac{d}{dr} + \frac{n^2 (1-\beta_n(r))^2}{r^2} + \frac{3 \lambda}{2} f_n^2(r) - \frac{\lambda}{2} 
\end{array} \right)\,. \label{operatorhbar0}
\end{equation}

\item The multipolar eigenfunctions have the form
{\small
\begin{equation}
\xi_\nu(\vec{x},n,\overline{k}) = \left( \begin{array}{c} \cos(\overline{k} \theta) \sin\theta [v'(r) -r f_n(r) w(r)] - \sin (\overline{k}\theta) \cos \theta \frac{\overline{k} v(r)}{r}  \\
- \cos(\overline{k} \theta) \cos\theta [v'(r) -r f_n(r) w(r)] - \sin (\overline{k}\theta) \sin \theta \frac{\overline{k} v(r)}{r} \\
\cos(\overline{k}\theta)\cos(n\theta) u(r) + \overline{k} \sin(\overline{k}\theta) \sin(n\theta) w(r) \\
\cos(\overline{k}\theta)\sin(n\theta) u(r) - \overline{k} \sin(\overline{k}\theta) \cos(n\theta) w(r) \end{array} \right)\,,\label{genericform04b}  
\end{equation}}
\hspace{-0.13cm}where $ \overline{k}=1,2,\dots $ and the radial functions $\zeta_\nu(r)= (v(r),u(r),w(r))^t$ are determined by the reduced eigenvalue problem $\overline{\cal H} \, \zeta_\nu(x)= \omega_\nu^2 \, \zeta_\nu(x)$, 
where 
{\scriptsize 
\begin{equation}
\overline{\cal H}_k =\left( \begin{array}{ccc} - \frac{d^2}{dr^2} - \frac{1}{r} \frac{d}{dr}+  \frac{\overline{k}^2}{r^2} + f_n^2(r) & 0 & 2(f_n(r) +r f_n'(r)) \\  
\frac{2n(1-\beta_n(r))f_n(r)}{r} \frac{d}{dr} &  \begin{array}{c} - \frac{d^2}{dr^2} - \frac{1}{r} \frac{d}{dr}+ \frac{\overline{k}^2}{r^2} + \\ + \frac{n^2(1-\beta_n(r))^2}{r^2} + \frac{3\lambda}{2} f_n^2(r) - \frac{\lambda}{2} \end{array} & - \frac{2n(1-\beta_n(r))(\overline{k}^2 + r^2 f_n^2(r))}{r^2} \\
\frac{2f_n'(r)}{r} & - \frac{2n(1-\beta_n(r))}{r^2}  & \begin{array}{c} - \frac{d^2}{dr^2} - \frac{1}{r} \frac{d}{dr}+ \frac{\overline{k}^2}{r^2} + \\ + \frac{n^2(1-\beta_n(r))^2}{r^2} + f_n^2(r) + \frac{\lambda}{2} f_n^2(r) - \frac{\lambda}{2} \end{array} \end{array} \right) \,.\label{hbarrak}
\end{equation}
}

For $\overline{k}\geq 1$, a second type of eigenfunctions $\chi_\nu(\vec{x},n,\overline{k})$ (orthogonal to (\ref{genericform04b})) can be constructed simply by making the changes $\cos(\overline{k}\theta)\rightarrow \sin(\overline{k}\theta)$ and $\sin(\overline{k}\theta)\rightarrow -\cos(\overline{k}\theta)$ in (\ref{genericform04b}), which satisfies the same spectral problem. Therefore, eigenvalues with $\overline{k}\geq 1$ are doubly degenerate.
\end{enumerate}
%\vspace{0.1cm}
\end{proposition}
\begin{proof}
We need to restrict the previous admissible eigenfunctions (\ref{genericform02}) and (\ref{genericform03}) such that they satisfy the background gauge conditions (\ref{backgroundgauge}) and check that no inconsistencies are introduced in the equations.

In the case of Derrick-type modes, we notice that (\ref{genericform03b}) follows the same form as (\ref{genericform03}) simply changing $v(r)=-\widetilde{g}(r)$ and $u(r)=\widetilde{t}(r)$. This is done to get a better fitting between this situation and the general situation, explained below. In this case, the background gauge condition is automatically satisfied and the differential equations (\ref{radialedo01}) read now
{
\begin{eqnarray}
	\hspace{-0.77cm}&& - \widetilde{v}''(r) - \frac{1}{r} \widetilde{v}'(r) + \Big( \frac{1}{r^2} + f_n^2(r) - \omega_n^2 \Big) \widetilde{v}(r) + \frac{2n}{r} (1-\beta_n(r)) f_n(r) \widetilde{u}(r) =0, \label{edoDerrick} \\
	\hspace{-0.77cm}&& - \widetilde{u}''(r) - \frac{1}{r} \widetilde{u}'(r) + \Big( \frac{n^2 (1-\beta_n(r))^2}{r^2} + \frac{3 \lambda}{2} f_n^2(r) - \frac{\lambda}{2} - \omega_n^2 \Big) \widetilde{u}(r) + \frac{2n}{r} (1-\beta_n(r)) f_n(r) \widetilde{v}(r) =0, \nonumber
\end{eqnarray}}
which define the spectral problem associated to the operator (\ref{operatorhbar0}). 

For the so called multipolar eigenfunctions, the situation is more complex. It is convenient to define the functions
\[
\overline{g}_+= -\overline{g}_1+\overline{g}_2 \hspace{0.0cm},\hspace{0.5cm}\overline{g}_-= -\overline{g}_1-\overline{g}_2 \hspace{0.cm},\hspace{0.5cm} \overline{t}_+= \overline{t}_1+\overline{t}_2 \hspace{0.cm},\hspace{0.5cm} \overline{t}_-= \overline{t}_1-\overline{t}_2,
\]
such that the equations (\ref{radialedo02}) now become
{\small
\begin{eqnarray}
	&& - \frac{d^2 \overline{g}_+(r)}{dr^2} - \frac{1}{r} \frac{d\overline{g}_+(r)}{dr}+ \Big[ \frac{\overline{k}^2+1}{r^2} +f_n^2(r) \Big] \overline{g}_+(r) - 
	\frac{2 \overline{k}}{r^2} \overline{g}_- (r)+2 f_n'(r) \, \overline{t}_-(r) = \omega_n^2 \overline{g}_+(r), \nonumber \\
	&&\hspace{-0.77cm} - \frac{d^2 \overline{g}_-(r)}{dr^2} - \frac{1}{r} \frac{d\overline{g}_-(r)}{dr}+ \Big[ \frac{\overline{k}^2+1}{r^2} +f_n^2(r) \Big] \overline{g}_-(r) - 
	\frac{2 \overline{k}}{r^2} \overline{g}_+ +\frac{2 n (1-\beta_n(r)) f_n(r)}{r} \overline{t}_+(r) = \omega_n^2 \overline{g}_-(r),\nonumber \\
	&& \hspace{-0.77cm}- \frac{d^2 \overline{t}_+(r)}{dr^2} - \frac{1}{r} \frac{d\overline{t}_+(r)}{dr}+ \Big[ \frac{\overline{k}^2+n^2(1-\beta_n(r))^2}{r^2} + \frac{3\lambda}{2} f_n^2(r) - \frac{\lambda}{2} \Big] \overline{t}_+(r) - 
	\frac{2 n \overline{k} (1-\beta_n(r))}{r^2} \overline{t}_- (r)\nonumber \\
	&&  \hspace{0.5cm} + \frac{2n(1-\beta_n(r))f_n(r)}{r} \, \overline{g}_-(r) = \omega_n^2 \overline{t}_+ (r),\label{radialedo03} \\
	&&\hspace{-0.77cm} - \frac{d^2 \overline{t}_-(r)}{dr^2} - \frac{1}{r} \frac{d\overline{t}_-(r)}{dr}+ \Big[ \frac{\overline{k}^2+n^2(1-\beta_n(r))^2}{r^2} + f_n^2(r)+ \frac{\lambda}{2} f_n^2(r) - \frac{\lambda}{2} \Big] \overline{t}_-(r) - 
	\frac{2 n \overline{k} (1-\beta_n(r))}{r^2} \overline{t}_+ (r)\nonumber \\
	&& \hspace{-0.77cm} \hspace{0.5cm} + 2 f_n'(r) \, \overline{g}_+(r) = \omega_n^2 \overline{t}_-(r). \nonumber
\end{eqnarray}
}
The introduction of the functions $\overline{g}_\pm(r)$ and $\overline{t}_\pm(r)$ is motivated because now the background gauge condition (\ref{backgroundgauge}) is simply expressed as
\[
\frac{\partial \overline{g}_+(r)}{\partial r} + \frac{1}{r} \overline{g}_+(r) - \frac{\overline{k}}{r} \overline{g}_-(r) - f_n(r)\, \overline{t}_- (r)=0,
\]
which allows us to solve for $\overline{g}_-$
\begin{equation}
\overline{g}_-(r)=\frac{r}{\overline{k}} \Big( \frac{\partial \overline{g}_+(r)}{\partial r} + \frac{1}{r} \overline{g}_+(r) - f_n(r) \, \overline{t}_-(r) \Big). \label{gmenos}
\end{equation}
If (\ref{gmenos}) is substituted into (\ref{genericform02}) it is easy to see that the eigenfunctions of this class have the form
\begin{eqnarray}
	a_1(r,\theta) \!\! \!\!&=& \!\! \!\! \cos(\overline{k} \theta) \sin(\theta) [v'(r) -r f_n(r) w(r)] - \sin (\overline{k}\theta) \cos \theta \frac{\overline{k}\, v(r)}{r}, \nonumber  \\
	a_2(r,\theta) \!\! \!\!&=& \!\! \!\!- \cos(\overline{k} \theta) \cos(\theta) [v'(r) -r f_n(r) w(r)] - \sin (\overline{k}\theta) \sin \theta \frac{\overline{k}\, v(r)}{r} ,\nonumber \\
	\varphi_1(r,\theta) \!\! \!\!&=& \!\! \!\!\cos(\overline{k}\theta)\cos(n\theta) u(r) + \overline{k} \sin(\overline{k}\theta) \sin(n\theta) w(r), \label{genericform04c}  \\
	\varphi_2(r,\theta) \!\! \!\!&=& \!\! \!\!\cos(\overline{k}\theta)\sin(n\theta) u(r) - \overline{k} \sin(\overline{k}\theta) \cos(n\theta) w(r), \nonumber
\end{eqnarray}
where we have introduced $\overline{g}_+(r)= \frac{\overline{k}}{r} v(r)$, $\overline{t}_+(r) = u(r)$ and $\overline{t}_-(r)=\overline{k} \,w(r)$. Notice that by construction (\ref{genericform04c}), automatically satisfies  the gauge condition (\ref{backgroundgauge}). It remains to verify the consistency of the procedure. The first, third and fourth equations in (\ref{radialedo03}) provide the relations
{\small
\begin{eqnarray}
	&&\hspace{-0.75cm} - \frac{d^2 v(r)}{dr^2} - \frac{1}{r} \frac{dv(r)}{dr}+ \Big[ \frac{\overline{k}^2}{r^2} + f_n^2(r) \Big] v(r) + 2(f_n(r) +r f_n'(r)) w(r) = \omega_n^2 v(r), \nonumber \\
	&& \hspace{-0.75cm}- \frac{d^2 u(r)}{dr^2} - \frac{1}{r} \frac{du(r)}{dr}+ \Big[ \frac{\overline{k}^2}{r^2} + \frac{n^2(1-\beta_n(r))^2}{r^2} + \frac{3\lambda}{2} f_n^2(r) - \frac{\lambda}{2} \Big] u(r)- \frac{2n(1-\beta_n(r))(\overline{k}^2 + r^2 f_n^2(r))}{r^2} w(r) \nonumber\\
	&&\hspace{-0.75cm} \hspace{0.5cm}+ \frac{2n(1-\beta_n(r))f_n(r)}{r} \frac{dv(r)}{dr} = \omega_n^2 u(r), \label{genericform04d} \\
	&& \hspace{-0.75cm}- \frac{d^2 w(r)}{dr^2} - \frac{1}{r} \frac{dw(r)}{dr}+ \Big[ \frac{\overline{k}^2}{r^2} + \frac{n^2(1-\beta_n(r))^2}{r^2} + f_n^2(r) + \frac{\lambda}{2} f_n^2(r) - \frac{\lambda}{2} \Big] w(r)- \frac{2n(1-\beta_n(r))}{r^2} u(r) + \nonumber\\
	&&\hspace{-0.75cm} \hspace{0.5cm}+ \frac{2f_n'(r)}{r} v(r) = \omega_n^2 w(r), \nonumber
\end{eqnarray}}
\hspace{-0.17cm}which can be used to determine the profiles of the three radial functions $v(r)$, $u(r)$ and $w(r)$. This establishes the spectral problem for the operator $\overline{\mathcal{H}}_k$, as stated in the proposition. Notably, the second equation derived from (\ref{radialedo03}) corresponds to a linear combination of the remaining equations. Specifically, it can be expressed as the sum of $r/\overline{k}$ times the derivative of the first equation in (\ref{genericform04d}), $1/\overline{k}$ times the first equation itself, and $-r f_n(r)/\overline{k}$ times the fourth equation. Note that, in this calculation, the differential equation (\ref{eqI3:ProfEq1}), which governs the scalar radial profile of the vortex, must be employed. This ensures the consistency of the procedure. 
\end{proof}
%\vspace{0.2cm}
\vspace{-0.2cm}
Finally, all the previous information regarding the angular dependence of the eigenfunctions under the background gauge must be complemented by an analysis of their radial dependence at the origin and at infinity. This is necessary to determine, for instance, possible restrictions on the value of the parameter $\overline{k}$. This analysis is summarized in the following result:

%\vspace{0.2cm}

\begin{theorem}
    
The spectrum of the eigenvalue problem (\ref{spectralproblem}) associated with the fluctuation operator (\ref{hessianoperator}) for non self-dual rotationally invariant $n$-vortices comprises three types of eigenfunctions: 
\begin{enumerate}
    \item \textit{Derrick-type modes:} These modes are characterized by the form (\ref{genericform03b}), which depends on two radial functions $\widetilde{v}(r)$ and $\widetilde{u}(r)$. These functions are determined by the reduced spectral problem associated to the 1D $2\times 2$ matrix differential operator (\ref{operatorhbar0}). The number of these modes is only restricted by the strength of the potential well arising in the spectral problem.

    \item \textit{Multipolar modes:} They follow the expression (\ref{genericform04b}) and involve three radial functions $v(r)$, $u(r)$ and $w(r)$, which correspond to the eigenfunctions of the 1D $3\times 3$ matrix differential operator (\ref{hbarrak}). There are two classes of the multipolar modes:
    \begin{enumerate}
        \item \textit{Type A:} There are eigenfunctions $\xi_\nu^{\mathrm{ A}}(\vec{x},n,\overline{k})$ with radial characteristic
\[
{\mathrm{ch}}_r[\xi_\nu^{\mathrm{ A}}(\vec{x},n,\overline{k})] = {\overline{k}-1 \choose n-\overline{k}}\,,
\]
which implies that the vector and scalar fields has opposite dependence on the parameter $\overline{k}$ in the power series with respect to $r$ near the origin. As a consequence, the number of eigenfunctions of this type are reduced to the range $\overline{k}\in \{1, \dots, n \}$.

{ \item \textit{Type B:} The second class of multipolar eigenfunctions $\xi_\nu^{\mathrm{ B}}(\vec{x},n,\overline{k})$ has radial characteristic
\[
{\mathrm{ch}}_r[\xi_\nu^{\mathrm{ B}}(\vec{x},n,\overline{k})] = {\overline{k}-1 \choose n+\overline{k}}\,,
\]
such that the vector and scalar fields exhibit the same powers of $r$ with respect to the parameter $\overline{k}$ in the expansion series near the origin. Now, the number of the discrete modes is only fixed by the potential well strength of the corresponding eigenvalue problem.}

    \end{enumerate}
\end{enumerate}

\end{theorem}

%\vspace{0.1cm}

%\noindent \textbf{Proof:}
\begin{proof}
It will be shown that the shape modes following the expressions (\ref{genericform03b}) and (\ref{genericform04b}) are square integrable eigenfunctions, $\xi_\nu (\vec{x})\in L^2(\mathbb{R}^2)\oplus \mathbb{R}^4$ of the operators (\ref{operatorhbar0}) and (\ref{hbarrak}). It only remains to show that the norm of these modes is well-defined. We distinguish two different scenarios:

\vspace{0.2cm}

\noindent 1. For Derrick-type modes the norm of the eigenfunction is written as
\begin{equation}
\| \xi(\vec{x},n,0)\|^2 = 2 \pi \int_0^\infty r [\widetilde{v}^2(r) + \widetilde{u}^2(r) ] \, dr ,\label{norma0}
\end{equation}
in terms of the radial functions $\widetilde{v}(r)$ and $\widetilde{u}(r)$. These functions satisfy the ordinary differential equations (\ref{edoDerrick}). In order to have $\| \xi(\vec{x},n,0)\|< \infty$  regularity conditions must be imposed on the functions $\widetilde{v}(r)$ and $\widetilde{u}(r)$ together with a fast enough decay when $r\rightarrow \infty$. First, we shall study the behavior of these functions in the neighborhood of $r=0$. In order to do that, we plug the expansion 
\begin{equation}
\widetilde{v}(r) = r^s \sum_{i=0}^\infty \widetilde{v}_i \, r^i \hspace{0.0cm}, \hspace{0.5cm} \widetilde{u}(r) = r^t \sum_{i=0}^\infty \widetilde{u}_i r^i, \label{powerseries0}
\end{equation}
into the equations which characterize the spectral problem for (\ref{operatorhbar0}). By construction, the coefficients $\widetilde{v}_0$ and $\widetilde{u}_0$ do not vanish and $s$ and $t$ are the orders of the power series for $\widetilde{v}(r)$ and $\widetilde{u}(r)$ respectively. Explicitly, at lowest order we obtain the conditions
\begin{eqnarray}
    (1-s^2) \widetilde{v}_0 r^{s-2} + [1-(s+1)^2] \widetilde{v}_1 r^{s-1} + O(r^{n_1})  \!\!\!\!&=& \!\!\!\! 0 ,\label{order0a} \\
    (n^2-t^2)\widetilde{u}_0 r^{t-2} + [n^2-(t+1)^2]\widetilde{u}_1 r^{t-1} + O(r^{n_2}) \!\!\!\!&=& \!\!\!\! 0 ,\label{order0b}
\end{eqnarray}
where $n_1=\min\{s,n+t-1\}$ and $n_2=\min\{t,n+s-1\}$.

%where several big $O$ terms have been introduced because the powers depending on $s$ and $t$ are unknown, making it impossible to determine the dominant order in the preceding expressions.

The existence of a non-trivial solution demands that the order of the power series (\ref{powerseries0}) must be set as
\[
s=1 \hspace{0.cm}, \hspace{0.5cm} t=n \hspace{0.5cm} \mbox{with} \hspace{0.5cm} \widetilde{v}_0,\widetilde{u}_0\in \mathbb{R}\hspace{0.5cm} \mbox{and} \hspace{0.5cm} \widetilde{v}_1,\widetilde{u}_1=0,
\]
which in addition maintains consistency in the order of the powers shown in (\ref{order0a}) and (\ref{order0b}). The previous condition implies that the radial characteristic of the Derrick-type modes is fixed as
\[
{\mathrm{ch}}_r[\xi_\mu(\vec{x},n,0)] = {1\choose n}\,,
\]
which, in turn,  means that this solution is regular at the origin. As a consequence, the integrand of the norm (\ref{norma0}) has a good behavior at $r=0$: $2\pi r[\widetilde{v}^2(r)+\widetilde{u}^2(r)] \sim 2\pi r^3$.

On the other hand, the asymptotic behavior of the eigenfunctions can be obtained form the differential equations
\[
-\frac{d^2 \widetilde{v}(r)}{d r^2} - \frac{1}{r} \frac{d \widetilde{v}(r)}{d r} + \Big[ \frac{1}{r^2} + 1- \omega_n^2 \Big] \widetilde{v}(r)= 0,\hspace{0.3cm} -\frac{d^2 \widetilde{u}(r)}{d r^2} - \frac{1}{r} \frac{d \widetilde{u}(r)}{d r} + \Big[ \lambda - \omega_n^2 \Big] \widetilde{v}(r)= 0,
\]
which are derived from the general equations (\ref{edoDerrick}) in this approximation. It can be shown that
\[
\widetilde{v}(r) \stackrel{r\rightarrow \infty}{\longrightarrow} \left\{ \begin{array}{lll}  \widehat {C}_1^{(v)} \, I_1(\sqrt{1-\omega_n^2}\, r) + \widehat{C}_2^{(v)} \, K_1(\sqrt{1-\omega_n^2}\, r), & \mbox{ if } \omega_n^2 <1, \\ C_1^{(v)} \, J_1(\sqrt{\omega_n^2-1}\, r) + C_2^{(v)} \, Y_1(\sqrt{\omega_n^2-1}\, r), & \mbox{ if } \omega_n^2 \geq 1,  \end{array} \right.
\]
and
\[
\widetilde{u}(r) \stackrel{r\rightarrow \infty}{\longrightarrow} \left\{ \begin{array}{lll}  \widehat {C}_1^{(u)} \, I_0(\sqrt{\lambda-\omega_n^2}\, r) +\widehat{C}_2^{(u)} \, K_0(\sqrt{\lambda-\omega_n^2}\, r), & \mbox{ if } \omega_n^2 <\lambda ,\\ C_1^{(u)} \, J_0(\sqrt{\omega_n^2-\lambda}\, r) +  C_2^{(u)} \, Y_0(\sqrt{\omega_n^2-\lambda}\, r), & \mbox{ if } \omega_n^2 \geq \lambda,  \end{array} \right.
\]
where $J_i$, $Y_i$, $I_i$ and $K_i$ stand for the usual modified Bessel functions. The shape modes (belonging to the discrete spectrum) are attained when the values of $v_0$ and $u_0$ are suitably chosen such that the constants $\widehat {C}_1^{(v)}$ and $\widehat {C}_1^{(u)}$ vanish in the asymptotic behavior. Therefore, for these eigenfunctions the norm (\ref{norma0}) is well-defined. 

\vspace{0.2cm}

\noindent 2. The case of the multipolar modes is more complicated. Now, the norm for these modes (\ref{genericform04b}) is given by
\begin{equation}
\| \xi(\vec{x},n,\overline{k})\|^2 = \int_0^\infty \pi r \Big[ u^2(r) + \overline{k}^2 w^2(r) + \left(v'(r)- r f_n(r) w(r)\right)^2 + \frac{\overline{k}^2 v^2(r)}{r^2}  \Big] \, dr, \label{norma1}
\end{equation}
in terms of the radial functions $v(r)$, $u(r)$ and $w(r)$. These functions satisfy the ordinary differential equations (\ref{genericform04d}).  Now, we substitute the power expansion near $r=0$
\[
v(r)=\sum_{i=0}^\infty v_i r^{s+i} \hspace{0.cm},\hspace{0.5cm} u(r)=\sum_{i=0}^\infty u_i r^{t+i} \hspace{0.cm},\hspace{0.5cm} w(r)=\sum_{i=0}^\infty w_i r^{m+i} ,
\]
into the equations which characterize the spectral problem. The first equation that we find is 
\begin{equation}
(\overline{k}^2 - s^2) v_0 r^{s-2}  + 2 d_0 w_0 (1+n) r^{n+m} + [\overline{k}^2 - (s+1)^2] v_1 r^{s-1} + O(r^{\widetilde{n}_1})  = 0, \label{indicial1a}
\end{equation}
with $\widetilde{n}_1= \min\{s,n+m+1\}$. Note that this relation involves the lowest power $r^{s-2}$ and $r^{n+m}$. Since $s$ and $m$ are at this point unknown, it cannot yet be determined which term dominates at the lowest order. The second equation in this process reads as 
{
\begin{eqnarray}
&&(-t^2+ \overline{k}^2+n^2) u_0 r^{t-2} - 2n \overline{k}^2 w_0 r^{m-2} +2 n s d_0 v_0 r^{n+s-2} + (-(t+1)^2+ \overline{k}^2+n^2) u_1 r^{t-1} - \nonumber \\
&& \hspace{1cm} - 2n \overline{k}^2 w_1 r^{m-1}+ 2 nd_0 v_1(1+s) r^{n+s-1}  + O(r^{\widetilde{n}_2})  = 0, \label{indicial1b}
\end{eqnarray}
}
whereas the third one becomes
\begin{eqnarray}
	&& (-m^2 + \overline{k}^2 + n^2) w_0 r^{m-2} - 2n u_0 r^{t-2}  + 2 n d_0 v_0 r^{n+s-2} + (-(m+1)^2 + \overline{k}^2 + n^2) w_1 r^{m-1} - \nonumber \\
    &&  \hspace{1cm} - 2n u_1 r^{t-1} + 2 n d_0 v_1 r^{n+s-1} + O(r^{\widetilde{n}_2})  =0 , \label{indicial1c}
\end{eqnarray}
being $\widetilde{n}_2=\min \{t,m,n+2\}$. Note that (\ref{indicial1b}) and (\ref{indicial1c}) involve the lowest power $r^{t-2}$, $r^{m-2}$ and $r^{n+s-2}$, leading to the same situation as before, where it is not predetermined which term dominates at the lowest order. For this reason, we have to distinguish two different cases:

\noindent \textbf{Type A:} A first choice of the orders of the series that allows for non-trivial solutions to the recurrence relations (\ref{indicial1a}), (\ref{indicial1b}) and (\ref{indicial1c}) is given by 
\begin{equation}
s=\overline{k} \hspace{0.cm} , \hspace{0.5cm} t=n-\overline{k} \hspace{0.cm} , \hspace{0.5cm} m=n-\overline{k}. \label{optionAA}
\end{equation}
With this choice, the lowest order term in (\ref{indicial1a}) vanishes, and the subsequent term leads to the condition $v_1=0$. The remaining terms  determine then the higher-order coefficients starting from $v_0\in \mathbb{R}$ (and also $w_0$ from the expansion of the function $w(r)$). On the other hand, the equations (\ref{indicial1a}) and (\ref{indicial1b}) now take the form 
\begin{eqnarray*}
2 \overline{k}n(u_0-\overline{k}\, w_0)r^{n-\overline{k}-2}+[(-(n-\overline{k}+1)^2+\overline{k}^2+n^2)u_1-2n \overline{k}^2 w_1]r^{n-\overline{k}-1}+O(r^{n-\overline{k}}) \!\!\!\! &=& \!\!\!\!0, \\
	-2n(u_0-\overline{k}\, w_0)^{n-\overline{k}-2} + [(-(n-\overline{k}+1)^2+\overline{k}^2+n^2)w_1-2n \overline{k}^2 u_1] r^{n-\overline{k}-1} + O(r^{n-\overline{k}}) \!\!\!\! &=& \!\!\!\!0.
\end{eqnarray*}
These equations yield non-trivial solutions under the conditions $v_0,w_0\in \mathbb{R}$ and $u_0= \overline{k} \, w_0$. Additionally, the next terms in the previous relation imply that $u_1=w_1=0$ and the following coefficients in the power series are determined by the values of $u_0$ and $w_0$ in the recurrence equations. 

As a consequence of the previous analysis, the eigenfunctions constructed with the choice (\ref{optionAA}) exhibit the following radial characteristic
\[
{\mathrm{ch}}_r[\xi_\nu^{\mathrm{ A}}(\vec{x},n,\overline{k})] = {\overline{k}-1 \choose n- \overline{k}}\,,
\]
Clearly, this implies that the value of the parameter $\overline{k}$ is restricted to $\overline{k}=1,\dots, n$, since otherwise, one of the components of the vortex solution would fail to be regular at the origin. Note that these Type A multipolar modes in the self-dual case give rise to zero modes. 

\vspace{0.2cm}

\noindent \textbf{Type B:} A second choice of the orders of the series leading to non-trivial solutions is given by
\begin{equation}
s=\overline{k} \hspace{0.cm} , \hspace{0.5cm} t=n+\overline{k} \hspace{0.cm} , \hspace{0.5cm} m=n+\overline{k}, \label{optionBB}
\end{equation}
In this case the lowest order term in (\ref{indicial1a}) automatically vanishes, and the following one leads to $v_1=0$. The rest of the terms determine, as before, the higher-order coefficients starting from $v_0,w_0\in \mathbb{R}$. On the other hand, (\ref{indicial1b}) and (\ref{indicial1c}) read now 
\begin{eqnarray*}
&&\hspace{-0.4cm} -2 n \overline{k}(u_0+\overline{k} w_0-d_0 v_0) r^{n+\overline{k}-2}+\left[ \left(-(n+\overline{k}+1)^2+\overline{k}^2+n^2\right)u_1-2n \overline{k}^2 w_1+2 n d_0 v_1(1+\overline{k})\right]  r^{n+\overline{k}-1}+ \\
&& + O(r^{n+\overline{k}}) = 0,
\end{eqnarray*}
and
\begin{eqnarray*}
&&\hspace{-0.4cm} -2n(u_0 +\overline{k} w_0- d_0 v_0) r^{n+\overline{k}-2}+ \left[ \left(-(n+\overline{k}+1)^2+\overline{k}^2+n^2 \right) w_1-2 n u_1+2n d_0 v_1\right] r^{n+\overline{k}-1} + \\
&& + O(r^{n+\overline{k}})= 0.
\end{eqnarray*}
Non-trivial regular modes can be obtained if we fix $v_0,w_0\in \mathbb{R}$ under the restriction $u_0=d_0v_0-\overline{k}w_0$. In addition, it can be checked that $u_1,w_1=0$. This analysis shows that the radial characteristic of the Type B multipolar eigenmodes is given by
\[
{\mathrm{ch}}_r[\xi_\nu^{\mathrm{ B}}(\vec{x},n,\overline{k})] = {\overline{k}-1\choose n+\overline{k}}\,,
\]
which does not introduce any restriction on the value of $\overline{k}\geq 1$. In the self-dual case this type of mode reproduces the shape modes with positive eigenvalues presented in \cite{AlonsoIzquierdo2016}.

The asymptotic behavior of the multipolar eigenfunctions are governed by the differential equation
\begin{eqnarray*}
	 - \frac{d^2 v}{dr^2} - \frac{1}{r} \frac{dv}{dr}+ \Big[ \frac{\overline{k}^2}{r^2} + 1 \Big] v(r) + 2  w(r) \!\!\!\!&=&\!\!\!\! \omega_n^2 v(r),   \\
	 - \frac{d^2 u}{dr^2} - \frac{1}{r} \frac{du}{dr}+ \Big[ \frac{\overline{k}^2}{r^2} + \lambda \Big] u(r)\!\!\!\! &=& \!\!\!\!\omega_n^2 u(r),  \\
	 - \frac{d^2 w}{dr^2} - \frac{1}{r} \frac{dw}{dr}+ \Big[ \frac{\overline{k}^2}{r^2} + 1 \Big] w(r)\!\!\!\! &=& \!\!\!\!\omega_n^2 w(r)  ,
\end{eqnarray*}
The second and the third equations leads to 
\[
u(r) \stackrel{r\rightarrow \infty}{\longrightarrow} \left\{ \begin{array}{lll}  \widehat {C}_1^{(u)} \, I_{\overline{k}}(\sqrt{\lambda-\omega_n^2}\, r) +\widehat{C}_2^{(u)} \, K_{\overline{k}}(\sqrt{\lambda-\omega_n^2}\, r), & \mbox{ if } \omega_n^2 <\lambda, \\ C_1^{(u)} \, J_{\overline{k}}(\sqrt{\omega_n^2-\lambda}\, r) +  C_2^{(u)} \, Y_{\overline{k}}(\sqrt{\omega_n^2-\lambda}\, r), & \mbox{ if } \omega_n^2 \geq \lambda,  \end{array} \right.
\]
and
\[
w(r) \stackrel{r\rightarrow \infty}{\longrightarrow} \left\{ \begin{array}{lll}  \widehat {C}_1^{(w)} \, I_{\overline{k}}(\sqrt{1-\omega_n^2}\, r) + \widehat{C}_2^{(w)} \, K_{\overline{k}}(\sqrt{1-\omega_n^2}\, r), & \mbox{ if } \omega_n^2 <1, \\ C_1^{(w)} \, J_{\overline{k}}(\sqrt{\omega_n^2-1}\, r) + C_2^{(w)} \, Y_{\overline{k}}(\sqrt{\omega_n^2-1}\, r), & \mbox{ if } \omega_n^2 \geq 1.  \end{array} \right.
\]
{
Finally, the first of the equations provides the expression
\[
v(r) \stackrel{r\rightarrow \infty}{\longrightarrow} \left\{ \begin{array}{lll}  \widehat {C}_1^{(v)} \, I_{\overline{k}}(\sqrt{1-\omega_n^2}\, r) + \widehat{C}_2^{(v)} \, K_{\overline{k}}(\sqrt{1-\omega_n^2}\, r) + v_p(r),& \mbox{ if } \omega_n^2 <1, \\ C_1^{(v)} \, J_{\overline{k}}(\sqrt{\omega_n^2-1}\, r) + C_2^{(v)} \, Y_{\overline{k}}(\sqrt{\omega_n^2-1}\, r) + v_p(r), & \mbox{ if } \omega_n^2 \geq 1 , \end{array} \right.
\]
where the particular solution $v_p(r)$ involves the presence of Meijer functions whose contribution is negligible compared to the contribution of the homogeneous solution.  It can be clearly seen from the previous expressions that this problem has two different mass thresholds, one starting at $\omega^2=1$ and the other one starting at $\omega^2=\lambda$. In other words, if $\omega_n^2>\lambda$, $u(r)$ becomes an oscillatory function at infinity. The same phenomenon occurs for $w(r)$ and $v(r)$ for $\omega_n^2>1$. 

} As before, it is possible to find values for the constants $v_0$ and $w_0$ such that the solution exhibits asymptotic behavior where the constants $\widehat{C}_1^{u,u,w}$ are zero, ensuring that the eigenfunctions vanish at infinity.
\end{proof}

At this stage, we have outlined the complete spectrum of fluctuations for the non-BPS vortices. Specifically, we have shown that the modes with $\overline{k}=0$ correspond to Derrick-type modes. Additionally, the multipolar modes with $\overline{k}=1$ have a clear interpretation as the translational modes. This interpretation is confirmed by observing that an infinitesimal gauge-invariant translation along the $x$ and $y$ directions generates the general form of the two fundamental translational modes \cite{Manton2004,Tong2014}
\begin{equation}\label{ZeroModes}
\xi_{0,x}(\vec{x}) = \left( 
\begin{array}{c} 0  \\
F_{12}(\vec{x})  \\
\widetilde{D}_1\psi_1(\vec{x})\\
\widetilde{D}_1\psi_2(\vec{x}) \end{array} \right)\,,  \qquad
\xi_{0,y}(\vec{x}) = \left( 
\begin{array}{c} -F_{12}(\vec{x})  \\
0  \\
\widetilde{D}_2\psi_1(\vec{x})\\
\widetilde{D}_2\psi_2(\vec{x})  \end{array} \right)\,. 
\end{equation}
Expressing \eqref{ZeroModes} in terms of the vortex profiles $f_n(r)$ and $\beta_n(r)$, these modes become
    \begin{equation}\label{ZeroModesx}
\xi_{0,x}(\vec{x}) = \left( 
\begin{array}{c} 0  \\
\frac{n}{r}\frac{d \beta_n(r)}{dr} \\
\frac{n f_n(r)}{r}(1-\beta_n(r))\sin(\theta)\sin(n \theta)+ \frac{d f_n(r)}{dr}\cos(\theta)\cos(n \theta)\\
-\frac{n f_n(r)}{r}(1-\beta_n(r))\sin(\theta)\cos(n \theta)+ \frac{d f_n(r)}{dr}\cos(\theta)\sin(n \theta)\end{array} \right)\,\,,  \qquad
\end{equation}
    \begin{equation}\label{ZeroModesy}
\xi_{0,y}(\vec{x}) = \left( 
\begin{array}{c} -\frac{n}{r}\frac{d \beta_n(r)}{dr}  \\
 0\\
-\frac{n f_n(r)}{r}(1-\beta_n(r))\cos(\theta)\sin(n \theta)+ \frac{d f_n(r)}{dr}\sin(\theta)\cos(n \theta)\\
\frac{n f_n(r)}{r}(1-\beta_n(r))\cos(\theta)\cos(n \theta)+ \frac{d f_n(r)}{dr}\sin(\theta)\sin(n \theta)\end{array} \right)\,\,.  \qquad
\end{equation}
The eigenfunction $\xi_{0,x}(\vec{x})$, given by \eqref{ZeroModesx}, can be written in the form \eqref{genericform04b} by simply choosing
\begin{equation}\label{uuww}
    \overline{k}=1, \quad v(r)=-n\frac{d \beta_n(r)}{dr}, \quad u(r)=\frac{d f_n (r)}{d r}, \quad w(r)=\frac{n f_n(r)}{r}(1-\beta_n(r)).
\end{equation}
On the other hand, the eigenmode $\xi_{0,y}(\vec{x})$, determined by \eqref{ZeroModesy}, can be identified with the orthogonal eigenfunction of (\ref{genericform04b}), which  arise when the changes  $\cos(\overline{k}\theta)\rightarrow \sin(\overline{k}\theta)$ and $\sin(\overline{k}\theta)\rightarrow -\cos(\overline{k}\theta)$ are considered in its expression. By direct substitution, it can be verified that the functions given by (\ref{uuww}) satisfy the spectral conditions \eqref{genericform04d}. To this end, it is necessary to use the differential equations \eqref{eqI3:ProfEq1} and \eqref{eqI3:ProfEq2}, which, for instance, lead to the relation $\frac{d v(r)}{d r}-r f_n(r) w(r)=\frac{v(r)}{r}$, which is useful for simplifying the resulting expressions in this calculation. It is worth mentioning that these two zero eigenmodes can be directly constructed from the vortex profiles $f_n(r)$ and $\beta_n(r)$ for any value of the coupling constant $\lambda$. 

Now, we explicitly show how the fluctuation spectrum of BPS $n$-vortices (in the self-dual regime $\lambda=1$) emerges from the general framework presented herein. 
\begin{itemize}
\item Firstly, it is clear that the $2n$ Type A multipolar modes become the $2n$ zero modes in the BPS case. Recall that these multipolar modes with $\overline{k}=1$ correspond to translational zero modes for any value of the coupling constant $\lambda$. The remaining $2n-2$ multipolar modes, whose eigenvalues are generally nonzero, become degenerate at zero when $\lambda=1$. This behavior is illustrated in the graphical representations of the spectra displayed later on in Figures \ref{fig:espectra1}-\ref{fig:espectra5}. The correspondence between the BPS and non-BPS formulations (\ref{bpszeromode4}) and (\ref{genericform04b}) is established by applying the following identifications: 
\[
k=n-\overline{k}, \hspace{0.2cm} v(r)=- \frac{r^{n-k} }{n-k} h_{nk}(r), \hspace{0.2cm} u(r)=- \frac{r^{n-k-1}}{f_n(r)}  h_{nk}'(r), \hspace{0.2cm}  w(r)=- \frac{r^{n-k-1}}{(n-k)f_n(r)}  h_{nk}'(r) \, .
\]
\item The non-degenerate Derrick-type mode $\overline{k}=0$ arising in the general case corresponds to the shape mode (\ref{bpsexcitedmode1}) with $k=0$ in the BPS regime. To establish this correspondence, we must identify
\[
\widetilde{v}(r)=\frac{d v_{n k}(r)}{d r} \hspace{0.4cm} \text{and } \hspace{0.4cm} \widetilde{u}(r)=f_n(r) v_{n k}(r) \, .
\]

\item Finally, the Type B multipolar modes determine the degenerate shape modes with $k=1,2,\dots$ in the self-dual case. Now, we have
\[
k= \overline{k}, \hspace{0.4cm}v(r)= v_{n k}(r), \hspace{0.4cm} u(r)=f_n(r) v_{n k}(r) \hspace{0.4cm} \text{and } \hspace{0.4cm} w(r)=0 \, .
\]
    
\end{itemize}

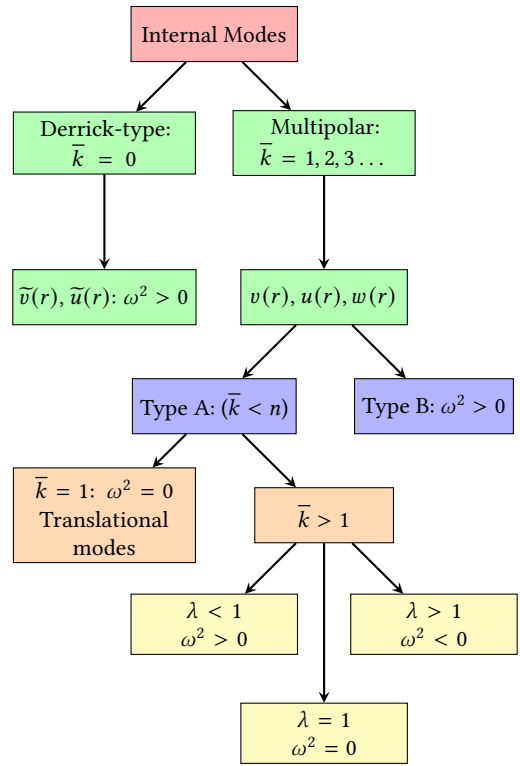
\begin{figure}[ht]
    \centering
\begin{tikzpicture}[node distance=2.8cm, scale=0.73, transform shape]

    % Nodos con colores
    \node (top) [block1] {Internal Modes };
    \node (left) [block2, below left of=top,text width=3cm] {Derrick-type:\\
    $\overline{k}=0$};
    \node (left2) [block2, below  of=left] {$\widetilde{v}(r)$, $\widetilde{u}(r)$: $\omega^2>0$};
    \node (right) [block2, below right of=top,text width=3cm] {Multipolar:
    \\ $\overline{k}=1,2,3\dots$};
    \node (right2) [block2, below  of=right] {$v(r)$, $u(r)$, $w(r)$};
    \node (rightA) [block4, below left of=right2] {Type A:  ($\overline{k}<n$)};
    \node (rightB) [block4, below right of=right2] {Type B: $\omega^2>0$};
    \node (rightAA) [block5, below left of=rightA, text width=3cm] {
        $\overline{k}=1$: $\omega^2=0$\\
        Translational modes
    };
    \node (rightBB) [block5, below right of=rightA] {
        $\overline{k}>1$
    };
    \node (rightBB1) [block3, below right of=rightBB, text width=1.5cm]{
    $\lambda>1$ $\omega^2<0$
    };
    \node (rightBB2) [block3, below left of=rightBB1, text width=1.5cm]{$\lambda=1$ $ \omega^2=0$};
    \node (rightBB3) [block3, below left  of=rightBB, text width=1.5cm]{$\lambda<1$ $\omega^2>0$};

    % Flechas
    \draw [arrow] (top) -- (left);
    \draw [arrow] (top) -- (right);
    \draw [arrow] (right) -- (right2);
    \draw [arrow] (left) -- (left2);
    \draw [arrow] (right2) -- (rightA);
    \draw [arrow] (right2) -- (rightB);
    \draw [arrow] (rightA) -- (rightAA);
    \draw [arrow] (rightA) -- (rightBB);
    \draw [arrow] (rightBB) -- (rightBB1);
    \draw [arrow] (rightBB) -- (rightBB2);
    \draw [arrow] (rightBB) -- (rightBB3);
 
    %\label{fig:internal_modes}
\end{tikzpicture}
\vspace{-0.2cm}
\caption{\textit{Classification of the internal modes of the $n$-vortex in the Abelian-Higgs model.}}
\label{Fig:Diagram}
\end{figure}

\vspace{-0.6cm}
  Finally we summarize all the results of this section. First of all, all internal modes can be classified into two categories: Derrick-type modes and multipolar modes. Every multipolar mode is double degenerate. Derrick-type modes preserve the angular symmetry of the vortex and depend on two functions $\widetilde{u}(r)$ and $\widetilde{v}(r)$, while multipolar do not preserve the aforementioned symmetry and depend on three functions $v(r)$, $u(r)$ and $w(r)$.  Among multipolar modes, these can be of Type $A$ or Type $B$, the former ones are always associated with a positive eigenvalue. The Type $A$ eigenfunctions with $\overline{k}=1$ correspond to translational eigenmodes for any value of $\lambda$. The rest of the modes of this type corresponds to unstable modes for $\lambda>1$, zero modes for the BPS limit and stable modes for $\lambda<1$, see Figure \ref{Fig:Diagram}. 
\vspace{-0.2cm}

%%%%%%%%%%%%%%%%%%%%%%%%
%%%%%%%%%%%%%%%%%%%%%%%%
%%%%%%%%%%%%%%%%%%%%%%%%
%%%%%%%%%%%%%%%%%%%%%%%%
%%%%%%%%%%%%%%%%%%%%%%%%
\vspace{-0.75cm}
\section{Vortex fluctuation spectra: numerical results}\label{num_spec}
\vspace{-0.2cm}
In this section, we explicitly determine the small fluctuation spectrum for non-BPS vortices with the lowest vorticities. The method we employ is general and can be extended to higher vorticities. The analysis introduced in Section \ref{corepaperinternalmodes} reduces the problem to solving the eigenvalue equation associated with the operators (\ref{operatorhbar0}) and (\ref{hbarrak}). In fact, it can be verified that the operator \eqref{hbarrak} can also be used to obtain the spectrum of \eqref{operatorhbar0}. Specifically, by setting $\overline{k}=0$ in the first of these operators and applying the transformations $\widetilde{v}(r)=\frac{d v(r)}{dr}- r f_n(r) w(r)$ and $\widetilde{u}(r)= u(r)$, the spectrum of the second operator can be recovered. Consequently, the entire vortex fluctuation spectrum can be obtained by solving the eigenvalue problem for the operator \eqref{hbarrak}.

For the numerical analysis, we adopt the procedure described in \cite{AlonsoIzquierdo2016}, now adapted to account for the fact that the relevant operator in our case has a $3\times 3$ matrix structure. The details of the numerical algorithm used in this work are explicitly shown in Appendix \ref{appen}. It is worth noting that the approach presented in this article significantly reduces computational demands. For instance, using a $2000$-point mesh per dimension, a brute-force diagonalization of the original spectral problem (\ref{hessianoperator}) would require handling matrices of size $16,000,000\times 16,000,000$. However, our analysis reduces this to matrices of size $6000\times 6000$, which are computationally feasible on a conventional computer. The eigenvalue problem has been solved on the interval $[0,r_{max}]$ using $N=2000$ points, with $r_{max}=20$.

\begin{figure}[h!]
\centering
{\footnotesize\begin{tabular}{c}\includegraphics[height=4.8cm]{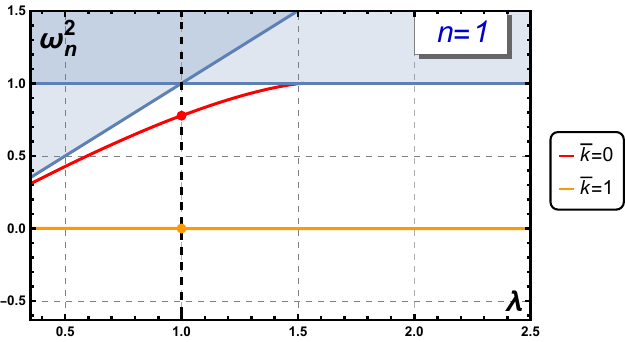} \end{tabular}\\ 
\begin{tabular}{|c|c|c|c|c|c|} \hline
\multicolumn{6}{|c|}{Case $n=1$} \\ \hline\hline
$\overline{k}$ & $\lambda=0.6$ & $\lambda=0.8$ & $\lambda=1.0$ & $\lambda=1.2$ & $\lambda=1.4$ \\ \hline\hline
$0$ & $0.50432$ & $0.64895$ & $0.77741$ & $0.88686$ & $0.97062$ \\ \hline
$1$ & $\sim 0.0$ & $\sim 0.0$ & $\sim 0.0$ & $\sim 0.0$ & $\sim 0.0$ \\ \hline
\end{tabular}}
\vspace{-0.2cm}
\caption{\textit{Spectral structure for the rotationally invariant $1$-vortex as a function of the coupling constant $\lambda$:  Graphical representation and  table of data for some values of $\lambda$.} }  
\label{fig:espectra1}
\end{figure}

\begin{figure}[h!]
\centering
{\footnotesize\begin{tabular}{c}\includegraphics[height=4.8cm]{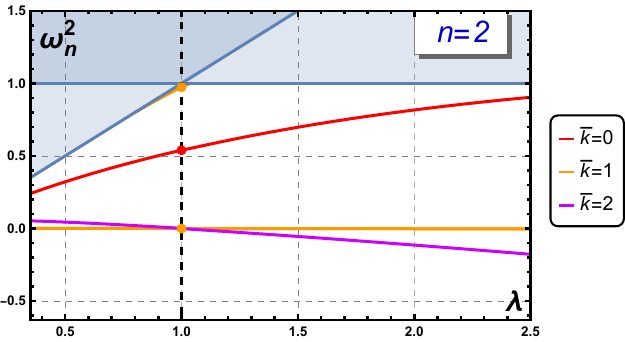} \end{tabular}\\
\begin{tabular}{|c|c|c|c|c|c|} \hline
\multicolumn{6}{|c|}{Case $n=2$} \\ \hline\hline
$\overline{k}$ & $\lambda=0.6$ & $\lambda=0.8$ & $\lambda=1.0$ & $\lambda=1.2$ & $\lambda=1.4$ \\ \hline\hline
$0$ &  $0.37130$ &  $0.46095$ & $0.53926$  & $0.60734$  &  $0.66932$  \\ \hline
$1$ & $\sim 0.0$ & $\sim 0.0$ & $\sim 0.0$ & $\sim 0.0$ & $\sim 0.0$ \\ 
    & - & $0.79905$ & $0.97338$ & - & - \\ \hline
$2$ & $0.037040$  & $0.020285$ &  $\sim 0.0$ &  $-0.021487$ & $-0.043283$  \\ \hline
\end{tabular}}
\caption{\textit{Spectral structure for the rotationally invariant $2$-vortex as a function of the coupling constant $\lambda$: Graphical representation and table of data for some values of $\lambda$.} }  
\label{fig:espectra2}
\end{figure}

\begin{figure}[h!]
\vspace{1cm}
\centering
{\footnotesize\begin{tabular}{c}\includegraphics[height=4.8cm]{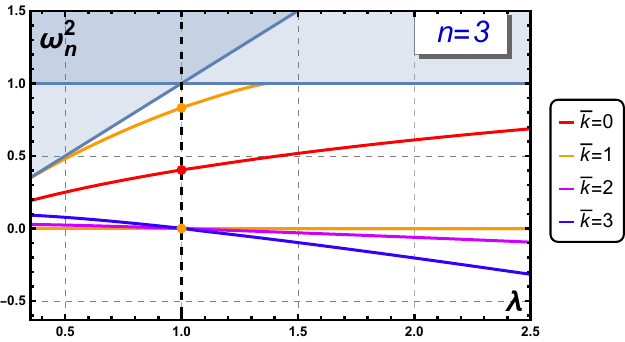} \end{tabular}\\
\begin{tabular}{|c|c|c|c|c|c|} \hline
\multicolumn{6}{|c|}{Case $n=3$} \\ \hline\hline
$\overline{k}$ & $\lambda=0.6$ & $\lambda=0.8$ & $\lambda=1.0$ & $\lambda=1.2$ & $\lambda=1.4$ \\ \hline\hline
$0$ & $0.28543$ & $0.34829$ &  $0.40373$  &  $0.45147$ & $0.49647$  \\ \hline
$1$ & $\sim 0.0$ & $\sim 0.0$ & $\sim 0.0$ & $\sim 0.0$ & $\sim 0.0$ 
\\ 
  & $0.56129$ & $0.70650$ & $0.83152$ & $0.93405$ &  - \\ \hline
$2$ & $0.019978$  & $0.010882$ & $\sim 0.0$ & $-0.011997$ & $-0.023537$ \\ \hline
$3$ & $0.064644$ & $0.035084$ & $\sim 0.0$ & $-0.037734$ & $-0.076565$ \\ \hline
\end{tabular}}
\caption{\textit{Spectral structure for the rotationally invariant $3$-vortex as a function of the coupling constant $\lambda$:  Graphical representation and  table of data for some values of $\lambda$. }}  
\label{fig:espectra3}
\end{figure}

\begin{figure}[h!]
\centering
{\footnotesize\begin{tabular}{c}\includegraphics[height=4.8cm]{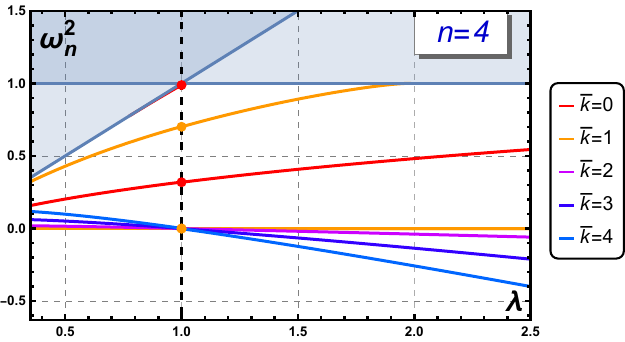} \end{tabular}\\
\vspace{-0.2cm}
\begin{tabular}{|c|c|c|c|c|c|} \hline
\multicolumn{6}{|c|}{Case $n=4$} \\ \hline\hline
$\overline{k}$ & $\lambda=0.6$ & $\lambda=0.8$ & $\lambda=1.0$ & $\lambda=1.2$ & $\lambda=1.4$ \\ \hline\hline
$0$ & $0.22982$ & $0.27775$ & $0.31893$  & $0.35670$ &  $0.39156$ \\
    & - & $0.79934$ & $0.98820$ & - & -  \\ \hline
$1$ & $\sim 0.0$ & $\sim 0.0$ & $\sim 0.0$ & $\sim 0.0$ & $\sim 0.0$ 
\\ 
  & $0.49289$ & $0.60572$ & $0.70103$ & $0.78497$ &  $0.85835$ \\ \hline
$2$ & $0.013308$  & $0.0073629$ & $\sim 0.0$ & $-0.0076856$ & $-0.015048$ \\ \hline
$3$ & $0.043027$ & $0.023293$ & $\sim 0.0$ & $-0.025528$ & $-0.051394$ \\ \hline
$4$ & $0.082528$  & $0.044591$ & $\sim 0.0$ & $-0.048017$ & $-0.097455$ \\ \hline
\end{tabular}}
\vspace{-0.2cm}
\caption{\textit{Spectral structure for the rotationally invariant $4$-vortex as a function of the coupling constant $\lambda$:  Graphical representation and table of data for some values of $\lambda$. }}  
\label{fig:espectra4}
\end{figure}

\begin{figure}[h!]
\centering
{\footnotesize\begin{tabular}{c}\includegraphics[height=4.8cm]{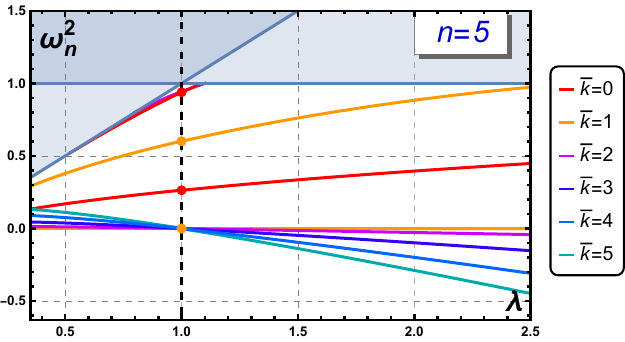} \end{tabular}\\
\vspace{-0.2cm}
\begin{tabular}{|c|c|c|c|c|c|} \hline
\multicolumn{6}{|c|}{Case $n=5$} \\ \hline\hline
$\overline{k}$ & $\lambda=0.6$ & $\lambda=0.8$ & $\lambda=1.0$ & $\lambda=1.2$ & $\lambda=1.4$ \\ \hline\hline
$0$ & $0.19169$  & $0.23024$  &  $0.26394$  & $0.29389$  &  $0.32217$  \\
    & $0.59629$ & $0.77644$  & $0.93860$  & - & -  \\ \hline
$1$ & $\sim 0.0$ & $\sim 0.0$ & $\sim 0.0$ & $\sim 0.0$ & $\sim 0.0$ \\  
& $0.43131$  & $0.52295$  &  $0.60156$ &  $0.67009$ &  $0.73243$  \\ \hline
$2$ &  $0.0097114$ & $0.0053945$ & $\sim 0.0$ & $-0.0055940$ & $-0.010833$ \\
    & - & -  &  $0.94291$ & - &  - \\ \hline
$3$ &  $0.031756$ & $0.017279$ & $\sim 0.0$ &  $-0.018512$ &  $-0.037296$ \\ \hline
$4$ &  $0.063050$  & $0.034015$  & $\sim 0.0$ & $-0.037165$  & $-0.075204$  \\ \hline
$5$ &  $0.092745$  & $0.050039$  & $\sim 0.0$ &  $-0.053929$ & $-0.10943$  \\ \hline
\end{tabular}}
\vspace{-0.2cm}
\caption{\textit{Spectral structure for the rotationally invariant $5$-vortex as a function of the coupling constant $\lambda$: Graphical representation and table of data for some values of $\lambda$. }}  
\label{fig:espectra5}
\end{figure}

The spectrum obtained for vortices with winding number $n=1,2,3,4,5$ can be found in Figures \ref{fig:espectra1}-\ref{fig:espectra5}. As expected, a vortex with $n=1$ does not posses any unstable eigenmode. Nevertheless, for $n>1$ and $\lambda>1$ every vortex has $n-1$ negative eigenvalues, which are responsible for their instability, as mentioned in Chapter \ref{Intro2}. It should be noted that for $\overline{k}>0$, every eigenvalue is doubly degenerate. Also, recall that for every value of $\lambda$ every vortex has two translational eigenmodes with $\overline{k}=1$.

Additionally, it can be seen that the unstable eigenmodes become zero modes in the BPS limit for every vortex with $\overline{k}>0$. It is also important to note that, as the vorticity grows, the potentials found in the spectral problem \eqref{numeric}-\eqref{numericbc} become deeper, which is the reason why it is possible to find more Type B eigenmodes for higher vorticities. For example, for $n=1$ only one Type B eigenmode can be found (see Figure \ref{fig:espectra1}). However, for $n=5$ five different Type B internal modes are present in the spectrum (see Figure \ref{fig:espectra5}).

{ In the next subsections, the eigenfunctions corresponding to the spectra found will be displayed. The effects of these eigenfunctions on the vortex and its corresponding energy density will also be discussed. }
\vspace{-0.2cm}

\vspace{-1cm}
\vspace{-0.5cm}

%\vspace{-0.5cm}
\subsection{Type A internal modes}

In Tables \ref{Tab1:AN2} and \ref{Tab2:AN3}, all Type A modes for $n=2,3$ and  several values of $\lambda$ are shown. In particular, we plot the functions $v(r)$, $u(r)$ and $w(r)$ as well as $|\Phi(x,y)|$ and $\mathrm{Arg}(\Phi(x,y))$ for the linearly excited vortex. 

As mentioned in Section \ref{non-bps}, Type A modes with $\overline{k}=1$ correspond to the translational modes of the vortex. This implies that these modes are zero modes for any value of $\lambda$. The effect of this mode on the vortex profile is just the splitting of one single vortex zero,  leaving a zero of multiplicity $n-1$ at the origin. It may seem that the translational mode does not move the entire  vortex as a whole, but, in reality, once these modes are triggered, the ``moving'' center attracts the rest of them translating the entire vortex. 

Regarding the remaining Type A modes, these separate the vortex centers and are responsible for the instabilities of vortex configurations with $\lambda>1$. For example, modes with $\overline{k}=2$ separate two vortex centers leaving the rest of them at the origin, forming a line. In the  case of a 2-vortex, this mode splits the condiguration into two 1-vortices. For a 3-vortex, the effect of the mode is a symmetrical spliting in a line of three 1-vortices. 

The effect of modes with higher $\overline{k}$ is similar to the behavior previously described. For example, modes with $\overline{k}=4$ would generate a configuration in which four vortices with $n=1$ form a square leaving a vortex with vorticity $n-4$ at the center. This behaviour can be seen in Table \ref{Tab2:AN3}, where modes with $\overline{k}=3$ generate a configuration in which the vortex splits into three individual vortices forming a triangle. It is worth noting that linear combinations of these internal modes can be used in order to construct multivortex configurations in which vortices are situated near each other, but not on top of each other, as mentioned in Section \ref{SecBPSlIMIT}.

%With respect to the profile functions $v(r)$, $u(r)$ and $w(r)$, it may seem that for every case $u(r)=\overline{k} \, w(r)$ but, as explained before, this relation is only fulfilled in the BPS limit. On the other hand, outside the BPS limit, what happens is that the first coefficients of the expansions of both functions near the origin are identical. But this is only true for the first few coefficients. Additionally, the asymptotic of both functions is also identical, which explains why these two profile functions are very similar. 

%As a last comment, it can be seen that in every plot the behavior of the three mode functions near the origin $r=0$ corresponds to the one described in Section \ref{non-bps}. 

%%%%%%%%%%%%%%%%%%%%%%%%%%%%%%%%%%%%%
%%%%%%%%%%%%%%%%%%%%%%%%%%%%%%%%%%%%%
%%%%%%%%%%%%%%%%%%%%%%%%%%%%%%%%%%%%%
%\vspace{-1.5cm}
\vspace{0.5cm}
\subsection{Derrick-type modes and Type B internal modes}
In Table \ref{Tab3} the plots corresponding to the Derrick-type modes for $n=1,2$ and several values of $\lambda$ are depicted. Additionally, we also show how the energy density changes once the internal modes have been triggered with a certain amplitude.  It can be seen that, when the corresponding eigenvalue approaches the mass threshold, profile functions exhibit a slower decay at large $r$. That is, they are less localized with respect to the vortex core. This behavior is clearly observed for $\lambda=1.4$ and $n=1$ in Table \ref{Tab4}. 

On the other hand, in Table \ref{Tab4}, the profile functions for Type B modes that arises for a vortex with $n=3$ are depicted along with the plots corresponding to the variations of the energy density once the internal mode has been triggered with a certain amplitude. It can be clearly seen that now, the variations of the energy density are not circularly symmetric; instead, the maximum deviation with respect to the unaltered vortex is located along the $x$ axis. Note that for the the mode corresponding to the form \eqref{genericform02b} these deviations would lie over the $y$ axis.

%Another remarkable phenomenon that can be extracted from the aforementioned figures is that, in the BPS limit, the function $w(r)$ is zero, as previously described in Section \ref{non-bps}. For the rest of the values of $\lambda$ it may seem that the previous statement also holds, but, although generally the value of $w(r)$ is much smaller  than the ones of $v(r)$ and $u(r)$, $w(r)$ is identically zero for Type B multipolar modes only when $\lambda=1$.

%As a last comment, it is worth mentioning that as, for $\lambda=0.6$ the eigenvalue corresponding to the Type B multipolar mode with $\overline{k}=1$ is close to the mass threshold, then, the functions $v(r)$, $u(r)$ and $w(r)$ become more spread, as also describe in Section \ref{non-bps}.

%As a last comment, it is worth mentioning that as the modes approach the mass threshold, they become more spread, that is, they are less localized with respect to the vortex core. This behavior is clearly observed is Table \ref{Tab4}. 

%%%%%%%%%%%%%%%%%%%%%%%%%%%%%%%%%%%%%
%%%%%%%%%%%%%%%%%%%%%%%%%%%%%%%%%%%%%
%%%%%%%%%%%%%%%%%%%%%%%%%%%%%%%%%%%%%
%%%%%%%%%%%%%%%%%%%%%%%%%%%%%%%%%%%%%

%%%%%%%%%%%%%%%%%%%%%%1
\begin{table}[htb!]
    \centering
        %\hspace{-1.cm}
    \begin{tabular}{|c|c|c|}
 
      \hline
 \multirow{3}{0.3cm}{ \rotatebox{90}{$n=2$ \hspace{2cm}$\overline{k}=1$ } } 
              &
          \rotatebox{90}{\hspace{-0.5cm}$\lambda=0.6$} &
        \begin{minipage}{0.74\textwidth}
        \centering          
        \includegraphics[width=0.31\textwidth]{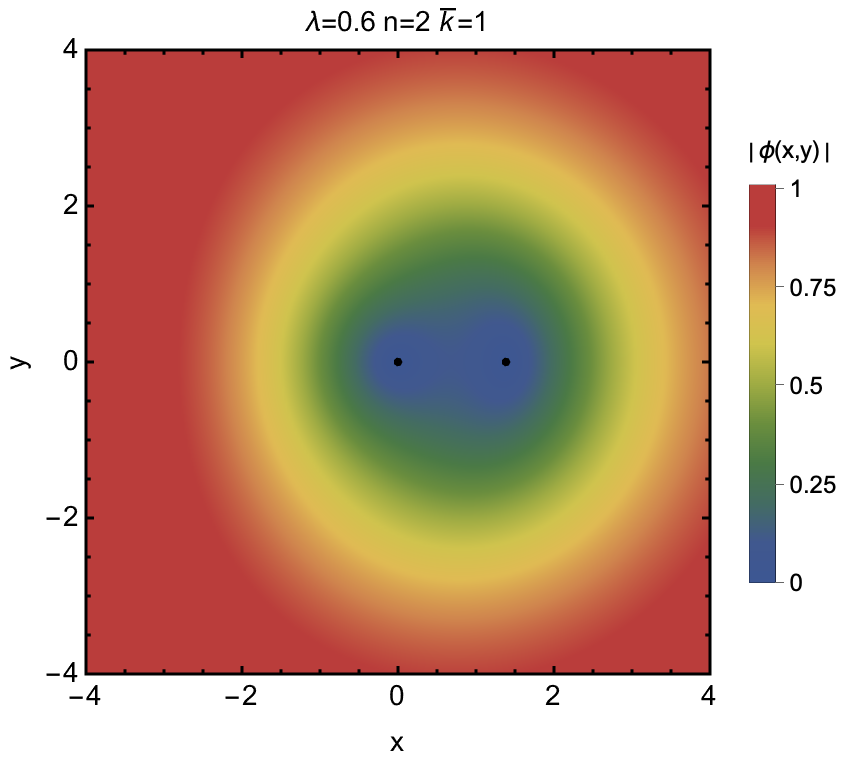}  \includegraphics[width=0.32\textwidth]{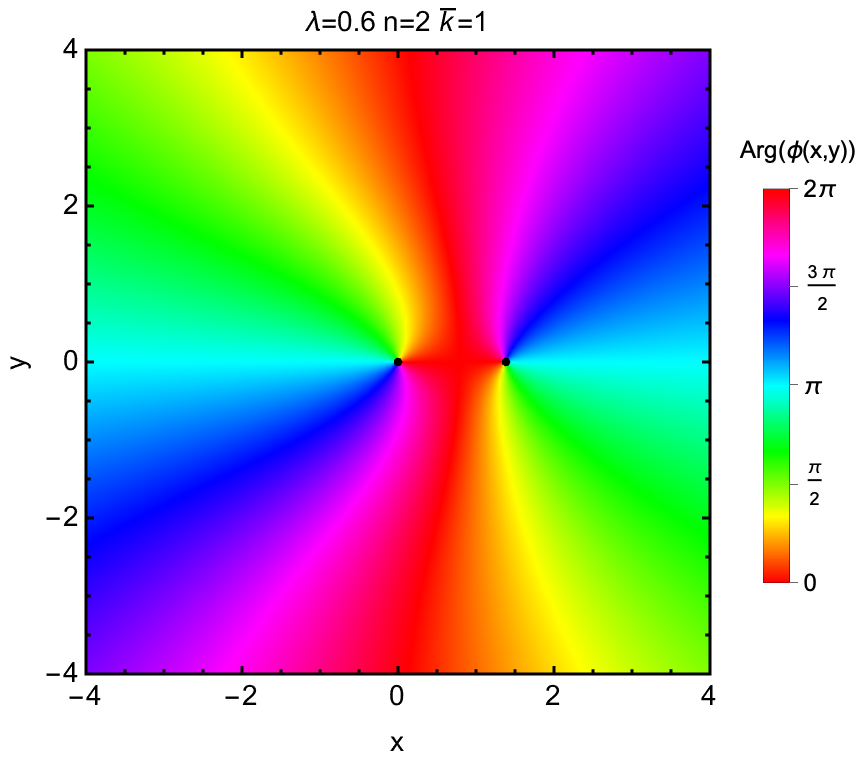}
        \raisebox{0.3\height}{\includegraphics[width=0.32\textwidth]{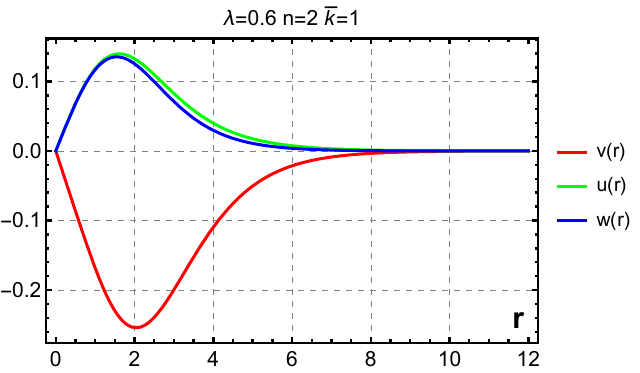}}
              %  \vspace*{0.1cm}
        \end{minipage}
     
\\
 %   \tabularnewline \cline{2-3}    
    &

          \rotatebox{90}{\hspace{-0.3cm}$\lambda=1$} &
        \begin{minipage}{0.74\textwidth}
        \centering                     
        \includegraphics[width=0.31\textwidth]{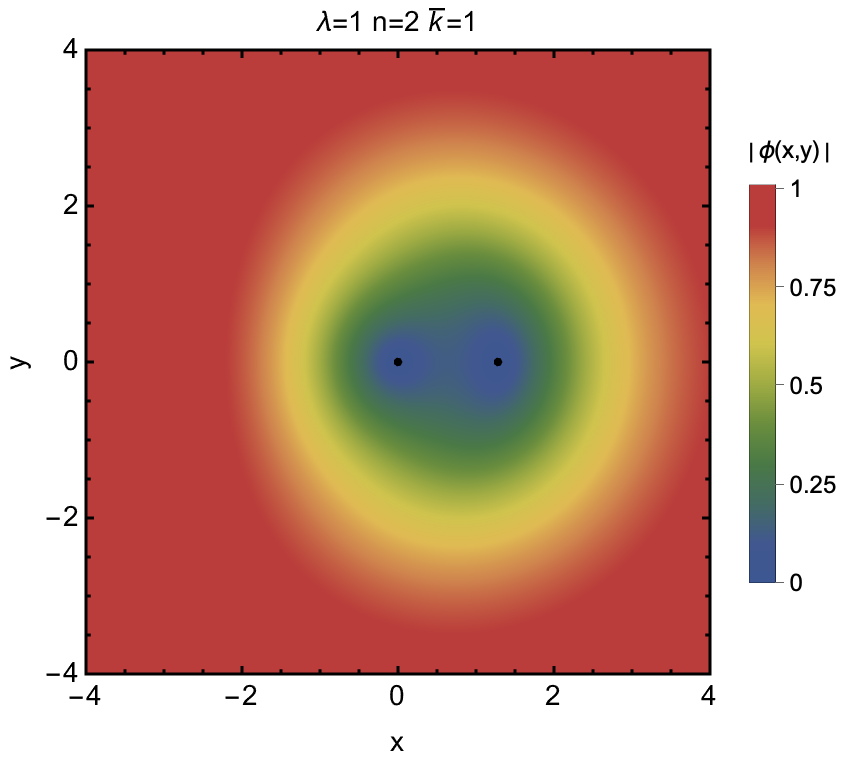}  \includegraphics[width=0.32\textwidth]{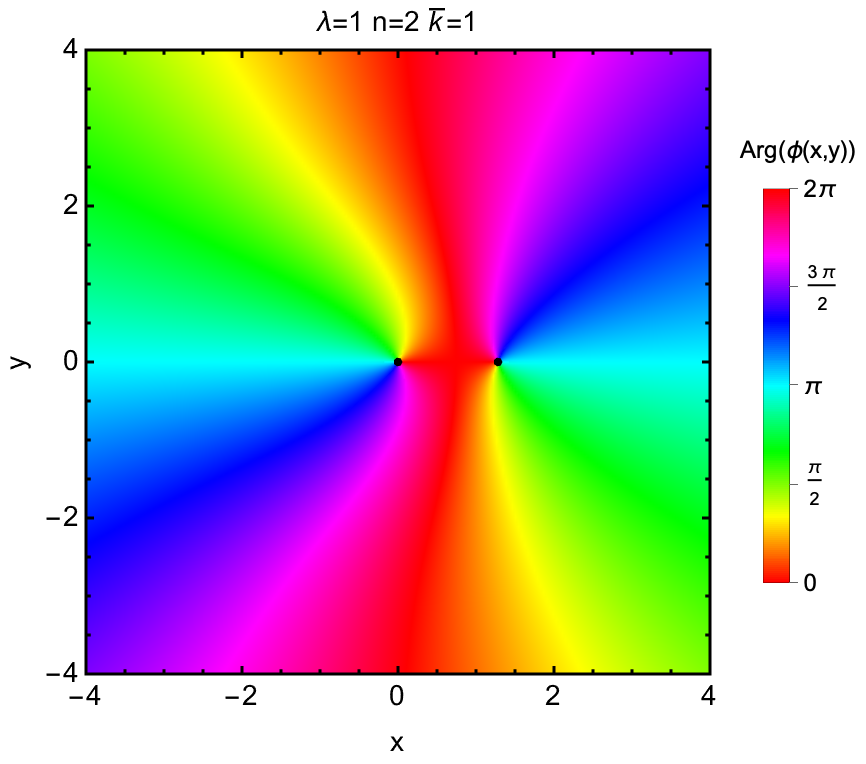}
        \raisebox{0.3\height}
       {\includegraphics[width=0.32\textwidth]{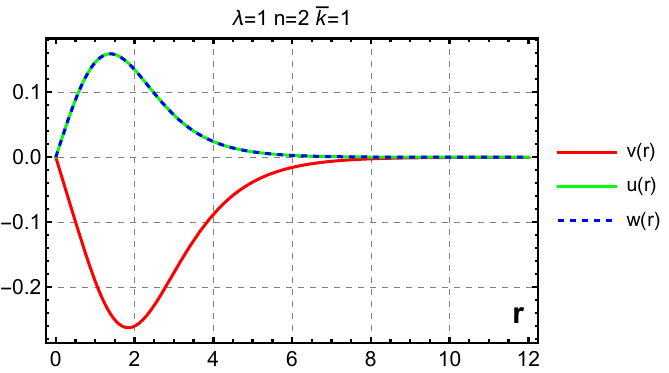}}
              %  \vspace*{0.1cm}
        \end{minipage}
 
\\
      % \tabularnewline \cline{2-3} 
      %\hline
       &

          \rotatebox{90}{\hspace{-0.5cm}$\lambda=1.4$} &
        \begin{minipage}{0.74\textwidth}
        \centering                         
        \includegraphics[width=0.31\textwidth]{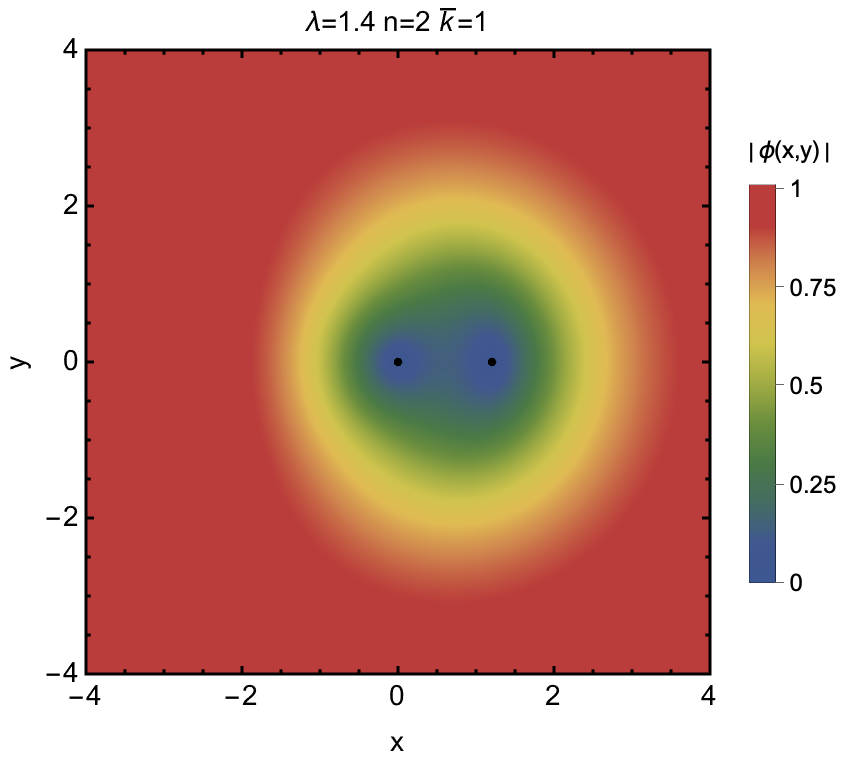}  \includegraphics[width=0.32\textwidth]{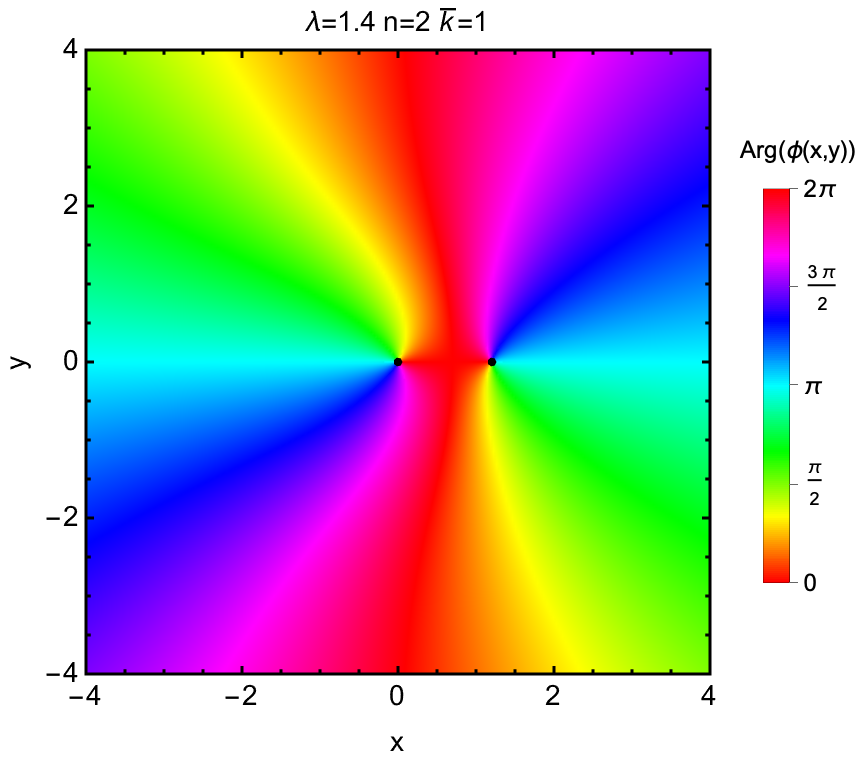}
        \raisebox{0.3\height}{\includegraphics[width=0.32\textwidth]{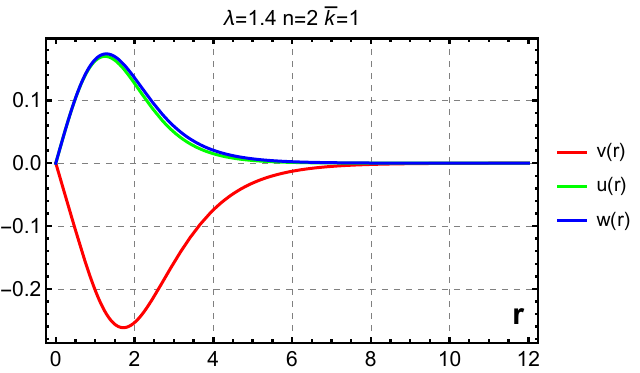}}
              %  \vspace*{0.1cm}
        \end{minipage}
 
\\

      \hline
 \multirow{3}{0.3cm}{ \rotatebox{90}{$n=2$ \hspace{2cm}$\overline{k}=2$ }} &      
          \rotatebox{90}{\hspace{-0.5cm}$\lambda=0.6$} &
        \begin{minipage}{0.74\textwidth}
        \centering
 \includegraphics[width=0.31\textwidth]{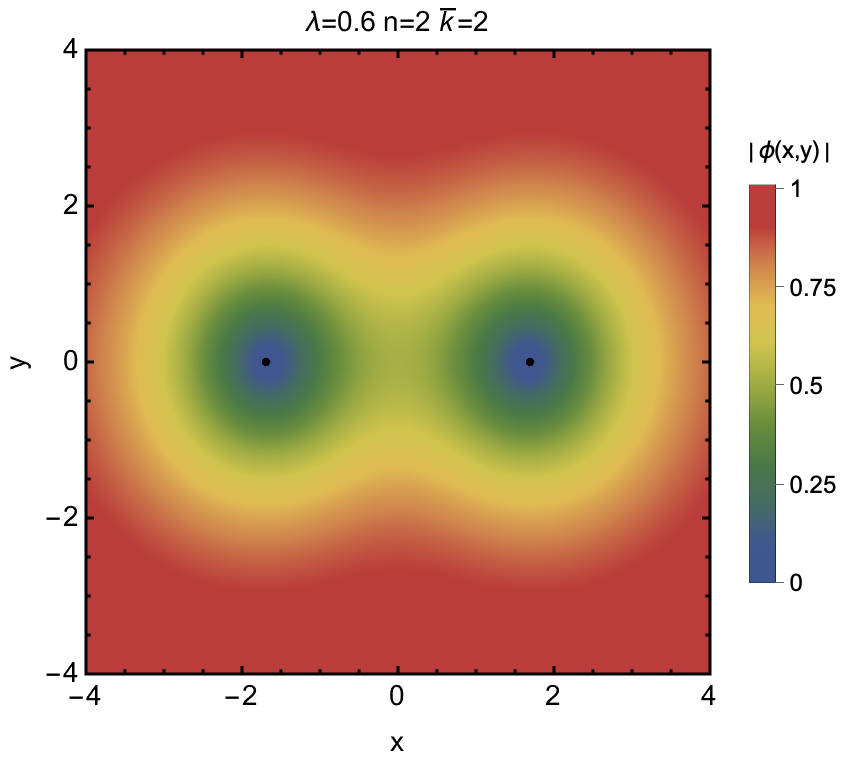}
        \includegraphics[width=0.32\textwidth]{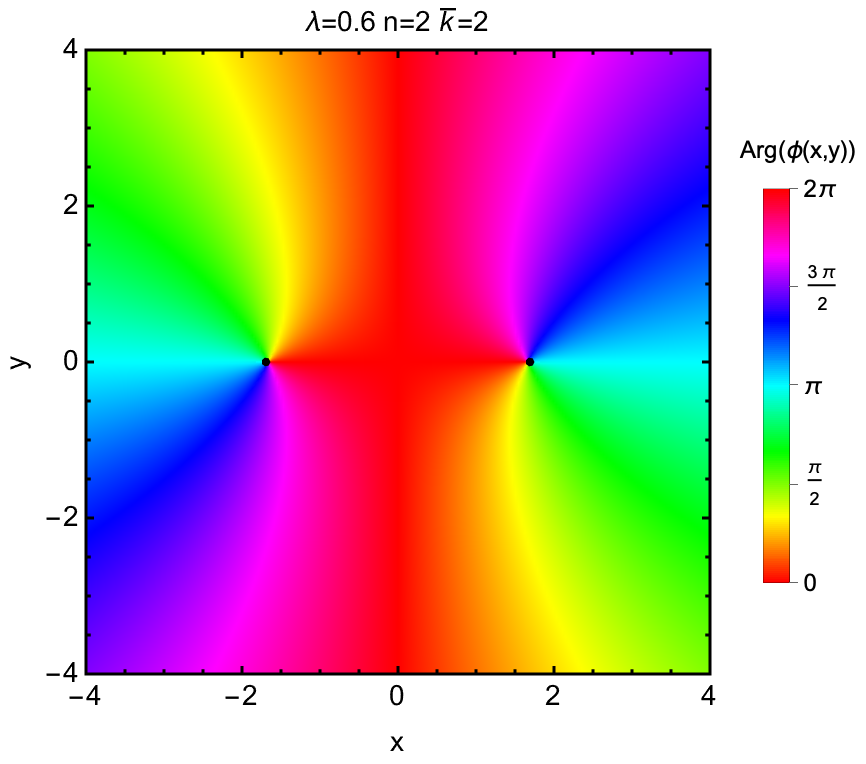}
    \raisebox{0.3\height}{\includegraphics[width=0.32\textwidth]{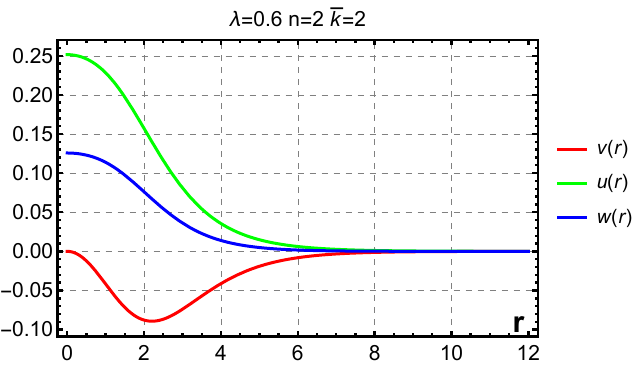}}
                      %  \vspace*{0.1cm}
        \end{minipage}
\\
          %   \tabularnewline \cline{2-3}
             &

          \rotatebox{90}{\hspace{-0.3cm}$\lambda=1$} &

        \begin{minipage}{0.74\textwidth}
        \centering
 \includegraphics[width=0.31\textwidth]{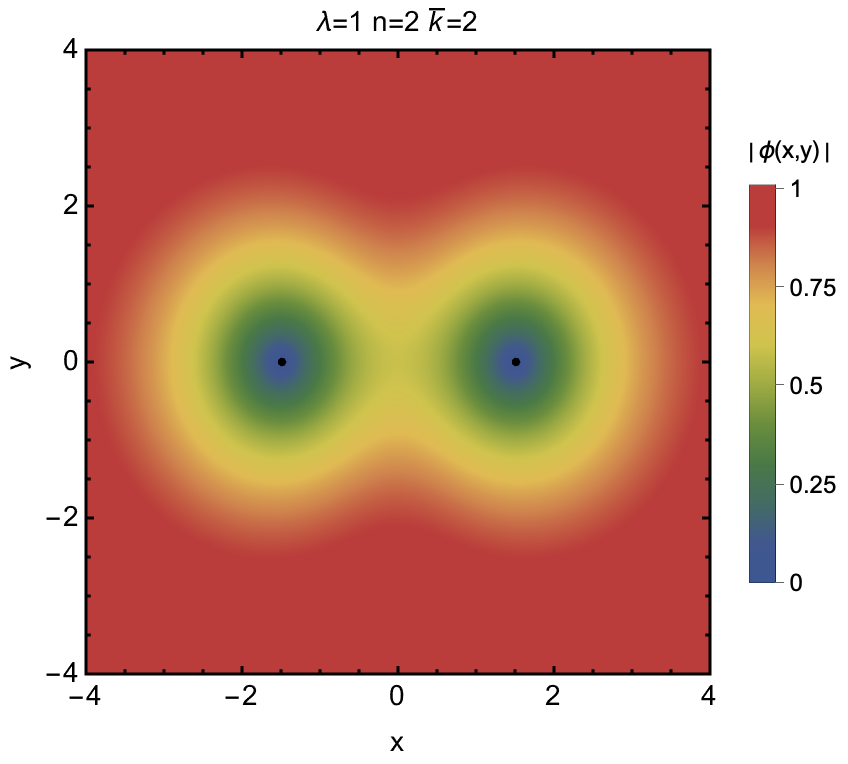}
        \includegraphics[width=0.32\textwidth]{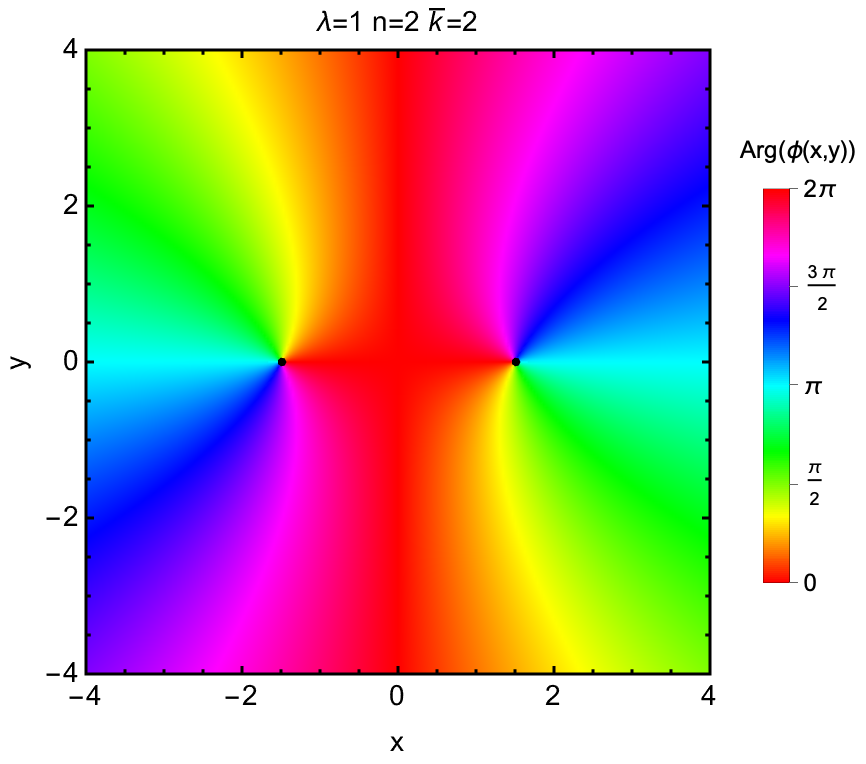}
    \raisebox{0.3\height}{\includegraphics[width=0.32\textwidth]{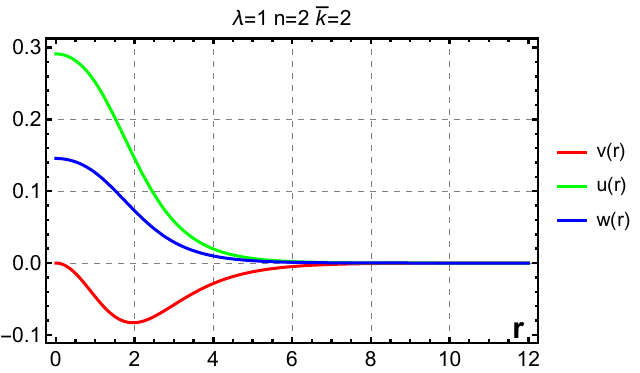}}
                      %  \vspace*{0.1cm}
        \end{minipage}
\\
       % \tabularnewline \cline{2-3} 
        &

          \rotatebox{90}{\hspace{-0.5cm}$\lambda=1.4$} &

        \begin{minipage}{0.74\textwidth}
        \centering
 \includegraphics[width=0.31\textwidth]{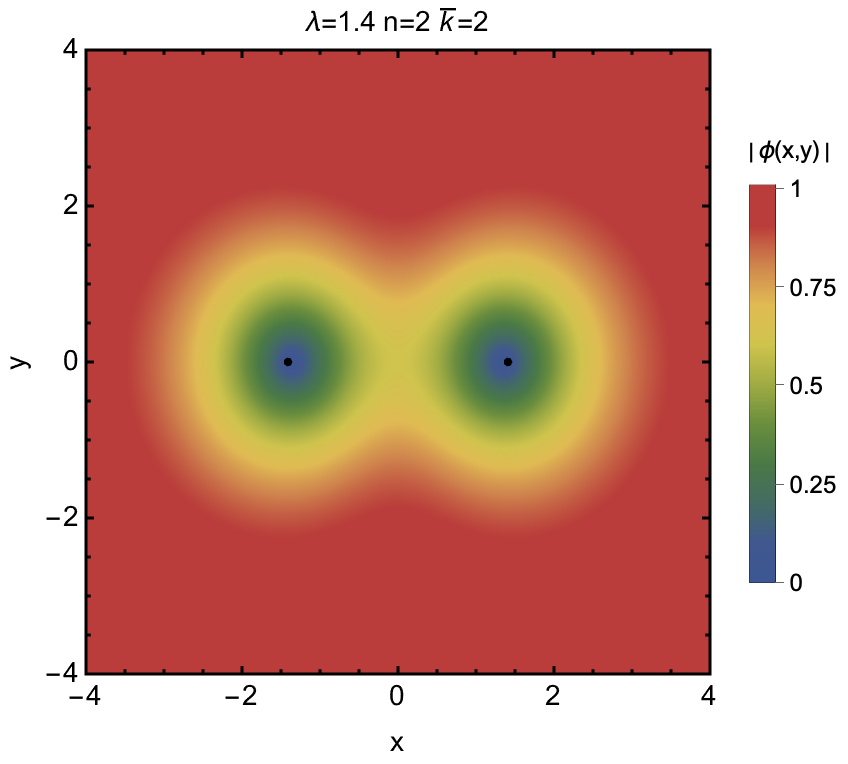}
        \includegraphics[width=0.32\textwidth]{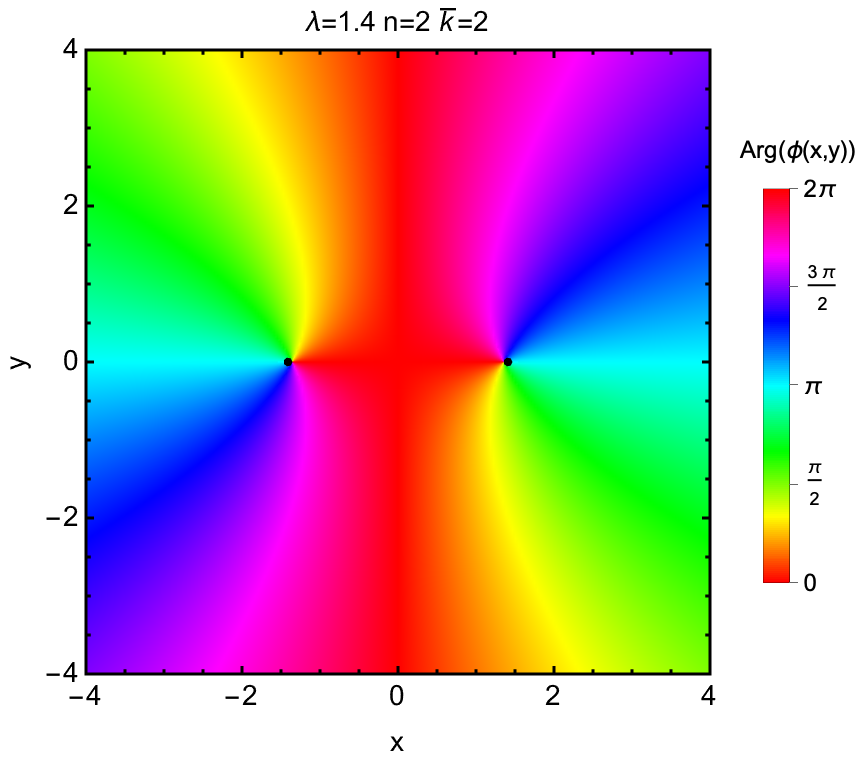}
    \raisebox{0.3\height}{\includegraphics[width=0.32\textwidth]{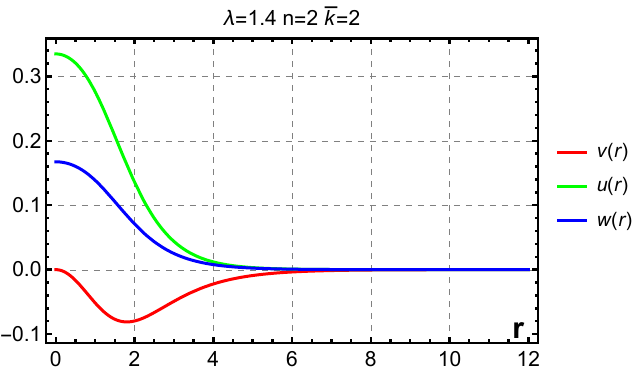}}
                      %  \vspace*{0.1cm}
        \end{minipage}
\\
        \hline
    \end{tabular}
\caption{\textit{Type A internal modes for different values of $\lambda$ and $n=2$. The plots correspond to $|\Phi(x,y)|$, $\mathrm{Arg}(\Phi(x,y))$ once the internal modes have been added to the static vortex configurations and the profile functions $v(r)$, $u(r)$ and $w(r)$.}}
    \label{Tab1:AN2}
\end{table}

\begin{table}[h!]
    \centering
        %\vspace{-1cm}
    \begin{tabular}{|c|c|c|}
    \hline
 \multirow{3}{0.3cm}{ \rotatebox{90}{$n=3$ \hspace{2cm}$\overline{k}=1$ } } 
              &

      \rotatebox{90}{\hspace{-0.5cm}$\lambda=0.6$} &
        \begin{minipage}{0.53\textwidth}
        \centering
                            
        \includegraphics[width=0.31\textwidth]{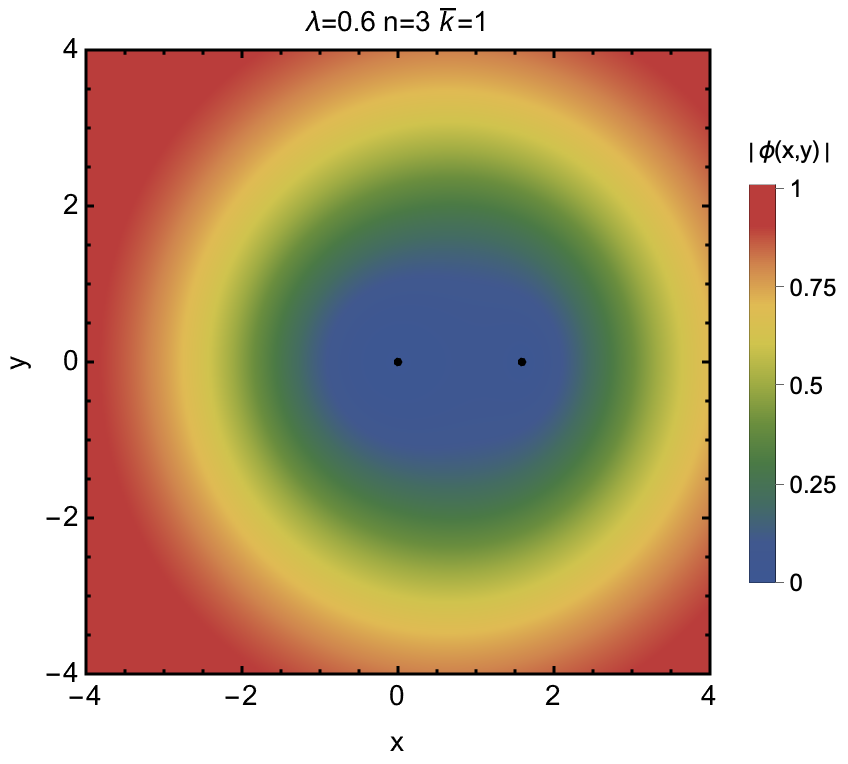}  \includegraphics[width=0.32\textwidth]{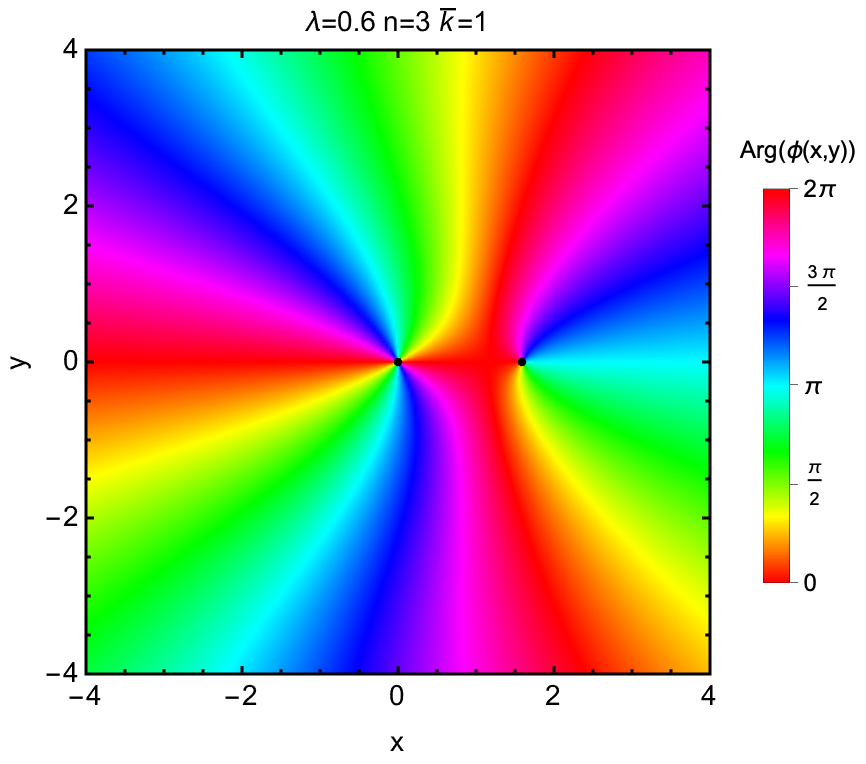}
        \raisebox{0.3\height}{\includegraphics[width=0.32\textwidth]{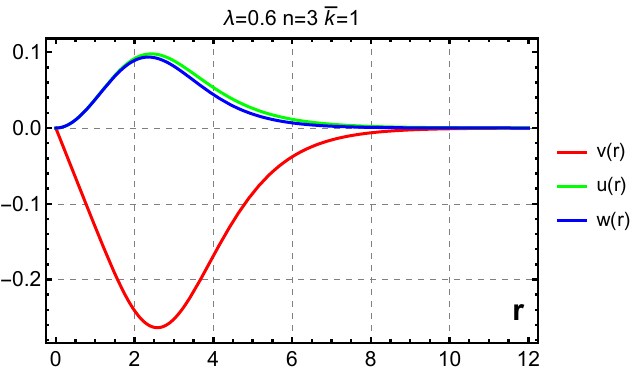}}
              %  \vspace*{0.1cm}
        \end{minipage}

\\
            & % \hline

      \rotatebox{90}{\hspace{-0.3cm}$\lambda=1$} &
        \begin{minipage}{0.53\textwidth}
        \centering
                            
        \includegraphics[width=0.31\textwidth]{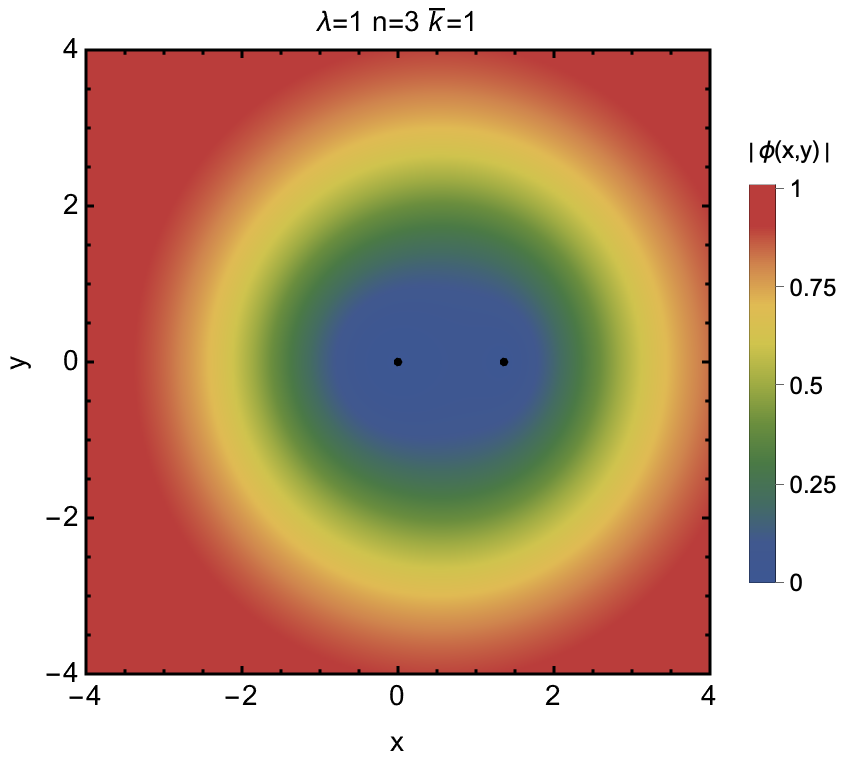}  \includegraphics[width=0.32\textwidth]{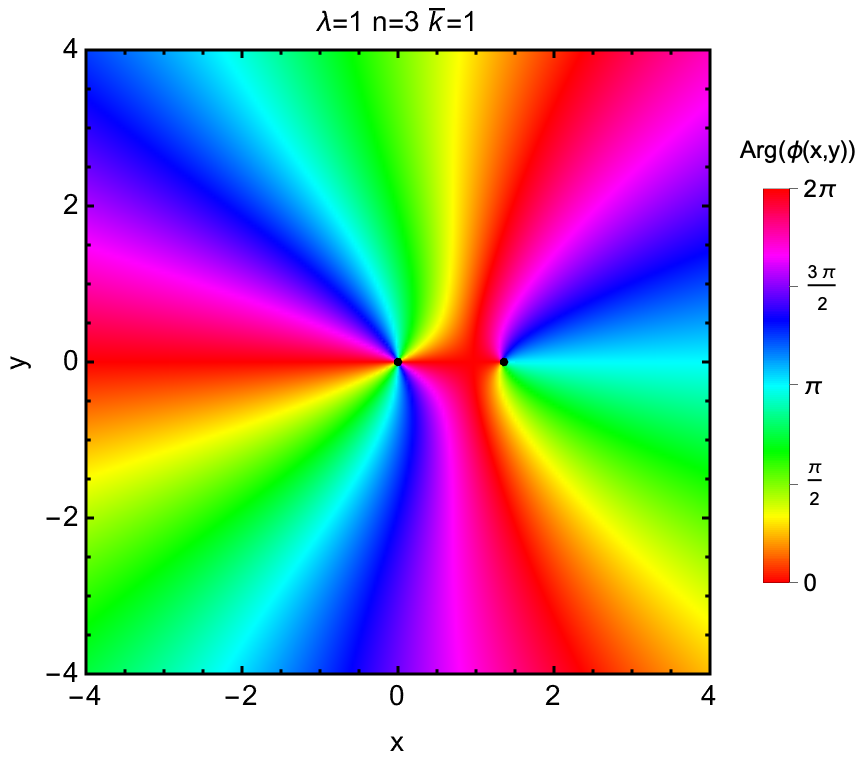}
        \raisebox{0.3\height}{\includegraphics[width=0.32\textwidth]{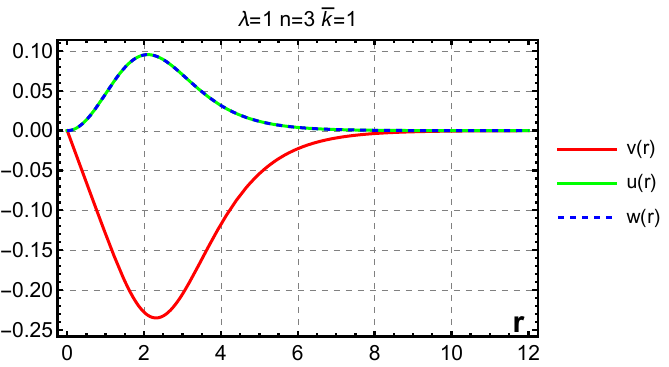}}
              %  \vspace*{0.1cm}
        \end{minipage}
 
\\
     & %  \hline

      \rotatebox{90}{\hspace{-0.5cm}$\lambda=1.4$} &
        \begin{minipage}{0.53\textwidth}
        \centering
                            
        \includegraphics[width=0.31\textwidth]{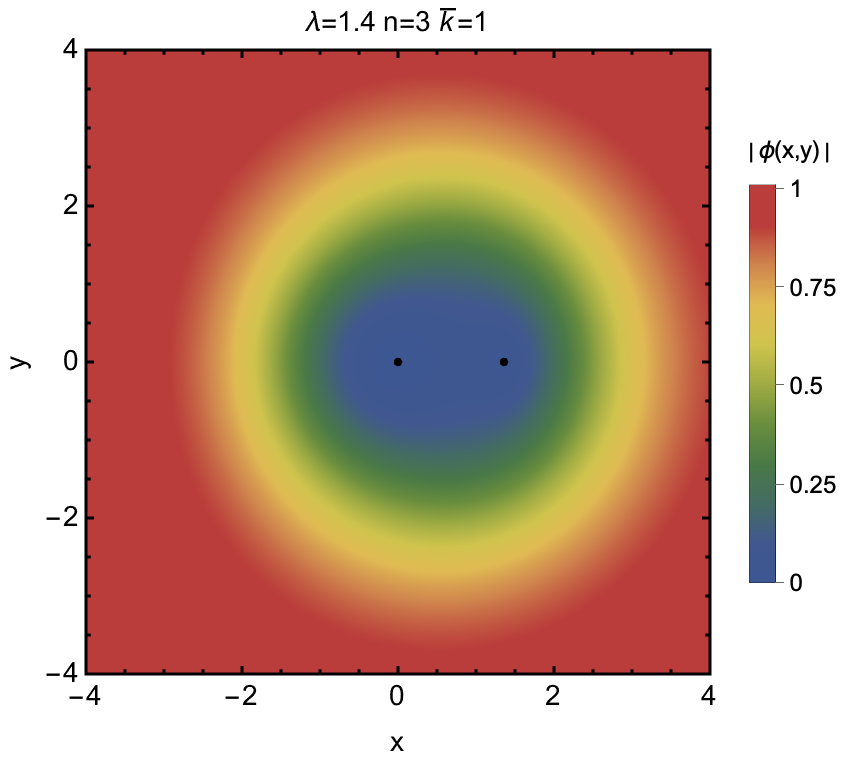}  \includegraphics[width=0.32\textwidth]{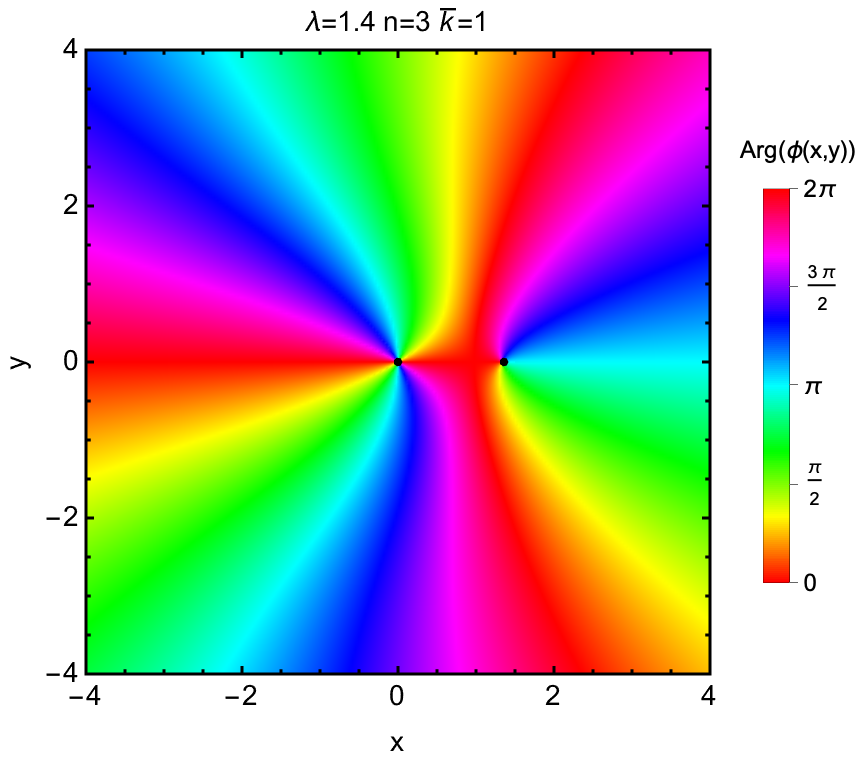}
        \raisebox{0.3\height}{\includegraphics[width=0.32\textwidth]{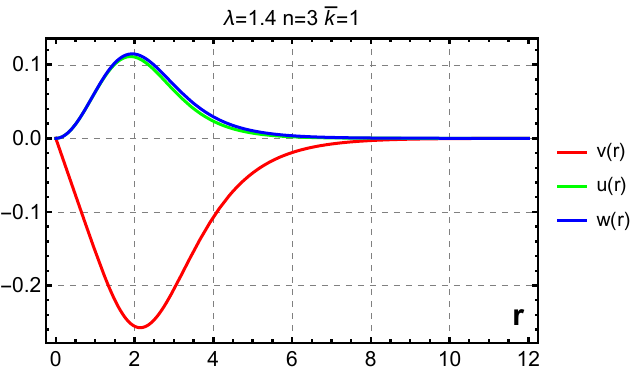}}
              %  \vspace*{0.1cm}
        \end{minipage}

\\
    %   & %\hline
           \hline
 \multirow{3}{0.3cm}{ \rotatebox{90}{$n=3$ \hspace{2cm}$\overline{k}=2$ } } 
            %  \hline
&
      \rotatebox{90}{\hspace{-0.5cm}$\lambda=0.6$} &

        \begin{minipage}{0.53\textwidth}
        \centering
 \includegraphics[width=0.31\textwidth]{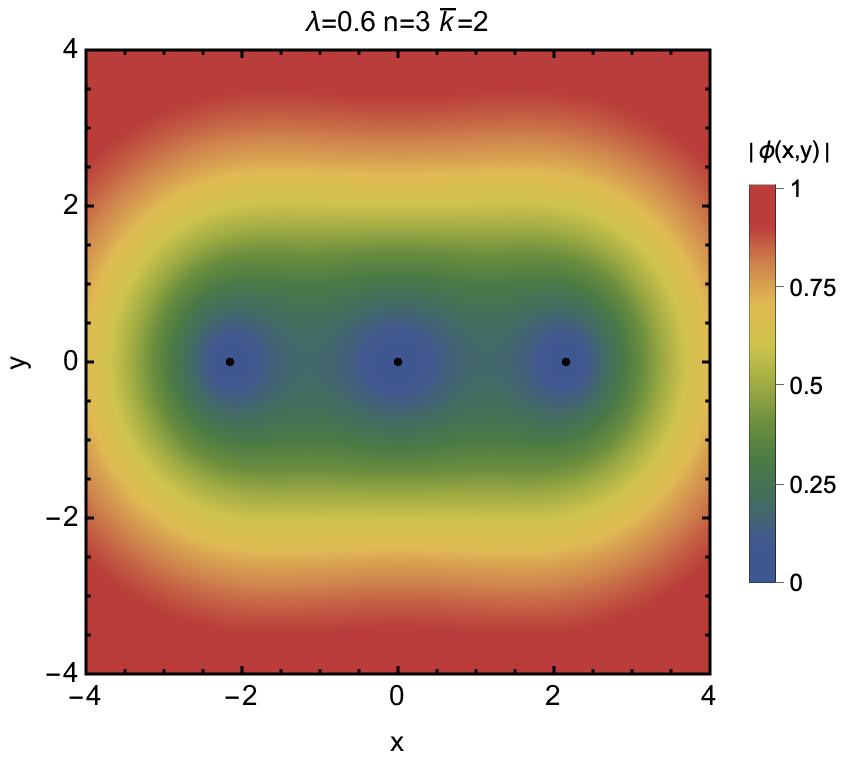}
        \includegraphics[width=0.32\textwidth]{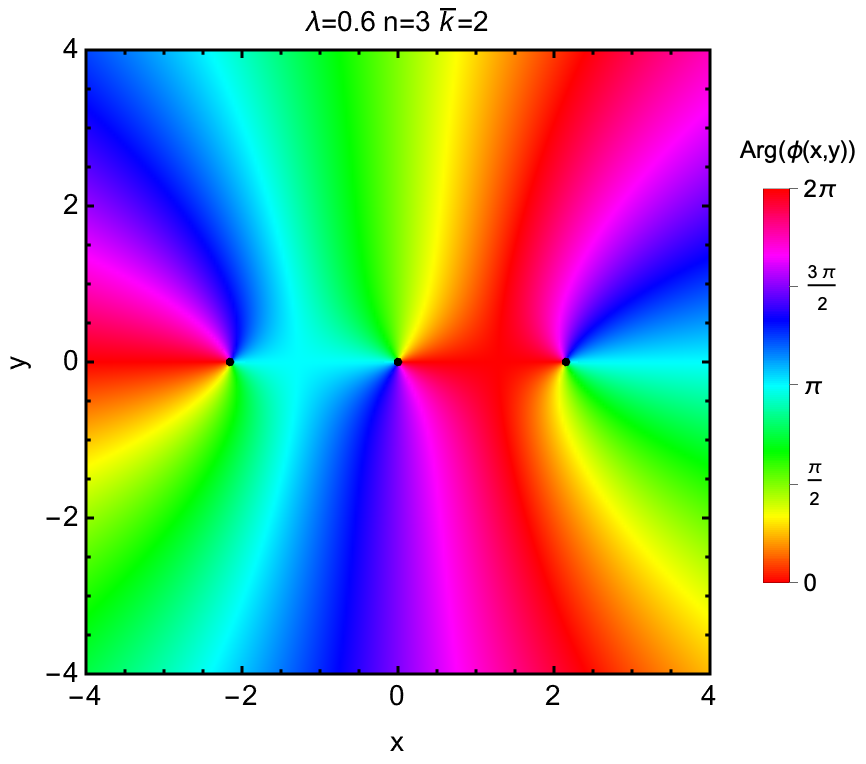}
    \raisebox{0.3\height}{\includegraphics[width=0.32\textwidth]{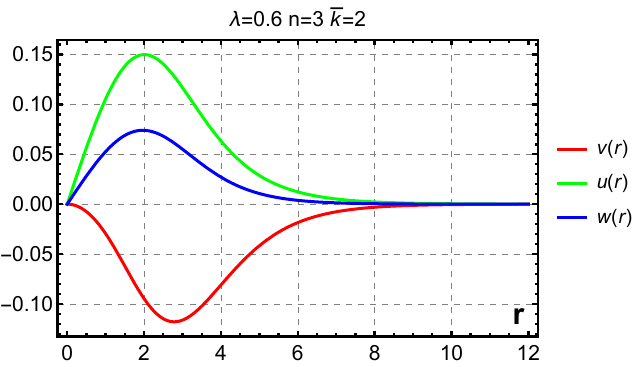}}
                      %  \vspace*{0.1cm}
        \end{minipage}
\\
        &     % \hline

      \rotatebox{90}{\hspace{-0.3cm}$\lambda=1$} &

        \begin{minipage}{0.53\textwidth}
        \centering
 \includegraphics[width=0.31\textwidth]{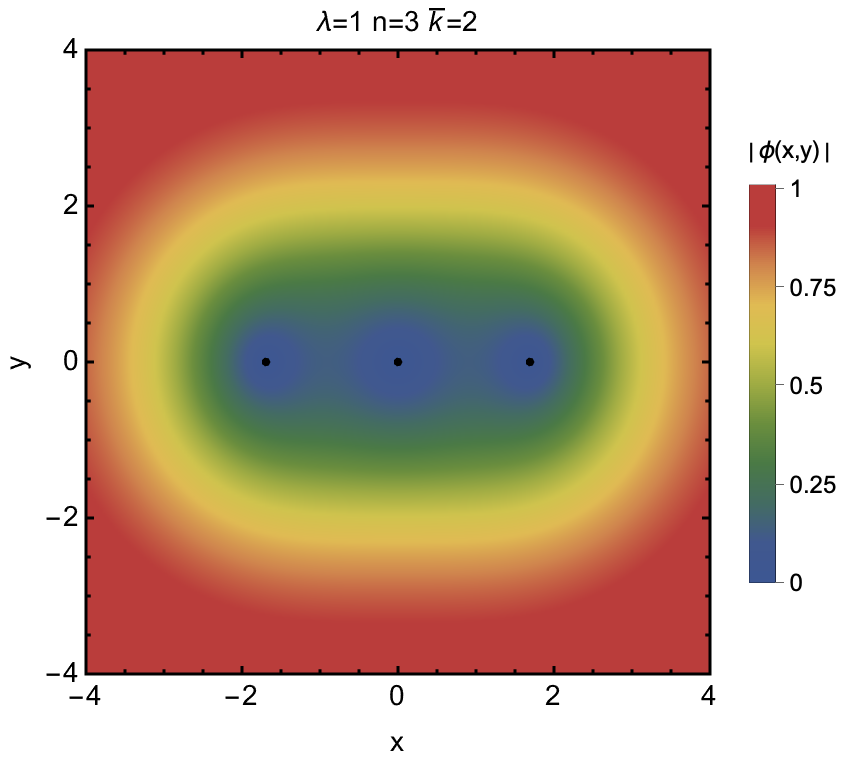}
        \includegraphics[width=0.32\textwidth]{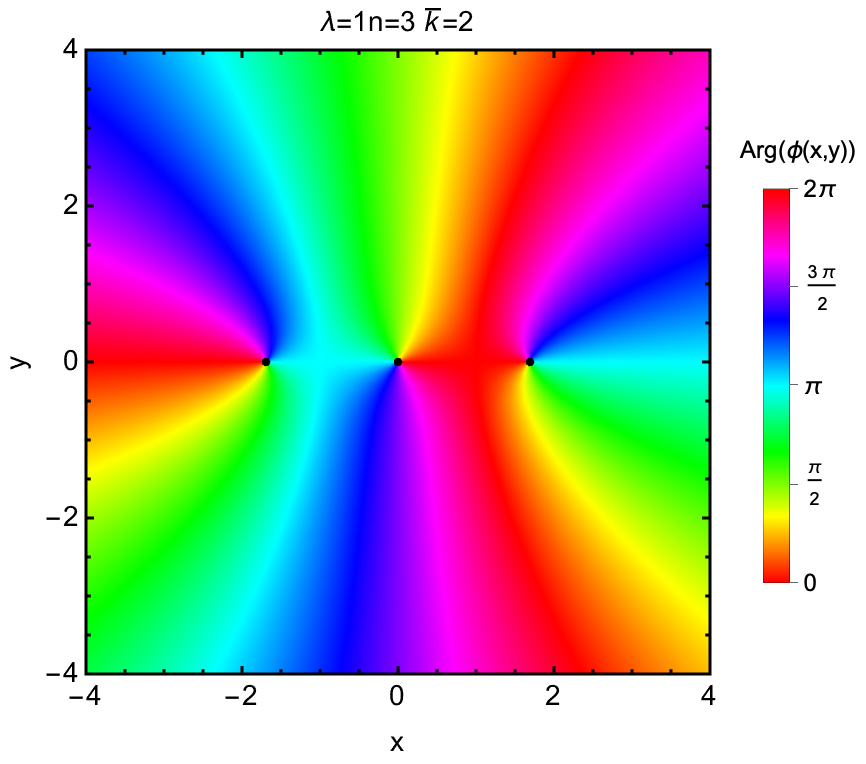}
    \raisebox{0.3\height}{\includegraphics[width=0.32\textwidth]{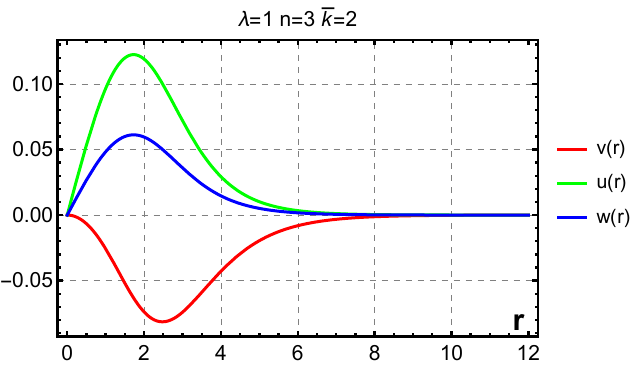}}
                      %  \vspace*{0.1cm}
        \end{minipage}
\\
      & % \hline

      \rotatebox{90}{\hspace{-0.5cm}$\lambda=1.4$} &

        \begin{minipage}{0.53\textwidth}
        \centering
 \includegraphics[width=0.31\textwidth]{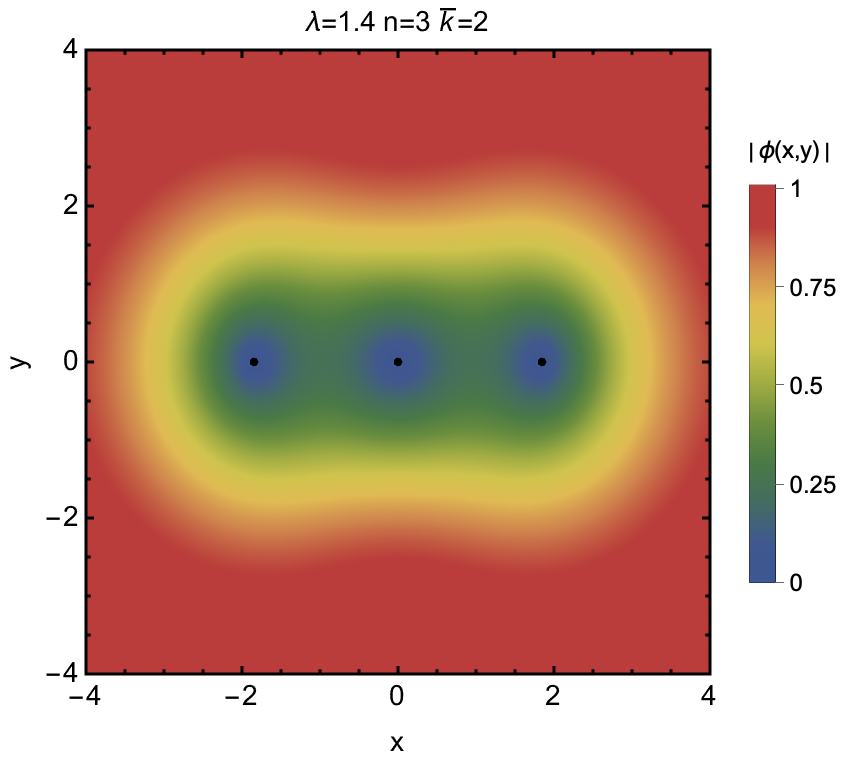}
        \includegraphics[width=0.32\textwidth]{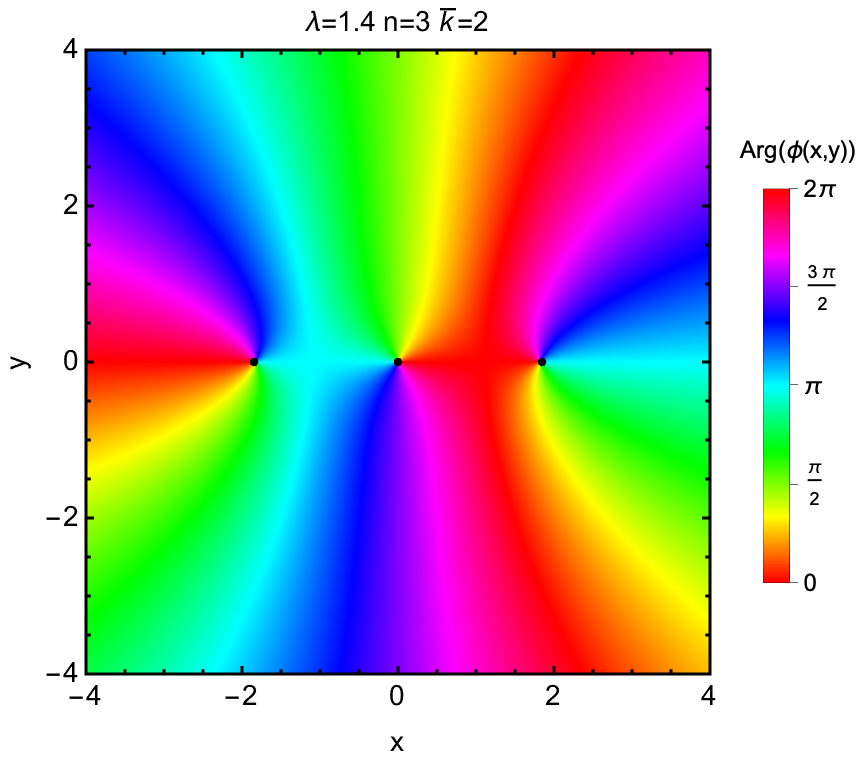}
    \raisebox{0.3\height}{\includegraphics[width=0.32\textwidth]{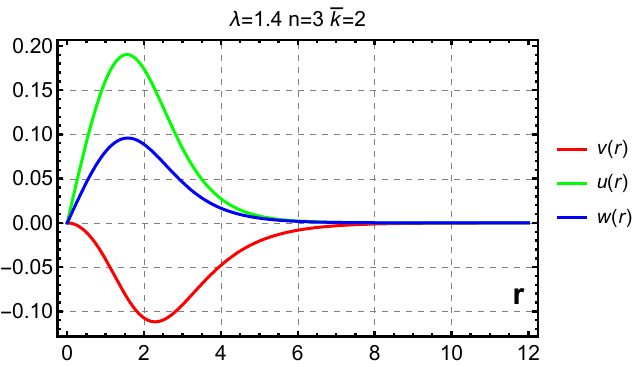}}
                      %  \vspace*{0.1cm}
        \end{minipage}
\\
        \hline

    %   & %\hline
           %\hline
 \multirow{3}{0.3cm}{ \rotatebox{90}{$n=3$ \hspace{2cm}$\overline{k}=3$ } } 
            %  \hline
&
      \rotatebox{90}{\hspace{-0.5cm}$\lambda=0.6$} &

        \begin{minipage}{0.53\textwidth}
        \centering
 \includegraphics[width=0.31\textwidth]{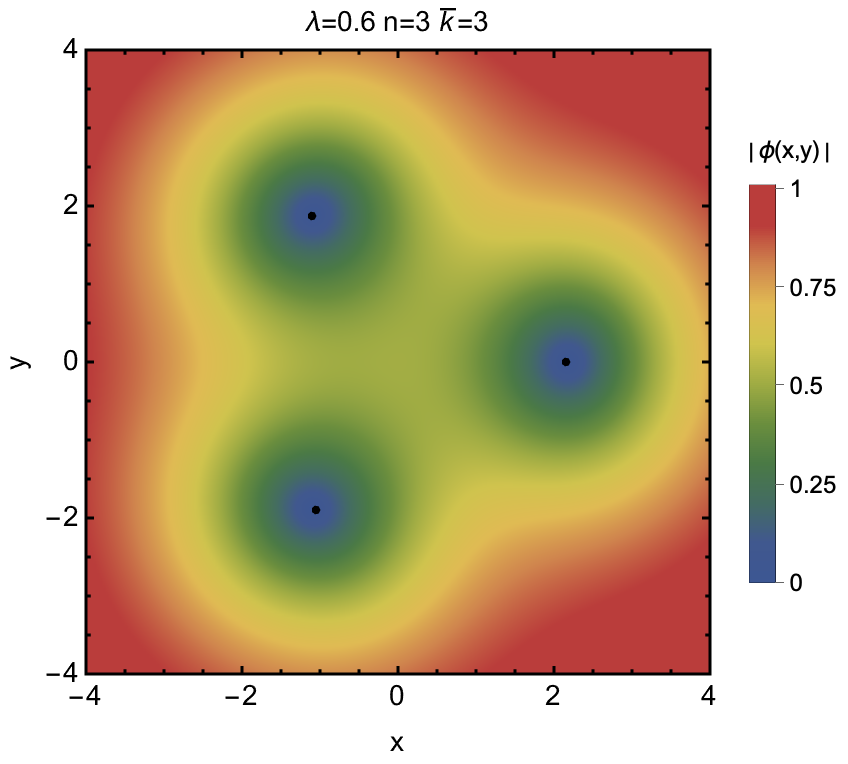}
        \includegraphics[width=0.32\textwidth]{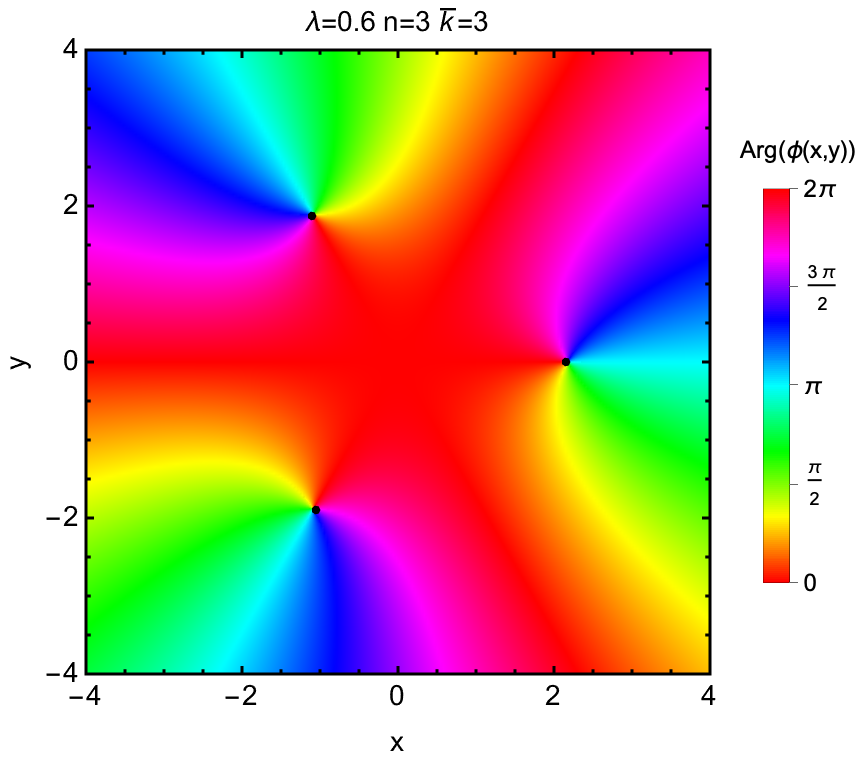}
    \raisebox{0.3\height}{\includegraphics[width=0.32\textwidth]{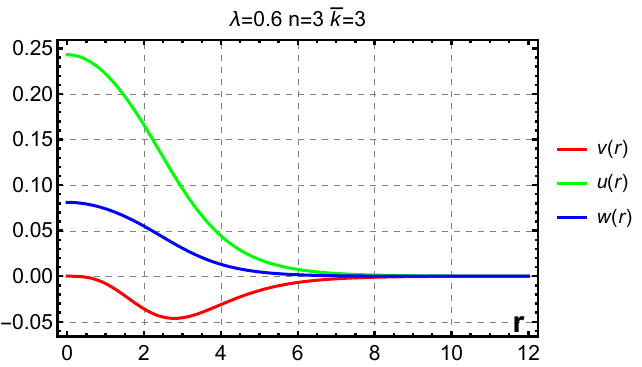}}
                      %  \vspace*{0.1cm}
        \end{minipage}
\\
        &     % \hline

      \rotatebox{90}{\hspace{-0.3cm}$\lambda=1$} &

        \begin{minipage}{0.53\textwidth}
        \centering
 \includegraphics[width=0.31\textwidth]{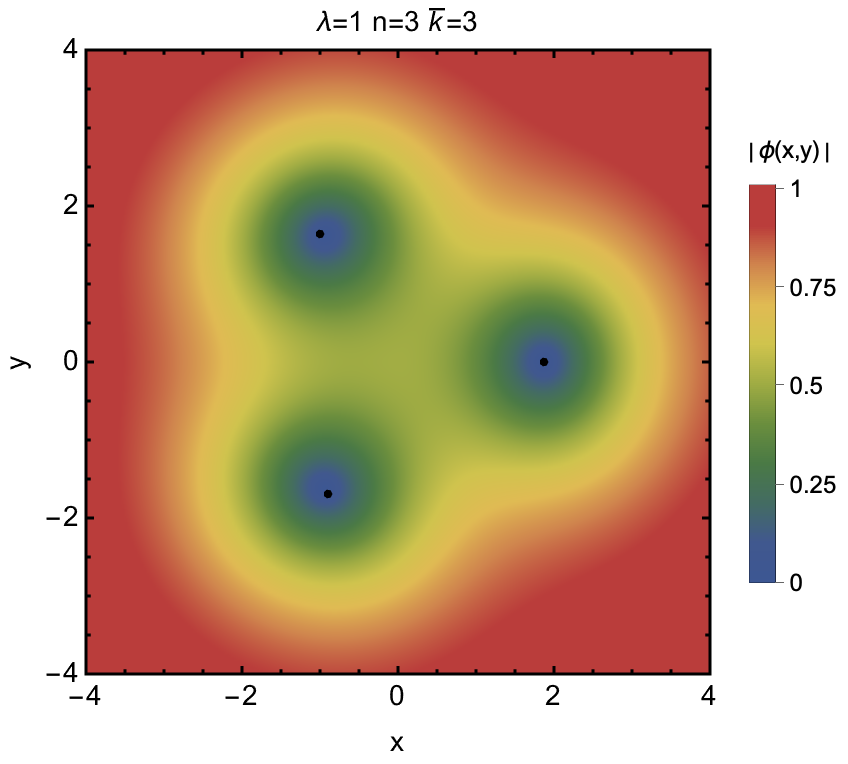}
        \includegraphics[width=0.32\textwidth]{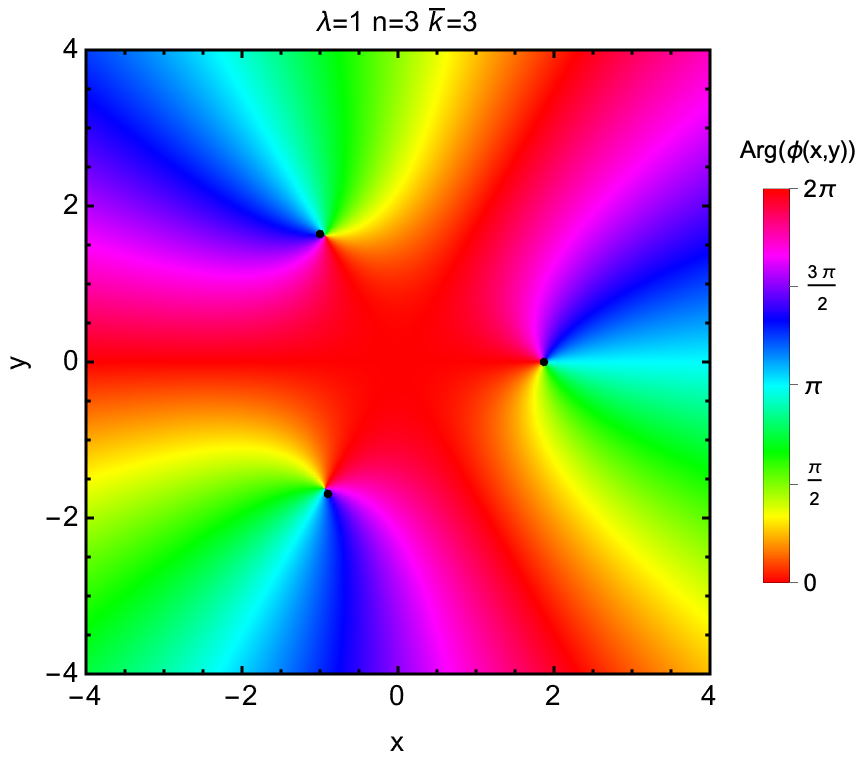}
    \raisebox{0.3\height}{\includegraphics[width=0.32\textwidth]{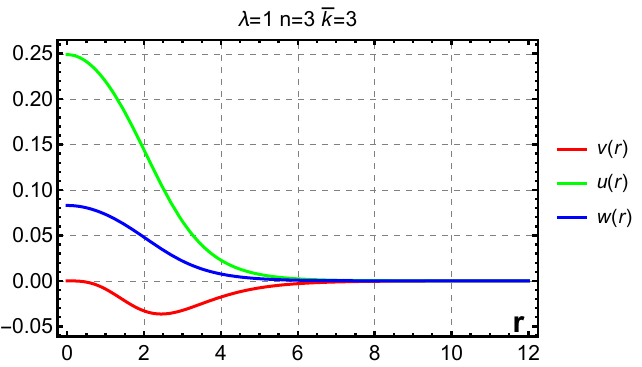}}
                      %  \vspace*{0.1cm}
        \end{minipage}
\\
      & % \hline

      \rotatebox{90}{\hspace{-0.5cm}$\lambda=1.4$} &

        \begin{minipage}{0.53\textwidth}
        \centering
 \includegraphics[width=0.31\textwidth]{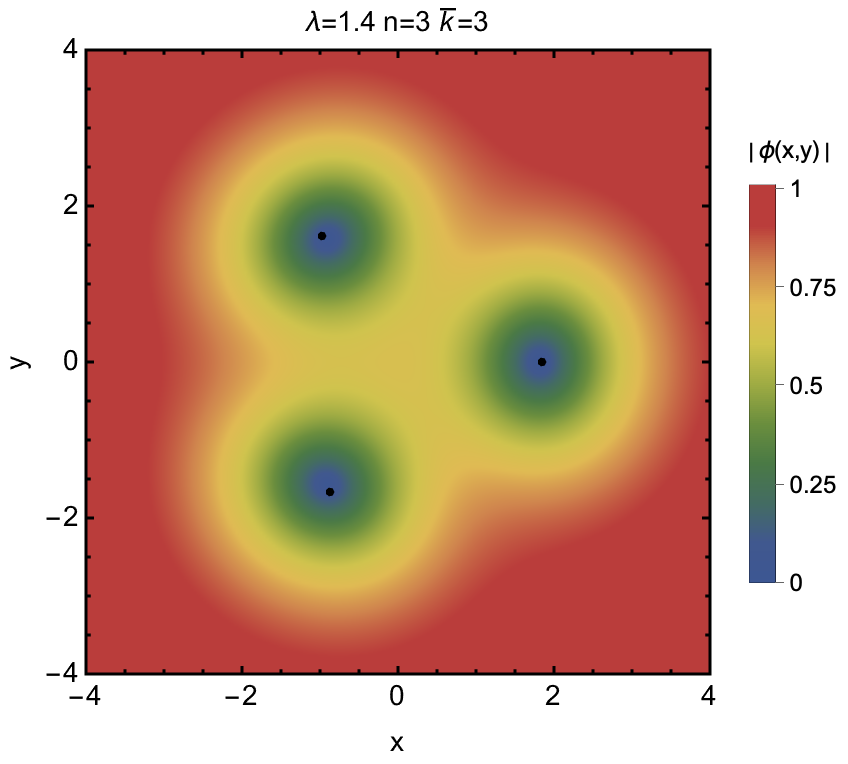}
        \includegraphics[width=0.32\textwidth]{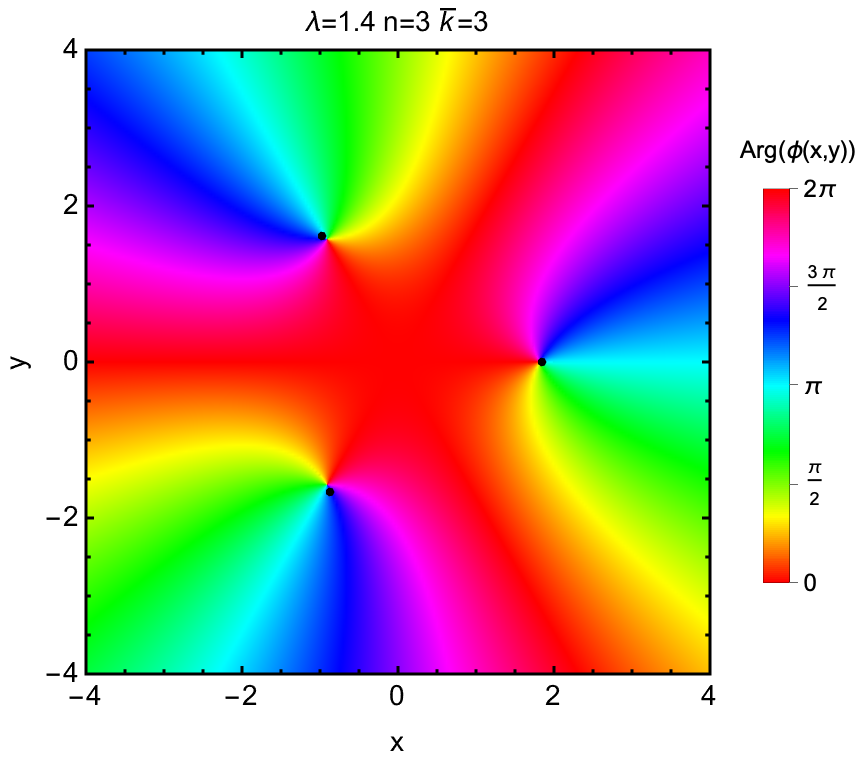}
    \raisebox{0.3\height}{\includegraphics[width=0.32\textwidth]{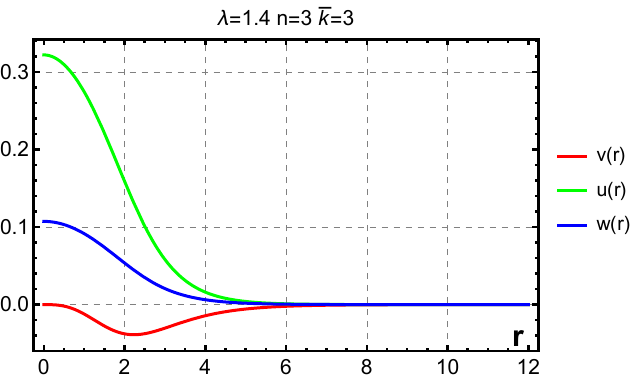}}
                      %  \vspace*{0.1cm}
        \end{minipage}
\\
        \hline
    \end{tabular}
    \caption{\textit{Type A internal modes for different values of $\lambda$ and $n=3$. The plots correspond to $|\Phi(x,y)|$, $\mathrm{Arg}(\Phi(x,y))$ once the internal modes have been added to the static vortex configurations and the profile functions $v(r)$, $u(r)$ and $w(r)$.}}
    \label{Tab2:AN3}
\end{table}

\begin{table}[htb]
    \centering
        %\hspace{1.cm}
        \vspace{-1.4cm}
    \begin{tabular}{|c|c|c|}
 
      \hline
 \multirow{3}{0.3cm}{ \rotatebox{90}{$n=1$ \hspace{2cm}$\overline{k}=0$ } } 
              &
          \rotatebox{90}{\hspace{-0.5cm}$\lambda=0.6$} &
            \begin{minipage}{0.5\textwidth}
        \centering
                            
        \includegraphics[width=0.43\textwidth]{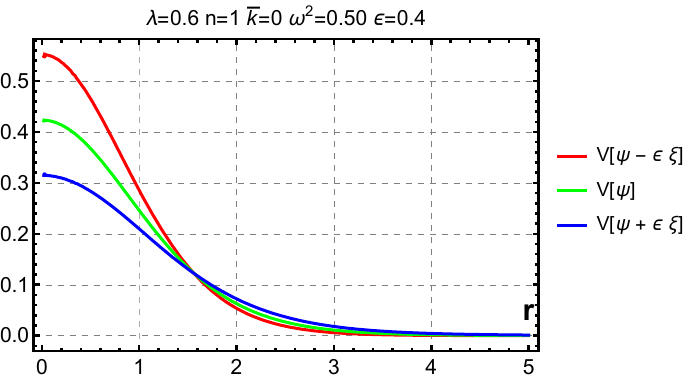}  
\includegraphics[width=0.43\textwidth]{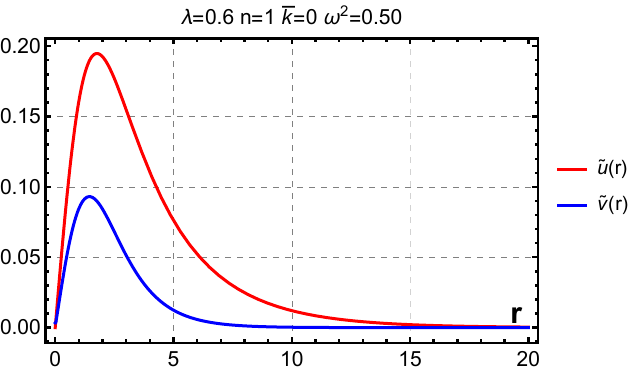}
              %  \vspace*{0.1cm}
        \end{minipage}
     
\\
 %   \tabularnewline \cline{2-3}    
   & \rotatebox{90}{\hspace{-0.3cm}$\lambda=1$}
    &

            \begin{minipage}{0.5\textwidth}
        \centering
        \includegraphics[width=0.43\textwidth]{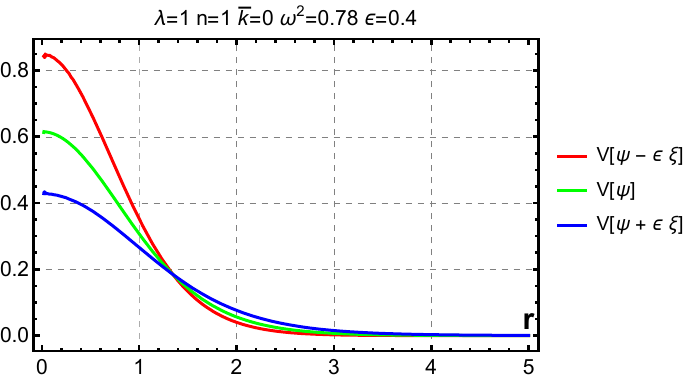}  
 \includegraphics[width=0.43\textwidth]{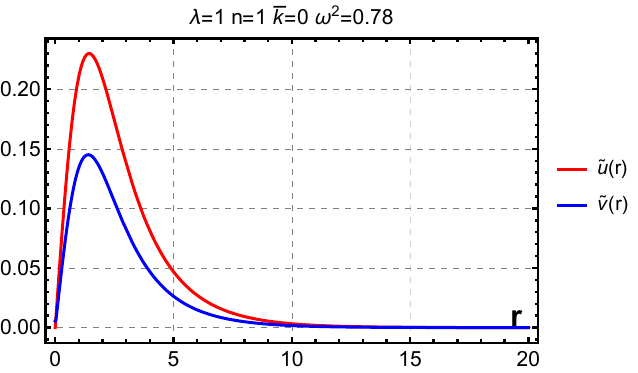}
                      %  \vspace*{0.1cm}
        \end{minipage}
 
\\
      % \tabularnewline \cline{2-3} 
      %\hline
   &   \rotatebox{90}{\hspace{-0.3cm}$\lambda=1.4$} &

            \begin{minipage}{0.5\textwidth}
        \centering
        \includegraphics[width=0.43\textwidth]{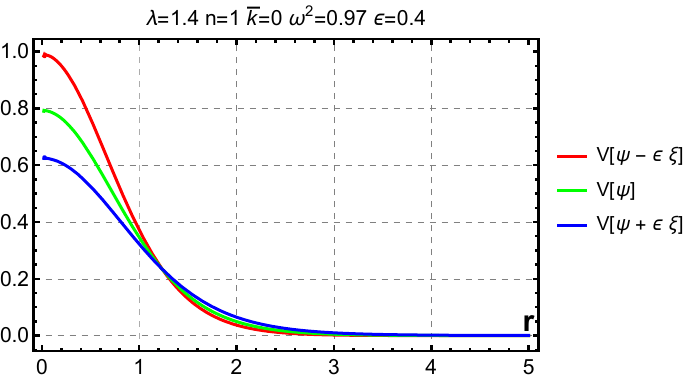}  
 \includegraphics[width=0.43\textwidth]{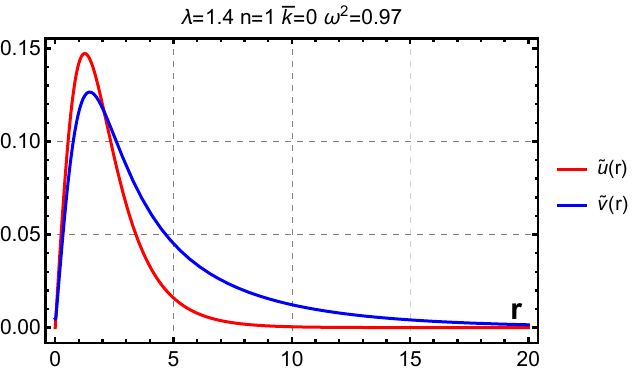}
                      %  \vspace*{0.1cm}
        \end{minipage}

\\

      \hline
 \multirow{3}{0.3cm}{ \rotatebox{90}{$n=2$ \hspace{2cm}$\overline{k}=0$ }} &      \rotatebox{90}{\hspace{-0.3cm}$\lambda=0.6$} &
            \begin{minipage}{0.5\textwidth}
        \centering
        \includegraphics[width=0.43\textwidth]{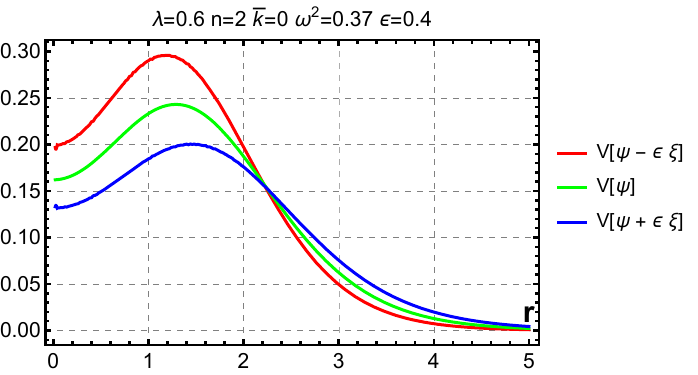}  
 \includegraphics[width=0.43\textwidth]{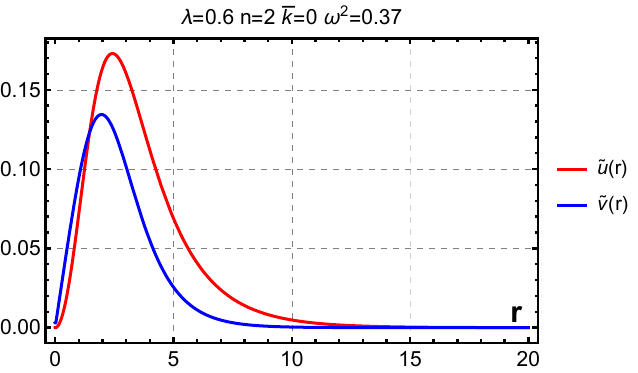}
                      %  \vspace*{0.1cm}
        \end{minipage}
\\
          %   \tabularnewline \cline{2-3}
             &  \rotatebox{90}{\hspace{-0.3cm}$\lambda=1$} &

          \begin{minipage}{0.5\textwidth}
        \centering
        \includegraphics[width=0.43\textwidth]{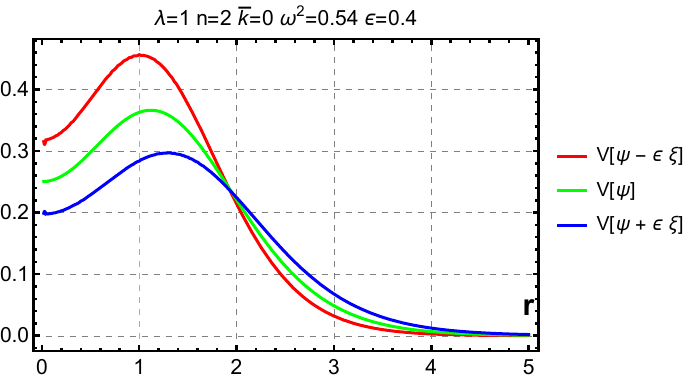}  
 \includegraphics[width=0.43\textwidth]{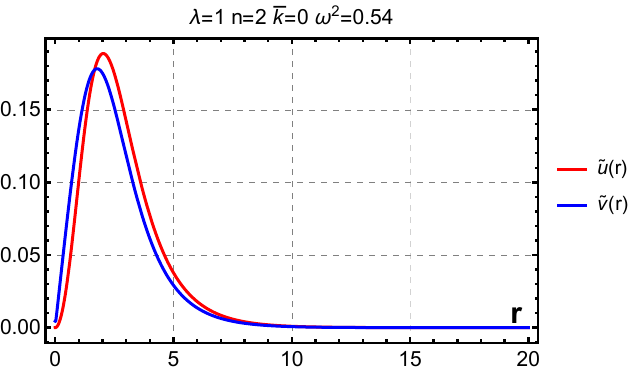}
                      %  \vspace*{0.1cm}
        \end{minipage}
\\
       % \tabularnewline \cline{2-3} 
          &
    \rotatebox{90}{\hspace{-0.3cm}$\lambda=1.4$} &

     \begin{minipage}{0.5\textwidth}
        \centering
        \includegraphics[width=0.43\textwidth]{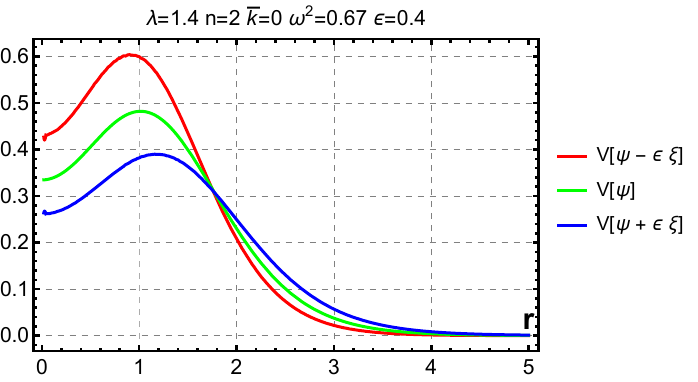}  
 \includegraphics[width=0.43\textwidth]{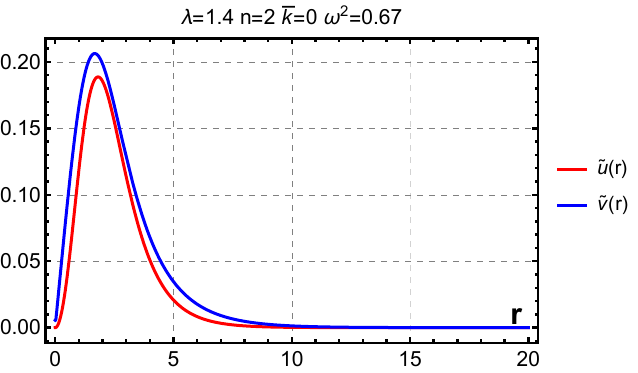}
                      %  \vspace*{0.1cm}
        \end{minipage}
\\
        \hline
    \end{tabular}
\caption{\textit{Profile functions $\widetilde{v}(r)$ and $\widetilde{u }(r)$ and variation of the energy density for Derrick mode types for a vortex with $\lambda=0.6, 1, 1.4$ and vorticity $n=1,2$. }}
    \label{Tab3}
\end{table}
\begin{table}[htb]
    %\centering
    \vspace{-1cm}
   %\vfill
        \hspace{3cm}
    \begin{tabular}{|c|c|}
      \hline   
   & $n=3$,  $\overline{k}=1$ \\
              \hline

      \rotatebox{90}{ \hspace{-0.3cm} $\lambda=0.6$} &
        \begin{minipage}{0.55\textwidth}
        \centering
                        
        \includegraphics[width=0.32\textwidth]{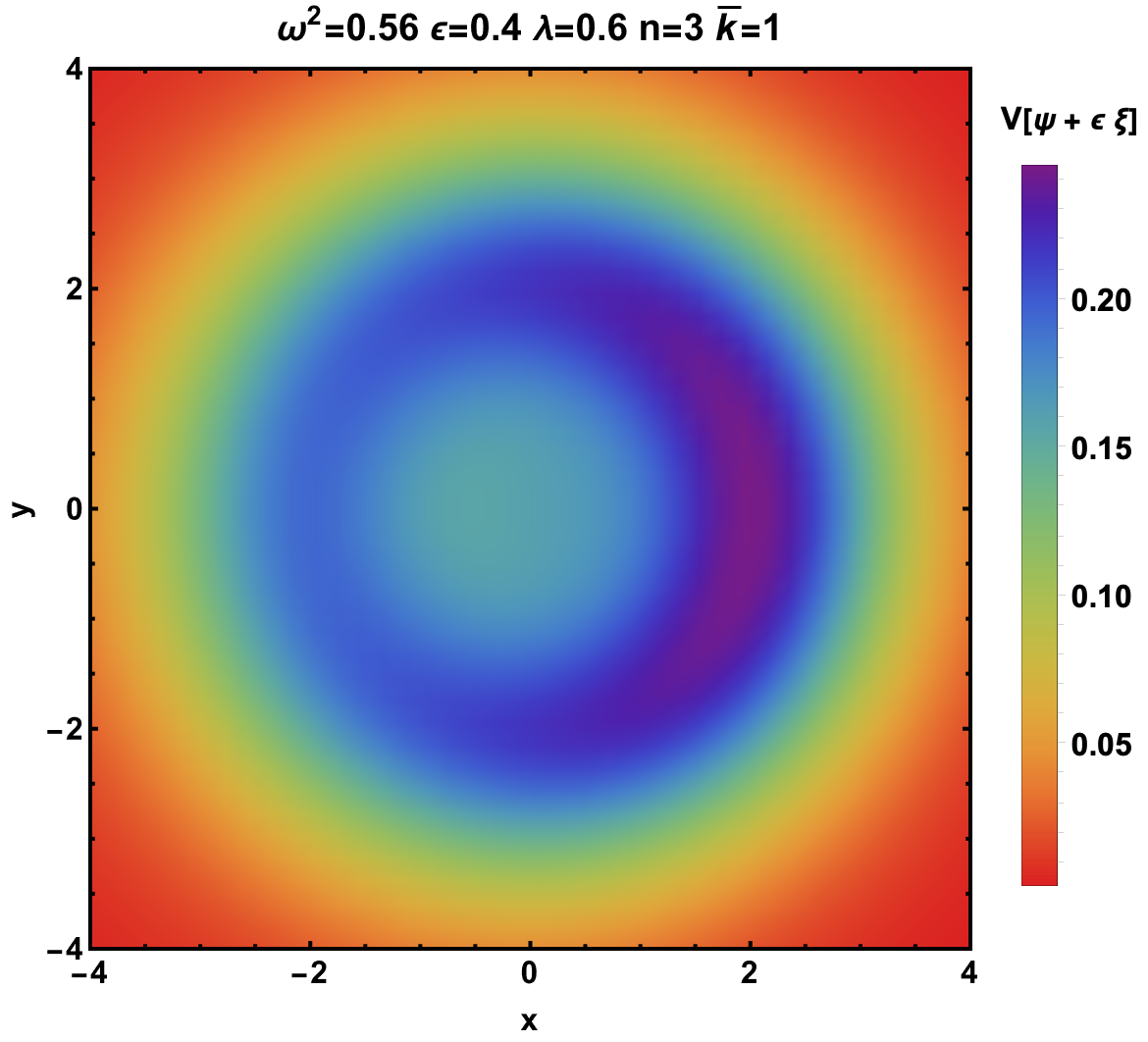}  
     \raisebox{0.25\height}{\includegraphics[width=0.41\textwidth]{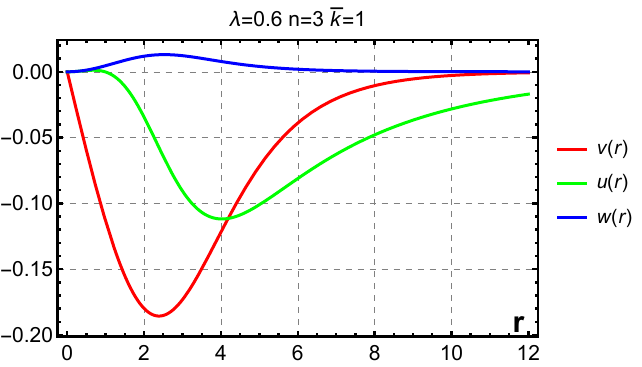}}
              %  \vspace*{0.1cm}
        \end{minipage}

\\
              \hline

      \rotatebox{90}{ \hspace{-0.3cm} $\lambda=1$} &
        \begin{minipage}{0.55\textwidth}
        \centering
                       
        \includegraphics[width=0.32\textwidth]{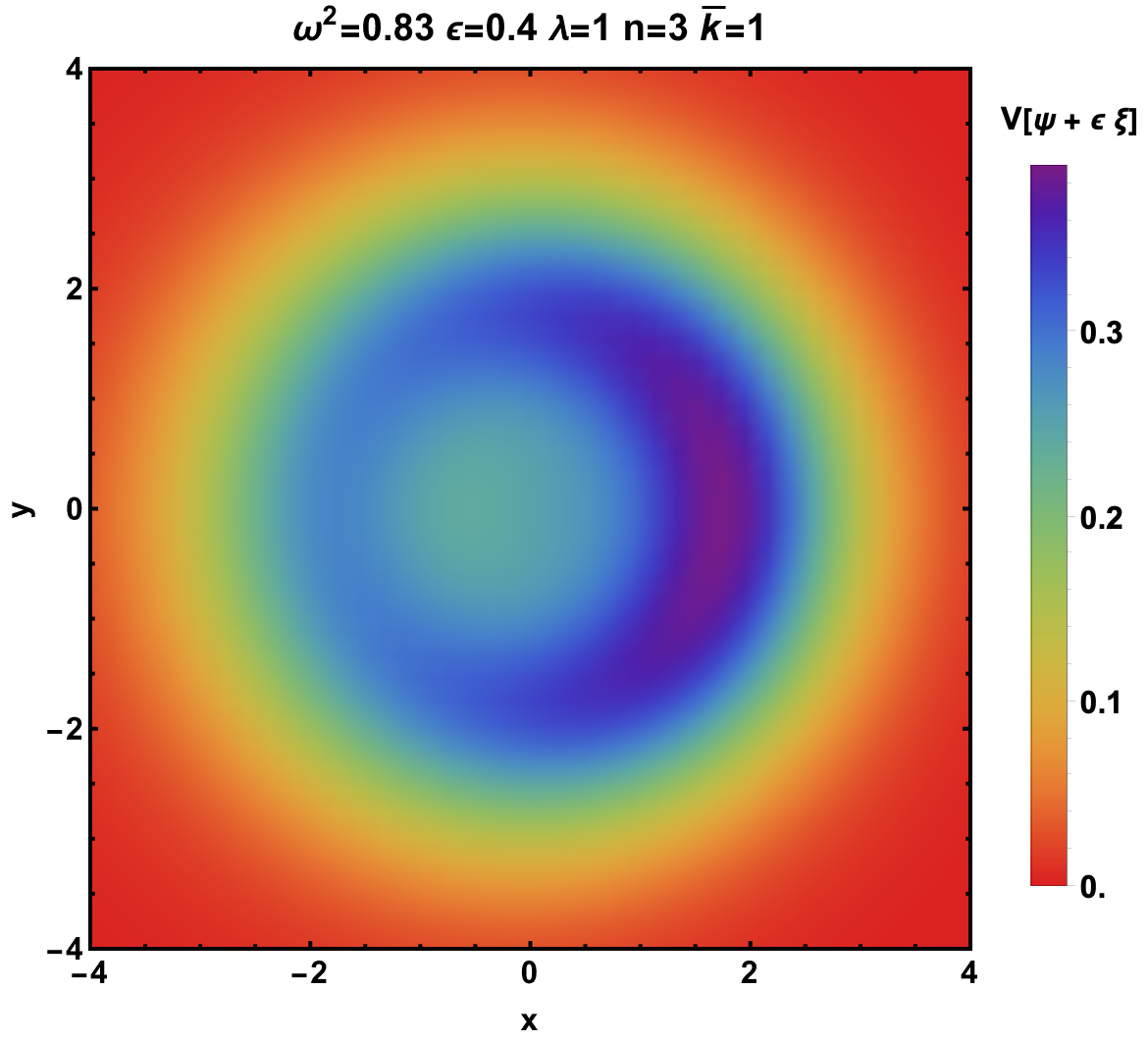}  
    \raisebox{0.25\height}{ \includegraphics[width=0.41\textwidth]{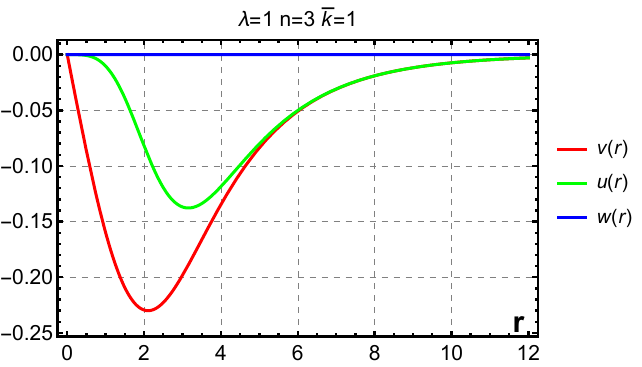}}
              %  \vspace*{0.1cm}
        \end{minipage}

\\
        \hline

       \rotatebox{90}{ \hspace{-0.5cm}$\lambda=1.2$ }&
        \begin{minipage}{0.55\textwidth}
        \centering

        \includegraphics[width=0.32\textwidth]{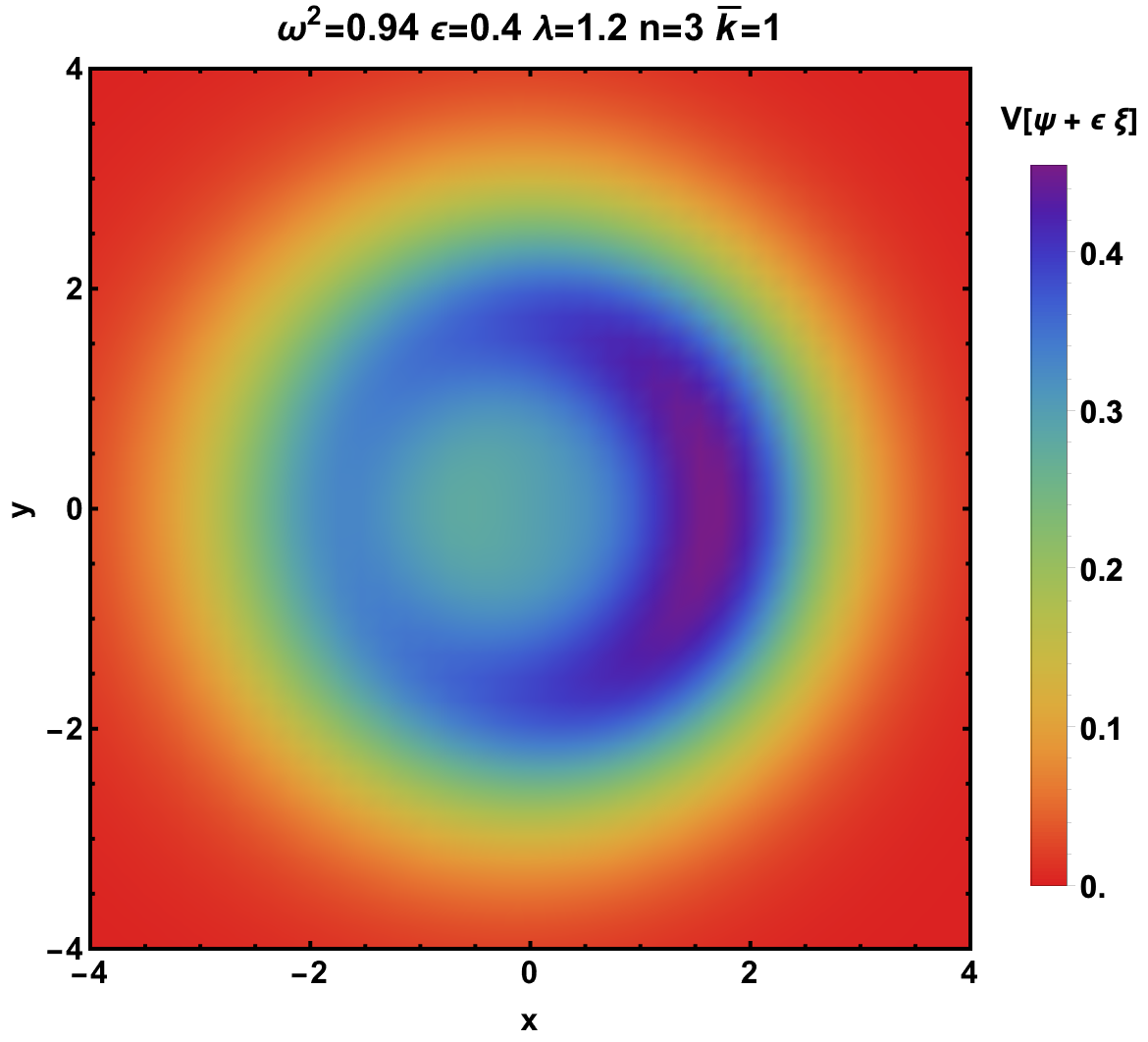}  
  \raisebox{0.25\height}{   \includegraphics[width=0.41\textwidth]{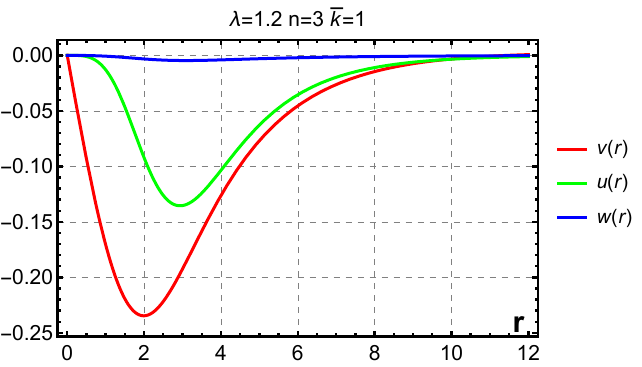}}
              %  \vspace*{0.1cm}
        \end{minipage}\\
        \hline
    \end{tabular}
  %  \caption{a}
     \caption{\textit{Profile functions $v(r)$, $u(r)$ and $w(r)$ for Type B multipolar modes for a vortex with $n=3$ and $\lambda=0.6,1,1.2$.}}
     \label{Tab4}
\end{table}

\clearpage
\newpage
\mbox{~}

\vspace{-1.5cm}
\section{Concluding remarks}\label{conclusions}

We have studied the internal structure of higher-charge $U(1)$ symmetric non-self-dual Abelian-Higgs vortices in $2+1$ dimensions. We have started our discussion analyzing the second order fluctuation operator and we have reviewed the internal fluctuations of the self-dual vortices. These fluctuations are classified into two sets: the zero and shape modes. The zero modes are responsible for the translations of the individual vortices and for a vortex of charge $n$ there are $2n$ of them. The shape modes correspond to fluctuations around the vortex and their number depends nontrivially on the vortex charge.  

The spectral structure of the non-self-dual vortices is richer and depends on the vortex type. For Type $I$ vortices we have found two zero modes. They correspond to the rigid translations of the vortex in two independent directions. The positive modes can be organized into two sets: the Derrick-type modes and the multipolar modes. The number of Derrick-type modes depends on both, the self-coupling $\lambda$ and the charge $n$. They correspond to axially symmetric fluctuations around the vortex center. 

On the other hand, the number of multipolar modes can be classified into two sets, Type A and Type B. The scalar and vector Type A modes have opposite dependence of the parameter $\bar{k}$ near the vortex center. Their frequency decreases monotonically with the self-coupling $\lambda$. At critical coupling, Type A modes become the zero modes of the BPS vortex. Therefore, there are in general $2n-2$ positive frequency Type A modes and $2$ zero modes, which gives a total of $2n$ zero modes in the critical case. Type B modes exhibit the same dependence in $\bar{k}$ near the origin and their frequency increases monotonically with $\lambda$. The number of Type B modes depends on both, the topological charge $n$ and the self-coupling constant $\lambda$.

 For Type $II$ vortices the situation is similar. Again, the number Derrick modes depends on both, the topological charge and the self-coupling. Their frequency grows monotonically with $\lambda$, and eventually, they dissappear into the continuum. As for Type $I$ vortices, the multipolar modes are continuously connected to the self-dual zero modes, but this time their frequency is negative. As a consequence, they become unstable modes that split the vortex into vortex configuration of smaller charge. 

 The zero modes (zero frequency multipolar modes) deserve a separate comment. For $n=1$ vortex, an infinitesimal excitation in the direction of the zero mode, corresponds, as expected, to a rigid translation of the vortex center. For $n>1$ vortices the situation is different. The position of the vortex is a zero of the scalar field with multiplicity $n$. An infinitesimal excitation in the direction of the zero mode only translates a zero multiplicity 1 from the origin, leaving a zero of multiplicity $n-1$ in the original position of the vortex. As a consequence, the linear excitation of the zero mode does not correspond to a rigid translation of the vortex. Instead, the translation of a zero of multiplicity $n$ requires a $n$-th order excitation in the direction of the zero mode.

Another interesting aspect, regarding the internal structure of the non-self-dual vortices in the existence of quasibound modes. As argued in \cite{AlonsoIzquierdo2024c}, and as it will be addressed in Chapter \ref{Chap4}, for large $\lambda$ the vortex core does not feel the magnetic field. This implies that the vortex behaves effectively as a global vortex. In this situation, the scalar modes of the Abelian-Higgs vortex remain bounded and approach the modes of the global vortex \cite{BlancoPillado2021, blanco2023}, while the vector modes become scattering states. As a consequence, the scalar-vector modes behave as quasibound modes. Therefore, for large $\lambda$ there is a correspondence between normal modes of the global vortex and quasibound modes of the Abelian-Higgs vortex. 

The decay of Derrick-type internal modes will be analyzed in detail in Chapter \ref{Chap4}, but the analysis of the dynamical properties of multipolar modes is left for a future investigation.
%The consequences of bound and quasibound modes for non-self-dual vortex scattering,  as well as their decay, are left for a future investigation. 

    \chapter{ Radiation emission of excited vortices}\label{Chap4}
    
This chapter is an adaptation from Reference \cite{AlonsoIzquierdo2024c}: 

\begin{figure}[htb]
    \centering
   \includegraphics[width=1\linewidth]{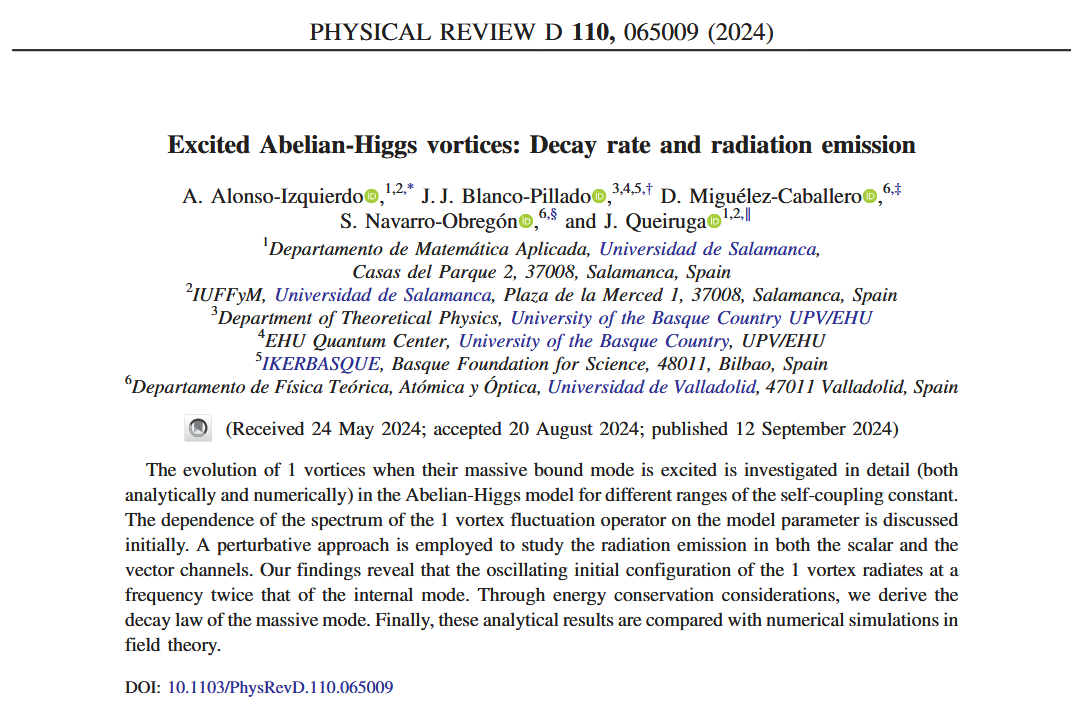}
\end{figure}

\section{Introduction}\label{Sec:1}

As stated in Chapter \ref{Chap3}, internal  modes may become particularly relevant when studying the dynamics of vortices
and their $3+1$ dimensional extensions, cosmic strings. The excitation of these solitonic structures may alter the mechanical properties of the strings, changing their energy and/or their tension. This suggests that further investigation into the dynamics of realistic strings extending beyond the thin wall approximation may be
necessary in some scenarios.

In fact, this idea has been recently suggested in \cite{Hindmarsh:2021mnl} as a possible way to reconcile the numerical results obtained in field theory simulations of cosmic string networks \cite{Hindmarsh:2017qff} and their Nambu-Goto counterparts in \cite{Blanco-Pillado:2013qja}. An initial examination of the dynamics of several loops from field theory simulations does not appear to indicate important deviations from the Nambu-Goto dynamics \cite{Blanco-Pillado:2023sap}. Therefore, the significance of these excitation modes for the conclusions of field theory simulations is not presently clear.

However, before we can understand the cosmological implications of these excited modes on strings 
one needs to investigate their most basic properties including their stability and potential decay rates via radiation emission. Note that understanding these properties is critical before extrapolating our results from a necessary limited numerical simulations to a cosmological context. 

In this chapter, we will thoroughly describe the evolution of a 1-vortex when its internal mode has been initially excited for different values of the self-coupling constant. For this purpose, we will employ the perturbative approach used in Chapter \ref{Chap1} and introduced in this manuscript in Section \ref{I2Manton}. %This analytical method has also been successfully employed to examine the evolution of kinks numerically \cite{Blanco-Pillado:2020smt, Navarro2023} as well as global vortices \cite{BlancoPillado2021,Blanco-Pillado:2022axf} and wobblers in two component scalar field theories \cite{AlonsoIzquierdo2023, AlonsoIzquierdo2024}. 
The resulting expressions of this perturbative analysis enable us to confirm that a 1-vortex radiates through scalar and gauge fields, with a predominant frequency which is twice that of the excited internal mode. It is clear that the coupling between the massive bound mode and the radiation modes will lead to the subsequent decay of the internal shape mode amplitude. Employing perturbation theory, we derive a temporal decay law for this mode which is well approximated by an inverse square root expression which is in very good agreement with our numerical simulations.

The backreaction of excitations of the longitudinal component of the $U(1)$ gauge field was studied in a 3+1 dimensional model in \cite{Arodz1996, Arodz1997}. Note that these are physically different modes than the ones studied here. In fact, they are not present in a $2+1$ dimensional vortex we study in the present paper. Furthermore, the mathematical approach and our results differ from the ones used in the aforementioned articles. We leave the discussion of these other modes for a future publication.

This chapter is organized as follows: In Section \ref{Sec:2}, a brief overview of the internal mode structure of the Abelian-Higgs 1-vortex is provided. Section \ref{Sec:3} offers a detailed perturbative study of the evolution of a 1-vortex whose internal mode has been initially excited. In Section \ref{Sec:4}, the validity of these results will be compared with those obtained from numerical simulations. Finally, Section \ref{Sec:5} summarizes the main findings of this work and outlines potential future prospects.

%\section{Some details about Derrick type modes }\label{Sec:2}
\section{Introduction to the problem}\label{Sec:2}
As previously stated, in this chapter we will only address the dynamics of excited Abelian-Higgs vortices for vortices with charge $n=1$ when they are triggered via a Derrick-type mode. We recall that these modes possess the same symmetry as the vortex itself, which implies that once triggered these modes change the size of the vortex. 
\begin{comment}
In order to identify the normal modes of the vortex we need to analyze the second-order small fluctuations operator ${\cal H}$. It has been demonstrated \cite{Izquierdo2016,AlonsoIzquierdo2016} that the lowest normal modes exhibit radial symmetry for vortex configurations with at least $n\leq 5$. Therefore, we will restrict our spectral analysis to fluctuations of the form $(\varphi(r),a_\theta(r))$\footnote{A more general ansatz for the internal mode structure along with the general eigenvalue problem can be found in \cite{Goodband1995}.} where $\varphi(r)$ and $a_\theta(r)$ denote respectively the fluctuations of the complex scalar field $\Phi$ and of the angular component of the vector field $A_\theta$. These perturbations have the same symmetry as the solutions derived from (\ref{Eq:betanfnEquations}) and satisfy the radial gauge condition, setting their radial component to zero, $a_r=0$. Therefore, the perturbed solution can be written as
\end{comment}

As mentioned in Chapter \ref{Chap3}, these modes can be found using the following ansatz:
\begin{equation}\label{Eq:Internalmodek0}
\Phi(r, \theta, t) = f_{n}(r)e^{i n \theta} + C_0\, \varphi(r)e^{i n \theta} \, e^{i \omega_n t} \hspace{0.0cm}, \hspace{0.4cm}  A_\theta(r, \theta, t) = \dfrac{n\,\beta_{n}(r)}{r} + C_0\,  a_\theta(r) \, e^{i \omega_n t} ,
\end{equation}
where $C_0$ is a small real number, and where we have used the convection $A_r=A_1 \cos\,\theta + A_2\sin\, \theta$ and $A_\theta= - A_1 \sin \, \theta + A_2\cos \,\theta$. 

Note that these fluctuations automatically satisfy the so-called background gauge condition \cite{Goodband1995,AlonsoIzquierdo2016}. If we plugged (\ref{Eq:Internalmodek0}) into  \eqref{eqI3:FE1}-\eqref{eqI3:FE2}, the equations of motion at linear order in $C_0$ lead to the spectral problem 
\begin{equation}\label{Eq:SecondOrderOperator}
\mathcal{H}
\begin{pmatrix}
    \varphi(r) \\
    a_\theta (r) \\
\end{pmatrix}
= \omega_{n,j}^2
\begin{pmatrix}
   \varphi(r) \\
    a_\theta (r) \\
\end{pmatrix}\,,
\end{equation}
where the subscript $n$ labels the vorticity of the configuration and $j$ labels the mode and where $\varphi(r)$ and $a_\theta(r)$ denote respectively the fluctuations of the complex scalar field $\Phi$ and of the angular component of the vector field $A_\theta$. The $n$-vortex fluctuation operator $\mathcal{H}$ reads
{
\begin{equation} \label{Eq:OperatorH}
    \mathcal{H}=
    \begin{pmatrix}
         -\dfrac{d^2}{dr^2}  - \dfrac{1}{r}\dfrac{d}{dr}  + \left(\dfrac{3}{2}\lambda f_{n}(r)^{2} -  \dfrac{\lambda}{2} + \dfrac{n^{2}}{r^{2}} - \dfrac{n^{2}\beta_{n}(r)}{r^{2}}\left( 2 - \beta_{n}(r) \right)\right) &  - \dfrac{2 n f_{n}(r)}{r}\left( 1 - \beta_{n}(r)\right)\\
         - \dfrac{2 n f_{n}(r)}{r}\left( 1 - \beta_{n}(r)\right) & - \dfrac{d^2}{dr^2}  - \dfrac{1}{r}\dfrac{d}{dr}  + \left(f_{n}(r)^2 + \dfrac{1}{r^2}\right)  
    \end{pmatrix}
    .
\end{equation}
}

The spectral problem (\ref{Eq:SecondOrderOperator}) couples the scalar and vector fluctuations. However, for large $r$ the fluctuation operator trivially decouples 
\begin{equation} \label{Eq:OperatorHInfinity}
    \mathcal{H}\big|_{\infty}=
    \begin{pmatrix}
         -\dfrac{d^2}{dr^2}  - \dfrac{1}{r}\dfrac{d}{dr}  + \lambda     &  0\\
        0 & - \dfrac{d^2}{dr^2}  - \dfrac{1}{r}\dfrac{d}{dr}  + \left(1 + \dfrac{1}{r^2}\right)  
    \end{pmatrix}
    \,.
\end{equation}

 The spectral problem given by the operator (\ref{Eq:OperatorHInfinity}) is analytically solvable. The (asymptotic) modes read
{
\begin{eqnarray}
    \varphi(r)&\hspace{-0.2cm}\xrightarrow[]{r\rightarrow\infty}\hspace{-0.2cm}&A_\varphi H_{0}^{(1)}(k_\phi r)+B_\varphi H_{0}^{(2)}(k_\phi r)\approx \sqrt{\dfrac{2}{\pi k_\phi r}}\left(A_\varphi e^{i(k_\phi r-\frac{\pi}{4})}+B_\varphi e^{-i(k_\phi r-\frac{\pi}{4})} \right), \label{Eq:AsymptoticPhi} \\
    a_\theta(r)&\hspace{-0.2cm}\xrightarrow[]{r\rightarrow\infty}\hspace{-0.2cm}&A_A H_{1}^{(1)}(k_A r)+B_A H_{1}^{(2)}(k_A r)\approx \sqrt{\dfrac{2}{\pi k_A r}}\left(A_A e^{i(k_A r-\frac{3\pi}{4})}+B_A  e^{-i(k_A r-\frac{3\pi}{4})} \right),\label{Eq:AsymptoticA}
\end{eqnarray}}
\hspace{-0.165cm}where $k_\phi^2=\omega_{n,j}^2-\lambda$, $k_A^2=\omega_{n,j}^2-1$, and $H_n^{(1)}$, $H_n^{(2)}$ are the Hankel functions of first and second kind respectively \cite{Abramowitz1972,NIST2010}. From \eqref{Eq:OperatorHInfinity}, it is clear that the continuum spectrum for the vector fluctuations starts at the threshold value $\omega_c^A= 1$ while for the scalar component depends on the coupling constant and starts at $\omega_c^\phi=\sqrt{\lambda}$, as exposed in Chapter \ref{Chap3}. Both of them coincide at critical coupling  ($\lambda=1$) where the self-dual vortices arise. As a consequence, bound states of (\ref{Eq:OperatorH}) must have eigenvalues $\omega_{n,j}^2<\lambda$ when $\lambda < 1$ and $\omega_{n,j}^2<1$ when $\lambda>1$. The numerical techniques applied to numerically solve \eqref{Eq:SecondOrderOperator} are identical to the ones that can be found in Appendix \ref{appen}.
%Due to the complexity of the spectral problem (\ref{Eq:SecondOrderOperator}) the discrete spectrum of (\ref{Eq:OperatorH}) must be obtained by employing numerical methods. The numerical scheme used in this paper is described in \ref{Sec:Appendix2}. 

In Figure \ref{fig:espectra1} we show the spectrum of the 1-vortex fluctuation operator (\ref{Eq:OperatorH}) as a function of the model parameter $\lambda$. We find that there is a single bound mode which ceases to exist at $\lambda\approx 1.5$.  
\begin{comment}
\begin{figure}[ht!]
    \centering{
   \includegraphics[width=0.55\linewidth]{Images/Chap4/Spectrumn1_V2_1.png}
    }
    \caption{\textit{Spectrum of the 1-vortex fluctuation operator $\mathcal{H}$, defined in \eqref{Eq:OperatorH}, as a function of the model parameter $\lambda$. The discrete eigenvalue corresponds to the blue curve, while the shaded red area represents the continuous spectrum.}}
    \label{Fig:Spectrumn1}
\end{figure}
\end{comment}

\begin{figure}[ht!]
    \centering
    
    \begin{subfigure}{0.489\textwidth}
        \includegraphics[width=\linewidth]{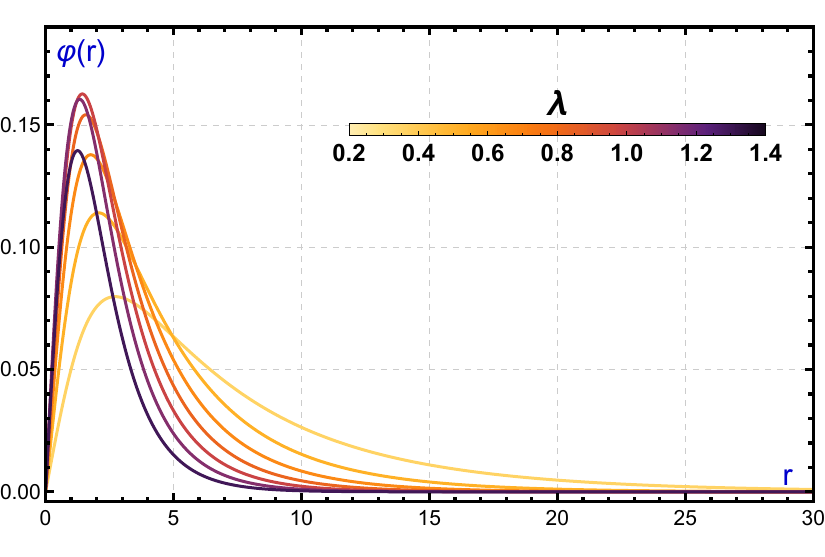}
    \end{subfigure}
    \hfill
    \begin{subfigure}{0.489\textwidth}
        \includegraphics[width=\linewidth]{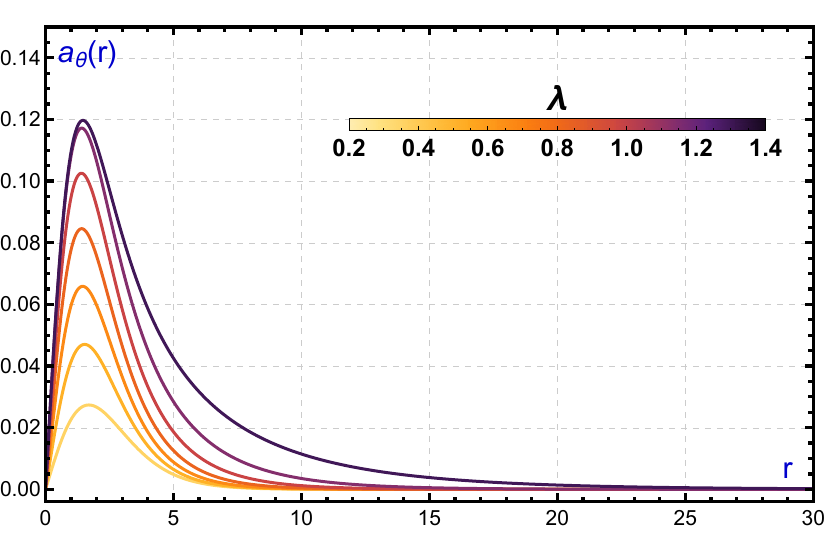}
    \end{subfigure}
  
    \caption{\textit{Scalar (left) and vector (right) component profiles of the 1-vortex bound mode for different values of the self-coupling constant $\lambda$.}}
    \label{Fig:Profiles_BoundMode}
\end{figure}

Finally, in Figure \ref{Fig:Profiles_BoundMode} we show the profiles of the 1-vortex bound mode for different values of the self-coupling constant $\lambda$.

\section{Internal mode evolution: Analytical approach}\label{Sec:3}

In this section we will determine the decay law for a circularly symmetry vortex of charge $n$ with a single internal mode excited. Then, for concreteness, we will compare our analytical results with field theory for the 1-vortex. 

%The perturbative approach used to analyze the decay law of the wobbling amplitude for a kink was originally introduced in \cite{Manton1997}, and later extended to study kinks numerically in \cite{Blanco-Pillado:2020smt} () as well as global vortices in \cite{BlancoPillado2021}. Here we employ similar techniques to derive the radiation emitted by an excited vortex. 

The perturbative approach that will be used was already explained in detail in Chapter \ref{I2Manton}. This method was originally introduced in \cite{Manton1997}, and later extended to study kinks numerically in \cite{Blanco-Pillado:2020smt}  as well as global vortices in \cite{BlancoPillado2021}. This technique was also used in Chapter \ref{Chap1} to analyze the decay and radiation emission of wobbling kinks in the \textit{double $\phi^4$ model}. 
Then, by energy considerations, we will compute the decay law for the internal modes.

\subsection{Perturbative approach}

We will employ the following circularly symmetric ansatz
\begin{eqnarray}
\Phi(r, \theta, t) &=& f_{n}(r)e^{i n \theta} + C(t)\varphi(r)e^{i n \theta} + \eta(r, t)e^{i n \theta}, \label{eq:phi_expansion} \\ 
A_{\theta}(r, t) &=& \dfrac{n\,\beta_{n}(r)}{r} + C(t) a_{\theta}(r) + \xi(r, t), \label{eq:a2_expansion}
\end{eqnarray}
where $\eta(r,t)$ and $\xi(r,t)$ represent the scalar and vector radiation field, $(\varphi(r), a_\theta(r))$  are the bound mode profiles and $C(t)$ is their time-dependent amplitude. A straightforward computation shows that $(\ref{eq:phi_expansion})$-$(\ref{eq:a2_expansion})$ still verify the temporal gauge and the background gauge conditions. 

To investigate the asymptotic radiation emitted due to the excitation of the internal mode, we insert the radially perturbed solution (\ref{eq:phi_expansion})-(\ref{eq:a2_expansion}) into the field equations \eqref{eqI3:FE1}-\eqref{eqI3:FE2}. The massive bound mode satisfies by definition the equations at first order in $C(t)$. However, the modes couple at higher order to radiation. Expanding at second order in $C(t)$ we derive the following equation of motion for the scalar field radiation {
\begin{eqnarray}\label{eq:diff_eta_k0}
& & \Ddot{\eta}(r,t) - \eta''(r,t) - \dfrac{1}{r}\eta'(r,t)  + \left(\dfrac{3}{2}\lambda f_{n}(r)^{2} -  \dfrac{\lambda}{2} + \dfrac{n^{2}}{r^{2}} - \dfrac{n^{2}\beta_{n}(r)}{r^{2}}\left( 2 - \beta_{n}(r) \right)\right)\eta(r,t) - \nonumber \\
& &  - \dfrac{2 n f_{n}(r)}{r}\left( 1 - \beta_{n}(r)\right)\xi(r,t) = \frac{2n C(t)^2 \varphi(r)a_{\theta}(r)}{r}(1-\beta (r)) - C(t)^2 f_{n}(r)a_{\theta}(r)^2 -  \\
& & - \frac{3}{2}\lambda \,C(t)^2 f_{n}(r)\varphi(r)^2 - \left(\Ddot{C}(t) + \omega_s^2 C(t)\right)\varphi(r).\nonumber
\end{eqnarray}}
\vspace{-0.2cm}
For the gauge field radiation we get{
\begin{eqnarray}\label{eq:diff_xi_k0}
& & \Ddot{\xi}(r,t) - \xi''(r,t) - \dfrac{1}{r}\xi'(r,t) + \left(f_{n}(r)^2 + \dfrac{1}{r^2}\right) \xi(r,t) - \dfrac{2n f_{n}(r)}{r}\left(1 - \beta_{n}(r) \right)\eta(r,t) = \nonumber\\ 
& & = \dfrac{n \,C(t)^2 \varphi^2(r)}{r}\left(1 - \beta_{n}(r)\right) - 2 C(t)^2 f_{n}(r)\varphi(r)a_{\theta}(r) - \left(\Ddot{C}(t) + \omega_s^2 C(t)\right) a_{\theta}(r).
\end{eqnarray}}
\vspace{-0.1cm}
Using the fact that the discrete modes are orthogonal to the radiation modes we can now project (\ref{eq:diff_eta_k0}) and (\ref{eq:diff_xi_k0}) onto $\varphi(r)$ and $a_\theta(r)$ to obtain,
\begin{equation}\label{eq:C_02}
\Ddot{C}(t) + \omega_{n,j}^2 C(t) + C(t)^2\gamma = 0,
\end{equation}
\vspace{-0.6cm}
where
\begin{equation} \label{eq:norma}
\gamma = \dfrac{2 \pi}{\mathcal{N}}\int_{0}^{\infty}\bigg( \dfrac{3}{2}\lambda f_{n}(r)\varphi(r)^3 + 3f_{n}(r)\varphi(r)a_{\theta}(r)^2 - \dfrac{3 n \varphi(r)^2 a_{\theta}(r)}{r}(1 - \beta_{n}(r)) \bigg)\,r dr,
\end{equation}
and $\mathcal{N}$ in (\ref{eq:norma}) is the normalization factor of the shape mode
\begin{equation}\label{eq:factor_N}
\mathcal{N} = 2\pi \int_{0}^{\infty}\left(\varphi(r)^2 + a_{\theta}(r)^2\right) r\, dr .
\end{equation} 
\vspace{-0.1cm}
The solution to the second order equation \eqref{eq:C_02} can be approximated as
\begin{equation}\label{eq:first}
C(t) = C_0\cos(\omega_{n,j} t). 
\end{equation}

\vspace{-0.6cm}
Furthermore, we will assume the following ansatz for the radiation
modes
\begin{eqnarray}\label{eq:rad_1}
\eta(r, t) \!\!\!\!&=&\!\!\!\! \eta_{r}(r)\,e^{i \omega t}~,\\ \label{eq:rad_2}
\xi(r,t) \!\!\!\!&=&\!\!\!\! \xi_{r}(r)\,e^{i \omega t}~,
\end{eqnarray}
which, together with first order solution (\ref{eq:first}) leads to the condition $\omega = 2\omega_{n,j}$, i.e. the radiation frequency in twice that of the discrete mode. After substituting (\ref{eq:rad_1}), (\ref{eq:rad_2}) and (\ref{eq:C_02}) into (\ref{eq:diff_eta_k0}) and (\ref{eq:diff_xi_k0}) we get the following equations
{\small
\begin{eqnarray}\label{eq:diff_eta_2}
& &\hspace{-0.6cm} - \eta_r''(r) - \dfrac{1}{r}\eta_r'(r)  + \left(\dfrac{3}{2}\lambda f_{n}(r)^{2} -  \dfrac{\lambda}{2} + \dfrac{n^{2}}{r^{2}} - 4\omega_{n,j}^2 - \dfrac{n^{2}\beta_{n}(r)}{r^{2}}\left( 2 - \beta_{n}(r) \right)\right)\eta_{r}(r) - \dfrac{2 n f_{n}(r)}{r}\left( 1 - \beta_{n}(r)\right)\xi_r(r) = \nonumber \\
& &\hspace{-0.6cm} = \frac{n C_0^2 \varphi(r)a_{\theta}(r)}{r}(1-\beta_{n}(r)) - \dfrac{1}{2}C_0^2 f_{n}(r)a_{\theta}(r)^2 - \frac{3}{4}\lambda C_0^2 f_{n}(r)\varphi(r)^2 + \dfrac{1}{2}C_0^2\gamma \varphi(r) \equiv C_0^2 F_{\phi}(r)~,
\end{eqnarray}
}and 
\begin{eqnarray}\label{eq:diff_xi_2}
& & - \xi_r''(r) - \dfrac{1}{r}\xi_r'(r) + \left(f_{n}(r)^2 + \dfrac{1}{r^2} - 4\omega_{n,j}^2\right) \xi_r(r) - \dfrac{2n f_{n}(r)}{r}\left(1 - \beta_{n}(r) \right)\eta_{r}(r) = \nonumber\\ 
& & = \dfrac{n C_0^2 \varphi^2(r)}{2r}\left(1 - \beta_{n}(r)\right) - C_0^2 f_{n}(r)\varphi(r)a_{\theta}(r) + \dfrac{1}{2}C_0^2\gamma a_{\theta}(r) \equiv C_0^2 F_{A}(r)~.
\end{eqnarray}

The equations (\ref{eq:diff_eta_2})-(\ref{eq:diff_xi_2}) constitute a coupled system of non-homogeneous linear ordinary differential equations. It should be noted, however, that the cross-terms exponentially vanish for large $r$. In order to solve the system we may use an iterative approach  similar to Bohr's approximation. The procedure works as follows: First, we apply the method of variation of parameters as if the coupling terms were part of the inhomogeneous terms. Then, a particular solution is given by{
\begin{eqnarray}
\eta_{r}^{(m)}(r) \!\!\!\!&=& \!\!\!\!- C_0^2\, z_{2\phi}(r) \int_{0}^{r} \left(F_{\phi}(r') - \dfrac{2 n f_{n}(r')}{r'}\left( 1 - \beta_{n}(r')\right)\xi_r^{(m - 1)}(r') \right) \dfrac{z_{1\phi}(r')}{W_{\phi}(r')}dr'\nonumber \\ 
& & - C_0^2\, z_{1\phi}(r) \int_{r}^{\infty} \left(F_{\phi}(r') - \dfrac{2 n f_{n}(r')}{r'}\left( 1 - \beta_{n}(r')\right)\xi_r^{(m - 1)}(r') \right) \dfrac{z_{2\phi}(r')}{W_{\phi}(r')}dr',\label{eq:VarPar_eta_k0}\\
\xi_r^{(m)}(r) \!\!\!\!&=&\!\!\!\! - C_0^2\, z_{2 A}(r) \int_{0}^{r} \left(F_{A}(r') - \dfrac{2 n f_{n}(r')}{r'}\left( 1 - \beta_{n}(r')\right)\eta_{r}^{(m - 1)}(r') \right) \dfrac{z_{1 A}(r')}{W_{A}(r')}dr'\nonumber\\ 
& & - C_0^2\, z_{1 A}(r) \int_{r}^{\infty} \left(F_{A}(r') - \dfrac{2 n f_{n}(r)}{r}\left( 1 - \beta_{n}(r)\right)\eta_{r}^{(m - 1)}(r') \right) \dfrac{z_{2 A}(r')}{W_{A}(r')}dr', \label{eq:VarPar_xi_k0}
\end{eqnarray}}
where  $z_{j\phi}$ and $z_{j A}$ ($j=1,2$) respectively denote the two linearly independent homogeneous solutions for the scalar and vector components $\eta$ and $\xi$ of (\ref{eq:diff_eta_2})-(\ref{eq:diff_xi_2}), and $F_{\phi}$ and $F_{A}$ account for their non-homogeneous terms as indicated in (\ref{eq:diff_eta_2})-(\ref{eq:diff_xi_2}). Besides, $W_{\phi}$ and $W_{A}$ are respectively the Wronskians associated to the homogeneous solutions $z_{j\phi}$ and $z_{j A}$, $j=1,2$. The index $m$ in (\ref{eq:VarPar_eta_k0})-(\ref{eq:VarPar_xi_k0}) indicates the iteration step, with $\eta_{r}^{(0)}(r)=0$ and $\xi_r^{(0)}(r)=0$. Although the solutions of the homogeneous system associated to (\ref{eq:diff_eta_2})-(\ref{eq:diff_xi_2}) cannot be analytically identified, it is possible to find their asymptotic behaviors (the equations decouple in this limit as shown in Section \ref{Sec:2}). We have that
\begin{eqnarray}
& & z_{\phi}(r) \xrightarrow{r \longrightarrow \infty} c_1 J_0(q_{\phi} \, r) + c_2 Y_0(q_{\phi}\, r ),\\  
& & z_{A}(r) \xrightarrow{r \longrightarrow \infty} d_1 J_1(q_{A}\, r ) + d_2 Y_1(q_{A}\, r )~,
\end{eqnarray}
where $J_n$ and $Y_n$ are the Bessel $J$ and Bessel $Y$ functions, respectively, and 
\begin{equation}
q_{\phi} = \sqrt{4 \,\omega_{n,j}^2 - \lambda}, \quad q_{A} = \sqrt{4 \, \omega_{n,j}^2 - 1}~.
\end{equation}
We choose the asymptotic radiation to be outgoing waves in the radial direction, then 
\begin{eqnarray}
z_{1\phi}(r) & \xrightarrow{r \longrightarrow \infty} & \Tilde{c}_1\,J_0(q_{\phi} \,r )~, \hspace{0.3cm} z_{2\phi}(r)  \xrightarrow{r \longrightarrow \infty} H^{(2)}_{0}(q_{\phi} \, r)\label{eq:asym_eta}~, \\
z_{1 A}(r) & \xrightarrow{r \longrightarrow \infty} & \Tilde{d}_2\,Y_1( q_{A}\,r)~, \hspace{0.3cm} z_{2 A}(r) \xrightarrow{r \longrightarrow \infty} H^{(2)}_{1}(q_{A}\,r)\label{eq:asym_xi}~,
\end{eqnarray}
where, once again, $H_n^{(2)}$ denote the Hankel function of second kind. The asymptotic expansion of (\ref{eq:VarPar_eta_k0})-(\ref{eq:VarPar_xi_k0}) reduces to 
\begin{eqnarray}
\eta_{r}^{(m)}(r) \!\!\!\!&=& \!\!\!\!- C_0^2\,\sqrt{\dfrac{2}{\pi r q_{\phi}}}  \cdot \, I^{(m)}_{\phi} \cdot  \, e^{-i r q_{\phi}+\frac{i \pi }{4}} , \label{eq:int_eta}\\ 
\xi_r^{(m)}(r)\!\!\!\! &=& \!\!\!\!- C_0^2\,\sqrt{\dfrac{2}{\pi r q_{A}}}  \cdot \, I^{(m)}_{A} \cdot  \, e^{-i r q_{A}+\frac{3 i \pi }{4}} \label{eq:int_xi},
\end{eqnarray}
where the factors $I^{(m)}_{\phi}$ and $I^{(m)}_{A}$ are defined by the following integrals
\begin{eqnarray}
I^{(m)}_{\phi} \!\!\!\!&=&\!\!\!\! \int_{0}^{\infty} \left(F_{\phi}(r') - \dfrac{2 n f_{n}(r')}{r'}\left( 1 - \beta_{n}(r')\right)\xi_r^{(m - 1)}(r') \right) \dfrac{z_{1\phi}(r')}{W_{\phi}(r')}dr', \label{integral1}\\
I^{(m)}_{A}\!\!\!\! &=&\!\!\!\! \int_{0}^{\infty} \left(F_{A}(r') - \dfrac{2 n f_{n}(r')}{r'}\left( 1 - \beta_{n}(r')\right)\eta_{r}^{(m - 1)}(r') \right) \dfrac{z_{1 A}(r')}{W_{A}(r')}dr'. \label{integral2}
\end{eqnarray}

To compute the integrals (\ref{integral1}) and (\ref{integral2}), it is necessary to identify numercally the homogeneous solutions $z_{1\phi}$ and $z_{1 A}$ with the boundary conditions $z_{1\phi}(0) = z_{1 A}(0) = 0$ and those given by the asymptotic behavior (\ref{eq:asym_eta}) and (\ref{eq:asym_xi}) (see Reference \cite{BlancoPillado2021}). After some straightforward algebraic manipulations we finally get
\begin{eqnarray}
\hspace{-1cm}\eta^{(m)}(r,t) \hspace{-0.2cm}&\hspace{-0.1cm}\approx\hspace{-0.1cm}&\hspace{-0.3cm} C_0^2\, {\mathrm{ Re}} \, \left[\dfrac{C^{(m)}_{\phi}}{\sqrt{r}}e^{i2\omega_{n,j} t - i r q_{\phi}+\frac{i \pi }{4}} \right] = C_0^2\dfrac{C^{(m)}_{\phi}}{\sqrt{r}}\,\cos \Big(2\omega_{n,j} t - r q_{\phi} + \dfrac{\pi}{4} + \zeta_{\phi} \Big) \label{eq:rad_eta}~, \\
\hspace{-1cm}\xi^{(m)}(r,t) \hspace{-0.3cm}&\approx& \hspace{-0.3cm}C_0^2\, {\mathrm{ Re}} \, \left[\dfrac{C^{(m)}_{A}}{\sqrt{r}}e^{i2\omega_{n,j} t - i r q_{A}+\frac{3i \pi }{4}} \right] = C_0^2\dfrac{C^{(m)}_{A}}{\sqrt{r}}\,\cos \Big (2\omega_{n,j} t - r q_{A} + \dfrac{3\pi}{4} + \zeta_A \Big) \label{eq:rad_xi}~,
\end{eqnarray}
where the normalized radiation amplitudes $C^{(m)}_{\phi}$ and $C^{(m)}_{A}$ are given by
\begin{equation} \label{eq:C_amplitudes} 
C^{(m)}_{\phi} = \sqrt{\dfrac{2}{\pi q_{\phi}}} |I^{(m)}_{\phi}| \hspace{0.5cm} \mbox{and} \hspace{0.5cm} C^{(m)}_{A} = \sqrt{\dfrac{2}{\pi  q_{A}}}|I^{(m)}_{A}| ~.
\end{equation}
The quantities $\zeta_{\phi}$ and $\zeta_A$ in (\ref{eq:rad_eta}) and (\ref{eq:rad_xi}) are the phases of the scalar and vector radiation irrelevant to our purposes. From equations $(\ref{eq:rad_eta})$-$(\ref{eq:rad_xi})$, it can be observed that the radiation amplitudes are proportional to the square of the shape mode amplitude $C_0^2$, consistent with our initial assumption. The asymptotic radiation profiles formally resemble the analogous expression for the global vortex \cite{BlancoPillado2021}. However, unlike in that case, there are two radiation channels (scalar and vector) which are sensitive to the self-coupling $\lambda$. We will discuss this is detail in the following sections.

\subsection{Internal mode decay law}

The asymptotic form of the radiation (\ref{eq:rad_eta})-(\ref{eq:rad_xi}), allows us to determine the decay law of the internal mode excitations. This can be done by comparing the average energy flux carried away by the radiation with the rate of change of the energy of the excited vortex. To begin, we compute the energy flux in the radial direction, represented by the $T_{0r}$ component of the energy momentum tensor
\begin{equation}
T_{0r} = \dot{\eta}(r,t)\partial_{r}\eta(r,t) + \dot{\xi}(r,t)\partial_{r}\xi(r,t) + \dot{\xi}(r,t)\dfrac{\xi(r,t)}{r}~.
\end{equation}
The average energy flux over one period is given by
\begin{equation}
\langle T_{0r} \rangle = - \dfrac{2 C_0^4 \omega_{n,j}}{\pi r}\left(|I^{(m)}_{\phi}|^2 + |I^{(m)}_{A}|^2\right)~.
\end{equation}
Thus, the power radiated to infinity is
\begin{equation}
\int_{0}^{2\pi}\, \langle T_{0r} \rangle r\,d\theta = - 4 C_0^4 \omega_{n,j}\left(|I^{(m)}_{\phi}|^2 + |I^{(m)}_{A}|^2\right)~.
\end{equation}
On the other hand, the energy associated to the vibrating vortex (excited only by the internal mode) is given by
\begin{equation}
E = M_V + \dfrac{1}{2}\mathcal{N}\omega_{n,j}^2 C_0^2~,
\end{equation}
where $M_V$ denotes the rest mass of the gauged vortex and $\mathcal{N}$ is the normalization factor previously defined in $(\ref{eq:factor_N})$. Therefore, the energy associated to the vibrational mode depends on the squared amplitude $C_0^2$, so while the vibrating vortex emits radiation, the amplitude must decrease over time to maintain the energy balance. This implies that
\begin{equation}  \label{edoCdet}
\dfrac{\mathcal{N}}{2}\omega_{n,j}\dfrac{d C_0(t)^2}{dt} = - 4C_0(t)^4 \left(|I^{(m)}_{\phi}|^2 + |I^{(m)}_{A}|^2\right).
\end{equation}
The solution of (\ref{edoCdet}) provides the decay law for the internal mode amplitude 
\begin{equation}\label{eq:decay_law}
C_0(t) = \dfrac{1}{\sqrt{C_0^{-2}(0) + \Gamma_{n,j}^{(m)}  t}}~,
\end{equation}
where $C_0(0)=C_0$ is the initial shape mode amplitude (at $t=0$) and 
\begin{equation}\label{eq:Gamma}
\Gamma_{n,j}^{(m)} = \dfrac{8 \left(|I^{(m)}_{\phi}|^2 + |I^{(m)}_{A}|^2\right)}{\mathcal{N}\omega_{n,j}}~,
\end{equation}
is the decay rate. The relation $(\ref{eq:decay_law})$ is formally identical to the decay law for the amplitude of a kink \cite{Manton1997} or for a global vortex \cite{BlancoPillado2021}, but here both the scalar and vector radiation field contribute to the final expression.

\begin{figure}[ht!]
    \centering{
   \includegraphics[width=0.62\linewidth]{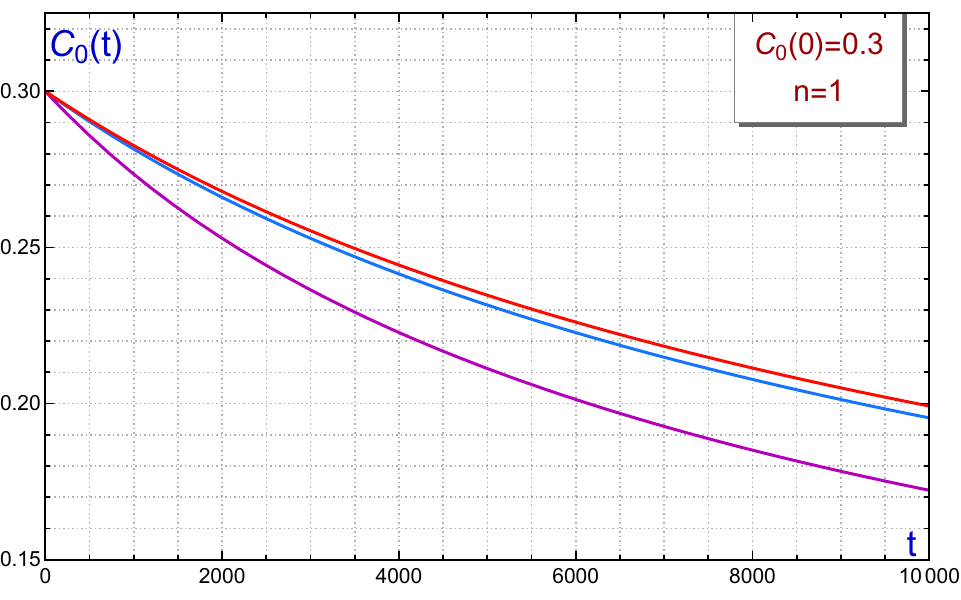}}
    \caption{\textit{Graphical representation for the theoretical decay of $C_0$ described by equation \eqref{eq:decay_law}  for $\lambda = 0.7$ (blue), $\lambda = 1$ (purple), and $\lambda = 1.4$ (red) .} }\label{Fig:AmpVsTimeTeor}
\end{figure}

\begin{figure}[ht!]
    \centering{
   \includegraphics[width=0.62\linewidth]{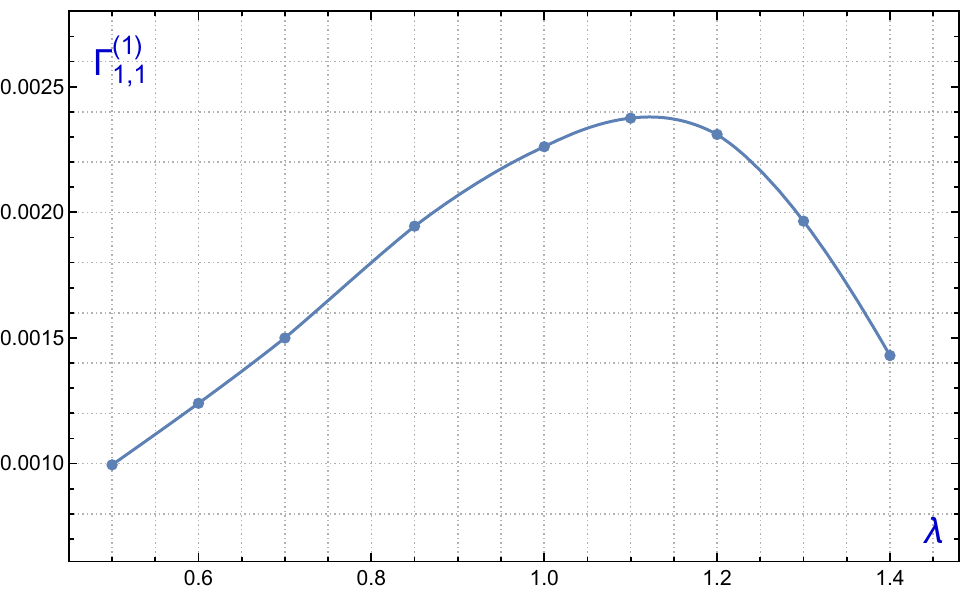}
    }
    \caption{\textit{Graphical representation of the decay rate $\Gamma_{1,1}^{(1)}$ as a function of the coupling constant $\lambda$ for 1-vortices. The points depict the analytical results derived from equation $(\ref{eq:Gamma})$, while the solid line illustrates a numerical interpolation.}}
    \label{Fig:Gamma_Lambda}
\end{figure}

Figure \ref{Fig:AmpVsTimeTeor} displays the evolution of the shape mode amplitude $C_0(t)$ for 1-vortices for the values of $\lambda = 0.7$, $\lambda = 1$ and $\lambda = 1.4$. To obtain a comprehensive pattern of the decay of the shape mode amplitude (\ref{eq:decay_law}) a representation illustrating the dependence of the decay rate $\Gamma$ on $\lambda$ can be found in Figure \ref{Fig:Gamma_Lambda}. 
We have taken $\lambda > 0.5$
since reducing $\lambda$ further would increase the size of the vortex core making necessary to use a substantially larger simulation box.
For this range of values the decay rate reaches a maximum for $\lambda$ close to $1$, that is, close to the critical value. This is related to the finite size of the source. Around the maximum of the decay rate the vortex size and the masses of the radiated particles are of the same order. Far from this point the difference in scales of the size of the object and the masses of the radiated particles should lead to radiation suppression \cite{BlancoPillado2021}. 

\section{Internal mode evolution: Numerical approach}  \label{Sec:4}
 
In the preceding section, we have employed perturbation theory to derive the asymptotic behavior of radiation in both the scalar and vector channels, enabling us to obtain an analytical expression that determines the decay of the shape mode amplitude. In this section, we carry out a similar investigation using numerical methods. The outcomes obtained through this approach will be compared with those from the preceding section, allowing us to validate the accuracy of the previously derived expressions.

\subsection{Numerical setup and radiation power spectrum analysis}\label{Sec:4.2}

As mentioned in Section \ref{Sec:2}, we will restrict our analysis to circularly symmetric configurations. Therefore, the system of differential equations to be numerically solved is as follows:
\begin{eqnarray}
\dfrac{\partial^2 F_n}{\partial t^2} - \dfrac{\partial^2 F_n}{\partial r^2} - \frac{1}{r}\dfrac{\partial F_n}{\partial r } + \frac{n^2}{r^2}(1-B_{n})^2 F_n - \dfrac{\lambda}{2}(1-F_n^2)F_n\!\!\!\!&=0, \label{eq:TemporalEvolution1}\\
\dfrac{\partial^2 B_{n}}{\partial t^2} - \dfrac{\partial^2 B_{n}}{\partial r^2} + \frac{1}{r}\frac{\partial B_{n}}{\partial r} - (1-B_{n}) F_n^2 \!\!\!\!&=0, \label{eq:TemporalEvolution2}
\end{eqnarray}
where $F_n(r,t)$ denotes the vortex profile of vorticity $n$ and $B_n(r,t) = \frac{r}{n} A_{\theta}(r,t)$ denotes the angular component of the gauge field. To solve this system, a second-order finite difference scheme in both space and time has been implemented with $\Delta r = 0.01$ and $\Delta t = 0.001$. The simulations were performed up to $t=10000$ over the radial interval $[0,L]$, where $L = 70$. We note that using larger boxes did not yield substantial differences in our simulation. To prevent radiation reflecting from the boundary at $r = L$, Mur absorbing boundary conditions have been introduced \cite{Mur}. Additionally, we introduced damping terms $-\epsilon(r) \frac{d F_n}{dt}$ and $-\epsilon(r) \frac{d B_n}{dt}$ respectively in (\ref{eq:TemporalEvolution1}) and (\ref{eq:TemporalEvolution2}) where

\begin{equation}
\epsilon(r) = \begin{cases}
0, & \text{if } 0 \leq r < r_{\text{cut}}, \\
\left[\dfrac{r - r_{\text{cut}}}{20}\right]^4, & \text{if } r_{\text{cut}} < r < L.
\end{cases}
\end{equation}
 Typical values have been chosen around $r_{cut} = 5L/6$, but we have performed several tests at different values of $r_{cut}$ to confirm the numerical stability of our numerical setup. A large number of simulations have been carried out to examine the evolution of a vibrating 1-vortex. Moreover, the eigenfunctions are assumed normalized with respect to the norm of $L^2(\mathbb{R}^2)\oplus\mathbb{R}^2$. In all numerical simulations we have focussed on the $n=1$ vortex. The evolution of $n-$vortices with $n>1$ will be explored in future studies, as they introduce some subtleties that require special analysis.

\begin{figure}[h!]
    \centering
    \begin{subfigure}{0.489\textwidth}
        \includegraphics[width=\linewidth]{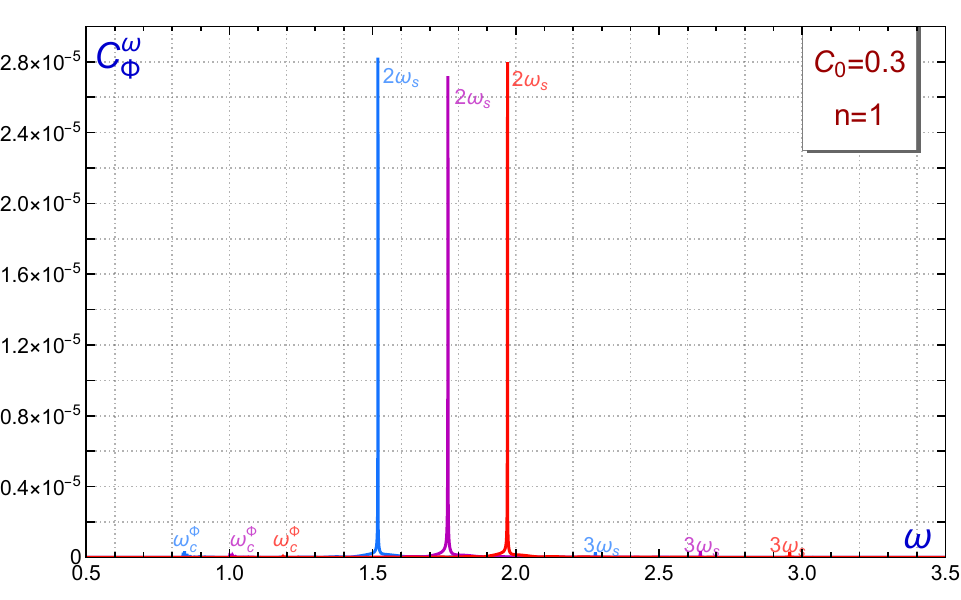}
        \caption{\centering  \textit{Scalar channel.}}
        \label{Fig:Power_Phi}
    \end{subfigure}
    \hfill
    \begin{subfigure}{0.489\textwidth}
        \includegraphics[width=\linewidth]{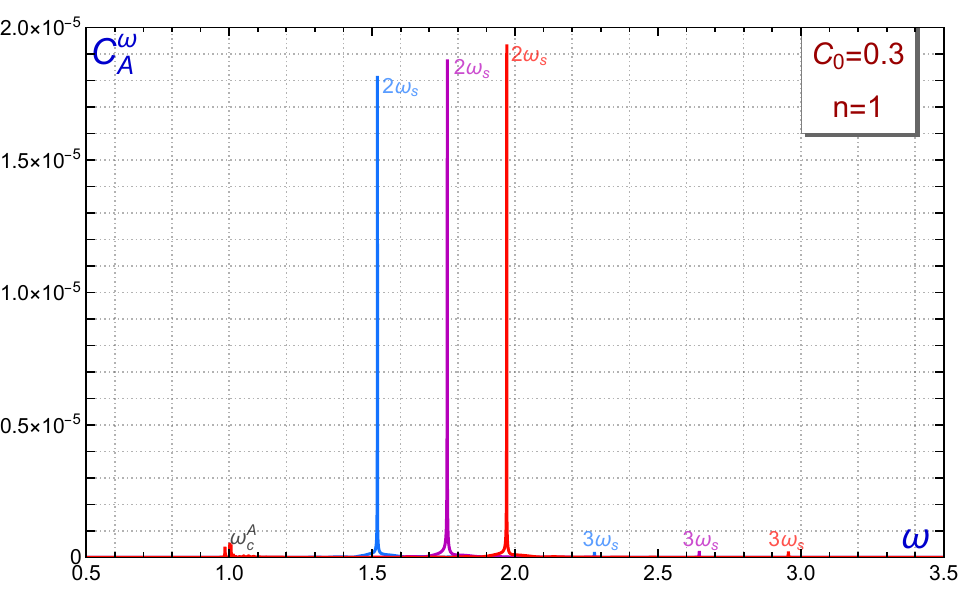}
        \caption{\centering\textit{ Vector channel.}}
        \label{Fig:Power_A}
    \end{subfigure}
  
    \caption{\textit{Power spectra of the scalar (left) and vector (right) radiation fields for $\lambda = 0.7$ (blue), $\lambda = 1$ (purple), and $\lambda = 1.4$ (red). The peak labels indicate the corresponding frequencies in each case, where $\omega_c^{\phi} = \sqrt{\lambda}$ denotes the scalar threshold frequency and $\omega_c^{A} = 1$ represents the gauge mass threshold frequency. The initial amplitude for the massive bound mode is set to $C_0 = 0.3$.} }
    \label{Fig:Power_Spectrum}
\end{figure}

The first analysis of the data extracted from these simulations is aimed at studying the frequencies of the radiation emitted by the 1-vortex at a point far away from the vortex core both in the scalar and vector channels. The results found for the model parameters $\lambda = 0.7, \lambda = 1.0$, and $\lambda = 1.4$ are illustrated in Figure \ref{Fig:Power_Spectrum}. The power spectrum of the radiation emitted by the gauge vortex in the scalar channel has been depicted in Figure \ref{Fig:Power_Phi} . As anticipated, the predominant peak occurs at $\omega(\lambda) = 2\omega_s(\lambda)$. Additionally, other peaks arise although comparatively suppressed. For example, small peaks around $3\omega_s(\lambda)$ are observed, due to the coupling between the internal and radiation modes at higher orders. 

The presence of these other peaks is a higher order effect and therefore is not included in our perturbative approach based on the dominant second order expansion. Nevertheless, they might be captured if we expanded up to the third order. Lastly, there are other minor peaks denoted by $\omega_c^{\phi} = \sqrt{\lambda}$ caused by random numerical noise. Analogously, Figure \ref{Fig:Power_A} shows the power spectrum of the radiation emitted in the vector channel, which exhibits a similar behaviour.

\begin{figure}[h!]
    \centering
    
    \begin{subfigure}{0.489\textwidth}
        \includegraphics[width=\linewidth]{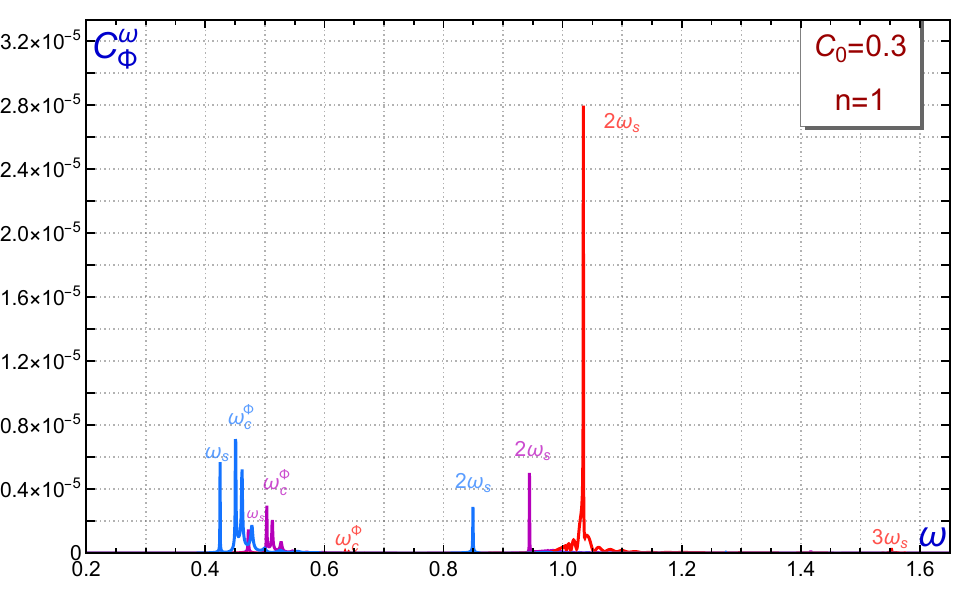}
        \caption{\centering \textit{Scalar channel.}}
        \label{Fig:Power_Phi_low}
    \end{subfigure}
    \hfill
    \begin{subfigure}{0.489\textwidth}
        \includegraphics[width=\linewidth]{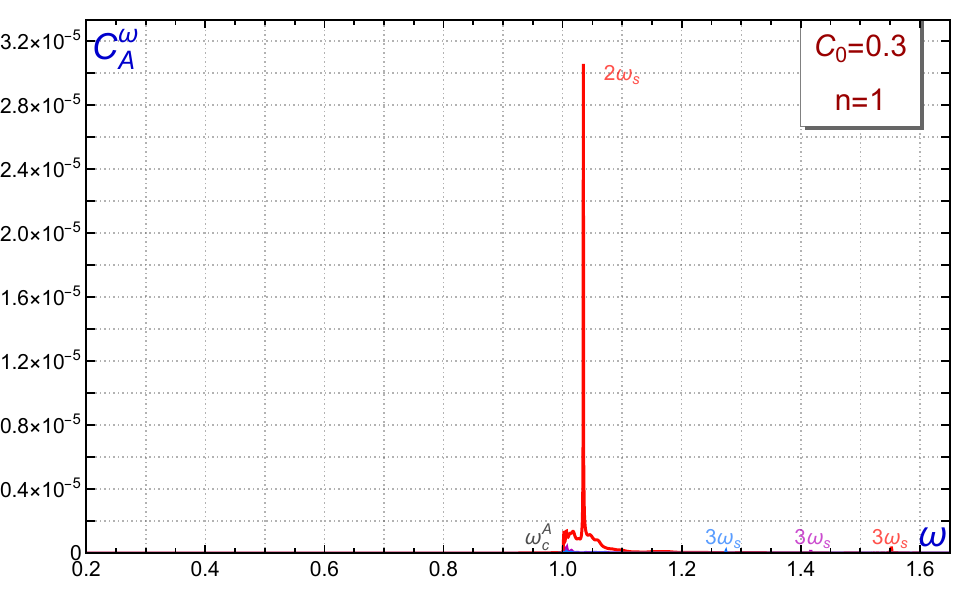}
        \caption{\centering \textit{ Vector channel.}}
        \label{Fig:Power_A_low}
    \end{subfigure}
  
    \caption{\textit{Power spectra of the scalar (left) and vector (right) radiation fields for $\lambda = 0.2$ (blue), $\lambda = 0.24$ (purple), and $\lambda = 0.4$ (red). The peak labels indicate the corresponding frequencies in each case, where $\omega_c^{\phi} = \sqrt{\lambda}$ denotes the scalar threshold frequency and $\omega_c^{A} = 1$ represents the gauge mass threshold frequency. The initial amplitude for the massive bound mode is set to $C_0 = 0.3$.}}
    \label{Fig:Power_Spectrum_2}
\end{figure}

The scheme depicted by the power spectrum in Figure \ref{Fig:Power_Spectrum} is valid in the regime where $\lambda > 0.282$ and $\lambda < 1.5$, in which there exists only one discrete mode and its first harmonic belongs to the doubly degenerate continuous spectrum. A qualitative change occurs for $\lambda \lesssim 0.282$. In this case, the first harmonic $\omega(\lambda) = 2 \omega_s(\lambda)$ is higher than the Higgs mass but lower than the gauge mass (see Figure \ref{fig:espectra1}). Thus, it is expected that only the scalar channel is available to emit radiation. In Figure \ref{Fig:Power_Spectrum_2}, we display the power spectrum for $\lambda = 0.2$, $\lambda = 0.25$, and $\lambda = 0.4$. Note that the cases $\lambda = 0.2$ and $\lambda = 0.25$ belong to the regime where only the scalar channel is available at the lowest order while for the case $\lambda = 0.4$ both channels can radiate. As mentioned in Section \ref{Sec:2} we anticipate a different behavior in the radiation emission for these two scenarios. As expected, in Figure \ref{Fig:Power_A_low} the peak at twice the frequency of the corresponding shape mode in the vector channel for $\lambda = 0.2$ and $\lambda = 0.25$ is highly suppressed. Note that in Figure \ref{Fig:Power_Phi_low}, the presence of shape mode oscillations for the cases $\lambda = 0.2$ and $\lambda = 0.25$ are noticeable at large distances (where the power spectrum analysis is considered). This is because the shape mode for these cases has a significant width, and the power spectrum in the simulations captures these frequencies. However, it is interesting to describe in detail the situation arising in this case. It should be noted that since the eigenvalue of the discrete mode is very close to the threshold value of the continuous spectrum associated to the scalar component and comparatively far from that of the vector component, the scalar component of the shape mode will be highly delocalized and will have a dominant behavior over the vector component. This explains the fact that only Figure \ref{Fig:Power_Phi_low} shows frequencies associated to the discrete mode when $\lambda < 0.282$.

In summary, the numerical simulations described in this section establish that the dominant frequency for the radiation emitted when a 1-vortex vibrates via its shape mode is precisely twice the natural frequency of the shape mode.
\vspace{0.3cm}

\subsection{Decay rate of the internal mode and radiation emission}
\vspace{0.3cm}

In this section, we will analyze the numerical decay of the shape mode amplitude and compare it with the analytical expression (\ref{eq:decay_law}) obtained in Section \ref{Sec:3}. 
The numerical amplitude of the shape mode at each time $t$ is computed in our simulations by projecting the difference between the evolving and the static 1-vortex onto the theoretical shape mode, as follows:
\begin{equation}
C(t) \approx 2 \pi \int_{0}^{\infty}\left[\left(F_n(r,t) - f_n(r)\right)\varphi(r) + \frac{n}{r}\left(B_{n}(r,t) - \,\beta_n(r)\right)a_{\theta}(r)\right]r\, dr,
\end{equation}
which is justified taking into account the expressions (\ref{eq:phi_expansion}) and (\ref{eq:a2_expansion}).

\begin{figure}[htb]
    \centering
    % --- Subfigura (a): lambda = 1 ---
    \begin{subfigure}{\textwidth}
        \centering
        \includegraphics[width=0.8\textwidth]{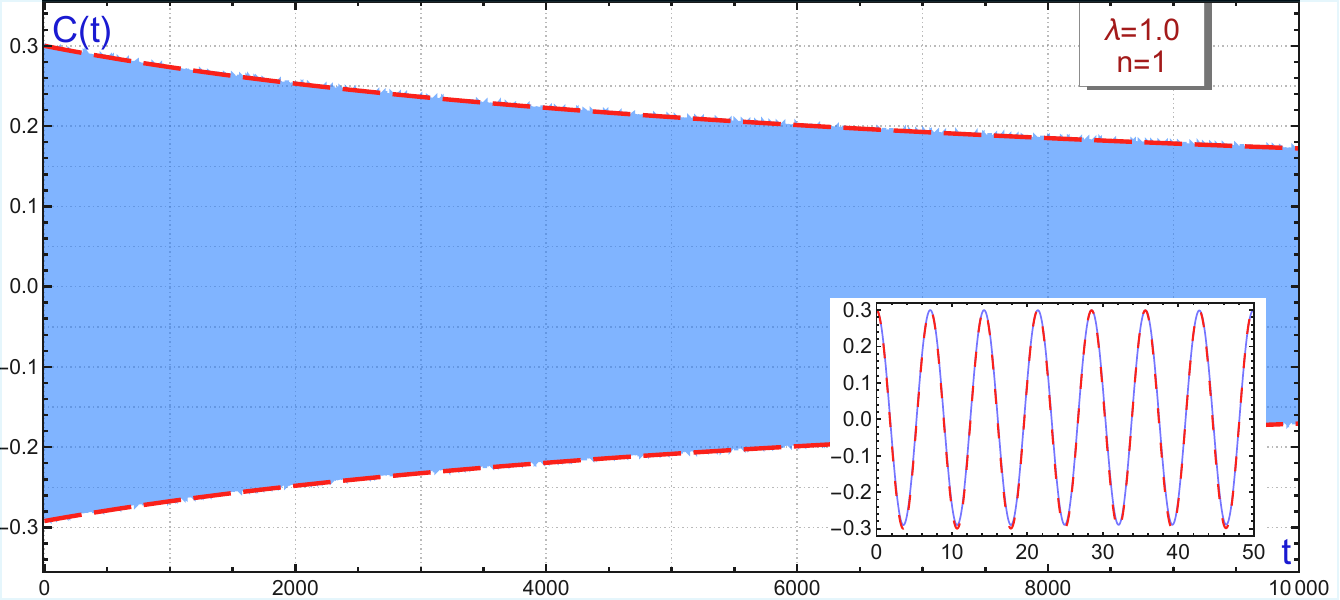}
        \caption{\centering$\lambda = 1$.}
    \end{subfigure}
  %  \vspace{0.4cm} % espacio vertical entre filas
    % --- Subfigura (b): lambda = 0.7 ---
    \begin{subfigure}{0.47\textwidth}
        \centering
        \includegraphics[width=\linewidth]{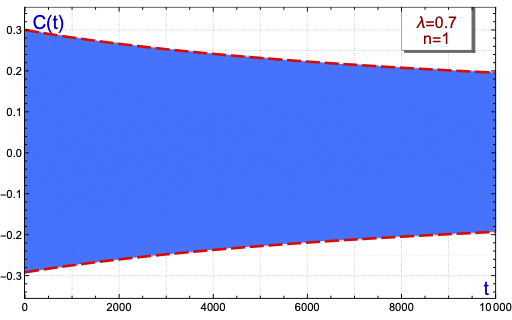}
        \caption{\centering$\lambda = 0.7$.}
    \end{subfigure}
    \hfill
    % --- Subfigura (c): lambda = 1.4 ---
    \begin{subfigure}{0.47\textwidth}
        \centering
        \includegraphics[width=\linewidth]{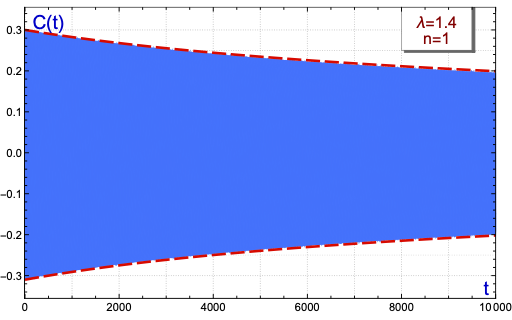}
        \caption{\centering$\lambda = 1.4$.}
    \end{subfigure}
    % --- Caption general ---
    \vspace{-0.3cm}
    \caption{\textit{Evolution of the numerical shape mode amplitude (blue solid curve) and the analytical decay law $(\ref{eq:decay_law})$ (red dashed curve) for the coupling constants $\lambda=0.7$, $\lambda=1.0$ and $\lambda=1.4$. The shape mode amplitude decays due to the coupling with scattering modes. All the simulations have been performed for an initial shape mode amplitude $C_0 = 0.3$. The inset shows the first oscillations of the internal mode amplitude.}}
    \label{Fig:Decay_Law}
\end{figure}

The numerical amplitude of the shape mode as a function of $t$ exhibits a large number of oscillations during our simulations. These oscillations are depicted by the blue curves in Figure \ref{Fig:Decay_Law} for the values $\lambda=0.7$, $\lambda=1$, and $\lambda=1.4$. However, due to the ratio between their oscillation periods and the total simulation time, they appear as a continuous blue area. On the other hand, the analytical response of the shape mode amplitudes (\ref{eq:decay_law}) are represented by the dashed red curves for the same cases in Figure \ref{Fig:Decay_Law}. This figure shows a close match between the envelope of the numerical oscillations of the shape mode and the analytical amplitude.
%\vspace{-0.2cm}

%\vspace{-0.2cm}

\begin{figure}[ht!]
    \centering
    
    \begin{subfigure}{0.328\textwidth}
        \includegraphics[width=\linewidth]{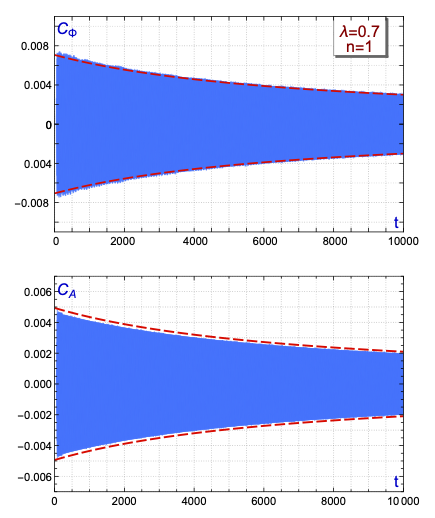}
        \caption{\centering$\lambda = 0.7$.}
    \end{subfigure}
    \hfill
    \begin{subfigure}{0.328\textwidth}
        \includegraphics[width=\linewidth]{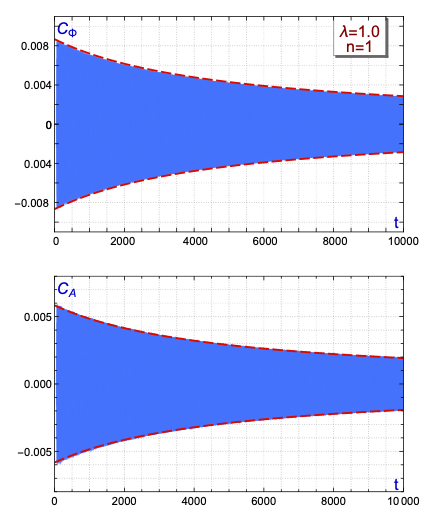}
        \caption{\centering$\lambda = 1.0$.}
    \end{subfigure}
    \hfill
    \begin{subfigure}{0.328\textwidth}
        \includegraphics[width=\linewidth]{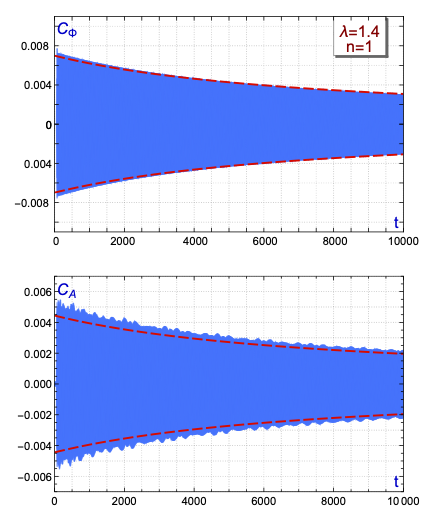}
        \caption{\centering$\lambda = 1.4$.}
    \end{subfigure}
    \vspace{-0.3cm} 
        
    \caption{\textit{Radiation field at $r_{rad} = 50$ for different self-coupling constants $\lambda$. The upper plots account for the scalar component and the lower plots the vector component. The simulations have been performed for an initial amplitude $C_0 = 0.3$. The red dashed line represent our analytical approximation.}}
    \label{Fig:Decay_Law_rad}
\end{figure}

In Figure \ref{Fig:Decay_Law_rad}, the radiation amplitudes in the scalar and vector channels at $r_{rad} = 50$ as a function of time $t$ are displayed. The dashed red lines represent the theoretical evolution of the radiation profile $(\ref{eq:decay_law_rad})$. A brief examination reveals a good agreement for the scalar component in all cases. The slight deviations in the vector channel are related to numerical precision. In order to compare the radiation amplitudes given by the formulas $(\ref{eq:rad_eta})$-$(\ref{eq:rad_xi})$ with the numerical data, we have multiplied the values from field theory by the factor $\sqrt{r_{rad}}/C_0^2$ to have a clear comparison independent of the particular position. Then, we have compared the resulting amplitudes with $C_{\phi}^{(m)}$ and $C_{A}^{(m)}$, which are given through the formulas $(\ref{eq:C_amplitudes})$. Those amplitudes and the corresponding decay rates shown in Figure \ref{Fig:Gamma_Lambda}, together with the following expressions\begin{equation}\label{eq:decay_law_rad}
C_{\phi}^{(m)}(t) = \dfrac{1}{\sqrt{(C_{\phi}^{(m)}(0))^{-2} + \Gamma^{(m)}  t}}, \hspace{1.5cm} C_{A}^{(m)}(t) = \dfrac{1}{\sqrt{(C_{A}^{(m)}(0))^{-2} + \Gamma^{(m)}  t}},
\end{equation}
have been used to predict the internal mode decay.

\subsection{The global string limit. Quasi-bound states}

\begin{figure}[ht!]
    \centering
    
    \begin{subfigure}{0.47\textwidth}
        \includegraphics[width=\linewidth]{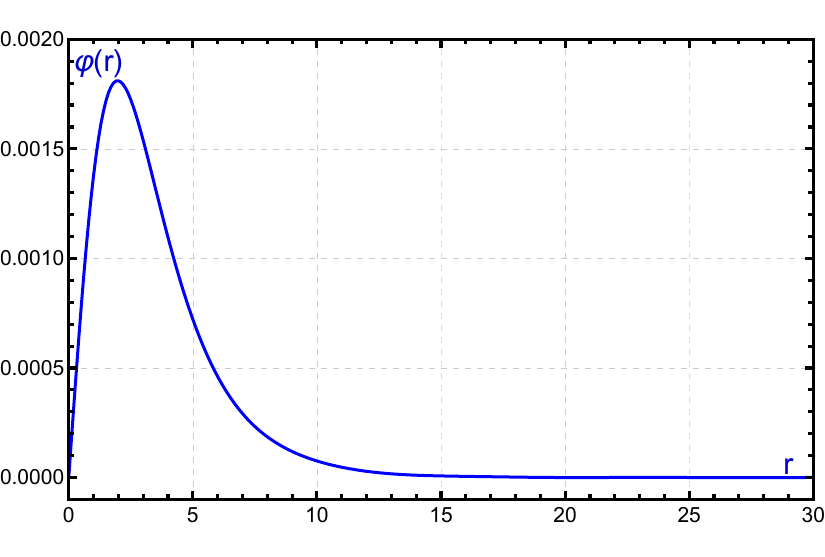}
                \vspace{-0.4cm}

        \caption{\centering \textit{Scalar channel.}}
       
    \end{subfigure}
    \hfill
    \begin{subfigure}{0.47\textwidth}
        \includegraphics[width=\linewidth]{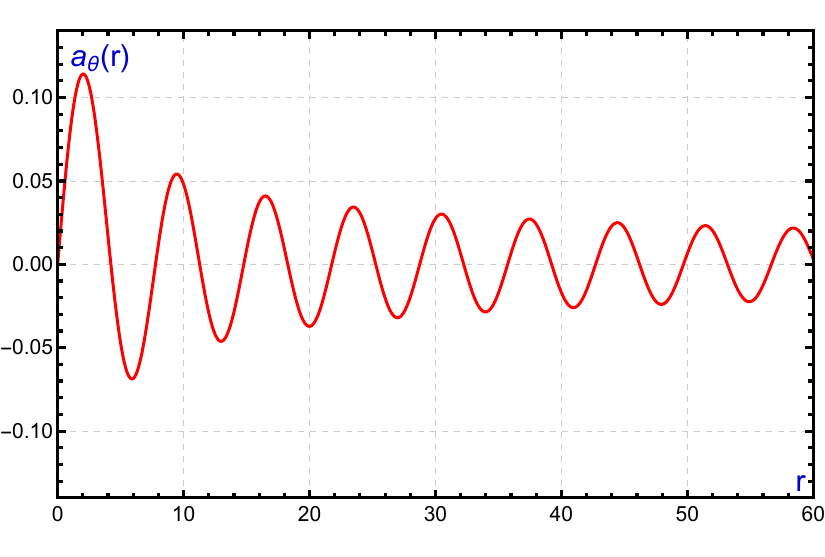}
        \vspace{-0.4cm}
        \caption{\centering \textit{Vector channel.}}
        
    \end{subfigure}
  
    \caption{\centering\textit{ Quasi-bound mode for the gauged vortex with $\lambda\gg1$}. }
    \label{Fig:GlobalMode}
\end{figure}

As we mentioned earlier, the bound state solutions of our coupled system of equations
ceases to exist when the parameter $\lambda>1.5$. The reason for this is clear, the
frequency of the bound state becomes in this case higher than the continuum for the 
vector field. This makes it impossible to have a bound state for the vector
component part of the perturbation in the analogue Schrödinger problem. However,
the frequency of the last bound state at around $\lambda=1.5$ is still below the
mass threshold for the scalar component. This suggests the possibility that
the lowest scattering states for $\lambda>1.5$ would be composed of a 
wave function similar to a bound state mode for the scalar part and
a radiative mode for the vector field part. This is indeed what one can
find in the numerical solutions presented in Figure \ref{Fig:GlobalMode}.

On the other hand, this also suggests that even though these modes 
are not truly bound states they may have a behavior that shares some similarities 
with them. In particular, it seems likely that excitations at this large coupling constants that
resemble the scalar part of these modes would have a long lifetime, comparable to the  ``bona fide'' bound states described earlier. One should therefore consider
these {\textit{ quasi-bound modes}} to be qualitatively in the same family of solutions
as the genuine bound states.

We can have an intuitive understanding of the reason for the long lifetime of these
{ \textit{quasi-bound modes}} if one considers the extreme Type II regime where $\lambda \gg 1$. In this
limit, the scalar field has a much larger mass than the vector field and the background
vortex configuration resembles a global vortex core for $r<m_A^{-1}$. It is therefore
reasonable to expect that the scalar field excitations of the global string could
be well approximated by bound states of the complete system. 

We have investigated this idea by initializing our numerical evolution for 
a purely global string bound state in our Abelian-Higgs model with several values $\lambda >1.5 $. The 
results indicate that indeed the system behaves like a global string for a long time
where the scalar core oscillations produced massive radiation much in the same
 way as it was previously observed in \cite{BlancoPillado2021}. Indeed, the mode displayed in Figure \ref{Fig:GlobalMode} is almost identical to that found in Figure \ref{figI3:GVInternalModesn1} for the global vortex.   Furthermore, we also note the presence
of a small amount of massive vector radiation but this does not have a big effect
on the system. Presumably the coupling between the oscillating scalar core and
the vector scattering states is low enough to allow for this possibility.

\vspace{-0.4cm}

\section{Concluding remarks}\label{Sec:5}

In this chapter, we have analyzed in detail the evolution of a vortex excited by the lowest internal bound mode. This discrete mode exhibits rotational symmetry, enabling analytical studies based on perturbation theory. Through this approach, we have demonstrated that vortices excited by Derrick-type shape modes emit radiation with radial symmetry at a frequency twice that of the shape mode, owing to quadratic nonlinear terms in the field equations. Additionally, we have provided an iterative procedure to analytically identify the decay of the shape mode amplitude, which follows an inverse square law, similar to that found in the case of the $\phi^4$ kink \cite{Manton1997}, the global vortex \cite{BlancoPillado2021} and the wobbling kinks introduced in Chapter \ref{Chap1}. We have performed numerical simulations that show very good agreement with our analytical predictions.

Depending of the selfcoupling values we have found different regimes. For $0.28 \leq \lambda\leq 1.5$ the excited vortex is able to decay by emitting radiation in both vector and scalar channels. For $\lambda\leq 0.28$ the vortex emits only though the scalar channel since the corresponding radiation frequency is below the vector mass threshold. Finally, for $\lambda\geq 1.5$ there are no ``proper'' bound modes. However, as we have argued, if $\lambda$ is large enough the local vortex resembles a global vortex at distances $r< m_A^{-1}$. In this regime, the scalar component profiles approach the global vortex bound modes and the excited configuration is able to store energy for large times.

There are two natural extensions of these results: the study of the decay of excited vortices with $n>1$ and the decay of excited local strings in $3+1$ dimensions. In both cases, the richer spectral structure associated to static configurations requires a detailed analysis. Both lines of research are  currently under investigation.

    %%%%%%%%%%%%%%%%%%%%%%%%%%%%%%%%%%%%%%%%%%%%%%%%%
    %% Please add the content of your thesis here. %%
    %%%%%%%%%%%%%%%%%%%%%%%%%%%%%%%%%%%%%%%%%%%%%%%%%

  %  \part{Wobbling kinks}\label{Part2}

  % \chapter{Demo Chapter}
   %\input{DELETE/demo}

    %%%%%%%%%%%%%%%%%%%%%%%%%%%%%%%%%
    %% End of adding your content. %%
    %%%%%%%%%%%%%%%%%%%%%%%%%%%%%%%%%

    % Add the following chapters not to the current ›part‹ but one level above instead.
    \makeatletter
        \def\toclevel@chapter{-1}
        \def\toclevel@section{0}
    \makeatother

    \addchap{Conclusions \& Further work}
Throughout this manuscript, we have discussed the main results obtained in \cite{AlonsoIzquierdo2024, AlonsoIzquierdo2025, AlonsoIzquierdo2023c, AlonsoIzquierdo2024b, AlonsoIzquierdo2025b}, which constitute this doctoral thesis. We have analyzed in detail novel results regarding the dynamics of kinks and Abelian-Higgs vortices. In these final pages, we summarize the main findings drawn from this document.

\begin{itemize}
\item In Chapter \ref{Intro0}, we introduced some historical remarks about the emergence of the concept of kink. Then, we gave a precise definition of a topological kink. The importance of these solutions in several areas of physics was also discussed. In this first chapter, we also addressed some details of the Ginzburg-Landau model and its relation to the Abelian-Higgs model. The importance of these models in cosmology, superconductivity, and superfluidity was also highlighted.

\item In Chapter \ref{Intro1}, we introduced the basic properties that characterize topological kinks. First, we presented the models that support the existence of these solutions and the basic internal mode structure that arises when analyzing their stability. Then, we introduced the $\phi^4$ and $\phi^6$ models. Kink solutions in these models were obtained, and their internal mode structure was detailed. The appearance of spectral walls in kink-antikink solutions was also discussed. Next, the dynamics of a wobbling kink in the $\phi^4$ model up to second order in perturbation theory were addressed.

We then generalized these results to analyze topological kinks arising in two-component field theories. The vacuum structure, kink solutions, and their stability were studied in the context of the MSTB and double $\phi^4$ models.

\item In Chapter \ref{Chap1}, we discussed the results presented in reference \cite{AlonsoIzquierdo2024}. In this work, the dynamics of a wobbling kink in the double $\phi^4$ model are addressed. First, the internal mode structure is analyzed, allowing us to determine that the number of possible internal modes depends on the constant coupling both field components. Then, the radiation emission of a wobbling kink is studied, and the radiation frequencies and amplitudes are found. Later, the results obtained were compared with those from numerical simulations, showing good agreement.

\item In Chapter \ref{Chap2}, kink/antikink collisions are studied in the context of the MSTB model. Here, we analyzed how the presence of a second internal mode affects the energy transfer mechanism between the shape modes and the translational mode. It was found that the amplitudes of both shape modes after the last collision are highly dependent on the coupling between the two field components. Moreover, a special value of the coupling constant was found for which the internal mode corresponding to the second field component transfers all its energy to the longitudinal eigenmode before the first collision takes place. This chapter is an adaptation of reference \cite{AlonsoIzquierdo2025}.

\item In Chapter \ref{Intro2}, we introduced the Abelian-Higgs model as a modification of the global vortex model, required to obtain static vortex solutions with finite energy. The topology associated with these vortex solutions and the reduced equations needed to obtain the profiles of these field configurations were presented. The BPS structure for the special value $\lambda_c = 1$ was then exploited to derive the corresponding BPS equations. Additionally, we analyzed several methods of generating multivortex configurations, both in the BPS limit and beyond. In Section \ref{SecIntBetVort}, the first notions about the stability of $n$-vortices are introduced, along with details about vortex/antivortex scattering for different values of the coupling constant $\lambda$. Finally, in Section \ref{spectralwallVort}, the appearance of spectral walls between two vortices separated by a certain distance is discussed.

\item In Chapter \ref{Chap3}, we analyzed the internal mode structure associated with Abelian-Higgs vortices. To do so, we reduced the original eigenvalue problem—consisting of four second-order partial differential equations in two spatial dimensions—to a system of three second-order ordinary differential equations. This was achieved by appropriately choosing the angular dependence of the eigenfunctions. In doing so, we identified all internal modes and classified them according to their angular dependence. On the one hand, Derrick-type modes shared the same symmetry as the vortex itself. On the other hand, multipolar modes exhibited different angular dependencies and were doubly degenerate. Among all multipolar eigenmodes, Type A eigenfunctions correspond to the zero modes of the vortex and its unstable eigenmodes in the $\lambda > 1$ limit, while Type B multipolar modes were always associated with positive eigenvalues. This chapter is an adaptation of references \cite{AlonsoIzquierdo2024b, AlonsoIzquierdo2025b}.

\item Finally, in Chapter \ref{Chap4}, the dynamics of an excited vortex up to second order in perturbation theory were studied. It was found that when a vortex is excited via a Derrick-type internal mode, it can emit radiation at twice the frequency of the internal mode. Using an extension of the perturbation theory employed in Chapter \ref{Chap1}, we computed the theoretical radiation amplitudes. Additionally, we derived an analytical law describing the decay of the internal mode amplitude. The analytical results were compared with data from numerical simulations, showing strong agreement. This chapter is an adaptation of reference \cite{AlonsoIzquierdo2024c}.
\end{itemize}

\begin{comment}
    The following points are proposed as future lines of research:
\begin{itemize}
    \item With respect to kink solutions, a natural continuation of the results obtained in Chapter \ref{Chap1} would consist of the analysis of the values of the coupling constant between field components for which a resonance between frequencies existed. On the other hand, regarding the results extracted from Chapter \ref{Chap2}, a possible research line would consist on the study of the scattering between wobbling kinks whose internal mode structure is more complex. This generalization would shed more light on the energy resonant mechanism between modes that is responsible for the fractal structure of the velocity diagrams presented in this study.

    \item Regarding vortex solutions, a possible continuation of the results provided in Chapter \ref{Chap3} would be analyzing the internal mode structure associated with vortices in theories that generalize the Abelian-Higgs model. With respect to the results exposed in Chapter \ref{Chap4}, a natural continuation would consist of analysis of the decay of internal modes not only for Derrick type modes, but also for multipolar ones and for $n$-vortices with $n>1$. Another possibility would be in generlization of this problem in order to study the excitation of cosmic strings. 
\end{itemize}
\end{comment}

The following points are proposed as future lines of research:
\begin{itemize}
\item Regarding kink solutions, a natural continuation of the results obtained in Chapter \ref{Chap1} would be the analysis of the values of the coupling constant between the field components for which resonances between frequencies occur. On the other hand, based on the results presented in Chapter \ref{Chap2}, a potential research direction would be the study of scattering processes between wobbling kinks with more complex internal mode structures. This generalization could provide further insight into the resonant energy exchange mechanisms responsible for the fractal structure observed in the velocity diagrams discussed in this study.

\item Regarding vortex solutions, one possible continuation of the results presented in Chapter \ref{Chap3} would be the analysis of the internal mode structure of vortices in models that generalize the Abelian-Higgs theory. As for the results shown in Chapter \ref{Chap4}, a natural extension would involve studying the decay of internal modes not only for Derrick-type modes but also for multipolar modes and for $n$-vortices with $n > 1$. Another possible direction would be the generalization of this problem to investigate the excitation of cosmic strings.
\end{itemize}

\addchap{Conclusiones y líneas futuras de trabajo}

   A lo largo del presente manuscrito, hemos analizado los principales resultados obtenidos en \cite{AlonsoIzquierdo2024, AlonsoIzquierdo2025, AlonsoIzquierdo2023c, AlonsoIzquierdo2024b, AlonsoIzquierdo2025b}, artículos que componen esta tesis doctoral. Hemos estudiado en detalle resultados originales relacionados con la dinámica asociada a kinks y vórtices de Abelian-Higgs. En estas últimas páginas, resumimos los principales hallazgos extraídos del presente manuscrito.

\begin{itemize}
\item En el Capítulo \ref{Intro0}, introdujimos detalles históricos relacionados con el surgimiento del concepto de kink. Esto nos permitió dar una definición precisa de kink topológico. También discutimos en detalle la importancia de estas soluciones en diversas áreas de la Física. En este primer capítulo, también presentamos algunos aspectos del modelo de Ginzburg-Landau y su relación con el modelo de Abelian-Higgs. Adicionalmente, destacamos la importancia de estos modelos en cosmología, superconductividad y el estudio de superfluidos.

\item En el Capítulo \ref{Intro1}, presentamos las propiedades básicas que caracterizan a los kinks topológicos. Primero, describimos la estructura de los modelos que admiten la existencia de dichas soluciones, para luego ofrecer una clasificación de los modos internos que surgen al analizar la estabilidad lineal de estos objetos. A continuación, introdujimos los modelos $\phi^4$ y $\phi^6$, a partir de los cuales obtuvimos las soluciones tipo kink y analizamos su estructura interna. Adicionalmente, se discutió la aparición de spectral walls en configuraciones kink/antikink. Posteriormente, se estudió la dinámica hasta segundo orden de un kink excitado en el contexto del modelo $\phi^4$.

Estos resultados fueron también generalizados para estudiar los kinks que surgen de teorías de campos escalares con dos componentes. Se analizó en detalle la estructura de vacíos, las soluciones tipo kink y su estabilidad lineal en los modelos MSTB y doble $\phi^4$.

\item En el Capítulo \ref{Chap1}, se introdujeron los resultados obtenidos en \cite{AlonsoIzquierdo2024}. En este trabajo, se analizó en detalle la dinámica asociada a un kink excitado en el modelo doble $\phi^4$. Primero, se estudió su estructura interna, lo cual permitió concluir que el número de modos internos posibles depende directamente de la constante de acoplamiento entre ambos campos. Posteriormente, se analizó la emisión de radiación por parte de un kink excitado, encontrando tanto las frecuencias como las amplitudes asociadas a dicha radiación. Finalmente, los resultados analíticos se compararon con los datos extraídos de simulaciones numéricas, mostrando buena concordancia.

\item En el Capítulo \ref{Chap2}, se estudiaron las colisiones kink/antikink en el modelo MSTB. Se analizó cómo la presencia de un segundo modo interno afecta el mecanismo de transferencia de energía entre los modos de forma y el modo translacional. Se encontró que las amplitudes de los modos internos después de la última colisión dependen fuertemente del acoplamiento entre las dos componentes del campo. Además, existe un valor específico de la constante de acoplamiento para el cual el modo interno correspondiente a la segunda componente del campo transfiere gran parte de su energía al modo presente en la primera componente antes de que el par kink/antikink colisione por primera vez. Este capítulo es una adaptación de la referencia \cite{AlonsoIzquierdo2025}.

\item En el Capítulo \ref{Intro2}, introdujimos el modelo de Abelian-Higgs como una modificación de un modelo con vórtices globales, necesaria para poder encontrar soluciones tipo vórtice con energía finita. Se describió la topología asociada a estas soluciones y las ecuaciones reducidas que deben resolverse para obtener los perfiles que caracterizan a los vórtices. Se explicó la estructura BPS del modelo para el caso particular $\lambda_c = 1$, lo cual permitió derivar las ecuaciones BPS. Adicionalmente, analizamos distintos métodos para generar configuraciones multivórtice, tanto en el límite BPS como fuera de él. En la Sección \ref{SecIntBetVort}, se introdujeron las primeras nociones sobre la estabilidad de $n$-vórtices, así como algunos detalles sobre colisiones vórtice/antivórtice para diferentes valores de la constante de acoplamiento $\lambda$. Finalmente, en la Sección \ref{spectralwallVort}, se discutió la aparición de spectral walls, analizando configuraciones de campo en las cuales dos vórtices se encuentran separados una cierta distancia dentro del límite BPS.

\item En el Capítulo \ref{Chap3}, analizamos la estructura interna asociada a los vórtices de Abelian-Higgs para cualquier valor de la constante de acoplamiento $\lambda$. Para ello, reducimos el problema original —la resolución de un sistema de cuatro ecuaciones diferenciales parciales definidas en dos dimensiones espaciales— a un sistema de tres ecuaciones diferenciales ordinarias definido en una sola coordenada espacial. Esto fue posible al escoger correctamente la dependencia angular de los modos internos. Así, identificamos todos los modos internos y los clasificamos según su dependencia angular. Por un lado, los modos tipo Derrick comparten la misma simetría que el vórtice. Por otro lado, los modos multipolares presentan una dependencia angular distinta y son doblemente degenerados. Entre todos los modos internos, los modos Tipo A se corresponden con los modos traslacionales y los modos inestables en el límite $\lambda > 1$, mientras que los modos Tipo B están siempre asociados con autovalores positivos. Este capítulo es una adaptación de las referencias \cite{AlonsoIzquierdo2024b, AlonsoIzquierdo2025b}.

\item Finalmente, en el Capítulo \ref{Chap4}, se analizó en detalle, hasta segundo orden en teoría de perturbaciones, la dinámica de un vórtice excitado. Se encontró que, cuando un vórtice es excitado mediante un modo tipo Derrick, es capaz de emitir radiación cuya frecuencia es justo el doble de la del modo interno. Usando una extensión de las técnicas analíticas empleadas en el Capítulo \ref{Chap1}, fuimos capaces de encontrar dichas amplitudes de radiación. Adicionalmente, derivamos la ley de decaimiento que describe la disminución de la amplitud del modo interno. Todos los resultados analíticos se compararon con los datos extraídos de simulaciones numéricas, mostrando un gran acuerdo. Este capítulo es una adaptación de la referencia \cite{AlonsoIzquierdo2024c}.
\end{itemize}

Como continuación de los resultados expuestos en este manuscrito, se proponen los siguientes puntos:  
\begin{itemize}
    \item Con respecto a las soluciones tipo \textit{kink}, una continuación natural de los resultados presentados en el Capítulo~\ref{Chap1} consistiría en el análisis detallado de los valores de la constante de acoplamiento para los cuales existen resonancias entre frecuencias. Por otro lado, como extensión de los resultados expuestos en el Capítulo~\ref{Chap2}, se propone el estudio de colisiones \textit{kink/antikink} en modelos que posean una estructura interna más compleja. Esta generalización permitiría comprender mejor el funcionamiento del mecanismo de transferencia de energía entre modos internos, responsable de la aparición de la estructura fractal característica de los diagramas de velocidad.

    \item Con respecto a las soluciones tipo vórtice, una posible línea de investigación, como continuación de los resultados presentados en el Capítulo~\ref{Chap3}, sería el análisis de la estructura interna de vórtices en modelos que generalicen la teoría de Abelian-Higgs. Otra posible extensión de los resultados de esta tesis consistiría en la generalización del estudio presentado en el Capítulo~\ref{Chap4}, con el fin de analizar el decaimiento y la emisión de radiación por parte de modos multipolares y de vórtices con carga topológica $n > 1$. Finalmente, una vía adicional de investigación sería la extensión de estos resultados para estudiar la excitación de cuerdas cósmicas.
\end{itemize}

\appendix

\chapter{Numerical methods}\label{appen}

The numerical method employed in Chapter \ref{Chap3} to solve the eigenvalue problem \eqref{radialedo03} is based on a second-order finite difference scheme, which discretizes the problem as follows:
\vspace{-0.2cm}
{\footnotesize \begin{eqnarray}
	&&\hspace{-.5cm} - \frac{v_{i+1}-2 v_i+v_{i-1}}{(\Delta r)^2 } -  \frac{v_{i+1}-v_{i-1}}{2 i (\Delta r)^2}+ \Big[ \frac{\overline{k}^2}{ (i \Delta r)^2} + f_{n,i}^2 \Big] v_i+ 2\left[f_{n,i} + \frac{i}{2}(f_{n,i+1}-f_{n,i-1}) \right] w_i = \omega_n^2 v_i, \nonumber \\
	&& \hspace{-.5cm}- \frac{u_{i+1}-2 u_i+u_{i-1}}{(\Delta r)^2} -  \frac{u_{i+1}-u_{i-1}}{2 i (\Delta r)^2}+ \Big[ \frac{\overline{k}^2}{(i \Delta r)^2} + \frac{n^2(1-\beta_{n,i})^2}{(i \Delta r)^2} + \frac{3\lambda}{2} f_{n,i}^2 - \frac{\lambda}{2} \Big] u_{i}- \frac{2n(1-\beta_{n,i})(\overline{k}^2 +   (i \Delta r)^2 f_{n,i}^2)}{ (i \Delta r)^2} w_i \nonumber\\
	&& \hspace{-.5cm}\hspace{0.5cm}+ \frac{n\,f_{n,i}\,(1-\beta_{n,i})(v_{i+1}-v_{i-1})}{ i (\Delta r)^2} = \omega_n^2 u_i, \label{numeric} \\
	&& \hspace{-.5cm}- \frac{w_{i+1}-2 w_i+w_{i-1}}{(\Delta r)^2} -  \frac{w_{i+1}-w_{i-1}}{2 i (\Delta r)^2}+ \Big[ \frac{\overline{k}^2}{(i \Delta r)^2} + \frac{n^2(1-\beta_{n,i})^2}{(i \Delta r)^2} + f_{n,i}^2 + \frac{\lambda}{2} f_{n,i}^2 - \frac{\lambda}{2} \Big] w_i- \frac{2n(1-\beta_{n,i})}{(i \Delta r)^2} u_i + \nonumber\\
	&& \hspace{-.5cm}\hspace{0.5cm}+ \frac{f_{n,i+1}-f_{n,i-1} }{i (\Delta r)^2} v_i= \omega_n^2 w_i, \nonumber
\end{eqnarray}
}
\hspace{-0.175cm} where $f_{n,i}=f_n(i\Delta r)$ and $\beta_{n,i} = \beta_n(i\Delta r)$ are respectively the values of the radial profiles of the scalar and vector field at the mesh points and $v_{i}= v( i\Delta r )$,  $u_{i}=u( i\Delta r )$, $w_{i}=w(i\Delta r )$ are the eigenfunction components at these points.  $\Delta r = r_{max}/N$ denotes the spatial step. 

The eigenfunctions asymptotically vanish so that the boundary conditions $v_N=u_N=w_N=0$ will be assumed in our problem. On the other hand, we shall consider the regularity conditions $u'(0)=u'(0)=w'(0)=0$. By using the progressive numerical first derivative this leads to the relations
\begin{equation}
v_0 = \frac{4}{3} v_1 - \frac{1}{3} v_2 ,\hspace{0.5cm}
u_0 = \frac{4}{3} u_1 - \frac{1}{3} u_2 ,\hspace{0.5cm}
w_0 = \frac{4}{3} w_1 - \frac{1}{3} w_2 \label{regularity01}
\end{equation}
between the discretized values of the fields. The previous relations allow us to avoid the singularities at $r=0$ arising in the discretized equations (\ref{numeric}) with $i=0$ (notice that the spectral problem is well-defined at $r=0$, as analytically demonstrated in Section \ref{corepaperinternalmodes}). Using (\ref{regularity01}), we can construct the discretization of the Hessian operator without explicitly considering the point $r=0$; instead, we simply take the discretization with values $v_i,u_i,w_i$, with $i=1,2,\dots,N$, replacing $v_0, u_0$ and $w_0$ with the value given above whenever necessary. For example, the first equation in this scheme reduces to 
{\small
\begin{eqnarray}
	&& \hspace{-0.8cm}- \frac{4\,v_2 }{3(\Delta r)^2} +\Big[ \frac{1}{(\Delta r)^2} \Big(\frac{4}{3}+ \overline{k}^2\Big)  +f_{n,1}^2 \Big] v_1  + 2[f_{n,1}+\frac{1}{2} (f_{n,2}-f_{n,0})]w_1 = \omega_n^2 v_1 \nonumber \\
    && \hspace{-0.8cm}- \frac{4\,u_2 }{3(\Delta r)^2} + \Big[ \frac{1}{(\Delta r)^2} \Big(\frac{4}{3}+\overline{k}^2 + n^2 (1-\beta_{n,1})^2 \Big)  + \frac{3\lambda}{2} f_{n,1}^2  -\frac{\lambda}{2} \Big] u_1  + \frac{4 n (1-\beta_{n,1})f_{n,1}}{3(\Delta r)^2} v_2  - \nonumber\\
    && \hspace{0.8cm}- \frac{4 n (1-\beta_{n,1})f_{n,1}}{3(\Delta r)^2} v_1 - 2n(1-\beta_{n,1}) \Big[ \frac{\overline{k}^2}{(\Delta r)^2} + f_{n,1}^2 \Big]w_1 = \omega_n^2 u_1 \label{numericbc} \\
    &&\hspace{-0.8cm} - \frac{4\,w_2 }{3(\Delta r)^2} +\Big[ \frac{1}{(\Delta r)^2} \Big(\frac{4}{3}+ \overline{k}^2 + n^2 (1-\beta_{n,1})^2\Big) +f_{n,1}^2 + \frac{\lambda}{2} f_{n,1}^2 - \frac{\lambda}{2} \Big] w_1  + \frac{f_{n,2}-f_{n,0}}{(\Delta r)^2} v_1  - \frac{2n(1-\beta_{n,1})}{(\Delta r)^2} u_1 = \omega_n^2 w_1 \nonumber \\
    && \hspace{0.8cm} \nonumber
\end{eqnarray}}
and the rest of entries are given by (\ref{numeric}) with $i=2,3,\dots,N$ as usual. There is no dependence on $g_0$ in our discretized spectral problem. 

The choice of the regularity boundary conditions are analytically justified. It has been shown that the eigenfunction profiles $v(r)$, $u(r)$ and $w(r)$ behave as $C r^j$, with $j\in \mathbb{N}$. Then, if $j > 1$, the functions clearly satisfies the imposed condition. When some $j = 0$, the functions takes the form $C_1 + C_2 r^2$, which also satisfies the condition. When some of the functions follow the previous behavior with $j = 1$ the regularity condition is not verified, however the imposed condition is very flexible in the sense that, although the exact eigenfunction does not analytically satisfy this condition, when computed numerically, it quickly adapts to the real eigenfunction, provided that a sufficiently high resolution is used. That is, the difference between the numerical and the real eigenfunction is only noticeable in a very small range.

We have tested other boundary conditions at the origin, leveraging the analytical insights provided in the article, but the obtained eigenvalues are the same. 

\cleardoublepage   % o \clearpage si no usas 'book'
\phantomsection
\specialannex[ \hspace{0.15cm}First paper]{Appendix B:\\
First paper}{\begin{itemize}
    \item \textbf{Authors:} A. Alonso-Izquierdo, D. Miguélez-Caballero and L.M. Nieto.
    \item \textbf{Title:} {Wobbling kinks and shape mode interactions in a coupled two-component} $\phi^4$ {theory}.
    \item \textbf{Journal:} Chaos, Solitons and Fractals.
    \item \textbf{DOI:} \href{https://doi.org/10.1016/j.chaos.2023.114373}{ 10.1016/j.chaos.2023.114373}
\end{itemize}}\label{Eppen1}
\afterpage{\blankpage}

%\phantomsection
\cleardoublepage   % o \clearpage si no usas 'book'
\phantomsection
    \specialannex[ \hspace{0.12cm}Second paper]{Appendix C:\\
Second paper}{\begin{itemize}
    \item \textbf{Authors:} A. Alonso-Izquierdo, D. Miguélez-Caballero and L.M. Nieto.
    \item \textbf{Title:} Scattering between orthogonally wobbling kinks.
    \item \textbf{Journal:} Physica D. 
    \item \textbf{DOI:} \href{https://doi.org/10.1016/j.physd.2024.134438}{10.1016/j.physd.2024.134438}
\end{itemize}}\label{Eppen2}

\afterpage{\blankpage}

\cleardoublepage   % o \clearpage si no usas 'book'
\phantomsection
    \specialannex[  \hspace{0.12cm}Third paper]{Appendix D:\\
Third paper}{\begin{itemize}
    \item \textbf{Authors:} A. Alonso-Izquierdo and D. Miguélez-Caballero.
    \item \textbf{Title:} Dissecting normal modes of vibration on vortices in Ginzburg-Landau superconductors.
    \item \textbf{Journal:} Physical Review D.
    \item \textbf{DOI:} \href{https://doi.org/10.1103/PhysRevD.110.125026}{10.1103/PhysRevD.110.125026}
\end{itemize}}\label{Eppen3}

\afterpage{\blankpage}

\cleardoublepage   % o \clearpage si no usas 'book'
\phantomsection
    \specialannex[ \hspace{0.15cm}Fourth paper]{Appendix E:\\
Fourth paper}{\begin{itemize}
    \item \textbf{Authors:} A. Alonso-Izquierdo, J.J. Blanco-Pillado, D. Miguélez-Caballero, S. Navarro-Obregón and J. Queiruga.
    \item \textbf{Title:} Excited Abelian-Higgs vortices: Decay rate and radiation emission.
    \item \textbf{Journal:} Physical Review D.
    \item \textbf{DOI:} \href{ https://doi.org/10.1103/PhysRevD.110.065009}{10.1103/PhysRevD.110.065009} 
    
\end{itemize}}\label{Eppen4}
\afterpage{\blankpage}

    \pagestyle{plain}
  %  \renewcommand*{\bibfont}{\small}
   % \printbibheading
    %\addcontentsline{toc}{chapter}{Bibliography}
    %\printbibliography[heading = none]
    
    %\renewcommand*{\bibfont}{\normalsize}
    %\include{core/bibliography/my_publications}

\end{document}